\documentclass[11pt,english]{book}
\usepackage[T1]{fontenc}
\usepackage[latin1]{inputenc}
\usepackage{geometry}
\usepackage{amsmath,amssymb}
\usepackage{feynmp}
\usepackage[dvips]{graphicx}
\usepackage{fancyhdr}
\usepackage{rotating}
\usepackage{subfigure}
\usepackage{babel}
\usepackage{graphicx}
\usepackage{setspace}
\usepackage{amssymb}

\makeatletter

\newcommand{\noun}[1]{\textsc{#1}}

\providecommand{\tabularnewline}{\\}

 \newcommand{\lyxaddress}[1]{
   \par {\raggedright #1
   \vspace{1.4em}
   \noindent\par}
 }
 \newcommand{\lyxrightaddress}[1]{
   \par {\raggedleft \begin{tabular}{l}\ignorespaces
   #1
   \end{tabular}
   \vspace{1.4em}
   \par}
 }
 \usepackage{verbatim}

\usepackage{framed}

\usepackage{babel}

\def    \be             {\begin{equation}}
\def    \ee             {\end{equation}}
\def    \ba             {\begin{eqnarray}}
\def    \ea             {\end{eqnarray}}

\def\beq{\begin{equation}}
\def\eeq{\end{equation}}
\def\beqn{\begin{eqnarray}}
\def\ba{\begin{eqnarray}}
\def\eeqn{\end{eqnarray}}
\def\ea{\end{eqnarray}}

\def\half{{\textstyle{1\over 2}}}

\def\atp{\frac{\alpha_s(Q^2)}{2\pi}}

\def\slash#1{#1\hskip-6pt/\hskip6pt}

\def\GeV{\,{\rm GeV}}

\def\Q{{\bf Q}}

\def\z{{\zeta}}

\def\A{\mathcal{A}}

\newcommand{\beqa}{\begin{eqnarray}}
\newcommand{\eeqa}{\end{eqnarray}}
\newcommand{\eps}{\epsilon}

\makeatother
\begin{document}
\begin{center}{\huge \thispagestyle{empty}}\end{center}{\huge \par}

\begin{center}{\large UNIVERSIT\`A DEGLI STUDI DI LECCE}\end{center}{\large \par}

\begin{center}{\large DIPARTIMENTO DI FISICA}
\end{center}{\large \par}

\begin{center}\vspace{1cm}\end{center}

\begin{center}\includegraphics[%
width=4cm]{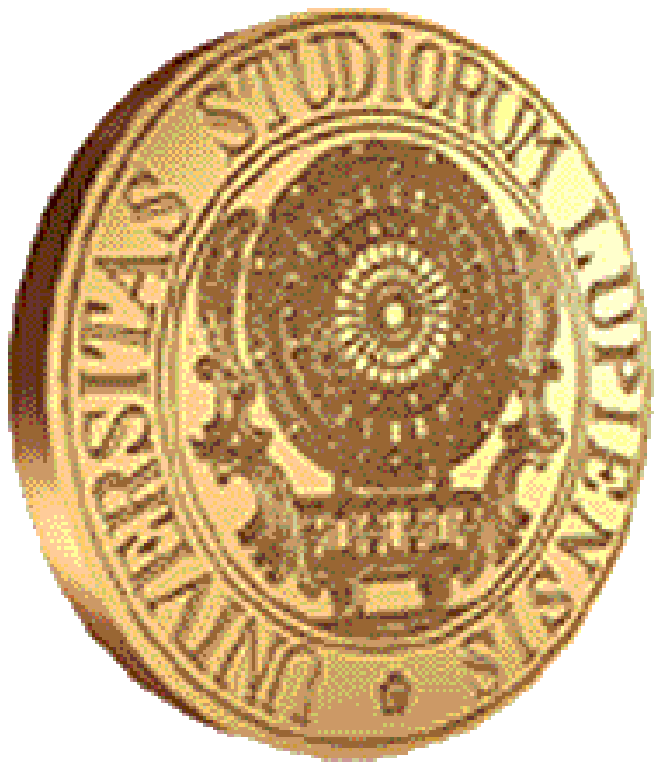}\vspace{2cm}\end{center}

\begin{center}\textbf{\huge QCD Studies at Hadron Colliders}\\
\textbf{\huge and in Deeply Virtual Neutrino Scattering}\vspace{3cm}\end{center}

\lyxaddress{\textbf{Advisor}\\
Claudio Corian\`o}

\lyxrightaddress{\textbf{Candidate}\\
Marco Guzzi}

\lyxrightaddress{\vspace{0.5cm}}

\begin{center}\noindent

\rule{16cm}{0.1mm}\renewcommand{\headrulewidth}{0pt}\end{center}

\begin{center}\noun{\large Tesi di Dottorato -- Anno Accademico
2005 2006 -- XVIII Ciclo}\fancyhf{}

\renewcommand{\headrulewidth}{0pt}

\fancyhead[LE,RO]{\thepage}

\fancyhead[RE]{\nouppercase\rightmark}

\fancyhead[LO]{\nouppercase\leftmark}\end{center}

\chapter*{Acknowledgements}

\addcontentsline{toc}{chapter}{Acknowledgements}\markboth{}{Acknowledgements}

This work is dedicated to my family for trusting and supporting me.

A big ``thank you'' goes to my advisor Dr. Claudio Corian\`o,
for all the patience, all the efforts he has done for introducing me to the art of Physics, for all the stimulating discussions and for his love for Physics.

To my friend and collaborator Dr. Alessandro Cafarella I want
to say ``thank you'', for all the nice time we spent together and for the help he gave me during this
thesis-period.

I'm very grateful to Prof. Phil Ratcliffe and to Dr. Enzo Barone
for giving me the opportunity to work with them in a very interesting
field of research, for their passion in Physics and for the
great support.

I cannot miss to thank Prof. Kyriakos Tamvakis
for the illuminating discussions and
for the nice time we spent together in Yoannina and in Lecce during the
``LHC School 2005''.

Of course, I cannot forget to thank my colleagues
Andrea, Karen and Iris, for the incredibly nice time we spent together
during these last years. Finally I thank Daniela, my love and inspiration,
for her patience and for her support in every situation.

\tableofcontents{}

\chapter*{List of publications}
\addcontentsline{toc}{chapter}{List of publications}\markboth{}{List of publications}

\section*{Research papers}

\begin{enumerate}
\item A.~Cafarella, C.~Corian\`o and M.~Guzzi,
\emph{An $x$-space analysis of evolution equations: Soffer's inequality
and the nonforward evolution}.
Published in J.High Energy Phys.\textbf{~11} 059, (2003).

\item P.~Amore, C.~Corian\`o and M.~Guzzi,
\emph{Deeply virtual neutrino scattering (DVNS)}.
Published in JHEP \textbf{0502} 038, (2005).

\item C.~Corian\`o, M.~Guzzi,
\emph{Leading twist amplitudes for exclusive neutrino
interactions in the deeply virtual limit}.
Published in Phys. Rev. D \textbf{71} 053002, (2005)

\item A.~Cafarella, C.~Corian\`o, M.~Guzzi, J.~Smith,
\emph{On the scale variation of the total cross section for Higgs production
at the LHC and at the Tevatron}, Eur. Phys. J. {\bf C 47}: 703, (2006).

\item M.~Guzzi, V.~Barone, A.~Cafarella, C.~Corian\`o, P.~Ratcliffe,
\emph{Double transverse-spin asymmetries in Drell-Yan processes with antiprotons},
Phys. Lett. {\bf B 639}: 483, (2006).

\item A.~Cafarella, C.~Corian\`o and M.~Guzzi,
\emph{NNLO logarithmic expansions and exact solutions of the DGLAP equations from $x$-space:
New algorithms for precision studies at the LHC}, Nucl. Phys. \textbf{B 748} (2006) 253.
\end{enumerate}

\section*{Conference proceedings}
\begin{enumerate}
\item A.~Cafarella, C.~Corian\`o, M.~Guzzi and D.~Martello,
\emph{Superstring relics, supersymmetric fragmentation and UHECR},
hep-ph/0208023, published in the proceedings of the
\emph{1st International Conference on String Phenomenology,
Oxford, England, 6-11 July 2002},
editors S.~Abel, A.~Faraggi, A.~Ibarra and M.~Plumacher, World Scientific (2003)

\item A.~Cafarella, C.~Corian\`o and M.~Guzzi,
\emph{Solving renormalization group equations by recursion relations},
hep-ph/0209149, published in
\emph{Proceedings of the Workshop: Nonlinear Physics: Theory and
Experiment. II}, editors M.J.~Ablowitz, M.~Boiti, F.~Pempinelli, B.~Prinari, World Scientific (2003)

\item G.~Chirilli, C.~Corian\`o and M.~Guzzi,
\emph{Using and constraining nonforward parton distributions:
Deeply virtual neutrino scattering in cosmic rays and light dark matter searches.},
hep-ph/0309069, based on a talk given at
\emph{QCD@Work 2003: 2nd International Workshop on Quantum Chromodynamics:
Theory and Experiment, Conversano, Italy, 14-18 Jun 2003.}
Published in \emph{eConf C030614}:\textbf{~023}, (2003), also in *Conversano 2003, QCD at work 2003* 163-167

\item M.~Guzzi, V.~Barone, A.~Cafarella, C.~Corian\`o, P.~Ratcliffe,
\emph{Double transverse-spin asymmetries in Drell-Yan and $J/\psi$
production from proton-antiproton collisions},
hep-ph/0604176, published in the proceedings of the
\emph{International Workshop on Transverse Polarisation Phenomena in
Hard Processes ``Transversity 200", Villa Olmo (Como), 7-10th. September 2005}
editor World Scientific (2006).

\item C.~Corian\`o and M.~Guzzi,
\emph{Deeply Virtual Neutrino Scattering at Leading Twist},
hep-ph/0612025, based on a talk given at ``NOW 2006'', Neutrino Oscillation Workshop,
Conca Specchiulla (Otranto, Lecce, Italy), September 9-16, (2006).
To be published in the Conference Proceedings on Nucl. Phys. B (Proc. Suppl.)

\end{enumerate}

\section*{Other papers}

\begin{enumerate}
\item Vincenzo Barone \it{et al},
\emph{Antiproton-proton scattering experiments with polarization},
hep-ex/0505054, (2005).

\item A.~Cafarella, C.~Corian\`o and M.~Guzzi,
\emph{Parton distributions, logarithmic expansions and kinetic evolution},
hep-ph/0602173, (2006).
\end{enumerate}

\chapter*{Introduction\addcontentsline{toc}{chapter}{Introduction}\markboth{}{Introduction}}
\fancyhead[LO]{\nouppercase{Introduction}}

The study of the behaviour of the hadronic interactions at colliders is 
a fascinating subject and a specialized field of research that is essential 
in order to widen our knowledge on the fundamental constituents of matter. 

Elementary particle physics, needless to say, is both a theoretical and an experimental science,
and requires a continuous interplay between the two 
approaches. Therefore it is important, 
from this viewpoint, to develope, from the theorist's side, theoretical 
tools and strategies that can be used by the experimental community, and this 
requires considerable effort.  

This requires the investigation of specific processes which can be tested 
by experiments, developing formalism that can be used for further 
theoretical elaborations, but this also points toward the need to develope 
software that can help the experimental collaborations to proceed with their complex phenomenological analysis. 

This thesis is the result of this philosophy, and collects several applications of Quantum Chromodynamics, the accepted theory of the strong interactions, 
to hadron colliders both at high and intermediate energies, and on the study of a specific process, termed by 
us ``deeply virtual neutrino scattering'', where we generalize the formalism of deeply virtual Compton scattering to 
neutral currents. The process is potentially relevant for neutrino detection at neutrino factories.

This dissertation is composed of two parts.
In the first section we focus our attention on the study of the initial state scaling violations and the evolution of 
the unpolarized parton distributions through Next-to-Next to Leading Order (NNLO) in $\alpha_s$, the strong coupling, suitable for precision studies of the parton model at the LHC. Specifically, we will analize the methods available to solve the equations and develope the theory that underlines a new proposed method which is shown to be highly accurate through NNLO. The theoretical analysis is accompanied by the developement of professional software partly documented in 
this thesis. 

The motivations of this study are in the need to compare traditional strategies in the solutions of the DGLAP 
equations with an independent theoretical 
approach which allows to reorganize all the scaling violations at a fixed perturbative order in terms of logarithms 
of the coupling constant times some scale invariant functions, introduced via factorization. 

We recall that high energy collisions can be well understood by the use of the factorization theorems,
which allow to separate the perturbative information contained in the hard scatterings of a given process 
from the non-perturbative information
which is contained in the parton distribution functions. These tell us what is the
probability density for finding a given parton with a fraction $x$ of the total initial momentum of the 
incoming hadron at an energy $Q$, and have a formal definition as light-cone correlation functions.

The knowledge of the PDFs with higher and higher accuracy is an important task for future experiments,
for the correct interpretation of the experimental data at the LHC, but  
they are also essential for a better description of the internal structure of the hadrons. 

In the first chapter of the thesis we discuss the solutions of the NNLO DGLAP equations using 
x-space methods and for this purpose we show how our approach allows to obtained accurate and exact solutions 
of the evolution equations for the pdf's using an analytical ansatz. We have applied the formal developements contained in this analysis is a numerical program, {\bf XSIEVE}
written in collaboration with A. Cafarella and C. Corian\`o that has been used for precision studies of 
two specific processes at hadron colliders.  

The first application 
has been in the NLO study of the double spin asymmetries in DY proton-antiproton
collisions, near the kinematical region of the $J/\psi$ resonance. 
This study, relevant for the proposed PAX experiment on the measurement of the transverse spin distribution, 
is analized in detail. We present result for the asymmetries of the process and elaborate on their possible 
measurements in the near future. 
A second application that we discuss is the study of the total cross section for Higgs production at the Tevatron and at the LHC, focusing in particular on the renormalization/factorization scale dependence of the predictions. While 
in other applications presented in the previous literature this study has been more limited, we perform a numerical analysis of the stability of the predictions on a large range of variability. These studies 
have been performed on a cluster.  

The second section of the thesis will be based on applications of perturbative
QCD to exclusive processes at intermediate energy in presence of weak interactions.

In particular we will analyse some theoretical aspects of the Generalized Parton Distribution
functions (GPDs) which are the natural generalization of the PDFs when perturbative QCD
is applied to exclusive reactions at intermediate energies. We introduce the notion of electroweak GPD's 
and develope the relevant formalism. 
Then, we will develope a phenomenological analysis of the neutrino-nucleon reaction mediated by
a $Z_0$ boson exchange and we will calculate the leading twist amplitudes for the charged current interaction as well, 
involving $W^{\pm}$ bosons.

\subsection*{The Relevance of the Perturbative QCD to the Precision Studies of the Physics at LHC:
Motivations}

Precision studies of some hadronic processes in the perturbative regime are going to be very important
in order to confirm the validity of
the mechanism of mass generation in the Standard Model
at the new collider, the LHC. This program involves a rather complex analysis of the
QCD background, with the corresponding radiative corrections taken into account to higher orders.
Studies of these corrections for specific processes have been performed by various groups,
at a level of accuracy which has reached the next-to-next-to-leading order (NNLO)
in $\alpha_s$, the QCD coupling constant. The quantification of the impact of these corrections
requires the determination of the hard scattering of the partonic cross sections up to
order $\alpha_s^3$, with the matrix of the
anomalous dimensions of the DGLAP kernels determined at the same perturbative order.

The study of the evolution of the parton distributions may include both a NNLO analysis and,
possibly, a resummation of the large logs which may appear in certain kinematic regions of specific
processes \cite{Sterman1}.
The questions that we address in this thesis concern the types of approximations which
are involved when we try to solve the DGLAP equations to higher
orders and the differences among the various methods proposed for their solution.

The clarification of these issues is important, since a chosen method has a direct
impact on the structure of the evolution codes and on their phenomenological predictions.
We address these questions by going over a discussions
of these methods and, in particular, we compare those based in Mellin space and
the analogous ones based in $x$-space. Mellin methods have been the most popular and have been implemented
up to NLO and, very recently, also at NNLO \cite{Vogt3}.
We remark that $x$-space methods based on logarithmic expansions have never been
thoroughly justified in the previous literature
even at NLO, in the case of the QCD parton distributions (pdf's) \cite{cafacor,nostri,gordon}.

We fill this gap and present exact proofs of the equivalence of these methods - in the
case of the evolution of the QCD pdf's - extending a proof which had been outlined
by Rossi \cite{Rossi} at LO and by Da Luz Vieira and Storrow \cite{Storrow} at NLO in their
study of the parton distributions of the photon.

In more recent times, these studies on the pdf's of the photon have triggered
similar studies also for the QCD pdf's. The result of these efforts was the proposal of new
expansions for the quark and gluon parton distributions \cite{cafacor,nostri,gordon}
which had to capture the logarithmic behaviour of the solution up to NLO.
Evidence of the consistency of the ansatz was,
in part, based on a comparative study of the generic structures of the logarithms that appear in the solution
using Mellin moments, since the same logarithms of the coupling could be reobtained by recursion relations.

In this thesis we are going to clarify - using the exact solutions of the corresponding recursion relations -
the role of these previous expansions and present their generalizations.
In particular we will show that they can be extended to retain higher logarithmic corrections and how they
can be made exact. It is shown that by a suitable extension of this analysis, all
that has been known so far in moment space can be reobtained directly from $x$-space.
In the photon case the Da Luz Vieira-Storrow solution \cite{Storrow} can be now understood
simply as a {\em first truncated ansatz} of the general truncated solutions that we analize.

\subsection*{Applications of the Perturbative Technique to the Neutrino Physics Regime}

There is no doubt that the study of neutrino masses and of flavour mixing
in the leptonic sector will play a crucial role for uncovering new physics
beyond the Standard Model and to test Unification.
In fact, the recent discovery of neutrino oscillations in atmospheric and solar neutrinos
(see \cite{Valle} for an overview) has raised the puzzle of the origin of the mass hierarchy
among the various neutrino flavours, a mystery which, at the moment, remains unsolved.
The study of the mixing among the leptons also raises the possibility of detecting
possible sources of CP violation in this sector as well.
It seems then obvious that the study of these aspects of flavour physics
requires the exploration of a new energy range for neutrino production and
detection beyond the one which is accessible at this time.

For this purpose, several proposals have been presented recently for neutrino factories,
where a beam, primarily made of muon neutrinos produced at an accelerator facility, is directed
to a large volume located several hundreds kilometers away at a second facility. The goal
of these experimental efforts is to uncover various
possible patterns of mixings among flavours - using the large distance
between the points at which neutrinos are produced and detected -
in order to study in a more detailed and ``artificial'' way the phenomenon of oscillations.
Detecting neutrinos at this higher energies is an aspect that deserves special attention since several
of these experimental proposals \cite{minos,Marciano1,Mangano} require a nominal energy
of the neutrino beam in the few GeV region.
We recall that the incoming neutrino beam, scattering off deuteron or other heavier targets
at the detector facility, has an energy which covers, in the various proposals,
both the resonant, the quasi-elastic (in the GeV range) and the deep inelastic region (DIS) at higher energy.
In the past, neutrino scattering on nucleons has been observed over a wide interval of energy,
ranging from few MeV up to 100 GeV, and these studies have been of significant help for
uncovering the structure of the fundamental interactions in the Standard Model.
Generally, one envisions contributions to the scattering cross section either in the
low energy region, such as in neutrino-nucleon elastic scattering, or in the deep inelastic
scattering (DIS) region.
Recent developments in perturbative QCD have emphasized that exclusive
(see \cite{Sterman}) and inclusive processes can be unified under a general
treatment using a factorization approach in a generalized kinematical domain.
The study of this domain, termed deeply virtual Compton scattering, or DVCS,
is an area of investigation of wide theoretical interest, with experiments planned in the
next few years at JLAB and at DESY.
The key constructs of the DVCS domain are the non-forward parton distributions,
where the term {\em non-forward} is there to indicate the asymmetry between the
initial and final state typical of a true Compton process,
in this case appearing not through unitarity, such as in DIS, but at amplitude level.

\chapter{The Logarithmic Expansions and Exact Solutions
of the DGLAP Equations from $x$-Space  \label{chap1}}
\fancyhead[LO]{\nouppercase{Chapter 1. The Logarithmic Expansions and Exact Solutions}}

\section{New Algorithms for Precision Studies at the LHC}

The chapter is organized as follows.
After defining the conventions, we bring in a simple example
that shows how a non-singlet LO solution of the DGLAP is obtained by an $x$-space ansatz.
Then we move to NLO and introduce the notion of truncated logarithmic solutions at this order,
moving afterwards to define exact recursive solutions from $x$-space. In all these cases, we show that
these solutions contain exactly
the same information of those obtained in Mellin space, to which they turn out to be equivalent.
The same analysis is then extended to NNLO.
The approach lays at the foundation of a numerical
method -based on $x$-space - that solves the NNLO DGLAP with great accuracy down to very small-$x$ $(10^{-5})$.
The method, therefore, not only does not suffer from the usual well known
inaccuracy of $x$-space based approaches at small-$x$ values \cite{CorianoSavkli}, but is,
a complementary way to look at the evolution of the pdf's in an extremely simple fashion.
We conclude with some comments concerning the timely issue of defining benchmarks for the evolution
of the pdf's, obtained by comparing solutions extracted by Mellin methods against
those derived from our approach, in particular
for those solutions which retain accuracy of a given order in $\alpha_s$ ($O(\alpha_s)$ accurate
solutions), relevant for precise determination of certain NNLO observables at the LHC.

\section{Definitions and conventions}

Before we start the analysis it is convenient to define
here the notations and conventions that we will use in the rest of the thesis.

We introduce the 3-loop evolution of the coupling via its $\beta$-function
\begin{equation}
\beta(\alpha_{s})\equiv\frac{\partial\alpha_{s}(Q^{2})}{\partial\log Q^{2}},
\label{eq:beta_def}
\end{equation}
and its three-loop expansion is
\begin{equation}
\beta(\alpha_{s})=-\frac{\beta_{0}}{4\pi}\alpha_{s}^{2}-\frac{\beta_{1}}{16\pi^{2}}\alpha_{s}^{3}-
\frac{\beta_{2}}{64\pi^{3}}\alpha_{s}^{4}+O(\alpha_{s}^{5}),
\label{eq:beta_exp}
\end{equation}
where
\ba
&&\beta_{0}=\frac{11}{3}N_{C}-\frac{4}{3}T_{f},\nonumber\\
&&\beta_{1}=\frac{34}{3}N_{C}^{2}-\frac{10}{3}N_{C}n_{f}-2C_{F}n_{f},\nonumber\\
&&\beta_{2}=\frac{2857}{54}N_{C}^{3}+2C_{F}^{2}T_{f}-\frac{205}{9}C_{F}N_{C}T_{f}-
\frac{1415}{27}N_{C}^{2}T_{f}+\frac{44}{9}C_{F}T_{f}^{2}+\frac{158}{27}N_{C}T_{f}^{2},\nonumber\\
\ea
are the coefficients of the beta function. In particular,
$\beta_2$ \cite{betafunction1,betafunction2} and $\beta_3$ \cite{betafunction3} are in the $\overline{MS}$ scheme.
We have set
\begin{equation}
N_{C}=3,\qquad C_{F}=\frac{N_{C}^{2}-1}{2N_{C}}=\frac{4}{3},\qquad T_{f}=T_{R}n_{f}=
\frac{1}{2}n_{f},
\end{equation}
where $N_{C}$ is the number of colors, $n_{f}$ is the number of
active flavors, that is fixed by the number of quarks with $m_{q}\leq Q$.
One can obtain either an exact or an accurate (truncated)
solution of this equation.
An exact solution includes higher order effects in $\alpha_s$, while a
truncated solution retains contributions only up to a given (fixed) order in a certain expansion parameter.
The structure of the NLO exact solution of the RGE for the coupling is well known
and relates $\alpha_s(\mu_1^2)$ in terms of $\alpha_s(\mu_2^2)$
via an implicit solution
\ba
\label{implicit}
\frac{1}{a_s(\mu_1^2)}=\frac{1}{a_s(\mu_2^2)}
+\beta_0 \ln \left(\frac{\mu_1^2}{\mu_2^2} \right)
-b_1\ln\left\{\frac{a_s(\mu_1^2) \, [ 1 + b_1 a_s(\mu_2^2) ]}
{a_s(\mu_2^2) \, [ 1 + b_1 a_s(\mu_1^2) ]} \right\},
\ea
where $a_s(\mu^2)=\alpha_s(\mu^2)/(4\pi)$.
The truncated solution is obtained by expanding up to a given order in a small
variable
\ba
\label{alphas}
\alpha_s(\mu_1^2)=\alpha_s(\mu_2^2)-\left[\frac{\alpha_s^2(\mu_2^2)}{4\pi}
+\frac{\alpha_s^3(\mu_2^2)}{(4\pi)^2}(-\beta_0^2 L^2+\beta_1 L)\right],
\ea
where the $\mu_1^2$ dependence is shifted into the factor $L=\ln(\mu_1^2/\mu_2^2)
$, and we have used a $\beta$-function expanded up to NLO,
involving $\beta_0$ and $\beta_1$. Exact solutions of the RGE for the running coupling are not
available (analytically) beyond NLO, while they can be obtained numerically.
Truncated solutions instead can be obtained quite easily, for instance
expanding in terms of the logarithm of a specific scale ($\Lambda$)
\begin{equation}
\alpha_{s}(Q^{2})=\frac{4\pi}{\beta_{0}L_{\Lambda}}\left\{ 1-\frac{\beta_{1}}{\beta_{0}^{2}}
\frac{\log L_{\Lambda}}{L_{\Lambda}}+
\frac{1}{\beta_{0}^{3}L_{\Lambda}^{2}}\left[\frac{\beta_{1}^{2}}{\beta_{0}}
\left(\log^{2}L_{\Lambda}-\log L_{\Lambda}-1\right)+
\beta_{2}\right]+O\left(\frac{1}{L_{\Lambda}^{3}}\right)\right\},
\label{eq:alpha_s_nnlo}
\end{equation}
where
\begin{equation}
L_{\Lambda}=\log\frac{Q^{2}}{\Lambda_{\overline{MS}}^{2}},
\end{equation}

and where $\Lambda_{\overline{MS}}^{(n_{f})}$ is calculated using the known
value of $\alpha_{s}(m_{Z})$ and imposing the continuity of $\alpha_{s}$
at the thresholds identified by the quark masses.

\section{General Issues}

For an integro-differential equation of DGLAP type, which is defined in a perturbative fashion,
the kernel $P(x)$ is known  perturbatively up to the first few orders
in $\alpha_s$, approximations which are commonly known as LO, NLO, NNLO \cite{vogt1}.

The equation is of the form
\ba
\label{dglap}
\frac{\partial f(x,Q^2)}{\partial \ln{Q^2}}=P(x,Q^2)\otimes f(x,Q^2),
\ea
with
\beq
a(x)\otimes b(x)\equiv \int_0^1\frac{dy}{y} a(y) b(x/y),
\eeq
and the expansion of the kernel at LO, for instance,  is given by
\ba
\label{kernLO}
&&{P(x,Q^2)}^{LO}=\left(\atp\right) P^{(0)}(x).
\ea
In the case of QCD one equation is scalar, termed non-singlet, the other equation involves 2-by-2 matrices,
the singlet. In other cases, for instance in supersymmetric QCD, both the singlet and the
non-singlet equations have a matrix structure \cite{Coriano}. Except for the LO case, exact analytic solutions of
the singlet equations are not known. However, various methods are available in order to
obtain a numerical solution with a good accuracy. These methods are of two types: 
{\em brute force} approaches based in $x$-space and those based on the inversion of the Mellin moments.

A brute force method involves a numerical solution of the PDE based on finite differences schemes.
One can easily find a stability scheme in which the differential 
operator on the left-hand-side of the equation gets replaced by its finite difference expression.
This method has the advantage that it allows to obtain the so called ``exact'' solution of the
equation at a given order (LO, NLO, NNLO). The only approximation involved in this numerical solution
comes from the perturbative expansion of the kernels. Solutions of this type are not
accurate to the working order of the expansion of the kernels,
since they retain higher order terms in $\alpha_s$. 
\footnote{It is common, however, to refer to
these solutions as to the ``exact'' ones, though they have no better status than the accurate (truncated) ones. 
In principle, large cancelations between contributions of 
higher order in the perturbative expansion of the kernels beyond NNLO, which are not available, 
and the known contributions, could take place at higher orders and this possibility 
remains unaccounted for in these ``exact'' solution. 
The term ``exact'', though being a misnomer, is however wide spread 
in the context of perturbative applications and 
for this reason we will use it throughout our work.} A short-come of brute force methods is the lack of an ansatz 
for the solution, which could instead be quite useful in order to understand the role of the retained perturbative logarithms. 
The use of Mellin inversion allows to extract, in the non singlet case, the exact solution quite immediately up to NNLO. 

The Mellin moments are defined as

\begin{equation}
a(N)=\int_{0}^{1}a(x)x^{N-1}\textrm{d}x
\end{equation}
and the basic advantage of working in moment space is to reduce the convolution product $\otimes$
into an ordinary product. For instance, at leading order we obtain
the LO DGLAP equation (\ref{dglap}) in moment space
\begin{equation}
\frac{\partial f(N,\alpha_{s})}{\partial\alpha_{s}}
=-\frac{\left(\frac{\alpha_{s}}{2\pi}\right)
P^{(0)}(N)}{\frac{\beta_{0}}{4\pi}\alpha_{s}^{2}}f(N,\alpha_{s}),
\label{dg1}
\end{equation}
which is solved by
\begin{equation}
f(N,\alpha_s)=f(N,\alpha_{0})\left(
\frac{\alpha_s}{\alpha_{0}}\right)^{-\frac{2P^{(0)}(N)}{\beta_{0}}}
=f(N,\alpha_{0})\exp\left\{-\frac{2P^{(0)}(N)}{\beta_{0}}
\log\left(\frac{\alpha_s}{\alpha_{0}}\right)\right\},
\label{solLO}
\end{equation}
where we have used the notation $\alpha\equiv\alpha(Q^{2})$ and $\alpha_{0}\equiv\alpha(Q_{0}^{2})$.
At this point, to contruct the solution in $x$-space, we need to perform a numerical
inversion of the moments, following a contour in the complex plane. This method is widely used in the numerical construction of the solutions and various optimization of this technique have been proposed 
\cite{kosower}.
We will show below how one can find solutions of any desired accuracy 
by using a set of recursion relations without the need of using numerical inversion of the 
Mellin moments.

\subsection{The logarithmic ansatz in LO}
To illustrate how the logarithmic expansion works and why it can reproduce the same solutions obtained from
moment space, it is convenient, for simplicity, to work at LO.
We try, in the ansatz, to organize the logarithmic behaviour of the solution
in terms of $\alpha_s$ and its logarithmic powers, times some
scale-invariant functions $A_n(x)$, which depend only on Bjorken $x$. As we are going to see, this re-arrangement of 
the scale dependent terms is rather general for evolution equations in QCD.
The number of the scale-invariant functions $A_n$ is actually infinite, and they are obtained recursively from a given initial condition.

The expansion that summarizes the logarithmic behaviour of the solution at LO is chosen
of the form
\beq
{f(x,Q^2)}^{LO}=\sum_{n=0}^{\infty}\frac{A_{n}(x)}{n!}
\left[\ln{\left(\frac{\alpha_s(Q^2)}{\alpha_s(Q^2_0)}\right)}\right]^{n}.
\label{an1}
\eeq

To determine $A_{n}(x)$ for every $n$ we introduce the simplified notation
\begin{equation}
L \equiv\log\frac{\alpha_{s}(Q^{2})}{\alpha_{s}(Q_{0}^{2})},
\end{equation}
and insert our ansatz (\ref{an1}) into the DGLAP equation
together with the LO expansion of the $\beta$-function to get

\begin{equation}
-\sum_{n=0}^{\infty}\frac{A_{n+1}}{n!}L^{n}
\frac{\beta_{0}}{4\pi}\alpha_{s}=\sum_{n=0}^{\infty}\frac{L^{n}}{n!}
\frac{\alpha_{s}}{2\pi}P^{(0)}\otimes A_{n}.
\end{equation}
Equating term by term in powers of $L$ we find the recursion relation
\begin{equation}
A_{n+1}=-\frac{2}{\beta_{0}}P^{(0)}\otimes A_{n}.
\label{LOrecursion}
\end{equation}
At this point we need to show that these recursion relations can be solved in terms of some
initial condition and that they reproduce the exact LO solution in moment space. This can be done by
taking Mellin moments of the recursion relations and solving the chain of these relations in terms of
the initial condition $A_0(x)$.
At LO the solution of (\ref{LOrecursion}) in moment space is simply given by  
\begin{equation}
A_{n}(N)=\left(-\frac{2}{\beta_{0}}P^{(0)}\right)^n q {(N,\alpha_s(Q_0^2))},
\end{equation}
having imposed the initial condition $A_0=q(x,\alpha_s(Q_0^2))$. 
At this point we plug in this solution into (\ref{an1}) to obtain 
\begin{equation}
f(N,Q^{2})=\sum_{n=0}^{\infty}\frac{A_{n}(N)}{n!}\log^{n}
\frac{\alpha_s(Q^{2})}{\alpha_s(Q_{0}^{2})},
\label{eq:LOansatz}
\end{equation}

which clearly coincides with (\ref{solLO}), after a simple expansion of the latter
\begin{equation}
f(N,\alpha_s)=f(N,\alpha_{0})\sum_{n=0}^{\infty}
\left\{\frac{1}{n!}\left[-\frac{2P^{(0)}(N)}{\beta_{0}}\right]^{n}
\log^{n}\left(\frac{\alpha_s}{\alpha_{0}}\right)\right\}.
\end{equation}
Notice that this non-singlet solution is an exact one. In the singlet case the same approach will succeed 
at the same order and there is no need to introduce truncated solution at this order. As expected, 
however, things will 
get more involved at higher orders, especially in the singlet case. 

The strategy that we follow in order to construct solutions of the DGLAP
equations is all contained in this trivial example, and we can summarize
our systematic search of logarithmic solutions at any order as follows: we

1) define the logarithmic ansatz up to a certain perturbative order and we insert it
into the DGLAP equation, appropriately expanded at that order;

2) derive recurrency relations for the scale invariant coefficients of the expansion;

3) take the Mellin moments of the recurrency relations and
{\em solve} them in terms of the moments of the initial conditions;

4) we show, finally, that the solution of the recursion relations, so obtained, is {\em exactly}
the solution of the original evolution equation firstly given in moment space.
This approach is sufficient to solve all the equations at any desired order of accuracy in the strong 
coupling, 
as we are going to show in the following sections. Ultimately, the success of the logarithmic ansatz lays on the fact that the solution
of the DGLAP equations in QCD resums only logarithms of the coupling constant.

\section{Truncated solution at NLO. Non-singlet}

The extension of our procedure to NLO (non-singlet) is more involved, but also in this
case proofs of consistency of the logarithmic ansatz can be formulated. 
However, before starting our technical analysis, we define the notion of  ``truncated solutions'' of the DGLAP
equations, expanding our preliminary discussion of the previous sections. We start with some definitions. 

A {\em truncated solution} retains only contributions up to a certain order in the expansion in the coupling.
We could define a $1$-$st$ truncated solution, a $2$-$nd$ truncated 
solution and so on. The sequence of truncated
solutions is expected to converge toward the exact solution of the DGLAP as the number of truncates 
increases. This can be done at any order in the 
expansion of the DGLAP kernels (NLO,NNLO,NNNLO,...). For instance, at NLO, we can build an exact solution
in moment space (this is true only in the non-singlet case) but we can also build the sequence of
truncated solutions. It is convenient to illustrate 
the kind of approximations which are involved in order 
to obtain these solutions and for this reason we try to detail the derivations.

Let's consider the NLO non-singlet DGLAP equation, written directly in moment space
\beq
\frac{\partial f(N,\alpha_{s})}{\partial\alpha_{s}}=
-\frac{\left(\frac{\alpha_{s}}{2\pi}\right)P^{(0)}(N)+
\left(\frac{\alpha_{s}}{2\pi}\right)^{2}P^{(1)}(N)}
{\frac{\beta_{0}}{4\pi}\alpha_{s}^{2}
+\frac{\beta_{1}}{16\pi^{2}}\alpha_{s}^{3}}f(N,\alpha_{s}),
\label{nontrunc}
\eeq
and search for its exact solution, which is given by
\ba
f(N,\alpha_s) & = & f(N,\alpha_{0})\left(
\frac{\alpha_s}{\alpha_{0}}\right)^{-\frac{2P^{(0)}(N)}{\beta_{0}}}
\left(\frac{4\pi\beta_{0}+\alpha_s\beta_{1}}{4\pi\beta_{0}+
\alpha_{0}\beta_{1}}\right)^{\frac{2P^{(0)}(N)}{\beta_{0}}-
\frac{4P^{(1)}(N)}{\beta_{1}}}.
\label{exactsol}
\ea
Notice that equation (\ref{nontrunc}) is the exact NLO equation. In particular we have preserved the
structure of the right-hand side, that involves both the beta function and the NLO kernels and is
given as a ratio of two polynomials in $\alpha_s$

\beq
\frac{P^{NLO}(x,\alpha_s)}{\beta^{NLO}(\alpha_s)},
\eeq
where
\beq
\label{kernNLO}
{P(x,Q^2)}^{NLO}=\left(\atp\right) P^{(0)}(x)+\left(\atp\right)^2 P^{(1)}(x)
\eeq
is the NLO kernel. The factorization of the LO solution from the NLO equation can be obtained 
expanding the ratio $P/\beta$ in $\alpha_s$, which allows the factorization of a $1/\alpha_s$ 
contribution. Equivalently, one can redefine the integral of the solution in moment space 
by subtraction of the LO part
\beq
\int_{\alpha_0}^{\alpha_s}d\alpha \left(\frac{P^{NLO}(x,\alpha)}{\beta^{NLO}(\alpha)} - \frac{{P}_{LO}(\alpha)}
{\beta_{LO}(\alpha)}\right).
\label{evolint}
\eeq
Denoting by $b_1=\beta_1/\beta_0$, the truncated differential
equation can be written as
\ba
\frac{\partial f(N,\alpha_s)}{\partial \alpha_s}=-\frac{2}{\beta_0 \alpha_s}
\left[P^{(0)}(N)+\atp \left(P^{(1)}-\frac{b_1}{2} P^{(0)}\right)\right]f(N,\alpha_s),
\label{tron}
\ea
which has the solution
\ba
f(N,\alpha_s)=\left[\frac{\alpha_s}{\alpha_0}\right]^{-\frac{2 P^{(0)}}{\beta_0}}
\times \exp{\left\{\frac{\left(\alpha_s-\alpha_0\right)}{\pi\beta_0}\left(
\frac{b_1}{2} P^{(0)}-P^{(1)}\right)\right\}}f(N,\alpha_0).
\label{TTR}
\ea
Notice that this solution of the truncated equation, exactly as in the exact solution (\ref{exactsol}),
contains as a factor the LO solution and therefore can be rewritten in the form
\ba
f(N,\alpha_s)=\exp{\left\{\frac{\left(\alpha_s-\alpha_0\right)}{\pi\beta_0}\left(
\frac{b_1}{2} P^{(0)}-P^{(1)}\right)\right\}}f^{LO}(N,\alpha_s),
\label{expansol}
\ea
where $f^{LO}(N,\alpha_s)$ is given by
\ba
f^{LO}(N,\alpha_s)=\left[\frac{\alpha_s}{\alpha_0}\right]^{-\frac{2 P^{(0)}}{\beta_0}}f(N,\alpha_0).
\ea
Eq. (\ref{expansol}) exemplifies a typical mathematical encounter in the search of solutions of PDE's 
of a certain accuracy: if we allow a perturbative expansion of the defining equation arrested at a given order,
the solution, however, is still affected
by higher order terms in the expansion parameter (in our case $\alpha_s$). 
To identify the expansion which converges to (\ref{expansol}) proceeds as follows. We start 
from the $1$-$st$ truncated solution.

Expanding (\ref{expansol}) to first order around the LO solution we obtain

\ba
\label{trunc1}
f(N,\alpha_s)=f^{LO}(N,\alpha_s)\times
\left\{1+\frac{(\alpha_s-\alpha_0)}{\pi\beta_0}\left(\frac{b_1}{2} P^{(0)}-P^{(1)}\right)\right\},
\ea
which is the expression of the $1$-$st$ truncated solution, accurate at order $\alpha_s$.
One can already see from (\ref{trunc1}) that the ansatz which we are looking for should involve a double expansion in two values of the coupling
constant: $\alpha_s$ and $\alpha_0$. This point will be made more clear below. 
For this reason 
we are naturally lead to study the logarithmic expansion
\beq
{f(x,Q^2)}^{NLO}=\sum_{n=0}^{\infty}\frac{A_{n}(x)}{n!}
\left[\ln{\left(\frac{\alpha_s}{\alpha_0}\right)}\right]^{n}
+\alpha_s\sum_{n=0}^{\infty}\frac{B_{n}(x)}{n!}
\left[\ln{\left(\frac{\alpha_s}{\alpha_0}\right)}\right]^{n},\,
\label{logexp}
\eeq
which is the obvious generalization of the analogous LO expansion (\ref{an1}).

Inserting this ansatz in the NLO DGLAP equation, we derive the following recursion relations
for $A_{n}$ and $B_{n}$
\ba
&&A_{n+1}=-\frac{2}{\beta_0}P^{(0)}(x)\otimes A_{n}(x),
\nonumber\\
&&B_{n+1}=-B_{n}(x)-\frac{\beta_1}{4\pi\beta_0}A_{n+1}(x)
-\frac{2}{\beta_0}P^{(0)}(x)\otimes B_{n}(x)-\frac{1}{\pi\beta_0}P^{(1)}(x)\otimes A_{n}(x),
\nonumber\\
\label{rr}
\ea
together with the initial condition
\ba
f(x,Q^2_{0})=A_0+\alpha_s(Q_0^2)B_0.
\ea

At this point we need to prove that the recursion relations (\ref{rr}) reproduce in moment
space (\ref{trunc1}).
To do so we rewrite the recursion relations in Mellin-space
\ba
&&A_{n+1}(N)=-\frac{2}{\beta_0}P^{(0)}(N)A_{n}(N),\nonumber\\
&&B_{n+1}(N)=-B_{n}(N)-\frac{\beta_1}{4\pi\beta_0}A_{n+1}(N)
-\frac{2}{\beta_0}P^{(0)}(N)B_{n}(N)-
\frac{1}{\pi\beta_0}P^{(1)}(N)A_{n}(N),\nonumber\\
\ea
and search for their solution over $n$.
After solving these relations with respect to $A_{0}$ and $B_{0}$,
it is simple to realize that our ansatz (\ref{logexp}) exactly
reproduces the truncated solution (\ref{trunc1}) only if the condition $B_{0}=0$
is satisfied. In fact, denoting by
\ba
&&R_0=-\frac{2}{\beta_0}P^{(0)}(N),\nonumber\\
&&R_1=\left(\frac{b_1}{2\pi\beta_0}P^{(0)}-
\frac{1}{\pi\beta_0}P^{(1)}\right),
\ea
the recursive coefficients, we can rewrite the recursion relations as
\ba
&&A_{n+1}=R_0 A_n,\nonumber\\
&&B_{n+1}=(R_0-1)B_n + R_1 A_n .
\label{rec1}
\ea
Then, observing that
\ba
&&A_n=R_0^{n}A_0,\nonumber\\
&&B_1=(R_0-1)B_0+R_1 A_0,\nonumber\\
&&B_2=(R_0-1)^2 B_0+R_1 A_0(2R_0-1),\nonumber\\
&&B_3=(R_0-1)^3 B_0+R_1 A_0\left[(2R_0-1)(R_0-1)+R_0^2\right],\nonumber\\
&&\vdots\nonumber\\
\ea
we identify the structure of the $n_{th}$ iterate in close form
\beq
B_n=(R_0-1)^n B_0+R_1 A_0\left[R_0^n-(R_0-1)^n\right].\nonumber
\label{rec2}
\eeq
Substituting the expressions for $A_n$ and $B_n$ so obtained in terms of $A_0$ and $B_0$
in the initial ansatz, and summing the logarithms (a procedure that we call ``exponentiation ``) we obtain
\beqa
\sum_{n=0}^\infty  \frac{A_n(N)}{n!} L^n &=& A_0\left(\frac{\alpha_s}{\alpha_0}\right)^{R_0},
\nonumber \\
\sum_{n=0}^\infty  \alpha_s \frac{B_n(N)}{n!} L^n &=& \sum_{n=0}^\infty
\alpha_s \frac{1}{n!}\left\{(R_0-1)^n B_0+R_1 A_0\left[R_0^n-(R_0-1)^n\right]\right\} \nonumber \\
&=&
\alpha_s B_0
\left(\frac{\alpha_s}{\alpha_0}\right)^{R_0-1}+
\alpha_s R_1 A_0\left(\frac{\alpha_s}{\alpha_0}\right)^{R_0}-
\alpha_s R_1 A_0\left(\frac{\alpha_s}{\alpha_0}\right)^{R_0-1},
\nonumber \\
\eeqa
expression that can be rewritten as
\ba
f(N,\alpha_s)=A_0\left(\frac{\alpha_s}{\alpha_0}\right)^{R_0} +\alpha_s B_0
\left(\frac{\alpha_s}{\alpha_0}\right)^{R_0-1}+
\alpha_s R_1 A_0\left(\frac{\alpha_s}{\alpha_0}\right)^{R_0}-
\alpha_s R_1 A_0\left(\frac{\alpha_s}{\alpha_0}\right)^{R_0-1}.
\ea
This expression after a simple rearrangement becomes 
\ba
f(N,\alpha_s)=\left(\frac{\alpha_s}{\alpha_0}\right)^{R_0}\left[1+
\left(\alpha_s-\alpha_0\right)R_1\right]f(N,Q^2_0)-
\left(\frac{\alpha_s}{\alpha_0}\right)^{R_0}\left(\alpha_s-\alpha_0\right)
(R_1\alpha_0 B_0).
\ea
This solution in moment space exactly coincides with the truncated solution (\ref{trunc1}) if we impose 
the condition $B_0=0$.
 It is clear that the solution gets organized in the form of a double expansion in the
two variables $\alpha_s$ and $\alpha_0$. While $\alpha_s$ appears explicitely
in the ansatz (\ref{logexp}), $\alpha_0$ appears only after the logarithmic summation
and the factorization of the leading order solution.
An obvious question to ask is how should we modify our ansatz if we want to reproduce
the exact solution of the truncated DGLAP equation in moment space, given by eq.~(\ref{TTR}).
The answer comes from a simple extension of our recursive method.

\subsection{Higher order truncated solutions}

We start by expanding the solution of the truncated equation (\ref{expansol}),
whose exponential factor is approximated by its double expansion in 
$\alpha_s$ and $\alpha_0$ to second order, thereby identifying the approximate solution 
\ba
f(N,\alpha_s) &=&\exp{\left\{\frac{\left(\alpha_s-\alpha_0\right)}{\pi\beta_0}\left(
\frac{b_1}{2} P^{(0)}-P^{(1)}\right)\right\}}\times f^{LO}(N,\alpha_s)\nonumber \\
&\simeq& \left(\frac{\alpha_s}{\alpha_{0}}\right)^{R_0}
\left[1- R_1 (\alpha_0-\alpha_s)
+\frac{1}{2} R_1^2(\alpha_0-\alpha_s)^2
+R_1(\alpha_0^2 -\alpha_s^2)\frac{b_1}{8\pi}\right]f(N,\alpha_{0}).
\label{newapp}
\nonumber\\
\ea
To generate this solution with the recursive method it is sufficient to introduce 
the higher order (2nd order)  ansatz 
\ba
\label{anss1}
\tilde{f}(x,\alpha_s)=\sum_{n=0}^{+\infty}\frac{L^n}{n!}\left[
A_n(x) +\alpha_s B_n(x)+\alpha_s^2 C_n(x)\right],
\ea
where we have included some new coefficients $C_n(x)$ that will take care of the higher order 
terms we aim to include.
Inserted into the NLO DGLAP equation, this ansatz
generates an appropriate chain of recursion relations
\ba
&&A_{n+1}(x)=-\frac{2}{\beta_0}P^{(0)}(x)\otimes A_{n}(x),
\nonumber\\
&&B_{n+1}(x)=-B_{n}(x)-\frac{2}{\beta_0}P^{(0)}(x)\otimes B_{n}(x)
-\frac{b_1}{(4\pi)}A_{n+1}(x)-\frac{1}{\pi\beta_0}P^{(1)}(x)\otimes A_{n}(x),
\nonumber\\
&&C_{n+1}(x)=-2 C_{n}(x)-\frac{2}{\beta_0}P^{(0)}(x)\otimes C_{n}(x)
-\frac{b_1}{(4\pi)}B_{n+1}(x)-\frac{b_1}{(4\pi)}B_{n}(x)
-\frac{1}{\pi\beta_0}P^{(1)}(x)\otimes B_{n}(x),
\nonumber\\
\ea
that we solve by going to Mellin space and obtain
\ba
\label{highNLO}
&&A_n=R_0^n A_0,\nonumber\\
&&B_n=R_1\left[R_0^n-(R_0-1)^n\right]A_0,
\nonumber\\
&&C_n=\left[\frac{1}{2}\left(R_0 -2\right)^n -\left(R_0 -1\right)^n
\right]R_1^2 A_0 + \nonumber\\
&&\hspace{1cm}\frac{1}{2}R_1^2 R_0^n A_0 +
\frac{1}{8\pi} R_1 b_1 \left(R_0 -2\right)^n A_0
-\frac{1}{8\pi}R_1 R_0^n b_1 A_0,
\ea
where the initial conditions are $\tilde{f}(N,\alpha_0)=A_0$ and $B_0=C_0=0$. It is a trivial exercise
to show that the solution of the recursion relation, inserted into (\ref{anss1}), coincides with (\ref{newapp}), after exponentiation.

Capturing more and more logs of the truncated logarithmic equation at this point is as easy as never before.
We can consider, for instance, a higher order ansatz accurate to $O(\alpha_s^3)$
for the NLO non-singlet solution
\ba
\label{anss2}
\tilde{f}(x,\alpha_s)=\sum_{n=0}^{\infty}\frac{L^n}{n!}\left[
A_n(x) +\alpha_s B_n(x)+\alpha_s^2 C_n(x)+\alpha_s^3 D_n(x)\right],
\ea
that generates four independent recursion relations. The relations for $A_{n+1}$,
$B_{n+1}$, $C_{n+1}$ are, for this extension, the same as in the previous case and are listed
in (\ref{highNLO}). Hence, we are left with an additional relation for the
$D_{n+1}$ coefficient which reads
\ba
D_{n+1}(x)=-3D_{n}(x)-\frac{2}{\beta_0}P^{(0)}\otimes D_{n}(x)
-\frac{b_1}{(4\pi)}C_{n+1}(x)-\frac{b_1}{(2\pi)}C_{n}(x)
-\frac{1}{\pi\beta_0}P^{(1)}\otimes C_{n}(x).
\nonumber\\
\ea
These are solved in Mellin space with respect to
$A_0,B_0,C_0,D_0$ (with the condition $B_0=C_0=D_0=0$). We obtain
\ba
&&A_n=R_0^n A_0,\nonumber\\
&&B_n=R_1\left[R_0^n- (R_0-1)^n\right]A_0,
\nonumber\\
&&C_n=\left[\frac{1}{2}\left(R_0 -2\right)^n -\left(R_0 -1\right)^n
\right]R_1^2 A_0 + \nonumber\\
&&\hspace{1cm}\frac{1}{2}R_1^2 R_0^n A_0 +
\frac{1}{8\pi} R_1 b_1 \left(R_0 -2\right)^n A_0
-\frac{1}{8\pi}R_1 R_0^n b_1 A_0,\nonumber\\
&&D_n=\left[-\frac{1}{6}\left(R_0 -3\right)^n
+\frac{1}{2}\left(R_0 -2\right)^n
-\frac{1}{2}\left(R_0 -1\right)^n
+\frac{1}{6}R_0^n \right]R_1^3 A_0\nonumber\\
&&\hspace{1cm}\left[-\frac{1}{8\pi}\left(R_0 -3\right)^n b_1
+\frac{1}{8\pi}\left(R_0 -2\right)^n b_1
+\frac{1}{8\pi}\left(R_0 -1\right)^n b_1
-\frac{1}{8\pi}R_0^n b_1\right]R_1^2 A_0\nonumber\\
&&\hspace{1cm}\left[-\frac{1}{48\pi^2}\left(R_0 -3\right)^n b_1^2
+\frac{1}{48\pi^2}R_0^n b_1^2\right]R_1 A_0.
\label{chain2}
\ea
Exponentiating we have
\ba
\label{hanss2}
&&\tilde{f}(x,\alpha_s)=
\left\{1+\alpha_s\left(1-\frac{\alpha_0}{\alpha_s}\right)R_1\right\}
A_0\left(\frac{\alpha_s}{\alpha_0}\right)^{R_0}
\nonumber\\
&&\hspace{1.5cm}+\alpha_s^2\left\{\left[\frac{1}{2}\left(\frac{\alpha_0^2}{\alpha_s^2}
-2~\frac{\alpha_0}{\alpha_s}+1\right)R_1^2
+\frac{b_1}{8\pi}\frac{\alpha_0^2}{\alpha_s^2}R_1
-\frac{b_1}{8\pi}R_1\right]\right\}A_0\left(\frac{\alpha_s}{\alpha_0}\right)^{R_0}
\nonumber\\
&&\hspace{1.5cm}+\alpha_s^3\left\{\left(-\frac{1}{6}\frac{\alpha_0^3}{\alpha_s^3}
+\frac{1}{2}\frac{\alpha_0^2}{\alpha_s^2}-\frac{1}{2}\frac{\alpha_0}{\alpha_s}
+\frac{1}{6}\right)R_1^3+\left(-\frac{b_1}{8\pi}\frac{\alpha_0^3}{\alpha_s^3}
+\frac{b_1}{8\pi}\frac{\alpha_0^2}{\alpha_s^2}
+\frac{b_1}{8\pi}\frac{\alpha_0}{\alpha_s}-\frac{b_1}{8\pi}\right)R_1^2
\right.\nonumber\\
&&\hspace{2.5cm}\left.+\left(\frac{b_1^2}{48\pi^2}\frac{\alpha_0^3}{\alpha_s^3}
+\frac{b_1^2}{48\pi^2}\right)R_1\right\}A_0\left(\frac{\alpha_s}{\alpha_0}\right)^{R_0}\,,
\ea
which is the solution of the truncated
equation computed with an $O(\alpha_s^3)$ accuracy.

\section{Non-singlet truncated solutions at NNLO}

The generalization of the method that takes to the truncated solutions
at NNLO is more involved, but to show the equivalence of these solutions to those in Mellin space
one proceeds as for the lower orders. As we have already pointed out, one has first to expand the ratio $P/\beta$ at a certain
order in $\alpha_s$, then solve the equation in moment space - solution that will bring in automatically
higher powers of $\alpha_s$ - and then reconstruct this solution via iterates.

At NNLO the kernels are given by
\ba
\label{ansatzeNNLO}
&&{P(x,Q^2)}^{NNLO}=\left(\atp\right) P^{(0)}(x)+\left(\atp\right)^2 P^{(1)}(x)
+\left(\atp\right)^3P^{(2)}(x),
\nonumber\\
\ea
and the equation in Mellin-space is given by
\ba
\frac{\partial f(N,\alpha_s)}{\partial\alpha_s}&=& \frac{P^{NNLO}(N)}{\beta^{NNLO}}f(N,\alpha_s).
\ea
We search for solutions of this equation of a given accuracy in $\alpha_s$ and for this purpose
we truncate the evolution integral of the ratio $P/\beta$ to $O(\alpha_s^2)$. 
This is the first order at which the $P^{(2)}$ component of the kernels appear. As we are going to see, 
this will generate the first truncate for the NNLO case. Therefore, while the first truncate at NLO 
is of $O(\alpha_s)$, the first truncate at NNLO is of $O(\alpha_s^2)$.
We obtain  
\beq
I_{NNLO}=\int_{\alpha_0}^{\alpha_s}d\alpha \left(\frac{P_{NNLO}(x,\alpha)}{\beta_{NNLO}(\alpha)}
- \frac{{P}_{LO}(\alpha)}{\beta_{LO}(\alpha)}\right)\approx
-R_1\alpha_0 -\frac{1}{2}R_2\alpha_0^2+R_1\alpha_s+\frac{1}{2}R_2\alpha_s^2.
\label{Intevol1}
\eeq

At this retained accuracy of the evolution integral,
the exact solution of the corresponding (truncated) DGLAP equation can be found, in moment space, as in (\ref{ltrunc})
\ba
&&f(N,\alpha_s)=f(N,\alpha_0)\left(\frac{\alpha_s}{\alpha_0}\right)^{-2 \frac{P^{(0)}}{\beta_0}} \left\{1 + \left(\alpha_s-\alpha_0\right)\left[-\frac{P^{(1)}}{\pi\beta_0}
+\frac{P^{(0)}\beta_1}{2\pi\beta_0^2}\right] \right.\nonumber\\
&&\left.\hspace{1.6cm}+\alpha_s^2\left[\frac{{P^{(1)}}^2}{2\pi^2\beta_0^2}
-\frac{P^{(2)}}{4\pi^2\beta_0}-\frac{P^{(0)}P^{(1)}\beta_1}{2\pi^2\beta_0^2}
+\frac{P^{(1)}\beta_1}{8\pi^2\beta_0^2}+\frac{{P^{(0)}}^2\beta_1^2}{8\pi^2\beta_0^4}
-\frac{P^{(0)}\beta_1^2}{16\pi^2\beta_0^3}+\frac{P^{(0)}\beta_2}{16\pi^2\beta_0^2}
\right]\right.\nonumber\\
&&\left.\hspace{1.6cm}+\alpha_0^2\left[\frac{{P^{(1)}}^2}{2\pi^2\beta_0^2}
+\frac{P^{(2)}}{4\pi^2\beta_0}-\frac{P^{(0)}P^{(1)}\beta_1}{2\pi^2\beta_0^2}
-\frac{P^{(1)}\beta_1}{8\pi^2\beta_0^2}+\frac{{P^{(0)}}^2\beta_1^2}{8\pi^2\beta_0^4}
+\frac{P^{(0)}\beta_1^2}{16\pi^2\beta_0^3}-\frac{P^{(0)}\beta_2}{16\pi^2\beta_0^2}
\right]\right.\nonumber\\
&&\left.\hspace{1.6cm}+\alpha_0\alpha_s\left[-\frac{{P^{(1)}}^2}{\pi^2\beta_0^2}
+\frac{P^{(0)}P^{(1)}\beta_1}{\pi^2\beta_0^3}-
\frac{{P^{(0)}}^2\beta_1^2}{4\pi^2\beta_0^4}\right]\right\}\,.
\label{para}
\ea
At this order this solution coincides with the exact NNLO solution of the DGLAP
equation, obtained from an exact evaluation of the integral (\ref{Intevol1}), followed by a 
double epansion in the couplings. Therefore, similarly to (\ref{newapp}), 
the solution is organized effectively as a double expansion
in $\alpha_s$ and $\alpha_0$. This approach remains valid also in the singlet case, when the equations assume a matrix form. As we have already pointed
out above, all the known solutions of the singlet equations in moment space are obtained after a truncation of the corresponding PDE, having retained a given accuracy of the ratio $P/\beta$. For this reason, and to compare with the
previous literature, it is convenient to rewrite (\ref{para}) in a form that parallels the analogous
singlet result \cite{ellis}. It is not difficult to perform the match of our result with that previous one, which takes the form 
\cite{ellis}, \cite{Vogt3}, \cite{Buras}
\ba
f(N,\alpha_s)&=&U(N,\alpha_s)f_{LO}(N,\alpha_s,\alpha_0)U^{-1}(N,\alpha_0)
\nonumber\\
&=&\left[1+\sum_{\kappa=1}^{+\infty}U_{\kappa}(N)\alpha_s^{\kappa}\right]
f_{LO}(N,\alpha_s,\alpha_0)\left[1+\sum_{\kappa=1}^{+\infty}U_{\kappa}(N)\alpha_0^{\kappa}\right]^{-1},
\ea
which becomes, after some manipulations 
\ba
\label{solNNLO}
&&f(N,\alpha_s)=\left(\frac{\alpha_s}{\alpha_0}\right)^{-\frac{2}{\beta_0}P^{(0)}}
\left[1+\left(\alpha_s-\alpha_0\right)U_1(N) +\alpha_s^2 U_2(N)\right.\nonumber\\
&&\hspace{4.5cm}\left.-\alpha_s\alpha_0 U_1^2(N)
+\alpha_0^2\left(U_1^2(N)-U_2(N)\right)\right]f(N,\alpha_0)\,,\nonumber\\
\ea
where the functions $U_{i}(N)$ are defined as
\ba
&&U_1(N)=\frac{1}{\pi\beta_0}\left[\frac{b_1 P^{(0)}(N)}{2}-P^{(1)}(N)\right]
\equiv R_1(N),\nonumber\\
&&U_2(N)=\frac{1}{2}\left[R_1^2(N)-R_2(N)\right],\nonumber\\
&&R_2(N)=\left[\frac{P^{(2)}(N)}{2\pi^2\beta_0}+
\frac{b_1}{4\pi}R_1(N)+\frac{b_2}{(4\pi)^2}R_0(N)\right],\nonumber\\
&&R_0(N)=-\frac{2}{\beta_0}P^{(0)}(N),
\,
\ea
where $\beta_1/\beta_0=b_1$, $\beta_2/\beta_0=b_2$. 

We intend to show rigorously that this solution is generated by a simple logarithmic ansatz
arrested at a specific order. 
For this purpose we simplify (\ref{solNNLO}) obtaining 

\ba
\label{arrangedNNLO}
&&f(N,Q^2)=\left(\frac{\alpha_s}{\alpha_0}\right)^{R_0(N)}
\left[1-R_1(N)\alpha_0+\frac{1}{2}R_1^2(N)\alpha_0^2+
\frac{1}{2}R_2(N)\alpha_0^2+R_1(N)\alpha_s\right.\nonumber\\
&&\hspace{4cm}\left.-R_1^2(N)\alpha_s\alpha_0+\frac{1}{2}R_1^2(N)\alpha_s^2-
\frac{1}{2}R_2(N)\alpha_s^2\right]f(N,Q^2_0),\nonumber\\
\ea
and the ansatz that captures its logarithmic behaviour can be easily
found and is given by
\ba
\label{nnloans}
&&{f(x,Q^2)}^{NNLO}=\sum_{n=0}^{\infty}\frac{A_{n}(x)}{n!}
\left[\ln{\left(\frac{\alpha_s(Q^2)}{\alpha_s(Q^2_0)}\right)}\right]^{n}
+\alpha_s(Q^2)\sum_{n=0}^{\infty}\frac{B_{n}(x)}{n!}
\left[\ln{\left(\frac{\alpha_s(Q^2)}{\alpha_s(Q^2_0)}\right)}\right]^{n}\nonumber\\
&&\hspace{3cm}+\alpha_s^2(Q^2)\sum_{n=0}^{\infty}\frac{C_{n}(x)}{n!}
\left[\ln{\left(\frac{\alpha_s(Q^2)}{\alpha_s(Q^2_0)}\right)}\right]^{n}\,.
\ea
Setting the initial conditions as
\ba
f(x,Q^2_{0})=A_0(x)+\alpha_0 B_0(x) + \alpha_0^2 C_0(x),
\ea
and introducing the 3-loop expansion of the $\beta$-function, we derive the following recursion relations

\ba
\label{NNLOrecurrence}
&&A_{n+1}(x)=-\frac{2}{\beta_0}P^{(0)}(x)\otimes A_{n}(x),
\nonumber\\
&&B_{n+1}(x)=-B_{n}(x)-\frac{\beta_1}{4\pi\beta_0}A_{n+1}(x)
-\frac{2}{\beta_0}P^{(0)}(x)\otimes B_{n}(x)-\frac{1}{\pi\beta_0}P^{(1)}(x)\otimes A_{n}(x),
\nonumber\\
&&C_{n+1}(x)=-2C_{n}(x)-\frac{\beta_{1}}{4\pi\beta_{0}}B_{n}(x)-
\frac{\beta_{1}}{4\pi\beta_{0}}B_{n+1}(x)-
\frac{\beta_{2}}{16\pi^{2}\beta_{0}}A_{n+1}(x)\nonumber \\
&&\hspace{2cm} -\frac{2}{\beta_{0}}P^{(0)}(x)\otimes C_{n}(x)-
\frac{1}{\pi\beta_{0}}P^{(1)}(x)\otimes B_{n}(x)\nonumber \\
&&\hspace{2cm}-\frac{1}{2\pi^{2}\beta_{0}}P^{(2)}(x)\otimes A_{n}(x).
\ea

We need to show that the solution of the NNLO recursion relations
reproduces (\ref{solNNLO}) in Mellin-space, once we have chosen appropriate initial conditions for
$A_0(N)$, $B_0(N)$ and $C_0(N)$ \footnote{It can be shown that the infinite set of recursion relations have internal symmetries 
and different choices of initial conditions can bring to the same solution. The choice that we make in our analysis is the simplest one.}.

At NLO we have already seen that $B_0(N)$ has to vanish for any $N$,
i.e. $B_0(x)=0$, and we try to impose the same condition on $C_0(N)$.
In this case we obtain the recursion relation for the moments
\ba
\label{NNLOrec_nonsing}
&&C_{n+1}(N)=-2 C_{n}(N)-\frac{\beta_{1}}{4\pi\beta_{0}}B_{n}(N)-
\frac{\beta_{1}}{4\pi\beta_{0}}B_{n+1}(N)-
\frac{\beta_{2}}{16\pi^{2}\beta_{0}}A_{n+1}(N)\nonumber \\
&&\hspace{2cm} -\frac{2}{\beta_{0}}P^{(0)}(N) C_{n}(N)-
\frac{1}{\pi\beta_{0}}P^{(1)}(N)  B_{n}(N)\nonumber \\
&&\hspace{2cm}-\frac{1}{2\pi^{2}\beta_{0}}P^{(2)}(N)  A_{n}(N),
\ea
which combined with the relations for $A_n(N)$ and $B_n(N)$
give
\ba
&&C_{n}=\left\{-R_1^2 \left(R_0-1\right)^n -\frac{1}{2}R_2 R_0^{n}
+\frac{1}{2}\left[\left(R_0-2\right)^n R_1^2+
\left(R_0-2\right)^n R_2+R_1^2 R_0^n\right]\right\}A_0,\nonumber\\
&&B_{n}= \left[R_0^n-\left(R_0-1\right)^n \right]R_1 A_0,\nonumber\\
&&A_{n}=R_0^n A_0\,,
\ea
where the $N$ dependence in the coefficients $R_i$ has been
suppressed for simplicity.
The solution is determined exactly as in (\ref{rec1})
and (\ref{rec2}) and can be easily brought to the form
\ba
f(N,Q^2)=\frac{1}{2}\left(\frac{\alpha_s}{\alpha_0}\right)^{R_0(N)}
\left[2-2 R_1 (\alpha_s -\alpha_0)+R_1^2(\alpha_s-\alpha_0)^2+
R_2(\alpha_0^2-\alpha_s^2)\right]A_0,
\ea
which is the result quoted in eq.~(\ref{arrangedNNLO}) with $A_0=f(N,\alpha_0)$.

We have therefore identified the correct logarithmic expansion at NNLO that solves in $x$-space the
DGLAP equation with an accuracy of order $\alpha_s^2$.

\section{Generalizations to all orders : exact solutions of truncated equations built
recursively}

We have seen that the order of approximation in $\alpha_s$ of the truncated solutions is in direct
correspondence with the order of the approximation used in the computation of the integral on the right-hand-side of the evolution equation (\ref{evolint}).
This issues is particularly important in the singlet case, as we are going to investigate next,
since any singlet solution involves a truncation. We have also seen that all the known solutions
obtained in moment space can be
easily reobtained from a logarithmic ansatz and therefore there is complete equivalence between the two
approaches. We will also seen that the structure of the ansatz is insensitive to whether the equations
that we intend to solve are of matrix forms or are scalar equations, since the ansatz and the
recursion relations are linear in the unknown matrix coefficients (for the singlet) and generate in both cases 
the same recursion relations. 

We pause here and try to describe the patterns that we have investigated in some generality.
We work at a generic order $N^mLO$, with m=1 denoting the NLO, $m=2$ the NNLO and so on.
We have already seen that
one can factorize the LO solution, having defined the evolution integral
\beq
I_{N^mLO}(\alpha_s,\alpha_0)=\int_{\alpha_0}^{\alpha_s}d\alpha
 \left(\frac{P_{N^mLO}(x,\alpha)}{\beta_{N^mLO}(\alpha)}
-\frac{{P}_{LO}(\alpha)}
{\beta_{LO}(\alpha)}\right),
\label{Intevolm}
\eeq
and the exact solution can be formally written as
\ba
f(N,\alpha_s)=f^{LO}(N,\alpha_s)\times e^{I_{N^mLO}(\alpha_s,\alpha_0)}.
\ea
A Taylor expansion of the integrand in the (\ref{Intevolm})
around $\alpha_s=0$ at order $\kappa=(m-1)$ gives, in moment space,
\ba
&&\left(\frac{P_{N^mLO}(N,\alpha_s)}{\beta_{N^mLO}(\alpha_s)}-\frac{{P}_{LO}(\alpha_s)}
{\beta_{LO}(\alpha_s)}\right)\approx R_1(P^{(0)},P^{(1)},N)+
R_2(P^{(0)},P^{(1)},P^{(2)},N)\alpha_s
\nonumber\\
&&\hspace{5cm}
+R_3(P^{(0)},P^{(1)},P^{(3)},N)\alpha_s^2
\nonumber\\
&&\hspace{5cm}
+\dots+R_{\kappa+1}(P^{(0)},P^{(1)},\dots,P^{(m)},N)\alpha_s^{\kappa},\,
\ea
which at NNLO becomes
\ba
&&\left(\frac{P_{NNLO}(N,\alpha_s)}{\beta_{NNLO}(\alpha_s)}-\frac{{P}_{LO}(\alpha_s)}
{\beta_{LO}(\alpha_s)}\right)\approx R_1(P^{(0)},P^{(1)},N)+
R_2(P^{(0)},P^{(1)},P^{(2)},N)\alpha_s.
\ea
Thus, integrating between $\alpha_s$ and $\alpha_0$ we, obtain the
following expression for $I_{N^mLO}(\alpha_s,\alpha_0)$ at 
$O(\alpha_s^{\kappa})$
\ba
I_{N^mLO}^{(\kappa)}(\alpha_s,\alpha_0)=-R_1\alpha_0 -\frac{1}{2}R_2\alpha_0^2-
\dots -\frac{1}{\kappa}R_{\kappa}\alpha_0^{\kappa}+R_1\alpha_s
+\frac{1}{2}R_2\alpha_s^2+\dots+\frac{1}{\kappa}R_{\kappa}\alpha_s^{\kappa},
\ea
where the $(P^{(0)},P^{(1)},\dots,P^{(m)},N)$ dependence in the $R_{\kappa}$
coefficients has been omitted.
To summarize: in order to solve the
$\kappa$-truncated version of the N$^{m}$LO DGLAP equation
\ba
\frac{\partial f(N,\alpha_{s})}{\partial\alpha_{s}}=
\frac{P_{\rm{N}^{m}\rm{LO}}(N,\alpha_s)}{\beta_{\rm{N}^{m}\rm{LO}}(\alpha_s)}f(N,\alpha_{s}),
\ea
which is obtained by a Taylor expansion - around $\alpha_s=0$ -
of the ratio $P_{\rm{N}^{m}\rm{LO}}(N,\alpha_s)/\beta_{\rm{N}^{m}\rm{LO}}(\alpha_s)$, we need to solve the equation
\ba
\label{trunc_eq}
\frac{\partial f(N,\alpha_{s})}{\partial\alpha_{s}}=\frac{1}{\alpha_s}\left[R_0+
\alpha_s R_1+\alpha_s^2 R_2+\dots+\alpha_s^{\kappa} R_{\kappa}\right]f(N,\alpha_{s}),
\ea
where the coefficients $R_{\kappa}$ have a dependence on $P^{(0)}$ and $P^{(1)}$
in the NLO case, and on $P^{(0)}$, $P^{(1)}$ and $P^{(2)}$ in the NNLO case.
Eq. (\ref{trunc_eq}) admits an exact solution of the form
\ba
f(N,\alpha_{s})=\left(\frac{\alpha_s}{\alpha_0}\right)^{R_0}\exp\left\{R_1\left(\alpha_s-\alpha_0\right)+
\frac{1}{2}R_2\left(\alpha_s^2-\alpha_0^2\right)+\dots +
\frac{1}{\kappa}R_{\kappa}\left(\alpha_s^{\kappa}-\alpha_0^{\kappa}\right)\right\}f(N,\alpha_0),
\nonumber\\
\ea
having factorized the LO solution.

At this stage, the Taylor expansion of the exponential
around $(\alpha_s,\alpha_0)=(0,0)$ generates an expanded solution of the form
\beqa
f_{N^mLO}(N,\alpha_s) &\approx& f_{LO}(N,\alpha_0) e^{I_{N^mLO}^{(\kappa)}}\nonumber \\
&=& f_{LO}(N,\alpha_0) \left(1 + I_{N^mLO}^{(\kappa)} + \frac{1}{2!}\left(I_{N^mLO}^{(\kappa)}\right)^2
+ \dots \right),
\label{ltrunc}
\eeqa

which can be also written as
\ba
&&f_{LO}(N,\alpha_0) \left(1 + I_{N^mLO}^{(\kappa)} + \frac{1}{2!}\left(I_{N^mLO}^{(\kappa)}\right)^2
+ \dots \right)= \nonumber\\
&&f^{LO}(N,\alpha_s)\times\left[c_0 + \alpha_s \left(c_{1,0} + c_{1,1}\alpha_0
+ c_{1,2}\alpha_0^2 + \dots + c_{1,\kappa-1}\alpha_0^{\kappa-1}+c_{1,\kappa}\alpha_0^{\kappa}\right)
\right. \nonumber \\
&&\hspace{3cm} \left.+ \alpha_s^2 \left( c_{2,0} + c_{2,1}\alpha_0 + c_{2,2}\alpha_0^2
+ \dots + c_{2,\kappa}\alpha_0^{\kappa}\right) + \dots c_{\kappa,0}\alpha_s^{\kappa} + \cdots\right],
\nonumber \\
\ea
where the coefficients $c_{ij}$ are defined in moment space.
The $\kappa$-th truncated solution of the equation above 
\beq
f_{N^m LO}(N,\alpha_s)\big|_{O(\alpha_s^{\kappa})}=f^{LO}(N,\alpha_s)
\left({\sum_{i,j=0}}^{i+j \leq \kappa} \alpha_s^i \alpha_0^j c_{i,j}\right),
\eeq
is therefore accurate to $O(\alpha_s^{\kappa})$, and clearly does not retain all the powers of the coupling constant which are, instead, part of (\ref{ltrunc}). However, as far as we are interested in an accurate solution of order $\kappa$, we can reobtain exactly the same expression from $x$-space using the ansatz

\beq
f_{N^mLO}(x,\alpha_s)\big|_{O(\alpha_s^{\kappa'})}=
\sum_{n=0}^\infty\left(A^0_n(x) + \alpha_s A^1_n(x)+ \alpha_s^2 A^2_n(x) + \dots +
\alpha_s^{\kappa'} A^{\kappa'}_n(x)\right)
\left[\ln{\left(\frac{\alpha_s(Q^2)}{\alpha_s(Q^2_0)}\right)}\right]^{n},
\label{truncatedseries}
\eeq
which can be correctly defined to be a {\em truncated solution of order $\kappa'$} of the $(\kappa)$-truncated equation.
Since we have truncated the evolution integral at order $\kappa$, this is also the maximum order at
which the truncated logarithmic expansion (\ref{truncatedseries}) coincides with the exact solution of the
full equation. This corresponds to the 
choice $\kappa'=\kappa$. Notice, however, that 
the number of coefficients $A_n^{\kappa'}$ that one introduces in the ansatz is unrelated to 
$\kappa$ and can be {\em larger} than this specific value. This implies that
we obtain an improved accuracy as we let $\kappa'$ in (\ref{truncatedseries}) grow. 

If we choose the accuracy of the evolution integral to be $\kappa$, while 
sending the index $\kappa'$
in the logarithmic expansion of (\ref{truncatedseries}) 
to infinity, then the ansatz that accompanies this choice becomes 
\beq
f_{N^mLO}(N,\alpha_s)=\sum_{n=0}^\infty\left(\sum_{l=0}^{\infty}
\alpha_s^{l} A_n^{l}(x)\right)
\left[\ln{\left(\frac{\alpha_s(Q^2)}{\alpha_s(Q^2_0)}\right)}\right]^{n},
\label{truncatedseries1}
\eeq
and converges to the exact solution of the (order $\kappa$) truncated equation (\ref{ltrunc}). 
Obviously, this exact solution starts to differ from the exact
solution of the exact DGLAP equation at $O(\alpha_s^{\kappa+1})$. Also in this case, as before, one should notice that the double expansion in
$\alpha_s$ and $\alpha_0$ of the exact solution of the $(\kappa)$-truncated equation 
can be reobtained after using our exponentiation, and not before. We remark, if not obvious, that 
the recursion relations, in this case, 
need to be solved at the chosen order $\kappa'$, as widely shown in the examples discussed before.

We remark that, as done for the LO, we could also factorize the NLO solution and determine the $N^mLO$ solution using
the integral
\beq
I_{N^mLO}=\int_{\alpha_0}^{\alpha_s}d\alpha \left(\frac{P_{N^mLO}(x,\alpha)}{\beta_{N^mLO}(\alpha)} 
- \frac{{P}_{NLO}(\alpha)}{\beta_{NLO}(\alpha)}\right),
\label{Intevol2}
\eeq
and then restart the previous procedure. Obviously, the two approaches imply a resummation of the logarithmic behaviour
of the pdf's in the two cases. 

It is convenient to summarize what we have achieved up to now.
A truncation of the evolution integral introduces an approximation in the search for
solutions, which is controlled by the accuracy ($\kappa$)
in the expansion of the same integral. The
exact solution of the corresponding truncated equation, as we have seen from
the previous examples, involves all the powers of $\alpha_s$ and $\alpha_0$ and, obviously,
a further expansion around
the point $\alpha_s=\alpha_0=0$ is needed in order to identify a set of truncated solutions which can be
reobtained by a logarithmic ansatz. This is possible because of 
the property of analiticity of the solution. 
Therefore two types of truncations
are involved in the approximation of the solutions: 1) truncation of the equation and
2) truncation ($\kappa'$) of the corresponding solution. In the non-singlet
case, which is particularly simple, one can therefore identify a wide choice 
of solutions (by varying $\kappa$ and $\kappa'$) that retain
higher order effects in quite different fashions. Previous studies of the evolution 
using an ansatz a la Rossi-Storrow \cite{Rossi,Storrow}, borrowed from the pdf's of the photon, 
were therefore quite limited in accuracy. Our generalized procedure is the logical step forward 
in order to equate the accuracy of solutions obtained in moment space to those in $x$-space, without having to rely on purely 
numerical brute force methods.\footnote{In \cite{Vogt3} a flag variable called
IMODEV allows to switch among the exact solution (IMODEV=1), the exact solution of
the $O(\alpha_s^2)$ truncated equation (IMODEV=2). A third option (IMODEV=0)
involves at NLO and at NNLO $O(\alpha_s)$ and $O(\alpha_s^2)$, respectively, truncated ansatz that,
in our cases, are reconstructed logarithmically.}

\subsection{Recursion relations beyond NNLO and for all $\kappa$'s}

In the actual numerical implementations, if we intend to use a generic truncate of the non-singlet equation
(the result is actually true also for the singlet), it is convenient
to work with implementations of the
recursion relations that are valid at any order. In fact it is not practical to rederive them at each new
order $\kappa'$ of approximation. Next we are going to show how to do it, identifying generic
relations that are easier to implement numerically.

The expression to all orders of the DGLAP kernels is given by
\ba
&&P(N,\alpha_s)=\sum_{l=0}^{\infty}\left(\frac{\alpha_s}{2\pi}\right)^{l+1} P^{(l)}(N)
\nonumber\\
&&\frac{\partial \alpha_s}{\partial \ln{Q^2}}=-\sum_{i=0}^{\infty}\alpha_s^{i+2}
\frac{\beta_i}{(4\pi)^{i+1}},
\ea
and the equation in Mellin-space in Melin space is given by
\ba
\label{tot1}
\frac{\partial f(N,\alpha_s)}{\partial \alpha_s}=-\frac{
\sum_{l=0}^{\infty}\left(\frac{\alpha_s}{2\pi}\right)^{l+1} P^{(l)}(N)}
{\sum_{i=0}^{\infty}\alpha_s^{i+2}
\frac{\beta_i}{(4\pi)^{i+1}}}f(N,\alpha_s),
\ea
whose exact solution can be formally written as
\ba
f(N,\alpha_s)=\left(\frac{\alpha_s}{\alpha_0}\right)^{-\frac{2}{\beta_0}P^{(0)}}
e^{{\cal F}(\alpha_s,\alpha_0,P^{(0)},P^{(1)},P^{(2)},...,
\beta_0,\beta_1,...)}f(N,\alpha_0)\,,
\ea
where ${\cal F}$ is obtained from the evolution integral and whose specific form is not relevant 
at this point.

We will use the notation $(\vec{P},\vec{\beta})$ to indicate the sequence of 
components of the kernels and the coefficients of the $\beta$-function 
$(P^{(0)},P^{(1)},P^{(2)},...,\beta_0,\beta_1,...)$.

Then, the Taylor expansion around $\alpha_s=\alpha_0$ of the solution is formally given by 
\ba
e^{{\cal F}(\alpha_s,\alpha_0,\vec{P},\vec{\beta})}=\sum_{n=0}^{\infty}
\Phi_n(\partial{\cal F},\partial^2{\cal F},...,\partial^n{\cal F})
|_{\alpha_s=\alpha_0}(\alpha_s-\alpha_0)^n\frac{1}{n!}
\ea
for an appropriate $\Phi_n(\partial{\cal F},\partial^2{\cal F},...,\partial^n{\cal F})$.
$\Phi_n$ is a function that depends over all the partial derivative obtained
by the Taylor expansion. Since it is
calculated for $\alpha_s=\alpha_0$, it has a parametric dependence only on $\alpha_0$,
and we can perform a further expansion around the value $\alpha_0=0$ obtaining
\ba
&&e^{{\cal F}(\alpha_s,\alpha_0,\vec{P},\vec{\beta})}=
\sum_{m=0}^{\infty}
\frac{\alpha_0^{m}}{m!}
\frac{{\partial}^m}{{\partial \alpha_0}^m}
\left[\sum_{n=0}^{\infty}
\Phi_n(\alpha_0,\vec{P},\vec{\beta})\left(\alpha_s-\alpha_0\right)^n
\frac{1}{n!}\right]_{\alpha_0=0}.\,
\ea
 This expression can always be arranged and simplified as follows
\ba
&&e^{{\cal F}(\alpha_s,\alpha_0,\vec{P},\vec{\beta})}=
\sum_{n=0}^{\infty}\left[\frac{\alpha_s^{n}}{n!}\Phi_n(0,\vec{P},\vec{\beta})
+\left(-n\frac{\alpha_s^{n-1}}{n!}\Phi_n(0,\vec{P},\vec{\beta})
+\frac{\alpha_s^{n}}{n!}\partial
\Phi_n(\alpha_0,\vec{P},\vec{\beta})\bigg|_{\alpha_0=0}\right)\alpha_0
+\right.\nonumber\\
&&\hspace{3cm}\left.\left(\frac{1}{2}\frac{(n-1)n\alpha_s^{n-2}}{n!}
\Phi_n(0,\vec{P},\vec{\beta})-n\frac{\alpha_s^{n-1}}{n!}
\partial\Phi_n(\alpha_0,\vec{P},\vec{\beta})\bigg|_{\alpha_0=0}
\right.\right.\nonumber\\
&&\hspace{3cm}\left.\left.
+\frac{1}{2}\alpha_s^n\partial^2\Phi_n(\alpha_0,\vec{P},\vec{\beta})
\bigg|_{\alpha_0=0}\right)\alpha_0^2+(\dots)\alpha_0^3+\dots\right]
\nonumber\\
&&\hspace{2cm}=\left[
\left(1+\alpha_s{\cal \xi}_{1}^{(0)}(\vec{P},\vec{\beta})+\dots
+\alpha_s^n{\cal \xi}_{n}^{(0)}(\vec{P},\vec{\beta})\right)
\right.\nonumber\\
&&\hspace{3cm}\left.
+\alpha_0\left({\cal \xi}_{0}^{(1)}(\vec{P},\vec{\beta})
+\alpha_s{\cal \xi}_{1}^{(1)}(\vec{P},\vec{\beta})
+\alpha_s^2{\cal \xi}_{2}^{(1)}(\vec{P},\vec{\beta})+\dots
+\alpha_s^n{\cal \xi}_{n}^{(1)}(\vec{P},\vec{\beta})\right)
\right.\nonumber\\
&&\hspace{3cm}\left.
+\alpha_0^2\left({\cal \xi}_{0}^{(2)}(\vec{P},\vec{\beta})
+\alpha_s{\cal \xi}_{1}^{(2)}(\vec{P},\vec{\beta})
+\alpha_s^2{\cal \xi}_{2}^{(2)}(\vec{P},\vec{\beta})+\dots
+\alpha_s^n{\cal \xi}_{n}^{(2)}(\vec{P},\vec{\beta})\right)
\right.\nonumber\\
&&\hspace{3cm}\left.
\vdots
\right.\nonumber\\
&&\hspace{3cm}\left.
+\alpha_0^m\left({\cal \xi}_{0}^{(m)}(\vec{P},\vec{\beta})
+\alpha_s{\cal \xi}_{1}^{(m)}(\vec{P},\vec{\beta})
+\alpha_s^2{\cal \xi}_{2}^{(m)}(\vec{P},\vec{\beta})+\dots+
\alpha_s^n{\cal \xi}_{n}^{(m)}(\vec{P},\vec{\beta})\right)
\right]\nonumber\\
&&\hspace{2cm}=\sum_{n=0}^{\infty}\sum_{m=0}^{\infty}
\alpha_0^{m}\alpha_s^{n}{\cal \xi}_{n}^{(m)}(\vec{P},\vec{\beta}),
\nonumber\\
\ea
where we are formally absorbing all the dependence on both the kernels $\vec{P}$
and the $\beta$-function $\vec{\beta}$, coming from
the functions $\partial^m\Phi_n$ calculated at the point $\alpha_0=0$,
in the coefficients ${\cal \xi}_{n}^{(m)}$. 
Finally, we can reorganize the solution to all orders as

\ba
\label{mar1}
f(N,\alpha_s)=\left(\frac{\alpha_s}{\alpha_0}\right)^{-\frac{2}{\beta_0}P^{(0)}}
f(N,\alpha_0)\sum_{n=0}^{\infty}\sum_{m=0}^{\infty}
\alpha_0^m\alpha_s^{n}{\cal \xi}_{n}^{(m)}(\vec{P},\vec{\beta}).
\ea
With the help of the general notation
\ba
&&\vec{P}_{NLO}=(P^{(0)},P^{(1)}),\nonumber\\
&&\vec{P}_{NNLO}=(P^{(0)},P^{(1)},P^{(2)}),\nonumber\\
&&\vec{\beta}_{NLO}=(\beta_{0},\beta_{1}),\nonumber\\
&&\vec{\beta}_{NNLO}=(\beta_{0},\beta_{1},\beta_{2})\,,
\ea
we can try to indentify, by a formal reasoning, the exact solution up to a fixed - but generic -
perturbative order of the expansion of the
kernels.

To obtain the NLO/NNLO exact solution it is sufficient to
take as null the components $(P^{(2)},P^{(3)},...)$ and
$(\beta_2,\beta_3,...)$ for the NLO and $(P^{(3)},...)$ and
$(\beta_3,...)$ for the NNLO case.
Since eq.(\ref{mar1}) contains all the powers of $\alpha_0 \alpha_s$ up to
$\alpha_0^m \alpha_s^n$ (i.e. it is a polynomial expression of order
$\alpha^{n+m}$), if we aim at an accuracy of order $\alpha_s^{\kappa}$,
we have to arrange the exact expanded solution as
\ba
\label{mar2}
f(N,\alpha_s)=\left(\frac{\alpha_s}{\alpha_0}\right)^{-\frac{2}{\beta_0}P^{(0)}}
f(N,\alpha_0)\sum_{n=0}^{\kappa}\sum_{j=0}^{\infty}\alpha_0^j
\alpha_s^{n-j} {\cal \xi}_{n}^{(j)}(\vec{P},\vec{\beta})+ O(\alpha_s^{\kappa}),
\ea
with $O(\alpha_s^{\kappa})$ indicating all the
higher-order terms containing powers of the type
$\alpha_0\alpha_s^{\kappa}+\dots+\alpha_0^{\kappa}\alpha_s^{\kappa}$.
Hence, the index $\kappa$ represents the order at which we truncate the
solution.

As a natural generalization of the cases discussed in the previous sections, we introduce
the higher-order ansatz ($\kappa$-truncated solution)
\ba
\tilde{f}(N,\alpha_s)=\sum_{n=0}^{\infty}\left[\sum_{m=0}^{\kappa}
\frac{O_{n}^m(N)}{n!}\alpha_s^m \right]\log^n\left(\frac{\alpha_s}{\alpha_0}\right)
\label{higher}
\ea
that reproduces the exact solution (\ref{mar2}) expanded at order $\kappa$.

In fact, inserting this last ansatz in (\ref{tot1}) we generate a generic 
chain of recursion relations of the form

\ba
\label{chain1}
&&O_{n+1}^0 (N)=F^0 (O_n^0(P^{(0)},\beta_0)),\nonumber\\
&&O_{n+1}^1 (N)=F^1 (O_n^0,O_{n+1}^0,O_n^1,\vec{P},\vec{\beta}),\nonumber\\
&&O_{n+1}^2 (N)=F^2 (O_n^0,O_n^1,O_n^2,O_{n+1}^0,O_{n+1}^1,\vec{P},\vec{\beta}),\nonumber\\
&&\vdots\nonumber\\
&&O_{n+1}^{\kappa} (N)=F^{\kappa}(O_n^0,...,O_n^{\kappa},O_{n+1}^0,...,
O_{n+1}^{\kappa-1},\vec{P},\vec{\beta})\,.
\ea

A deeper look at the explicit structure of (\ref{chain1}),
for the NLO non-singlet case, reveals the following structures for the generic iterates

\ba
\label{genrecnlo}
&&O_{n+1}^{0}(N)=-\frac{2}{\beta_{0}}\left[P^{(0)}(N)\, O_{n}^{0}(N)\right],
\nonumber\\
&&O_{n+1}^{\kappa}(N)= -\frac{2}{\beta_{0}}\left[P^{(0)}\, O_{n}^{\kappa}\right](N)
-\frac{1}{\pi\beta_{0}}\left[P^{(1)}(N)\, O_{n}^{\kappa-1}(N)\right]
\nonumber \\
&&\hspace{2cm}-\frac{\beta_{1}}{4\pi\beta_{0}}
O_{n+1}^{\kappa-1}(N)-\kappa O_{n}^{\kappa}(N)-(\kappa-1)
\frac{\beta_{1}}{4\pi\beta_{0}} O_{n}^{\kappa-1}(N),\,
\ea
at NLO, while for the NNLO case we obtain
\ba
\label{genrecnnlo}
&&O_{n+1}^{0}(N)=-\frac{2}{\beta_{0}}\left[P^{(0)}(N)\, O_{n}^{0}(N)\right],
\nonumber\\
&& O_{n+1}^{1}(N)= -\frac{2}{\beta_{0}}\left[P^{(0)}(N)\, O_{n}^{1}(N)\right]
-\frac{1}{\pi\beta_{0}}\left[P^{(1)}(N)\, O_{n}^{0}(N)\right]\nonumber \\
&&\hspace{2cm} -\frac{\beta_{1}}{4\pi\beta_{0}}O_{n+1}^{0}(N)-O_{n}^{1}(N),
\nonumber\\
&& O_{n+1}^{\kappa}(N) =  -\frac{2}{\beta_{0}}\left[P^{(0)}
(N)\, O_{n}^{\kappa}(N)\right]
-\frac{1}{\pi\beta_{0}}\left[P^{(1)}(N)\, O_{n}^{\kappa-1}(N)\right]\nonumber \\
&&\hspace{2cm} -\frac{1}{2\pi^{2}\beta_{0}}
\left[P^{(2)}(N)\, O_{n}^{\kappa-2}(N)\right]\nonumber \\
&&\hspace{2cm} -\frac{\beta_{1}}{4\pi\beta_{0}}O_{n+1}^{\kappa-1}(N)
-\frac{\beta_{2}}{16\pi^{2}\beta_{0}} O_{n+1}^{\kappa-2}(N)\nonumber \\
&&\hspace{2cm} -\kappa O_{n}^{\kappa}(N)-(\kappa-1)
\frac{\beta_{1}}{4\pi\beta_{0}} O_{n}^{\kappa-1}(N)-(\kappa-2)
\frac{\beta_{2}}{16\pi^{2}\beta_{0}} O_{n}^{\kappa-2}(N).
\ea
Hence, one is able to determine the structure of the $\kappa$-$th$
recursion relation when the $\kappa=0$ and $\kappa=1$ cases are known.
This property is very useful from the computational point of view.
\footnote{The relations (\ref{genrecnlo}) and (\ref{genrecnnlo}) hold
also in the NLO/NNLO singlet case and can be generalized to any
perturbative order in the expansion of the kernels.}

Passing to the resolution of the recursion relations in moment space, we get the
formal expansion of each $O_{n}^{\kappa}(N)$
in terms of the initial condition $O_{0}^0(N)$, which reads
\ba
\tilde{f}(N,\alpha_0)=O_{0}^0(N)+\slash{0}\alpha_0+\slash{0}\alpha_0^2+...
+\slash{0}\alpha_0^{\kappa}\,.
\ea
Here we have set to zero all the higher order terms, as a natural 
generalization of $B_0=C_0=0$..., according to what has been discussed above.

These relations can be solved as we have shown in previous examples,
and the generic structure of their solution can be identified.
If we define $R_0=-\frac{2}{\beta_0}P^{(0)}$, the expressions of all the
$O_{n}^{\kappa}(N)$ in terms of $\tilde{f}(N,\alpha_0)\equiv\tilde{f}_0$ become
\ba
&&O_{n}^0 (N)=R_0^n\tilde{f}_0\nonumber\\
&&O_{n}^1 (N)=G^1 (R_0^n,(R_0 -1)^n,\vec{P},\vec{\beta})
\tilde{f}_0,\nonumber\\
&&O_{n}^2 (N)=G^2 (R_0^n,(R_0 -1)^n,(R_0 -2)^n,\vec{P},\vec{\beta})
\tilde{f}_0,\nonumber\\
&&\vdots\nonumber\\
&&O_{n}^{\kappa} (N)=G^{\kappa} (R_0^n,(R_0 -1)^n,(R_0 -2)^n,...,(R_0 -{\kappa})^n,\vec{P},\vec{\beta})
\tilde{f}_0\,,
\ea
and in particular, by an explicit calculation of $O_{n}^{\kappa}(N)$, one can work out
the structure of these special functions $G^m$. For instance, we get for $m=\kappa$ the expression
\ba
\label{Gform}
G^{\kappa} (R_0^n,(R_0 -1)^n,(R_0 -2)^n,...,(R_0 -{\kappa})^n,\vec{P},\vec{\beta})=\sum_{j=0}^{\kappa}
(R_0 -j)^n \xi_{\kappa}^{(j)}(\vec{P},\vec{\beta})\,,
\ea
for suitable coefficients $\xi_{\kappa}^{(j)}$.
Substituting the $O_{n}^{m}(N)$ functions with $m=0,\dots\kappa$  in the higher-order ansatz
(\ref{higher}) and performing our exponentiation, we get an expression of the form
\ba
&&\tilde{f}(N,\alpha_s)=G^{0}\Bigg(\left(\frac{\alpha_s}{\alpha_0}\right)^{R_0}\Bigg)
+\alpha_s \,G^{1}\Bigg(\left(\frac{\alpha_s}{\alpha_0}\right)^{R_0},
\left(\frac{\alpha_s}{\alpha_0}\right)^{R_0-1}\Bigg)
\nonumber\\
&&\hspace{2cm}
+\alpha_s^2 \,G^{2}\Bigg(\left(\frac{\alpha_s}{\alpha_0}\right)^{R_0},
\left(\frac{\alpha_s}{\alpha_0}\right)^{R_0-1},\left(\frac{\alpha_s}{\alpha_0}\right)^{R_0-2}\Bigg)
\nonumber\\
&&\hspace{2cm}
+\dots \alpha_s^{\kappa} \,G^{\kappa}\Bigg(\left(\frac{\alpha_s}{\alpha_0}\right)^{R_0},
\left(\frac{\alpha_s}{\alpha_0}\right)^{R_0-1},\left(\frac{\alpha_s}{\alpha_0}\right)^{R_0-2},\dots,
\left(\frac{\alpha_s}{\alpha_0}\right)^{R_0-\kappa}\Bigg)\,,
\nonumber\\
\ea
which can be written as follows by the use of eq.~(\ref{Gform})
\ba
\tilde{f}(N,\alpha_s)=\left(\frac{\alpha_s}{\alpha_0}\right)^{R_0}\tilde{f}(N,\alpha_0)
\sum_{m=0}^{\kappa}\sum_{j=0}^{m} \alpha_s^{m-j}\alpha_0^{j}~\xi_{m}^{(j)}(\vec{P},\vec{\beta})\,.
\ea
This is the exact solution expanded up to $O(\alpha_s^\kappa)$ in accuracy.

\section{The Search for the exact non-singlet NLO solution}
We have shown in the previous sections how to construct exact solutions of truncated equations
using logarithmic expansions. We have also shown the equivalence of these approaches with the Mellin method, since the recursion relations for the unknown
coefficient functions
of the expansions can be solved to all orders and so reproduce the solution in Mellin space
of the truncated equation. The question that we want to address in this section is whether we
can search for exact solutions of the exact (untruncated) equations as well.
These solutions are known exactly in the non singlet case up to NLO. It is not difficult also to obtain the
exact NNLO solution in Mellin space, and we will reconstruct the same solutions 
using modified recursion relations. 
The expansions that we will be using at NLO are logarithmic and solve the untruncated equation. 
The NNLO case, instead, will be treated in a following section, where, again, we will use
recursion relations to build the exact solution but with a non-logarithmic ansatz \footnote{In PEGASUS 
\cite{Vogt3} the NNLO non-singlet solution is built by truncation in Mellin space of the evolution equation,
while the NLO solution is implemented as an exact solution.}.

The exact NLO non-singlet solution has been given in eq.~(\ref{exactsol}).
The identification of an expansion that allows to reconstruct in moment space eq.~(\ref{exactsol}) follows quite naturally once the typical
properties of the convolution product $\otimes$ are identified.
For this purpose we define the serie of convolution products
\beqa
e^{F_A P_A(x)\otimes} &\equiv& \sum_{n=0}^\infty \frac{F_A^n}{n!}\left(P_A(x)\otimes\right)^n
\ea
that acts on a given initial function as
\beqa
e^{F_A P_A(x)\otimes}\phi(x)&=&
\left( \delta(1-x)\otimes + F_A P_A(x)\otimes + \frac{1}{2!}F_A^2 P_A\otimes P_A\otimes
+ \dots\right)\phi(x) \nonumber \\
&=&
\phi(x) +  F_A \left(P_A \otimes \phi\right) + \frac{1}{2!}F_A^2 \left(P_A\otimes P_A\otimes \right)\phi(x)
\,+ \dots.
\eeqa

The functions $F_A$ and $F_B$ are parametrically dependent on any other variable except the variable $x$.
The proof of the associativity, distributivity and commutativity of the $\otimes$ product is easily obtained after mapping these products in Mellin space. For instance, for generic functions $a(x)$ $b(x)$ and $c(x)$ for which the $\otimes$ product is a regular function one has
\beqa
\mathcal{M}\left[\left(a\otimes b\right)\otimes c\right](N)&=&\mathcal{M}\left[ a\otimes \left(b\otimes c\right)\right](N) \nonumber \\
&=& a(N) B(N) C(N),
 \eeqa
where $\mathcal{M}$ denotes the Mellin transform and $N$ is the moment variable.
Also one obtains
\beq
e^{F_A P_A(x)\otimes}\,e^{F_B P_B(x)\otimes}\,\phi(x)=e^{\left(F_A P_A(x) + F_B P_B(x)\right)\otimes}\,\phi(x),
\label{expo}
\eeq
and
\beq
\mathcal{M}\left[e^{M a(x)\otimes}\,\phi\right](N)=e^{M a(N)}\phi(N),
\eeq
with $M$ $x$-independent,
since both left-hand-side and right-hand-side of (\ref{expo}) can be mapped to the same function in Mellin space.
Notice that the role of the identity in $\otimes$-space is taken by the function $\delta(1-x)$. We will also use the notation
\beq
\left( \sum_{n=0}^\infty {A'}_n(x) F_A^n\right)_\otimes \phi(x)\equiv \left( A_0(x)\otimes + F_A A_1(x)\otimes + \dots\right)\phi(x),
\eeq
where the ${A'}_n(x)$ and the $A_n(x)$ capture the operatorial and the functional expansion 
- respectively - and are trivially related
\beq
A'_n(x)\otimes \phi(x)= A_n(x).
\eeq
To identify the $x$-space ansatz we rewrite (\ref{exactsol}) as
\ba
f(N,\alpha)& = & f(N,\alpha_{0})e^{a(N)L}e^{b(N)M}\nonumber \\
& = & f(N,\alpha_{0})\left(\sum_{n=0}^{\infty}
\frac{a(N)^{n}}{n!}L^{n}\right)\left(\sum_{m=0}^{\infty}
\frac{b(N)^{m}}{m!}M^{m}\right),
\label{labelE}
\ea
where we have introduced the notations
\ba
L & = & \log\frac{\alpha_s}{\alpha_{0}},\nonumber \\
M & = & \log\frac{4\pi\beta_{0}+\alpha_s\beta_{1}}{4\pi\beta_{0}+\alpha_{0}\beta_{1}},\nonumber \\
a(N) & = & -\frac{2P^{(0)}(N)}{\beta_{0}},\nonumber \\
b(N) & = & \frac{2P^{(0)}(N)}{\beta_{0}}-\frac{4P^{(1)}(N)}{\beta_{1}}.
\label{labelA}
\ea
Our ansatz for the exact solution in $x$-space is chosen of the form
\ba
f(x,Q^{2}) & = & \left(\sum_{n=0}^{\infty}\frac{A'_{n}(x)}{n!}L^{n}\right)_\otimes
\left(\sum_{m=0}^{\infty}\frac{B'_{m}(x)}{m!}M^{m}\right)_\otimes f(x,Q_0^2)\nonumber \\
& = & \sum_{s=0}^{\infty}\sum_{n=0}^{s}L^{n}M^{s-n} \frac{A'_{n}(x)\otimes B'_{s-n}(x)}{n!(s-n)!}
\otimes f(x,Q_0^2)\nonumber \\
& = & \sum_{s=0}^{\infty}\sum_{n=0}^{s}\frac{C_{n}^{s}(x)}{n!(s-n)!}L^{n}M^{s-n},
\label{eq:NLOansatz}
\ea
where in the first step we have turned the product of two series into
a single series of a combined exponent $s=n+m$, and in the last step
we have introduced the functions
\begin{equation}
C_{n}^{s}(x)=A'_{n}(x)\otimes B'_{s-n}(x)\otimes f(x,Q_0^2),\qquad(n\leq s).
\end{equation}
Setting $Q=Q_{0}$ in (\ref{eq:NLOansatz}) we get the initial condition $A'_0(x)=B'_0(x)=\delta(1-x)$ or,
equivalently,
\begin{equation}
f(x,Q_{0}^{2})=C_{0}^{0}(x).
\end{equation}
Inserting the ansatz (\ref{eq:NLOansatz}) into the NLO DGLAP equation, with the expressions of
the kernel and beta function included at the corresponding order, we obtain the identity
\begin{eqnarray}
\sum_{s=0}^{\infty}\sum_{n=0}^{s}\left\{\left(-\frac{\beta_{0}}{4\pi}
\alpha-\frac{\beta_{1}}{16\pi^{2}}\alpha^{2}\right)C_{n+1}^{s+1}-
\frac{\beta_{1}}{16\pi^{2}}\alpha^{2}C_{n}^{s+1}\right\}
\frac{L^{n}M^{s-n}}{n!(s-n)!}\nonumber \\
=\sum_{s=0}^{\infty}\sum_{n=0}^{s}\left\{ \frac{\alpha}{2\pi}P^{(0)}
\otimes C_{n}^{s}+\frac{\alpha^{2}}{4\pi^{2}}P^{(1)}\otimes C_{n}^{s}\right\}
\frac{L^{n}M^{s-n}}{n!(s-n)!}.
\end{eqnarray}
Equating term by term the coefficients of $\alpha$ and
$\alpha^{2}$, we find from this identity the new exact recursion relations
\begin{eqnarray}
C_{n+1}^{s+1} & = & -\frac{2}{\beta_{0}}P^{(0)}\otimes C_{n}^{s},\\
C_{n}^{s+1} & = & -C_{n+1}^{s+1}-\frac{4}{\beta_{1}}P^{(1)}\otimes C_{n}^{s},
\end{eqnarray}
or, equivalently,  
\begin{eqnarray}
C_{n}^{s} & = & -\frac{2}{\beta_{0}}P^{(0)}\otimes C_{n-1}^{s-1}
\label{eq:NLO_rec_diagonale},\\
C_{n}^{s} & = & -C_{n+1}^{s}-\frac{4}{\beta_{1}}P^{(1)}\otimes C_{n}^{s-1}.
\label{eq:NLO_rec_verticale}
\end{eqnarray}
Notice that although the coefficients $C_n^s$ are convolution products of two functions,
the recursion relations do not let these products appear explicitely.
These relations just written down allow to compute all the coefficients $C_{n}^{s}$
$(n\leq s)$ up to a chosen $s$ starting from $C_{0}^{0}$, which is
given by the initial conditions. In particular eq.~(\ref{eq:NLO_rec_diagonale}) allows us to move along the diagonal
arrow according to the diagram reproduced in Table \ref{cap:schemaNLO};
eq.~(\ref{eq:NLO_rec_verticale}) instead allows us to compute a coefficient
in the table once we know the coefficients at its right and the coefficient
above it (horizontal and vertical arrows).
To compute $C_{n}^{s}$ there is a certain freedom, as illustrated
in the diagram.  For instance, to determine $C_{s}^{s}$
we can only use (\ref{eq:NLO_rec_diagonale}),
and for the coefficients $C_{0}^{s}$ we can only use (\ref{eq:NLO_rec_verticale}).
For all the other coefficients one can prove that using (\ref{eq:NLO_rec_diagonale})
or (\ref{eq:NLO_rec_verticale}) brings to the same determination of the coefficients, 
and in our numerical studies we
have chosen to implement (\ref{eq:NLO_rec_diagonale}), being this relation 
less time consuming since it involves $P^{(0)}$ instead of $P^{(1)}$.

The recursion relations defining the iterated solution can be solved as follows.
From the first relation (\ref{eq:NLO_rec_diagonale}), keeping the $s$-index fixed, we have
\ba
\label{exasolve1}
&&C_{n}^{s} = -\frac{2}{\beta_{0}}P^{(0)}\otimes C_{n-1}^{s-1}\Rightarrow
\nonumber\\
&&C_{n}^{s} = \left[-\frac{2}{\beta_{0}}P^{(0)}\right]^{n}\otimes C_{0}^{s-n-1}\,,
\ea
then, since the second relation (\ref{eq:NLO_rec_verticale}) also holds for $n=0$,
we can write (using eq. (\ref{exasolve1}))
\ba
&&C_{n}^{s}=-C_{n+1}^{s}-\frac{4}{\beta_{1}}P^{(1)}\otimes C_{n}^{s-1} \Rightarrow
\nonumber\\
&&C_{0}^{s}=\left[\frac{2}{\beta_0}P^{(0)}-\frac{4}{\beta_{1}}P^{(1)}\right]
\otimes C_{0}^{s-1}\Rightarrow
\nonumber\\
&&C_{0}^{s}=\left[\frac{2}{\beta_0}P^{(0)}-\frac{4}{\beta_{1}}P^{(1)}\right]^{s}
\otimes C_{0}^{0}.
\ea
Finally, inserting the above relation in (\ref{exasolve1}) we can write
\ba
C_{n}^{s} =\left[\frac{2}{\beta_0}P^{(0)}\right]^{n}\otimes
\left[\frac{2}{\beta_0}P^{(0)}-\frac{4}{\beta_{1}}P^{(1)}\right]^{s-n}
\otimes C_{0}^{0}\,,
\ea
which is the solution we have been searching for. The last step in the proof consists in
taking the Mellin transform of this operatorial solution and summing the corresponding series

\ba
f(N,\alpha_s)&=&\sum_{s=0}^{\infty}\sum_{n=0}^{s}\frac{C_{n}^{s}(N)}{n!(s-n)!}L^{n}M^{s-n}
\nonumber\\
&=&\sum_{s=0}^{\infty}\sum_{n=0}^{s}\frac{L^{n}M^{s-n}}{n!(s-n)!}
\left[\frac{2}{\beta_0}P^{(0)}\right]^{n}
\left[\frac{2}{\beta_0}P^{(0)}-\frac{4}{\beta_{1}}P^{(1)}\right]^{s-n}
 C_{0}^{0}(N),\,
\ea
that after summation gives  
\ba
f(N,\alpha_s)=e^{-\frac{2}{\beta_0}P^{(0)}(N)\log\left(\frac{\alpha_s}{\alpha_0}\right)}
\exp\left\{\left[\frac{2}{\beta_0}P^{(0)}(N)-\frac{4}{\beta_{1}}P^{(1)}(N)\right]
\log{\left(\frac{4\pi\beta_0+\alpha_s\beta_1}{4\pi\beta_0+\alpha_0\beta_1}\right)}\right\}
C_{0}^{0}(N),
\nonumber\\
\ea
which is exactly the expression in eq. (\ref{labelE}).
Hence, it is obvious that the exact solution of the DGLAP 
equation (\ref{nontrunc}) can be written in $x$-space as
\ba
f(x,\alpha_s(Q^2))=e^{-\log\left(\frac{\alpha_s}{\alpha_0}\right)\frac{2}{\beta_0}P^{(0)}(x)\otimes}
e^{\log{\left(\frac{4\pi\beta_0+\alpha_s\beta_1}{4\pi\beta_0+\alpha_0\beta_1}\right)}
\left[\frac{2}{\beta_0}P^{(0)}(x)-\frac{4}{\beta_{1}}P^{(1)}(x)\right]\otimes}C_{0}^{0}(x),
\nonumber\\
\ea
therefore proving that the ansatz (\ref{eq:NLOansatz}) reproduces the exact solution of the
NLO DGLAP equation from $x$-space.

A second version of the same ansatz for the NLO exact solution can be built using a
factorization of the NLO DGLAP equation. This strategy is analogous to the method of
factorization for ordinary PDE's. For this purpose we define a modified LO DGLAP equation,
involving $\beta^{NLO}$
\beq
\frac{\partial \tilde{f}_{LO}(x,\alpha_{s})}{\partial\alpha_{s}}
=\left(\frac{\alpha_{s}}{2\pi \beta^{NLO}}\right)
P^{(0)}(x)\otimes \tilde{f}_{LO}(x,\alpha_{s}),
\eeq
whose solution is given by
\ba
\tilde{f}_{LO}(x,\alpha)&=& e^{M \left(\frac{2 P^{(0)}}{\beta_0}\right)\otimes} f_{LO}(x,\alpha)
\nonumber \\
f_{LO}(x,\alpha)&=& e^{L \left(\frac{-2 P^{(0)}}{\beta_0}\right)\otimes} f(x,\alpha_0),
\ea
where we have introduced the ordinary LO solution $f_{LO}$, expressed in terms of a typical initial
condition $f(x,\alpha_0)$, and the NLO recursion relations can be obtained from the expansion
\beq
f_{NLO}(x,\alpha)=\left(\sum_{n=0}^{\infty}\frac{B_{n}(x)}{n!}M^{n}\right)_\otimes
\tilde{f}_{LO}(x,\alpha).
\label{nlosolve}
\eeq
Inserting this relation into (\ref{nontrunc}) we obtain the recursion relations
\beqa
B_{n+1}&=& \left(-\frac{4}{\beta_1} P^{(1)}\right)\otimes B_n \nonumber \\
B_0(x) &=&\delta(1-x)\,,
\eeqa
which is solved in moment space by
\beqa
B_n(N)&=&\left(-\frac{4}{\beta_1} P^{(1)}\right)^n B_0(N)\nonumber \\
B_0(N)&=&1.
\eeqa
The solution eq.~(\ref{nlosolve}) can be re-expressed in the form
\ba
f_{NLO}(x,\alpha) &=& e^{M \left(-\frac{1}{4 \beta_1} P^{(1)}\right)\otimes}e^{M \left(\frac{2 P^{(0)}}{\beta_0}\right)\otimes}e^{L \left(\frac{-2 P^{(0)}}{\beta_0}\right)\otimes} f(x,\alpha_0) \nonumber \\
&=& e^{M \left(-\frac{1}{4 \beta_1} P^{(1)} +\frac{2 P^{(0)}}{\beta_0}\right)\otimes}e^{L \left(\frac{-2 P^{(0)}}{\beta_0}\right)\otimes} f(x,\alpha_0)
\ea
which agrees with (\ref{labelE}) once $a(N)$ and $b(N)$ have been defined as in (\ref{labelA}).

We have therefore proved that the exact NLO
solution of the DGLAP equation can be described by an exact ansatz. Since the ansatz is built
by inspection, it is obvious that one needs to know the solution in moment space in order
to reconstruct the coefficients. Though the recursive scheme used to construct the solution
in $x$-space is more complex compared to the recursion relations for the truncated solution,
its numerical implementation is still very stable and very precise, reaching the same level of
accuracy of the traditional methods based on the inversion of the Mellin
moments.
\begin{table}
\begin{center}$\begin{array}{ccccccccc}
C_{0}^{0}\\
\downarrow & \searrow\\
C_{0}^{1} & \leftarrow & C_{1}^{1}\\
\downarrow & \searrow &  & \searrow\\
C_{0}^{2} & \leftarrow & C_{1}^{2} &  & C_{2}^{2}\\
\downarrow & \searrow &  & \searrow &  & \searrow\\
C_{0}^{3} & \leftarrow & C_{1}^{3} &  & C_{2}^{3} &  & C_{3}^{3}\\
\downarrow & \searrow &  & \searrow &  & \searrow &  & \searrow\\
\ldots &  & \ldots &  & \ldots &  & \ldots &  & \ldots\end{array}$
\end{center}
\caption{Schematic representation of the procedure followed to compute each
coefficient $C_{n}^{s}$.\label{cap:schemaNLO}}
\end{table}

\section{Finding the exact non-singlet  NNLO solution}

To identify the NNLO exact solution we proceed similarly to the NLO case and start
from the DGLAP equation in moment space at the corresponding perturbative order

\beq
\frac{\partial f(N,\alpha_{s})}{\partial\alpha_{s}}=
-\frac{\left(\frac{\alpha_{s}}{2\pi}\right)P^{(0)}(N)+
\left(\frac{\alpha_{s}}{2\pi}\right)^{2}P^{(1)}(N)+
\left(\frac{\alpha_{s}}{2\pi}\right)^{3}P^{(2)}(N)}{\frac{\beta_{0}}{4\pi}
\alpha_{s}^{2}+\frac{\beta_{1}}{16\pi^{2}}\alpha_{s}^{3}+
\frac{\beta_{2}}{64\pi^{3}}\alpha_{s}^{4}}f(N,\alpha_{s}).
\label{eqnn}
\eeq
After a separation of variables, all the new logarithmic/non logarithmic and dependences
come from the integral
\beq
\int_{\alpha_s(Q_0^2)}^{\alpha_s(Q^2)} d\alpha \frac{P^{NNLO}(\alpha_s)}{\beta^{NNLO}(\alpha)},
\label{integer}
\eeq
and the solution of (\ref{eqnn}) is
\begin{eqnarray}
f(N,\alpha) & = & f(N,\alpha_{0})e^{a(N)\mathcal{L}}e^{b(N)\mathcal{M}}e^{c(N)\mathcal{Q}}\nonumber \\
& = & f(N,\alpha_{0})\left(\sum_{n=0}^{\infty}\frac{a(N)^{n}}{n!}\mathcal{L}^{n}\right)
\left(\sum_{m=0}^{\infty}\frac{b(N)^{m}}{m!}\mathcal{M}^{m}\right)\left(\sum_{p=0}^{\infty}
\frac{c(N)^{p}}{p!}\mathcal{Q}^{p}\right),
\label{nnlotest}
\end{eqnarray}
where we have defined
\begin{eqnarray}
\mathcal{L} & = & \log\frac{\alpha}{\alpha_{0}},\\
\mathcal{M} & = & \log\frac{16\pi^{2}\beta_{0}+4\pi\alpha\beta_{1}
+\alpha^{2}\beta_{2}}{16\pi^{2}\beta_{0}+4\pi\alpha_{0}\beta_{1}+\alpha_{0}^{2}\beta_{2}},\\
\mathcal{Q} & = & \frac{1}{\sqrt{4\beta_{0}\beta_{2}-\beta_{1}^{2}}}
\arctan\frac{2\pi(\alpha-\alpha_{0})\sqrt{4\beta_{0}\beta_{2}-
\beta_{1}^{2}}}{2\pi(8\pi\beta_{0}+(\alpha+\alpha_{0})\beta_{1})+
\alpha\alpha_{0}\beta_{2}},\\
a(N) & = & -\frac{2P^{(0)}(N)}{\beta_{0}},\\
b(N) & = & \frac{P^{(0)}(N)}{\beta_{0}}-\frac{4P^{(2)}(N)}{\beta_{2}},\\
c(N) & = & \frac{2\beta_{1}}{\beta_{0}}P^{(0)}(N)-8P^{(1)}(N)
+\frac{8\beta_{1}}{\beta_{2}}P^{(2)}(N).
\end{eqnarray}
Notice that for $n_{f}=6$ the solution has a branch point since
$4\beta_{0}\beta_{2}-\beta_{1}^{2}<0$. If we increase $n_f$ as we
step up in the factorization scale then, for $n_f=6$,
$\mathcal{Q}$ is replaced by its analytic continuation
\begin{equation}
\mathcal{Q}=\frac{1}{\sqrt{\beta_{1}^{2}-4\beta_{0}\beta_{2}}}
\textrm{arctanh}\frac{2\pi(\alpha-\alpha_{0})\sqrt{\beta_{1}^{2}
-4\beta_{0}\beta_{2}}}{2\pi(8\pi\beta_{0}+(\alpha+
\alpha_{0})\beta_{1})+\alpha\alpha_{0}\beta_{2}}.
\end{equation}
Eq.~(\ref{nnlotest}) incorporates all the nontrivial dependence on the coupling
constant $\alpha_s$ (now determined at 3-loop level)
into $\mathcal{L}, \mathcal{M}$ and $\mathcal{Q}$.

As a side remark we emphasize that it is also possible to obtain various  NNLO exact recursion
relations using the formalism of the convolution series introduced above.
For this purpose it is convenient to define suitable operatorial expressions, for instance
\beqa
\mathcal{E}_1 &\equiv& e^{\int_{\alpha_0}^{\alpha_s} d\alpha
 \frac{P^{NLO}(x,\alpha)}{\beta^{NNLO}(\alpha)}_\otimes}, \nonumber \\
\mathcal{E}_2 &\equiv& e^{\int_{\alpha_0}^{\alpha_s} d\alpha
 \left(\frac{\alpha}{2 \pi}\right)^3 \frac{P^{(2)}(x,\alpha)}{\beta^{NNLO}(\alpha)}_\otimes},
\label{seq1}
\ea

which are manipulated under the prescription that the integral in $\alpha$ is evaluated before that any
convolution product acts on the initial conditions. The re-arrangement of these operatorial
expressions is therefore quite simple and one can use simple identities such as
\beqa
J_0 &=&\int_{\alpha_0}^{\alpha_s} d\alpha
 \left(\frac{\alpha}{2 \pi}\right) \frac{P^{(0)}(x,\alpha)}{\beta^{NNLO}(\alpha)}_\otimes
= 2\frac{\beta_1}{\beta_0}\mathcal{Q}P^{(0)}_\otimes -\frac{2}{\beta_0}\mathcal{L} P^{(0)}_\otimes
+ \frac{1}{\beta_0}\mathcal{M}P^{(0)}_\otimes, \nonumber\\
J_1 &=&\int_{\alpha_0}^{\alpha_s} d\alpha
 \left(\frac{\alpha}{2 \pi}\right)^2 \frac{P^{(1)}(x,\alpha)}{\beta^{NNLO}(\alpha)}_\otimes
= - 8 \mathcal{Q} P^{(1)}\otimes, \nonumber \\
J_2&=&\int_{\alpha_0}^{\alpha_s} d\alpha
 \left(\frac{\alpha}{2 \pi}\right)^3 \frac{P^{(2)}(x,\alpha)}{\beta^{NNLO}(\alpha)}_\otimes =
\left(-\frac{4}{\beta_2}\mathcal{M} + 8 \frac{\beta_1}{\beta_2}\mathcal{Q}\right) P^{(2)}\otimes,
\nonumber \\
J_{NNLO}&=&\int_{\alpha_0}^{\alpha_s} d\alpha
 \frac{P^{NLO}(x,\alpha)}{\beta^{NNLO}(\alpha)}_\otimes \nonumber \\
&=&  \mathcal{Q}\left(2 \frac{\beta_1}{\beta_0} P^{(0)}\otimes - 8 P^{(1)}\otimes \right)
-\frac{2}{\beta_0}\mathcal{L}P^{(0)}\otimes + \frac{1}{\beta_0}\mathcal{M} P^{(0)}\otimes,
\nonumber \\
\ea
to build the NNLO exact solution using a suitable recursive algorithm. For instace, using
(\ref{seq1})
one can build an intermediate solution of the equation
\beq
\frac{\partial \tilde{f}_{NLO}(x,\alpha_s)}{\partial \alpha_s}=
\frac{P^{NLO}}{\beta^{NNLO}}\tilde{f}_{NLO}(x,\alpha_s)
\eeq
given by
\ba
\tilde{f}_{NLO} &=& \mathcal{E}_1 f(x,\alpha_0),
\ea
and then constructs with a second recursion the exact solution
\ba
f(x,\alpha_s) &=& \mathcal{E}_2 \tilde{f}_{NLO}.
\ea
A straightforward approach, however, remains the one described in the previous section,
that we are going now to extend to NNLO.
In this case, in the choice of the recursion relations,
one is bound to equate 3 independent logarithmic powers of
$\mathcal{L}$, $\mathcal{M}$ and $\mathcal{Q}$ that appear in the symmetric ansazt
\ba
f(x,Q^{2}) & = & \left(\sum_{n=0}^{\infty}\frac{A'_{n}(x)}{n!}\mathcal{L}^{n}\right)_\otimes
\left(\sum_{m=0}^{\infty}\frac{B'_{m}(x)}{m!}\mathcal{M}^{m}\right)_\otimes
\left(\sum_{p=0}^{\infty}\frac{C'_{p}(x)}{p!}\mathcal{Q}^{p}\right)_\otimes f(x,Q_0^2)\nonumber \\
& = & \sum_{s=0}^{\infty}\sum_{t=0}^{s}\sum_{n=0}^{t}\frac{A'_{n}(x)\otimes
B'_{t-n}(x)\otimes {C'_{s-t}(x)}}{n!(t-n)!(s-t)!}\otimes f(x,Q_0^2)\,\mathcal{L}^{n}\mathcal{M}^{t-n}\mathcal{Q}^{s-t}\nonumber \\
& = & \sum_{s=0}^{\infty}\sum_{t=0}^{s}\sum_{n=0}^{t}
\frac{ D_{t,n}^{s}(x)}{n!(t-n)!(s-t)!}\mathcal{L}^{n}\mathcal{M}^{t-n}\mathcal{Q}^{s-t},
\label{eq:NNLOansatz}
\ea
and where
\beq
D_{t,n}^{s}(x)= A'_{n}(x)\otimes
B'_{t-n}(x)\otimes C'_{s-t}(x)\otimes f(x,Q_0^2).
\eeq
The ansatz is clearly identified quite simply by inspection, once
the structure of the solution in moment space (\ref{nnlotest}) is known explicitely.
In (\ref{eq:NNLOansatz}) we have at a first step re-arranged the product of the three series
into a single series with a given total exponent $s=n+m+p$, and we have
introduced an index $t=n+m$. The triple-indexed function $D_{t,n}^{s}(x)$ can be defined also as
an ordinary product
\begin{equation}
D_{t,n}^{s}(x)=A_{n}(x)\otimes B_{t-n}(x)\otimes C_{s-t}(x),\qquad(n\leq t\leq s),
\end{equation}
where we have absorbed the $\otimes$ operator into the definition of $A$, $B$ and $C$,
\beq
A(x) B(x) C(x)=A'(x)\otimes \bigg[\Big(B'(x)\otimes \Big(C'(x)\otimes f(x,Q_0^2)\Big)\Big)\bigg].
\eeq
Setting $Q=Q_{0}$ in (\ref{eq:NNLOansatz}) we get the initial condition
\begin{equation}
f(x,Q_{0}^{2})=D_{0,0}^{0}(x).
\end{equation}
Inserting the ansatz (\ref{eq:NNLOansatz}) into the 3-loop DGLAP equation
together with the beta function determined at the same order and equating
the coefficients of $\alpha$, $\alpha^{2}$ and $\alpha^{3}$, we
find the recursion relations satisfied by the unknown coefficients $D_{t,n}^{s}(x)$
\begin{eqnarray}
D_{t+1,n+1}^{s+1} & = & -\frac{2}{\beta_{0}}P^{(0)}\otimes D_{t,n}^{s},\\
D_{t+1,n}^{s+1} & = & -\frac{1}{2}D_{t+1,n+1}^{s+1}-
\frac{4}{\beta_{2}}P^{(2)}\otimes D_{t,n}^{s},\\
D_{t,n}^{s+1} & = & -2\beta_{1}\left(D_{t+1,n}^{s+1}+
D_{t+1,n+1}^{s+1}\right)-8P^{(1)}\otimes D_{t,n}^{s},
\end{eqnarray}
or equivalently
\begin{eqnarray}
D_{t,n}^{s} & = & -\frac{2}{\beta_{0}}P^{(0)}
\otimes D_{t-1,n-1}^{s-1}
\label{eq:NNLO_rec_diagonale},\\
D_{t,n}^{s} & = & -\frac{1}{2}D_{t,n+1}^{s}-
\frac{4}{\beta_{2}}P^{(2)}\otimes D_{t-1,n}^{s-1}
\label{eq:NNLO_rec_verticale},\\
D_{t,n}^{s} & = & -2\beta_{1}\left(D_{t+1,n}^{s}+
D_{t+1,n+1}^{s}\right)-8P^{(1)}\otimes D_{t,n}^{s-1}.
\label{eq:NNLO_rec_orizzontale}
\end{eqnarray}
In the computation of a given coefficient $D_{t,n}^{s}$, if more
than one recursion relation is allowed to determine that specific coefficient,
we will choose to implement the less time consuming path,
i.e.~in the order (\ref{eq:NNLO_rec_diagonale}),
(\ref{eq:NNLO_rec_orizzontale}) and (\ref{eq:NNLO_rec_verticale}).
At a fixed integer $s$ we proceed as follows: we

\begin{enumerate}
\item compute all the coefficients $D_{t,n}^{s}$ with $n\neq0$ using (\ref{eq:NNLO_rec_diagonale});
\item compute the coefficient $D_{s,0}^{s}$ using (\ref{eq:NNLO_rec_verticale});
\item compute the coefficient $D_{t,0}^{s}$ with $t\neq s$ using (\ref{eq:NNLO_rec_orizzontale}),
in decreasing order in $t$.
\end{enumerate}
This computational strategy is exemplified in the diagram in Table \ref{cap:schemaNNLO}
for $s=4$, where the various paths are highlighted.

\begin{table}
\begin{center}$\begin{array}{ccccccccc}
 &  &  &  &  &  &  &  & \underline{D_{4,4}^{4}}\\
\\ &  &  &  &  &  & \underline{D_{3,3}^{4}} &  & \underline{D_{4,3}^{4}}\\
\\ &  &  &  & \underline{D_{2,2}^{4}} &  & \underline{D_{3,2}^{4}} &  &
\underline{D_{4,2}^{4}}\\
\\ &  & \underline{D_{1,1}^{4}} &  & \underline{D_{2,1}^{4}} &  &
\underline{D_{3,1}^{4}} &  & \underline{D_{4,1}^{4}}\\
 & \swarrow &  & \swarrow &  & \swarrow &  & \swarrow & \downarrow\\
D_{0,0}^{4} & \longleftarrow & D_{1,0}^{4} & \longleftarrow & D_{2,0}^{4} &
\longleftarrow & D_{3,0}^{4} & \longleftarrow & D_{4,0}^{4}\end{array}$
\end{center}

\caption{Schematic representation of the procedure followed to compute each
coefficient $D_{t,n}^{s}$ for $s=4$. The underlined coefficients are
computed via eq.~(\ref{eq:NNLO_rec_diagonale}).\label{cap:schemaNNLO}}
\end{table}

Following a procedure similar to the one used for the NLO case, we can solve
the recursion relations for the NNLO ansatz with the initial
conditions $D_{0,0}^{0}(x)$.
Solving the relations (\ref{eq:NNLO_rec_diagonale}-\ref{eq:NNLO_rec_orizzontale}), 
we obtain the chain conditions
\ba
&&D_{t,n}^{s}=-\frac{2}{\beta_{0}}P^{(0)}
\otimes D_{t-1,n-1}^{s-1}\Rightarrow
\nonumber\\
&&D_{t,n}^{s}= \left[-\frac{2}{\beta_{0}}P^{(0)}\right]^{n}\otimes D_{t-n,0}^{s-n}.
\ea
Then, from the second relation we get the additional ones
\ba
&&D_{t,n}^{s}=-\frac{1}{2}D_{t,n+1}^{s}-
\frac{4}{\beta_{2}}P^{(2)}\otimes D_{t-1,n}^{s-1}\Rightarrow
\nonumber\\
&&D_{t-n,0}^{s-n}=\left[\frac{P^{(0)}}{\beta_{0}}
-\frac{4P^{(2)}}{\beta_2}\right]^{t-n}\otimes D_{0,0}^{s-t-2n}.
\ea
From the last relation we also obtain the relations 
\ba
&&D_{t,n}^{s}=-2\beta_{1}\left(D_{t+1,n}^{s}+
D_{t+1,n+1}^{s}\right)-8P^{(1)}\otimes D_{t,n}^{s-1}\Rightarrow
\nonumber\\
&&D_{0,0}^{s-t-2n}=\left[-8 P^{(1)}+\frac{2\beta_1}{\beta_0}P^{(0)}
+\frac{8\beta_1}{\beta_2}P^{(2)}\right]^{s-t-2n}\otimes D_{0,0}^{0}\,,
\ea
which solve the recursion relations in $x$-space in terms of the initial condition $D_{0,0}^{0}$.
Finally, the explicit expression of the $D_{t,n}^{s}$ coefficient
will be given by 
\ba
D_{t,n}^{s}(x)=\left[-\frac{2}{\beta_0}P^{(0)}\right]^{n}
\otimes\left[\frac{P^{(0)}}{\beta_0}-\frac{4P^{(2)}}{\beta_2}\right]^{t-n}
\otimes\left[-8 P^{(1)} +\frac{2\beta_1}{\beta_0}P^{(0)}+
8\frac{\beta_1}{\beta_2}P^{(2)}\right]^{s-t-2n}\otimes D_{0,0}^{0}(x).
\nonumber\\
\ea
The solution of the NNLO DGLAP equation reproduced by (\ref{eq:NNLOansatz})
in Mellin space will then be written in the form 
\ba
f(N,\alpha_s)&=&\sum_{s=0}^{\infty}\sum_{t=0}^{s}\sum_{n=0}^{t}
\frac{D_{t,n}^{s}(N)}{n!(t-n)!(s-t)!}\mathcal{L}^{n}\mathcal{M}^{t-n}\mathcal{Q}^{s-t}
\nonumber\\
&=&\sum_{s=0}^{\infty}\sum_{t=0}^{s}\sum_{n=0}^{t}
\frac{\mathcal{L}^{n}\mathcal{M}^{t-n}\mathcal{Q}^{s-t}}{n!(t-n)!(s-t)!}
\left[-\frac{2}{\beta_0}P^{(0)}\right]^{n}
\left[\frac{P^{(0)}(N)}{\beta_0}-\frac{4P^{(2)}(N)}{\beta_2}\right]^{t-n}
\nonumber\\
&\times&\left[-8 P^{(1)}(N) +\frac{2\beta_1}{\beta_0}P^{(0)}(N)
+ 8\frac{\beta_1}{\beta_2}P^{(2)}(N)\right]^{s-t-2n}D_{0,0}^{0}(N)\,,
\ea
which is equivalent to
\ba
&&f(N,\alpha_s)= e^{\left[-\frac{2}{\beta_0}P^{(0)}\right]
\log{\left(\frac{\alpha_s}{\alpha_0}\right)}}
\exp\left\{\left[\frac{P^{(0)}(N)}{\beta_{0}}-\frac{4P^{(2)}(N)}{\beta_{2}}\right]
\log\frac{16\pi^{2}\beta_{0}+4\pi\alpha_s\beta_{1}
+\alpha_s^{2}\beta_{2}}{16\pi^{2}\beta_{0}+4\pi\alpha_{0}\beta_{1}+\alpha_{0}^{2}\beta_{2}}
\right\}\times
\nonumber\\
&&\hspace{2cm}\exp\left\{\left[\frac{2\beta_{1}}{\beta_{0}}P^{(0)}(N)-8P^{(1)}(N)
+\frac{8\beta_{1}}{\beta_{2}}P^{(2)}(N)\right]\times
\right.\nonumber\\
&&\left.\hspace{2cm}\Bigg(\frac{1}{\sqrt{4\beta_{0}\beta_{2}-\beta_{1}^{2}}}
\arctan\frac{2\pi(\alpha_s-\alpha_{0})\sqrt{4\beta_{0}\beta_{2}-
\beta_{1}^{2}}}{2\pi(8\pi\beta_{0}+(\alpha_s+\alpha_{0})\beta_{1})+
\alpha_s\alpha_{0}\beta_{2}}\Bigg)
\right\}D_{0,0}^{0}(N),
\ea
and it reproduces the result in (\ref{nnlotest}).
In $x$-space the above solution can be simply written as
\ba
&&f(x,\alpha_s(Q^2))= e^{\left[\log{\left(\frac{\alpha_s}{\alpha_0}\right)}
-\frac{2}{\beta_0}P^{(0)}\right]\otimes}
\exp\left\{\log\left(\frac{16\pi^{2}\beta_{0}+4\pi\alpha_s\beta_{1}
+\alpha_s^{2}\beta_{2}}{16\pi^{2}\beta_{0}+4\pi\alpha_{0}\beta_{1}+\alpha_{0}^{2}\beta_{2}}\right)
\left[\frac{P^{(0)}(N)}{\beta_{0}}-\frac{4P^{(2)}(N)}{\beta_{2}}\right]\otimes\right\}
\nonumber\\
&&\hspace{2cm}\exp\left\{
\Bigg(\frac{1}{\sqrt{4\beta_{0}\beta_{2}-\beta_{1}^{2}}}
\arctan\frac{2\pi(\alpha_s-\alpha_{0})\sqrt{4\beta_{0}\beta_{2}-
\beta_{1}^{2}}}{2\pi(8\pi\beta_{0}+(\alpha_s+\alpha_{0})\beta_{1})+
\alpha_s\alpha_{0}\beta_{2}}\Bigg)\right.\nonumber\\
&&\left.\hspace{2cm}
\left[\frac{2\beta_{1}}{\beta_{0}}P^{(0)}(N)-8P^{(1)}(N)
+\frac{8\beta_{1}}{\beta_{2}}P^{(2)}(N)\right]\otimes\right\}D_{0,0}^{0}(x).
\ea
We have shown how to obtain exact NNLO solutions of the non-singlet equations using recursion relations.
It is clear that the solution shown above conceals all the logarithms of the coupling constant
into more complicated functions of $\alpha_s$ and therefore
it performs an intrinsic resummation of
all these contributions, as obvious, being the exact solution of the non-singlet
equation at NNLO.
A numerical implementation of the recursion relations associated to these new
functions of the coupling constants, in this case, is no
different from the previous cases, when only functions of the form
$\log(\alpha/\alpha_0)$ have been considered, but with a faster convergence rate.

\section{Truncated solutions at LO and NNLO in the singlet case}

The proof of the existence of a valid logarithmic ansatz that reproduces the
truncated solution of the singlet DGLAP equation at NLO is far
more involved compared to the non singlet
case. Before we proceed with this discussion, it is important to clarify
some points regarding some known results concerning these equations in moment space.
First of all, as we have widely remarked before, there are no exact solutions
of the singlet equations in moment space beyond those known at LO,
due to the matrix structure of the equations.
Therefore, it is no surprise that there is no logarithmic ansatz that can't do better
than to reproduce the truncated solution, since only these ones are available
analytically in moment space.
If we knew the structure of the exact solution in moment space we could
construct an ansatz that would generate by recursion relations all the moments of that solution,
following the same strategy outlined for the non-singlet equation. Therefore,
inverting numerically the equations for the moments has no advantage whatsoever compared
to the numerical implementation of the logarithmic series using the algorithm that we have developed here.
However, we can arbitrarily improve the logarithmic series in order to capture higher
order contributions in the truncated solution, a feature that can be very appealing
for phenomenological purposes.

The proof that a suitable logarithmic ansatz reproduces the truncated
solution of the moments of the singlet pdf's at NLO goes as follows.
\footnote{Based on the paper~\cite{CCG}}

\subsection{The exact solution at LO}

We start from the singlet matrix equation
\begin{equation}
\label{NLOsinglet1}
\frac{\partial}{\partial\log Q^{2}}\left(\begin{array}{c}
q^{(+)}(x,Q^{2})\\
g(x,Q^{2})\end{array}\right)=\left(\begin{array}{cc}
P_{qq}(x,\alpha_{s}(Q^{2})) & P_{qg}(x,\alpha_{s}(Q^{2}))\\
P_{gq}(x,\alpha_{s}(Q^{2})) & P_{gg}(x,\alpha_{s}(Q^{2}))
\end{array}\right)\otimes\left(\begin{array}{c}
q^{(+)}(x,Q^{2})\\
g(x,Q^{2})\end{array}\right),
\end{equation}
whose well known LO solution in Mellin space can be easily identified
\ba
\vec{f}(N,\alpha_s)=\hat{L}(\alpha_s,\alpha_0,N)\vec{f}(N,\alpha_0),
\label{easily}
\ea
and where $\hat{L}(\alpha_s,\alpha_0,N)=
\left(\frac{\alpha_s}{\alpha_0}\right)^{\hat{R}_{0}(N)}$ is the evolution
operator.

Diagonalizing the $\hat{R}_{0}$ operator, in the equation above, we can write
the evolution operator $\hat{L}(\alpha_s,\alpha_0,N)$ as
\ba
\hat{L}(\alpha_s,\alpha_0,N)=
e_{+}\left(\frac{\alpha_s}{\alpha_0}\right)^{r_{+}}
+e_{-}\left(\frac{\alpha_s}{\alpha_0}\right)^{r_{-}},
\ea
where $r_{\pm}$ are the eigenvalues of the matrix
$\hat{R}_0=(-2/\beta_0) \hat{P}_0$ and $e_{+}$
and $e_{-}$ are projectors \cite{Petronzio1}, \cite{Petronzio2} defined as
\ba
e_{\pm}=\frac{1}{r_{\pm}-r_{\mp}}\left[\hat{R}_0-
r_{\mp}\hat{I}\right].
\ea
Since the $e_{\pm}$ are projection operators, the following properties
hold
\ba
e_+ e_+=e_+\hspace{1.5cm} e_- e_-=e_-\hspace{1.5cm} e_+ e_-=e_- e_+=0
\hspace{1.5cm} e_+ + e_-=1 \,.
\label{project}
\ea
Hence it is not difficult to see that
\ba
\hat{R}_{0}(N)=e_{+}r_{+}+e_{-}r_{-}\,.
\ea
It is important to note that one can write a solution of the singlet
DGLAP equation in a closed exponential form only at LO.

It is quite straightforward to reproduce this exact matrix solution at 
LO using a logarithmic expansion and the associated recursion relations. 
These are obtained from the ansatz (here written directly in moment space)
\ba
\label{RossiLO}
\vec{f}(N,\alpha_s)=\sum_{n=0}^{\infty}\frac{\vec{A}_{n}(N)}{n!}
\left[\ln{\left(\frac{\alpha_s}{\alpha_0}\right)}\right]^{n}\,,
\ea
subject to the initial condition
\ba
\vec{f}(N,\alpha_0)=\vec{A}_0(N).
\ea
Then the recursion relations become
\ba
\vec{A}_{n+1}(N)=-\frac{2}{\beta_0}\hat{P}^{(0)}(N)
\vec{A}_{n}(N)\equiv\hat{R}_{0}(N)\vec{A}_{n}(N),
\ea
and can be solved as
\ba
\vec{A}_{n}&=&\left[\hat{R}_{0}(N)\right]^n \vec{A}_{0}(N) \nonumber \\
&=&\left( e_+ r_+^n + e_- r_-^n\right) \vec{f}(N,\alpha_0),
\ea
having used eq.~(\ref{project}). Inserting this expression into eq.~(\ref{RossiLO}) we easily
obtain the relations
\ba
\vec{f}(N,\alpha_s)=\sum_{n=0}^{\infty}\frac{\left[\hat{R}_{0}(N)\right]^n}{n!}
\left[\ln{\left(\frac{\alpha_s}{\alpha_0}\right)}\right]^{n}\vec{A}_{0}(N)=
\left(\frac{\alpha_s}{\alpha_0}\right)^{\hat{R}_{0}(N)}\vec{f}(N,\alpha_0),
\ea
in agreement with eq.~(\ref{easily}).
\subsection{The standard NLO solution from moment space}
Moving to NLO, one can build a truncated
solution in moment space of eq.~(\ref{NLOsinglet1})
by a series expansion around the lowest order solution.

We start from the truncated version of the
vector equation (\ref{NLOsinglet1})
\ba
\frac{\partial{\vec{f}(N,\alpha_s)}}{\partial\alpha_s}=
-\frac{2}{\beta_0 \alpha_s}
\left[\hat{P}^{(0)}
+\frac{\alpha_s}{2\pi}\left(\hat{P}^{(1)}-
\frac{b_1}{2}\hat{P}^{(0)}\right)\right]\vec{f}(N,\alpha_s)\,,
\ea
that we re-express in the form
\ba
\label{NLOsinglet2}
\frac{\partial{\vec{f}(N,\alpha_s)}}{\partial\alpha_s} &=&
-\frac{1}{\alpha_s}\left[-\hat{R_0}+\alpha_s
\left(\frac{\hat{P}^{(1)}}{\pi\beta_0}+
\frac{\hat{R}_0 b_1}{4\pi}\right)\right]\vec{f}(N,\alpha_s) \nonumber \\
&=&
\frac{1}{\alpha_s}\left[\hat{R_0}+\alpha_s \hat{R}_1\right]
\vec{f}(N,\alpha_s),\, \nonumber\\
\ea
and with the $\hat{R_1}$ operator defined as
\ba
\hat{R_1}(N)=-\frac{1}{\pi}\left(\frac{b_1}{4}\hat{R}_0(N)+
\frac{\hat{P}^{(1)}(N)}{\beta_0}\right).
\ea
We use \cite{Petronzio1,Petronzio2} a truncated vector
solution of (\ref{NLOsinglet2}) - accurate at $O(\alpha_s)$ - of the form
\ba
\label{NLOans_sing1}
\vec{f}(N,\alpha_s)&=&\hat{U}(\alpha_s,N)\,\hat{L}(\alpha_s,\alpha_0,N)\,
\hat{U}^{-1}(\alpha_0,N)\,\vec{f}(N,\alpha_0)\nonumber\\
&=&\left[1+\alpha_s
\hat{U}_{1}(N)\right]
\hat{L}(\alpha_s,\alpha_0,N)\left[1+\alpha_0
\hat{U}_{1}(N)\right]^{-1}\vec{f}(N,\alpha_0)\,,\nonumber\\
\ea
where we have expanded in powers of $\alpha_s$ the operators
$\hat{U}(\alpha_s,N)$ and $\hat{U}^{-1}(\alpha_0,N)$.
Inserting (\ref{NLOans_sing1}) in eq.~(\ref{NLOsinglet2}),
we obtain the commutation relations involving the operators $\hat{U_1}$, $\hat{R}_0$ and $\hat{R}_1$
\ba
\label{com1}
\left[\hat{R}_0,\hat{U}_1\right]=\hat{U}_1-\hat{R}_1,
\ea
which appear in the solution in Mellin space \cite{Petronzio1,Petronzio2}.
Then, using the properties of the projection operators
\ba
\hat{U}_1=e_+\hat{U}_1 e_+ + e_+\hat{U}_1 e_- + e_-\hat{U}_1 e_+
+e_-\hat{U}_1 e_- = \hat{U}_1^{++} +\hat{U}_1^{-+}+\hat{U}_1^{+-}+\hat{U}_1^{--}\,,
\ea
and inserting this relation in the commutator (\ref{com1})
we easily derive the relation
\ba
\hat{U}_1=\left[e_+\hat{R}_1 e_+ + e_-\hat{R}_1 e_-\right]-
\frac{e_+\hat{R}_1 e_-}{r_+ -r_- -1}-\frac{e_-\hat{R}_1 e_+}{r_- -r_+ -1}.
\ea
Finally, expanding the term $\left[1+\alpha_0 \hat{U}_1\right]^{-1}$ in eq.~(\ref{NLOans_sing1}) 
we arrive at the solution \cite{Petronzio1,Petronzio2}
\ba
\label{NLOtrsolsin}
\vec{f}(N,\alpha_s)=\left[\hat{L}+\alpha_s\hat{U}_1\hat{L}
-\alpha_0\hat{L}\hat{U}_1\right]\vec{f}(N,\alpha_0)\,,
\ea
where the $(\alpha_s,\alpha_0,N)$ dependence has been dropped.
Such solution can be put in a more readable form as
\ba
\label{solution}
&&\vec{f}(N,\alpha_s)=\left\{
\left(\frac{\alpha_s}{\alpha_0}\right)^{r_+}\left[e_{+}
+(\alpha_s-\alpha_0)\,e_{+}\hat{R_1}e_{+}+\right.\right.\nonumber\\
&&\hspace{2cm}\left.\left.\left(\alpha_0-\alpha_s
\left(\frac{\alpha_s}{\alpha_0}\right)^{r_- -r_+}\right)
\frac{e_+\hat{R_1}e_{-}}{r_+ -r_- -1}\right]
+(+\leftrightarrow -)\right\}\vec{f}(N,\alpha_0)\,,
\ea
which can be called the {\em standard} NLO solution, having been introduced in the literature
about 20 years ago \cite{Petronzio2}. It is obvious that this solution is a
(first) truncated solution of the NLO singlet DGLAP equation, with the equation truncated
at the same order.

\subsection{Reobtaining the standard NLO solution using the logarithmic expansion}

Having worked out the well-known NLO singlet solution in moment space, our aim is
to show that the
same solution can be reconstructed using a logarithmic ansatz. This fills a gap in the
previous literature on this types of ansatze for the QCD pdf's.
To facilitate our duty, we stress once more that the type of recursion relations
obtained in the non-singlet and singlet cases are similar.
In fact the matrix structure of the equations doesn't play any role
in the derivation due to the linearity of the ansatz in the (vector) coefficient
functions that appear in it.

Our NLO singlet ansatz has the form
\ba
\label{ansatzevec}
{\vec{f}(x,Q^2)}^{NLO}=\sum_{n=0}^{\infty}\frac{\vec{A}_{n}(x)}{n!}
\left[\ln{\left(\frac{\alpha_s}{\alpha_0}\right)}\right]^{n}
+\alpha_s\sum_{n=0}^{\infty}\frac{\vec{B}_{n}(x)}{n!}
\left[\ln{\left(\frac{\alpha_s}{\alpha_0}\right)}\right]^{n}\,,
\ea

with $A_n$ and $B_n$ now being vectors involving the
singlet components. The recursion relations are
\ba
\label{vecrec}
&&\vec{A}_{n+1}(N)=\hat{R}_{0}(N)\vec{A}_{n}(N),\nonumber\\
&&\vec{B}_{n+1}(N)=-\vec{B}_{n}(N)-\frac{b_1}{4\pi}\vec{A}_{n+1}(N)
+\hat{R}_{0}(N)\vec{B}_{n}(N)-
\frac{1}{\pi\beta_0}\hat{P}^{(1)}(N)\vec{A}_{n}(N),
\ea
subjected to the initial condition
\ba
\vec{f}(N,\alpha_0)=\vec{A}_0(N)+\alpha_0 \vec{B}_0(N).\,
\ea
The solution of (\ref{vecrec}) in moment space can be easily found and is given by
\ba
\label{vecrec1}
&&\vec{A}_{n}(N)=\left[e_+(r_+)^n +e_-(r_-)^n\right]\vec{A}_{0}(N).
\ea
It has to be pointed out that the ${\vec{B}}_{n}(N)$ vectors are bidimensional column vectors and they
can be decomposed in $\mathbf{R}^{2}$ orthonormal basis $\left\{{\bf e}_{1},{\bf e}_{2}\right\}$ as
\ba
{\vec{B}}_{n}(N)={\bf e}_{1} B_{n}^{(1)}(N)+{\bf e}_{2} B_{n}^{(2)}(N)=
{\vec{B}}_{n}^{+}(N)+{\vec{B}}_{n}^{-}(N)\,.
\ea
Also, we can re-arrange the $\vec{B}_{n+1}(N)$ relation
into the form
\ba
&&\vec{B}_{n+1}(N)=\left(\hat{R}_0-1\right)\vec{B}_{n}(N)+
\hat{R}_1\hat{R}_0^n\vec{A}_{0}(N).
\label{brel1}
\ea
The last step to follow in order
to construct the truncated solution involves a projection of the recursion relations
(\ref{brel1}) in the basis of the projectors $\hat{e}_\pm$, and in the basis
$\left\{{\bf e}_1,{\bf e}_2\right\}$.
To do this, we separate the equations as
\ba
&&\vec{B}_{n+1}=\left(e_+ r_+ +e_- r_- -1\right)\left[
\vec{B}_{n}^{++}+\vec{B}_{n}^{+-}+\vec{B}_{n}^{-+}+
\vec{B}_{n}^{--}\right]+\nonumber\\
&&\hspace{1.5cm}\left[\hat{R}_1^{++}+
\hat{R}_1^{+-}+\hat{R}_1^{-+}+\hat{R}_1^{--}\right]
\left(e_+ r_+^n +e_- r_-^n\right)\vec{A}_0,
\ea
where we have used the notation (not to be confused with the previous one)
\ba
&&e_{+}B^{(1)}_{n}(N){\bf e}_{1}=\vec{B}_{n}^{++}\nonumber\\
&&e_{+}B^{(2)}_{n}(N){\bf e}_{2}=\vec{B}_{n}^{+-}\nonumber\\
&&e_{-}B^{(1)}_{n}(N){\bf e}_{1}=\vec{B}_{n}^{-+}\nonumber\\
&&e_{-}B^{(2)}_{n}(N){\bf e}_{2}=\vec{B}_{n}^{--}\,.
\ea
Then, we can split the relation (\ref{brel1}) in four recursion relations
\ba
&&\vec{B}_{n+1}^{++}=\left(r_+ -1\right)\vec{B}_{n}^{++}+
\hat{R}_1^{++}r_+^n\vec{A}_0,\nonumber\\
&&\vec{B}_{n+1}^{+-}=\left(r_+ -1\right)\vec{B}_{n}^{+-}+
\hat{R}_1^{+-}r_-^n\vec{A}_0,\nonumber\\
&&\vec{B}_{n+1}^{-+}=\left(r_- -1\right)\vec{B}_{n}^{-+}+
\hat{R}_1^{-+}r_+^n\vec{A}_0,\nonumber\\
&&\vec{B}_{n+1}^{--}=\left(r_- -1\right)\vec{B}_{n}^{--}+
\hat{R}_1^{--}r_-^n\vec{A}_0\,.
\ea
Finally, it is an easy task to verify that the solutions of the recursion relations
at NLO are given by
\ba
\label{Bproj}
&&\vec{B}_{n}^{++}=\left[r_+^n -(r_+ -1)^n \right]\hat{R}_1^{++}
\vec{A}_0,\nonumber\\
&&\vec{B}_{n}^{--}=\left[r_-^n -(r_- -1)^n \right]\hat{R}_1^{--}
\vec{A}_0,\nonumber\\
&&\vec{B}_{n}^{+-}=\left[-r_-^n +(r_+ -1)^n \right]
\frac{\hat{R}_1^{+-}}{r_+ -r_- -1}\vec{A}_0,\nonumber\\
&&\vec{B}_{n}^{-+}=\left[-r_+^n +(r_- -1)^n \right]
\frac{\hat{R}_1^{-+}}{r_- -r_+ -1}\vec{A}_0,\,\nonumber\\
\ea
where we have expressed the $n_{th}$ iterate in terms of the initial conditions,
and we have taken $\vec{B}_0=\vec{0}$.
Summing over all the projections, we arrive at the following expression for the
NLO truncated solution of the singlet parton distributions
\ba
\vec{f}(N,\alpha_s)=\sum_{n=0}^{\infty}\frac{L^n}{n!}
\left[\vec{A}_{n}+\alpha_s
\left(\vec{B}_{n}^{++}+\vec{B}_{n}^{--}
+\vec{B}_{n}^{-+}+\vec{B}_{n}^{+-}\right)\right]\,,
\ea
which can be easily exponentiated to give
\ba
&&\vec{f}(N,\alpha_s)=e_{+}\vec{A}_0 \left(\frac{\alpha_s}{\alpha_0}\right)^{r_+}+
e_{-}\vec{A}_0 \left(\frac{\alpha_s}{\alpha_0}\right)^{r_-}+\nonumber\\
&&\hspace{1.5cm}\alpha_s\left\{
e_{+}\hat{R}_1e_{+}
\left(\frac{\alpha_s}{\alpha_0}\right)^{r_+}-
e_{+}\hat{R}_1e_{+}
\left(\frac{\alpha_s}{\alpha_0}\right)^{(r_+ -1)}+\right.\nonumber\\
&&\hspace{2cm}\left.
e_{-}\hat{R}_1e_{-}
\left(\frac{\alpha_s}{\alpha_0}\right)^{r_-}-
e_{-}\hat{R}_1e_{-}
\left(\frac{\alpha_s}{\alpha_0}\right)^{(r_- -1)}+\right.\nonumber\\
&&\hspace{2cm}\left.\frac{1}{(r_+ -r_- -1)}\left[
-e_{+}\hat{R}_1e_{-}
\left(\frac{\alpha_s}{\alpha_0}\right)^{r_-}+
e_{+}\hat{R}_1e_{-}
\left(\frac{\alpha_s}{\alpha_0}\right)^{(r_+ -1)}\right]+\right.\nonumber\\
&&\hspace{2cm}\left.\frac{1}{(r_- -r_+ -1)}\left[
-e_{-}\hat{R}_1e_{+}
\left(\frac{\alpha_s}{\alpha_0}\right)^{r_+}+
e_{-}\hat{R}_1e_{+}
\left(\frac{\alpha_s}{\alpha_0}\right)^{(r_- -1)}\right]\right\}\vec{A}_0\,.
\nonumber\\
\ea
Finally, organizing the various pieces we obtain exactly the solution in
eq.~(\ref{solution}). We have therefore shown that the logarithmic ansatz coincides with the 
solution of the singlet DGLAP equation at NLO known from the previous literature 
and reported in the previous section. It is intuitively obvious that we can 
build with this approach truncated solutions of higher orders improving on the standard 
solution (\ref{solution}) known from moment space, and we can do this with any accuracy. 
However, before discussing this point in one of the following sections, 
we want to show how the same strategy works at NNLO.

\subsection{Truncated Solution at NNLO}

At this point, to complete our investigation, we need to discuss the generalization
of the procedure illustrated above to the
NNLO case.
As usual, we start from a truncated version of eq.~(\ref{NLOsinglet1}), that at
NNLO can be written as
\ba
\frac{\partial{\vec{f}(N,\alpha_s)}}{\partial\alpha_s}=
\frac{1}{\alpha_s}\left[\hat{R_0}+\alpha_s \hat{R}_1 +\alpha_s^2
\hat{R}_2\right]\vec{f}(N,\alpha_s),\,\nonumber\\
\label{NNLOsinglet2}
\ea
where
\ba
\hat{R}_2=-\frac{1}{\pi}\left(\frac{\hat{P^{(2)}}}{2\pi\beta_0}
+\frac{\hat{R}_1 b_1}{4}+\frac{\hat{R}_0 b_2}{16\pi}\right),
\ea
whose solution is expected to be of the form \cite{ellis}
\ba
\label{NNLOans_sing1}
\vec{f}(N,\alpha_s)=\left[1+\alpha_s\hat{U}_{1}(N)+\alpha_s^2
\hat{U}_{2}(N)\right]\hat{L}(\alpha_s,\alpha_0,N)
\left[1+\alpha_0\hat{U}_{1}(N)+\alpha_0^2\hat{U}_{2}(N)\right]^{-1}
\vec{f}(N,\alpha_0),\,\nonumber\\
\ea
where
\ba
&&\left[\hat{R}_0,\hat{U}_1\right]=\hat{U}_1-\hat{R}_1,\nonumber\\
&&\left[\hat{R}_0,\hat{U}_2\right]=-\hat{R}_2 -\hat{R}_1\hat{U}_1 +2\hat{U}_2.\,
\ea
Using the projectors $e_{+},e_{-}$ in the $\pm$ subspaces, one can remove the commutators, obtaining
 \ba
&&\hat{U}_2^{++}=\frac{1}{2}\left[\hat{R}_1^{++}\hat{R}_1^{++}
+\hat{R}_2^{++}-\frac{\hat{R}_1^{+-}
\hat{R}_1^{-+}}{r_- -r_+ -1}\right],\nonumber\\
&&\hat{U}_2^{--}=\frac{1}{2}\left[\hat{R}_1^{--}\hat{R}_1^{--}
+\hat{R}_2^{--}-\frac{\hat{R}_1^{-+}
\hat{R}_1^{+-}}{r_+ -r_- -1}\right],\nonumber\\
&&\hat{U}_2^{+-}=\frac{1}{r_+ -r_- -2}\left[-\hat{R}_1^{+-}\hat{R}_1^{--}
-\hat{R}_2^{+-}+\frac{\hat{R}_1^{++}\hat{R}_1^{+-}}{r_+ -r_- -1}
\right],\nonumber\\
&&\hat{U}_2^{-+}=\frac{1}{r_- -r_+ -2}\left[-\hat{R}_1^{-+}\hat{R}_1^{++}
-\hat{R}_2^{-+}+\frac{\hat{R}_1^{--}\hat{R}_1^{-+}}{r_- -r_+ -1}
\right],\,
\ea
and the formal solution from Mellin space can be simplified to
\ba
\label{NNLOtrsolsing}
&&\vec{f}(N,\alpha_s)=\left[\hat{L}+\alpha_s\hat{U}_1\hat{L}
-\alpha_0\hat{L}\hat{U}_1\right.\nonumber\\
&&\hspace{2cm}\left.+\alpha_s^2 \hat{U}_2\hat{L}
-\alpha_s\alpha_0\hat{U}_1\hat{L}\hat{U}_1
+\alpha_0^2\hat{L}\left(\hat{U}_1^2-\hat{U}_2\right)
\right]\vec{f}(N,\alpha_0)\,.
\ea
At this point we introduce our ($1$-$st$ truncated) logarithmic ansatz that is expected to reproduce
(\ref{NNLOtrsolsing}). Now it includes also an infinite set of new coefficients $\vec{C_n}$,
similar to the non-singlet NNLO case
\ba
\vec{f}(N,\alpha_s)=\sum_{n=0}^{\infty}\frac{L^n}{n!}\left[\vec{A_n} +\alpha_s\vec{B_n}+
\alpha_s^2\vec{C_n}\right].
\label{altroans}
\ea
Inserting the NNLO logarithmic ansatz into (\ref{NNLOsinglet2}), we obtain in moment space the
recursion relations
\ba
\label{vecrecNNLO}
&&\vec{A}_{n+1}=\hat{R}_{0}\vec{A}_{n},\nonumber\\
&&\vec{B}_{n+1}=\left(\hat{R}_0-1\right)\vec{B}_{n}+
\hat{R}_1\hat{R}_0^n\vec{A}_{0},\nonumber\\
&&\vec{C}_{n+1}=\left(\hat{R}_0-2\right)\vec{C}_{n}
-\frac{b_{1}}{4\pi}\left(\vec{B}_{n}+\vec{B}_{n+1}\right)
+\left[\frac{b_{1}}{4\pi}\hat{R}_{0}+\hat{R}_{1}\right]\vec{B}_{n},
\nonumber \\
&&\hspace{1cm}+\left[\hat{R}_2+\frac{b_1}{4\pi}\hat{R}_1\right]\hat{R}_0^n \vec{A}_0,
\ea
whose solution has to coincide with (\ref{NNLOtrsolsing}). Also in this case, as before,
we use the $e_{\pm}$ projectors and notice that the structure of the recursion relations for the
coefficients $A_n$ and $B_n$ remain the same as in NLO.
Therefore, the solutions of the recursion relations for $\vec{A}_n$ and $\vec{B}_n$ are still given by
(\ref{vecrec1}) and (\ref{Bproj}). 
We then have to find only an explicit solution of the relations for the new coefficients $\vec{C}_{n+1}(N)$.

These relations can be solved in terms of $\vec{A}_0$, $\vec{B}_0$ and
$\vec{C}_0$ with the help of (\ref{Bproj}).
Finally, taking $\vec{B}_0 =0$ and $\vec{C}_0 =0$ after a lengthy computation we obtain the explicit solutions
for the projected components
\ba
&&\vec{C}_{n}^{++}=-\frac{1}{2}\frac{\hat{R}_1^{+-}\hat{R}_1^{-+}}
{(r_+ -r_- -1)(r_- -r_+ -1)}\times\nonumber\\
&&\hspace{1cm}\left[2(r_- -1)^n -(r_+ -2)^n
-(r_+ -2)^n r_+ -r_+^n +r_+^{n+1}
+ r_- \left((r_+ -2)^n -r_+^n \right)\right]\vec{A}_0
\nonumber\\
&&\hspace{1cm}+\frac{1}{2}\hat{R}_1^{++}\hat{R}_1^{++}\left[
r_+^n -2(r_+ -1)^n +(r_+ -2)^n\right]\vec{A}_0
\nonumber\\
&&\hspace{1cm}+\frac{1}{2}\hat{R}_2^{++}\left[r_+^n
-(r_+ -2)^n\right]\vec{A}_0,
\nonumber\\
%%%%%%%%%%%%%%%%%%%%%%%%
&&\vec{C}_{n}^{--}=-\frac{1}{2}\frac{\hat{R}_1^{-+}\hat{R}_1^{+-}}
{(r_+ -r_- -1)(r_- -r_+ -1)}\times\nonumber\\
&&\hspace{1cm}\left[2(r_+ -1)^n -(r_- -2)^n
-(r_- -2)^n r_- -r_+^n +r_+^{n+1}
+ r_+ \left((r_- -2)^n -r_-^n \right)\right]\vec{A}_0
\nonumber\\
&&\hspace{1cm}+\frac{1}{2}\hat{R}_1^{--}\hat{R}_1^{--}\left[
r_-^n -2(r_- -1)^n +(r_- -2)^n\right]\vec{A}_0
\nonumber\\
&&\hspace{1cm}+\frac{1}{2}\hat{R}_2^{--}\left[r_-^n
-(r_- -2)^n\right]\vec{A}_0,
\nonumber\\
%%%%%%%%%%%%%%%%%%%%
&&\vec{C}_{n}^{+-}=\frac{1}{2+r_-^2 +r_-(3-2r_+)-3r_+ +r_+^2}
\left[\hat{R}_2^{+-}\left(r_-^n -(r_+-2)^n\right)\left(1+ r_- -r_+\right)
\right.\nonumber\\
&&\hspace{1cm}\left.-\hat{R}_1^{+-}\hat{R}_1^{--}\left(2(r_- -1)^n
+(r_--1)^n r_- -r_-^{n+1}\right.\right.\nonumber\\
&&\hspace{1cm}\left.\left.-(r_+ -2)^n +r_-^n(r_+ -1) -(r_- -1)^n r_+\right)
\right.\nonumber\\
&&\hspace{1cm}\left.+\hat{R}_1^{++}\hat{R}_1^{+-}
\left(r_-^n+(r_+ -2)^n +r_-(r_+ -2)^n
\right.\right.\nonumber\\
&&\hspace{1cm}\left.\left.-2(r_+ -1)^n -r_-(r_+ -1)^n
-r_+(r_+ -2)^n + r_+(r_+ -1)^n \right)
\right]\vec{A}_0,
\nonumber\\
%%%%%%%%%%%%%%%%%%%%%%%%%
&&\vec{C}_{n}^{-+}=\frac{1}{2+r_-^2+3r_+ +r_+^2 -r_-(3+2r_+)}
\left[\hat{R}_2^{-+}(r_- -r_+ -1)\left((r_- -2)^n -r_+^n\right)
\right.\nonumber\\
&&\hspace{1cm}\left.-\hat{R}_1^{--}\hat{R}_1^{-+}\left(
-(r_- -2)^n +2(r_- -1)^n +((r_- -2)^n -(r_- -1)^n)r_-
\right.\right.\nonumber\\
&&\hspace{1cm}\left.\left.-(r_- -2)^n r_+ +(r_- -1)^n r_+
-r_+^n\right)
\right.\nonumber\\
&&\hspace{1cm}\left.+\hat{R}_1^{-+}\hat{R}_1^{++}\left(
(r_- -2)^n -2(r_+ -1)^n +r_-(r_+ -1)^n\right.\right.\nonumber\\
&&\hspace{1cm}\left.\left.-(r_+ -1)^n r_+
+r_+^n -r_- r_+^n +r_+^{n+1}\right)
\right]\vec{A}_0.
\nonumber\\
\ea
Re-inserting these solutions into the NNLO ansatz (\ref{altroans}) and after exponentiation
one can show explicitely that the logarithmic solution so obtained coincides
with (\ref{NNLOtrsolsing}). Details can be found in appendix A.
In the practical implementations of these solutions, there are two obvious strategies
that can be followed. One consists in the implementation of the recursion relations
as we have done in various cases above: a sufficiently large number of iterates will converge
to the truncated solution (\ref{NNLOtrsolsing}). This is obtained by implementing eqs.~(\ref{vecrecNNLO})
and incorporating them into (\ref{altroans}). A second method consists in the direct computation of
(\ref{NNLOtrsolsing}) in $x$-space which becomes
\ba
\label{NNLOtrsolsingx}
&&\vec{f}(x,\alpha_s)=\left[\hat{L}+\alpha_s\hat{U}_1\otimes\hat{L}
-\alpha_0\hat{L}\otimes\hat{U}_1\right.\nonumber\\
&&\hspace{2cm}\left.+\alpha_s^2 \hat{U}_2\otimes\hat{L}
-\alpha_s\alpha_0\hat{U}_1\otimes\hat{L}\otimes \hat{U}_1
+\alpha_0^2\hat{L}\otimes \left(\hat{U}_1\otimes \hat{U}_1-\hat{U}_2\right)
\right]\otimes \vec{f}(x,\alpha_0),\nonumber \\
\ea
and with $\hat{L}$ now replaced by its operatorial $(\otimes)$ form 
\beq
\hat{L}\to e^{L^n R_0 \otimes}= \left(\sum_{n=0}^{\infty} \frac{R_0^n}{n!} L^n\right)_\otimes.
\eeq
An implementation of (\ref{NNLOtrsolsingx}) would reduce 
its numerical evaluation to that of a sequence of LO solutions built around ``artificial''
initial conditions given by
$\hat{U}_1\otimes f(x,\alpha_0)$, $\hat{U}_1\otimes \hat{U}_1\otimes f(x,\alpha_0)$ and so on.

\section{Higher order logarithmic approximation of the NNLO singlet solution}

The procedure studied in the previous section can be
generalized and applied to obtain solutions that retain higher order
logarithmic contributions in the NLO/NNLO singlet cases.
The same procedure is also the one that
has been implemented in all the existing codes for the singlet:
one has to truncate the equation and then try to reach the exact solution by a sufficiently
high number of iterates.
On the other hand, $x$-space (non brute force) implementations are, from this respect,
still lagging since they are only based
on the Rossi-Storrow formulation \cite{cafacor,gordon}, which we have analized thoroughly
and largely extended in this section.

Therefore, the only way at our disposal to reach from $x$-space the exact solution
is by using higher order truncates. This is of practical relevance since our algorithm
allows to perform separate
checks between truncated solutions of arbitrary high orders built either from
Mellin or from $x$-space
\footnote{ The current benchmarks available at NNLO are limited to exact solutions and
do not involve comparisons
between truncated solutions}. We recall, if not obvious, that in the analysis
of hadronic processes the two criteria of using
either truncated or exact solutions are both acceptable.

To summarize: since in the singlet case is not possible
to write down a solution of the DGLAP equation in a
closed exponential form because of the non-commutativity
of the operators $\hat{R}_i$, the best thing we can do is to arrange
the singlet DGLAP equation in the truncated form,
as in eqs.~(\ref{NLOsinglet2}) and (\ref{NNLOsinglet2}).
The truncated vector solutions
(\ref{NNLOtrsolsing}) and (\ref{NLOtrsolsin}) are equivalent to those obtained using the
vector recursion relations at NLO/NNLO.

Now, working at NNLO, we will show how the basic NNLO solution
can be improved and the higher truncates identified.
Clearly, it is important to show explicitely that these truncates,
generated after solving the recursion relations,
can be rewritten exactly in the form previously known from Mellin space.
We are going to show here that this
is in fact the case, although some of the explicit expressions for the higher order
coefficient functions $C_n, D_n'$s  will
be given explicitely only in part. The expressions are in fact slightly lengthy
\footnote{They will be included in
a file that will be made available in the same distribution of our code, which is in preparation.}.
Therefore, here we will just outline the procedure and
illustrate the proof up to the second truncate of the NNLO singlet only for the sake
of clarity.

The exact singlet NNLO equation in Mellin space is given by
\beqa
\frac{\partial\vec{f}(N,\alpha_s)}{\partial\alpha_s} &=&
-\frac{\left(\frac{\alpha_{s}}{2\pi}\right)\hat{P}^{(0)}(N)+
\left(\frac{\alpha_{s}}{2\pi}\right)^{2}\hat{P}^{(1)}(N)+
\left(\frac{\alpha_{s}}{2\pi}\right)^{3}
\hat{P}^{(2)}(N)}{\frac{\beta_{0}}{4\pi}
\alpha_{s}^{2}+\frac{\beta_{1}}{16\pi^{2}}\alpha_{s}^{3}+
\frac{\beta_{2}}{64\pi^{3}}\alpha_{s}^{4}}\vec{f}(N,\alpha_{s}) \nonumber \\
&=&
\frac{\hat{P}^{NNLO}(N,\alpha_s)}{\beta^{NNLO}(\alpha_s)}\vec{f}(N,\alpha_{s}),
\nonumber\\
\ea
where we have introduced the singlet kernels.
After a Taylor expansion of $\hat{P}^{NNLO}(N,\alpha_s)/\beta^{NNLO}(\alpha_s)$
up to $\alpha_s^3$ it becomes
\ba
\frac{\partial\vec{f}(N,\alpha_s)}{\partial\alpha_s}=
\frac{1}{\alpha_s}\left[\hat{R}_0-\frac{b_2}{(4\pi)^2}\hat{R}_1
+\alpha_s\hat{R}_1+\alpha_s^2\hat{R}_2-\frac{b_1}{4\pi}\alpha_s^3\hat{R}_2
\right]\vec{f}(N,\alpha_{s}),\,
\ea
which is the truncated equation of order $\alpha_s^3$. The $\hat{R}_i$ $(i=0,1,2)$ operators are listed below
\ba
&&\hat{R}_0=-\frac{2}{\beta_0}\hat{P}^{(0)},
\nonumber\\
&&\hat{R}_1=-\frac{\hat{P}^{(1)}}{\pi\beta_0}+\frac{b_1}{2\pi\beta_0}\hat{P}^{(0)},
\nonumber\\
&&\hat{R}_2=-\frac{\hat{P}^{(2)}}{2\pi^2\beta_0}+\frac{b_1}{4\pi^2\beta_0}\hat{P}^{(1)}
+\left[-\frac{b_1^2}{8\pi^2\beta_0}+\frac{b_2}{8\pi^2\beta_0}\right]\hat{P}^{(0)}.
\ea
The formal solution of this equation can be written as \cite{Buras},\cite{kosower}
\ba
\vec{f}(N,\alpha_s)&=&T_{\alpha}\left[\exp\left\{\int_{\alpha_0}^{\alpha_s} d\,\alpha_s'\,\frac{1}{\alpha_s'}\left(\hat{R}_0
-\frac{b_2}{(4\pi)^2}\hat{R}_1+\alpha_s'\hat{R}_1+{\alpha_s'}^2\hat{R}_2
-\frac{b_1}{4\pi}{\alpha_s'}^3\hat{R}_2\right)\right\}\right]\vec{f}(N,\alpha_0)\nonumber\\
&=&\hat{U}(N,\alpha_s)\hat{L}(\alpha_s,\alpha_0)
\hat{U}^{-1}(N,\alpha_0)\vec{f}(N,\alpha_0),
\ea
where the $T_{\alpha}$ operator acts on the exponential similarly to a
time-ordered product, but this time in the space of the
couplings. Again, expanding the $\hat{U}(N,\alpha_s)$ and
$\hat{U}^{-1}(N,\alpha_0)$ operators in the formal solution
around $\alpha_s=0$ and $\alpha_0=0$, we have
\ba
\label{NNLOhigher1}
&&\vec{f}(N,\alpha_s)=\left[\hat{L}+\alpha_s\hat{U}_1\hat{L}
-\alpha_0\hat{L}\hat{U}_1+\alpha_s^2\hat{U}_2\hat{L}
-\alpha_s\alpha_0\hat{U}_1\hat{L}\hat{U}_1
+\alpha_0^2\hat{L}\left(\hat{U}_1^2-\hat{U}_2\right)\right.
\nonumber\\
&&\hspace{2cm}\left.+\alpha_s^3\hat{U}_3\hat{L}
+\alpha_s\alpha_0^2\hat{U}_1\hat{L}\left(\hat{U}_1^2-\hat{U}_2\right)
-\alpha_s^2\alpha_0\hat{U}_2\hat{L}\hat{U}_1
\right.\nonumber\\
&&\hspace{2cm}\left.-\alpha_0^3\hat{L}\left(\hat{U}_1^3
-\hat{U}_1\hat{U}_2-\hat{U}_2\hat{U}_1+\hat{U}_3\right)
\right]\vec{f}(N,\alpha_0).
\ea
Inserting the expanded solution into eq.~(\ref{NNLOhigher1}) and equating the various power
of $\alpha_s$ we arrive at the following chain of commutation relations
\ba
&&\left[\hat{R}_0,\hat{U}_1\right]=\hat{U}_1-\hat{R}_1,
\nonumber\\
&&\left[\hat{R}_0,\hat{U}_2\right]=-\hat{R}_2-\hat{R}_1\hat{U}_1+2\hat{U}_2,
\nonumber\\
&&\left[\hat{R}_0,\hat{U}_3\right]=\frac{b_2}{(4\pi)^2}\hat{R}_1
+\frac{b_1}{(4\pi)}\hat{R}_2-\hat{R}_1\hat{U}_2-\hat{R}_2\hat{U}_1
+3\hat{U}_3.
\ea
Removing the commutators by the $e_\pm$ projectors one obtains
\ba
&&\hat{U}_1^{++}=\hat{R}_1^{++},\nonumber\\
&&\hat{U}_1^{--}=\hat{R}_1^{--},\nonumber\\
&&\hat{U}_1^{+-}=-\frac{\hat{R}_1^{+-}}{r_+ -r_- -1},\nonumber\\
&&\hat{U}_1^{-+}=-\frac{\hat{R}_1^{-+}}{r_- -r_+ -1},\nonumber\\
&&\hat{U}_2^{++}=\frac{1}{2}\left[\hat{R}_1^{++}\hat{R}_1^{++}
+\hat{R}_2^{++}-\frac{\hat{R}_1^{+-}
\hat{R}_1^{-+}}{r_- -r_+ -1}\right],\nonumber\\
&&\hat{U}_2^{--}=\frac{1}{2}\left[\hat{R}_1^{--}\hat{R}_1^{--}
+\hat{R}_2^{--}-\frac{\hat{R}_1^{-+}
\hat{R}_1^{+-}}{r_+ -r_- -1}\right],\nonumber\\
&&\hat{U}_2^{+-}=\frac{1}{r_+ -r_- -2}\left[-\hat{R}_1^{+-}\hat{R}_1^{--}
-\hat{R}_2^{+-}+\frac{\hat{R}_1^{++}\hat{R}_1^{+-}}{r_+ -r_- -1}
\right],\nonumber\\
&&\hat{U}_2^{-+}=\frac{1}{r_- -r_+ -2}\left[-\hat{R}_1^{-+}\hat{R}_1^{++}
-\hat{R}_2^{-+}+\frac{\hat{R}_1^{--}\hat{R}_1^{-+}}{r_- -r_+ -1}
\right],\nonumber\\
&&\hat{U}_3^{++}=\frac{1}{3}\left[-\frac{b_1}{(4\pi)}\hat{R}_1^{++}
-\frac{b_2}{(4\pi)^2}\hat{R}_2^{++}+\hat{R}_1^{+-}\hat{U}_2^{-+}
+\hat{R}_1^{++}\hat{U}_2^{++}+\hat{R}_2^{+-}\hat{U}_1^{-+}
+\hat{R}_2^{++}\hat{U}_1^{++}\right],\nonumber\\
&&\hat{U}_3^{--}=\frac{1}{3}\left[-\frac{b_1}{(4\pi)}\hat{R}_1^{--}
-\frac{b_2}{(4\pi)^2}\hat{R}_2^{--}+\hat{R}_1^{-+}\hat{U}_2^{+-}
+\hat{R}_1^{--}\hat{U}_2^{--}+\hat{R}_2^{-+}\hat{U}_1^{+-}
+\hat{R}_2^{--}\hat{U}_1^{--}\right],
\nonumber\\
&&\hat{U}_3^{+-}=\frac{1}{r_+ -r_- -3}\left[-\frac{b_1}{(4\pi)}\hat{R}_1^{+-}
-\frac{b_2}{(4\pi)^2}\hat{R}_2^{+-}-\hat{R}_1^{+-}\hat{U}_2^{--}
-\hat{R}_1^{++}\hat{U}_2^{+-}-\hat{R}_2^{+-}\hat{U}_1^{--}
-\hat{R}_2^{++}\hat{U}_1^{+-}\right],
\nonumber\\
&&\hat{U}_3^{-+}=\frac{1}{r_- -r_+ -3}\left[-\frac{b_1}{(4\pi)}\hat{R}_1^{-+}
-\frac{b_2}{(4\pi)^2}\hat{R}_2^{-+}-\hat{R}_1^{-+}\hat{U}_2^{++}
-\hat{R}_1^{--}\hat{U}_2^{-+}-\hat{R}_2^{-+}\hat{U}_1^{++}
-\hat{R}_2^{--}\hat{U}_1^{-+}\right].\,
\nonumber\\
\ea
As one can see, the $\hat{U}_{3}$ operator is expressed in terms 
of the kernels $P^{(0)},\hat{P}^{(1)}$ and $\hat{P}^{(2)}$.
One can prove that by imposing a higher order ansatz in Mellin space
of the form
\ba
\vec{\tilde{f}}(N,\alpha_s)=\sum_{n=0}^{\infty}\frac{L^n}{n!}
\left[\vec{A}_n(N)+\alpha_s\vec{B}_n(N)+\alpha_s^2\vec{C}_n(N)
+\alpha_s^3 \vec{D}_n(N)\right]\,,
\ea
 the solution (\ref{NNLOhigher1}) is generated. 
The vector recursion relations in this case become 
\ba
&&\vec{A}_{n+1}=-\frac{2}{\beta_0}\hat{P}^{(0)}\vec{A}_{n},\nonumber\\
&&\vec{B}_{n+1}=-\vec{B}_{n}-\frac{1}{\pi\beta_0}\hat{P}^{(1)}\vec{A}_{n}-
\frac{\beta_1}{\pi\beta_0}\vec{A}_{n+1}-\frac{2}{\beta_0}\hat{P}^{(0)}\vec{B}_{n},
\nonumber\\
&&\vec{C}_{n+1}=-\frac{1}{2\pi^2\beta_0}\hat{P}^{(2)}\vec{A}_{n}
-\frac{\beta_2}{(4\pi)^2\beta_0}\vec{A}_{n+1}
-\frac{1}{\pi\beta_0}\hat{P}^{(1)}\vec{B}_{n},\nonumber\\
&&\hspace{1.4cm}-\frac{\beta_1}{4\pi\beta_0}\vec{B}_{n}
-\frac{\beta_1}{4\pi\beta_0}\vec{B}_{n+1}
-\frac{2}{\beta_0}\hat{P}^{(0)}\vec{C}_{n}
-2\vec{C}_{n},\nonumber\\
&&\vec{D}_{n+1}=-\frac{1}{2\pi^2\beta_0}\hat{P}^{(2)}\vec{B}_{n}
-\frac{\beta_2}{(4\pi)^2\beta_0}\vec{B}_{n}
-\frac{\beta_2}{(4\pi)^2\beta_0}\vec{B}_{n+1}
-\frac{1}{\pi\beta_0}\hat{P}^{(1)}\vec{C}_{n}\nonumber\\
&&\hspace{1.4cm}-\frac{\beta_1}{2\pi\beta_0}\vec{C}_{n}
-\frac{\beta_1}{(4\pi)\beta_0}\vec{C}_{n+1}
-\frac{2}{\beta_0}\hat{P}^{(0)}\vec{D}_{n}
-3\vec{D}_{n}.\,
\ea
Applying the properties of the $e_\pm $ operators and decomposing into the 
$\left\{{\bf e}_1,{\bf e}_2\right\}$ basis,
we project out the $\vec{B}_{n+1}^\pm $, $\vec{C}_{n+1}^\pm\dots$ components of these relations,
and imposing the initial conditions $\vec{B}_0=\vec{C}_0=\vec{D}_0=0$ together with
\ba
\vec{f}(N,\alpha_0)=\vec{A}_0\,,
\ea
we obtain the explicit form of $\vec{\tilde{f}}(N,\alpha_s)$.
In order to construct the $\vec{\tilde{f}}(N,\alpha_s)$ solution,
the four ${\pm}$ projections of $\vec{D}_{n+1}$ must be solved
with respect to $\vec{A}_0(N)$, since the other projections
$\vec{B}_{n}^{\pm}$, $\vec{C}_{n}^{\pm}$
are already known. 
A direct computation shows that the structure of the solution
can be organized as follows in terms of the components of $R_i$
\ba
&&\vec{D}_{n}^{++}(N)=
{\bf W}_1(\hat{R}_1^{++\,3},r_+,r_-,N,\vec{A}_0)
+{\bf W}_2(\hat{R}_1^{+-}\hat{R}_1^{-+}\hat{R}_1^{++},r_+,r_-,N,\vec{A}_0)
\nonumber\\
&&\hspace{1.7cm}
+{\bf W}_3(\hat{R}_1^{+-}\hat{R}_1^{--}\hat{R}_1^{-+},r_+,r_-,N,\vec{A}_0)
+{\bf W}_4(\hat{R}_1^{++}\hat{R}_1^{+-}\hat{R}_1^{-+},r_+,r_-,N,\vec{A}_0)
\nonumber\\
&&\hspace{1.7cm}
+{\bf W}_5(\hat{R}_1^{++}\hat{R}_2^{++},r_+,r_-,N,\vec{A}_0)
+{\bf W}_6(\hat{R}_2^{++}\hat{R}_1^{++},r_+,r_-,N,\vec{A}_0)
\nonumber\\
&&\hspace{1.7cm}
+{\bf W}_7(\hat{R}_1^{+-}\hat{R}_2^{-+},r_+,r_-,N,\vec{A}_0)
+{\bf W}_8(\hat{R}_2^{+-}\hat{R}_1^{-+},r_+,r_-,N,\vec{A}_0)
\nonumber\\
&&\hspace{1.7cm}
+{\bf W}_9(\hat{R}_1^{++},r_+,r_-,N,\vec{A}_0)
+{\bf W}_{10}(\hat{R}_2^{++},r_+,r_-,N,\vec{A}_0),
\nonumber\\
\nonumber\\
&&\vec{D}_{n}^{--}(N)=
{\bf W}_1(\hat{R}_1^{--\,3},r_+,r_-,N,\vec{A}_0)
+{\bf W}_2(\hat{R}_1^{-+}\hat{R}_1^{+-}\hat{R}_1^{--},r_+,r_-,N,\vec{A}_0)
\nonumber\\
&&\hspace{1.7cm}
+{\bf W}_3(\hat{R}_1^{-+}\hat{R}_1^{++}\hat{R}_1^{+-},r_+,r_-,N,\vec{A}_0)
+{\bf W}_4(\hat{R}_1^{--}\hat{R}_1^{-+}\hat{R}_1^{+-},r_+,r_-,N,\vec{A}_0)
\nonumber\\
&&\hspace{1.7cm}
+{\bf W}_5(\hat{R}_1^{--}\hat{R}_2^{--},r_+,r_-,N,\vec{A}_0)
+{\bf W}_6(\hat{R}_2^{--}\hat{R}_1^{--},r_+,r_-,N,\vec{A}_0)
\nonumber\\
&&\hspace{1.7cm}
+{\bf W}_7(\hat{R}_1^{-+}\hat{R}_2^{+-},r_+,r_-,N,\vec{A}_0)
+{\bf W}_8(\hat{R}_2^{-+}\hat{R}_1^{+-},r_+,r_-,N,\vec{A}_0)
\nonumber\\
&&\hspace{1.7cm}
+{\bf W}_9(\hat{R}_1^{--},r_+,r_-,N,\vec{A}_0)
+{\bf W}_{10}(\hat{R}_2^{--},r_+,r_-,N,\vec{A}_0),
\nonumber\\
\nonumber\\
&&\vec{D}_{n}^{+-}(N)=
{\bf Z}_1(\hat{R}_1^{++}\hat{R}_1^{+-}\hat{R}_1^{--},r_+,r_-,N,\vec{A}_0)
+{\bf Z}_2(\hat{R}_1^{++}\hat{R}_1^{++}\hat{R}_1^{+-},r_+,r_-,N,\vec{A}_0)
\nonumber\\
&&\hspace{1.7cm}
+{\bf Z}_3(\hat{R}_1^{+-}\hat{R}_1^{--}\hat{R}_1^{--},r_+,r_-,N,\vec{A}_0)
+{\bf Z}_4(\hat{R}_1^{+-}\hat{R}_1^{-+}\hat{R}_1^{+-},r_+,r_-,N,\vec{A}_0)
\nonumber\\
&&\hspace{1.7cm}
+{\bf Z}_5(\hat{R}_1^{+-}\hat{R}_2^{--},r_+,r_-,N,\vec{A}_0)
+{\bf Z}_6(\hat{R}_2^{+-}\hat{R}_1^{--},r_+,r_-,N,\vec{A}_0)
\nonumber\\
&&\hspace{1.7cm}
+{\bf Z}_7(\hat{R}_1^{++}\hat{R}_2^{+-},r_+,r_-,N,\vec{A}_0)
+{\bf Z}_8(\hat{R}_2^{++}\hat{R}_1^{+-},r_+,r_-,N,\vec{A}_0)
\nonumber\\
&&\hspace{1.7cm}
+{\bf Z}_9(\hat{R}_1^{+-},r_+,r_-,N,\vec{A}_0)
+{\bf Z}_{10}(\hat{R}_2^{+-},r_+,r_-,N,\vec{A}_0),
\nonumber\\
\nonumber\\
&&\vec{D}_{n}^{-+}(N)=
{\bf Z}_1(\hat{R}_1^{--}\hat{R}_1^{-+}\hat{R}_1^{++},r_+,r_-,N,\vec{A}_0)
+{\bf Z}_2(\hat{R}_1^{--}\hat{R}_1^{--}\hat{R}_1^{-+},r_+,r_-,N,\vec{A}_0)
\nonumber\\
&&\hspace{1.7cm}
+{\bf Z}_3(\hat{R}_1^{-+}\hat{R}_1^{++}\hat{R}_1^{++},r_+,r_-,N,\vec{A}_0)
+{\bf Z}_4(\hat{R}_1^{-+}\hat{R}_1^{+-}\hat{R}_1^{-+},r_+,r_-,N,\vec{A}_0)
\nonumber\\
&&\hspace{1.7cm}
+{\bf Z}_5(\hat{R}_1^{-+}\hat{R}_2^{++},r_+,r_-,N,\vec{A}_0)
+{\bf Z}_6(\hat{R}_2^{-+}\hat{R}_1^{++},r_+,r_-,N,\vec{A}_0)
\nonumber\\
&&\hspace{1.7cm}
+{\bf Z}_7(\hat{R}_1^{--}\hat{R}_2^{-+},r_+,r_-,N,\vec{A}_0)
+{\bf Z}_8(\hat{R}_2^{--}\hat{R}_1^{-+},r_+,r_-,N,\vec{A}_0)
\nonumber\\
&&\hspace{1.7cm}
+{\bf Z}_9(\hat{R}_1^{-+},r_+,r_-,N,\vec{A}_0)
+{\bf Z}_{10}(\hat{R}_2^{-+},r_+,r_-,N,\vec{A}_0),
\nonumber\\
\ea
where we used the notation
$\hat{R}_1^{++\,3}=\hat{R}_1^{++}\hat{R}_1^{++}\hat{R}_1^{++}$. The expressions of the functions W and Z's can
be extracted by a symbolic manipulation of the coefficients $\vec{D}_{n}^{++}$.
We have included one of the projections for completeness in an appendix for the interested reader.

In $x$-space, the structure of the (2nd) truncated (or $\alpha_s^3$) NNLO solution can
be expressed as a sequence of convolution products of the form
\ba
&&\vec{f}_{O(\alpha_s^3)}^{NNLO}(x,\alpha_s)=\left[\hat{L}(x)+
\alpha_s\hat{U}_1(x)\otimes\hat{L}(x)
-\alpha_0\hat{L}(x)\otimes\hat{U}_1(x)+
\alpha_s^2\hat{U}_2(x)\otimes\hat{L}(x)
\right.\nonumber\\
&&\hspace{2cm}\left.
-\alpha_s\alpha_0\hat{U}_1(x)\otimes\hat{L}(x)\otimes\hat{U}_1(x)
+\alpha_0^2\hat{L}(x)\otimes\left(\hat{U}_1(x)\otimes
\hat{U}_1(x)-\hat{U}_2(x)\right)
\right.\nonumber\\
&&\hspace{2cm}\left.
+\alpha_s^3\hat{U}_3(x)\otimes\hat{L}(x)
+\alpha_s\alpha_0^2\hat{U}_1(x)\otimes\hat{L}(x)\otimes\left(\hat{U}_1(x)
\otimes\hat{U}_1(x)-\hat{U}_2(x)\right)
\right.\nonumber\\
&&\hspace{2cm}\left.
-\alpha_s^2\alpha_0\hat{U}_2(x)\otimes\hat{L}(x)\otimes\hat{U}_1(x)
-\alpha_0^3\hat{L}(x)\otimes
\right.\nonumber\\
&&\hspace{2cm}\left.
\left(\hat{U}_1(x)\otimes\hat{U}_1(x)\otimes\hat{U}_1(x)
-\hat{U}_1(x)\otimes\hat{U}_2(x)-\hat{U}_2(x)\otimes\hat{U}_1(x)
+\hat{U}_3(x)\right)
\right]\otimes\vec{f}(x,\alpha_0),
\nonumber\\
\ea
and is reproduced by the $\alpha_s^3$ logarithmic expansion
\ba
\vec{\tilde{f}}(x,\alpha_s)=\sum_{n=0}^{\infty}\frac{L^n}{n!}
\left[\vec{A}_n(x)+\alpha_s\vec{B}_n(x)+\alpha_s^2\vec{C}_n(x)
+\alpha_s^3 \vec{D}_n(x)\right]\,,
\ea
with the initial condition
\ba
\vec{f}(x,\alpha_0)=\vec{A}_0(x).
\ea
The study of higher order truncates is performed numerically with the implementation of the
generalized recursion relations (\ref{genrecnnlo}) given in the previous sections.

\section{Comparison with existing programs}

In this section we present a numerical test of our solution algorithm.
For this aim, we compare the results of a computer program that implements our method
with the results of QCD-Pegasus \cite{Vogt3}, a PDF evolution program based on Mellin-space inversion,
which has been used by the QCD Working Group to set some
benchmark results \cite{leshouches,parton2005}. In the following,
we refer to these results as to the \emph{benchmark}.

In the tables we are going to show in this chapter, we set the renormalization and
factorization scales to be equal, and we adopt the fixed flavor number scheme. The
final evolution scale is $\mu_{F}^{2}=10^{4}\,\textrm{GeV}^{2}$.
We limit ourselves to this case because it is enough to test the reliability of our method.
A more lenghty and detailed analysis, which will take into account many 
other cases with renormalization scale dependence and variable flavor number scheme, 
is planned to be presented for the future.

As in the published benchmarks, we start the evolution at $\mu_{F,0}^{2}=2\,\textrm{GeV}^{2}$,
where the test input distributions, regardless of the perturbative order, are parametrized
by the following toy model
\begin{eqnarray}
xu_{v}(x) & = & 5.107200x^{0.8}(1-x)^{3}\nonumber \\
xd_{v}(x) & = & 3.064320x^{0.8}(1-x)^{4}\nonumber \\
xg(x) & = & 1.700000x^{-0.1}(1-x)^{5}\nonumber \\
x\bar{d}(x) & = & 0.1939875x^{-0.1}(1-x)^{6}\nonumber \\
x\bar{u}(x) & = & (1-x)x\bar{d}(x)\nonumber \\
xs(x)=x\bar{s}(x) & = & 0.2x(\bar{u}+\bar{d})(x)
\end{eqnarray}
and the running coupling has the value
\begin{equation}
\alpha_{s}(\mu_{R,0}^{2}=2\,\textrm{GeV}^{2})=0.35.
\end{equation}
We remind that $q_{v}=q-\bar{q}$, $q_{+}=q+\bar{q}$,
$L_{\pm}=\bar{d}\pm\bar{u}$. Our results are obtained using the exact solution method for the nonsinglet and the
LO singlet, and the $\kappa$-th truncate method for the NLO and NNLO singlet, with $\kappa=10$.
In each entry in the tables, the first number is our result and the second is the difference
between our results and the benchmark.

In Table \ref{LOtable} we compare our results at leading order with the results
reported in Table 2 of \cite{leshouches}. The agreement is excellent for any
value of $x$ for the nonsinglet ($xu_{v}$ and $xd_{v}$); regarding the singlet
($xg$ column), the agreement is excellent except at very high $x$: we have a sizeable
difference at $x=0.9$. In Table \ref{NLOtable} we analyze the next-to-leading order evolution; the results for the
proposed benchmarks are reported in Table 3 of \cite{leshouches}. The agreement is very good
for any value of $x$ for the nonsinglet; for the singlet the agreement is good,
except at very high $x$ ($x=0.9$).

Moving to the NNLO case (Table \ref{NNLOtable}, the benchmarks are shown in Table 14 of \cite{parton2005}),
some comments are in order.
We don't solve the non-singlet equation as in PEGASUS,
since (\ref{eqnn}) admits an exact solution
(\ref{nnlotest}), which in PEGASUS is obtained only by iteration of truncated solutions.
Our implementation is based on the exact solution presented in this thesis.
The discrepancy between our results and PEGASUS are of the order of few percent,
and they become large in the gluon case at $x=0.9$, as for the lower orders. Another comments
should be made for the sea asimmetry of the $s$ quark, which is nonvanishing at NNLO.
In this case we have a sizeable
relative discrepancy for any value of $x$ (reported in the column $xs_{v}$), but it is evident
that this asymmetry is quite small, especially if compared with the column $xs_{+}$, whose entries
are several orders of magnitude larger. This means that $xs$ and $x\bar{s}$ should be comparable and their
difference therefore pretty small.
We have to notice that $xs_{v}$
 cannot be computed directly by a single evolution equation.
Indeed, it is computed by a difference between very close numbers, a procedure
that amplifies the relative error.
\begin{sidewaystable}
\begin{small}
\begin{center}
\begin{tabular}{|c||c|c|c|c|c|c|c|c|}
\hline
\multicolumn{8}{|c|}{LO, $n_{f}=4$, $\mu_{F}^{2}=\mu_{R}^{2}=10^{4}\,\textrm{GeV}^{2}$}
\tabularnewline
\hline
$x$&
$xu_{v}$&
$xd_{v}$&
$xL_{-}$&
$2 xL_{+}$&
$xs_{+}$&
$xc_{+}$&
$xg$\tabularnewline
\hline
\hline
$10^{-7}$&
$5.7722\cdot10^{-5}$&
$3.4343\cdot10^{-5}$&
$7.6527\cdot10^{-7}$&
$9.9465\cdot10^{+1}$&
$4.8642\cdot10^{+1}$&
$4.7914\cdot10^{+1}$&
$1.3162\cdot10^{+3}$\tabularnewline
&
$0.0000\cdot10^{-5}$&
$0.0000\cdot10^{-5}$&
$0.0000\cdot10^{-7}$&
$0.0000\cdot10^{+1}$&
$0.0000\cdot10^{+1}$&
$0.0000\cdot10^{+1}$&
$0.0000\cdot10^{+3}$\tabularnewline
\hline
$10^{-6}$&
$3.3373\cdot10^{-4}$&
$1.9800\cdot10^{-4}$&
$5.0137\cdot10^{-6}$&
$5.0259\cdot10^{+1}$&
$2.4263\cdot10^{+1}$&
$2.3685\cdot10^{+1}$&
$6.0008\cdot10^{+2}$\tabularnewline
&
$0.0000\cdot10^{-4}$&
$0.0000\cdot10^{-4}$&
$0.0000\cdot10^{-6}$&
$0.0000\cdot10^{+1}$&
$0.0000\cdot10^{+1}$&
$0.0000\cdot10^{+1}$&
$0.0000\cdot10^{+2}$\tabularnewline
\hline
$10^{-5}$&
$1.8724\cdot10^{-3}$&
$1.1065\cdot10^{-3}$&
$3.1696\cdot10^{-5}$&
$2.4378\cdot10^{+1}$&
$1.1501\cdot10^{+1}$&
$1.1042\cdot10^{+1}$&
$2.5419\cdot10^{+2}$\tabularnewline
&
$0.0000\cdot10^{-3}$&
$0.0000\cdot10^{-3}$&
$0.0000\cdot10^{-5}$&
$0.0000\cdot10^{+1}$&
$0.0000\cdot10^{+1}$&
$0.0000\cdot10^{+1}$&
$0.0000\cdot10^{+2}$\tabularnewline
\hline
$10^{-4}$&
$1.0057\cdot10^{-2}$&
$5.9076\cdot10^{-3}$&
$1.9071\cdot10^{-4}$&
$1.1323\cdot10^{+1}$&
$5.1164\cdot10^{+0}$&
$4.7530\cdot10^{+0}$&
$9.7371\cdot10^{+1}$\tabularnewline
&
$0.0000\cdot10^{-2}$&
$0.0000\cdot10^{-3}$&
$0.0000\cdot10^{-4}$&
$0.0000\cdot10^{+1}$&
$0.0000\cdot10^{+0}$&
$0.0000\cdot10^{+0}$&
$0.0000\cdot10^{+1}$\tabularnewline
\hline
$10^{-3}$&
$5.0392\cdot10^{-2}$&
$2.9296\cdot10^{-2}$&
$1.0618\cdot10^{-3}$&
$5.0324\cdot10^{+0}$&
$2.0918\cdot10^{+0}$&
$1.8089\cdot10^{+0}$&
$3.2078\cdot10^{+1}$\tabularnewline
&
$0.0000\cdot10^{-2}$&
$0.0000\cdot10^{-2}$&
$0.0000\cdot10^{-3}$&
$0.0000\cdot10^{+0}$&
$0.0000\cdot10^{+0}$&
$0.0000\cdot10^{+0}$&
$0.0000\cdot10^{+1}$\tabularnewline
\hline
$10^{-2}$&
$2.1955\cdot10^{-1}$&
$1.2433\cdot10^{-1}$&
$4.9731\cdot10^{-3}$&
$2.0433\cdot10^{+0}$&
$7.2814\cdot10^{-1}$&
$5.3247\cdot10^{-1}$&
$8.0546\cdot10^{+0}$\tabularnewline
&
$0.0000\cdot10^{-1}$&
$0.0000\cdot10^{-1}$&
$0.0000\cdot10^{-3}$&
$0.0000\cdot10^{+0}$&
$0.0000\cdot10^{-1}$&
$0.0000\cdot10^{-1}$&
$0.0000\cdot10^{+0}$\tabularnewline
\hline
$0.1$&
$5.7267\cdot10^{-1}$&
$2.8413\cdot10^{-1}$&
$1.0470\cdot10^{-2}$&
$4.0832\cdot10^{-1}$&
$1.1698\cdot10^{-1}$&
$5.8864\cdot10^{-2}$&
$8.8766\cdot10^{-1}$\tabularnewline
&
$0.0000\cdot10^{-1}$&
$0.0000\cdot10^{-1}$&
$0.0000\cdot10^{-2}$&
$0.0000\cdot10^{-1}$&
$0.0000\cdot10^{-1}$&
$0.0000\cdot10^{-2}$&
$0.0000\cdot10^{-1}$\tabularnewline
\hline
$0.3$&
$3.7925\cdot10^{-1}$&
$1.4186\cdot10^{-1}$&
$3.3029\cdot10^{-3}$&
$4.0165\cdot10^{-2}$&
$1.0516\cdot10^{-2}$&
$4.1380\cdot10^{-3}$&
$8.2676\cdot10^{-2}$\tabularnewline
&
$0.0000\cdot10^{-1}$&
$0.0000\cdot10^{-1}$&
$0.0000\cdot10^{-3}$&
$0.0000\cdot10^{-2}$&
$0.0000\cdot10^{-2}$&
$+0.0001\cdot10^{-3}$&
$0.0000\cdot10^{-2}$\tabularnewline
\hline
$0.5$&
$1.3476\cdot10^{-1}$&
$3.5364\cdot10^{-2}$&
$4.2815\cdot10^{-4}$&
$2.8624\cdot10^{-3}$&
$7.3137\cdot10^{-4}$&
$2.6481\cdot10^{-4}$&
$7.9242\cdot10^{-3}$\tabularnewline
&
$0.0000\cdot10^{-1}$&
$0.0000\cdot10^{-2}$&
$0.0000\cdot10^{-4}$&
$0.0000\cdot10^{-3}$&
$-0.0001\cdot10^{-4}$&
$0.0000\cdot10^{-4}$&
$+0.0002\cdot10^{-3}$\tabularnewline
\hline
$0.7$&
$2.3123\cdot10^{-2}$&
$3.5943\cdot10^{-3}$&
$1.5868\cdot10^{-5}$&
$6.8970\cdot10^{-5}$&
$1.7730\cdot10^{-5}$&
$6.5593\cdot10^{-6}$&
$3.7301\cdot10^{-4}$\tabularnewline
&
$0.0000\cdot10^{-2}$&
$0.0000\cdot10^{-3}$&
$0.0000\cdot10^{-5}$&
$+0.0009\cdot10^{-5}$&
$+0.0005\cdot10^{-5}$&
$+0.0044\cdot10^{-6}$&
$-0.0010\cdot10^{-4}$\tabularnewline
\hline
$0.9$&
$4.3443\cdot10^{-4}$&
$2.2287\cdot10^{-5}$&
$1.1042\cdot10^{-8}$&
$3.3030\cdot10^{-8}$&
$8.5607\cdot10^{-9}$&
$3.2577\cdot10^{-9}$&
$1.3887\cdot10^{-6}$\tabularnewline
&
$0.0000\cdot10^{-4}$&
$0.0000\cdot10^{-5}$&
$0.0000\cdot10^{-8}$&
$-0.3263\cdot10^{-8}$&
$-0.1631\cdot10^{-8}$&
$-1.6316\cdot10^{-9}$&
$+0.2969\cdot10^{-6}$\tabularnewline
\hline
\end{tabular}
\end{center}

\caption{Comparison between our and the benchmark results at LO. In each entry, 
the first number is our result and the second one is the difference between our result and the benchmark. 
The benchmark values are reported in Table 2 of \cite{leshouches}}
\label{LOtable}
\end{small}
\end{sidewaystable}

\begin{sidewaystable}
\begin{small}
\begin{center}
\begin{tabular}{|c||c|c|c|c|c|c|c|c|}
\hline
\multicolumn{8}{|c|}{NLO, $n_{f}=4$, $\mu_{F}^{2}=\mu_{R}^{2}=10^{4}\,\textrm{GeV}^{2}$}
\tabularnewline
\hline
$x$&
$xu_{v}$&
$xd_{v}$&
$xL_{-}$&
$2 xL_{+}$&
$xs_{+}$&
$xc_{+}$&
$xg$\tabularnewline
\hline
\hline
$10^{-7}$&
$1.0620\cdot10^{-4}$&
$6.2353\cdot10^{-5}$&
$4.2455\cdot10^{-6}$&
$1.3710\cdot10^{+2}$&
$6.7469\cdot10^{+1}$&
$6.6750\cdot10^{+1}$&
$1.1517\cdot10^{+3}$\tabularnewline
&
$+0.0004\cdot10^{-4}$&
$+0.0025\cdot10^{-5}$&
$+0.0015\cdot10^{-6}$&
$+0.0112\cdot10^{+2}$&
$+0.0556\cdot10^{+1}$&
$+0.0555\cdot10^{+1}$&
$+0.0034\cdot10^{+3}$\tabularnewline
\hline
$10^{-6}$&
$5.4196\cdot10^{-4}$&
$3.1730\cdot10^{-4}$&
$1.9247\cdot10^{-5}$&
$6.8896\cdot10^{+1}$&
$3.3592\cdot10^{+1}$&
$3.3021\cdot10^{+1}$&
$5.4048\cdot10^{+2}$\tabularnewline
&
$+0.0019\cdot10^{-4}$&
$+0.0011\cdot10^{-4}$&
$+0.0006\cdot10^{-5}$&
$+0.0500\cdot10^{+1}$&
$+0.0250\cdot10^{+1}$&
$+0.0250\cdot10^{+1}$&
$+0.0137\cdot10^{+2}$\tabularnewline
\hline
$10^{-5}$&
$2.6878\cdot10^{-3}$&
$1.5682\cdot10^{-3}$&
$8.3598\cdot10^{-5}$&
$3.2936\cdot10^{+1}$&
$1.5788\cdot10^{+1}$&
$1.5335\cdot10^{+1}$&
$2.3578\cdot10^{+2}$\tabularnewline
&
$+0.0008\cdot10^{-3}$&
$+0.0005\cdot10^{-3}$&
$+0.0023\cdot10^{-5}$&
$+0.0208\cdot10^{+1}$&
$+0.0103\cdot10^{+1}$&
$+0.0104\cdot10^{+1}$&
$+0.0050\cdot10^{+2}$\tabularnewline
\hline
$10^{-4}$&
$1.2844\cdot10^{-2}$&
$7.4576\cdot10^{-3}$&
$3.4919\cdot10^{-4}$&
$1.4824\cdot10^{+1}$&
$6.8744\cdot10^{+0}$&
$6.5156\cdot10^{+0}$&
$9.3026\cdot10^{+1}$\tabularnewline
&
$+0.0003\cdot10^{-2}$&
$+0.0018\cdot10^{-3}$&
$+0.0008\cdot10^{-4}$&
$+0.0078\cdot10^{+1}$&
$+0.0389\cdot10^{+0}$&
$+0.0387\cdot10^{+0}$&
$+0.0154\cdot10^{+1}$\tabularnewline
\hline
$10^{-3}$&
$5.7937\cdot10^{-2}$&
$3.3343\cdot10^{-2}$&
$1.4164\cdot10^{-3}$&
$6.1899\cdot10^{+0}$&
$2.6783\cdot10^{+0}$&
$2.4001\cdot10^{+0}$&
$3.1540\cdot10^{+1}$\tabularnewline
&
$+0.0011\cdot10^{-2}$&
$+0.0006\cdot10^{-2}$&
$+0.0002\cdot10^{-3}$&
$+0.0251\cdot10^{+0}$&
$+0.0124\cdot10^{+0}$&
$+0.0123\cdot10^{+0}$&
$+0.0038\cdot10^{+1}$\tabularnewline
\hline
$10^{-2}$&
$2.3029\cdot10^{-1}$&
$1.2930\cdot10^{-1}$&
$5.3258\cdot10^{-3}$&
$2.2587\cdot10^{+0}$&
$8.4518\cdot10^{-1}$&
$6.5540\cdot10^{-1}$&
$8.1120\cdot10^{+0}$\tabularnewline
&
$+0.0003\cdot10^{-1}$&
$+0.0002\cdot10^{-1}$&
$+0.0007\cdot10^{-3}$&
$+0.0060\cdot10^{+0}$&
$+0.0298\cdot10^{-1}$&
$+0.0294\cdot10^{-1}$&
$+0.0054\cdot10^{+0}$\tabularnewline
\hline
$0.1$&
$5.5456\cdot10^{-1}$&
$2.7338\cdot10^{-1}$&
$1.0012\cdot10^{-2}$&
$3.9392\cdot10^{-1}$&
$1.1517\cdot10^{-1}$&
$6.0619\cdot10^{-2}$&
$8.9872\cdot10^{-1}$\tabularnewline
&
$+0.0004\cdot10^{-1}$&
$+0.0002\cdot10^{-1}$&
$+0.0001\cdot10^{-2}$&
$+0.0056\cdot10^{-1}$&
$+0.0028\cdot10^{-1}$&
$+0.0268\cdot10^{-2}$&
$+0.0005\cdot10^{-1}$\tabularnewline
\hline
$0.3$&
$3.5395\cdot10^{-1}$&
$1.3158\cdot10^{-1}$&
$3.0363\cdot10^{-3}$&
$3.5884\cdot10^{-2}$&
$9.2210\cdot10^{-3}$&
$3.4066\cdot10^{-3}$&
$8.3415\cdot10^{-2}$\tabularnewline
&
$+0.0002\cdot10^{-1}$&
$0.0000\cdot10^{-1}$&
$+0.0001\cdot10^{-3}$&
$+0.0036\cdot10^{-2}$&
$+0.0180\cdot10^{-3}$&
$+0.0176\cdot10^{-3}$&
$-0.0036\cdot10^{-2}$\tabularnewline
\hline
$0.5$&
$1.2271\cdot10^{-1}$&
$3.1968\cdot10^{-2}$&
$3.8266\cdot10^{-4}$&
$2.4149\cdot10^{-3}$&
$5.8539\cdot10^{-4}$&
$1.7068\cdot10^{-4}$&
$8.0412\cdot10^{-3}$\tabularnewline
&
$0.0000\cdot10^{-1}$&
$+0.0001\cdot10^{-2}$&
$+0.0001\cdot10^{-4}$&
$+0.0023\cdot10^{-3}$&
$+0.0115\cdot10^{-4}$&
$+0.0113\cdot10^{-4}$&
$-0.0061\cdot10^{-3}$\tabularnewline
\hline
$0.7$&
$2.0429\cdot10^{-2}$&
$3.1474\cdot10^{-3}$&
$1.3701\cdot10^{-5}$&
$5.3703\cdot10^{-5}$&
$1.2432\cdot10^{-5}$&
$2.8201\cdot10^{-6}$&
$3.8654\cdot10^{-4}$\tabularnewline
&
$0.0000\cdot10^{-2}$&
$+0.0001\cdot10^{-3}$&
$0.0000\cdot10^{-5}$&
$+0.0081\cdot10^{-5}$&
$+0.0039\cdot10^{-5}$&
$+0.0394\cdot10^{-6}$&
$-0.0067\cdot10^{-4}$\tabularnewline
\hline
$0.9$&
$3.6097\cdot10^{-4}$&
$1.8317\cdot10^{-5}$&
$8.9176\cdot10^{-9}$&
$1.6196\cdot10^{-8}$&
$1.6717\cdot10^{-9}$&
$-2.6084\cdot10^{-9}$&
$1.8308\cdot10^{-6}$\tabularnewline
&
$+0.0001\cdot10^{-4}$&
$0.0000\cdot10^{-5}$&
$-0.0054\cdot10^{-9}$&
$-0.4724\cdot10^{-8}$&
$-2.3673\cdot10^{-9}$&
$-2.3681\cdot10^{-9}$&
$+0.6181\cdot10^{-6}$\tabularnewline
\hline
\end{tabular}
\end{center}

\caption{Same as in Table \ref{LOtable} in the NLO case. The benchmark values are reported in 
Table 3 of \cite{leshouches}}
\label{NLOtable}
\end{small}
\end{sidewaystable}

\begin{sidewaystable}
\begin{small}
\begin{center}
\begin{tabular}{|c||c|c|c|c|c|c|c|c|}
\hline
\multicolumn{9}{|c|}{NNLO, $n_{f}=4$, $\mu_{F}^{2}=\mu_{R}^{2}=10^{4}\,\textrm{GeV}^{2}$}
\tabularnewline
\hline
$x$&
$xu_{v}$&
$xd_{v}$&
$xL_{-}$&
$2 xL_{+}$&
$xs_{v}$&
$xs_{+}$&
$xc_{+}$&
$xg$\tabularnewline
\hline
\hline
$10^{-7}$&
$1.4069\cdot10^{-4}$&
$9.0435\cdot10^{-5}$&
$5.5759\cdot10^{-6}$&
$1.4184\cdot10^{+2}$&
$1.9569\cdot10^{-5}$&
$6.9844\cdot10^{+1}$&
$6.9127\cdot10^{+1}$&
$1.0519\cdot10^{+3}$\tabularnewline
&
$-0.1218\cdot10^{-4}$&
$-0.1201\cdot10^{-4}$&
$-0.1259\cdot10^{-6}$&
$+0.0994\cdot10^{+2}$&
$-1.1868\cdot10^{-5}$&
$+0.4967\cdot10^{+1}$&
$+0.4966\cdot10^{+1}$&
$+0.0543\cdot10^{+3}$\tabularnewline
\hline
$10^{-6}$&
$6.5756\cdot10^{-4}$&
$4.0826\cdot10^{-4}$&
$2.4722\cdot10^{-5}$&
$7.1794\cdot10^{+1}$&
$5.8862\cdot10^{-5}$&
$3.5043\cdot10^{+1}$&
$3.4474\cdot10^{+1}$&
$5.1093\cdot10^{+2}$\tabularnewline
&
$-0.3420\cdot10^{-4}$&
$-0.3458\cdot10^{-4}$&
$-0.0688\cdot10^{-5}$&
$+0.3295\cdot10^{+1}$&
$-3.5417\cdot10^{-5}$&
$+0.1646\cdot10^{+1}$&
$+0.1646\cdot10^{+1}$&
$+0.1969\cdot10^{+2}$\tabularnewline
\hline
$10^{-5}$&
$3.0260\cdot10^{-3}$&
$1.8199\cdot10^{-3}$&
$1.0393\cdot10^{-4}$&
$3.4373\cdot10^{+1}$&
$1.4230\cdot10^{-4}$&
$1.6509\cdot10^{+1}$&
$1.6057\cdot10^{+1}$&
$2.2899\cdot10^{+2}$\tabularnewline
&
$-0.0721\cdot10^{-3}$&
$-0.0775\cdot10^{-3}$&
$-0.0326\cdot10^{-4}$&
$+0.0902\cdot10^{+1}$&
$-0.8560\cdot10^{-4}$&
$+0.0450\cdot10^{+1}$&
$+0.0450\cdot10^{+1}$&
$+0.0602\cdot10^{+2}$\tabularnewline
\hline
$10^{-4}$&
$1.3656\cdot10^{-2}$&
$8.0052\cdot10^{-3}$&
$4.1299\cdot10^{-4}$&
$1.5403\cdot10^{+1}$&
$2.2837\cdot10^{-4}$&
$7.1661\cdot10^{+0}$&
$6.8085\cdot10^{+0}$&
$9.2125\cdot10^{+1}$\tabularnewline
&
$-0.0066\cdot10^{-2}$&
$-0.0967\cdot10^{-3}$&
$-0.1259\cdot10^{-4}$&
$+0.0199\cdot10^{+1}$&
$-1.3807\cdot10^{-4}$&
$+0.0991\cdot10^{+0}$&
$+0.0988\cdot10^{+0}$&
$+0.1457\cdot10^{+1}$\tabularnewline
\hline
$10^{-3}$&
$5.9360\cdot10^{-2}$&
$3.4135\cdot10^{-2}$&
$1.5650\cdot10^{-3}$&
$6.3657\cdot10^{+0}$&
$8.9572\cdot10^{-5}$&
$2.7684\cdot10^{+0}$&
$2.4913\cdot10^{+0}$&
$3.1592\cdot10^{+1}$\tabularnewline
&
$+0.0200\cdot10^{-2}$&
$+0.0085\cdot10^{-2}$&
$-0.0358\cdot10^{-3}$&
$+0.0427\cdot10^{+0}$&
$-0.5522\cdot10^{-4}$&
$+0.0210\cdot10^{+0}$&
$+0.0209\cdot10^{+0}$&
$+0.0243\cdot10^{+1}$\tabularnewline
\hline
$10^{-2}$&
$2.3139\cdot10^{-1}$&
$1.2958\cdot10^{-1}$&
$5.5064\cdot10^{-3}$&
$2.2868\cdot10^{+0}$&
$-3.5702\cdot10^{-4}$&
$8.6094\cdot10^{-1}$&
$6.7224\cdot10^{-1}$&
$8.1503\cdot10^{+0}$\tabularnewline
&
$+0.0061\cdot10^{-1}$&
$+0.0039\cdot10^{-1}$&
$-0.0624\cdot10^{-3}$&
$+0.0116\cdot10^{+0}$&
$+2.1611\cdot10^{-4}$&
$+0.0592\cdot10^{-1}$&
$+0.0601\cdot10^{-1}$&
$+0.0122\cdot10^{+0}$\tabularnewline
\hline
$0.1$&
$5.5125\cdot10^{-1}$&
$2.7142\cdot10^{-1}$&
$9.9834\cdot10^{-3}$&
$3.9119\cdot10^{-1}$&
$-1.9045\cdot10^{-4}$&
$1.1453\cdot10^{-1}$&
$6.0520\cdot10^{-2}$&
$8.9909\cdot10^{-1}$\tabularnewline
&
$-0.0052\cdot10^{-1}$&
$-0.0023\cdot10^{-1}$&
$-0.0040\cdot10^{-2}$&
$+0.0100\cdot10^{-1}$&
$+1.1582\cdot10^{-4}$&
$+0.0067\cdot10^{-1}$&
$+0.0747\cdot10^{-2}$&
$-0.0654\cdot10^{-1}$\tabularnewline
\hline
$0.3$&
$3.5017\cdot10^{-1}$&
$1.3005\cdot10^{-1}$&
$3.0025\cdot10^{-3}$&
$3.4975\cdot10^{-2}$&
$-1.9830\cdot10^{-5}$&
$8.8758\cdot10^{-3}$&
$3.1421\cdot10^{-3}$&
$8.3041\cdot10^{-2}$\tabularnewline
&
$-0.0054\cdot10^{-1}$&
$-0.0020\cdot10^{-1}$&
$-0.0073\cdot10^{-3}$&
$-0.0383\cdot10^{-2}$&
$+1.2061\cdot10^{-5}$&
$-0.1722\cdot10^{-3}$&
$-0.1640\cdot10^{-3}$&
$-0.1145\cdot10^{-2}$\tabularnewline
\hline
$0.5$&
$1.2099\cdot10^{-1}$&
$3.1485\cdot10^{-2}$&
$3.7667\cdot10^{-4}$&
$2.1876\cdot10^{-3}$&
$-1.6924\cdot10^{-6}$&
$4.8155\cdot10^{-4}$&
$7.4120\cdot10^{-5}$&
$7.9784\cdot10^{-3}$\tabularnewline
&
$-0.0018\cdot10^{-1}$&
$-0.0043\cdot10^{-2}$&
$-0.0075\cdot10^{-4}$&
$-0.1991\cdot10^{-3}$&
$+1.0291\cdot10^{-6}$&
$-0.9810\cdot10^{-4}$&
$-0.9758\cdot10^{-4}$&
$-0.1342\cdot10^{-3}$\tabularnewline
\hline
$0.7$&
$2.0052\cdot10^{-2}$&
$3.0849\cdot10^{-3}$&
$1.3411\cdot10^{-5}$&
$1.5984\cdot10^{-5}$&
$-6.2854\cdot10^{-8}$&
$-6.1500\cdot10^{-6}$&
$-1.5545\cdot10^{-5}$&
$3.8226\cdot10^{-4}$\tabularnewline
&
$-0.0025\cdot10^{-2}$&
$-0.0037\cdot10^{-3}$&
$-0.0023\cdot10^{-5}$&
$-3.8260\cdot10^{-5}$&
$+0.3821\cdot10^{-7}$&
$-1.9084\cdot10^{-5}$&
$-1.9075\cdot10^{-5}$&
$-0.0722\cdot10^{-4}$\tabularnewline
\hline
$0.9$&
$3.5078\cdot10^{-4}$&
$1.7767\cdot10^{-5}$&
$8.6326\cdot10^{-9}$&
$-6.4293\cdot10^{-7}$&
$-9.1828\cdot10^{-11}$&
$-3.2776\cdot10^{-7}$&
$-3.3190\cdot10^{-7}$&
$1.9280\cdot10^{-6}$\tabularnewline
&
$-0.0033\cdot10^{-4}$&
$-0.0016\cdot10^{-5}$&
$-0.0184\cdot10^{-9}$&
$-6.6986\cdot10^{-7}$&
$+0.5579\cdot10^{-10}$&
$-3.3487\cdot10^{-7}$&
$-3.3487\cdot10^{-7}$&
$+0.7144\cdot10^{-6}$\tabularnewline
\hline
\end{tabular}
\end{center}

\caption{Same as in Table \ref{LOtable} in the NNLO case. 
The benchmark values are reported in Table 14 of \cite{parton2005}}
\label{NNLOtable}
\end{small}
\end{sidewaystable}

\section{Future objectives}

We have shown that logarithmic expansions identified in $x$-space and
implemented in this space carry the same information as the solution
of evolution equations in Mellin space. This has been obtained by the introduction of new and
generalized expansions that we expect to be very useful in order to establish benchmarks
for the evolution of the pdf's at the
LHC. Our analysis has been presented up to NNLO.
We have also shown how exact expansions can be derived.
We have presented analytical proofs of the
equivalence and clarified the role of previous similar analysis which were quite limited in their reach.
We have also presented a numerical comparison
of our results against those obtained using PEGASUS, for a specific setting.
The overall agreement, as we have seen, is very good down to very small x-values.
One future objective will be a more detailed analysis based on
the numerical implementation of our results with various comparisons between our
approach and other approaches with different set up.

There are several issues which are still unclear in this area and concern the role
of the NNLO effects in the evolution and in the hard scatterings, the role of the theoretical errors
in the determination of the pdf's, whether they dominate over the NNLO effects or not, and the
impact of the choices of various truncations in the determination of the numerical solution of the pdf's, along
the lines of our work. Similar analysis can be performed by other methods, but we think
that it is important, in the search for precise determination of cross sections at the LHC, to state clearly
which algorithm is implemented and what accuracy is retained, with a particular attention to the issues connected
to the resummation of the perturbative expansion \cite{Catani}. Our work, here, has been limited to a
(fixed order) NNLO
analysis. We hope that our analysis has shown that $x$-space approaches have a very solid
base and provide a simple view on the structure
of the solutions of the DGLAP equations, valid to all orders.

\section{Appendix A. Derivation of the recursion relations at NNLO}

As an illustration we have included here a derivation of the recursion relations
for the first truncated ansatz of $O(\alpha_s^2)$ that appears at NNLO.

Inserting the NNLO truncated ansatz for the solution into the
DGLAP equation we get at the left-hand-side of the defining equation
\begin{eqnarray}
&&\sum_{n=1}^{\infty}\left\{ \frac{A_{n}(x)}{n!}nL^{n-1}
\frac{\beta(\alpha_{s})}{\alpha_{s}}+\alpha_{s}\frac{B_{n}(x)}{n!}nL^{n-1}
\frac{\beta(\alpha_{s})}{\alpha_{s}}\right.\nonumber \\
&&\left.\qquad+\alpha_{s}^{2}\frac{C_{n}(x)}{n!}nL^{n-1}
\frac{\beta(\alpha_{s})}{\alpha_{s}}\right\} \nonumber \\
&&+\sum_{n=0}^{\infty}\left\{ \beta(\alpha_{s})\frac{B_{n}(x)}{n!}L^{n}+
2\alpha_{s}\beta(\alpha_{s})\frac{C_{n}(x)}{n!}L^{n}\right\} .
\label{eq:ansatz2}
\end{eqnarray}
Note that the first sum starts at $n=1$, because the $n=0$ term
in (\ref{eq:ansatz2}) does not have a $Q^{2}$ dependence. Sending
$n\rightarrow n+1$ in the first sum, using the three-loop expansion
of the beta function (\ref{eq:beta_exp}) and neglecting all the terms
of order $\alpha_{s}^{4}$ or higher, the previous formula becomes
\begin{eqnarray}
&&\sum_{n=0}^{\infty}\left\{ \frac{A_{n+1}(x)}{n!}L^{n}
\left(-\frac{\beta_{0}}{4\pi}\alpha_{s}-\frac{\beta_{1}}{16\pi^{2}}\alpha_{s}^{2}-
\frac{\beta_{2}}{64\pi^{3}}\alpha_{s}^{3}\right)\right.\nonumber \\
&&+\frac{B_{n+1}(x)}{n!}L^{n}\left(-\frac{\beta_{0}}{4\pi}\alpha_{s}^{2}-
\frac{\beta_{1}}{16\pi^{2}}\alpha_{s}^{3}\right)+\frac{C_{n+1}(x)}{n!}L^{n}
\left(-\frac{\beta_{0}}{4\pi}\alpha_{s}^{3}\right)\nonumber \\
&&\left.+\frac{B_{n}(x)}{n!}L^{n}\left(-\frac{\beta_{0}}{4\pi}\alpha_{s}^{2}-
\frac{\beta_{1}}{16\pi^{2}}\alpha_{s}^{3}\right)+2\frac{C_{n}(x)}{n!}L^{n}
\left(-\frac{\beta_{0}}{4\pi}\alpha_{s}^{3}\right)\right\} .\label{eq:recrelLHS}
\end{eqnarray}
At this point we use the NNLO expansion of the kernels.
We get at the right-hand-side of the defining equation
\begin{eqnarray}
&&\sum_{n=0}^{\infty}\frac{L^{n}}{n!}\left\{ \frac{\alpha_{s}}{2\pi}\left[P^{(0)}
\otimes A_{n}\right](x)+\frac{\alpha_{s}^{2}}{4\pi^{2}}\left[P^{(1)}\otimes A_{n}\right](x)
\right.\nonumber \\
&&\quad\quad\quad+\frac{\alpha_{s}^{3}}{8\pi^{3}}\left[P^{(2)}
\otimes A_{n}\right](x)+\frac{\alpha_{s}^{2}}{2\pi}\left[P^{(0)}
\otimes B_{n}\right](x)\nonumber \\
&&\left.\quad\quad\quad+\frac{\alpha_{s}^{3}}{4\pi^{2}}\left[P^{(1)}
\otimes B_{n}\right](x)+\frac{\alpha_{s}^{3}}{2\pi}\left[P^{(0)}
\otimes C_{n}\right](x)\right\}.
\label{eq:recrelRHS}
\end{eqnarray}
Equating (\ref{eq:recrelLHS}) and (\ref{eq:recrelRHS}) term by term
and grouping the terms proportional respectively to $\alpha_{s}$,
$\alpha_{s}^{2}$ and $\alpha_{s}^{3}$ we get the three desired recursion
relations (\ref{NNLOrec_nonsing}).
Setting $Q=Q_{0}$ in (\ref{nnloans}) we get
\begin{equation}
f(x,Q_{0}^{2})=A_{0}(x)+\alpha_{s}(Q_{0}^{2})B_{0}(x)+
\left(\alpha_{s}(Q^{2})\right)^{2}C_{0}(x).
\label{eq:boundary2}
\end{equation}
We have seen that the initial conditions should be chosen as
\beq
B_{0}(x)=C_{0}(x)=0,\qquad f(x,Q_{0}^{2})=A_{0}(x)
\eeq
in order to reproduce the moments of the truncated solution of the DGLAP equation.

\section{Appendix B. NNLO singlet truncated solution}

Putting all the projections of the coefficients $\vec{C}_n$
into the NNLO singlet ansatz we get
\ba
&&\vec{f}(N,\alpha_s)=\sum_{n=0}^{\infty}\frac{L^n}{n!}
\left[\vec{A}_{n}+\alpha_s
\left(\vec{B}_{n}^{++}+\vec{B}_{n}^{--}
+\vec{B}_{n}^{-+}+\vec{B}_{n}^{+-}\right)\right.\nonumber\\
&&\hspace{3.5cm}\left.+\alpha_s^2
\left(\vec{C}_{n}^{++}+\vec{C}_{n}^{--}
+\vec{C}_{n}^{-+}+\vec{C}_{n}^{+-}\right)\right]\,,
\ea
and exponentiating
\begin{small}
\ba
&&\vec{f}(N,\alpha_s)= e_{+}\vec{A}_0 \left(\frac{\alpha_s}{\alpha_0}\right)^{r_+}+
e_{-}\vec{A}_0 \left(\frac{a_s}{\alpha_0}\right)^{r_-}+\nonumber\\
&&\hspace{1.5cm}\alpha_s\left\{
e_{+}\hat{R}_1e_{+}
\left(\frac{\alpha_s}{\alpha_0}\right)^{r_+}-
e_{+}\hat{R}_1e_{+}
\left(\frac{\alpha_s}{\alpha_0}\right)^{(r_+ -1)}+\right.\nonumber\\
&&\hspace{2cm}\left.
e_{-}\hat{R}_1e_{-}
\left(\frac{\alpha_s}{\alpha_0}\right)^{r_-}-
e_{-}\hat{R}_1e_{-}
\left(\frac{\alpha_s}{\alpha_0}\right)^{(r_- -1)}+\right.\nonumber\\
&&\hspace{2cm}\left.\frac{1}{(r_+ -r_- -1)}\left[
-e_{+}\hat{R}_1e_{-}
\left(\frac{\alpha_s}{\alpha_0}\right)^{r_-}+
e_{+}\hat{R}_1e_{-}
\left(\frac{\alpha_s}{\alpha_0}\right)^{(r_+ -1)}\right]+\right.\nonumber\\
&&\hspace{2cm}\left.\frac{1}{(r_- -r_+ -1)}\left[
-e_{-}\hat{R}_1e_{+}
\left(\frac{\alpha_s}{\alpha_0}\right)^{r_+}+
e_{-}\hat{R}_1e_{+}
\left(\frac{\alpha_s}{\alpha_0}\right)^{(r_- -1)}\right]\right\}\vec{A}_0
\nonumber\\
%%%%%%%%%%%%%%%%%%%%%%%%%%%%%%%%%%
&&\hspace{1.5cm}+\alpha_s^2\left\{
\left(\frac{\alpha_s}{\alpha_0}\right)^{r_+}\left[
\frac{1}{2\alpha_s^2}
\left(e_{+}\hat{R}_1e_{+}\hat{R}_1e_{+}(\alpha_0-\alpha_s)^2
-e_{+}\hat{R}_2e_{+}(\alpha_0^2-\alpha_s^2)\right)
\right.\right.\nonumber\\
&&\hspace{4cm}\left.\left.
+\alpha_s\alpha_0\frac{e_{+}\hat{R}_1e_{-}\hat{R}_1e_{+}}
{\alpha_s^2((r_- -r_+)^2 -1)}
\left(\frac{\alpha_s}{\alpha_0}\right)^{r_--r_+}
\right.\right.\nonumber\\
&&\hspace{4cm}\left.\left.
+e_{+}\hat{R}_1e_{-}\hat{R}_1e_{+}\frac{\left((r_--r_+-1)\alpha_0^2
+(r_+-r_- -1)\alpha_s^2\right)}{2\alpha_s^2((r_- -r_+)^2 -1)}
\right]\right\}\vec{A}_0
\nonumber\\
%%%%%%%%%%%%%%%%%%%%%%%%%%%
&&\hspace{1.5cm}+\alpha_s^2\left\{
\left(\frac{\alpha_s}{\alpha_0}\right)^{r_-}\left[
\frac{1}{2\alpha_s^2}
\left(e_{-}\hat{R}_1e_{-}\hat{R}_1e_{-}(\alpha_0-\alpha_s)^2
-e_{-}\hat{R}_2e_{-}(\alpha_0^2-\alpha_s^2)\right)
\right.\right.\nonumber\\
&&\hspace{4cm}\left.\left.
+\alpha_s\alpha_0\frac{e_{-}\hat{R}_1e_{+}\hat{R}_1e_{-}}
{\alpha_s^2((r_- -r_+)^2 -1)}
\left(\frac{\alpha_s}{\alpha_0}\right)^{r_+-r_-}
\right.\right.\nonumber\\
&&\hspace{4cm}\left.\left.
+e_{-}\hat{R}_1e_{+}\hat{R}_1e_{-}\frac{\left((r_+-r_--1)\alpha_0^2
+(r_--r_+ -1)\alpha_s^2\right)}{2\alpha_s^2((r_- -r_+)^2 -1)}
\right]\right\}\vec{A}_0
\nonumber\\
%%%%%%%%%%%%%%%%%%%%%%%%%%%
&&\hspace{1.5cm}+\alpha_s^2\left\{
\left(\frac{\alpha_s}{\alpha_0}\right)^{r_+}
\left[
\frac{e_{+}\hat{R}_1e_{-}\hat{R}_1e_{-}\alpha_0^2}
{\alpha_s^2(1+r_--r_+)(2+r_- -r_+)}
+\frac{\left(e_{+}\hat{R}_1e_{+}\hat{R}_1e_{-}
-e_{+}\hat{R}_2e_{-}\right)\alpha_0^2}
{\alpha_s^2(2+r_- -r_+)}
\right.\right.\nonumber\\
&&\hspace{4cm}\left.\left.+
\frac{e_{+}\hat{R}_1e_{+}\hat{R}_1e_{-}\alpha_0}
{\alpha_s(1+r_- -r_+)}
\right]+
\right.\nonumber\\
&&\hspace{2.6cm}\left.
\left(\frac{\alpha_s}{\alpha_0}\right)^{r_-}
\left[
\frac{e_{+}\hat{R}_1e_{+}\hat{R}_1e_{-}}
{(1+r_--r_+)(2+r_- -r_+)}+
\frac{\left(e_{+}\hat{R}_2e_{-}
+e_{+}\hat{R}_1e_{-}\hat{R}_1e_{-}\right)}
{(2+r_- -r_+)}
\right.\right.\nonumber\\
&&\hspace{4cm}\left.\left.
-\frac{e_{+}\hat{R}_1e_{-}\hat{R}_1e_{-}\alpha_0}
{\alpha_s(1+r_--r_+)}\right]
\right\}\vec{A}_0
\nonumber\\
%%%%%%%%%%%%%%%%%%%%%%%%%%%
&&\hspace{1.5cm}+\alpha_s^2\left\{
\left(\frac{\alpha_s}{\alpha_0}\right)^{r_-}
\left[
\frac{e_{-}\hat{R}_1e_{+}\hat{R}_1e_{+}\alpha_0^2}
{(r_- -r_+-1)(r_--r_+-2)\alpha_s^2}
+\frac{\left(e_{-}\hat{R}_2e_{+}-
e_{-}\hat{R}_1e_{-}\hat{R}_1e_{+}\right)\alpha_0^2}
{(r_--r_+-2)\alpha_s^2}
\right.\right.\nonumber\\
&&\hspace{4cm}\left.\left.
+\frac{e_{-}\hat{R}_1e_{-}\hat{R}_1e_{+}\alpha_0}
{(r_--r_+-1)\alpha_s}
\right]
\right.\nonumber\\
&&\hspace{2.6cm}\left.
\left(\frac{\alpha_s}{\alpha_0}\right)^{r_+}
\left[
\frac{e_{-}\hat{R}_1e_{-}\hat{R}_1e_{+}}
{(r_--r_+-1)(r_--r_+-2)}
-\frac{\left(e_{-}\hat{R}_1e_{+}\hat{R}_1e_{+}+e_{-}\hat{R}_2e_{+}
\right)}{(r_--r_+-2)}
\right.\right.\nonumber\\
&&\hspace{4cm}\left.\left.
+\frac{e_{-}\hat{R}_1e_{+}\hat{R}_1e_{+}\alpha_0}
{(r_--r_+-1)\alpha_s}\right]
\right\}\vec{A}_0
\nonumber\\
\ea
\end{small}
This expression is equivalent to that obtained from eq. (\ref{NNLOtrsolsing}).
Projecting over all the $\pm$ components, and formally introducing the quantities
${\vec{q}(N,\alpha_s)}^{++},{\vec{q}(N,\alpha_s)}^{+-},\dots$, we can write
\begin{small}
\ba
\label{eqlhs}
&&{\vec{q}(N,\alpha_s)}^{++}=\left(\frac{\alpha_s}{\alpha_0}\right)^{r_+}
\left\{
e_+ + (\alpha_s -\alpha_0)e_+\hat{R}_1 e_+
+\alpha_s^2\frac{1}{2}\left[e_+\hat{R}_1 e_+\hat{R}_1 e_+
+e_+\hat{R}_2 e_+ -\frac{e_+\hat{R}_1 e_-\hat{R}_1 e_+}{(r_--r_+-1)}\right]
\right.\nonumber\\
&&\hspace{3.5cm}\left.-\alpha_s \alpha_0\left[
e_+\hat{R}_1e_+\hat{R}_1 e_+ +
\frac{e_+\hat{R}_1e_-\hat{R}_1 e_+}{(r_+-r_--1)(r_--r_+-1)}
\left(\frac{\alpha_s}{\alpha_0}\right)^{r_--r_+}\right]
\right.\nonumber\\
&&\hspace{4cm}\left.+\alpha_0^2 \left[
\frac{1}{2}e_+\hat{R}_1e_+\hat{R}_1 e_+ +
\frac{e_+\hat{R}_1e_-\hat{R}_1 e_+}{(r_+-r_--1)(r_--r_+-1)}
\right.\right.\nonumber\\
&&\hspace{5cm}\left.\left.
-\frac{1}{2}e_+\hat{R}_2 e_+ +\frac{1}{2}\frac{e_+\hat{R}_1e_-\hat{R}_1 e_+}{(r_--r_+-1)}
\right]\right\}\vec{q}(N,\alpha_0)\,,
\nonumber\\\\
%%%%%%%%%%%%%%%%%%%%%%%%%%%%%%%%%%%%%%%%
&&{\vec{q}(N,\alpha_s)}^{--}=\left(\frac{\alpha_s}{\alpha_0}\right)^{r_-}
\left\{
e_- + (\alpha_s -\alpha_0)e_-\hat{R}_1 e_-
+\alpha_s^2\frac{1}{2}\left[e_-\hat{R}_1e_-\hat{R}_1 e_-
+e_-\hat{R}_2e_- -\frac{e_-\hat{R}_1 e_+\hat{R}_1 e_-}{(r_+-r_--1)}\right]
\right.\nonumber\\
&&\hspace{3.5cm}\left.-\alpha_s \alpha_0\left[
e_-\hat{R}_1e_-\hat{R}_1e_- +
\frac{e_-\hat{R}_1e_+\hat{R}_1 e_-}{(r_--r_+-1)(r_+-r_--1)}
\left(\frac{\alpha_s}{\alpha_0}\right)^{r_+-r_-}\right]
\right.\nonumber\\
&&\hspace{4cm}\left.+\alpha_0^2 \left[
\frac{1}{2}e_-\hat{R}_1e_-\hat{R}_1 e_- +
\frac{e_-\hat{R}_1e_+\hat{R}_1 e_-}{(r_--r_+-1)(r_+-r_--1)}
\right.\right.\nonumber\\
&&\hspace{5cm}\left.\left.
-\frac{1}{2}e_-\hat{R}_2 e_- +\frac{1}{2}\frac{e_-\hat{R}_1e_+\hat{R}_1 e_-}{(r_+-r_--1)}
\right]\right\}\vec{q}(N,\alpha_0)\,,
\nonumber\\\\
%%%%%%%%%%%%%%%%%%%%%%%%%%%%%%%%%%%%%%%%
&&{\vec{q}(N,\alpha_s)}^{+-}=\left\{
-\alpha_s\frac{e_+\hat{R}_1e_-}{(r_+-r_--1)}\left(\frac{\alpha_s}{\alpha_0}\right)^{r_-}
+\alpha_0\frac{e_+\hat{R}_1e_-}{(r_+-r_--1)}\left(\frac{\alpha_s}{\alpha_0}\right)^{r_+}
\right.\nonumber\\
&&\hspace{2.5cm}\left.+\frac{\alpha_s^2}{(r_+-r_--2)}\left[
-e_+\hat{R}_1e_-\hat{R}_1e_- - e_+\hat{R}_2e_- +\frac{e_+\hat{R}_1e_+\hat{R}_1e_-}
{(r_+-r_--1)}\right]\left(\frac{\alpha_s}{\alpha_0}\right)^{r_-}
\right.\nonumber\\
&&\hspace{2.5cm}\left.
-\alpha_s\alpha_0\left[-\frac{e_+\hat{R}_1e_+\hat{R}_1e_-}{(r_+-r_--1)}
\left(\frac{\alpha_s}{\alpha_0}\right)^{r_+}
-\frac{e_+\hat{R}_1e_-\hat{R}_1e_-}{(r_+-r_--1)}
\left(\frac{\alpha_s}{\alpha_0}\right)^{r_-}\right]
\right.\nonumber\\
&&\hspace{2.5cm}\left.+\alpha_0^2\left(\frac{\alpha_s}{\alpha_0}\right)^{r_+}
\left[\left(-\frac{e_+\hat{R}_1e_+\hat{R}_1e_-}{(r_+-r_--1)}
-\frac{e_+\hat{R}_1e_-\hat{R}_1e_-}{(r_+-r_--1)}\right)
\right.\right.\nonumber\\
&&\hspace{4.8cm}\left.\left.
-\left(-\frac{e_+\hat{R}_1e_-\hat{R}_1e_-}{(r_+-r_--2)}
-\frac{e_+\hat{R}_2e_-}{(r_+-r_--2)}
\right.\right.\right.\nonumber\\
&&\hspace{5cm}\left.\left.\left.
+\frac{e_+\hat{R}_1e_+\hat{R}_1e_-}{(r_+-r_--2)(r_+-r_--1)}\right)
\right]\right\}\vec{q}(N,\alpha_0)\,,
\nonumber\\\\
%%%%%%%%%%%%%%%%%%%%%%%%%%%%%%%%%%%%%%%%
&&{\vec{q}(N,\alpha_s)}^{-+}=\left\{
-\alpha_s\frac{e_-\hat{R}_1e_+}{(r_--r_+ -1)}\left(\frac{\alpha_s}{\alpha_0}\right)^{r_+}
+\alpha_0\frac{e_-\hat{R}_1e_+}{(r_--r_+ -1)}\left(\frac{\alpha_s}{\alpha_0}\right)^{r_-}
\right.\nonumber\\
&&\hspace{2.5cm}\left.+\frac{\alpha_s^2}{(r_--r_+-2)}\left[
-e_-\hat{R}_1e_+\hat{R}_1e_+ - e_-\hat{R}_2e_+ +\frac{e_-\hat{R}_1e_-\hat{R}_1e_+}
{(r_--r_+-1)}\right]\left(\frac{\alpha_s}{\alpha_0}\right)^{r_+}
\right.\nonumber\\
&&\hspace{2.5cm}\left.
-\alpha_s\alpha_0\left[-\frac{e_-\hat{R}_1e_-\hat{R}_1e_+}{(r_--r_+-1)}
\left(\frac{\alpha_s}{\alpha_0}\right)^{r_-}
-\frac{e_-\hat{R}_1e_+\hat{R}_1e_+}{(r_--r_+-1)}
\left(\frac{\alpha_s}{\alpha_0}\right)^{r_+}\right]
\right.\nonumber\\
&&\hspace{2.5cm}\left.+\alpha_0^2\left(\frac{\alpha_s}{\alpha_0}\right)^{r_-}
\left[\left(-\frac{e_-\hat{R}_1e_-\hat{R}_1e_+}{(r_--r_+-1)}
-\frac{e_-\hat{R}_1e_+\hat{R}_1e_+}{(r_--r_+-1)}\right)
\right.\right.\nonumber\\
&&\hspace{4.8cm}\left.\left.
-\left(-\frac{e_-\hat{R}_1e_+\hat{R}_1e_+}{(r_--r_+-2)}
-\frac{e_-\hat{R}_2e_+}{(r_--r_+ -2)}
\right.\right.\right.\nonumber\\
&&\hspace{5cm}\left.\left.\left.
+\frac{e_-\hat{R}_1e_-\hat{R}_1e_+}{(r_--r_+-2)(r_--r_+-1)}\right)
\right]\right\}\vec{q}(N,\alpha_0)\,.
\nonumber\\
\ea
\end{small}
Using $\vec{A}_0=\vec{f}(N,\alpha_0)=\vec{q}(N,\alpha_0)$ it is simple
to obtain that
\ba
&&{\vec{q}(N,\alpha_s)}^{++}=\sum_{n=0}^{\infty}\frac{L^n}{n!}
\left[e_+ r^n_+\vec{A}_{0}+\alpha_s
\vec{B}_{n}^{++}+\alpha_s^2\vec{C}_{n}^{++}\right]
\nonumber\\
&&{\vec{q}(N,\alpha_s)}^{--}=\sum_{n=0}^{\infty}\frac{L^n}{n!}
\left[e_- r^n_-\vec{A}_{0}+\alpha_s
\vec{B}_{n}^{--}+\alpha_s^2\vec{C}_{n}^{--}\right]
\nonumber\\
&&{\vec{q}(N,\alpha_s)}^{+-}=\sum_{n=0}^{\infty}\frac{L^n}{n!}
\left[\alpha_s
\vec{B}_{n}^{+-}+\alpha_s^2\vec{C}_{n}^{+-}\right]
\nonumber\\
&&{\vec{q}(N,\alpha_s)}^{-+}=\sum_{n=0}^{\infty}\frac{L^n}{n!}
\left[\alpha_s
\vec{B}_{n}^{-+}+\alpha_s^2\vec{C}_{n}^{-+}\right]\,.
\ea
For example we can check the first of the relations above,
gives
\ba
&&\hspace{-1cm}\sum_{n=0}^{\infty}\frac{L^n}{n!}
\left[e_+ r^n_+\vec{A}_{0}+\alpha_s
\vec{B}_{n}^{++}+\alpha_s^2\vec{C}_{n}^{++}\right]=
\left(\frac{\alpha_s}{\alpha_0}\right)^{r_+}
\left\{e_{+}+(\alpha_s-\alpha_0)e_{+}\hat{R}_1e_{+}
\right\}\vec{f}(N,\alpha_0)
\nonumber\\
&&\hspace{5cm}+\left\{
\left(\frac{\alpha_s}{\alpha_0}\right)^{r_+}\left[
\frac{1}{2}
\left(e_{+}\hat{R}_1e_{+}\hat{R}_1e_{+}(\alpha_0-\alpha_s)^2
-e_{+}\hat{R}_2e_{+}(\alpha_0^2-\alpha_s^2)\right)
\right.\right.\nonumber\\
&&\hspace{7.5cm}\left.\left.
+\alpha_s\alpha_0\frac{e_{+}\hat{R}_1e_{-}\hat{R}_1e_{+}}
{(r_- -r_+)^2 -1}
\left(\frac{\alpha_s}{\alpha_0}\right)^{r_--r_+}
\right.\right.\nonumber\\
&&\hspace{5cm}\left.\left.
+e_{+}\hat{R}_1e_{-}\hat{R}_1e_{+}\frac{\left((r_--r_+-1)\alpha_0^2
+(r_+-r_- -1)\alpha_s^2\right)}{2((r_- -r_+)^2 -1)}
\right]\right\}\vec{f}(N,\alpha_0)\,.
\nonumber\\
\ea
Factorizing $(\alpha_s/\alpha_0)^{r_+}$ and expanding the power of $\alpha_s$
the previous expression becomes
\ba
&&{\vec{f}(N,\alpha_s)}^{++}=\left(\frac{\alpha_s}{\alpha_0}\right)^{r_+}
\left\{e_+ + (\alpha_s-\alpha_0)e_+\hat{R}_1 e_+
+\frac{1}{2}\alpha_s^2 \left[e_+\hat{R}_1 e_+\hat{R}_1 e_+
+e_+\hat{R}_2 e_+ -\frac{e_+\hat{R}_1 e_-\hat{R}_1 e_+}{(r_--r_+-1)}\right]
\right.\nonumber\\
&&\hspace{2cm}\left.+\frac{1}{2}\alpha_0^2
\left[-\frac{e_+\hat{R}_1 e_-\hat{R}_1 e_+}
{(r_+-r_--1)}+e_+\hat{R}_1 e_+\hat{R}_1 e_+ -
e_+\hat{R}_2 e_+\right]
\right.\nonumber\\
&&\hspace{2cm}\left.-\alpha_s\alpha_0\left[
e_+\hat{R}_1e_+\hat{R}_1 e_+ +
\frac{e_+\hat{R}_1e_-\hat{R}_1 e_+}{(r_+-r_--1)(r_--r_+-1)}
\left(\frac{\alpha_s}{\alpha_0}\right)^{r_--r_+}\right]
\right\}\vec{f}(N,\alpha_0)\,.
\ea
which agrees with the left hand side of eq.~(\ref{eqlhs}).

\section{Appendix C. Calculation of $\vec{D}_{n}^{++}$}

An explicit calculation of the vector coefficient $\vec{D}_{n}^{++}$ of the
$\kappa=4$ (4th truncated) solution of the NNLO singlet equation has been done
in this section. Since the expressions of the coefficients
$\vec{D}_{n}^{+-}$, $\vec{D}_{n}^{-+}$ and $\vec{D}_{n}^{--}$ have a
structure similar to $\vec{D}_{n}^{++}$, we omit them and
give only the explicit form of this one
\begin{small}
\ba
&&\vec{D}_{n}^{++}={\left( -3 + r_+ \right) }^n\,\left( 4\,b_2\,\hat{R}_1^{++}
- 32\,{\pi }^2\,\hat{R}_1^{++\,3} +
16\,b_1\,\pi \,\hat{R}_2^{++} - 5\,b_2\,\hat{R}_1^{++}\,{r_-}^2
+ 40\,{\pi }^2\,\hat{R}_1^{++\,3}\,{r_-}^2
\right.\nonumber\\
&&\left.
-20\,b_1\,\pi \,\hat{R}_2^{++}\,{r_-}^2 + b_2\,\hat{R}_1^{++}\,{r_-}^4 -
8\,{\pi }^2\,\hat{R}_1^{++\,3}\,{r_-}^4 + 4\,b_1\,\pi \,\hat{R}_2^{++}\,{r_-}^4 +
96\,{\pi }^2\,\hat{R}_1^{++\,3}\,{\left( \frac{-2 + r_+}{-3 + r_+} \right) }^n -
\right.\nonumber\\
&&\left.
120\,{\pi }^2\,\hat{R}_1^{++\,3}\,{r_-}^2\,{\left( \frac{-2 + r_+}{-3 + r_+} \right) }^n +
24\,{\pi }^2\,\hat{R}_1^{++\,3}\,{r_-}^4\,{\left( \frac{-2 + r_+}{-3 + r_+} \right) }^n -
96\,{\pi }^2\,\hat{R}_1^{++\,3}\,{\left( \frac{-1 + r_+}{-3 + r_+} \right) }^n +
\right.\nonumber\\
&&\left.
120\,{\pi }^2\,\hat{R}_1^{++\,3}\,{r_-}^2\,{\left( \frac{-1 + r_+}{-3 + r_+} \right) }^n -
24\,{\pi }^2\,\hat{R}_1^{++\,3}\,{r_-}^4\,{\left( \frac{-1 + r_+}{-3 + r_+} \right) }^n +
10\,b_2\,\hat{R}_1^{++}\,r_-\,r_+
- 80\,{\pi }^2\,\hat{R}_1^{++\,3}\,r_-\,r_+ +
\right.\nonumber\\
&&\left.
40\,b_1\,\pi \,\hat{R}_2^{++}\,r_-\,r_+ - 4\,b_2\,\hat{R}_1^{++}\,{r_-}^3\,r_+ +
32\,{\pi }^2\,\hat{R}_1^{++\,3}\,{r_-}^3\,r_+ - 16\,b_1\,\pi \,\hat{R}_2^{++}\,{r_-}^3\,r_+ +
\right.\nonumber\\
&&\left.
240\,{\pi }^2\,\hat{R}_1^{++\,3}\,r_-\,{\left( \frac{-2 + r_+}{-3 + r_+} \right) }^n\,r_+
-96\,{\pi }^2\,\hat{R}_1^{++\,3}\,{r_-}^3\,{\left( \frac{-2 + r_+}{-3 + r_+} \right) }^n\,r_+ -
240\,{\pi }^2\,\hat{R}_1^{++\,3}\,r_-\,{\left( \frac{-1 + r_+}{-3 + r_+} \right) }^n\,r_+
\right.\nonumber\\
&&\left.
+96\,{\pi }^2\,\hat{R}_1^{++\,3}\,{r_-}^3\,{\left( \frac{-1 + r_+}{-3 + r_+} \right) }^n\,r_+
-5\,b_2\,\hat{R}_1^{++}\,{r_+}^2 + 40\,{\pi }^2\,\hat{R}_1^{++\,3}\,{r_+}^2 -
20\,b_1\,\pi \,\hat{R}_2^{++}\,{r_+}^2 + 6\,b_2\,\hat{R}_1^{++}\,{r_-}^2\,{r_+}^2
\right.\nonumber\\
&&\left.
-48\,{\pi }^2\,\hat{R}_1^{++\,3}\,{r_-}^2\,{r_+}^2 +
24\,b_1\,\pi \,\hat{R}_2^{++}\,{r_-}^2\,{r_+}^2 -
120\,{\pi }^2\,\hat{R}_1^{++\,3}\,{\left( \frac{-2 + r_+}{-3 + r_+} \right) }^n\,{r_+}^2
\right.\nonumber\\
&&\left.
+144\,{\pi }^2\,\hat{R}_1^{++\,3}\,{r_-}^2\,{\left( \frac{-2 + r_+}{-3 + r_+} \right) }^n\,{r_+}^2 +
120\,{\pi }^2\,\hat{R}_1^{++\,3}\,{\left( \frac{-1 + r_+}{-3 + r_+} \right) }^n\,{r_+}^2 -
144\,{\pi }^2\,\hat{R}_1^{++\,3}\,{r_-}^2\,{\left( \frac{-1 + r_+}{-3 + r_+} \right) }^n\,{r_+}^2
\right.\nonumber\\
&&\left.
-4\,b_2\,\hat{R}_1^{++}\,r_-\,{r_+}^3 + 32\,{\pi }^2\,\hat{R}_1^{++\,3}\,r_-\,{r_+}^3 -
16\,b_1\,\pi \,\hat{R}_2^{++}\,r_-\,{r_+}^3
-96\,{\pi }^2\,\hat{R}_1^{++\,3}\,r_-\,{\left( \frac{-2 + r_+}{-3 + r_+} \right) }^n\,{r_+}^3 +
\right.\nonumber\\
&&\left.
96\,{\pi }^2\,\hat{R}_1^{++\,3}\,r_-\,{\left( \frac{-1 + r_+}{-3 + r_+} \right) }^n\,{r_+}^3 +
b_2\,\hat{R}_1^{++}\,{r_+}^4 - 8\,{\pi }^2\,\hat{R}_1^{++\,3}\,{r_+}^4 +
4\,b_1\,\pi \,\hat{R}_2^{++}\,{r_+}^4 +
\right.\nonumber\\
&&\left.24\,{\pi }^2\,\hat{R}_1^{++\,3}\,{\left( \frac{-2 + r_+}{-3 + r_+} \right) }^n\,{r_+}^4 -
24\,{\pi }^2\,\hat{R}_1^{++\,3}\,{\left( \frac{-1 + r_+}{-3 + r_+} \right) }^n\,{r_+}^4 -
4\,b_2\,\hat{R}_1^{++}\,{\left( \frac{r_+}{-3 + r_+} \right) }^n
\right.\nonumber\\
&&\left.
+32\,{\pi }^2\,\hat{R}_1^{++\,3}\,{\left( \frac{r_+}{-3 + r_+} \right) }^n -
16\,b_1\,\pi \,\hat{R}_2^{++}\,{\left( \frac{r_+}{-3 + r_+} \right) }^n +
5\,b_2\,\hat{R}_1^{++}\,{r_-}^2\,{\left( \frac{r_+}{-3 + r_+} \right) }^n
\right.\nonumber\\
&&\left.
-40\,{\pi }^2\,\hat{R}_1^{++\,3}\,{r_-}^2\,{\left( \frac{r_+}{-3 + r_+} \right) }^n +
20\,b_1\,\pi \,\hat{R}_2^{++}\,{r_-}^2\,{\left( \frac{r_+}{-3 + r_+} \right) }^n -
b_2\,\hat{R}_1^{++}\,{r_-}^4\,{\left( \frac{r_+}{-3 + r_+} \right) }^n +
\right.\nonumber\\
&&\left.
8\,{\pi }^2\,\hat{R}_1^{++\,3}\,{r_-}^4\,{\left( \frac{r_+}{-3 + r_+} \right) }^n -
4\,b_1\,\pi \,\hat{R}_2^{++}\,{r_-}^4\,{\left( \frac{r_+}{-3 + r_+} \right) }^n -
10\,b_2\,\hat{R}_1^{++}\,r_-\,r_+\,{\left( \frac{r_+}{-3 + r_+} \right) }^n
\right.\nonumber\\
&&\left.
+80\,{\pi }^2\,\hat{R}_1^{++\,3}\,r_-\,r_+\,{\left( \frac{r_+}{-3 + r_+} \right) }^n -
40\,b_1\,\pi \,\hat{R}_2^{++}\,r_-\,r_+\,{\left( \frac{r_+}{-3 + r_+} \right) }^n +
4\,b_2\,\hat{R}_1^{++}\,{r_-}^3\,r_+\,{\left( \frac{r_+}{-3 + r_+} \right) }^n
\right.\nonumber\\
&&\left.
-32\,{\pi }^2\,\hat{R}_1^{++\,3}\,{r_-}^3\,r_+\,{\left( \frac{r_+}{-3 + r_+} \right) }^n +
16\,b_1\,\pi \,\hat{R}_2^{++}\,{r_-}^3\,r_+\,{\left( \frac{r_+}{-3 + r_+} \right) }^n +
5\,b_2\,\hat{R}_1^{++}\,{r_+}^2\,{\left( \frac{r_+}{-3 + r_+} \right) }^n -
\right.\nonumber\\
&&\left.
40\,{\pi }^2\,\hat{R}_1^{++\,3}\,{r_+}^2\,{\left( \frac{r_+}{-3 + r_+} \right) }^n +
20\,b_1\,\pi \,\hat{R}_2^{++}\,{r_+}^2\,{\left( \frac{r_+}{-3 + r_+} \right) }^n -
6\,b_2\,\hat{R}_1^{++}\,{r_-}^2\,{r_+}^2\,{\left( \frac{r_+}{-3 + r_+} \right) }^n
\right.\nonumber\\
&&\left.
+48\,{\pi }^2\,\hat{R}_1^{++\,3}\,{r_-}^2\,{r_+}^2\,{\left( \frac{r_+}{-3 + r_+} \right) }^n -
24\,b_1\,\pi \,\hat{R}_2^{++}\,{r_-}^2\,{r_+}^2\,{\left( \frac{r_+}{-3 + r_+} \right) }^n +
4\,b_2\,\hat{R}_1^{++}\,r_-\,{r_+}^3\,{\left( \frac{r_+}{-3 + r_+} \right) }^n
\right.\nonumber\\
&&\left.
-32\,{\pi }^2\,\hat{R}_1^{++\,3}\,r_-\,{r_+}^3\,{\left( \frac{r_+}{-3 + r_+} \right) }^n +
16\,b_1\,\pi \,\hat{R}_2^{++}\,r_-\,{r_+}^3\,{\left( \frac{r_+}{-3 + r_+} \right) }^n -
b_2\,\hat{R}_1^{++}\,{r_+}^4\,{\left( \frac{r_+}{-3 + r_+} \right) }^n
\right.\nonumber\\
&&\left.
+8\,{\pi}^2\,\hat{R}_1^{++\,3}\,{r_+}^4\,{\left( \frac{r_+}{-3 + r_+} \right) }^n -
4\,b_1\,\pi \,\hat{R}_2^{++}\,{r_+}^4\,{\left( \frac{r_+}{-3 + r_+} \right) }^n
\right.\nonumber\\
&&\left.
+16\,{\pi }^2 \left( -2 + r_- + {r_-}^2 - r_+ - 2\,r_-\,r_+ + {r_+}^2 \right)\times
\right.\nonumber\\
&&\left.
\left( -2 + r_- + 3\,{\left(\frac{-2 + r_-}{-3 + r_+} \right)}^n -
{\left( \frac{r_+}{-3 + r_+} \right) }^n - r_-\,{\left( \frac{r_+}{-3 + r_+} \right)}^n
+r_+\,\left( -1 + {\left( \frac{r_+}{-3 + r_+} \right) }^n \right)  \right)
\,\hat{R}_1^{+-}\,\hat{R}_2^{-+}
\right.\nonumber\\
&&\left.
-8\,{\pi }^2\,\left( -2 + 3\,{\left( \frac{-2 + r_+}{-3 + r_+} \right) }^n -
{\left( \frac{r_+}{-3 + r_+} \right) }^n \right) \times\,
\right.\nonumber\\
\nonumber
%%%%%%%%%%%%%%%%
\ea
\end{small}
%%%%%%%%%%%%%%%%%%%%
\begin{small}
\ba
&&\left.
\left( 4 + {r_-}^4 - 4\,{r_-}^3\,r_+ - 5\,{r_+}^2 + {r_+}^4 +
{r_-}^2\,\left( -5 + 6\,{r_+}^2 \right)  + r_-\,\left( 10\,r_+ - 4\,{r_+}^3 \right)  \right) \,
\hat{R}_1^{++}\, \hat{R}_2^{++}
\right.\nonumber\\
&&\left.
+ 32\,{\pi }^2\,\hat{R}_2^{+-}\, \hat{R}_1^{-+} -
16\,{\pi }^2\,r_-\,\hat{R}_2^{+-}\, \hat{R}_1^{-+}
- 32\,{\pi }^2\,{r_-}^2\,\hat{R}_2^{+-}\,\hat{R}_1^{-+} +
16\,{\pi }^2\,{r_-}^3\,\hat{R}_2^{+-}\, \hat{R}_1^{-+}
\right.\nonumber\\
&&\left.
-96\,{\pi }^2\,{\left( \frac{-1 + r_-}{-3 + r_+} \right) }^n\,\hat{R}_2^{+-}\, \hat{R}_1^{-+} -
48\,{\pi }^2\,r_-\,{\left( \frac{-1 + r_-}{-3 + r_+} \right) }^n\,\hat{R}_2^{+-}\, \hat{R}_1^{-+} +
48\,{\pi }^2\,{r_-}^2\,{\left( \frac{-1 + r_-}{-3 + r_+} \right) }^n\,\hat{R}_2^{+-}\, \hat{R}_1^{-+}
\right.\nonumber\\
&&\left.
+16\,{\pi }^2\,r_+\,\hat{R}_2^{+-}\, \hat{R}_1^{-+} +
64\,{\pi }^2\,r_-\,r_+\,\hat{R}_2^{+-}\, \hat{R}_1^{-+}
-48\,{\pi }^2\,{r_-}^2\,r_+\,\hat{R}_2^{+-}\, \hat{R}_1^{-+}
\right.\nonumber\\
&&\left.
+48\,{\pi }^2\,{\left( \frac{-1 + r_-}{-3 + r_+} \right) }^n\,r_+\,\hat{R}_2^{+-}\, \hat{R}_1^{-+}
-96\,{\pi }^2\,r_-\,{\left( \frac{-1 + r_-}{-3 + r_+} \right) }^n\,r_+\,\hat{R}_2^{+-}\, \hat{R}_1^{-+}
\right.\nonumber\\
&&\left.
-32\,{\pi }^2\,{r_+}^2\,\hat{R}_2^{+-}\, \hat{R}_1^{-+} +
48\,{\pi }^2\,r_-\,{r_+}^2\,\hat{R}_2^{+-}\, \hat{R}_1^{-+}
\right.\nonumber\\
&&\left.
+48\,{\pi }^2\,{\left( \frac{-1 + r_-}{-3 + r_+} \right) }^n\,{r_+}^2\,\hat{R}_2^{+-}\, \hat{R}_1^{-+} -
16\,{\pi }^2\,{r_+}^3\,\hat{R}_2^{+-}\, \hat{R}_1^{-+}
\right.\nonumber\\
&&\left.
+64\,{\pi }^2\,{\left( \frac{r_+}{-3 + r_+} \right) }^n\,\hat{R}_2^{+-}\, \hat{R}_1^{-+} +
64\,{\pi }^2\,r_-\,{\left( \frac{r_+}{-3 + r_+} \right) }^n\,\hat{R}_2^{+-}\, \hat{R}_1^{-+} -
16\,{\pi }^2\,{r_-}^2\,{\left( \frac{r_+}{-3 + r_+} \right) }^n\,\hat{R}_2^{+-}\, \hat{R}_1^{-+}
\right.\nonumber\\
&&\left.
-16\,{\pi }^2\,{r_-}^3\,{\left( \frac{r_+}{-3 + r_+} \right) }^n\,\hat{R}_2^{+-}\, \hat{R}_1^{-+} -
64\,{\pi }^2\,r_+\,{\left( \frac{r_+}{-3 + r_+} \right) }^n\,\hat{R}_2^{+-}\, \hat{R}_1^{-+}
\right.\nonumber\\
&&\left.
+32\,{\pi }^2\,r_-\,r_+\,{\left( \frac{r_+}{-3 + r_+} \right) }^n\,\hat{R}_2^{+-}\, \hat{R}_1^{-+}
+48\,{\pi }^2\,{r_-}^2\,r_+\,{\left( \frac{r_+}{-3 + r_+} \right) }^n\,\hat{R}_2^{+-}\, \hat{R}_1^{-+}
\right.\nonumber\\
&&\left.
-16\,{\pi }^2\,{r_+}^2\,{\left( \frac{r_+}{-3 + r_+} \right) }^n\,\hat{R}_2^{+-}\, \hat{R}_1^{-+} -
48\,{\pi }^2\,r_-\,{r_+}^2\,{\left( \frac{r_+}{-3 + r_+} \right) }^n\,\hat{R}_2^{+-}\, \hat{R}_1^{-+}
\right.\nonumber\\
&&\left.
+16\,{\pi }^2\,{r_+}^3\,{\left( \frac{r_+}{-3 + r_+} \right) }^n\,\hat{R}_2^{+-}\, \hat{R}_1^{-+} +
32\,{\pi }^2\,\hat{R}_2^{++}\, \hat{R}_1^{++} - 40\,{\pi }^2\,{r_-}^2\,\hat{R}_2^{++}\, \hat{R}_1^{++} +
8\,{\pi }^2\,{r_-}^4\,\hat{R}_2^{++}\, \hat{R}_1^{++}
\right.\nonumber\\
&&\left.
-96\,{\pi }^2\,{\left( \frac{-1 + r_+}{-3 + r_+} \right) }^n\,\hat{R}_2^{++}\, \hat{R}_1^{++} +
120\,{\pi }^2\,{r_-}^2\,{\left( \frac{-1 + r_+}{-3 + r_+} \right) }^n\,\hat{R}_2^{++}\, \hat{R}_1^{++} -
24\,{\pi }^2\,{r_-}^4\,{\left( \frac{-1 + r_+}{-3 + r_+} \right) }^n\,\hat{R}_2^{++}\, \hat{R}_1^{++}
\right.\nonumber\\
&&\left.
+80\,{\pi }^2\,r_-\,r_+\,\hat{R}_2^{++}\, \hat{R}_1^{++} -
32\,{\pi }^2\,{r_-}^3\,r_+\,\hat{R}_2^{++}\, \hat{R}_1^{++} -
240\,{\pi }^2\,r_-\,{\left( \frac{-1 + r_+}{-3 + r_+} \right) }^n\,r_+\,\hat{R}_2^{++}\, \hat{R}_1^{++}
\right.\nonumber\\
&&\left.
+96\,{\pi }^2\,{r_-}^3\,{\left( \frac{-1 + r_+}{-3 + r_+} \right) }^n\,r_+\,
 \hat{R}_2^{++}\, \hat{R}_1^{++} - 40\,{\pi }^2\,{r_+}^2\,\hat{R}_2^{++}\, \hat{R}_1^{++} +
48\,{\pi }^2\,{r_-}^2\,{r_+}^2\,\hat{R}_2^{++}\, \hat{R}_1^{++}
\right.\nonumber\\
&&\left.
+120\,{\pi }^2\,{\left( \frac{-1 + r_+}{-3 + r_+} \right) }^n\,{r_+}^2\,\hat{R}_2^{++}\, \hat{R}_1^{++} -
144\,{\pi }^2\,{r_-}^2\,{\left( \frac{-1 + r_+}{-3 + r_+} \right) }^n\,{r_+}^2\,
\hat{R}_2^{++}\, \hat{R}_1^{++} - 32\,{\pi }^2\,r_-\,{r_+}^3\,\hat{R}_2^{++}\, \hat{R}_1^{++}
\right.\nonumber\\
&&\left.
+96\,{\pi }^2\,r_-\,{\left( \frac{-1 + r_+}{-3 + r_+} \right) }^n\,{r_+}^3\,
 \hat{R}_2^{++}\, \hat{R}_1^{++} + 8\,{\pi }^2\,{r_+}^4\,\hat{R}_2^{++}\, \hat{R}_1^{++}
\right.\nonumber\\
&&\left.
-24\,{\pi }^2\,{\left( \frac{-1 + r_+}{-3 + r_+} \right) }^n\,{r_+}^4\,\hat{R}_2^{++}\, \hat{R}_1^{++} +
64\,{\pi }^2\,{\left( \frac{r_+}{-3 + r_+} \right) }^n\,\hat{R}_2^{++}\, \hat{R}_1^{++}
\right.\nonumber\\
&&\left.
-80\,{\pi }^2\,{r_-}^2\,{\left( \frac{r_+}{-3 + r_+} \right) }^n\,\hat{R}_2^{++}\, \hat{R}_1^{++} +
16\,{\pi }^2\,{r_-}^4\,{\left( \frac{r_+}{-3 + r_+} \right) }^n\,\hat{R}_2^{++}\, \hat{R}_1^{++}
\right.\nonumber\\
&&\left.
+160\,{\pi }^2\,r_-\,r_+\,{\left( \frac{r_+}{-3 + r_+} \right) }^n\,\hat{R}_2^{++}\, \hat{R}_1^{++}
-64\,{\pi }^2\,{r_-}^3\,r_+\,{\left( \frac{r_+}{-3 + r_+} \right) }^n\,\hat{R}_2^{++}\, \hat{R}_1^{++}
\right.\nonumber\\
&&\left.
-80\,{\pi }^2\,{r_+}^2\,{\left( \frac{r_+}{-3 + r_+} \right) }^n\,\hat{R}_2^{++}\, \hat{R}_1^{++}
+96\,{\pi }^2\,{r_-}^2\,{r_+}^2\,{\left( \frac{r_+}{-3 + r_+} \right) }^n\,
\hat{R}_2^{++}\, \hat{R}_1^{++}
\right.\nonumber\\
&&\left.
- 64\,{\pi }^2\,r_-\,{r_+}^3\,{\left( \frac{r_+}{-3 + r_+} \right) }^n\,
\hat{R}_2^{++}\, \hat{R}_1^{++}
+ 16\,{\pi }^2\,{r_+}^4\,{\left( \frac{r_+}{-3 + r_+} \right) }^n\,
\hat{R}_2^{++}\, \hat{R}_1^{++} - 32\,{\pi }^2\,\hat{R}_1^{+-}\, \hat{R}_1^{--}\, \hat{R}_1^{-+}
\right.\nonumber\\
&&\left.
+48\,{\pi }^2\,r_-\,\hat{R}_1^{+-}\, \hat{R}_1^{--}\, \hat{R}_1^{-+}
-16\,{\pi }^2\,{r_-}^2\,\hat{R}_1^{+-}\, \hat{R}_1^{--}\, \hat{R}_1^{-+} +
96\,{\pi }^2\,{\left( \frac{-2 + r_-}{-3 + r_+} \right) }^n\,\hat{R}_1^{+-}\, \hat{R}_1^{--}\, \hat{R}_1^{-+}
\right.\nonumber\\
&&\left.
-48\,{\pi }^2\,r_-\,{\left( \frac{-2 + r_-}{-3 + r_+} \right) }^n\,
 \hat{R}_1^{+-}\, \hat{R}_1^{--}\, \hat{R}_1^{-+}
-48\,{\pi }^2\,{r_-}^2\,{\left( \frac{-2 + r_-}{-3 + r_+} \right) }^n\,
 \hat{R}_1^{+-}\, \hat{R}_1^{--}\, \hat{R}_1^{-+}
\right.\nonumber\\
&&\left.
-96\,{\pi }^2\,{\left( \frac{-1 + r_-}{-3 + r_+} \right) }^n\,\hat{R}_1^{+-}\, \hat{R}_1^{--}\, \hat{R}_1^{-+}
-48\,{\pi }^2\,r_-\,{\left( \frac{-1 + r_-}{-3 + r_+} \right) }^n\,
 \hat{R}_1^{+-}\, \hat{R}_1^{--}\, \hat{R}_1^{-+}
\right.\nonumber\\
\nonumber
\ea
\end{small}
%%%%%%%%%%%%%%%%%%%%%%%
\begin{small}
\ba
&&\left.
+48\,{\pi }^2\,{r_-}^2\,{\left( \frac{-1 + r_-}{-3 + r_+} \right) }^n\,
 \hat{R}_1^{+-}\, \hat{R}_1^{--}\, \hat{R}_1^{-+}
-48\,{\pi }^2\,r_+\,\hat{R}_1^{+-}\, \hat{R}_1^{--}\, \hat{R}_1^{-+} +
\right.\nonumber\\
&&\left.
32\,{\pi }^2\,r_-\,r_+\,\hat{R}_1^{+-}\, \hat{R}_1^{--}\, \hat{R}_1^{-+} +
48\,{\pi }^2\,{\left( \frac{-2 + r_-}{-3 + r_+} \right) }^n\,r_+\,
 \hat{R}_1^{+-}\, \hat{R}_1^{--}\, \hat{R}_1^{-+}
\right.\nonumber\\
&&\left.
+96\,{\pi }^2\,r_-\,{\left( \frac{-2 + r_-}{-3 + r_+} \right) }^n\,r_+\,
 \hat{R}_1^{+-}\, \hat{R}_1^{--}\, \hat{R}_1^{-+} +
48\,{\pi }^2\,{\left( \frac{-1 + r_-}{-3 + r_+} \right) }^n\,r_+\,
 \hat{R}_1^{+-}\, \hat{R}_1^{--}\, \hat{R}_1^{-+}
\right.\nonumber\\
&&\left.
-96\,{\pi }^2\,r_-\,{\left( \frac{-1 + r_-}{-3 + r_+} \right) }^n\,r_+\,
 \hat{R}_1^{+-}\, \hat{R}_1^{--}\, \hat{R}_1^{-+} -
16\,{\pi }^2\,{r_+}^2\,\hat{R}_1^{+-}\, \hat{R}_1^{--}\, \hat{R}_1^{-+}
\right.\nonumber\\
&&\left.
-48\,{\pi }^2\,{\left( \frac{-2 + r_-}{-3 + r_+} \right) }^n\,{r_+}^2\,
 \hat{R}_1^{+-}\, \hat{R}_1^{--}\, \hat{R}_1^{-+} +
48\,{\pi }^2\,{\left( \frac{-1 + r_-}{-3 + r_+} \right) }^n\,{r_+}^2\,
 \hat{R}_1^{+-}\, \hat{R}_1^{--}\, \hat{R}_1^{-+}
\right.\nonumber\\
&&\left.
+32\,{\pi }^2\,{\left( \frac{r_+}{-3 + r_+} \right) }^n\,\hat{R}_1^{+-}\, \hat{R}_1^{--}\,
\hat{R}_1^{-+} +
48\,{\pi }^2\,r_-\,{\left( \frac{r_+}{-3 + r_+} \right) }^n\,
 \hat{R}_1^{+-}\, \hat{R}_1^{--}\, \hat{R}_1^{-+}
\right.\nonumber\\
&&\left.
+16\,{\pi }^2\,{r_-}^2\,{\left( \frac{r_+}{-3 + r_+} \right) }^n\,
 \hat{R}_1^{+-}\, \hat{R}_1^{--}\, \hat{R}_1^{-+} -
48\,{\pi }^2\,r_+\,{\left( \frac{r_+}{-3 + r_+} \right) }^n\,
\hat{R}_1^{+-}\, \hat{R}_1^{--}\, \hat{R}_1^{-+}
\right.\nonumber\\
&&\left.
-32\,{\pi }^2\,r_-\,r_+\,{\left( \frac{r_+}{-3 + r_+} \right) }^n\,
 \hat{R}_1^{+-}\, \hat{R}_1^{--}\, \hat{R}_1^{-+} +
16\,{\pi }^2\,{r_+}^2\,{\left( \frac{r_+}{-3 + r_+} \right) }^n\,
 \hat{R}_1^{+-}\, \hat{R}_1^{--}\, \hat{R}_1^{-+}
\right.\nonumber\\
&&\left.
- 32\,{\pi }^2\,\hat{R}_1^{+-}\,
 \hat{R}_1^{-+}\, \hat{R}_1^{++}
+32\,{\pi }^2\,r_-\,\hat{R}_1^{+-}\, \hat{R}_1^{-+}\, \hat{R}_1^{++} +
8\,{\pi }^2\,{r_-}^2\,\hat{R}_1^{+-}\, \hat{R}_1^{-+}\, \hat{R}_1^{++}
\right.\nonumber\\
&&\left.
-8\,{\pi }^2\,{r_-}^3\,\hat{R}_1^{+-}\, \hat{R}_1^{-+}\, \hat{R}_1^{++} +
96\,{\pi }^2\,{\left( \frac{-2 + r_-}{-3 + r_+} \right) }^n\,\hat{R}_1^{+-}\,
\hat{R}_1^{-+}\, \hat{R}_1^{++} + 48\,{\pi }^2\,r_-\,{\left( \frac{-2 + r_-}{-3 + r_+} \right) }^n\,
 \hat{R}_1^{+-}\, \hat{R}_1^{-+}\, \hat{R}_1^{++}
\right.\nonumber\\
&&\left.
-96\,{\pi }^2\,{\left( \frac{-1 + r_+}{-3 + r_+} \right) }^n\,\hat{R}_1^{+-}\,
\hat{R}_1^{-+}\, \hat{R}_1^{++} -
96\,{\pi }^2\,r_-\,{\left( \frac{-1 + r_+}{-3 + r_+} \right) }^n\,
 \hat{R}_1^{+-}\, \hat{R}_1^{-+}\, \hat{R}_1^{++}
\right.\nonumber\\
&&\left.
+24\,{\pi }^2\,{r_-}^2\,{\left( \frac{-1 + r_+}{-3 + r_+} \right) }^n\,
 \hat{R}_1^{+-}\, \hat{R}_1^{-+}\, \hat{R}_1^{++}
+24\,{\pi }^2\,{r_-}^3\,{\left( \frac{-1 + r_+}{-3 + r_+} \right) }^n\,
 \hat{R}_1^{+-}\, \hat{R}_1^{-+}\, \hat{R}_1^{++}
\right.\nonumber\\
&&\left.
-32\,{\pi }^2\,r_+\,\hat{R}_1^{+-}\, \hat{R}_1^{-+}\, \hat{R}_1^{++}
-16\,{\pi }^2\,r_-\,r_+\,\hat{R}_1^{+-}\, \hat{R}_1^{-+}\, \hat{R}_1^{++} +
24\,{\pi }^2\,{r_-}^2\,r_+\,\hat{R}_1^{+-}\, \hat{R}_1^{-+}\, \hat{R}_1^{++}
\right.\nonumber\\
&&\left.
-48\,{\pi }^2\,{\left( \frac{-2 + r_-}{-3 + r_+} \right) }^n\,r_+\,
 \hat{R}_1^{+-}\, \hat{R}_1^{-+}\, \hat{R}_1^{++}
+96\,{\pi }^2\,{\left( \frac{-1 + r_+}{-3 + r_+} \right) }^n\,r_+\,
 \hat{R}_1^{+-}\, \hat{R}_1^{-+}\, \hat{R}_1^{++}
\right.\nonumber\\
&&\left.
-48\,{\pi }^2\,r_-\,{\left( \frac{-1 + r_+}{-3 + r_+} \right) }^n\,r_+\,
 \hat{R}_1^{+-}\, \hat{R}_1^{-+}\, \hat{R}_1^{++}
-72\,{\pi }^2\,{r_-}^2\,{\left( \frac{-1 + r_+}{-3 + r_+} \right) }^n\,r_+\,
 \hat{R}_1^{+-}\, \hat{R}_1^{-+}\, \hat{R}_1^{++}
\right.\nonumber\\
&&\left.
+8\,{\pi }^2\,{r_+}^2\,\hat{R}_1^{+-}\, \hat{R}_1^{-+}\, \hat{R}_1^{++}
-24\,{\pi }^2\,r_-\,{r_+}^2\,\hat{R}_1^{+-}\, \hat{R}_1^{-+}\, \hat{R}_1^{++}
\right.\nonumber\\
&&\left.
+24\,{\pi }^2\,{\left( \frac{-1 + r_+}{-3 + r_+} \right) }^n\,{r_+}^2\,
\hat{R}_1^{+-}\, \hat{R}_1^{-+}\, \hat{R}_1^{++} +
72\,{\pi }^2\,r_-\,{\left( \frac{-1 + r_+}{-3 + r_+} \right) }^n\,{r_+}^2\,
 \hat{R}_1^{+-}\, \hat{R}_1^{-+}\, \hat{R}_1^{++}
\right.\nonumber\\
&&\left.
+8\,{\pi }^2\,{r_+}^3\,\hat{R}_1^{+-}\, \hat{R}_1^{-+}\, \hat{R}_1^{++} -
24\,{\pi }^2\,{\left( \frac{-1 + r_+}{-3 + r_+} \right) }^n\,{r_+}^3\,
 \hat{R}_1^{+-}\, \hat{R}_1^{-+}\, \hat{R}_1^{++} +
32\,{\pi }^2\,{\left( \frac{r_+}{-3 + r_+} \right) }^n\,\hat{R}_1^{+-}\,
\hat{R}_1^{-+}\, \hat{R}_1^{++}
\right.\nonumber\\
&&\left.
+16\,{\pi }^2\,r_-\,{\left( \frac{r_+}{-3 + r_+} \right) }^n\,
 \hat{R}_1^{+-}\, \hat{R}_1^{-+}\, \hat{R}_1^{++}
-32\,{\pi }^2\,{r_-}^2\,{\left( \frac{r_+}{-3 + r_+} \right) }^n\,
 \hat{R}_1^{+-}\, \hat{R}_1^{-+}\, \hat{R}_1^{++}
\right.\nonumber\\
&&\left.
 -16\,{\pi }^2\,{r_-}^3\,{\left( \frac{r_+}{-3 + r_+} \right) }^n\,
 \hat{R}_1^{+-}\, \hat{R}_1^{-+}\, \hat{R}_1^{++}
-16\,{\pi }^2\,r_+\,{\left( \frac{r_+}{-3 + r_+} \right) }^n\,
 \hat{R}_1^{+-}\, \hat{R}_1^{-+}\, \hat{R}_1^{++}
\right.\nonumber\\
&&\left.
+64\,{\pi }^2\,r_-\,r_+\,{\left( \frac{r_+}{-3 + r_+} \right) }^n\,
 \hat{R}_1^{+-}\, \hat{R}_1^{-+}\, \hat{R}_1^{++} +
48\,{\pi }^2\,{r_-}^2\,r_+\,{\left( \frac{r_+}{-3 + r_+} \right) }^n\,
 \hat{R}_1^{+-}\, \hat{R}_1^{-+}\, \hat{R}_1^{++}
\right.\nonumber\\
&&\left.
 -32\,{\pi }^2\,{r_+}^2\,{\left( \frac{r_+}{-3 + r_+} \right) }^n\,
 \hat{R}_1^{+-}\, \hat{R}_1^{-+}\, \hat{R}_1^{++} -
48\,{\pi }^2\,r_-\,{r_+}^2\,{\left( \frac{r_+}{-3 + r_+} \right) }^n\,
 \hat{R}_1^{+-}\, \hat{R}_1^{-+}\, \hat{R}_1^{++}
 \right.\nonumber\\
&&\left.
+16\,{\pi }^2\,{r_+}^3\,{\left( \frac{r_+}{-3 + r_+} \right) }^n\,
 \hat{R}_1^{+-}\, \hat{R}_1^{-+}\, \hat{R}_1^{++}
 - 32\,{\pi }^2\,\hat{R}_1^{++}\, \hat{R}_1^{+-}\,
 \hat{R}_1^{-+} +16\,{\pi }^2\,r_-\,\hat{R}_1^{++}\, \hat{R}_1^{+-}\, \hat{R}_1^{-+}
 \right.\nonumber\\
\nonumber
\ea
\end{small}
%%%%%%%%%%%%%%%%%%%%%%%%%%%%
\begin{small}
\ba
&&\left.
 +32\,{\pi }^2\,{r_-}^2\,\hat{R}_1^{++}\, \hat{R}_1^{+-}\, \hat{R}_1^{-+} -
16\,{\pi }^2\,{r_-}^3\,\hat{R}_1^{++}\, \hat{R}_1^{+-}\, \hat{R}_1^{-+} -
96\,{\pi }^2\,{\left( \frac{-1 + r_-}{-3 + r_+} \right) }^n\,\hat{R}_1^{++}\,
\hat{R}_1^{+-}\, \hat{R}_1^{-+}
 \right.\nonumber\\
&&\left.
+48\,{\pi }^2\,r_-\,{\left( \frac{-1 + r_-}{-3 + r_+} \right) }^n\,
 \hat{R}_1^{++}\, \hat{R}_1^{+-}\, \hat{R}_1^{-+} +
96\,{\pi }^2\,{\left( \frac{-2 + r_+}{-3 + r_+} \right) }^n\,\hat{R}_1^{++}\,
\hat{R}_1^{+-}\, \hat{R}_1^{-+} -
\right.\nonumber\\
&&\left.
96\,{\pi }^2\,r_-\,{\left( \frac{-2 + r_+}{-3 + r_+} \right) }^n\,
 \hat{R}_1^{++}\, \hat{R}_1^{+-}\, \hat{R}_1^{-+} -
24\,{\pi }^2\,{r_-}^2\,{\left( \frac{-2 + r_+}{-3 + r_+} \right) }^n\,
 \hat{R}_1^{++}\, \hat{R}_1^{+-}\, \hat{R}_1^{-+}
\right.\nonumber\\
&&\left.
 +24\,{\pi }^2\,{r_-}^3\,{\left( \frac{-2 + r_+}{-3 + r_+} \right) }^n\,
 \hat{R}_1^{++}\, \hat{R}_1^{+-}\, \hat{R}_1^{-+} -
16\,{\pi }^2\,r_+\,\hat{R}_1^{++}\, \hat{R}_1^{+-}\, \hat{R}_1^{-+}
-64\,{\pi }^2\,r_-\,r_+\,\hat{R}_1^{++}\, \hat{R}_1^{+-}\, \hat{R}_1^{-+}
\right.\nonumber\\
&&\left.
+48\,{\pi }^2\,{r_-}^2\,r_+\,\hat{R}_1^{++}\, \hat{R}_1^{+-}\, \hat{R}_1^{-+} -
48\,{\pi }^2\,{\left( \frac{-1 + r_-}{-3 + r_+} \right) }^n\,r_+\,
 \hat{R}_1^{++}\, \hat{R}_1^{+-}\, \hat{R}_1^{-+} +
 \right.\nonumber\\
&&\left.
96\,{\pi }^2\,{\left( \frac{-2 + r_+}{-3 + r_+} \right) }^n\,r_+\,
 \hat{R}_1^{++}\, \hat{R}_1^{+-}\, \hat{R}_1^{-+} +
48\,{\pi }^2\,r_-\,{\left( \frac{-2 + r_+}{-3 + r_+} \right) }^n\,r_+\,
 \hat{R}_1^{++}\, \hat{R}_1^{+-}\, \hat{R}_1^{-+}
\right.\nonumber\\
&&\left.
-72\,{\pi }^2\,{r_-}^2\,{\left( \frac{-2 + r_+}{-3 + r_+} \right) }^n\,r_+\,
 \hat{R}_1^{++}\, \hat{R}_1^{+-}\, \hat{R}_1^{-+} +
32\,{\pi }^2\,{r_+}^2\,\hat{R}_1^{++}\, \hat{R}_1^{+-}\, \hat{R}_1^{-+}
\right.\nonumber\\
&&\left.
-48\,{\pi }^2\,r_-\,{r_+}^2\,\hat{R}_1^{++}\, \hat{R}_1^{+-}\, \hat{R}_1^{-+} -
24\,{\pi }^2\,{\left( \frac{-2 + r_+}{-3 + r_+} \right) }^n\,{r_+}^2\,
 \hat{R}_1^{++}\, \hat{R}_1^{+-}\, \hat{R}_1^{-+}
\right.\nonumber\\
&&\left.
+72\,{\pi }^2\,r_-\,{\left( \frac{-2 + r_+}{-3 + r_+} \right) }^n\,{r_+}^2\,
 \hat{R}_1^{++}\, \hat{R}_1^{+-}\, \hat{R}_1^{-+} +
16\,{\pi }^2\,{r_+}^3\,\hat{R}_1^{++}\, \hat{R}_1^{+-}\, \hat{R}_1^{-+}
 \right.\nonumber\\
&&\left.
-24\,{\pi }^2\,{\left( \frac{-2 + r_+}{-3 + r_+} \right) }^n\,{r_+}^3\,
 \hat{R}_1^{++}\, \hat{R}_1^{+-}\, \hat{R}_1^{-+} +
32\,{\pi }^2\,{\left( \frac{r_+}{-3 + r_+} \right) }^n\,\hat{R}_1^{++}\,
\hat{R}_1^{+-}\, \hat{R}_1^{-+}
 \right.\nonumber\\
&&\left.
+32\,{\pi }^2\,r_-\,{\left( \frac{r_+}{-3 + r_+} \right) }^n\,
 \hat{R}_1^{++}\, \hat{R}_1^{+-}\, \hat{R}_1^{-+} -
8\,{\pi }^2\,{r_-}^2\,{\left( \frac{r_+}{-3 + r_+} \right) }^n\,
 \hat{R}_1^{++}\, \hat{R}_1^{+-}\, \hat{R}_1^{-+}
 \right.\nonumber\\
&&\left.
 -8\,{\pi }^2\,{r_-}^3\,{\left( \frac{r_+}{-3 + r_+} \right) }^n\,
 \hat{R}_1^{++}\, \hat{R}_1^{+-}\, \hat{R}_1^{-+} -
32\,{\pi }^2\,r_+\,{\left( \frac{r_+}{-3 + r_+} \right) }^n\,
 \hat{R}_1^{++}\, \hat{R}_1^{+-}\, \hat{R}_1^{-+}
 \right.\nonumber\\
&&\left.
+16\,{\pi }^2\,r_-\,r_+\,{\left( \frac{r_+}{-3 + r_+} \right) }^n\,
 \hat{R}_1^{++}\, \hat{R}_1^{+-}\, \hat{R}_1^{-+}
 24\,{\pi }^2\,{r_-}^2\,r_+\,{\left( \frac{r_+}{-3 + r_+} \right) }^n\,
 \hat{R}_1^{++}\, \hat{R}_1^{+-}\, \hat{R}_1^{-+}
 \right.\nonumber\\
&&\left.
-8\,{\pi }^2\,{r_+}^2\,{\left( \frac{r_+}{-3 + r_+} \right) }^n\,
 \hat{R}_1^{++}\, \hat{R}_1^{+-}\, \hat{R}_1^{-+} -
24\,{\pi }^2\,r_-\,{r_+}^2\,{\left( \frac{r_+}{-3 + r_+} \right) }^n\,
 \hat{R}_1^{++}\, \hat{R}_1^{+-}\, \hat{R}_1^{-+}
 \right.\nonumber\\
&&\left.
 +8\,{\pi }^2\,{r_+}^3\,{\left( \frac{r_+}{-3 + r_+} \right) }^n\,
\hat{R}_1^{++}\, \hat{R}_1^{+-}\, \hat{R}_1^{-+} \right)\vec{A}_0
\nonumber\\
&&\left[48\,{\pi }^2\,\left( 4 + {r_-}^4 - 4\,{r_-}^3\,r_+ - 5\,{r_+}^2 + {r_+}^4 +
{r_-}^2\,\left( -5 + 6\,{r_+}^2 \right)  + r_-\,\left( 10\,r_+ - 4\,{r_+}^3
\right)\right) \right]^{-1}\,.
\nonumber\\
\ea
\end{small}

\chapter{Applications: Solving the $x$-space Evolution Equations for Transversity at NLO \label{chap2}}
\fancyhead[LO]{\nouppercase{Chapter 2. Solving the $x$-space Evolution Equations for Transversity at NLO}}

\section{Introduction}
One of the most fascinating aspects of the structure of the nucleon
is the study of the distribution of spin among its constituents, a topic of
remarkable conceptual complexity which has gained a lot of attention
in recent years. This study is entirely based on the classification and on the
phenomenological modeling of all the leading-twist parton distributions, used
as building blocks for further investigations in hadronic physics.

There are various theoretical
ways to gather information on these non-local matrix elements. One among the various
possibilities is to discover sum rules connecting moments of these distributions to other
fundamental observables. Another possibility is to discover
bounds - or inequalities - among them and use these results in the process of their modeling.
There are various bounds that can be studied, particularly in the context of the
new generalized parton dynamics typical of the skewed distributions
\cite{Ji,Radyushkin}.
All these relations can be analized in perturbation theory and studied using
the Renormalization Group (RG), although a complete description of their perturbative dynamics
is still missing. This study, we believe,
may require considerable theoretical effort since it involves a global understanding both of the (older) forward (DGLAP) dynamics and of the generalized
new dynamics encoded in the skewed distributions.

In this context, a program aimed at the study of various bounds in perturbation theory using primarily a parton dynamics in $x$-space has been outlined \cite{CafaCor}.
This requires accurate algorithms to solve the equations up to
next-to-leading order (NLO). Also, underlying this type of description is, in many cases, a
probabilistic approach \cite{Teryaev1} which has some interesting consequences worth
of a closer look . In fact, the DGLAP equation, viewed as a probabilistic process,
can be rewritten in a {\em master form} which is at the root of some interesting
formal developements. In particular, a wide set of results, available from the theory
of stochastic processes, find their way in the study of the evolution.
In a recent work \cite{CafaCor} it has been proposed a Kramers-Moyal expansion of the
DGLAP equation as an alternative way to describe the dynamics of parton evolution.
Here, this analysis will be extended to the case of the non-forward evolution.

With these objectives in mind, in this study we
test  $x$-space algorithms up to NLO developed in the previous chapter,
and verify their accuracy using a stringent test:
Soffer's inequality. As usual, we are bound to work with specific models of initial conditions.
The implementations on which our analysis are based are general,
with a varying flavour number $n_f$ at any threshold of intermediate quark mass
in the evolution.
Here, we address Soffer's inequality using an approach
based on the notion of ``superdistributions''
\cite{Teryaev},
which are constructs designed to have a simple (positive)
evolution thanks to the existence of an underlying master form \cite{Teryaev1,CafaCor}.
The original motivation for using such a master form
(also termed {\em kinetic} or {\em probabilistic})
to prove positivity has been presented in \cite{Teryaev},
while further extensions of these arguments have been presented in \cite{CafaCor}.
In a final section we propose the extension of the evolution algorithm
to the case of the skewed distributions, and illustrate its implementation
in the non-singlet case. As for the forward case,
numerical tests of the inequality are performed for
two different models. We show that even starting from a saturated inequality at the
lowest evolution scale, the various models differ significantly even for a moderate final
factorization scale of $Q=100$ GeV. Finally, we illustrate in another application the
evolution of the tensor charge and show that, in the models considered, differences
in the prediction of the tensor charge are large.

\section{Prelude to $x$-space: A Simple Proof of Positivity of $h_1$ to NLO}
There are some nice features of the parton dynamics, at least in the leading logarithmic approximation (LO), when viewed in $x$-space, once a suitable ``master form'' of the
parton evolution equations is identified.

The existence of such a master form, as firstly shown by Teryaev,
is a special feature of the evolution equation
itself. The topic has been addressed before
in LO \cite{Teryaev} and reanalized in more detail in
\cite{CafaCor} where, starting from a kinetic interpretation of the
evolution, a differential equation obtained
from the Kramers-Moyal expansion of the DGLAP equation
has also been proposed.

The arguments of refs.~\cite{Teryaev,CafaCor}
are built around a form of the evolution equation
which has a simple kinetic interpretation and is written
in terms of transition probabilities constructed from the kernels.

The strategy used, at least in leading order,
to demonstrate the positivity of the LO evolution for
special combinations of parton distributions
$\Q_\pm$ \cite{Teryaev}, to be defined below, or the NLO evolution for $h_1$,
which we are going to address, is based on some results of
ref.\cite{Teryaev}, briefly reviewed here, in order
to be self-contained.

 A master equation is typically given by
\beq
\frac{\partial }{\partial \tau}f(x,\tau)=\int dx'\left(
w(x|x') f(x',\tau) -w(x'|x) f(x,\tau)\right) dx'
\label{masterforms}
\eeq

and if through some manipulations, a DGLAP equation

\beq
\frac{d q(x,Q^2)}{d \log( Q^2)} = \int_x^1 \frac{dy}{y} P(x/y)q(y,Q^2),
\eeq
with kernels $P(x)$, is rewritten in such a way to resemble
eq. (\ref{masterforms})

\beq
\frac{d}{d \tau}q(x,\tau) = \int_x^1 dy \hat{P}\left(\frac{x}{y}\right)\frac{q(y,\tau)}{y}
-\int_0^x \frac{dy}{y}\hat{ P}\left(\frac{y}{x}\right)\frac{q(x,\tau)}{x},
\label{bolz}
\eeq
with a (positive) transition probability
\beq
w(x|y)= \frac{\alpha_s}{2 \pi} \hat{P}(x/y)\frac{\theta(y > x)}{y}
\eeq
then positivity of the evolution is established.

For equations of non-singlet type, such
as those evolving $q^{(-)}=q - \bar{q}$, the valence quark distribution,
or $h_1$, the transverse spin distribution,
this rewriting of the equation is possible, at least in LO.
NLO proofs are, in general, impossible to construct by this method,
since the kernels turn out, in many cases, to be negative. The only possible proof,
in these cases, is just a numerical one, for suitable (positive)
boundary conditions observed by the initial form of the parton distributions.
Positivity of the evolution is then a result of an unobvious interplay between
the various contributions to the kernels in various regions in $x$-space.

In order to discuss the probabilistic version of the DGLAP equation it
is convenient to separate the bulk contributions of the kernels $(x<1)$ from the
edge point contributions at $x=1$. For this purpose
we recall that the structure of the kernels is, in general, given by
\beq
P(z) = \hat{P}(z) - \delta(1-z) \int_0^1 \hat{P}(z)\, dz,
\label{form}
 \eeq
where the bulk contributions $(z<1)$ and the edge point contributions
$(\sim \delta(z-1))$ have been explicitely separated.
We focus on the transverse spin distributions as an example.
With these prerequisites,
proving the LO and NLO positivity of the transverse spin distributions
is quite straightforward, but requires a numerical inspection of the transverse
kernels. Since the evolutions for $\Delta_T q^{(\pm)}\equiv h^q_1$ are purely non-singlet,
diagonality in flavour of the subtraction terms $(\sim \int_0^x w(y|x)q(x,\tau))$
is satisfied, while the edge-point subtractions can be tested
to be positive numerically.
We illustrate the explicit construction of the master equation for $h_1$ in LO, since extensions to NLO of this construction are rather straighforward.

In this case the LO kernel is given by

\beqn
\Delta_{T}P^{(0)}_{qq}(x)= C_{F}\left[\frac{2}{(1-x)_{+}}-2 +\frac{3}{2}\delta(1-x)\right]
\eeqn
and by some simple manipulations we
can rewrite the corresponding evolution equation
in a suitable master form. That this is possible is an elementary fact
since the subtraction terms
can be written as integrals of a positive function. For instance,
a possibility is
to choose the transition probabilities
\beqa
w_1[x|y] &=& \frac{C_F}{y}\left(\frac{2}{1- x/y} - 2 \right)
\theta(y>x) \theta(y<1)\nonumber \\
w_2[y|x] &=& \frac{C_F}{x} \left(\frac{2}{1- y/x} - \frac{3}{2}\right)
\theta(y > -x)\theta(y<0)
\nonumber \\
\eeqa
which reproduce the evolution equation for $h_1$ in master form

\beq
\frac{d h_1}{d \tau}= \int_0^1 dy w_1(x|y)h_1(y,\tau)
-\int_0^1 dy w_2(y|x) h_1(x,\tau).
\label{masterix}
\eeq

\begin{figure}
{\centering \resizebox*{12cm}{!}{\rotatebox{-90}{\includegraphics{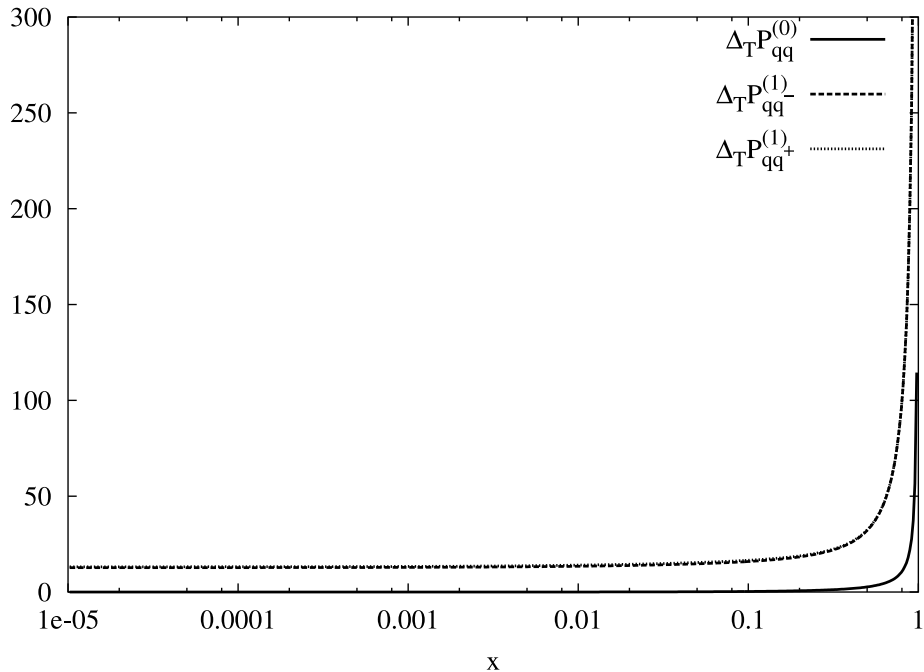}}} \par}
\caption{Plot of the transverse kernels.}
\label{transversekernels}
\end{figure}

The NLO proof of positivity is also rather straightforward.
For this purpose we have analized numerically the behaviour of the NLO kernels both
in their bulk region and at the edge-point.
We show in Table 1 of Appendix B results
for the edge point contributions to NLO for both of
the $\Delta_T P^{(1)}_\pm$ components,
which are numerically the same.
There we have organized these terms in the form $\sim C\delta(1-x)$ with
\beq
C=-\log(1- \Lambda) A + B
\label{sub},
\eeq
with A and B being numerical coefficients depending on the number
of flavours included in the kernels.
The (diverging) logarithmic contribution ($\sim \int_0^\Lambda dz/(1-z)$)
have been regulated by a cutoff. This divergence
in the convolution cancels when these terms are combined with the divergence at
$x=1$ of the first term of the master equation (\ref{masterix})
for all the relevant components
containing ``+'' distributions. As for the
bulk contributions $(x<1)$, positivity up to NLO of the transverse kernels
is shown numerically in Fig. (\ref{transversekernels}).
All the conditions of positivity are therefore satisfied and therefore
the $\Delta_{T\pm}q$ distributions evolve positively up to NLO.
The existence of a master form of the equation is then guaranteed.

Notice that the NLO positivity of $\Delta_{T\pm}q$ implies positivity of the
nucleon tensor charge \cite{JJ}
\beq
\delta q\equiv\int_0^1 dx \left( h_1^q(x) - h_1^{\bar{q}(x)}\right)
\eeq
for each separate flavour for positive initial conditions.
As we have just shown, this proof of positivity is very short, as far as one
can check numerically that both components of eq.(\ref{masterix})
are positive.

\section{Soffer's inequality}
Numerical tests of Soffer's inequality can be performed
either in moment space or, as we are going to illustrate
in the next section, directly in $x$-space, using suitable
algorithms to capture the perturbative nature of the evolution.
We recall that Soffer's inequality
\beq
|h_1(x)| < q^+(x)
\eeq
sets a bound on the transverse spin distribution $h_1(x)$ in terms of the
components of the positive helicity component of the quarks, for a given flavour.
An original proof of Soffer's inequality
in LO has been discussed in ref.\cite{Barone}, while
in \cite{Teryaev} an alternative proof was presented, based
on a kinetic interpretation of the evolution equations.

We recall that $h_1$, also denoted by the symbol
\begin{equation}
\Delta _{T}q(x,Q^{2})\equiv q^{\uparrow }(x,Q^{2})-q^{\downarrow }(x,Q^{2}),
\end{equation}
has the property
of being purely non-singlet and of appearing at leading twist. It is
identifiable in transversely polarized
hadron-hadron collisions and not in Deep Inelastic Scattering (from now on we will
omit sometime the x-dependence in the kernels and in the distributions when obvious).
In the following we will use interchangeably the notations $h_1\equiv h_1^q$
and $\Delta_T q$ to 
denote the transverse asymmetries. We introduce also the combinations 
\beqa
\Delta_T(q + \bar{q}) &=& h_1^q + h_1^{\bar{q}} \nonumber \\
\Delta_T q^{(-)}=\Delta_T(q - \bar{q}) &=& h_1^q - h_1^{\bar{q}} \nonumber \\
\Delta_T q^{(+)} &=& \sum_i \Delta_T(q_i + \bar{q}_i) \nonumber \\
\eeqa
where we sum over the flavor index $(i)$, and we have introduced singlet and non-singlet 
contributions for distributions of fixed helicities 
\beqa
q_+^{(+)}&=&\sum_i\left( q_{+ i} + \bar{q}_{+ i}\right)\nonumber \\
q_+^{(-)}&=& q_{+ i} -\bar{q}_{+ i}\equiv \Sigma. \nonumber \\
\eeqa
In our analysis we solve all the equations in the helicity basis and reconstruct 
the various helicities after separating singlet and non-singlet sectors. 
We mention that 
the non-singlet sector is now given by a set of 2 equations, each involving 
$\pm$ helicities and the singlet sector is given by a 4-by-4 matrix.   

In the singlet sector we have 

\begin{eqnarray}
{dq_+^{(+)} \over{dt}}=
{\alpha_s \over {2 \pi}} (P_{++}^{qq}\otimes q_+^{(+)}+
P_{+-}^{qq} \otimes q_-^{(-)}  \nonumber \\
+P_{++}^{qG} \otimes G_++
P_{+-}^{qG} \otimes G_-),
\nonumber \\
{dq_-^{(+)}(x) \over{dt}}=
{\alpha_s \over {2 \pi}} (P_{+-} \otimes q_+^{(+)} +
P_{++}  \otimes q_-^{(+)} \nonumber \\
+P_{+-}^{qG} \otimes G_+ +
P_{++}^{qG} \otimes G_-),  \nonumber \\
{dG_+(x) \over{dt}}=
{\alpha_s \over {2 \pi}} (P_{++}^{Gq} \otimes q_+^{(+)}+
P_{+-}^{Gq} \otimes q_-^{(+)} \nonumber \\
+P_{++}^{GG}\otimes G_+ +
P_{+-}^{GG} \otimes G_-),  \nonumber \\
{dG_-(x) \over{dt}}=
{\alpha_s \over {2 \pi}} (P_{+-}^{Gq} \otimes q_+^{(+)} +
P_{++}^{Gq} \otimes q_-^{(+)} \nonumber \\
+P_{+-}^{GG} \otimes G_+ +
P_{++}^{GG} \otimes G_-).
\label{hs}\end{eqnarray}

while the non-singlet (valence) analogue of this equation is also easy to
write down
\begin{eqnarray}
{dq_{+ i}^{(-)}(x) \over{dt}}=
{\alpha_s \over {2 \pi}} (P^{NS}_{++} \otimes q_{+ i}^{(-)}+
P^{NS}_{+-} \otimes q_{-}^{(-)}(y)), \nonumber \\
{dq_{- i}^{(-)}(x) \over{dt}}=
{\alpha_s \over {2 \pi}} (P^{NS}_{+-} \otimes q_{+}^{(-)}+
P^{NS}_{++} \otimes q_{- i}^{(-)}).
\label{h}\end{eqnarray}
Above, $i$ is the flavor index, $(\pm)$ indicate $q\pm \bar{q}$ components and the lower subsctipt $\pm$ stands for the helicity.

Similarly to the unpolarized case the flavour reconstruction is done by adding 
two additional equations for each flavour in the helicity $\pm$
\beq
\chi_{\pm,i}= q_{\pm i}^{(+)}- \frac{1}{n_f}q^{(+)}_\pm
\eeq
whose evolution is given by 
\beqa
{d \chi_{+ i}^{(-)}(x) \over{dt}} &=&
{\alpha_s \over {2 \pi}} (P^{NS}_{++} \otimes \chi_{+ i} +
P^{NS}_{+-} \otimes \chi_{- i}), \nonumber \\
{d \chi_{- i} (x) \over{dt}} &=&
{\alpha_s \over {2 \pi}} (P^{NS}_{+-} \otimes \chi_{+ i} +
P^{NS}_{++} \otimes \chi_{- i}). \nonumber \\
\label{h11}
\end{eqnarray}

The reconstruction of the various contributions in flavour space 
for the two helicities is finally done 
using the linear combinations 
\beq
q_{\pm i}=\frac{1}{2}\left( q_{\pm i}^{(-)} + \chi_{\pm i} +\frac{1}{n_f}q_{\pm}^{(+)}\right).
\eeq

We will be needing these equations below when we present
a proof of positivity up to LO, and we will thereafter proceed with a NLO implementation of these and other evolution equations. For this we will be needing some more notations. 

We recall that the following relations are also true to all orders 
\beqa
P(x) &=&\frac{1}{2}\left( P_{++}(x) + P_{+-}(x)\right)\nonumber \\
&=&\frac{1}{2}\left( P_{--}(x) + P_{-+}(x)\right)\nonumber 
\eeqa
between polarized and unpolarized $(P)$ kernels 
and 
\beq
P_{++}(x) =  P_{--}(x),\,\,\,P_{-+}(x)=P_{+-}(x)
\eeq
relating unpolarized kernels to longitudinally polarized ones. 
Generically, the kernels of various type are expanded up to NLO as 
\beq
P(x)= \frac{\alpha_s}{2 \pi} P^{(0)}(x) + \left(\frac{\alpha_s}{2 \pi}\right)^2 P^{(1)}(x),
\eeq
and specifically, in the transverse case we have

\begin{eqnarray} \label{pm}
\Delta_T P_{qq,\pm}^{(1)} &\equiv& \Delta_T P_{qq}^{(1)} \pm \Delta_T 
P_{q\bar{q}}^{(1)} \; , \\
\end{eqnarray}
with the corresponding evolution equations 

\begin{equation} \label{evol3}
\frac{d}{d\ln Q^2} \Delta_T q_{\pm} (Q^2) = \Delta_T P_{qq,\pm} 
(\alpha_s (Q^2))\otimes \Delta_T q_\pm (Q^2) \; .
\end{equation}

We also recall that the kernels in the helcity basis in LO are given by 
\beqa
P_{NS\pm,++}^{(0)} &=&P_{qq, ++}^{(0)}=P_{qq}^{(0)}\nonumber \\
P_{qq,+-}^{(0)}&=&P_{qq,-+}^{(0)}= 0\nonumber \\
P_{qg,++}^{(0)}&=& n_f x^2\nonumber \\
P_{qg,+-}&=& P_{qg,-+}= n_f(x-1)^2 \nonumber \\
P_{gq,++}&=& P_{gq,--}=C_F\frac{1}{x}\nonumber \\ 
P_{gg,++}^{(0)}&=&P_{gg,++}^{(0)}= N_c
\left(\frac{2}{(1-x)_+} +\frac{1}{x} -1 -x - x^2 \right) +{\beta_0}\delta(1-x) \nonumber \\
P_{gg,+-}^{(0)}&=& N_c
\left( 3 x +\frac{1}{x} -3 - x^2 \right). 
\label{stand1}
\eeqa

An inequality, such as Soffer's inequality, can be stated as positivity condition 
for suitable linear combinations of parton distributions \cite{Teryaev} 
and this condition can be analized - as we have just shown 
for the $h_1$ case -  in a most direct way using the master form.

For this purpose consider the linear valence combinations
\beqa
\Q_+ &=& q_+ + h_1 \nonumber \\
\Q_-  &=& q_+ - h_1 \nonumber \\
\eeqa
which are termed ``superdistributions'' in ref.\cite{Teryaev}. Notice that a proof 
of positivity of the $\Q$ distributions is equivalent to verify Soffer's inequality. 
However, given the mixing of singlet and non-singlet sectors, the analysis of 
the master form is, in this case, more complex. As we have just mentioned, what can spoil the proof of 
positivity, in general, is the negativity of the kernels to higher order. 
We anticipate here the result that we will illustrate below where we show 
that a LO proof of the positivity of the evolution for $\Q$ can be established using 
kinetic arguments, being the kernels are positive at this order. However 
we find 
that the NLO kernels do not satisfy 
this condition. 
In any case, let's see how the identification of such master form proceeds in general. 
We find useful to illustrate the result using the separation between singlet and non-singlet 
sectors. In this case we introduce the combinations

\beqa
\Q_\pm^{(-)} &=& q_+^{(-)} \pm h_1^{(-)} \nonumber \\
\Q_\pm^{(+)} &=& q_+^{(+)} \pm h_1^{(+)} \nonumber \\
\label{separation}
\eeqa
with $h_1^{(\pm)}\equiv \Delta_T q^{(\pm)}$.  

Differentiating these two linear combinations  (\ref{separation}) we get
\beqa
\frac{d \Q_\pm^{(-)}}{d\log(Q^2)}= P^{NS}_{++} q_+^{(-)} 
+ P^{NS}_{+ -} q_-^{(-)} \pm P_T h_1^{(-)} \nonumber \\
\eeqa
which can be rewritten as
\beqa
\frac{d \Q_+^{(-)}}{d\log(Q^2)} &=& \frac{1}{2}\left(P_{++}^{(-)} + P_T^{(-)}\right)\Q_+^{(-)}
 + \frac{1}{2}\left(P_{++}^{(-)} - P_T^{(-)}\right)\Q_-^{(-)} + P_{+ -}^{(-)}q_-^{(-)} \nonumber \\
\frac{d \Q_+^{(-)}}{d\log(Q^2)} &=& \frac{1}{2}\left(P_{++}^{(-)} - P_T^{(-)}\right)\Q_+^{(-)}
+ \frac{1}{2}\left(P_{++}^{(-)} + P_T\right)^{(-)}\Q_-^{(-)} + P_{+ -}^{(-)}q_-^{(-)} \nonumber \\
\eeqa
with $P^{(-)}\equiv P^{NS}$ being the non-singlet (NS) kernel. 

At this point we define the linear combinations
\beqa
{\bar{P}^Q}_{+\pm}= \frac{1}{2}\left(P_{++} \pm P_T\right)
\eeqa
and rewrite the equations above as
\beqa
\frac{d \Q_+ i}{d\log(Q^2)} &=& \bar{P}^Q_{ ++}\Q_{i+}
 + \bar{P}^Q_{+-}\Q_{i-} + P^{qq}_{+ -}q_{i-} \nonumber \\
\frac{d \Q_{i+}}{d\log(Q^2)} &=& \bar{P}^Q_{+-}\Q_{i+}
 + \bar{P}^Q_{++}\Q_{i-} + P_{+ -}^{qq}q_{i-} \nonumber \\
\label{pos1}
\eeqa
where we have reintroduced $i$ as a flavour index.
From this form of the equations it is easy to establish the leading order positivity of the evolution, after checking the positivity of the kernel and the existence of a master form.

The second non-singlet sector is defined via the variables 

\beq
\chi_{i\pm}=q_{i \pm}^{(+)} - \frac{1}{n_f}q_{i\pm}^{(+)}
\eeq
which evolve as non-singlets 
and the two additional distributions 
\beqa
\Q_{\chi i,\pm}= \chi_{i +} \pm h_1^{i(+)}.
\eeqa
Also in this case we introduce  the kernels 
\beqa
{\bar{P}^{Q_\chi}}_{+\pm} &=& \frac{1}{2}\left(P_{++} \pm 
\Delta_T P^{(+)}\right)
\eeqa
to obtain the evolutions 
\beqa
\frac{d \Q_{\chi i+}}{d\log(Q^2)} &=& \bar{P}^{Q_\chi}_{ ++}\Q_{\chi i+}
 + \bar{P}^{Q_\chi}_{+-}\Q_{\chi i-} + P^{qq}_{+ -}\chi_{i-} \nonumber \\
\frac{d \Q_{i+}}{d\log(Q^2)} &=& \bar{P}^Q_{\chi +-}\Q_{\chi i+}
 + \bar{P}^{Q_\chi}_{++}\Q_{\chi i-} + P_{+ -}^{qq}\chi_{i-}. \nonumber \\
\label{pos12}
\eeqa

For the singlet sector, we simply define $Q_+^{(+)}=q^{(+)}$, and the 
corresponding evolution is similar to the singlet equation of the helicity basis. 
Using the equations above, 
the distributions $\Q_{i\pm}$ are then reconstructed as 
\beq
\Q_{i\pm} = \frac{1}{2}\left(\Q_{i \pm}^{(-)} + 
\Q_{\chi i \pm}^{(-)} + \frac{1}{n_f}Q_+^{(+)}\right)
\eeq
and result positive for any flavour if the addends are positive as well. 
However, as we have just mentioned, positivity of all the kernels introduced above is easy to check numerically to LO, together with their diagonality in flavour which guarantees the existence 
of a master form.  

As an example, consider the LO evolution of $\Q\pm$. 
The proof of positivity is a simple consequence of the structure of eq.~(\ref{pos1}).
In fact the edge-point
contributions appear only in $P^Q_{++}$, i.e. they are diagonal in the evolution of
$\Q_{\pm}$. The inhomogenous terms on the right hand side of (\ref{pos1}), proportional to
$q_-$ are are harmless, since the $P_{+-}$ kernel has no edge-point contributions. Therefore under 1) diagonality in flavour of the subtraction terms and 
2) positivity of first and second term 
(transition probabilities) we can have positivity of the evolution. A 
refined arguments to support this claim has been presented in \cite{CafaCor}.    

\begin{figure}
{\centering \resizebox*{12cm}{!}{\rotatebox{-90}{\includegraphics{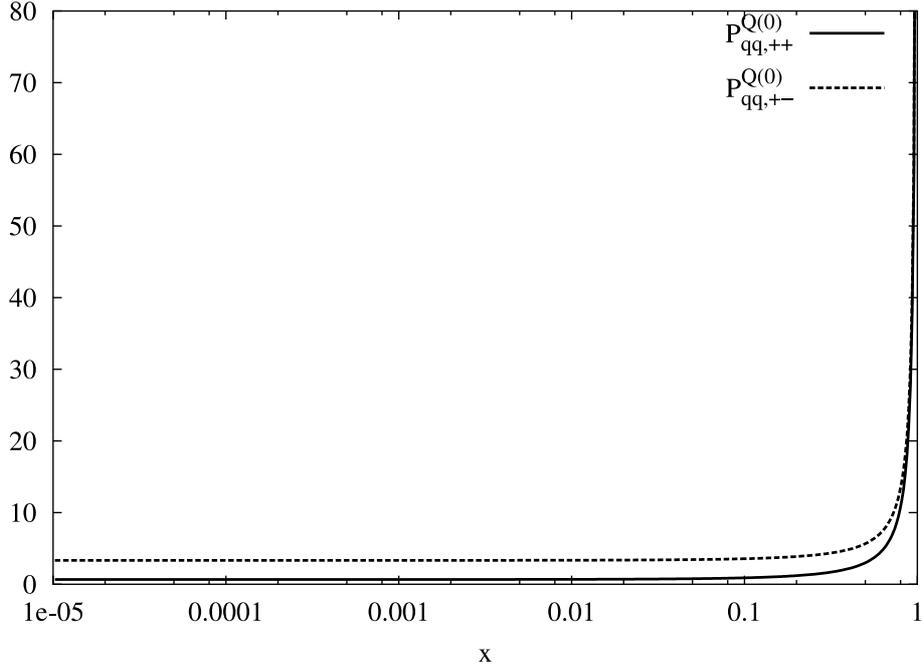}}} \par}
\caption{Plot of the LO kernels for the $\Q$ distributions}
\label{Qkernels}
\end{figure}

\begin{figure}
{\centering \resizebox*{12cm}{!}{\rotatebox{-90}{\includegraphics{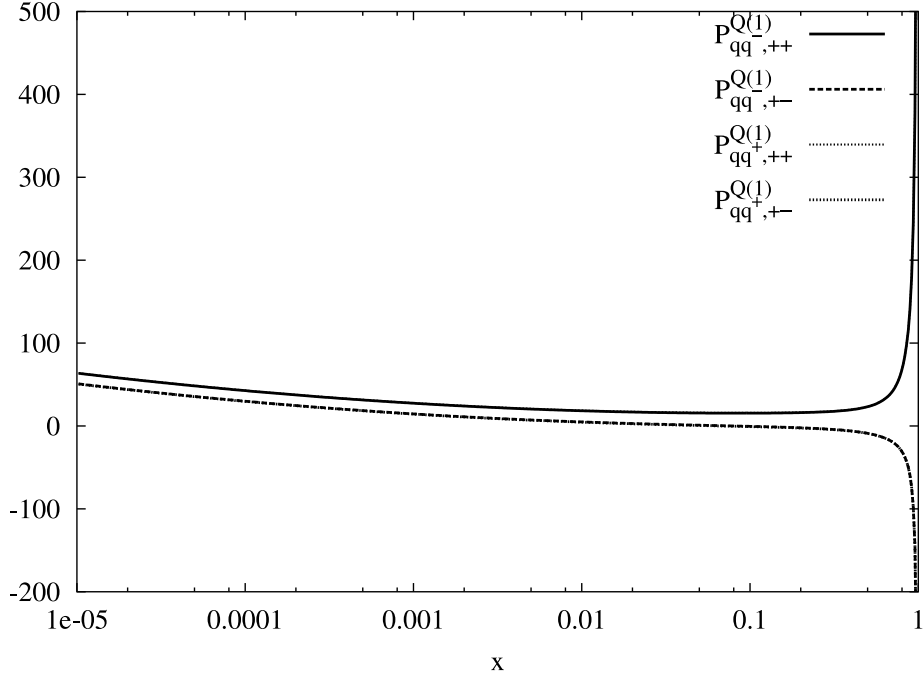}}} \par}
\caption{Plot of the NLO kernels for the $\Q$ distributions, 
showing a negative behaviour at large $x$}
\label{QNLOkernels}
\end{figure}

This construction is not valid to NLO. In fact, while the features of flavour 
diagonality of the 
master equation  are satisfied, the transition probabilities $w(x,y)$ 
are not positive in the whole $x,y$ range. The existence 
of a crossing from positive to negative values in $P^{\Q}_{++}$ 
can, in fact, be established quite easily using a numerical analysis. We illustrate in Figs. 
(\ref{Qkernels}) and (\ref{QNLOkernels}) plots of the $\Q$ kernels at LO and NLO, 
showing that, at NLO, the requirement of positivity of some components is violated.  
The limitations of this sort of proofs -based on kinetic arguments- are strictly 
linked to the positivity of the transition probabilities once a master form of the 
equation is identified.
\footnote{Based on the article published in JHEP 0311:059, (2003)}
\section{The numerical investigation}
We have seen that NLO proofs of positivity, can be -at least partially- obtained
only for suitable sets of boundary conditions. To this purpose, we choose to
investigate the numerical behaviour of the solution using $x$-space
based algorithms which need to be tested up to NLO.

Our study validates a method which can be used to solve evolution equations
with accuracy in leading and in next-to-leading order. The method is entirely
based on an expansion \cite{Rossi} used in the context of spin physics \cite{Gordon}
and in supersymmetry \cite{Coriano}. An interesting feature of the
expansion, once combined with Soffer's inequality, is to generate an infinite set of
relations among the scale invariant coefficients $(A_n, B_n)$ which characterize it.

In this approach, the NLO expansion of the distributions in the DGLAP equation is the one studied
in the previous section and it is given by
\beq
f(x,Q^2)=\sum_{n=0}^{\infty} \frac{A_n(x)}{n!}\log^n
\left(\frac{\alpha(Q^2)}{\alpha(Q_0^2)}\right) +
\alpha(Q^2)\sum_{n=0}^\infty \frac{B_n(x)}{n!}\log^n
\left(\frac{\alpha(Q^2)}{\alpha(Q_0^2)}\right)
\label{expansion}
 \eeq
where, to simplify the notation, we assume a short-hand matrix notation for all the convolution products.
Therefore $f(x,Q^2)$ stands for a vector having as components
any of the helicities of the various flavours  $(\Q_{\pm},q_\pm,G_\pm)$.
The ansatz implies a tower of recursion relations once the running coupling
is kept into account and implies that (see Eqns.~\ref{rr})
\beq
A_{n+1}(x) =  -\frac{2}{\beta_0}P^{(0)}\otimes A_n(x)
\label{recur}
\eeq
to leading order
and
\beqa
B_{n+1}(x) & = & - B_n(x)- \left(\frac{\beta_1}{4 \beta_0} A_{n+1}(x)\right)
- \frac{1}{4 \pi\beta_0}P^{(1)}\otimes A_n(x) -\frac{2}{\beta_0}P^{(0)}
\otimes B_n(x) \nonumber \\
 & = &  - B_n(x) + \left(\frac{\beta_1}{2 \beta_0^2}P^{(0)}\otimes A_n(x)\right)
\nonumber \\
&&- \frac{1}{4 \pi\beta_0}P^{(1)}\otimes A_n(x) -\frac{2}{\beta_0}P^{(0)}
\otimes B_n(x),  \nonumber \\
\label{recur1}
\eeqa
to NLO, relations which are solved with the initial condition $B_0(x)=0$.
The initial conditions for the coefficients $ A_0(x)$ and $B_0(x)$ are specified
with $q(x,Q_0^2)$ as a leading order ansatz for the initial
distribution
\beq
A_0(x)= \delta(1-x)\otimes q(x,Q_0^2)\equiv q_0(x)
\eeq
which also requires $B_0(x)=0$, since we have to
satisfy the boundary condition
\beq
A_0(x) + \alpha_0 B_0(x)= q_0(x).
\label{bdry}
\eeq

If we introduce Rossi's expansion for $h_1$, $q_+$, and the linear combinations $\Q_\pm$ (in short form)
\beqa
h_1 &\sim&\left(A^{h}_n,B^{h +}_n\right)\nonumber \\
q_\pm &\sim&\left(A^{q_\pm}_n,B^{q_\pm}_n\right)\nonumber \\
\Q_\pm &\sim&\left(A^{Q +}_n,B^{Q +}_n\right) \nonumber \\
\eeqa
we easily get the inequalities
\beq
(-1)^n\left(A^{q_+}_n + A^{h}_n\right) >0
\eeq
and
\beq
(-1)^n\left(A^{q_+}_n - A^{h}_n\right) >0
\eeq
valid to leading order,which we can check numerically. 
Notice that the signature factor has to be included due to the alternation
in sign of the expansion.
To next to leading order we obtain 

\beq
 (-1)^{n+1} \left(A^{q_+}_n(x) +\alpha(Q^2) B^{q_+}_n (x)\right)\, 
< (-1)^n \left(A^{h}_n(x)\,+\alpha(Q^2) B^{h}_n (x)\right)\,  <\, (-1)^n \left(A^{q_+}_n(x) +\alpha(Q^2) 
B^{q_+}_n (x)\right), 
\eeq
valid for $n\geq 1$, obtained after identification of the corresponding 
logarithmic powers $\log\left(\alpha(Q^2)\right)$ at any $Q$. 
In general, one can assume a saturation of the inequality at the initial evolution scale 
\beq
\Q_-(x,Q_0^2)=h_1(x,Q_0^2) -\frac{1}{2}q_+(x,Q_0^2)=0. 
\eeq
This initial condition has been evolved in $Q$ solving the equations 
for the $\Q_\pm$ distributions  to NLO.

\section{Relations among moments}
In this section we elaborate on the relation between the coefficients of the 
recursive expansion as defined above and the standard solution 
of the evolution equations in the space of Mellin moments. We will show that the 
two solutions can be related in an unobvious way. 

Of our concern here is the relation between 
the Mellin moments of the coefficients appearing in the expansion 
\beqa
A(N) &=& \int_0^1\,dx \, x^{N-1} A(x)\nonumber \\
B(N) &=&\int_0^1\,dx \, x^{N-1} B(x) \nonumber \\
\eeqa
and those of the distributions
\beq
\Delta_T q_{\pm} (N,Q^2)=\int_0^1\,dx \, x^{N-1} \Delta_T q_{\pm}(x,Q^2)).
\eeq
For this purpose we recall that the general (non-singlet) solution to NLO for the latter moments is given by
\begin{eqnarray} \label{evsol}
\nonumber
\Delta_T q_{\pm} (N,Q^2) &=& K(Q_0^2,Q^2,N)
\left( \frac{\alpha_s (Q^2)}{\alpha_s (Q_0^2)}\right)^{-2\Delta_T
P_{qq}^{(0)}(N)/ \beta_0}\! \Delta_T q_{\pm}(N, Q_0^2)
\label{qTsolution}
\end{eqnarray}
with the input distributions $\Delta_T q_{\pm}^n (Q_0^2)$ at the input scale
$Q_0$
and where we have set
\beq
K(Q_0^2,Q^2,N)= 1+\frac{\alpha_s (Q_0^2)-
\alpha_s (Q^2)}{\pi\beta_0}\!
\left[ \Delta_T P_{qq,\pm}^{(1)}(N)-\frac{\beta_1}{2\beta_0} \Delta_T
P_{qq\pm}^{(0)}(N) \right].
\eeq
In the expressions above we have introduced the corresponding moments for the LO and NLO kernels
($\Delta_T P_{qq}^{(0),N}$,
$ \Delta_T P_{qq,\pm}^{(1),N})$.

We can easily get the relation between the moments of the coefficients of the non-singlet
$x$-space expansion and those of the parton distributions at any $Q$, as expressed by eq.~(\ref{qTsolution})
\beq
A_n(N) + \alpha_s B_n(N)=\Delta_T q_\pm(N,Q_0^2)K(Q_0,Q,N)\left(\frac{-2 \Delta_T P_{qq}(N)}{\beta_0}\right)^n.
\label{relation}
\eeq

As a check of this expression, notice that the initial condition is easily obtained from
(\ref{relation}) setting $Q\to Q_0, n\to 0$, thereby obtaining
\beq
A_0^{NS}(N) + \alpha_s B_0^{NS} (N)= \Delta_T q_\pm(N,Q_0^2)
\eeq
which can be solved with $A_0^{NS}(N)=\Delta_T q_\pm(N,Q_0^2)$ and
$B_0^{NS} (N)=0$.

It is then evident that the expansion (\ref{expansion}) involves a resummation of the logarithmic contributions,
as shown in eq.~(\ref{relation}).

\section{An Example: The Evolution of the Transverse Spin Distributions}

LO and NLO recursion relations for the coefficients of the expansion
can be worked out quite easily, although the numerical implementation of these
equations is far from being obvious. Things are somehow
simpler to illustrate in the case of simple non-singlet evolutions, such
as those involving transverse spin distributions, as we are going to show below.
Some details and definitions can be found in the appendix.

For the first recursion relation (eq. (\ref{recur})) we have
\beqn
&&A^{\pm}_{n+1}(x)=-\frac{2}{\beta_{0}}\Delta_{T}P^{(0)}_{qq}(x)\otimes A^{\pm}_{n}(x)=\nonumber\\ 
&&C_{F}\left(-\frac{4}{\beta_{0}}\right)\left[\int^{1}_{x}\frac{dy}{y}\frac{y A^{\pm}_{n}(y) - x A^{\pm}_{n}(x)}{y-x} + A^{\pm}_{n}(x) \log(1-x)\right]+\nonumber\\
&&C_{F}\left(\frac{4}{\beta_{0}}\right) \left(\int_{x}^{1}\frac{dy}{y} A^{\pm}_{n}(y)\right) + C_{F}\left(-\frac{2}{\beta_{0}}\right)\frac{3}{2} A^{\pm}_{n}(x)\,.
\eeqn
As we move to NLO, it is convenient to summarize 
the structure of the transverse kernel $\Delta_{T}P^{\pm, (1)}_{qq}(x)$ as  

\beqn
&&\Delta_{T}P^{\pm, (1)}_{qq}(x)= K^{\pm}_{1}(x)\delta(1-x) + K^{\pm}_{2}(x)S_{2}(x) +K^{\pm}_{3}(x)\log(x)\nonumber\\
&&+ K^{\pm}_{4}(x)\log^{2}(x) +K^{\pm}_{5}(x)\log(x)\log(1-x) + K^{\pm}_{6}(x)\frac{1}{(1-x)_{+}} + K^{\pm}_{7}(x)\,.    
\eeqn

Hence, for the $(+)$ case we have 

\beqn
&&\Delta_{T}P^{+, (1)}_{qq}(x)\otimes A^{+}_{n}(x) = K^{+}_{1} A^{+}_{n}(x) + \int^{1}_{x}\frac{dy}{y}\left[K^{+}_{2}(z) S_{2}(z) + K^{+}_{3}(z)\log(z) \right.\nonumber\\
&& \left. + \log^{2}(z)K^{+}_{4}(z) + \log(z)\log(1-z)K^{+}_{5}(z)\right] A^{+}_{n}(y) +  \nonumber\\ 
&&K^{+}_{6}\left\{\int^{1}_{x}\frac{dy}{y} \frac{yA^{+}_{n}(y) - xA^{+}_{n}(x)}{y-x} + A^{+}_{n}(x)\log(1-x) \right\} + K^{+}_{7}\int^{1}_{x}\frac{dy}{y}A^{+}_{n}(y)\,, 
\eeqn

where $z={x}/{y}$. For the $(-)$ case we get a similar expression.
  
\begin{figure}[tbh]
{\centering \resizebox*{12cm}{!}{\rotatebox{-90}{\includegraphics{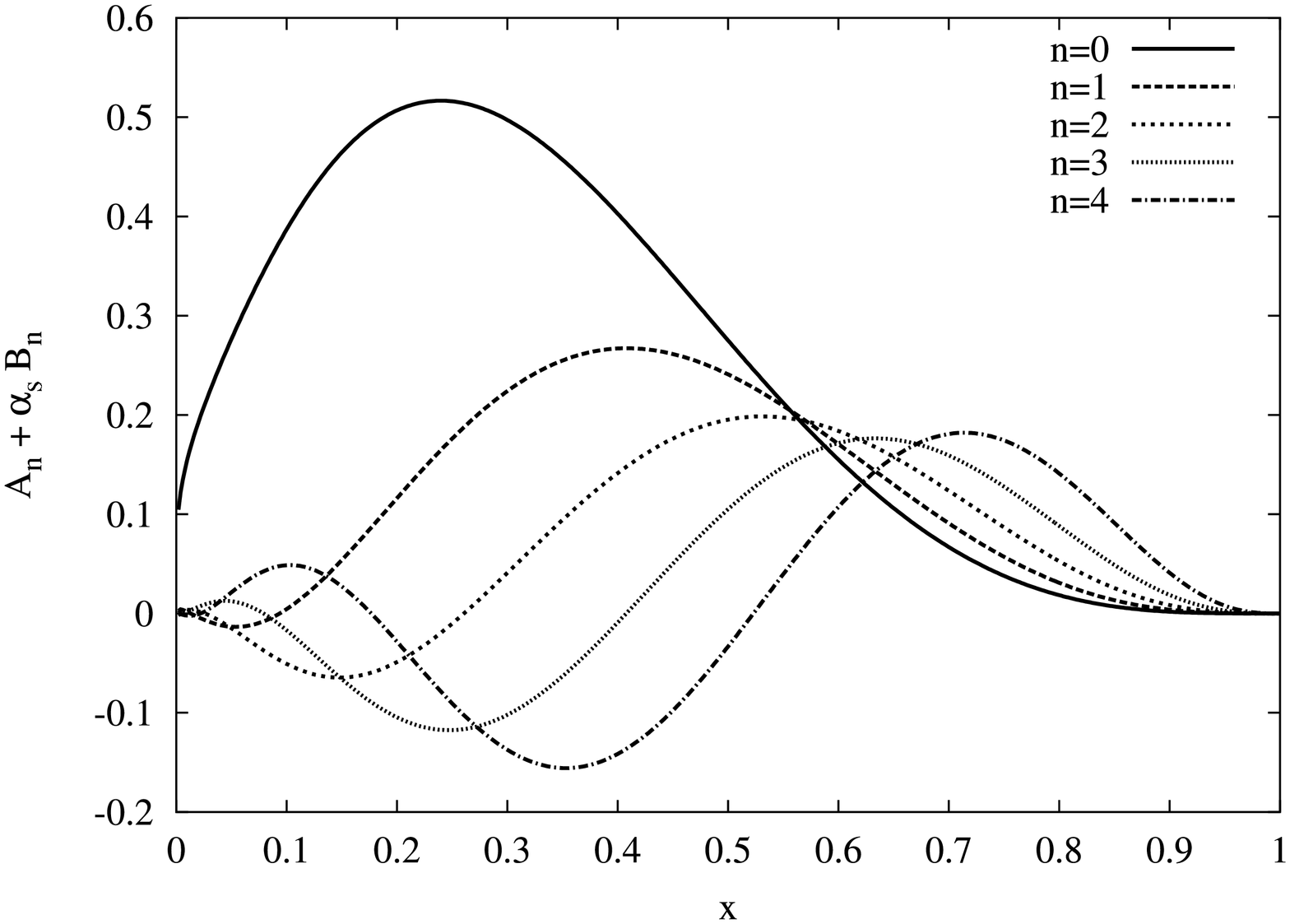}}} \par}
\caption{Coefficients \protect\( A_{n}(x)+\alpha _{s}(Q^{2})B_{n}\protect \),
with \protect\( n=0,\ldots ,4\protect \) for a final scale \protect\( Q=100\protect \)
GeV for the quark up.}
\label{an}
\end{figure}

\begin{figure}[tbh]
{\centering \resizebox*{12cm}{!}{\rotatebox{-90}{\includegraphics{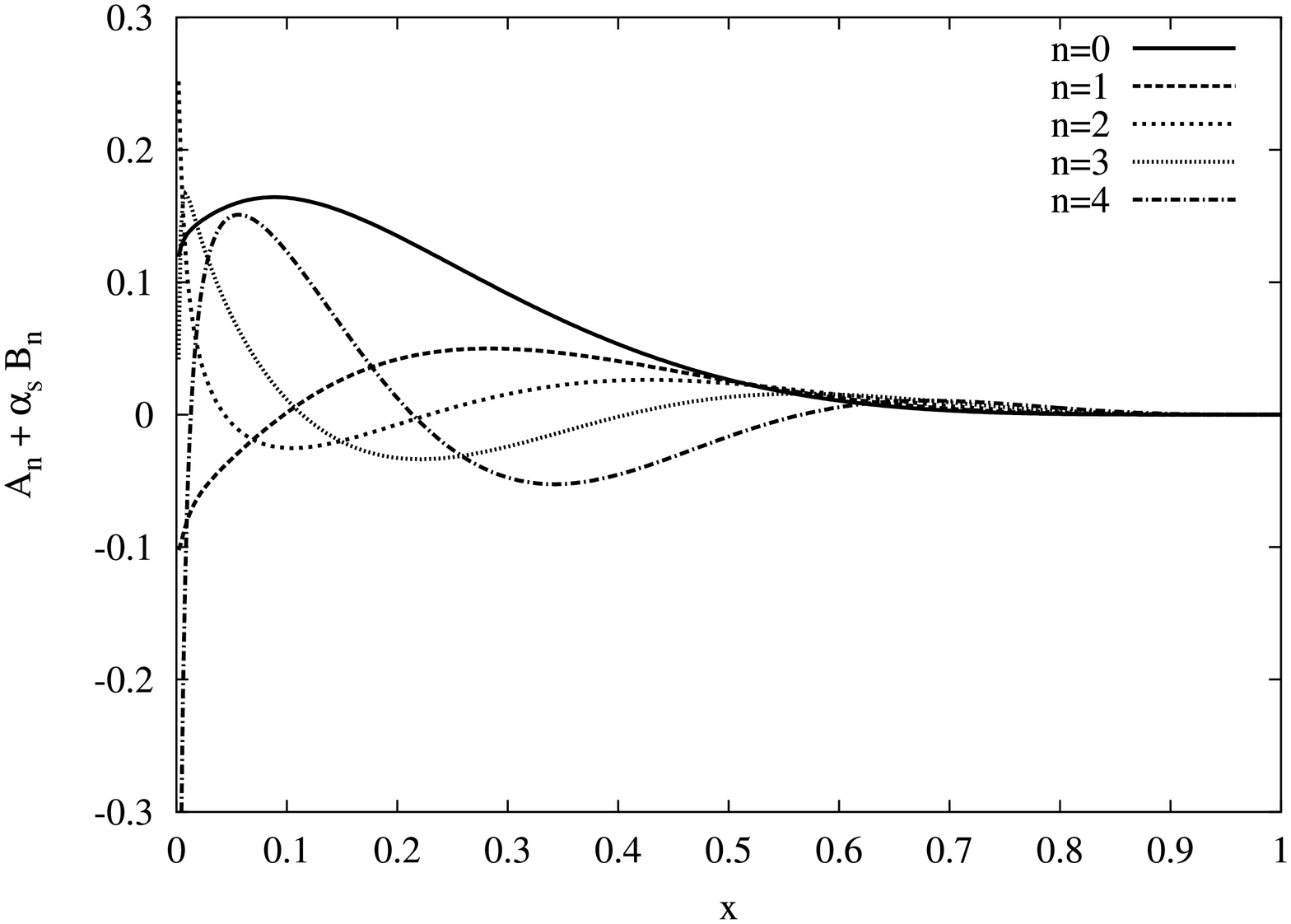}}} \par}

\caption{Coefficients \protect\( A_{n}(x)+\alpha _{s}(Q^{2})B_{n}\protect \),
with \protect\( n=0,\ldots ,4\protect \) for a final scale \protect\( Q=100\protect \)
GeV for the quark down.}
\label{anprime}
\end{figure}

Now we are ready to write down the expression for the $B^{\pm}_{n+1}(x)$ coefficient to NLO, 
similarly to eq. (\ref{recur1}). So we get (for the $(+)$ case) 

\ba
&&B^{+}_{n+1}(x) = - B^{+}_{n}(x) + \frac{\beta_{1}}{2\beta^{2}_{0}} \left\{2C_{F}\left[\int^{1}_{x}\frac{dy}{y}\frac{y A^{+}_{n}(y) - x A^{+}_{n}(x)}{y-x} + A^{+}_{n}(x) \log(1-x)\right]\right.+\nonumber\\
&&\left.-2C_{F}\left(\int_{x}^{1}\frac{dy}{y} A^{+}_{n}(y)\right) + C_{F}\frac{3}{2} A^{+}_{n}(x)\right\}-\frac{1}{4\pi\beta_{0}}K^{+}_{1} A^{+}_{n}(x)+ \int^{1}_{x}\frac{dy}{y}\left[ K^{+}_{2}(z) S_{2}(z) + \right.\nonumber\\
&&+ \left.K^{+}_{3}(z)\log(z)+\log^{2}(z)K^{+}_{4}(z) + \log(z)\log(1-z)K^{+}_{5}(z)\right]\left(-\frac{1}{4\pi\beta_{0}}\right)A^{+}_{n}(y)+\nonumber\\
&&K^{+}_{6}\left(-\frac{1}{4\pi\beta_{0}}\right)\left\{\left[\int^{1}_{x}\frac{dy}{y} \frac{yA^{+}_{n}(y) - xA^{+}_{n}(x)}{y-x} + A^{+}_{n}(x)\log(1-x) \right] + K^{+}_{7}\int^{1}_{x}\frac{dy}{y}A^{+}_{n}(y)\right\}-\nonumber\\
&&C_{F}\left(-\frac{4}{\beta_{0}}\right)\left[\int^{1}_{x}\frac{dy}{y}\frac{y B^{\pm}_{n}(y) - x B^{\pm}_{n}(x)}{y-x} + B^{\pm}_{n}(x) \log(1-x)\right]+\nonumber\\
&&C_{F}\left(\frac{4}{\beta_{0}}\right) \left(\int_{x}^{1}\frac{dy}{y} B^{\pm}_{n}(y)\right) + C_{F}\left(-\frac{2}{\beta_{0}}\right)\frac{3}{2} B^{\pm}_{n}(x)\,.\nonumber\\
\ea
As we have already mentioned, the implementation of these recursion relations require particular numerical care, since, as $n$ increases, numerical instabilities tend to add up 
unless high accuracy is used in the computation of the integrals. In particular we use finite 
element expansions to extract analitically the logarithms in the convolution 
(see the discussion in Appendix A). 
NLO plots of the coefficients $A_n(x) + \alpha(Q^2) B_n(x)$ are shown in figs. 
(\ref{an},\ref{anprime}) for a specific set of initial conditions (GRSV, as discussed below). 
As the index $n$ increases, the number of nodes also increases. A stable implementation 
can be reached for several thousands of grid-points and up to $n\approx 10$. 
Notice that the asymptotic expansion is suppressed by $n!$ and that additional 
contributions ($n>10$) are insignificant even at large ($> 200$ GeV) final evolution scales $Q$.

\section{Nonforward Extensions}
In this section we finally discuss the nonforward extension of the evolution 
algorithm. In the case of nonforward distributions a second scaling parameter $\zeta$ 
controls the asymmetry between the initial and the final nucleon momentum in the 
deeply virtual limit of nucleon Compton scattering. The solution of the evolution 
equations, in this case, are known in operatorial form. Single and double parton distributions are obtained sandwiching the operatorial solution with 4 possible types of initial/final states $<p|...|p>, <p|...|0>, <p'|...|p>$, corresponding, respectively, 
to the case of diagonal parton distributions, distribution amplitudes and, in the latter 
case, skewed and double parton distributions \cite{Radyushkin}. Here we will simply analize the non-singlet case and discuss the extension of the forward algorithm to this more general case. 
Therefore, given the off-forward distributions $H_q(x,\xi)$, in Ji's notation, 
we set up the expansion 
\beq 
H_q(x,\xi)=\sum_{k=0}^{\infty} \frac{A_k(x,\xi)}{k!}\log^k
\left(\frac{\alpha(Q^2)}{\alpha(Q_0^2)}\right) +
\alpha(Q^2)\sum_{k=0}^\infty \frac{B_k(x,\xi)}{k!}\log^k
\left(\frac{\alpha(Q^2)}{\alpha(Q_0^2)}\right),
\label{expansionx}
 \eeq
which is the natural extension of the forward algorithm discussed 
in the previous sections. We recall that in the light-cone gauge $H(x,\xi)$ 
is defined as 
\beq
H_q(x,\xi,\Delta^2))= \frac{1}{2}\int \frac{dy^-}{2 \pi}e^{-i x \bar{P}^+y^-}
\langle P'| \bar{\psi}_q(0,\frac{y^-}{2},{\bf 0_\perp})
\frac{1}{2}\gamma^+ \psi_q (0,\frac{y^-}{2},{\bf 0_\perp})|P\rangle
\eeq
with $\Delta= P' - P$, $\bar{P}^+=1/2(P + \bar{P})$ \cite{Ji} (symmetric choice) and 
$\xi \bar{P}=1/2\,\,\Delta^+$.

This distribution describes for $x>\xi$ and $x < -\xi$ the DGLAP-type region for the quark and the antiquark distribution respectively, and the ERBL 
\cite{ERBL} (see also \cite{CL} for an overview) distribution amplitude 
for $-\xi <x < \xi$. In the following we will omit the $\Delta$ dependence from $H_q$.  

Again, once we insert the ansatz (\ref{expansionx}) into the evolution equations we obtain an infinite set of recursion relations which we can solve numerically. In LO, it is rather simple to relate the Gegenbauer moments of the skewed distributions and those of the generalized 
scaling coefficients $A_n$.
We recall that in the 
nonforward evolution, the multiplicatively renormalizable operators appearing 
in the light cone expansion are given in terms of Gegenbauer polynomials 
\cite{Radyushkin}. 
The Gegenbauer moments of the coefficients $A_n$ of our expansion (\ref{expansionx}) 
can be easily related to those of the off-forward distribution 

\beq
C_n(\xi, Q^2) =\zeta^n\int_{-1}^{1} C_n^{3/2}(z/\xi)H(z,\xi, Q^2) dz. 
\eeq  
The evolution of these moments is rather simple  
\beq
C_n(\zeta, Q^2)=C_n(\zeta, Q_0^2) \left(\frac{\alpha(Q^2)}{\alpha(Q_0^2)}\right)^{\gamma_n/\beta_0} 
\eeq
with
\beq
\gamma_n= C_F \left(\frac{1}{2} - \frac{1}{(n+1)(n+2)} + 2 \sum_{j=2}^{n+1}\frac{1}{j}\right)
\eeq
being the non-singlet anomalous dimensions. If we define the Gegenbauer moments of 
our expansion 
\beq
A_k^{(n)}(\xi, Q^2) =\xi^n\int_{-1}^{1} C_n^{3/2}(z/\xi)H(z\,\xi, Q^2) dz 
\eeq  
we can relate the moments of the two expansions as 
\beq
A_k^{(n)}(\xi)=C_n(\zeta,Q_0^2)\left( \frac{\gamma_n}{\beta_0}\right)^k. 
\eeq
Notice that expansions similar to (\ref{expansionx}) hold also for other choices 
of kinematical variables, such as those defining the non-forward distributions 
\cite{Radyushkin}, where the t-channel longitudinal momentum exchange $\Delta^+$ is related to the longitudinal momentum of the incoming nucleon as $\Delta=\zeta P$. We recall 
that $H_q(x.\xi)$ as defined in \cite{Ji} can be mapped 
into two independent distributions $\hat{\mathcal{F}}_q(X,\zeta)$ and 
$\hat{\mathcal{F}}_{\bar{q}}(X,\zeta)$ through the mappings \cite{Golec}
\beqa
X_1 &=& \frac{(x_1+ \xi)}{(1 +\xi)} \nonumber \\
X_2 &=& \frac{\xi - x_2}{(1 +\xi)} \nonumber \\
\xi &=&\zeta/(2 - \zeta) \nonumber \\
\mathcal{F}_q(X_1,\zeta) &=& \frac{1}{1 - \zeta/2}H_q(x_1,\xi) \nonumber \\
\mathcal{F}_{\bar{q}}(X_2,\zeta) &=& \frac{-1}{1 - \zeta/2}H_q(x_2,\xi),\nonumber \\
\eeqa
in which the interval  $-1 \leq x \leq 1$ is split 
into two coverings, partially overlapping (for $-\xi\leq x \leq \xi$, or ERBL region) 
in terms of the two variables $-\xi \leq x_1 \leq 1$ ($0\leq X_1 \leq 1$) and 
$-1 \leq x_2 \leq \xi$ ($0\leq X_2 \leq 1$). In this new parameterization, the 
momentum fraction carried by the emitted quark is $X$, 
as in the case of ordinary distributions, where it is parametrized by Bjorken $x$. 
For definitess, we focus here on the DGLAP-like $(X> \zeta)$ region of 
the non-singlet evolution. The non-singlet kernel is given in this case by 
$(x\equiv X)$
\beqa
P_\zeta(x,\zeta)=\frac{\alpha}{\pi}C_F\left( \frac{1}{y - x}\left[1 + \frac{x x'}{y y'}\right] -
\delta(x - y)\int_0^1 dz \frac{1 + z^2}{1 - z}\right),
\label{nfkernel}
\eeqa
 we introduce a LO ansatz 
\beq 
\mathcal{F}_q(x,\zeta)=\sum_{k=0}^{\infty} \frac{\mathcal{A}_k(x,\zeta)}{k!}\log^k
\left(\frac{\alpha(Q^2)}{\alpha(Q_0^2)}\right)
\eeq
and insert it into the evolution of this region to obtain the very simple recursion relations 
\beqa
\A_{n+1}(X,\zeta) &=& -\frac{2}{\beta_0} C_F 
\int_X^1 \frac{dy}{y} \frac{y \A_n(y,\zeta) -
x \A_n(X,\zeta)}{y-X}  -\frac{2}{\beta_0} C_F 
\int_X^1 \frac{dy (X-\zeta)}{y(y-\zeta)} \frac{\left(y \A_n(X,\zeta) -X \A_n(y,\zeta)\right)}{y-X} 
\nonumber \\
&& -\frac{2}{\beta_0}C_F\hat{\A}_n(X,\zeta)\left[\frac{3}{2}+\ln\frac{(1 - X)^2(1 - x/\zeta)}{1 - \zeta}\right].
\eeqa
The recursion relations can be easily reduced to a weighted sum of contributions in which $\zeta$ is a spectator parameter. Here we will not make a complete implementation, but we will illustrate 
in an appendix the general strategy to be followed. There we show a very accurate analytical 
method to evaluate the logarithms generated by the expansion without having 
to rely on brute-force computations.  
 
\section{Positivity of the non-singlet Evolution}
Positivity of the non-singlet evolution is a simple consequence of the master-form associated to the 
non-forward kernel (\ref{nfkernel}). As we have already emphasized above, 
positivity of the initial conditions are sufficient to guarantee a positivity of the 
solution at any scale $Q$. The master-form of the equation allows to reinterpret the parton dynamics as a random walk biased toward small-x values as $\tau=\log(Q^2)$ 
increases.

In the non-forward case the identification of a transition probability 
for the random walk \cite{CafaCor} associated with the evolution of the parton distribution is obtained 
via the non-forward transition probability 
\beqa
w_\zeta(x|y) &=&\frac{\alpha}{\pi}C_F \frac{1}{y- x}\left[1 + \frac{x}{y}\frac{(x-\zeta)}{y - \zeta}\right]
\theta(y>x)\nonumber \\
w'_\zeta(y|x)&=&\frac{\alpha}{\pi}C_F \frac{x^2 + y^2}{x^2(x - y)}\theta(y<x)
\eeqa
and the corresponding master equation is given by 
\beq
\frac{d \mathcal{F}_q}{d\tau}=\int_x^1 dy\, w_\zeta(x|y)\mathcal{F}_q(y,\zeta,\tau)-
\int_0^x dy\, w'_\zeta(y|x)\mathcal{F}_q(x,\zeta,\tau),
\eeq
that can be re-expressed in a form which is a simple generalization of the formula for the 
forward evolution \cite{CafaCor}

\beqa
\frac{d \mathcal{F}_q}{d\log Q^2} &=& \int_x^1 dy\, w_\zeta(x|y)\mathcal{F}_q(y,\zeta,\tau) - 
\int_0^x dy \,w'_\zeta(y|x) \mathcal{F}_q(x,\zeta,\tau)
\nonumber \\
&=& -\int_0^{\alpha(x)} dy w_\zeta(x+y|x)* \mathcal{F}_q(x,\zeta,\tau)+ 
\int_0^{-x} dy\, w'_\zeta(x+y|x)\mathcal{F}_q(x,\zeta,\tau),
\eeqa
where a Moyal-like product appears 
\beq
w_\zeta(x+y|x)*\mathcal{F}_q(x,\zeta,\tau)\equiv w_\zeta(x+y|x) e^{-y \left(\overleftarrow{\partial}_x + 
\overrightarrow{\partial}_x\right)} \mathcal{F}_q(x,\zeta,\tau)
\eeq
and $\alpha(x) =x-1$. 
A Kramers-Moyal expansion of the equation allows to generate a differential equation 
of infinite order with a parametric dependence on $\zeta$ 

\beqa
\frac{d \mathcal{F}_q}{d\log Q^2} &=&\int_{\alpha(x)}^{0}dy\,  
w_\zeta(x+y|x)\mathcal{F}_q(x,\zeta,\tau) + 
\int_{0}^{-x}dy\,  
w'_\zeta(x+y|x)\mathcal{F}_q(x,\zeta,\tau) \nonumber \\
&& - \sum_{n=1}^{\infty}\int_0^{\alpha(x)}dy \frac{(-y)^n}{n!}{\partial_x}^n
\left(w_\zeta(x+y|x)\mathcal{F}_q(x,\zeta,\tau)\right).
\label{expans}
\eeqa
We define
\beqa
\tilde{a}_0(x,\zeta) &=& \int_{\alpha(x)}^{0} dy w_\zeta(x+y|x)\mathcal{F}_q(x,\zeta,\tau)
+ \int_{0}^{-x}dy\,  
w'_\zeta(x+y|x)\mathcal{F}_q(x,\zeta,\tau)
 \nonumber \\
a_n(x,\zeta) &=&\int_0^{\alpha(x)} dy \,y^n w_\zeta (x+y|x) \mathcal{F}_q(x,\zeta,\tau) \nonumber \\
\tilde{a}_n(x,\zeta)&=&\int_0^{\alpha(x)}
dy y^n {\partial_x}^n \left(w_\zeta(x+y|x)\mathcal{F}_q(x,\zeta,\tau)\right) \,\,\,n=1,2,...
\eeqa
If we arrest the expansion at the first two terms $(n=1,2)$ we are able to derive an 
approximate equation describing the dynamics of partons for non-diagonal transitions. 
The procedure is a slight generalization of the method presented in \cite{CafaCor}, 
to which we refer for further details. For this purpose we use the identities

\beqa
\tilde{a}_1(x,\zeta) &=&\partial_x a_1(x,\zeta) - \alpha(x) \partial_x \alpha(x)
w_\zeta(x + \alpha(x)|x)\mathcal{F}_q(x,\zeta,\tau) \nonumber \\
\tilde{a}_2(x,\zeta) &=&\partial_x^2 a_2(x,\zeta) - 
2 \alpha(x) (\partial_x \alpha(x))^2 w_\zeta(x+ \alpha(x)|x)\mathcal{F}_q(x,\zeta,\tau)
\nonumber \\
&&  - 
\alpha(x)^2 \partial_x\alpha(x)
\partial_x\left( w_\zeta(x+ \alpha(x)|x)\mathcal{F}_q(x,\zeta,\tau)\right)\nonumber \\ 
&& - \alpha^2(x)\partial_x \alpha(x) \partial_x\left( w_\zeta(x + y|x)
\mathcal{F}_q(x,\zeta,\tau)\right)|_{y=\alpha(x)}.
\eeqa
which allow to compute the first few coefficients of the expansion. 
Using these relations, the Fokker-Planck approximation to this equation 
can be worked out explicitely. We omit details on the derivation which is unobvious 
since particular care is needed to regulate the (canceling) divergences and just quote 
the result. 

A lengthy computation gives 
\beqa
\frac{d \mathcal{F}_q}{d\tau}&=& \frac{\alpha}{\pi}C_F\left(\frac{x_{0,-3}}{(x-\z)^3}
+ \frac{x_{0,-1}}{(x-\z)} + x_{0,0}\right)\mathcal{F}_q(x,\z,\tau) 
\nonumber \\
&& + \frac{\alpha}{\pi}C_F\left(\frac{x_{1,-3}}{(x-\z)^3}
+ \frac{x_{1,-1}}{(x-\z)}\right)\partial_x\mathcal{F}_q(x,\z,\tau) 
+ \frac{\alpha}{\pi}C_F\frac{x_{0,-3}}{(x-\z)^3}\partial_x^2\mathcal{F}_q(x,\z,\tau) 
\nonumber \\	
\eeqa
where we have defined  
\beqa
x_{0,-3}&=&\frac{-\left( {\left( -1 + x \right) }^3\,
      \left( 17 x^3 - \z^2 \left( 3 + 4\z \right)  + 
        3 x \z\left( 3 + 5 \z \right)  - 3 x^2\left( 3 + 7 \z \right) 
        \right)  \right) }{12\,x^3}\nonumber \\
x_{0,-1} &=&
\frac{-29 x^4 - 3 + x^2\,\left( -1 + \z \right)  + 2 \z - 
    2 x \left( 1 + 3 \z \right)  + x^3 \left( 12 + 23 \z \right) }{3 x^3}
\nonumber \\
x_{0,0} &=& 4 + \frac{1}{2 x^2} - \frac{3}{x} + 2 \log \frac{(1 - x)}{x}
\nonumber \\
x_{1,-1}&=& 
\frac{-\left( \left( -1 + 6 x - 15 x^2 + 14 x^3 \right) \,
      \left( x - \z \right)  \right) }{3 x^2}\nonumber \\
x_{1,-3}&=&
\frac{1}{2} - \frac{5 x}{3} + 5 x^3 - \frac{23 x^4}{6} + \frac{7 \z}{3} - 
  \frac{3 \z}{4 x} + \frac{5 x \z}{2} \nonumber \\
&& - 15 x^2 \z + 
  \frac{131 x^3 \z}{12} - \frac{5 \z^2}{2} + \frac{\z^2}{4 x^2} - 
  \frac{\z^2}{x} + 13 x \z^2 - \frac{39 x^2 \z^2}{4} - 3 \z^3 + 
  \frac{\z^3}{3 x^2} + \frac{8 x \z^3}{3} \nonumber \\
x_{2,-3}&=&\frac{-\left( {\left( -1 + x \right) }^2\,{\left( x - \z \right) }^2\,
      \left( 3 + 23 x^2 + 4 \z - 2 x\left( 7 + 8 \z \right)  \right) 
      \right) }{24 x}.
\eeqa

This equation and all the equations obtained by arresting the 
Kramers-Moyal expansion to higher order provide a complementary 
description of the non-forward dynamics in the DGLAP region, at least 
in the non-singlet case. Moving to higher order is straightforward 
although the results are slightly lengthier. A full-fledged 
study of these equations is under way and we expect that the DGLAP dynamics 
is reobtained - directly from these equations - as the order of the approximation increases.

\section{Model Comparisons, Saturation and the Tensor Charge}

In this last section we discuss some implementations of our methods to the 
standard (forward) evolution by doing a NLO model comparisons both in the 
analysis of Soffer's inequality and for the evolution of the tensor charge. 
We have selected two models, motivated quite independently 
and we have compared the predicted evolution of the Soffer bound at an 
accessable final evolution scale around $100$ GeV for the light quarks and 
around $200$ GeV for the heavier generations. At this point we recall that 
in order to generate suitable initial 
conditions for the analysis of Soffer's inequality, one needs an ansatz 
in order to quantify the difference between its left-hand side and right-hand side 
at its initial value.

The well known strategy to build reasonable initial conditions for the transverse 
spin distribution consists in generating polarized distributions (starting 
from the unpolarized ones) and then saturate the inequality at some lowest scale, 
which is the approach we have followed for all the models that we have implemented. 

Following Ref. \cite{GRSV} (GRSV model), 
we have used as input distributions - in the unpolarized case - the formulas
in Ref. \cite{GRV}, calculated to NLO in the \( \overline{\textrm{MS}} \)
scheme at a scale \( Q_{0}^{2}=0.40\, \textrm{GeV}^{2} \)
\begin{eqnarray}
x(u-\overline{u})(x,Q_{0}^{2}) & = & 0.632x^{0.43}(1-x)^{3.09}(1+18.2x)\nonumber \\
x(d-\overline{d})(x,Q_{0}^{2}) & = & 0.624(1-x)^{1.0}x(u-\overline{u})(x,Q_{0}^{2})\nonumber \\
x(\overline{d}-\overline{u})(x,Q_{0}^{2}) & = & 0.20x^{0.43}(1-x)^{12.4}(1-13.3\sqrt{x}+60.0x)\nonumber \\
x(\overline{u}+\overline{d})(x,Q_{0}^{2}) & = & 1.24x^{0.20}(1-x)^{8.5}(1-2.3\sqrt{x}+5.7x)\nonumber \\
xg(x,Q_{0}^{2}) & = & 20.80x^{1.6}(1-x)^{4.1}
\end{eqnarray}
and \( xq_{i}(x,Q_{0}^{2})=x\overline{q_{i}}(x,Q_{0}^{2})=0 \) for
\( q_{i}=s,c,b,t \).

We have then related the unpolarized input distribution to the longitudinally
polarized ones by as in Ref. \cite{GRSV} 
\begin{eqnarray}
x\Delta u(x,Q_{0}^{2}) & = & 1.019x^{0.52}(1-x)^{0.12}xu(x,Q_{0}^{2})\nonumber \\
x\Delta d(x,Q_{0}^{2}) & = & -0.669x^{0.43}xd(x,Q_{0}^{2})\nonumber \\
x\Delta \overline{u}(x,Q_{0}^{2}) & = & -0.272x^{0.38}x\overline{u}(x,Q_{0}^{2})\nonumber \\
x\Delta \overline{d}(x,Q_{0}^{2}) & = & x\Delta \overline{u}(x,Q_{0}^{2})\nonumber \\
x\Delta g(x,Q_{0}^{2}) & = & 1.419x^{1.43}(1-x)^{0.15}xg(x,Q_{0}^{2})
\end{eqnarray}
and \( x\Delta q_{i}(x,Q_{0}^{2})=x\Delta \overline{q_{i}}(x,Q_{0}^{2})=0 \)
for \( q_{i}=s,c,b,t \). 

Following \cite{MSSV}, we assume the saturation of
Soffer inequality:\begin{equation}
\label{eq:saturation}
x\Delta _{T}q_{i}(x,Q_{0}^{2})=\frac{xq_{i}(x,Q_{0}^{2})+x\Delta q_{i}(x,Q_{0}^{2})}{2}
\end{equation}
and study the impact of the different evolutions 
on both sides of Soffer's inequality at various final evolution scales $Q$. 

In the implementation of the second model (GGR model)
we have used as input distributions in the unpolarized case the 
CTEQ4 parametrization \cite{CTEQ}, calculated to NLO in the
\( \overline{\textrm{MS}} \) scheme at a scale \( Q_{0}=1.0\, \textrm{GeV} \)

\begin{eqnarray}
x(u-\overline{u})(x,Q_{0}^{2}) & = & 1.344x^{0.501}(1-x)^{3.689}(1+6.402x^{0.873})\nonumber \\
x(d-\overline{d})(x,Q_{0}^{2}) & = & 0.64x^{0.501}(1-x)^{4.247}(1+2.69x^{0.333})\nonumber \\
xs(x,Q_{0}^{2})=x\overline{s}(x,Q_{0}^{2}) & = & 0.064x^{-0.143}(1-x)^{8.041}(1+6.112x)\nonumber \\
x(\overline{d}-\overline{u})(x,Q_{0}^{2}) & = & 0.071x^{0.501}(1-x)^{8.041}(1+30.0x)\nonumber \\
x(\overline{u}+\overline{d})(x,Q_{0}^{2}) & = & 0.255x^{-0.143}(1-x)^{8.041}(1+6.112x)\nonumber \\
xg(x,Q_{0}^{2}) & = & 1.123x^{-0.206}(1-x)^{4.673}(1+4.269x^{1.508})
\end{eqnarray}
and \( xq_{i}(x,Q_{0}^{2})=x\overline{q_{i}}(x,Q_{0}^{2})=0 \) for
\( q_{i}=c,b,t \)
and we have related the unpolarized input distribution to the longitudinally
polarized ones by the relations \cite{GGR}

\begin{eqnarray}
x\Delta \overline{u}(x,Q_{0}^{2}) & = & x\eta _{u}(x)xu(x,Q_{0}^{2})\nonumber \\
x\Delta u(x,Q_{0}^{2}) & = & \cos \theta _{D}(x,Q_{0}^{2})\left[ x(u-\overline{u})-\frac{2}{3}x(d-\overline{d})\right] (x,Q_{0}^{2})+x\Delta \overline{u}(x,Q_{0}^{2})\nonumber \\
x\Delta \overline{d}(x,Q_{0}^{2}) & = & x\eta _{d}(x)xd(x,Q_{0}^{2})\nonumber \\
x\Delta d(x,Q_{0}^{2}) & = & \cos \theta _{D}(x,Q_{0}^{2})\left[ -\frac{1}{3}x(d-\overline{d})(x,Q_{0}^{2})\right] +x\Delta \overline{d}(x,Q_{0}^{2})\nonumber \\
x\Delta s(x,Q_{0}^{2})=x\Delta \overline{s}(x,Q_{0}^{2}) & = & x\eta _{s}(x)xs(x,Q_{0}^{2})
\end{eqnarray}
and \( x\Delta q_{i}(x,Q_{0}^{2})=x\Delta \overline{q_{i}}(x,Q_{0}^{2})=0 \)
for \( q_{i}=c,b,t \).

\begin{figure}[tbh]
{\centering \resizebox*{12cm}{!}{\rotatebox{-90}{\includegraphics{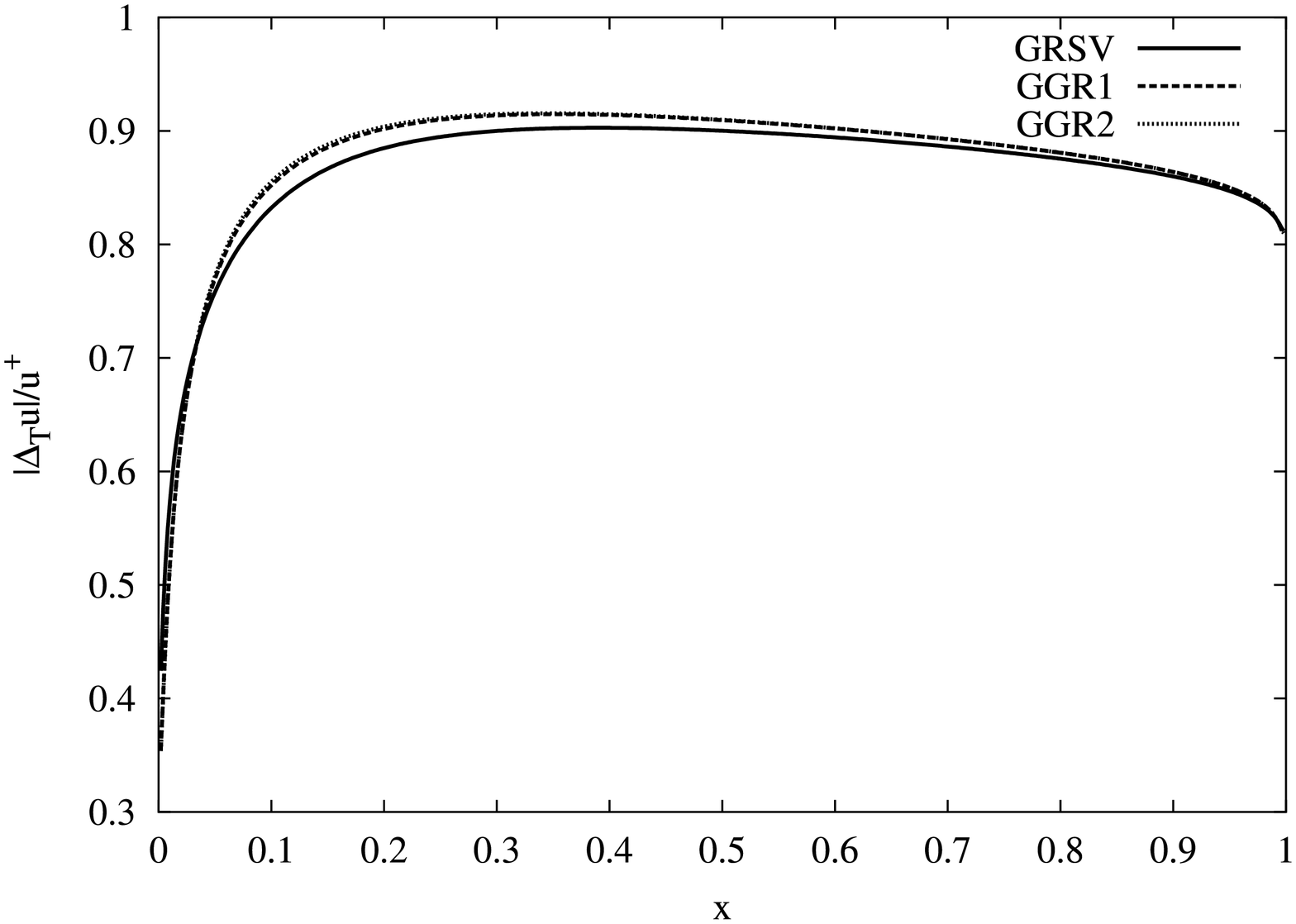}}} \par}
\caption{ Test of Soffer's inequality for quark up at \protect\( Q=100\protect \)
GeV for different models.}
\label{upsof}
\end{figure}

\begin{figure}[tbh]
{\centering \resizebox*{12cm}{!}{\rotatebox{-90}{\includegraphics{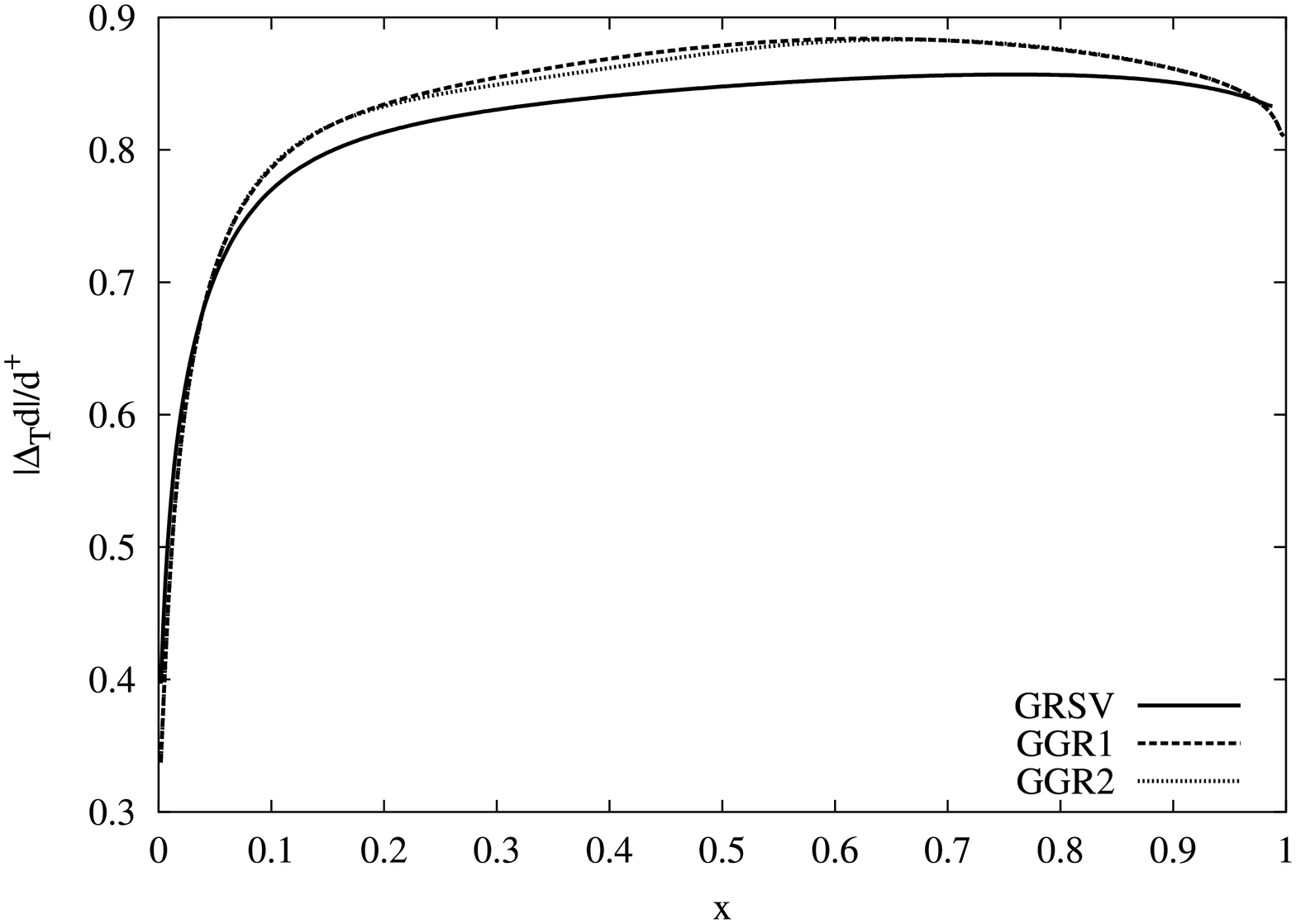}}} \par}
\caption{ Test of Soffer's inequality for quark down at \protect\( Q=100\protect \)
GeV for different models}
\label{downsof}
\end{figure}

\begin{figure}[tbh]
{\centering \resizebox*{12cm}{!}{\rotatebox{-90}{\includegraphics{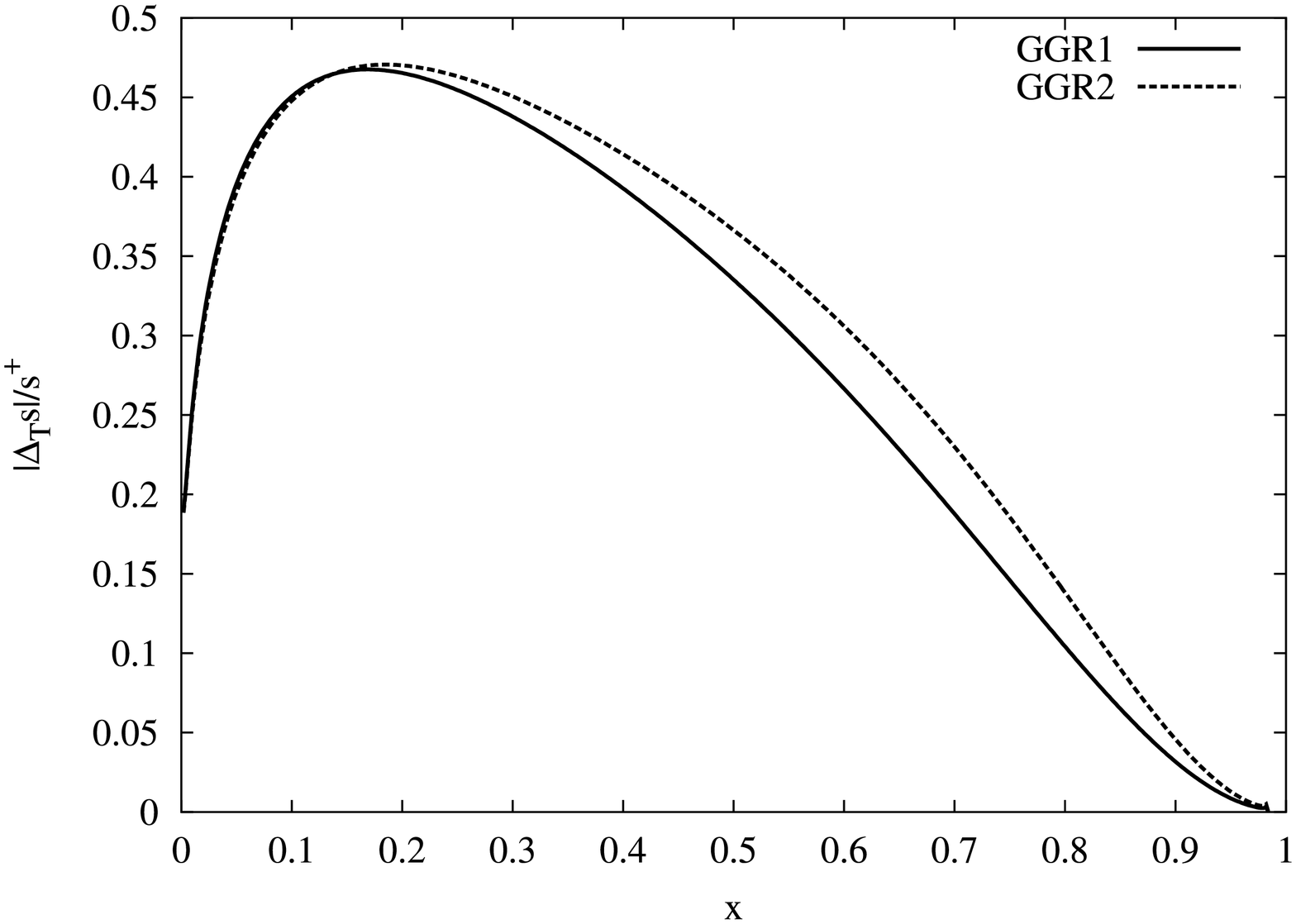}}} \par}
\caption{ Test od Soffer's inequality for quark strange at \protect\( Q=100\protect \)
GeV for different models}
\label{strangesof}
\end{figure}

\begin{figure}[tbh]
\resizebox*{9cm}{!}{\rotatebox{-90}{\includegraphics{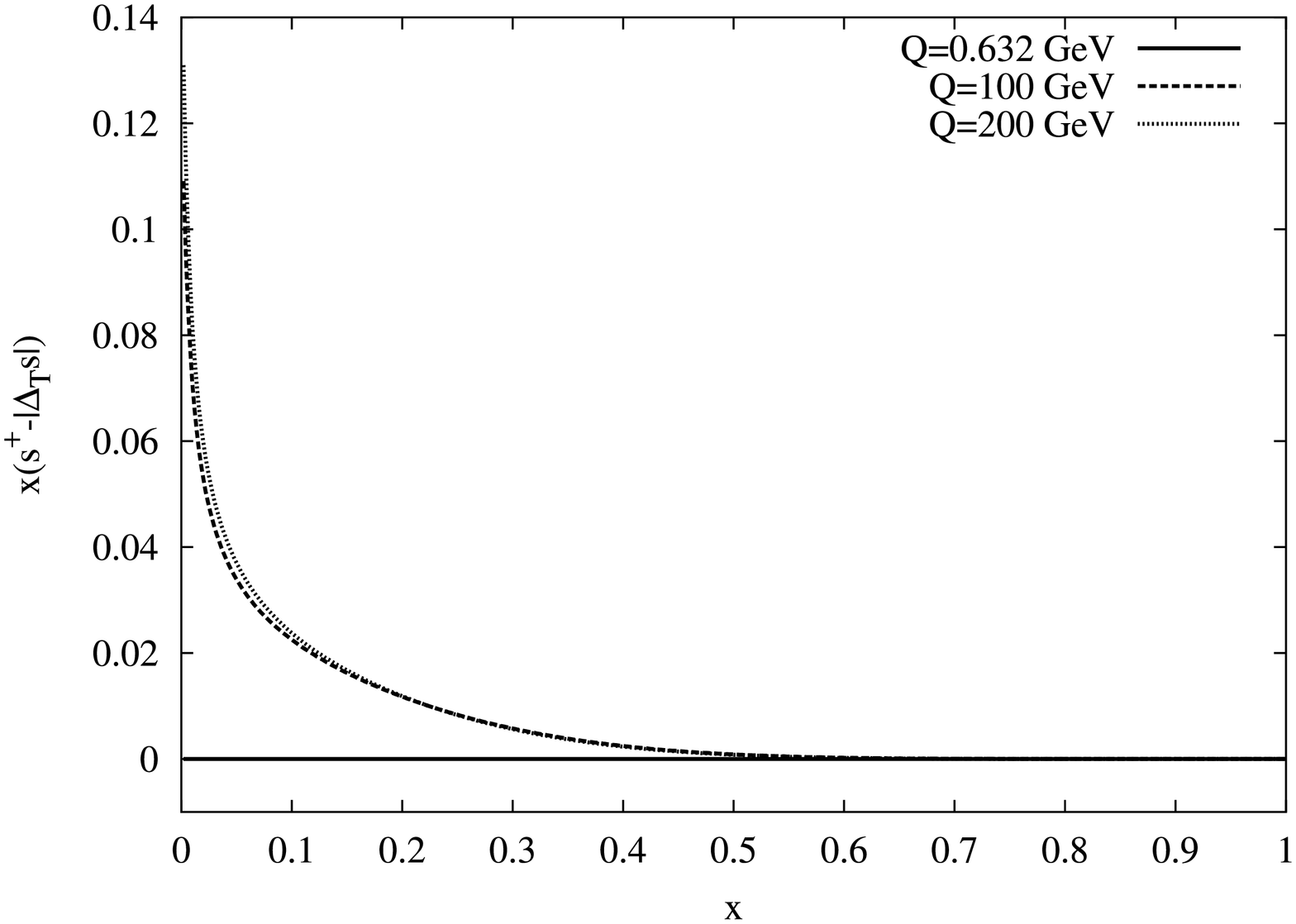}}}
\resizebox*{9cm}{!}{\rotatebox{-90}{\includegraphics{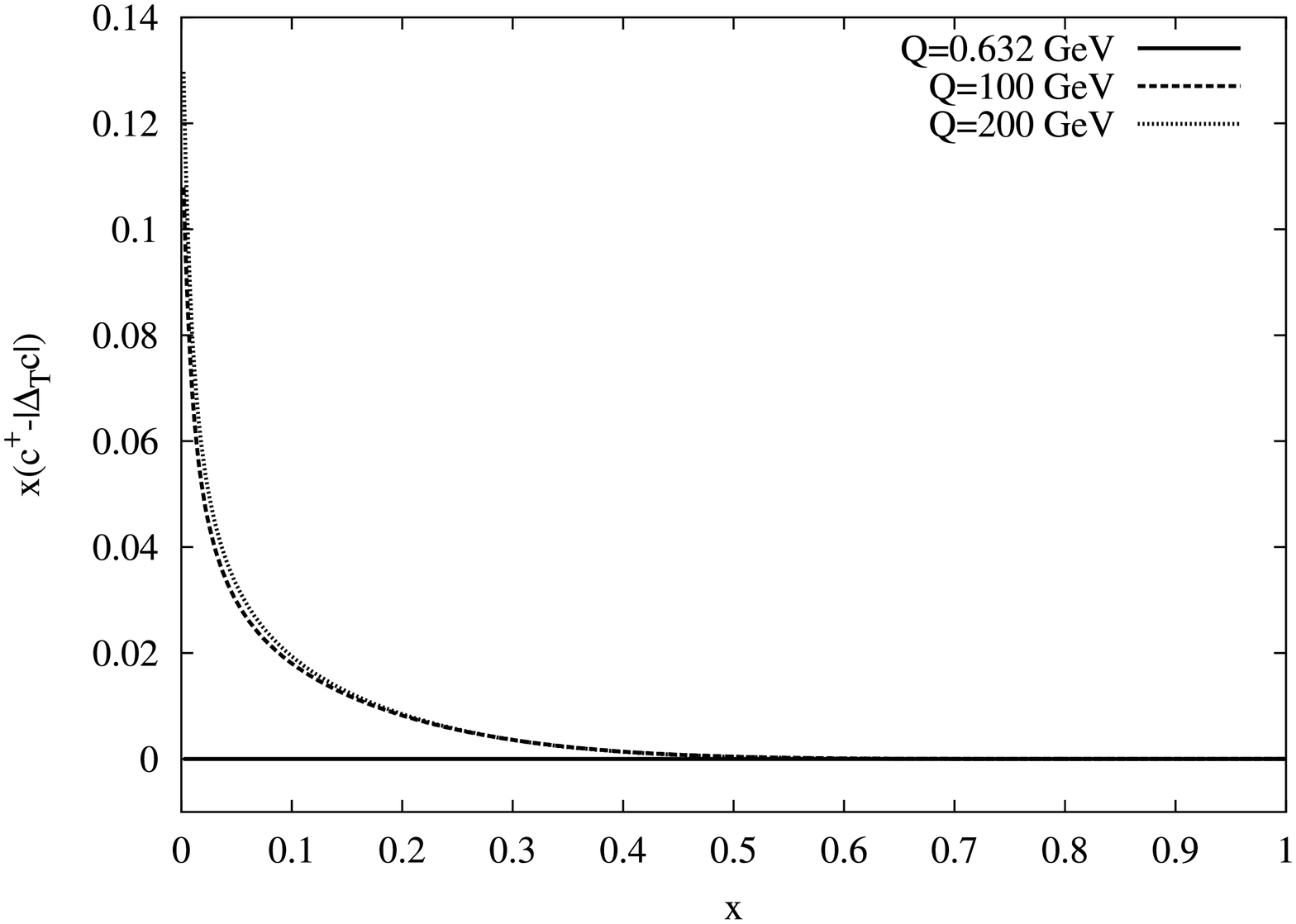}}}
\caption{Soffer's inequality for strange and charm in the GRSV model.}
\label{scsof}
\end{figure}

\begin{figure}[tbh]
\resizebox*{9cm}{!}{\rotatebox{-90}{\includegraphics{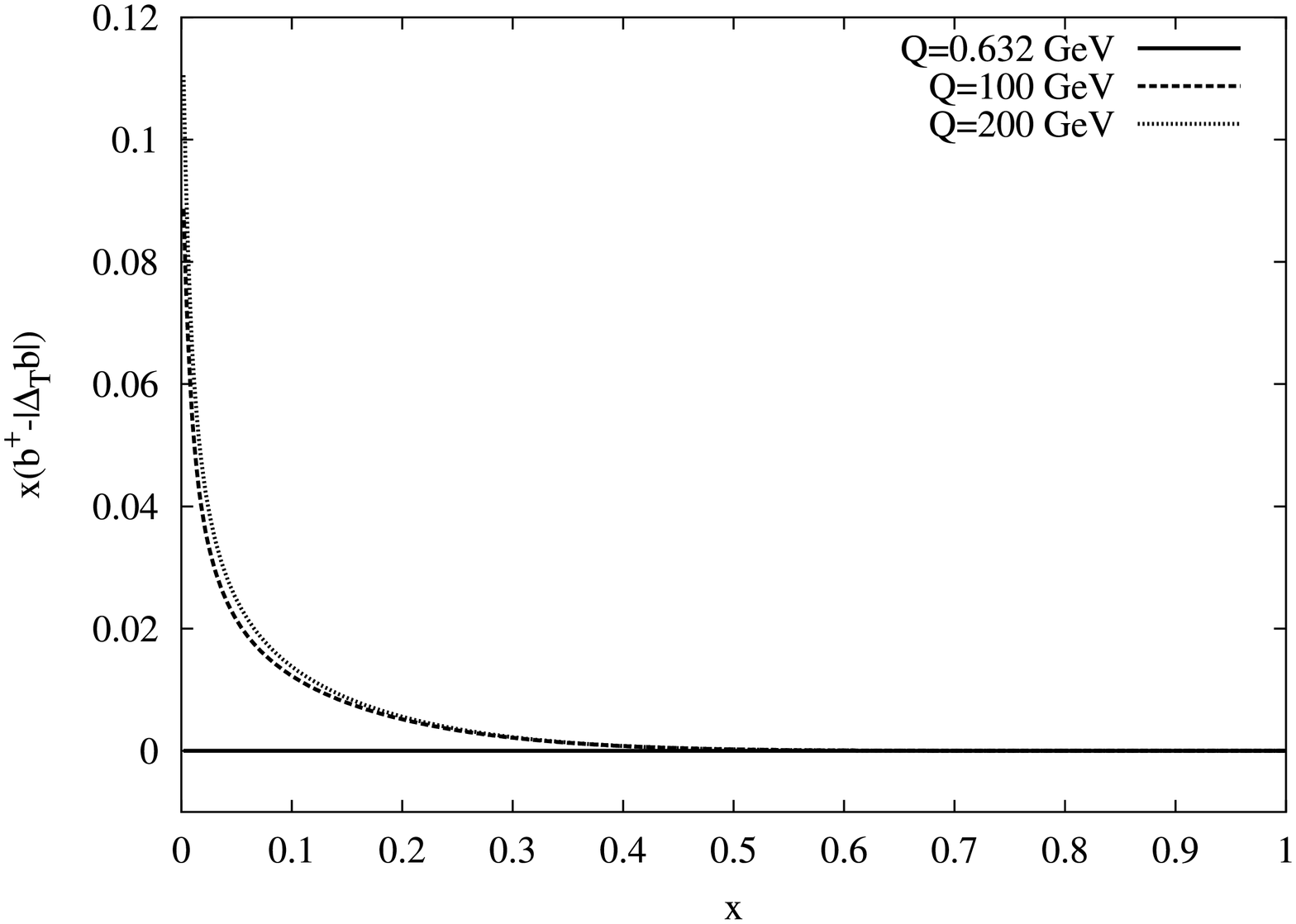}}}
\resizebox*{9cm}{!}{\rotatebox{-90}{\includegraphics{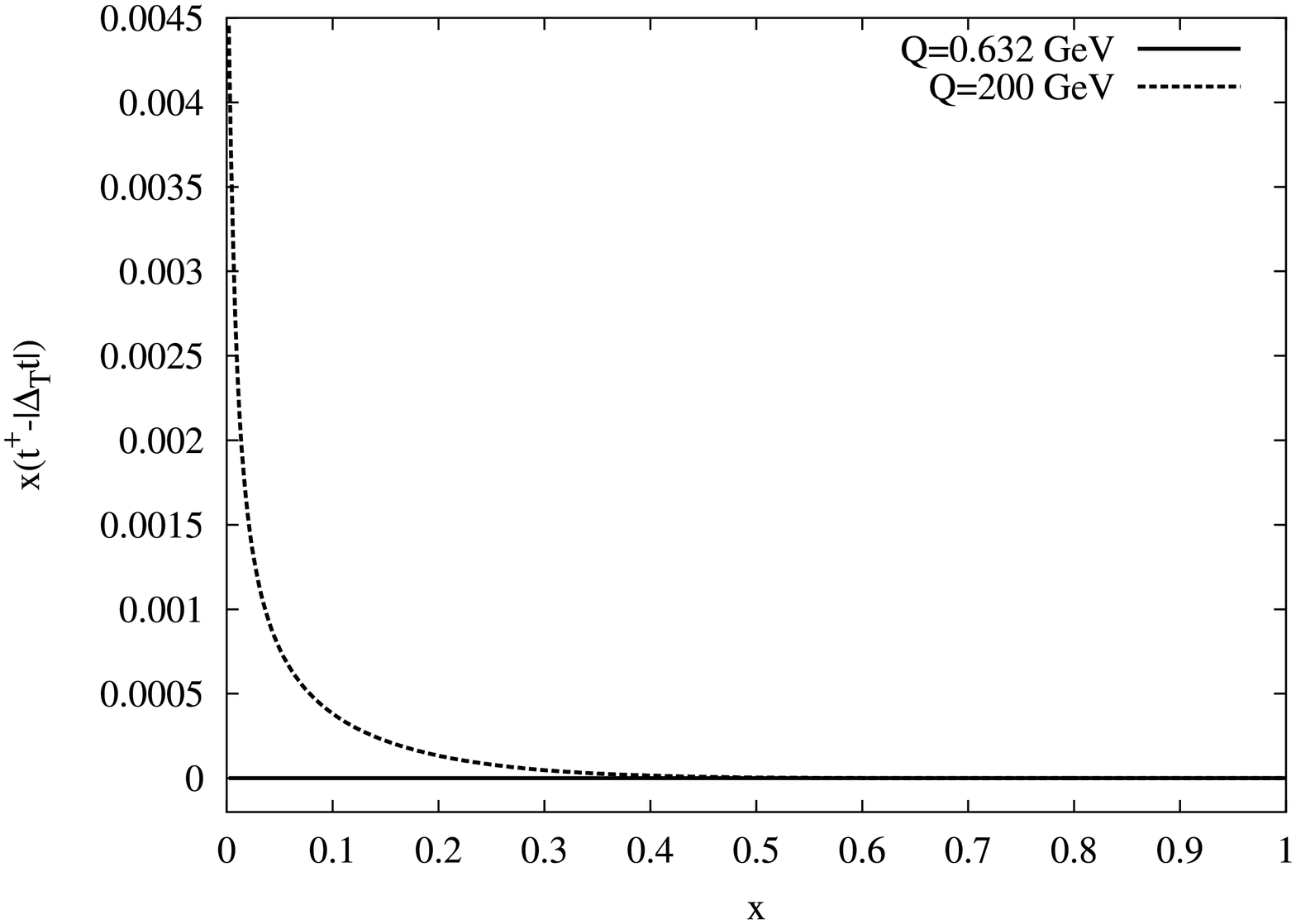}}}
\caption{Soffer's inequality for bottom and top
in the GRSV model.}
\label{btsof}
\end{figure}

A so-called ``spin dilution factor'' as defined in \cite{GGR}, which appears 
in the equations above is given by
\begin{equation}
\cos \theta _{D}(x,Q_{0}^{2})=\left[ 1+\frac{2\alpha _{s}(Q^{2})}{3}\frac{(1-x)^{2}}{\sqrt{x}}\right] ^{-1}.
\end{equation}
In this second (GGR) model, in regard to the initial conditions for the gluons, 
we have made use of two different options, characterized by a parameter 
\( \eta  \) dependent on the corresponding option. 
The first option, that
we will denote by GGR1, assumes that gluons are moderately polarized 
\begin{eqnarray}
x\Delta g(x,Q_{0}^{2}) & = & x\cdot xg(x,Q_{0}^{2})\nonumber \\
\eta _{u}(x)=\eta _{d}(x) & = & -2.49+2.8\sqrt{x}\nonumber \\
\eta _{s}(x) & = & -1.67+2.1\sqrt{x},
\end{eqnarray}
while the second option (GGR2) assumes that gluons are not polarized 
\begin{eqnarray}
x\Delta g(x,Q_{0}^{2}) & = & 0\nonumber \\
\eta _{u}(x)=\eta _{d}(x) & = & -3.03+3.0\sqrt{x}\nonumber \\
\eta _{s}(x) & = & -2.71+2.9\sqrt{x}.
\end{eqnarray}
We have plotted both ratios $\Delta_T/f^+$ and differences 
$(x f^+ - x\Delta_T f)$ for 
various flavours as a function of $x$. For the up quark, while the two models GGR1 and GGR2 
are practically overlapping, the difference between the GGR 
and the GRSV models in the the ratio $\Delta_T u/u^+$ 
is only slightly remarked in the intermediate x region $(0.1-0.5)$. In any case, it is just at the few percent level (Fig. (\ref{upsof})), while the inequality is satisfied 
with a ratio between the plus helicity distribution and transverse around 10 percent from the saturation value, and above. There is a wider gap in the inequality at small x, region characterized by larger transverse distribution, with values up to 40 percent from saturation. A similar trend is noticed for the x-behaviour of the inequality in the case 
of the down quark (Fig. \ref{downsof}). In this latter case the GGR 
and the GRSV model show a more remarked difference, especially for 
intermediate x-values. An interesting features appears in the corresponding 
plot for the strange quark (Fig.(\ref{strangesof})), showing a much 
wider gap in the inequality (50 percent and higher) compared 
to the other quarks. Here we have plotted results for the two GGR models (GGR1 and GGR2). 
Differently from the case of the other quarks, in this case we observe 
a wider gap between lhs and rhs at larger x values, increasing as $x\rightarrow 1$. 
In figs. (\ref{scsof})and (\ref{btsof}) we plot the differences $(x f^+ - x\Delta_T f)$ 
for strange and charm and for bottom and top quarks respectively, which 
show a much more reduced evolution from the saturation value up to the final corresponding 
evolving scales (100 and 200 GeV). 
As a final application we finally discuss the behaviour of the tensor charge 
of the up quark for the two models as a function of the final 
evolution scale $Q$. We recall that like the isoscalar and the isovector 
axial vector charges defined from the forward matrix element 
of the nucleon, the nucleon tensor charge is defined from the matrix 
element of the tensor current 
\beq
\langle P S_T|\bar{\psi}\sigma^{\mu\nu}\gamma_5 \lambda^a \psi|P,S_T\rangle 
=2 \delta q^a(Q_0^2)\left( P^\mu S_T^\nu - P^\nu S_T^\mu\right)
\eeq
where $\delta^a q(Q_0^2)$ denotes the flavour (a) contribution to the nucleon tensor charge at a scale $Q_0$ and $S_T$ is the transverse spin. 

\begin{figure}[tbh]
{\centering \resizebox*{12cm}{!}{\rotatebox{-90}{\includegraphics{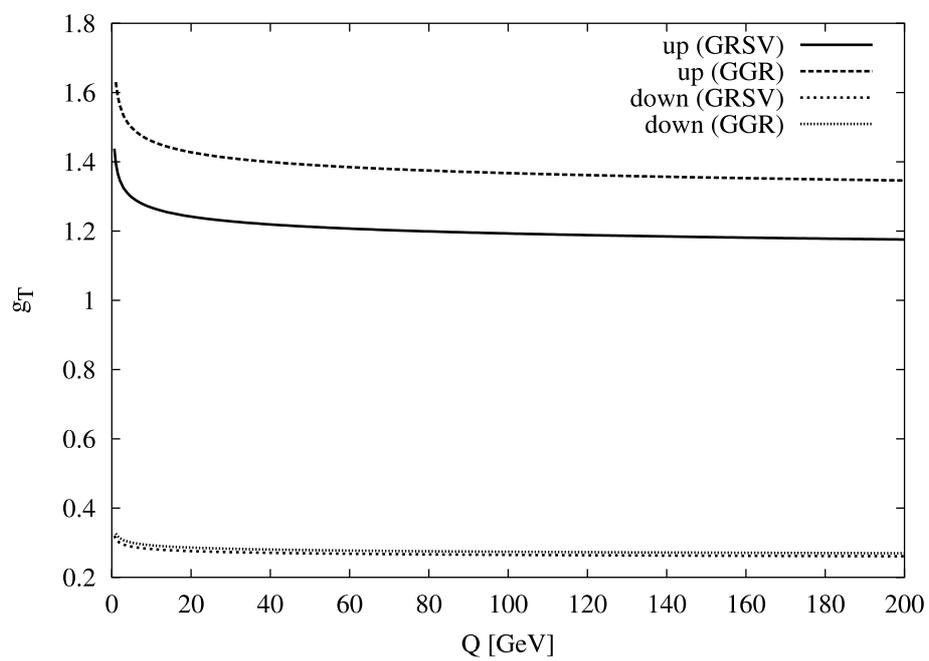}}} \par}
\caption{Tensor charge \protect\( g_{T}\protect \) as a function of \protect\( Q\protect \)
for up and down quark for the GRSV and GGR models.}
\label{figure}
\end{figure}

In fig. (\ref{figure}) we plot the evolution of the tensor charge for the models 
we have taken in exam. At the lowest evolution scales the charge is, in these models, 
above 1 and decreases slightly as the factorization scale $Q$ increases. 
We have performed an evolution up to 200 GeV as an illustration of this behaviour. 
There are substantial differences between these models, as one can easily observe, which are around 20 percent. From the analysis of these differences at various factorization 
scales we can connect low energy dynamics to observables at higher energy, thereby 
distinguishing between the various models. Inclusion of the correct evolution, 
up to subleading order is, in general, essential.

\section{Conclusions}
We have illustrated the use of $x$-space based algorithms for the solution 
of evolution equations in the leading and in the next-to-leading 
approximation and we have 
provided some applications of the method both in the analysis 
of Soffer's inequality and in the investigation 
of other relations, such as the evolution of 
the proton tensor charge, for various models. 
The evolution has been implemented using a suitable base, 
relevant for an analysis of positivity in LO, using kinetic arguments. 
The same kinetic argument 
has been used to prove the positivity of the evolution of 
$h_1$ and of the tensor charge up to NLO. 
In our implementations we have completely relied on recursion relations without 
any reference to Mellin moments. We have provided several illustrations of the 
recursive algorithm and extended it to the non-forward evolution up to NLO. 
Building on previous work
for the forward evolution, we have presented a master-form of the non-singlet
evolution of the skewed distributions, a simple proof of positivity
and a related Kramers Moyal expansion,
valid in the DGLAP region of the skewed evolution for any value
of the asymmetry parameter $\zeta$. We hope to return with a complete study of the
nonforward evolution and related issues not discussed here in the near future.

\section{Appendix A. Weighted Sums}
In this appendix we briefly illustrate the reduction of
recursion relations to analytic expressions based on finite element decompositions of the
corresponding integrals. The method allows to write in analytic forms the
most dangerous integrals thereby eliminating possible sources of instabilities in the
implementation of the recursion relations. The method uses a linear
interpolation formula for the coefficients $A_n$, $B_n$ which, in principle can also
be extended to higher (quadratic) order. However, enugh accuracy can be achieved by
increasing the grid points in the discretization. Notice that using this method
we can reach any accuracy since we have closed formulas for the integrals.
In practice these and similar equations are introduced analytically as functions in the
numerical integration procedures.

Below, we will use a simplified notation ($X\equiv x$ for simplicity).

We define $\bar{P}(x,\zeta)\equiv x P(x,\zeta)$ and $\bar{A}(x,\zeta)\equiv x A(x)$ and the
convolution products

\beq
J(x)\equiv\int_x^1 \frac{dy}{y}\left(\frac{x}{y}\right)
P\left(\frac{x}{y},\zeta\right)\bar{A}(y). \
\eeq
The integration interval in $y$ at any fixed x-value is
partitioned in an array of
increasing points ordered from left to right
$\left(x_0,x_1,x_2,...,x_n,x_{n+1}\right)$
with $x_0\equiv x$ and $x_{n+1}\equiv 1$ being the upper edge of the
integration
region. One constructs a rescaled array
$\left(x,x/x_n,...,x/x_2,x/x_1, 1 \right)$. We define
$s_i\equiv x/x_i$, and $s_{n+1}=x < s_n < s_{n-1}<... s_1 < s_0=1$.
We get
\beq
J(x,\zeta)=\sum_{i=0}^N\int_{x_i}^{x_{i+1}}\frac{dy}{y}
\left(\frac{x}{y}\right) P\left(\frac{x}{y},\zeta\right)\bar{A}(y,\zeta)
\eeq
At this point we introduce the linear interpolation
\beq
\bar{A}(y,\zeta)=\left( 1- \frac{y - x_i}{x_{i+1}- x_i}\right)\bar{A}(x_i,\zeta) +
\frac{y - x_i}{x_{i+1}-x_i}\bar{A,\zeta}(x_{i+1})
\label{inter}
\eeq
and perform the integration on each subinterval with a change of variable
$y->x/y$ and replace the integral $J(x,\zeta)$ with
its discrete approximation $J_N(x)$
to get
\beqa
J_N(x,\zeta) &=& \bar{A}(x_0)\frac{1}{1- s_1}\int_{s_1}^1 \frac{dy}{y}P(y,\zeta)(y - s_1)
\nonumber \\
&+& \sum_{i=1}^{N}\bar{A}(x_i,\zeta) \frac{s_i}{s_i - s_{i+1}}
\int_{s_{i+1}}^{s_i} \frac{dy}{y}P(y)(y - s_{i+1})\nonumber \\
& -& \sum_{i=1}^{N}\bar{A}(x_i,\zeta) \frac{s_i}{s_{i-1} - s_{i}}
\int_{s_{i}}^{s_{i-1}} \frac{dy}{y}P(y,\zeta)(y - s_{i-1}) \nonumber \\
\eeqa
with the condition $\bar{A}(x_{N+1},\zeta)=0$.
Introducing the coefficients  $W(x,x,\zeta)$ and $W(x_i,x,\zeta)$, the integral
is cast in the form
\beq
J_N(x,\zeta)=W(x,x,\zeta) \bar{A}(x,\zeta) + \sum_{i=1}^{n} W(x_i,x,\zeta)\bar{A}(x_i,\zeta)
\eeq
where
\beqa
W(x,x,\zeta) &=& \frac{1}{1-s_1} \int_{s_1}^1 \frac{dy}{y}(y- s_1)P(y,\zeta), \nonumber \\
W(x_i,x,\zeta) &=& \frac{s_i}{s_i- s_{i+1}}
\int_{s_{i+1}}^{s_i} \frac{dy}{y}\left( y - s_{i+1}\right) P(y,\zeta) \nonumber \\
& -& \frac{s_i}{s_{i-1} - s_i}\int_{s_i}^{s_{i-1}}\frac{dy}{y}\left(
y - s_{i-1}\right) P(y,\zeta).\nonumber \\
\eeqa
For instance, after some manipulations we get
\beq
\int_X^1 \frac{dy}{y} \frac{y \A_n(y,\zeta) -
x \A_n(X,\zeta)}{y-X}= {\bf In_0}(x)\A_n(x,\zeta) +\sum_{i+1}^{N}
\left({\bf Jn_i}(x)- {\bf Jnt_i}(x)\right) \A_n(x_i) - \ln( 1- x) \A_n(x,\zeta)
\eeq
where
\beqa
{\bf I_0}(x) & = &
 \frac{1}{1- s_1} \log(s_1) + \log(1-s_1) \nonumber \\
\nonumber \\
{\bf J_i}(x) & = & \frac{1}{s_i - s_{i +1}}
\left[ \log\left(\frac{1 - s_{i+1}}{1 - s_i}\right)
+ s_{i+1} \log\left(\frac{1- s_i}{1 - s_{i+1}}
\frac{s_{i+1}}{s_i}\right)\right]
\nonumber \\
{\bf J'_i}(x) & = & \frac{1}{s_{i-1}- s_i}\left[ \log\left(\frac{1 - s_i}{1 -
s_{i-1}}\right)
+ s_{i-1}\log\left( \frac{s_i}{s_{i-1}}\right) + s_{i-1}\left(
\frac{1 - s_{i-1}}{1 - s_i}\right)\right],   \,\,\,\,\,\ i=2,3,..N \nonumber \\
{\bf J_1}(x) &=& \frac{1}{1- s_1}\log s_1. \nonumber \\
\eeqa
These functions, as shown here, and similar ones, are computed once and for all
the kernels and allow to obtain very fast and extremely accurate implementations
for any $\zeta$.

\section{Appendix B}

\makeatother

\begin{tabular}{|c|c|c|}
\hline
\( n_{f} \)&
\( A \)&
\( B \)\\
\hline
\hline
3&
12.5302&
12.1739\\
\hline
4&
10.9569&
10.6924\\
\hline
5&
9.3836&
9.2109\\
\hline
6&
7.8103&
7.7249\\
\hline
\end{tabular}
Table 1. Coefficients A and B for various flavour, to NLO
for $\Delta_T P_{qq, \pm}$.

\section{Appendix C}
Here we define some notations in regard to the recursion relations used for the
NLO evolution of the transverse spin distributions.

For the $(+)$ case we have these expressions
\ba
&&K_{1}^{+}(x)=\frac{1}{72}C_{F} (-2 n_{f} (3+4\pi^{2}) + N_{C}(51 + 44\pi^{2} - 216 \zeta(3))+ 9C_{F}(3-4\pi^{2}+48\zeta(3))\nonumber\\
&&K_{2}^{+}(x)= \frac{2 C_{F}(-2C_{F}+N_{C})x}{1+x}\\
&&K_{3}^{+}(x)= \frac{C_{F}(9C_{F}-11 N_{C}+2n_{f})x}{3(x-1)}\\
&&K_{4}^{+}(x)=\frac{C_{F}N_{C}x}{1-x}\\
&&K_{5}^{+}(x)=\frac{4C_{F}^{2}x}{1-x}\\
&&K_{6}^{+}(x)=-\frac{1}{9}C_{F}(10n_{f}+N_{C}(-67 + 3\pi^{2}))\\
&&K_{7}^{+}(x)=\frac{1}{9}C_{F}(10n_{f}+N_{C}(-67 + 3\pi^{2}))\\
\ea
and for the $(-)$ case we have
\ba
&&K_{1}^{-}(x)=\frac{1}{72}C_{F} (-2 n_{f} (3+4\pi^{2}) + N_{C}(51 + 44\pi^{2} - 216 \zeta(3))+ 9C_{F}(3-4\pi^{2}+48\zeta(3))\nonumber\\
&&K_{2}^{-}(x)= \frac{2 C_{F}(+2C_{F}-N_{C})x}{1+x}\\
&&K_{3}^{-}(x)= \frac{C_{F}(9C_{F}-11 N_{C}+2n_{f})x}{3(x-1)}\\
&&K_{4}^{-}(x)=\frac{C_{F}N_{C}x}{1-x}\\
&&K_{5}^{-}(x)=\frac{4C_{F}^{2}x}{1-x}\\
&&K_{6}^{-}(x)=-\frac{1}{9}C_{F}(10n_{f}+N_{C}(-67 + 3\pi^{2}))\\
&&K_{7}^{-}(x)=-\frac{1}{9}C_{F}(10n_{f}-18 C_{F}(x-1)+N_{C}(-76 +3\pi^{2}+9x))\\
\ea

\chapter{Double Transverse-Spin Asymmetries in Drell--Yan Processes
with Antiprotons \label{chap3}}
\fancyhead[LO]{\nouppercase{Chapter 3. Double Transverse-Spin Asymmetries in Drell--Yan Processes
with Antiprotons}}

\section{Introduction}

In this chapter we will use what we learned about the evolution of the
transversely polarized parton distributions to make predictions about
the NLO asymmetry in DY processes with transversely polarized
antiprotons.
\footnote{Based on the article~\cite{BCCGR}}

The experiments with antiproton beams planned for the next decade in
the High-Energy Storage Ring at GSI will provide a variety of perturbative and
non-perturbative tests of QCD~\cite{brodsky}. In particular, the possible
availability of \emph{transversely polarised} antiprotons opens the way to
direct investigation of transversity, which is currently one of the main goals
of high-energy spin physics~\cite{bdr}. The quark transversity (\emph{i.e.}
transverse polarisation) distributions $\Delta_Tq$ were first introduced and
studied in the context of transversely polarised Drell--Yan (DY)
production~\cite{rs}; this is indeed the cleanest process probing these
quantities. In fact, whereas in semi-inclusive deep-inelastic scattering
transversity couples to another unknown quantity, the Collins fragmentation
function~\cite{collins}, rendering the extraction of $\Delta_Tq$ a not
straightforward task, the DY double-spin asymmetry
\begin{equation}
A_{TT}^{DY}\equiv\frac{d\sigma^{\uparrow \uparrow} - d\sigma^{\uparrow \downarrow}}
{d\sigma^{\uparrow \uparrow} - d\sigma^{\uparrow \downarrow}}=
\frac{\Delta_T \sigma}{\sigma_\text{unp}}
\label{att1}
\end{equation}
only contains combinations of transversity distributions. At leading order, for
instance, for the process $p^{\uparrow}p^{\uparrow}\to\ell^+\ell^-X$ one has
\begin{equation}
  A_{TT}^{DY} =
  a_{TT} \,
  \frac{\sum_q e_q^2
    [
      \Delta_T      q (x_1, M^2) \, \Delta_T \bar{q}(x_2, M^2) +
      \Delta_T \bar{q}(x_1, M^2) \, \Delta_T      q (x_2, M^2)
    ]
  }{\sum_q e_q^2
    [q(x_1, M^2) \, \bar{q}(x_2, M^2) + \bar{q}(x_1, M^2) \, q(x_2, M^2)]
  },
  \label{att2}
\end{equation}
where $M$ is the invariant mass of the lepton pair, $q(x,M^2)$ is the
unpolarised distribution function, and $a_{TT}$ is the spin asymmetry of the QED
elementary process $q\bar{q}\to\ell^+\ell^-$. In the dilepton centre-of-mass
frame, integrating over the production angle $\theta$, one has
\begin{equation}
  a_{TT} (\varphi) = \half \, \cos 2 \varphi \,,
\end{equation}
%%%
where $\varphi$ is the angle between the dilepton direction and the plane
defined by the collision and polarisation axes.
%%%

Measurement of $p^{\uparrow}p^{\uparrow}$ DY is planned at RHIC~\cite{rhic}. It
turns out, however, that $A_{TT}^{DY}(pp)$ is rather small at such
energies~\cite{bcd,mssv,ratcliffe}, no more than a few percent (similar values
are found for double transverse-spin asymmetries in prompt-photon
production~\cite{Mukherjee:2003pf} and single-inclusive hadron
production~\cite{Mukherjee:2005rw}). The reason is twofold: 1) $A_{TT}^{DY}(pp)$
depends on antiquark transversity distributions, which are most likely to be
smaller than valence transversity distributions; 2) RHIC kinematics
($\sqrt{s}=200\GeV$, $M<10\GeV$ and $x_1x_2=M^2/s\lesssim3\times10^{-3}$) probes
the low-$x$ region, where QCD evolution suppresses $\Delta_Tq(x,M^2)$ as
compared to the unpolarised distribution $q(x,M^2)$~\cite{bcd2,Barone:1997fh}.
The problem may be circumvented by studying transversely polarised
proton--\emph{anti}proton DY production at more moderate energies. In this case
a much larger asymmetry is expected~\cite{bcd,Anselmino:2004ki,Efremov:2004qs}
since $A_{TT}^{DY}(p\bar{p})$ is dominated by valence distributions at medium
$x$. The PAX collaboration has proposed the study of
$p^{\uparrow}\bar{p}^{\uparrow}$ Drell--Yan production in the High-Energy
Storage Ring (HESR) at GSI, in the kinematic region
$30\GeV^2\lesssim{s}\lesssim200\GeV^2$, $2\GeV\lesssim{M}\lesssim10\GeV$ and
$x_1x_2\gtrsim0.1$~\cite{pax}. An antiproton polariser for the PAX experiment is
currently under study~\cite{pax_prl}: the aim is to achieve a polarisation of
30--40\%, which would render the measurement of $A_{TT}^{DY}(p\bar{p})$ very
promising.

Leading-order predictions for the $p\bar{p}$ asymmetry at moderate $s$ were
presented in~\cite{Anselmino:2004ki}. It was also suggested there to access
transversity in the $J/\psi$ resonance production region, where the production
rate is much higher. The purpose of this study is to extend the calculations
of~\cite{Anselmino:2004ki} to next-to-leading order (NLO) in QCD.\footnote{The
results presented here were communicated at the QCD--PAC meeting at GSI (March
2005) and reported by one of us (M.G.) at the Int. Workshop ``Transversity 2005''
(Como, September 2005)~\cite{guzzi}.} This is a necessary check of the previous
conclusions, given the moderate values of $s$ in which we are interested. We
shall see that the NLO corrections are actually rather small and double
transverse-spin asymmetries are confirmed to be of order 20--40\%.

\section{The kinematics}

The kinematic variables describing the Drell--Yan process are (1 and 2
denote the colliding hadrons):
\begin{equation}
  \xi_1 = \sqrt{\tau} \, e^y \,, \qquad
  \xi_2 = \sqrt{\tau} \, e^{-y} \,, \qquad
  y = \frac{1}{2} \, \ln \frac{\xi_1}{\xi_2} \,,
  \label{kin}
\end{equation}
with $\tau=M^2/s$. We denote by $x_1$ and $x_2$ the longitudinal momentum
fractions of the incident partons. At leading order, $\xi_1$ and $\xi_2$
coincide with $x_1$ and $x_2$, respectively. The QCD factorisation formula for
the transversely polarised cross-section for the proton--antiproton Drell--Yan
process is
\begin{multline}
  \frac{d\Delta_T \sigma}{d M \, d y \,d\varphi}
  =
  \sum_q e_q^2 \int_{\xi_1}^1 d x_1 \int_{\xi_2}^1 d x_2
  \left[ \Delta_T q(x_1,\mu^2) \Delta_T q(x_2,\mu^2) \right.
\\
  \null +
  \left. \Delta_T \bar{q}(x_1,\mu^2) \Delta_T \bar{q}(x_2,\mu^2) \right]
  \frac{d\Delta_T \hat\sigma}{d M \, d y \, d\varphi}
  \,,
  \label{fact}
\end{multline}

where $\mu$ is the factorisation scale
%%%
and we take the quark (antiquark) distributions of the antiproton equal to the
antiquark (quark) distributions of the proton.
%%%
Note that, since gluons cannot be transversely polarised (there is no such thing
as a gluon transversity distribution for a spin one-half object like the
proton), only quark--antiquark annihilation subprocesses (with their radiative
corrections) contribute to $d\Delta_T\sigma$. In Eq.~\eqref{fact} we use the
fact that antiquark distributions in antiprotons equal quark distributions in
protons, and \emph{viceversa}. At NLO, \emph{i.e.} at order $\alpha_s$, the
hard-scattering cross-section $d\Delta_T\hat\sigma^{(1)}$, taking the diagrams
of Fig.~\ref{dyqqg2} into account, is given by~\cite{mssv}
\begin{align}
  \frac{
    d\Delta_T \hat\sigma^{(1),\overline{\text{MS}}}
  }{
    d M \, d y \, d\varphi
  }
  \hspace{-4em} & \hspace{4em} =
  \frac{2\alpha^2}{9 s M} \, C_F \,
  \frac{\alpha_s(\mu^2)}{2\pi} \,
  \frac{4\tau(x_1x_2+\tau)}{x_1x_2(x_1+\xi_1)(x_2+\xi_2)} \,
  \cos(2\varphi)
  \nonumber
  \\[1ex]
  & \hspace{-1em} \null \times
  \Bigg\{
  \delta(x_1-\xi_1)\delta(x_2-\xi_2) \! \left[
  \frac{1}{4}\ln^2\frac{(1-\xi_1)(1-\xi_2)}{\tau}
  +\frac{\pi^2}{4}-2 \right]
  \nonumber
  \\[1ex]
  & \null +
  \delta(x_1-\xi_1)
  \left[
    \frac{1}{(x_2-\xi_2)_{+}} \ln\frac{2x_2(1-\xi_1)}{\tau(x_2+\xi_2)}
    + \left(\frac{\ln(x_2-\xi_2)}{x_2-\xi_2}\right)_{\!+}
    + \frac{1}{x_2-\xi_2}\ln\frac{\xi_2}{x_2}
  \right]
  \nonumber
  \\[1ex]
  & \null
  + \frac{1}{2[(x_1-\xi_1)(x_2-\xi_2)]_{+}}
  + \frac{(x_1+\xi_1)(x_2+\xi_2)}{(x_1 \xi_2+x_2\xi_1)^2}
  - \frac{3\ln\!\left(\frac{x_1x_2+\tau}{x_1\xi_2+x_2\xi_1} \right)}
         {(x_1-\xi_1)(x_2-\xi_2)}
  \nonumber
  \\[1ex]
  & \null +
  \ln \frac{M^2}{\mu^2} \left[
  \delta(x_1-\xi_1)\delta(x_2-\xi_2) \!
  \left( \frac{3}{4}+\frac{1}{2}\ln \frac{(1-\xi_1)(1-\xi_2)}{\tau} \right)
  +
  \frac{\delta(x_1-\xi_1)}{(x_2-\xi_2)_{+}} \right] \! \Bigg\}
  \nonumber
  \\[1ex]
  & \hspace{25em} \null + \big[ 1 \leftrightarrow 2 \big],
  \label{sigmanlo}
\end{align}
%%%
where we have taken the factorisation scale $\mu$ equal to the renormalisation
scale. In our calculations we set $\mu = M$.
%%%

\begin{figure}[hbt]
  \centering
  \begin{fmffile}{attnlo-fmf}
    \setlength\unitlength{1.0mm}
    \fmfset{arrow_len}{2.5mm}
    \fmfset{curly_len}{3.0mm}
    \fmfset{wiggly_slope}{75}
    \begin{fmfgraph}(30,30)
      \fmfleft{i2,i1}
      \fmfright{o1}
      \fmf{fermion,tension=2.0}{i1,v1,v2,v3,i2}
      \fmf{photon,tension=1.5}{v2,o1}
      \fmffreeze
      \fmf{gluon,right=0.3}{v1,v3}
    \end{fmfgraph}%
    \qquad
    \begin{fmfgraph}(30,30)
      \fmfset{curly_len}{2.0mm}
      \fmfleft{i2,i1}
      \fmfright{o1}
      \fmf{fermion,tension=3.0}{i1,v1,v2,v3}
      \fmf{fermion}{v3,i2}
      \fmf{photon,tension=1.5}{v3,o1}
      \fmffreeze
      \fmf{gluon,right}{v2,v1}
    \end{fmfgraph}%
    \qquad
    \fmfframe(0,1)(0,1){%
    \begin{fmfgraph}(50,28)
      \fmfleft{i2,i1}
      \fmfright{o2,o1}
      \fmf{fermion}{i1,v1}
      \fmf{fermion,l.s=right,tension=0.0}{v1,v2}
      \fmf{fermion}{v2,i2}
      \fmf{photon}{v1,o1}
      \fmf{gluon}{o2,v2}
    \end{fmfgraph}%
    }%
  \end{fmffile}%
  \\[1ex]
  (a)\hspace{33mm}(b)\hspace{44mm}(c)\hspace{10mm}
  \caption{Elementary processes contributing to the transverse Drell--Yan
           cross-section at NLO: (a, b) virtual-gluon corrections and (c)
           real-gluon emission.}
  \label{dyqqg2}
\end{figure}

The unpolarised Drell--Yan differential cross-section can be found, for
instance, in~\cite{sutton}; besides the diagrams of Fig.~\ref{dyqqg2}, it also
includes the contribution of quark--gluon scattering processes.

\section{Drell--Yan Asymmetries}

To compute the Drell--Yan asymmetries we need an assumption for the
transversity distributions, which as yet are completely unknown. We might
suppose, for instance, that transversity equals helicity at some low scale, as
suggested by confinement models~\cite{bcd2} (this is exactly true in the
non-relativistic limit). Thus, one possibility is
\begin{equation}
  \Delta_T q(x,\mu_0^2) = \Delta q(x,\mu_0^2)\,,
  \label{minbound}
\end{equation}
where typically $\mu_0\lesssim1\GeV$. Another possible assumption for
$\Delta_Tq$ is the saturation of Soffer's inequality~\cite{soffer}, namely
\begin{equation}
  \big| \Delta_T q(x,\mu_0^2) \big| =
  \half \big[ q(x,\mu_0^2) + \Delta q(x,\mu_0^2) \big] ,
  \label{sofbound}
\end{equation}
which represents an upper bound on the transversity distributions.

Since Eqs.~\eqref{minbound} and \eqref{sofbound} make sense only at very low
scales, in practical calculations one has to resort to radiatively generated
helicity and number densities, such as those provided by the GRV
fits~\cite{grv}. The GRV starting scale is indeed (at NLO) $\mu_0^2=0.40\GeV^2$.
We should however bear in mind that in the GRV parametrisation there is a
sizeable gluon contribution to the nucleon's helicity already at the input scale
($\Delta g$ is of order $0.5$). On the other hand, as already mentioned, gluons
do not contribute to the nucleon's transversity. Thus, use of
Eq.~\eqref{minbound} with the GRV parametrisation may lead to substantially
underestimating the quark transversity distributions and hence is a sort of
``minimal bound'' for transversity. Incidentally, the experimental verification or
otherwise of the theoretical predictions of $A_{TT}$ based on the low-scale
constraints (\ref{minbound}, \ref{sofbound}) would represent an indirect test of
the ``valence glue'' hypothesis behind the GRV fits.
%%%%
Note too that, although the assumption (\ref{minbound}) may, in principle,
violate the Soffer inequality, we have explicitly checked that this is not the
case with all the distributions we use.
%%%%

\begin{figure}[hbt]
  \centering
  \includegraphics[width=7cm,angle=-90]{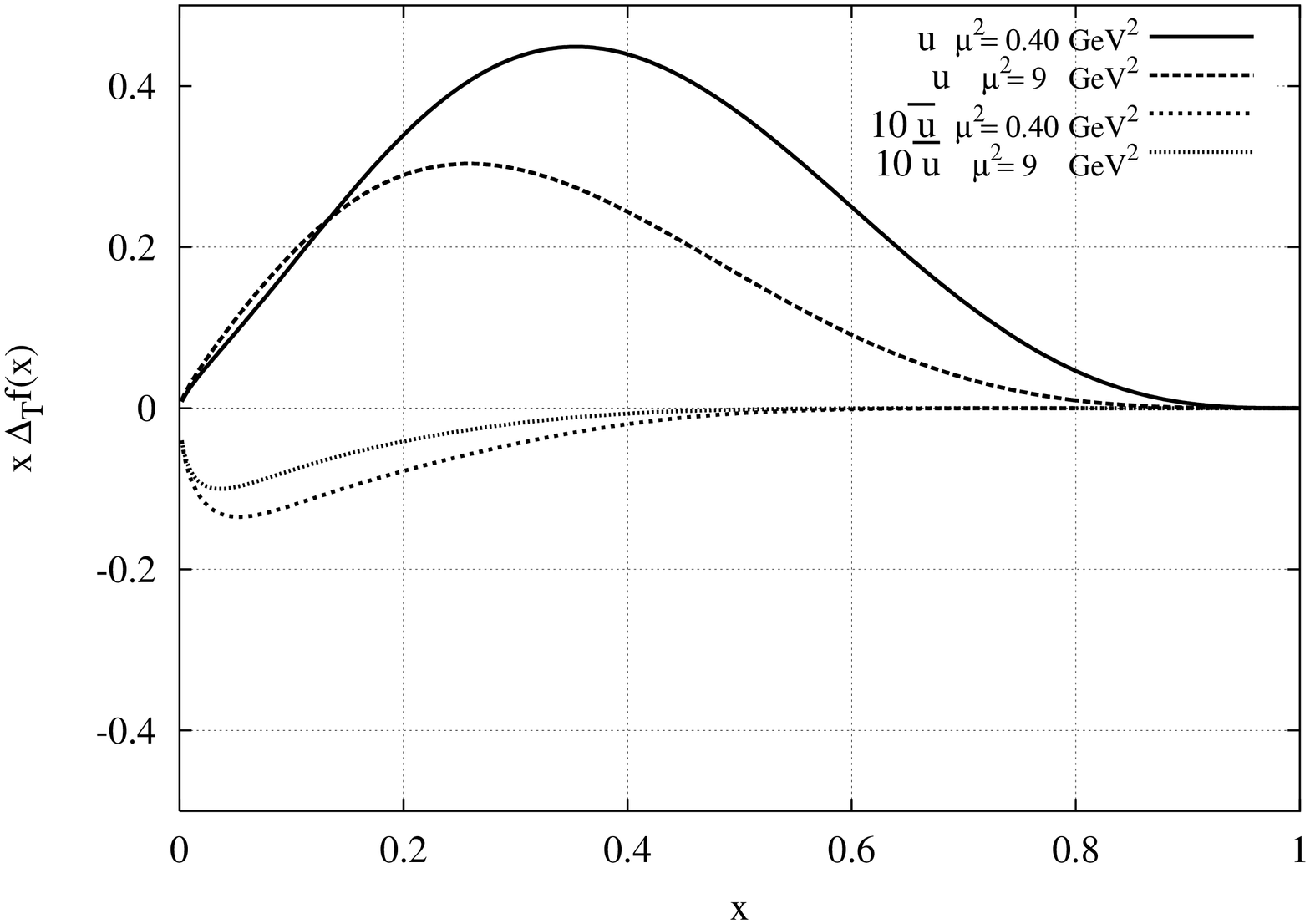}
  \includegraphics[width=7cm,angle=-90]{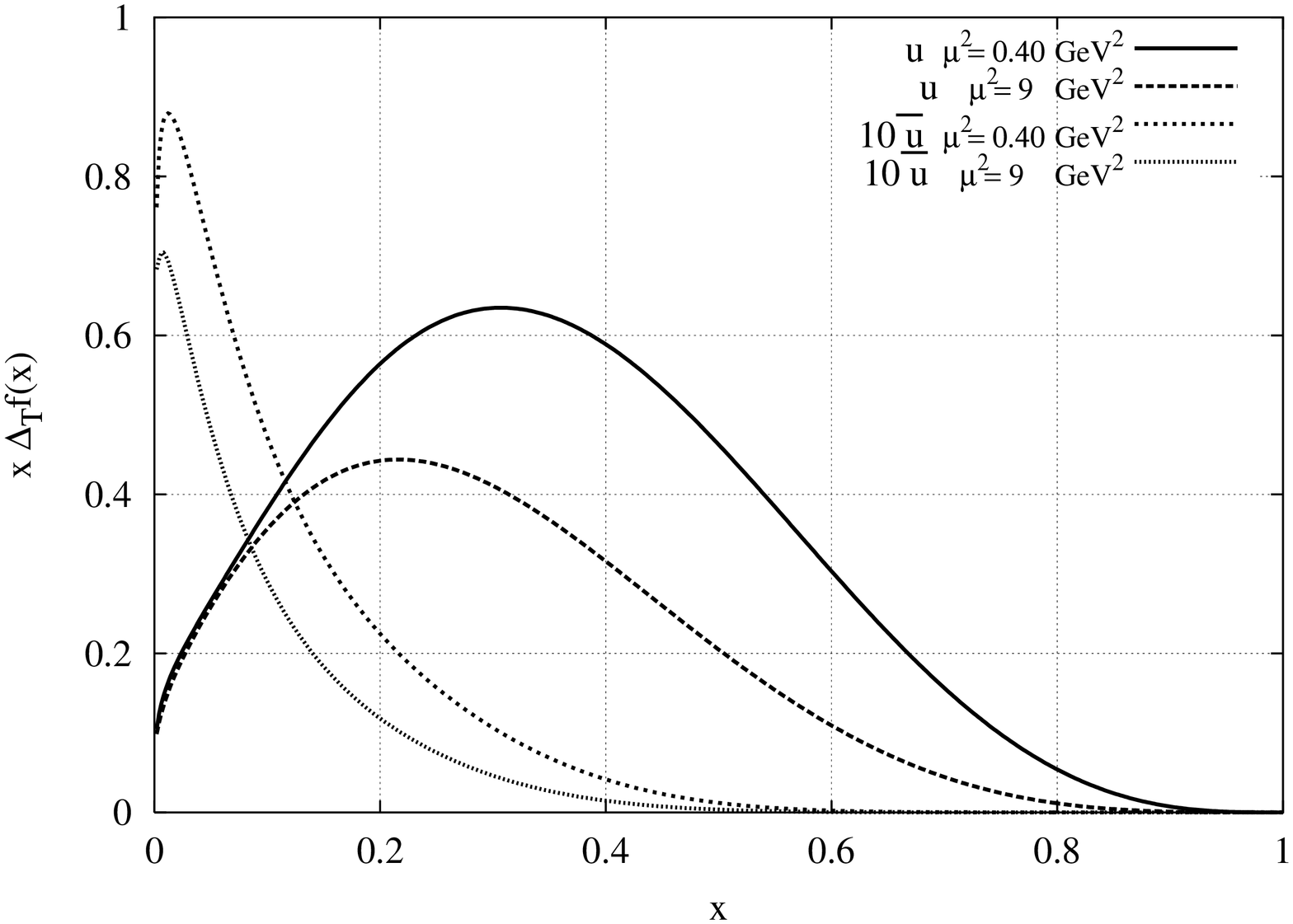}
  \caption{The $u$ and $\bar{u}$ transversity distributions, as obtained from
  the GRV parametrisation and Eq.~\eqref{minbound}, top panel, or
  Eq.~\eqref{sofbound}, bottom:
  $x\Delta_Tu$ at $\mu^2=\mu_0^2=0.40\GeV^2$ (dashed curve) and
  $\mu^2=9\GeV^2$ (solid curve);
  $x\Delta_T\bar{u}$ at $\mu^2=\mu_0^2=0.40\GeV^2$ (dotted curve) and
  $\mu^2=9\GeV^2$ (dot--dashed curve).
  Note that the $\bar{u}$ transversity distributions have been multiplied by a
  factor of $10$.}
  \label{umin}
\end{figure}

After setting the initial condition \eqref{minbound} or \eqref{sofbound}, all
distributions are evolved at NLO according to the appropriate DGLAP equations
(for transversity, see~\cite{h1evol}; the numerical codes we use to solve the
DGLAP equations are described in~\cite{cafacor}). The $u$ sector of
transversity is displayed in Fig.~\ref{umin} for the minimal bound
\eqref{minbound} and for the Soffer bound~\eqref{sofbound}.
%
%\begin{figure}[hbt]
%  \centering
%  \includegraphics[width=8cm,angle=-90]{tr_pdf_s_3}
%  \caption{The $u$ and $\bar{u}$ transversity distributions as obtained from
%  the GRV parametrisation and Eq.~\eqref{sofbound}. The meaning of the curves
%  is as in Fig.~\ref{umin}.}
%  \label{usof}
%\end{figure}

\begin{figure}[hbt]
  \centering
  \includegraphics[width=7cm,angle=-90]{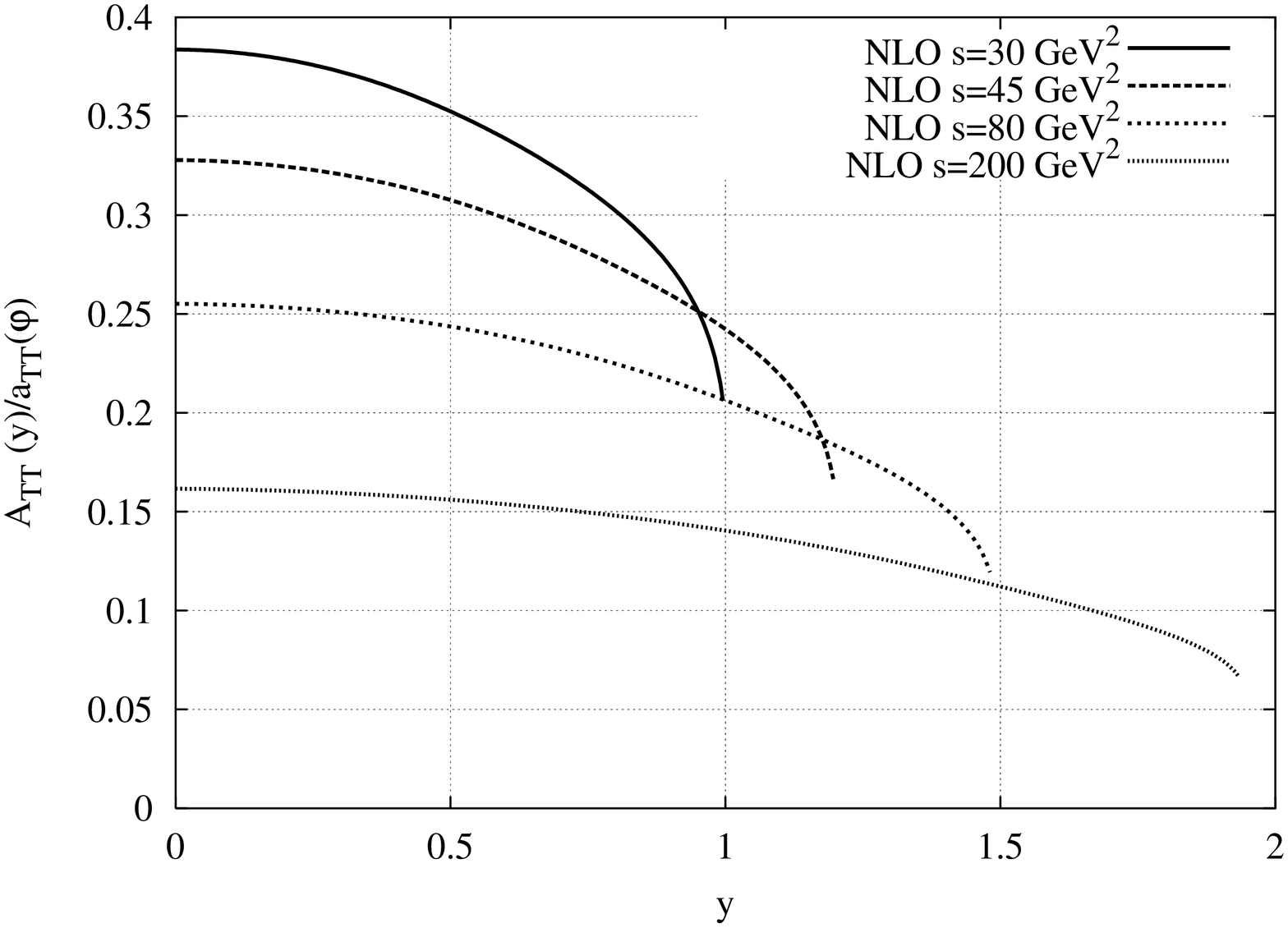}
  \caption{The NLO double transverse-spin asymmetry $A_{TT}(y)/a_{TT}$,
  integrated between $M=2\GeV$ and $M=3\GeV$, for various values of $s$; the
  minimal bound \eqref{minbound} is used for the input distributions.}
  \label{nlom23}
\end{figure}

\begin{figure}[hbt]
  \centering
  \includegraphics[width=7cm,angle=-90]{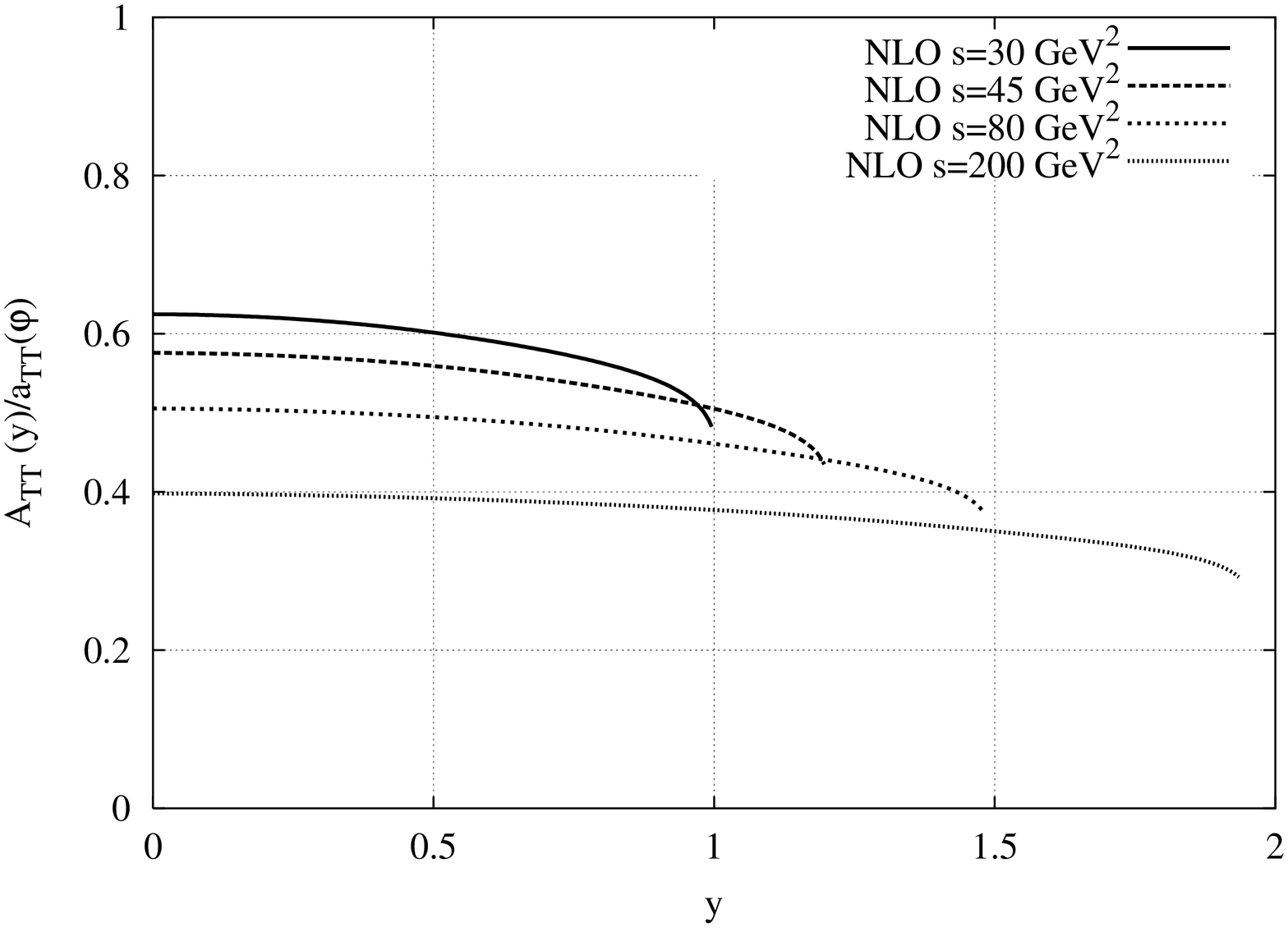}
  \caption{As Fig.~\ref{nlom23}, but with input distributions corresponding to
  the Soffer bound~\eqref{sofbound}.}
  \label{nlos23}
\end{figure}

The transverse Drell--Yan asymmetry $A_{TT}^{DY}/a_{TT}$, integrated over $M$
between $2\GeV$ and $3\GeV$ (\emph{i.e.} below the $J/\psi$ resonance region),
for various values of $s$ is shown in Fig.~\ref{nlom23}. As can be seen, the
asymmetry is of order of 30\% for $s=30\GeV^2$ (fixed-target option) and
decreases by a factor two for a centre-of-mass energy typical of the collider
mode ($s=200\GeV^2$). The corresponding asymmetry obtained by saturating the
Soffer bound, that is by using Eq.~\eqref{sofbound} for the input distributions,
is displayed in Fig.~\ref{nlos23}. As expected, it is systematically larger,
rising to over 50\% for fixed-target kinematics.

\begin{figure}[hbt]
  \centering
  \includegraphics[width=7cm,angle=-90]{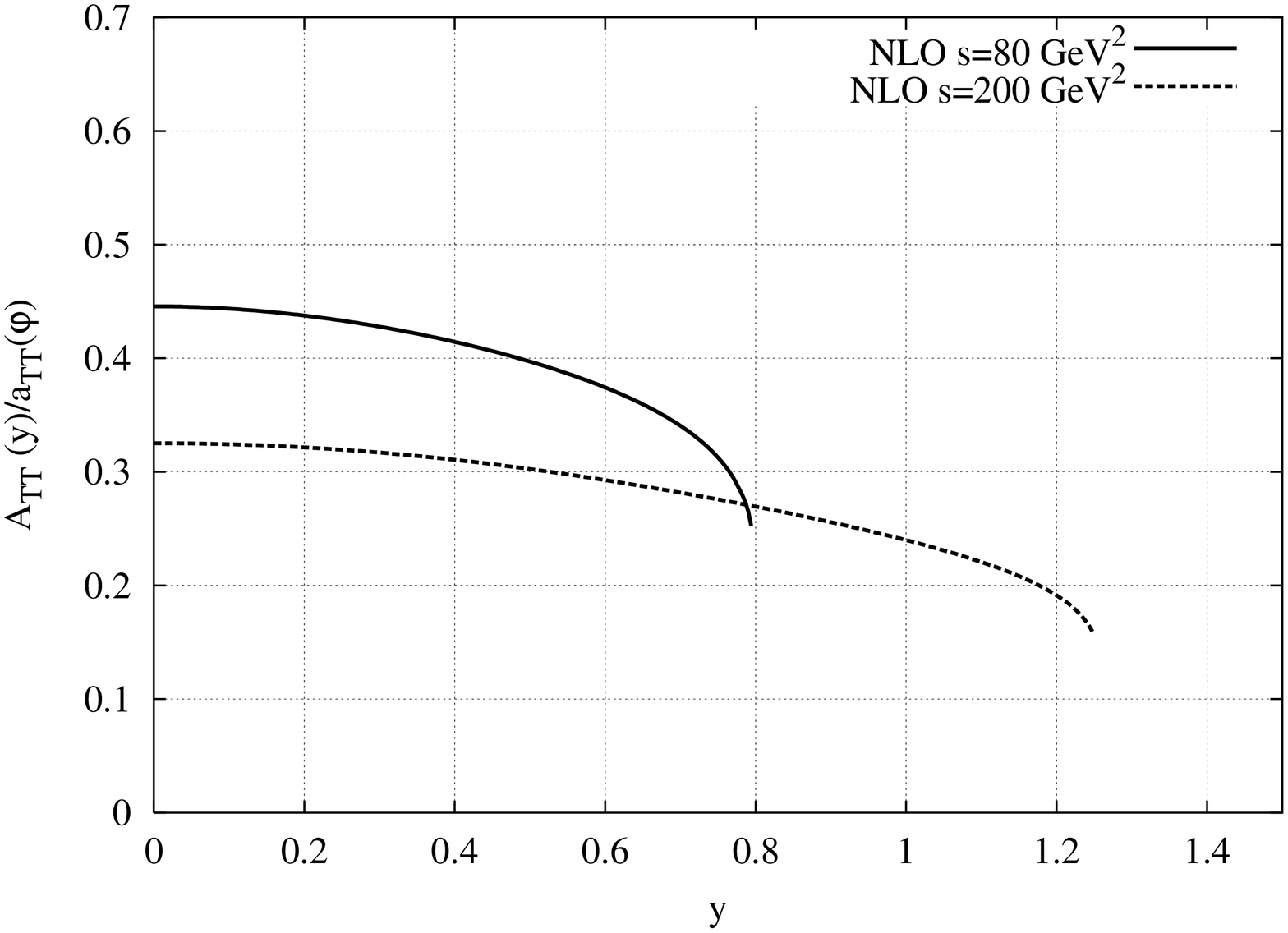}
  \caption{The NLO double transverse-spin asymmetry $A_{TT}(y)/a_{TT}$,
  integrated between $M=4\GeV$ and $M=7\GeV$, for various values of $s$; the
  minimal bound \eqref{minbound} is used for the input distributions.}
  \label{nlom47}
\end{figure}

Above the $J/\psi$ peak $A_{TT}^{DY}/a_{TT}$ appears as shown in
Fig.~\ref{nlom47}, where we present the results obtained with the minimal bound
\eqref{minbound}. Comparing Figs.~\ref{nlom23} and \ref{nlom47}, we see that the
asymmetry increases at larger $M$ (recall though that the cross-section falls
rapidly with growing $M$).

\begin{figure}[hbt]
  \centering
  \includegraphics[width=7cm,angle=-90]{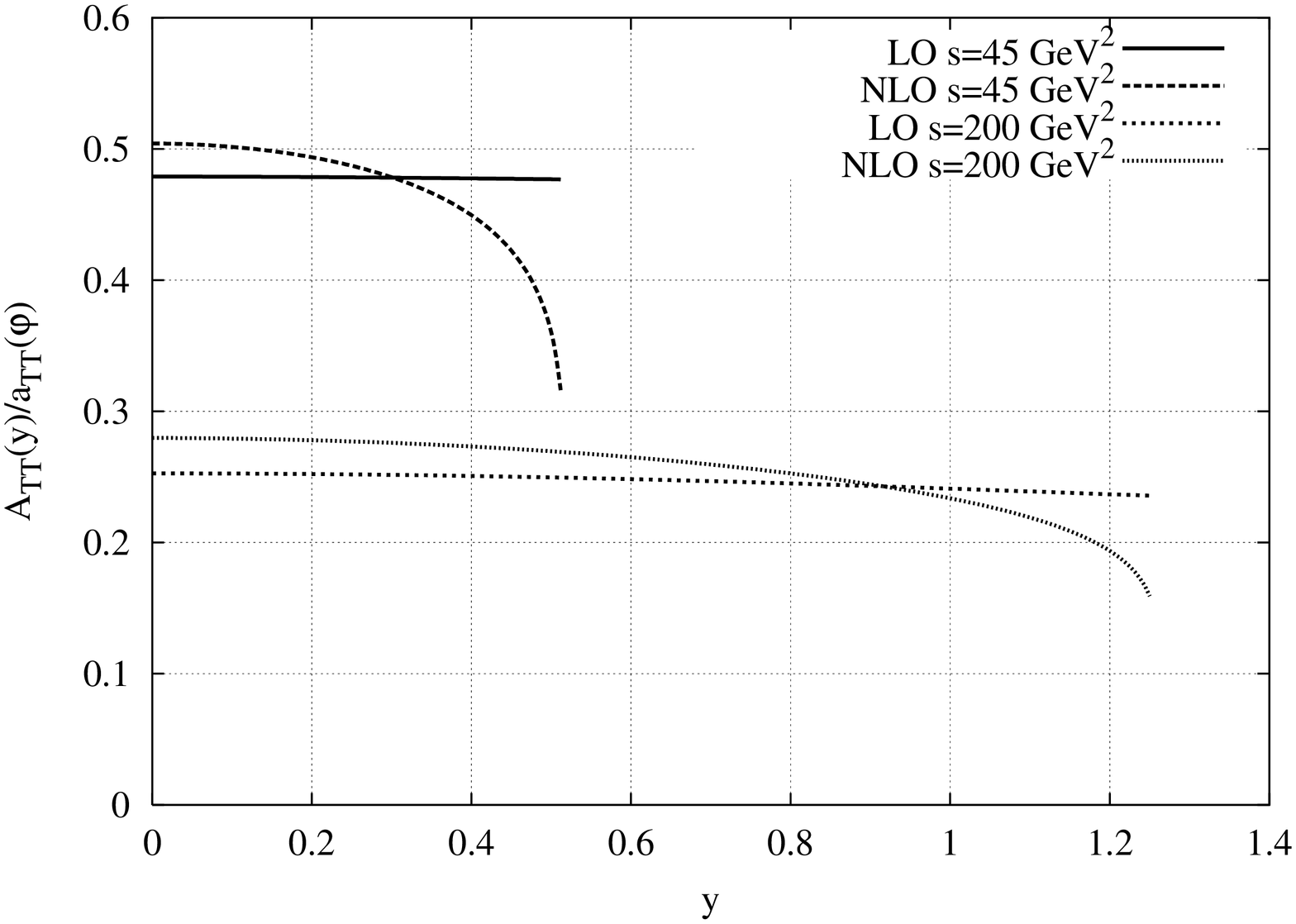}
  \caption{NLO \emph{vs.} LO double transverse-spin asymmetry $A_{TT}(y)/a_{TT}$
  at $M=4\GeV$ for $s=45\GeV^2$ and $s=200\GeV^2$; the minimal bound
  \eqref{minbound} is used for the input distributions.}
  \label{nlolo4}
\end{figure}

The importance of NLO QCD corrections may be appreciated from Fig.~\ref{nlolo4},
where one sees that the NLO effects hardly modify the asymmetry since the $K$
factors of the transversely polarised and unpolarised cross-sections are similar
to each other and therefore cancel out in the ratio.
%%%%%
As for the dependence on the factorisation scale $\mu$ (we recall that the
results presented in all figures are obtained setting $\mu = M$), we have
repeated the calculations with two other choices ($\mu = 2 M$ and $\mu = M/2$)
and found no sensible differences.
%%%%%%

A \emph{caveat} is in order at this point. The GSI kinematics is dominated by
the domain of large $\tau$ and large $z=\tau/x_1x_2$, where real-gluon emission
is suppressed and where there are powers of large logarithms of the form
$\ln(1-z)$, which need to be resummed to all orders in
$\alpha_s$~\cite{shimizu}. It turns out that the effects of threshold
resummation on the asymmetry $A_{TT}^{DY}$ in the regime we are considering,
although not irrelevant, are rather small (about 10\%) if somewhat dependent on
the infrared cutoff for soft-gluon emission.

%%%%
The feasibility of the $A_{TT}$ measurement at GSI has been thoroughly
investigated by the PAX Collaboration (see App.~F of~\cite{pax}). In collider
mode, with a luminosity of $5 \cdot 10^{30}\,\text{cm}^{-2}\,\text{s}^{-1}$, a
proton polarisation of 80\%, an antiproton polarisation of 30\% and considering
dimuon invariant masses down to $M = 2\GeV$, after one year's data taking one
expects a few hundred events per day and a statistical accuracy on $A_{TT}$ of
10--20\%.
%%%%

\section{The $J/\psi$ region}
Before concluding, we briefly comment on the possibility
of accessing transversity via $J/\psi$ production in $p\bar{p}$ scattering.
%Although the double transverse-spin asymmetries are quite sizeable,
%the Drell--Yan counting rate in the continuum region above
%$M=4\GeV$ falls down rapidly
%and therefore large statistics is required in order to determine
%$A_{TT}$ with high accuracy away from the low-$M$ region.
It is known that the dilepton production rate around $M=3\GeV$,
\emph{i.e.} at the $J/\psi$ peak, is two orders of magnitude higher than in the
region $M\simeq4\GeV$. Thus, with a luminosity of $5 \cdot
10^{30}\,\text{cm}^{-2}\,\text{s}^{-1}$, one expects a number of
$p\bar{p}\to{J}/\psi\to\ell^-\ell^+$ events of order $10^5$ per year at GSI
collider energies.
This renders the measurement of $A_{TT}$
in the $J/\psi$-resonance region extremely advantageous from a statistical
point of view.

As explained in~\cite{Anselmino:2004ki}, if $J/\psi$
formation is dominated by the $q\bar{q}$ annihilation channel,
at leading order the double
transverse-spin asymmetry at the $J/\psi$ peak has the same structure as the
asymmetry for Drell--Yan continuum production, since the $J/\psi$ is a vector
particle and the $q\bar{q}\,J/\psi$
coupling has the same helicity structure as
the $q\bar{q}\gamma^*$ coupling. The CERN SPS data~\cite{sps} show that the
$p\bar{p}$ cross-section for $J/\psi$ production at $s=80\GeV^2$ is about ten
times larger than the corresponding $pp$ cross-section, which is a strong
indication that the $q\bar{q}$-fusion mechanism is indeed dominant. Therefore,
at the $s$ values of interest here ($s\lesssim200\GeV^2$)
dilepton production in
the $J/\psi$ resonance region can be described in a manner analogous to
Drell--Yan continuum production, with the elementary subprocess
$q\bar{q}\to\gamma^*\to\ell^-\ell^+$ replaced by $q\bar{q}\to
J/\psi\to\ell^-\ell^+$~\cite{carlson}.
%Instead of the quark electric charges one
%has the $J/\psi$ vector couplings to $q\bar{q}$ and to $\ell^+\ell^-$,
%\begin{equation}
%  16 \pi^2 \alpha_\text{em}^2 e_q^2 \rightarrow (g_q^V)^2 (g_{\ell}^V)^2 \,,
%\end{equation}
%and the virtual-photon propagator is replaced by a Breit--Wigner function,
%\begin{equation}
%  \frac{1}{M^4} \rightarrow
%  \frac{1}{(M^2 - M_{J/\psi}^2)^2 + M_{J/\psi}^2 \Gamma_{J/\psi}^2} \,,
%\end{equation}
%where $\Gamma_{J/\psi}$ is the $J/\psi$ width.
Using this model, which successfully accounts
for the SPS $J/\psi$ production data at moderate values of~$s$,
it was found in \cite{Anselmino:2004ki} that the transverse asymmetry
at the $J/\psi$ peak is of the order of 25--30\%.

At next-to-leading order,
due to QCD radiative corrections,
one cannot use a point-like $q\bar{q}\,J/\psi$
coupling, and therefore
it is not possible to extend in a straightforward way the
model used to evaluate $A_{TT}$ at leading order. Were NLO effects
not dominant, as is the case for continuum production,
one could still expect the $J/\psi$ asymmetry to be quite sizeable,
but this is no more than an educated guess.
What we wish to emphasise, however, is the
importance of experimentally investigating
the $J/\psi$ double transverse-spin
asymmetry, which can
shed light both on the transversity content of the
nucleon and on the mechanism of $J/\psi$ formation
(since gluon-initiated hard processes do not contribute
to the transversely polarised scattering,
the study of $A_{TT}$
in the $J/\psi$ resonance region may give information
on the relative weight of gluon and quark-antiquark subprocesses
in $J/\psi$ production).

%Moreover, since the $u$ sector is largely dominant, there is only one
%$J/\psi$ coupling (to $u$ quarks), which thus cancels in the cross-section
%ratio. The conclusion is then that the transverse asymmetry for $J/\psi$
%production in $p\bar{p}$ scattering at GSI energies is essentially the same as
%the Drell--Yan transverse asymmetry at $M=3.1\GeV$ (the $J/\psi$ mass).
%The result is shown in Fig.~\ref{nlojpsi}.
%One sees that the asymmetry is again
%rather large but, compared to the continuum case,
%its measurement is made easier
%by the high counting rate of dileptons at the resonance peak.
%\begin{figure}[hbt]
%  \centering
%  \includegraphics[width=7cm,angle=-90]{asy_sc5}
%  \caption{The double transverse-spin asymmetry
%in the $J/\psi$ resonance region
%  for various c.m.\ energies. As usual, the minimal bound \eqref{minbound} is
%  used for the input distributions.}
%  \label{nlojpsi}
%\end{figure}

\section{Conclusions}
In conclusion, experiments with polarised antiprotons at GSI will
represent a unique opportunity to investigate the transverse polarisation
structure of hadrons. The present chapter, which confirms the results
of~\cite{Anselmino:2004ki}, shows that the double transverse-spin asymmetries
are large enough to be experimentally measured and therefore
represent the most promising observables to directly access
the quark transversity distributions.

%%%%%%%%%%%%%%%%%%%%%%%%%%%%%%%%%%%%%%%%
%%%%%%%%%%%%%%%%%%%%%%%%%%%%%%%%%%%%%%%%%

\chapter{On the Scale Variation of the Total Cross Section for
 Higgs Production at the LHC and at the Tevatron\label{chap4}}
\fancyhead[LO]{\nouppercase{Chapter 4. On the Scale Variation of the Total Cross Section for
 Higgs Production}}

The validity of the mechanism of mass generation in the Standard Model
will be tested at the new collider, the LHC.  For this we require precision
studies in the Higgs sector to confirm its existence. 
This program involves a rather complex analysis of the QCD backgrounds 
with the corresponding radiative corrections fully taken into account.
Studies of these corrections for specific processes have been performed 
by various groups, to an accuracy which has reached the 
next-to-next-to-leading order (NNLO) level in $\alpha_s$,
the QCD coupling constant. The quantification of the impact of these 
corrections requires the determination of the hard scattering partonic 
cross sections up to order $\alpha_s^3$, together with the DGLAP kernels 
controlling the evolution of the parton distributions 
determined at the same perturbative order. 
Therefore, the study of the evolution of the parton distributions, 
using the three-loop results on the anomalous dimensions \cite{vogt1}, is  
critical for the success of this program. Originally NNLO predictions 
for some particular processes such as total cross sections 
\cite{RSVN} have been obtained using the approximate expressions for 
these kernels \cite{VN1}. The completion of the exact computation of the 
NNLO DGLAP kernels motivates more detailed studies of the same 
observables based on these exact kernels and the investigation of the
factorization ($\mu_F$) and renormalization ($\mu_R$) scale dependences of the result,
which are still missing. In this work we are going to reanalyze 
these issues from a broader perspective.  Our analysis is here exemplified 
in the case of the total cross sections at the LHC ($pp$) and at the 
Tevatron ($p \bar{p}$) for Higgs production using the hard scatterings 
computed in \cite{RSVN} and their dependence on the factorization and 
renormalization scales.
Our study is based on the exact and well defined
NNLO computations of the hard scatterings for this process and we have
not taken into account any threshold re summation since this involves further
approximations. These effects have been considered in \cite{Nason}.
The DGLAP equation is solved directly in $x$-space
using a method which is briefly illustrated below
and which is accurate up to order $\alpha_s^2$. Our input distributions at
a small scale will be specified below. We also analyse the corresponding
K-factors and the region of stability of the perturbative expansion
by studying their variation under changes in all the relevant
scales. It is shown that the NNLO corrections are sizeable while the region
of reduced scale dependence is near the value $\mu_F=m_H$
with $\mu_R$ around the same value but slightly higher.

\section{Higgs production at LHC}

\begin{figure}
{\centering \resizebox*{7cm}{!}{\rotatebox{0}
{\includegraphics{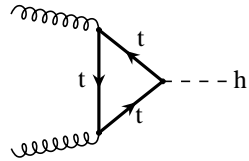}}}\par}
\caption{The leading order diagram for Higgs production by gluon fusion}
\label{g-fusion}
\end{figure}

\begin{figure}
{\centering \resizebox*{7cm}{!}{\rotatebox{0}
{\includegraphics{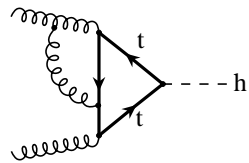}}}\par}
\caption{A typical NLO diagram for Higgs production by gluon fusion}
\label{g-fusion1}
\end{figure}
The Higgs field, being responsible for the mechanism of mass generation, 
can be radiated off by any massive state and its coupling is proportional 
to the mass of the same state. At the LHC one of the golden plated modes 
to search for the Higgs is its production via the mechanism of gluon fusion.
The leading order contribution is shown in Fig.~(\ref{g-fusion}) which
shows that dependence of the amplitude is through the quark loop. Most of the
contribution comes from the top quark, since this is the heaviest quark and has
the largest coupling to the Higgs field. NLO and NNLO corrections
have been computed in the last few years by various groups \cite{Harl&Kil}, \cite{Anas&Melni}.
A typical NLO correction is shown in Fig.~(\ref{g-fusion1}).
In the infinite mass limit of the quark mass in the loop (see \cite{dawson}
for a review), an effective description of the process is obtained in
leading order by the Lagrangian density
\beqn
{\cal L}_{\rm eff}&=&{\alpha_s \over 12 \pi} G_{\mu\nu}^A G^{A~\mu\nu}
\biggl({H\over v}\biggr)\nonumber \\
&=&{\beta_F \over g_s} G_{\mu\nu}^A G^{A~\mu\nu}
\biggl({H\over 2v}\biggr)(1-2 \alpha_s/\pi),\nonumber
\label{effth}
\eeqn
with
\beq
\beta_F={g_s^3 N_H\over 24 \pi^2}
\eeq
being the contribution of $N_H$ heavy fermion loops to the QCD beta function.
This effective Lagrangian can be used to compute the radiative corrections
in the gluon sector. A discussion of the NNLO approach to the computation
of the gluon fusion contributions to Higgs production has been presented
in \cite{RSVN}, to which we refer for more details. We recall that in
this chapter we presented a study for both scalar and pseudoscalar Higgs
production, the pseudoscalar appearing in 2-Higgs doublets models. 
The diagonalization of the mass matrix for the Higgs at the minimum 
introduces scalar and pseudoscalar interactions between the various Higgs 
and the quarks, as shown from the structure of the operator $O_2$ below 
in eq. (\ref{eqn2.2}).
In the large top-quark mass limit the Feynman rules
for scalar Higgs  production (${\rm H}$) can be derived from the
effective Lagrangian density \cite{RSVN1}, \cite{Stein}, \cite{Laenen&Spira},
\begin{eqnarray}
\label{eqn2.1}
{\cal L}^{\rm H}_{eff}=G_{\rm H}\,\Phi^{\rm H}(x)\,O(x) \quad
\mbox{with} \quad O(x)=-\frac{1}{4}\,G_{\mu\nu}^a(x)\,G^{a,\mu\nu}(x)\,,
\end{eqnarray}
whereas the production of a pseudo-scalar Higgs \cite{Chet&Bard},
(${\rm A}$) is obtained from
\begin{eqnarray}
\label{eqn2.2}
&&{\cal L}_{eff}^{\rm A}=\Phi^{\rm A}(x)\Bigg [G_{\rm A}\,O_1(x)+
\tilde G_{\rm A}\,O_2(x)\Bigg ] \quad \mbox{with} \quad
\nonumber\\[2ex]
&&O_1(x)=-\frac{1}{8}\,\epsilon_{\mu\nu\lambda\sigma}\,G_a^{\mu\nu}\,
G_a^{\lambda\sigma}(x) \,,
\nonumber\\[2ex]
&&O_2(x) =-\frac{1}{2}\,\partial^{\mu}\,\sum_{i=1}^{n_f}
\bar q_i(x)\,\gamma_{\mu}\,\gamma_5\,q_i(x)\,,
\end{eqnarray}
where $\Phi^{\rm H}(x)$ and  $\Phi^{\rm A}(x)$ represent the scalar and
pseudo-scalar fields respectively and $n_f$ denotes the number of light
flavours.
$G_a^{\mu\nu}$ is the field strength of QCD
and the quark fields are denoted by $q_i$.
We refer the reader to \cite{RSVN} for further details.

Using the effective Lagrangian one can calculate the
total cross section of the reaction

\begin{eqnarray}
\label{eqn2.10}
H_1(P_1)+H_2(P_2)\rightarrow {\rm B} + X\,,
\end{eqnarray}
where $H_1$ and $H_2$ denote the incoming hadrons and $X$ represents an
inclusive hadronic state and ${\rm B}$ denotes the scalar or the psudoscalar
particle produced in the reaction.
The total cross section is given by
\begin{eqnarray}
\label{eqn2.11}
&&\sigma_{\rm tot}=\frac{\pi\,G_{\rm B}^2}{8\,(N^2-1)}\,\sum_{a,b=q,\bar q,g}\,
\int_x^1 dx_1\, \int_{x/x_1}^1dx_2\,f_a(x_1,\mu^2)\,f_b(x_2,\mu^2)\,
\nonumber\\[2ex] && \qquad\qquad\times
\Delta_{ab,{\rm B}}\left ( \frac{x}{x_1\,x_2},\frac{m^2}{\mu^2} \right ) \,,
\nonumber\\[2ex]
&&\mbox{with}\quad x=\frac{m^2}{S} \quad\,,\quad S=(P_1+P_2)^2\quad
\,,
\end{eqnarray}
where the factor $1/(N^2-1)$ is due to the average over colour.
The parton distributions $f_a(y,\mu^2)$ ($a,b=q,\bar q,g$)
depend on the mass factorization/renormalization scale $\mu$.
$\Delta_{ab,{\rm B}}$ denotes the partonic hard scattering coefficient
computed with NNLO accuracy.

\section{The NNLO Evolution}
We summarize the main features of the NNLO DGLAP evolution. 
As usual we introduce singlet $(+)$ and non-singlet $(-)$ parton distributions

\begin{equation}
q_{i}^{(\pm)}=q_{i}\pm\overline{q}_{i},\qquad q^{(\pm)}=\sum_{i=1}^{n_{f}}q_{i}^{(\pm)}
\label{eq:definizioni}
\end{equation}
whose evolution is determined by the corresponding equations 

\begin{equation}
\frac{\textrm{d}}{\textrm{d}\log Q^{2}}\left(\begin{array}{c}
q^{(+)}(x,Q^{2})\\
g(x,Q^{2})\end{array}\right)=\left(\begin{array}{cc}
P_{qq}(x,\alpha_{s}(Q^{2})) & P_{qg}(x,\alpha_{s}(Q^{2}))\\
P_{gq}(x,\alpha_{s}(Q^{2})) & P_{gg}(x,\alpha_{s}(Q^{2}))\end{array}\right)\otimes\left(\begin{array}{c}
q^{(+)}(x,Q^{2})\\
g(x,Q^{2})\end{array}\right)
\label{eq:singlet}
\end{equation}
for the singlet combination 
and a scalar one for the non-singlet case

\begin{equation}
\frac{\textrm{d}}{\textrm{d}\log Q^{2}}q_i^{(-)}(x,Q^{2})=P_{NS}(x,\alpha_{s}(Q^{2}))\otimes q_i^-(x,Q^{2}).
\label{eq:DGLAP}
\end{equation}

The convolution product is defined by
\begin{equation}
\left[a\otimes b\right](x)=\int_{x}^{1}\frac{\textrm{d}y}{y}a\left(\frac{x}{y}\right)b(y)=
\int_{x}^{1}\frac{\textrm{d}y}{y}a(y)b\left(\frac{x}{y}\right).
\end{equation}

We recall that the perturbative expansion, up to NNLO, of the kernels
is
\begin{equation}
P(x,a_{s})= a_s P^{(0)}(x)+ a_s^2 P^{(1)}(x)+ a_s^3 P^{(2)}(x)+\ldots.
\label{kern1}
\end{equation}
where $a_s\equiv \alpha_s/(4 \pi)$.
In order to solve the evolution equations directly in $x$-space \cite{Rossi},
\cite{Giele}, (see \cite{cafacor} for an NLO implementation of the method), we
assume solutions of the form \cite{CCG}
\begin{eqnarray}
f(x,Q^{2}) & = & \sum_{n=0}^{\infty}\frac{A_{n}(x)}{n!}
\log^{n}\frac{a_{s}(Q^{2})}{a_{s}(Q_{0}^{2})}+a_s (Q^{2})
\sum_{n=0}^{\infty}\frac{B_{n}(x)}{n!}\log^{n}
\frac{a_{s}(Q^{2})}{a_{s}(Q_{0}^{2})}\nonumber \\
&&+ a_{s}^2(Q^{2})\sum_{n=0}^{\infty}
\frac{C_{n}(x)}{n!}\log^{n}\frac{a_{s}(Q^{2})}{a_{s}(Q_{0}^{2})}
\label{eq:ansatz}
\end{eqnarray}
for each parton distribution $f$, where $Q_{0}$ defines the initial
evolution scale. The ansatz is introduced into the evolution equations and
used to derive recurrence relations for its unknown coefficients
$A_n,B_n,C_n$, involving polylogarithmic functions
\cite{remiddi1,remiddi2} which are then implemented numerically.

This ansatz corresponds to a solution of the DGLAP equation accurate up to
order $a_s^2$ (truncated solution).
It can be shown that \cite{CCG} this ansatz reproduces the solution of the
DGLAP equation in (Mellin) moment space obtained with the same accuracy
in $a_s$.  Modifications of this ansatz also allow to obtain the
so-called ``exact'' solutions of the equations for the moments \cite{Vogt3}.
These second solutions include higher order terms in $a_s$ and can be
identified only in the non-singlet case. Exact approaches also include
an exact solution of the renormalization group equation
for the $\beta$-function, which embodies the effects of the coefficients
$\beta_0, \beta_1$ and $\beta_2$ to higher order in $a_s$.
The term ``exact'' is, however, a misnomer since the accuracy of the solution
is limited to the knowledge of the first three contributions to the
expansion in the beta function and in the kernels.
It can be shown both for exact and for the truncated solutions that solving
the equations by an ansatz in $x$-space is completely equivalent to
searching for the solution in moment space, since in moment space the
recursion relations can be solved exactly \cite{CCG}.

A numerical comparison of our approach
with that of \cite{Vogt3} has been presented in \cite{CCG} where
it is shown that at LO and NLO there is excellent agreement,
while at NNLO there are discrepancies of a few percent (mainly
in the Singlet case, and at very small and large $x$ values).
For a more detailed discussion we refer
the reader to section $11$ of \cite{CCG}.

\section{Renormalization scale dependence}

For a better determination of the dependence of the perturbative cross
section on the scales of a certain process it is important
to keep these scales independent and study the behaviour of the corresponding
hadronic cross section under their variation. In our case the two relevant
scales are the factorization scale $\mu_F$ and the renormalization scale
$\mu_R$ which can be both included in the evolution by a rearrangement
of the evolution kernels up to NNLO.

The study of the dependence of the solution upon the various scales is then
performed in great generality and includes also the logarithmic
contributions $\log(\mu_F/\mu_R)$ coming from the hard scatterings given
in \cite{RSVN}, where, however, only the specific
point $\mu_F=\mu_R=m_H$ was considered. The separation of the scales
should then appear not only in the hard scatterings but also in the evolution
equations. This issue has been addressed in \cite{Vogt3} and can be
reconsidered also from $x$-space \cite{CCG} using the $x$-space logarithmic
ansatz (\ref{eq:ansatz}).

The scale dependence of the parton distribution
functions is then expressed by a generalized DGLAP equation
\ba
\label{evolution}
\frac{\partial}{\partial \ln \mu_F^2}\, f_i(x,\mu_F^2,\mu_R^2)=
P_{ij}(x,\mu_F^2,\mu_R^2) \otimes f_j(x,\mu_F^2,\mu_R^2)\,,
\label{scales}
\ea
where $\mu_F$ is now a generic factorization scale.

Generally speaking, both the kernels and the PDF's have a dependence on
the scales $\mu_F$ and $\mu_R$, and formally, a comparison between these
scales is always possible up to a fixed order by using the renormalization
group equations for the running coupling $\alpha_s$.

The renormalization scale dependence of the ansatz (\ref{eq:ansatz}) that 
solves (\ref{scales}) is obtained quite straightforwardly 
by a Taylor expansion of the running coupling $\alpha_s(\mu_F^2)$ in terms 
of $\alpha_s(\mu_R^2)$ \cite{CCG}
\ba
\label{alphasmuf}
\alpha_s(\mu_F^2)=\alpha_s(\mu_R^2)-\left[\frac{\alpha_s^2(\mu_R^2)}{4\pi}
+\frac{\alpha_s^3(\mu_R^2)}{(4\pi)^2}(-\beta_0^2 L^2+\beta_1 L)\right]
\label{betaf}
\ea
where the $\mu_F^2$ dependence is included in the factor 
$L=\ln(\mu_F^2/\mu_R^2) $, and the coefficients of the $\beta$-function,
(the $\beta_i$) are listed below \cite{Tarasov}, \cite{Larin}, \cite{Ritbergen}
\ba
&&\beta_{0}=\frac{11}{3}N_{C}-\frac{4}{3}T_{f},\nonumber\\
&&\beta_{1}=\frac{34}{3}N_{C}^{2}-\frac{10}{3}N_{C}n_{f}-2C_{F}n_{f},\nonumber\\
&&\beta_{2}=\frac{2857}{54}N_{C}^{3}+2C_{F}^{2}T_{f}-\frac{205}{9}C_{F}N_{C}T_{f}
-\frac{1415}{27}N_{C}^{2}T_{f}+\frac{44}{9}C_{F}T_{f}^{2}+\frac{158}{27}N_{C}T_{f}^{2}.
\nonumber\\
\ea
As usual we have set
\begin{equation}
N_{C}=3,\qquad C_{F}=\frac{N_{C}^{2}-1}{2N_{C}}=\frac{4}{3},
\qquad T_{f}=T_{R}n_{f}=\frac{1}{2}n_{f},
\end{equation}
where $N_{C}$ is the number of colors and $n_{f}$ is the number of
active flavors. This number is varied as we step into a region 
characterized by an evolution scale 
$\mu$ larger than a specific quark mass ($\mu \geq m_{q}$). Also the
NNLO matching conditions across flavor thresholds \cite{bmsn},
\cite{cs} are implemented.

Since the perturbative expansion of eq.~(\ref{kern1}) contains powers 
of $\alpha_{s}(\mu_F^2)$ which can be related to the value of 
$\alpha_s(\mu_R^2)$ by (\ref{betaf}), from 
\ba
\label{kern2}
P_{ij}^{NNLO}(x,\mu_F^2)=
\sum_{k=0}^{2} \, \left(\frac{\alpha_s(\mu_F^2)}{4\pi}\right)^{k+1}
P_{ij}^{(k)}(x)\,,
\ea
substituting eq. (\ref{alphasmuf}) into (\ref{kern2}), we obtain the corresponding expression
of the kernels organized in powers of $\alpha_{s}(\mu_R^2)$ up to NNLO, and it reads \cite{Vogt3}
\ba
\label{kern3}
&&P_{ij}(x,\mu_F^2,\mu_R^2)=\frac{\alpha_s(\mu_R^2)}{4\pi}P_{ij}^{(0)}(x) \nonumber\\
&&\hspace{2.5cm}+\frac{\alpha_s^2(\mu_R^2)}{(4\pi)^2}\left(P_{ij}^{(1)}(x)
-\beta_0P_{ij}^{(0)}(x) L\right)
\nonumber\\
&&\hspace{2.5cm}+\frac{\alpha_s^3(\mu_R^2)}{(4\pi)^3}\left[P_{ij}^{(2)}(x)-
2\beta_0 L P_{ij}^{(1)}(x)
-\left(\beta_1 L - \beta_0^2 L^2 \right) P_{ij}^{(0)}(x)\right]\,.\nonumber\\
\ea

The implementation of the method in $x$-space is quite straightforward and
allows us to perform a separate study of the predictions in terms
of $\mu_F$ and $\mu_R$.

\section{Numerical Results}
The use of the NNLO evolution of the parton distributions together
with the results of \cite{RSVN} allows us to provide accurate
predictions for the total cross section for Higgs production.
Here we summarize and discuss our numerical results.
\footnote{Based on the article~\cite{CCG_Eur}}
We use as inital conditions at low scales the sets of distributions
given by MRST \cite{MRST} and Alekhin \cite{Alekhin}.
Our final plots refer to center-of-mass energies which are reachable
at the LHC, with 14 TeV being the largest one achievable in a not so
distant future, and at the Tevatron, where we have selected the corresponding
value as 2 TeV. We have also taken the Higgs mass $m_H$ as a parameter
in the prediction, with an interval of variability which goes from a light
to a heavy Higgs (100 GeV to 300 GeV). Therefore $\mu_F$, $\mu_R$ and $m_H$
are studied chosing various combinations of their possible values in the
determination of total cross sections at leading $(\sigma_{LO})$,
next-to-leading $(\sigma_{NLO})$, and next-to-next-to-leading
order $(\sigma_{NNLO})$. We present both standard two-dimensional
plots and also some three dimensional plots in order to characterize
in detail the structure of the region of stability of the
perturbative expansion. We have also evaluated the K-factors for the
total Higgs cross section  at NLO, defined by
\beq
K^{NLO}=\frac{\sigma_{NLO}}{\sigma_{LO}}
\eeq
and at NNLO
\beq
K^{NNLO}=\frac{\sigma_{NNLO}}{\sigma_{NLO}}.
\eeq
The study of the K-factors has been performed first by keeping the three
scales equal $(\mu_F=\mu_R=m_H)$ and then letting them vary around the
typical value $m_H$. A second set of studies has been performed by taking
typical values of $m_H$ and varying the value of the renormalization scale.
%%%%%%%%%%%%%%%%%%%%%%%%%%%%%%%%%%%%%%%%%%%%%%%%%%%%%%%%%%%%%%%%%%

\subsection{The errors on the cross sections}

\begin{figure}
\subfigure[LHC]{\includegraphics[%
  width=6cm,
  angle=-90]{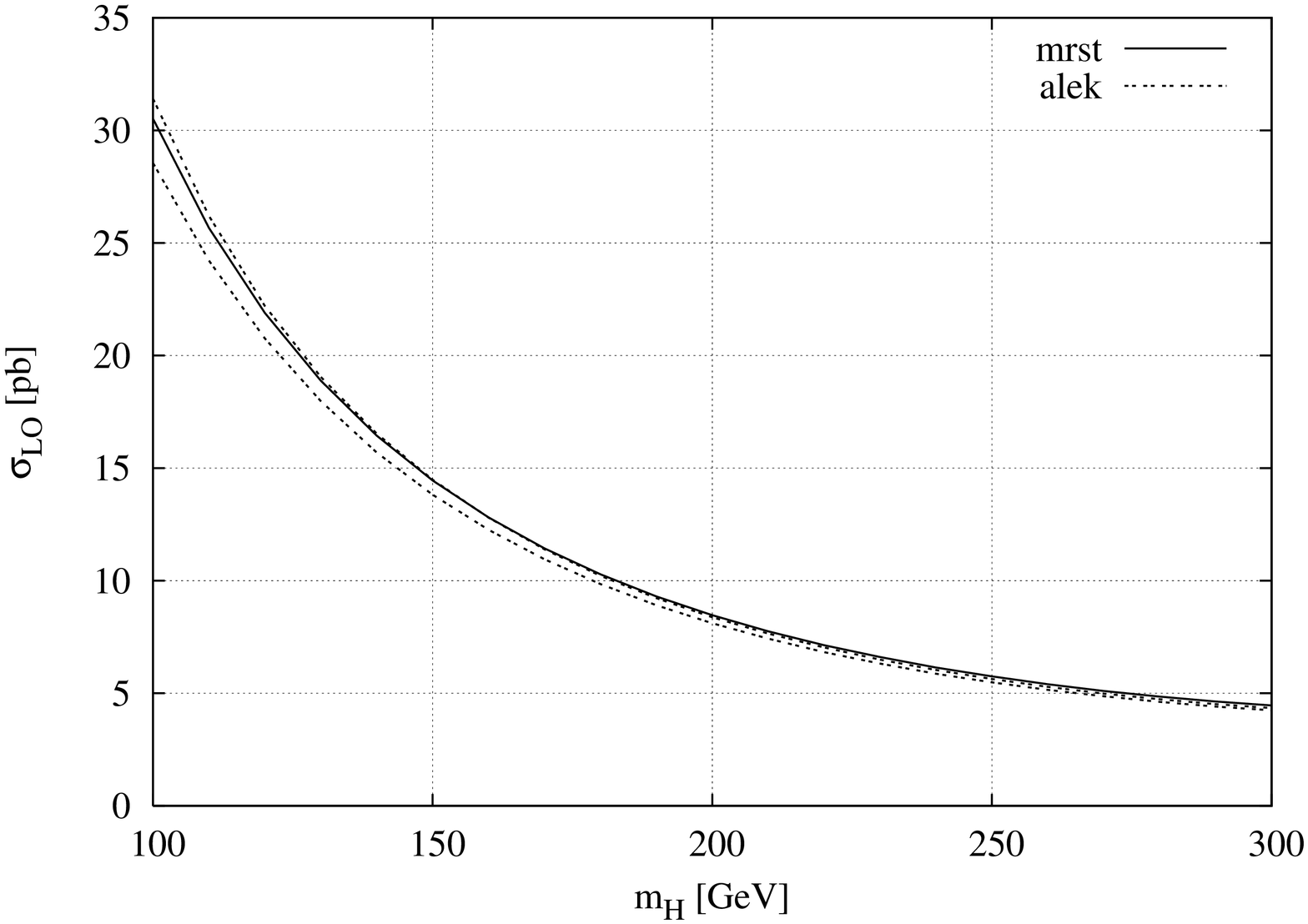}}
\subfigure[LHC]{\includegraphics[%
  width=6cm,
  angle=-90]{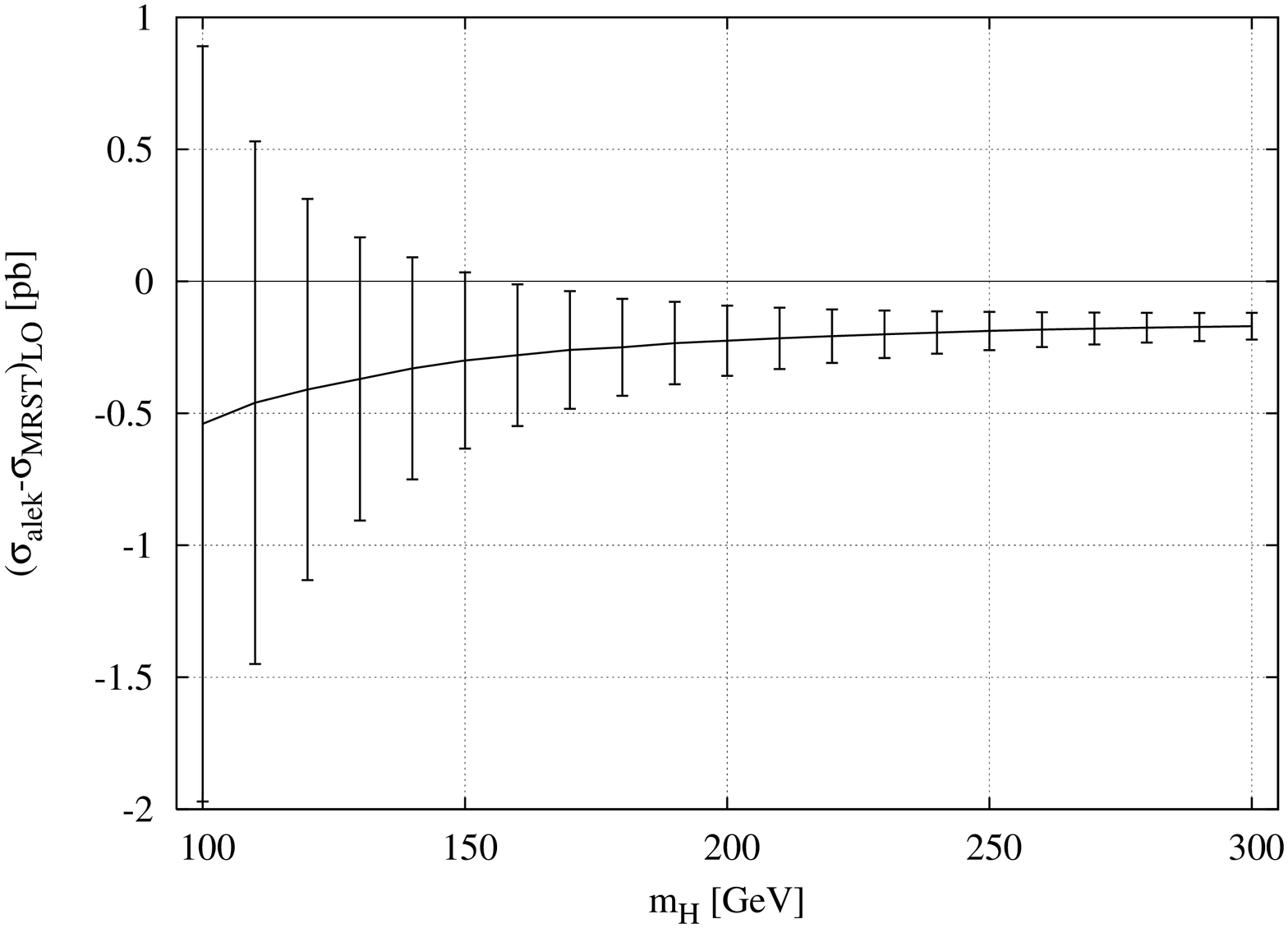}}
\subfigure[LHC]{\includegraphics[%
  width=6cm,
  angle=-90]{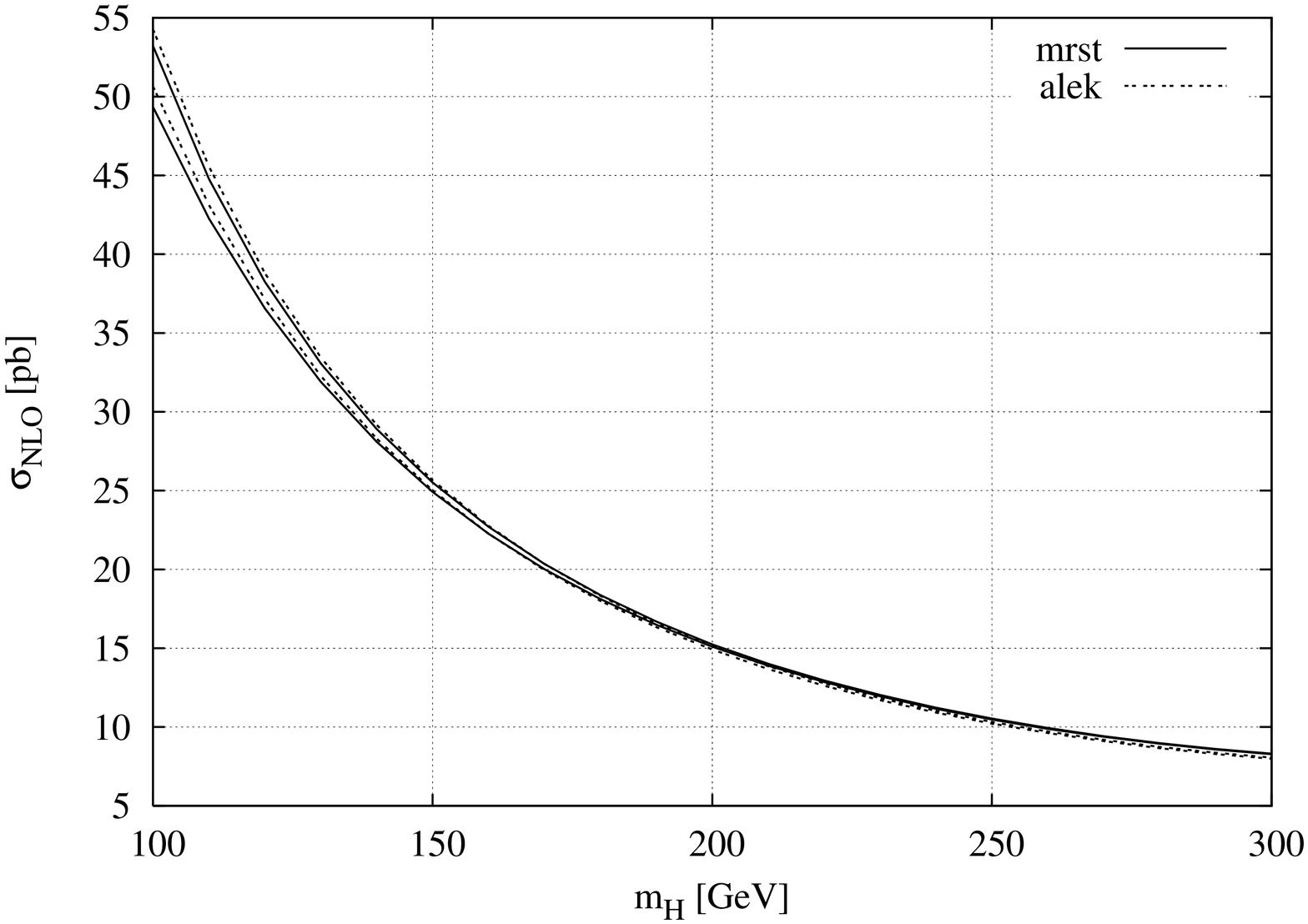}}
\subfigure[LHC]{\includegraphics[%
  width=6cm,
  angle=-90]{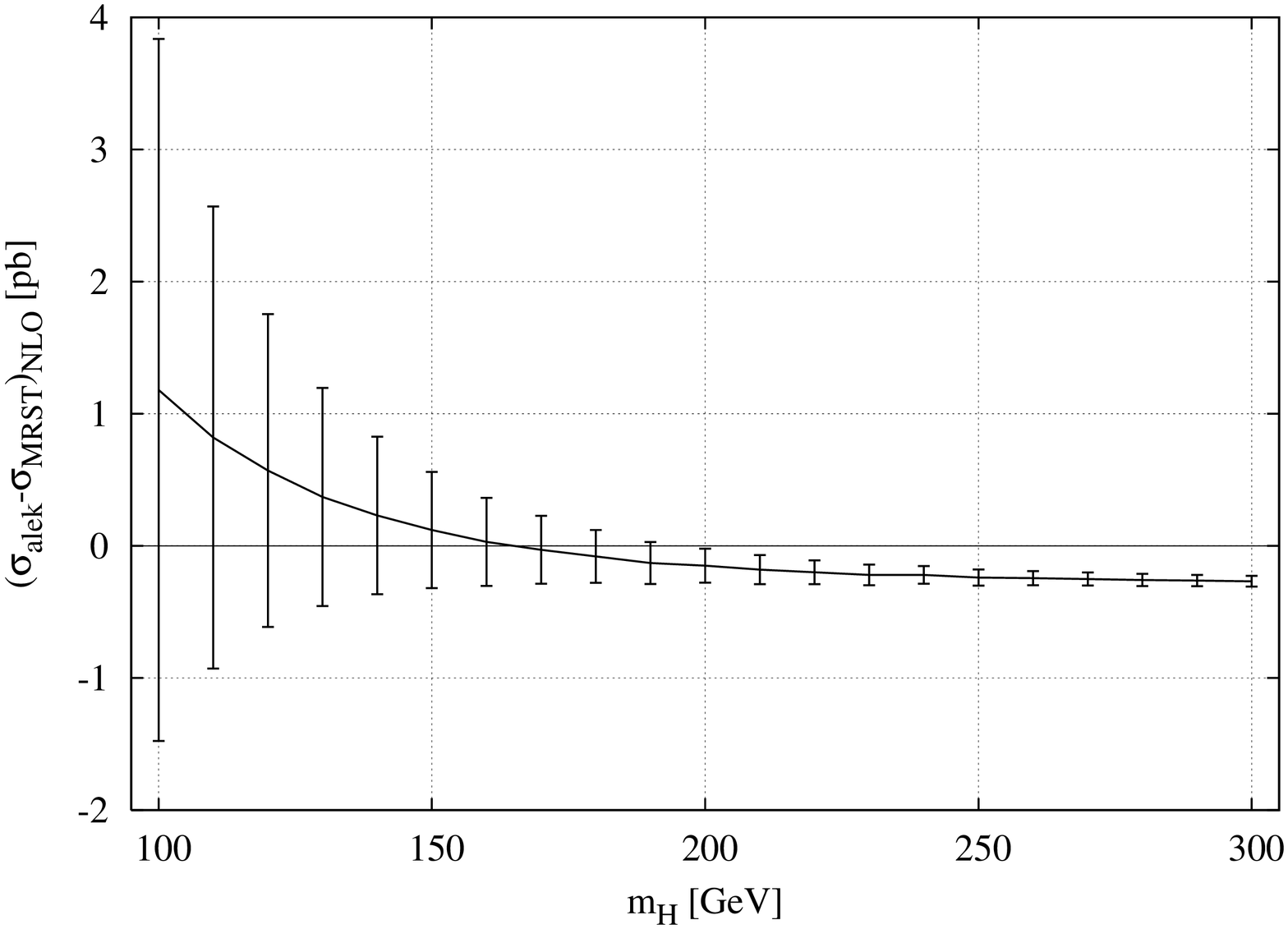}}
\subfigure[LHC]{\includegraphics[%
  width=6cm,
  angle=-90]{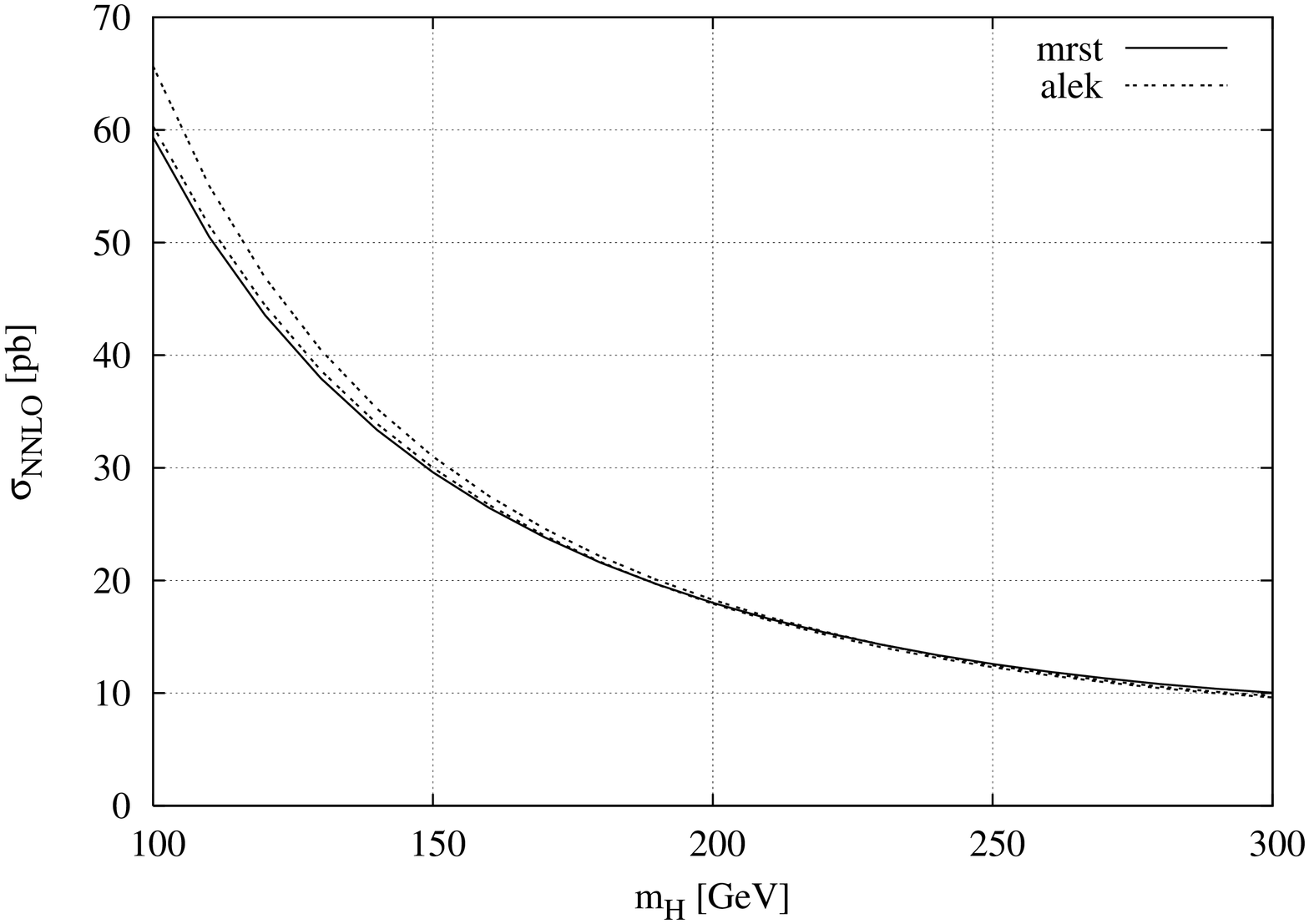}}
\subfigure[LHC]{\includegraphics[%
  width=6cm,
  angle=-90]{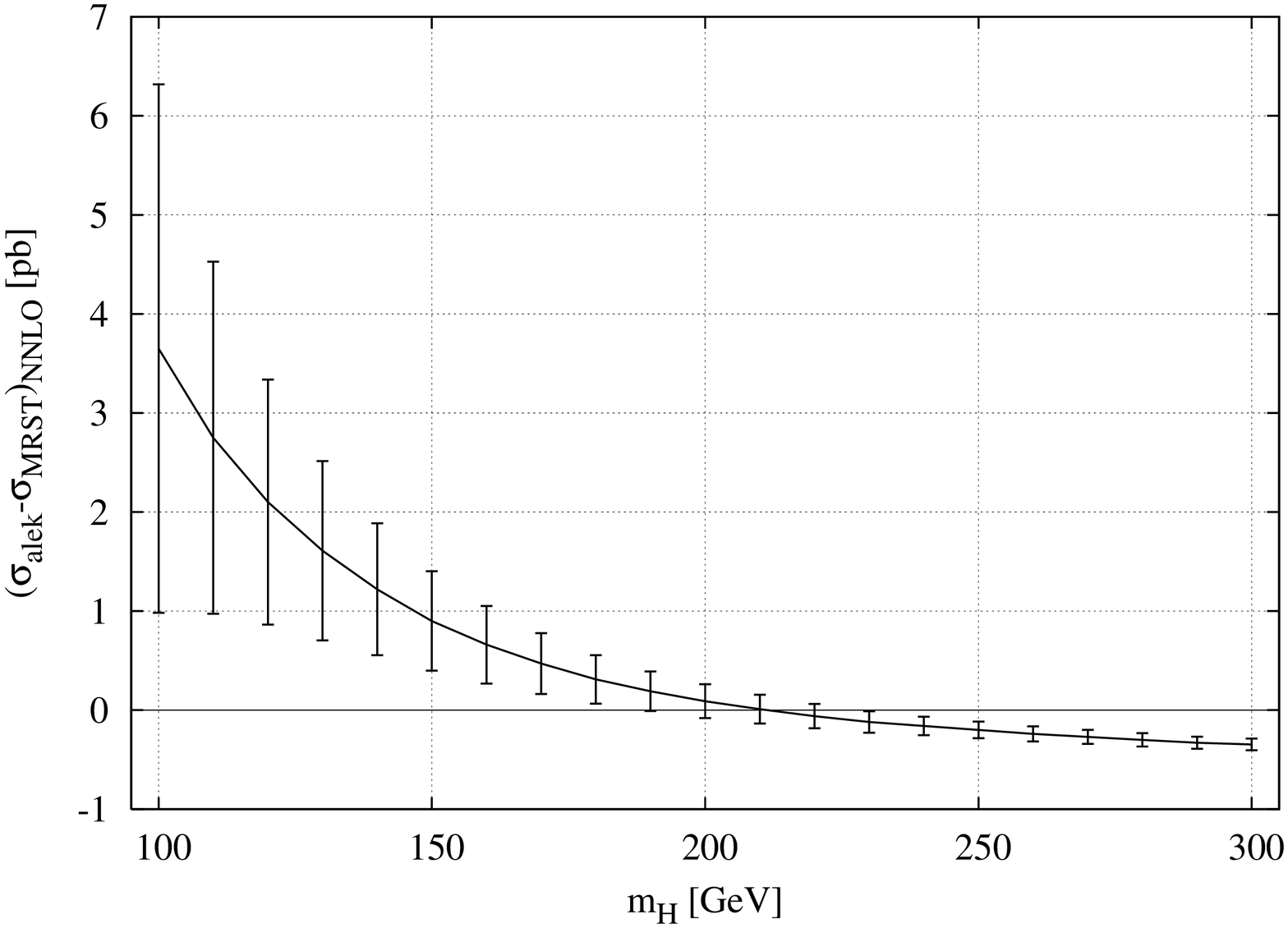}}
\caption{Cross sections for the scalar Higgs production at the LHC
with $\mu_R^2=\mu_F^2=m_H^2$.
When available (Alekhin at all orders and MRST at NLO) the error bands are shown.
See sec. 5 for comments.}
\label{S1P_lhc}
\end{figure}

\begin{figure}
\subfigure[Tevatron]{\includegraphics[%
  width=6cm,
  angle=-90]{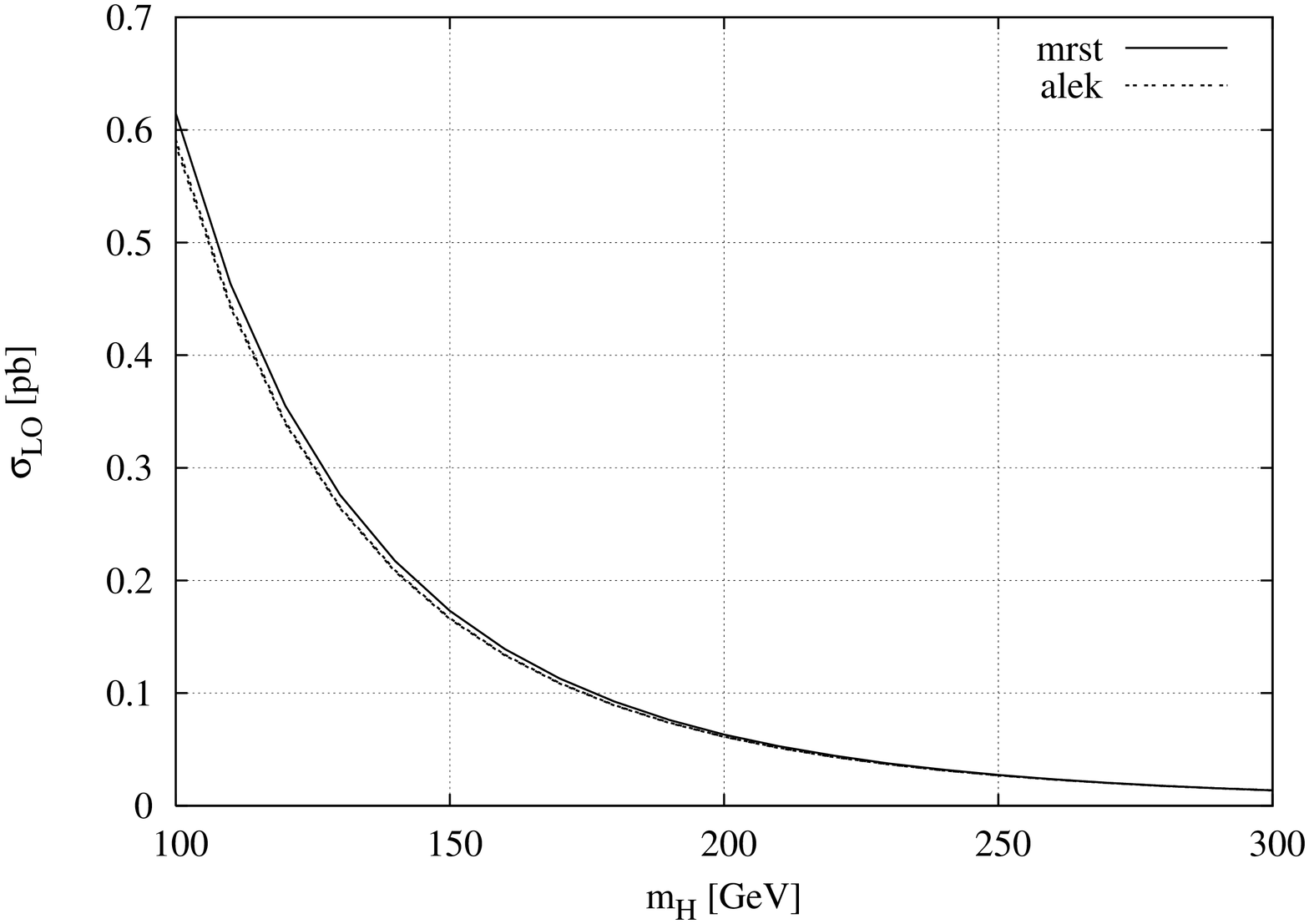}}
\subfigure[Tevatron]{\includegraphics[%
  width=6cm,
  angle=-90]{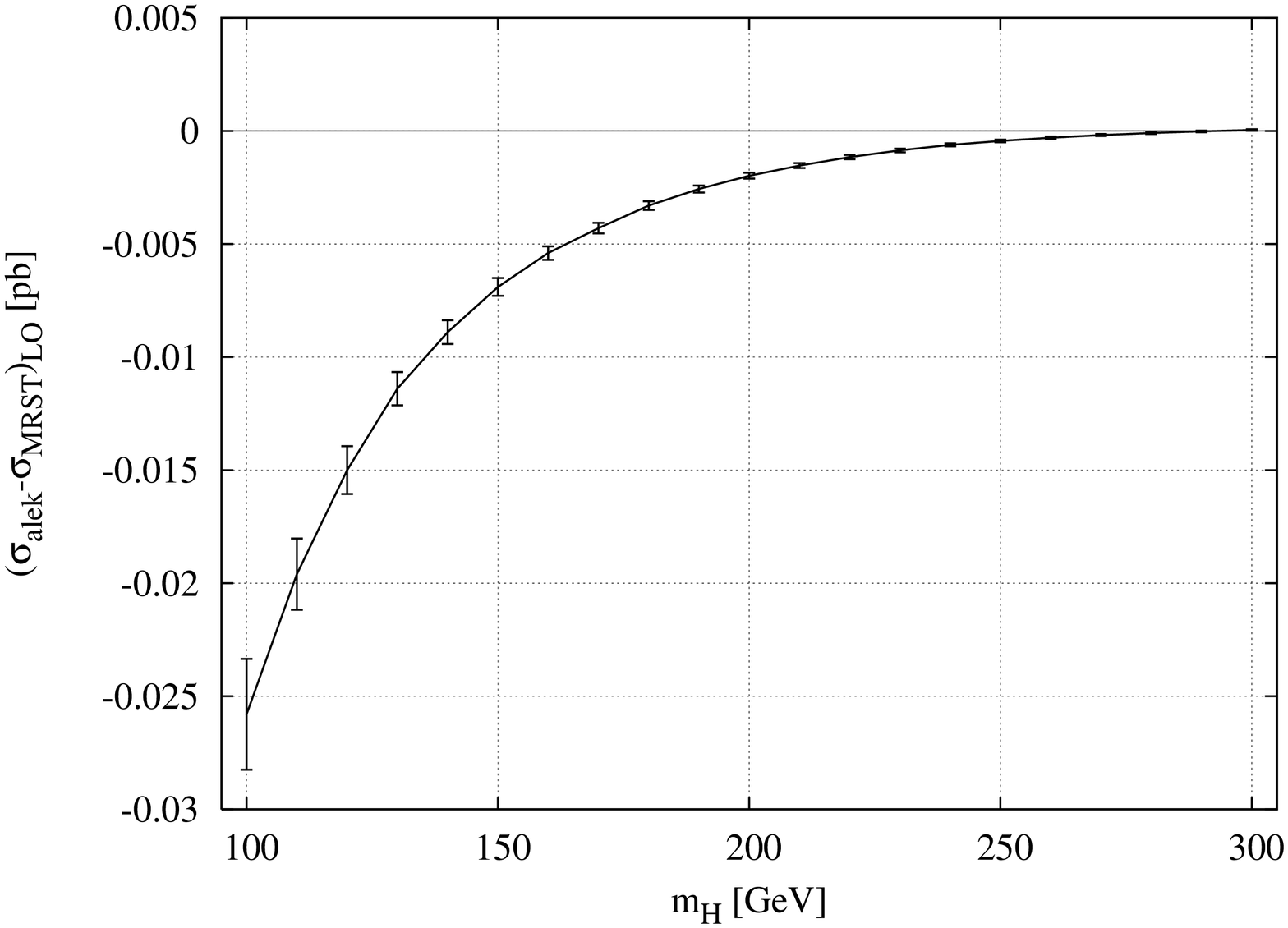}}
\subfigure[Tevatron]{\includegraphics[%
  width=6cm,
  angle=-90]{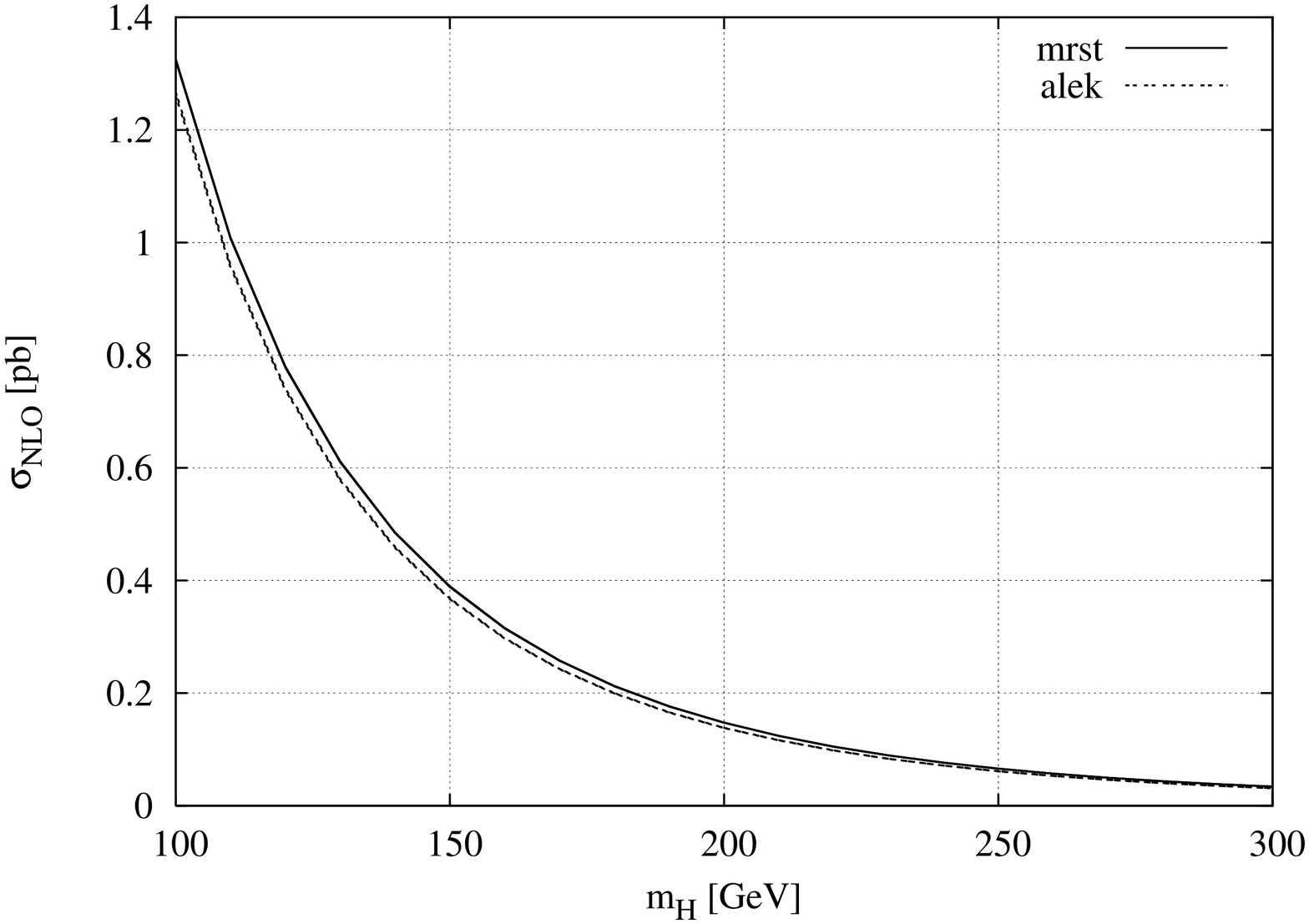}}
\subfigure[Tevatron]{\includegraphics[%
  width=6cm,
  angle=-90]{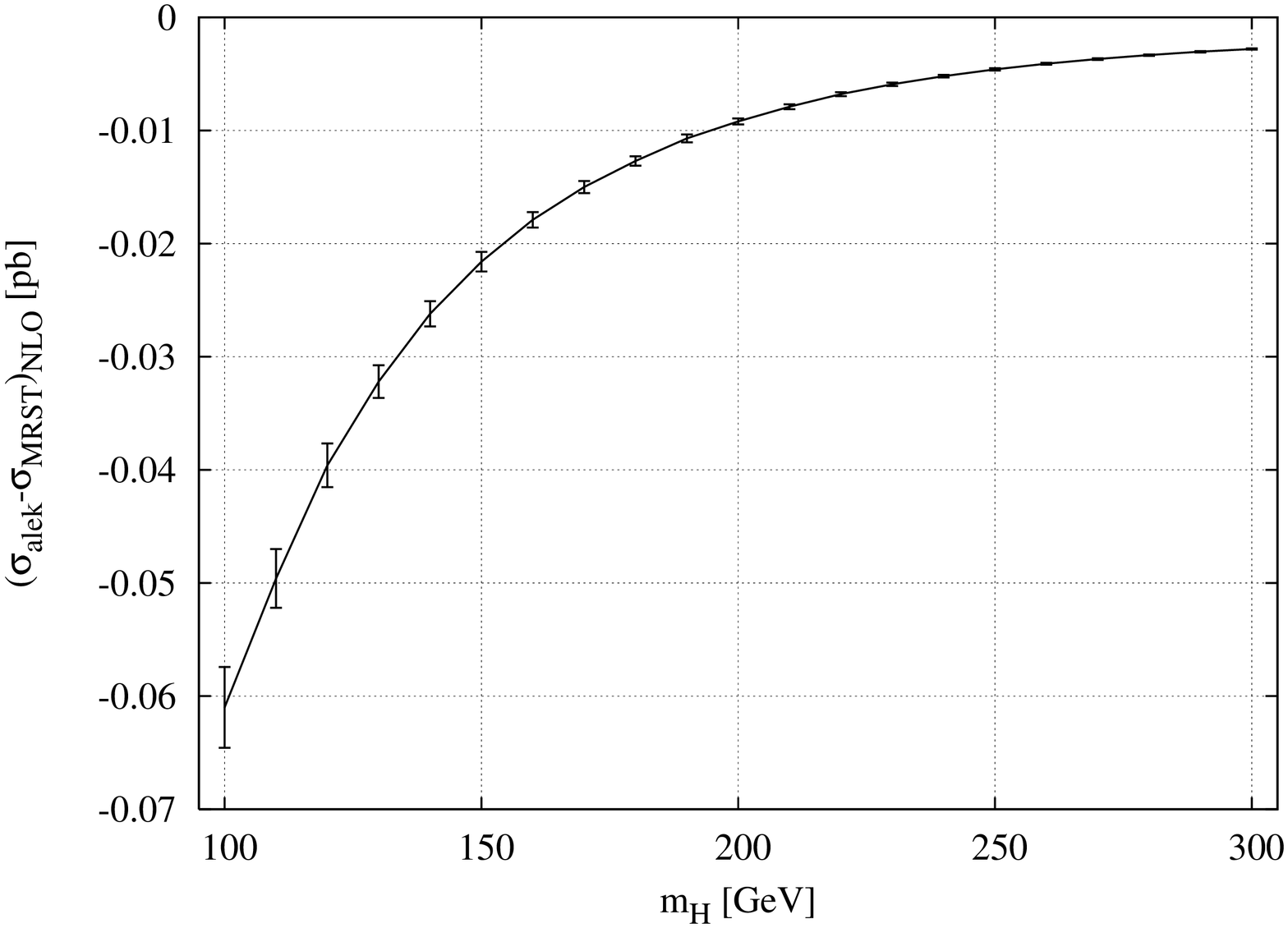}}
\subfigure[Tevatron]{\includegraphics[%
  width=6cm,
  angle=-90]{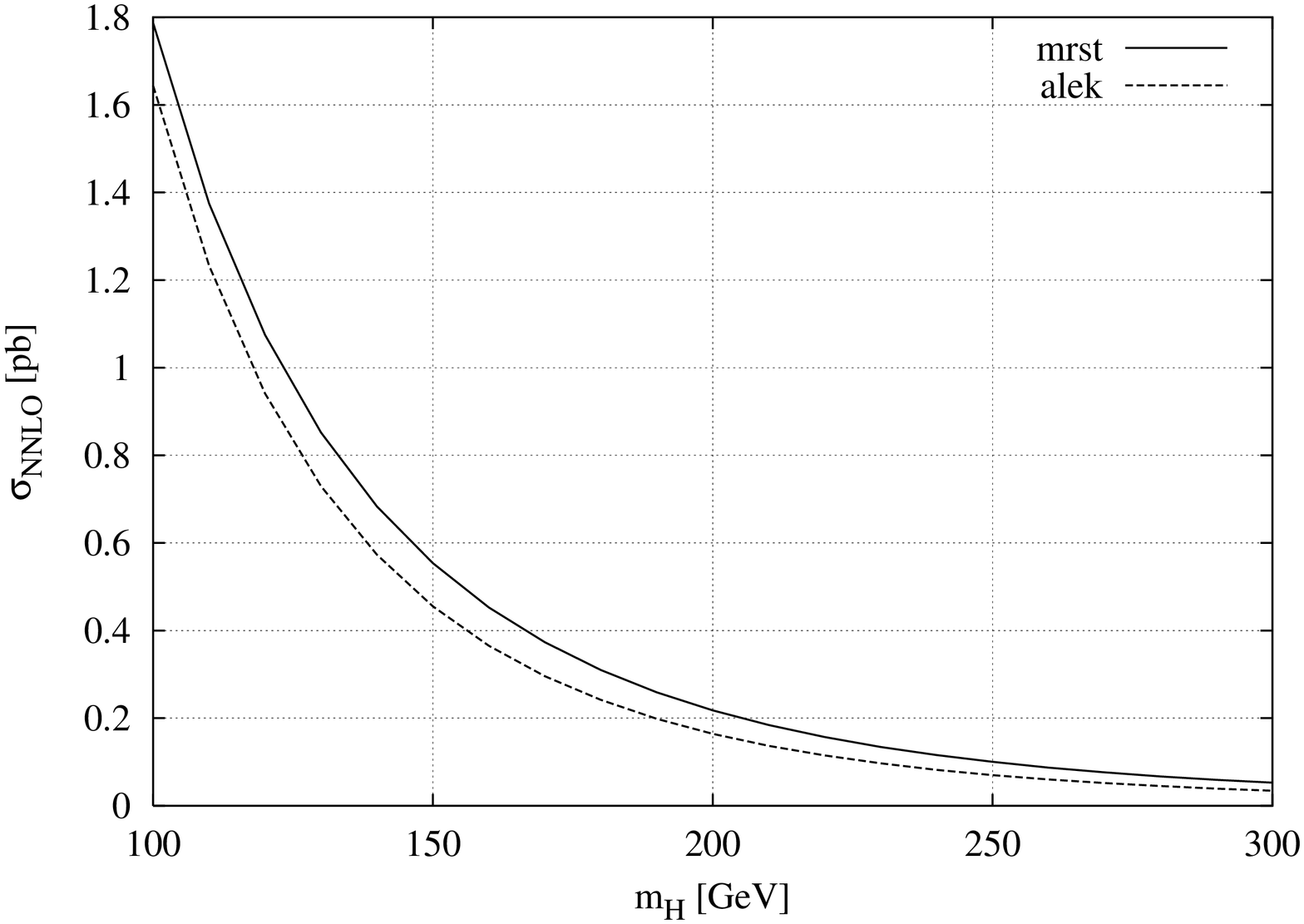}}
\subfigure[Tevatron]{\includegraphics[%
  width=6cm,
  angle=-90]{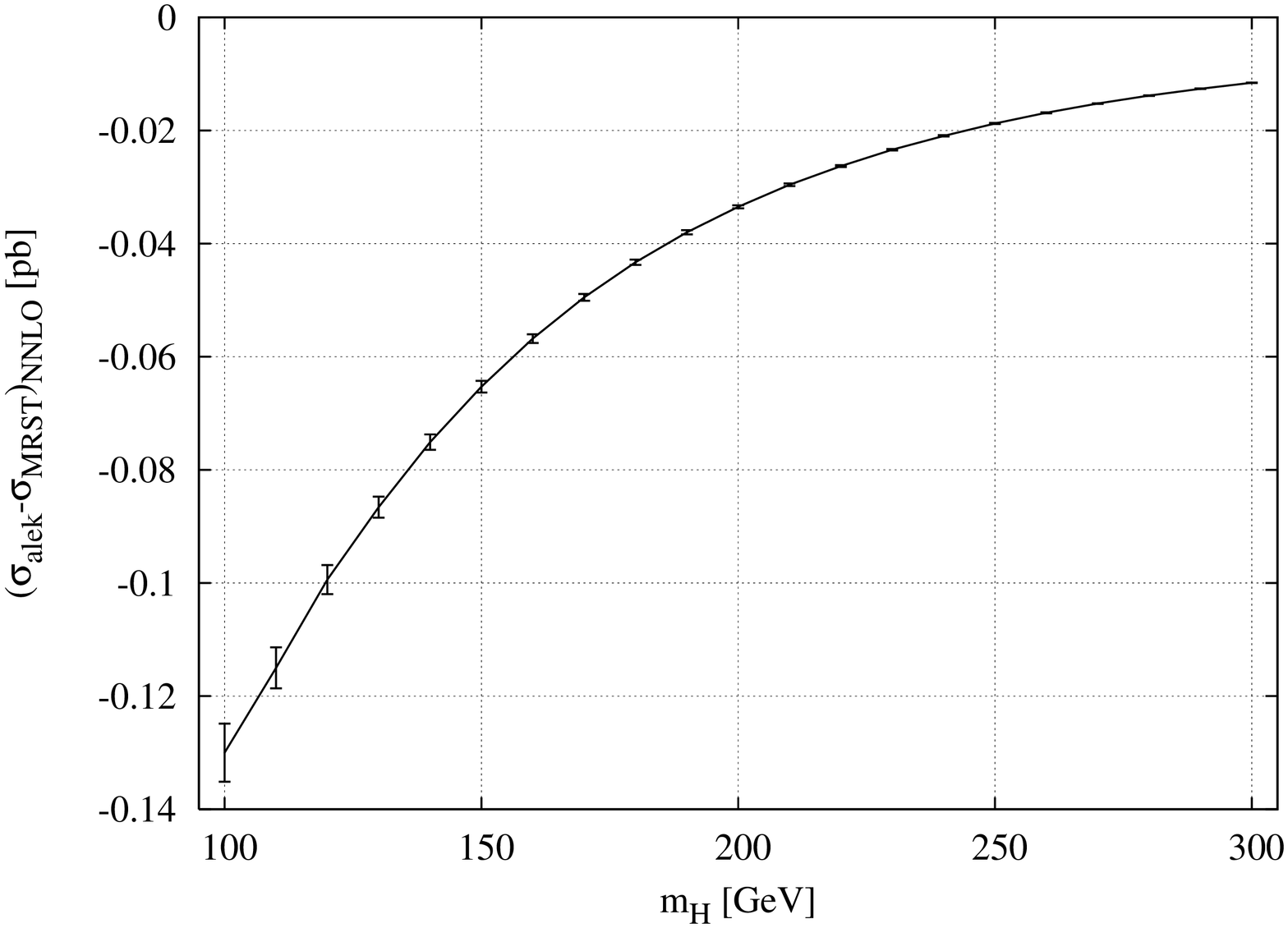}}
\caption{Like in Fig.~(\ref{S1P_lhc}) but for the Tevatron.
In Figs.~(a),~(c) and~(e) the error bands are so tiny that
the lines look superimposed.}
\label{S1P_tev}
\end{figure}

We present in Figs.~(\ref{S1P_lhc}) and (\ref{S1P_tev}) the LO, NLO and NNLO
results for the total Higgs cross sections at the LHC ($\sqrt{S}=14$ TeV) and
at the Tevatron ($\sqrt{S}=2$ TeV), with the corresponding errors, by setting
the condition $\mu_R=\mu_F$. We have chosen to compute these only for two figures,
as an illustration of the size of the errors on the parton distribution functions compared to the best fits,
since these are smaller than the variation induced when moving from one
perturbative order to the next. The numerical determination of the errors 
is computationally very intensive and has been performed on a cluster.

In the figures on the left the cross sections obtained using
MRST input are represented by a solid line, and the ones obtained using
Alekhin's input, by a dashed line.
In the figures on the right we present a plot of the difference between
the values of the MRST cross sections and Alekhin's cross sections
for each perturbative order, with the respective errors.
The calculation of the error bands has been done following the usual
theory of the linear propagation of the errors.
Starting from the errors on the PDFs known in the
literature (see \cite{Alekhin},\cite{MRST1}), we have generated different
sets of cross sections. Then, the error on the cross section has been
calculated using the formula
\ba
\Delta \sigma=\frac{1}{2}\sqrt{\sum_{k=1}^{N}\left[\sigma_{2k-1}-\sigma_{2 k}\right]^{2}},
\ea
where $\sigma_{k}$ is the $k$-th cross section belonging to a certain set,
and $N$ is the number of free parameters, which is 15 for MRST and 17 for Alekhin.

The PDFs with the related error analysis are available at all orders for the
Alekhin input but only at NLO for the MRST's input
(Figs.~(\ref{S1P_lhc})-c and (\ref{S1P_tev})-c).

When in Figs.~(\ref{S1P_lhc}), (\ref{S1P_tev}), and~(\ref{K2P}) we plot
more than one line for a single set the lower line is the minimal value (best fit minus error) and
the upper line is the maximal value (best fit plus error).

%%%%%%%%%%%%%%%%%%%%%%%%%%%%%%%%%%%%%%%%%%%%%%%%%%%%%%%%%%%
The LO cross sections increase by a factor of approximately 100 as
we change the energy from 2 TeV Figs.~(\ref{S1P_tev}), to 14 TeV Figs.~(\ref{S1P_lhc}),
and sharply decrease as we raise the mass of the Higgs boson.
At 14 TeV the range of variation of $\sigma_{LO}$
is between 30 and 5 pb, with the highest value reached for $m_H=100$ GeV.

In the same figures we compare LO, NLO and NNLO cross
sections at these two typical energies. It is quite evident that the role
of the NLO corrections is to increase by a factor of approximately 2
the LO cross section bringing the interval of variation of $\sigma_{NLO}$
between 60 and 10 pb, for an increasing value of $m_H$. NNLO corrections
at 14 TeV increase these values by an additional 10 per cent compared
to the NLO prediction, with a growth which is more pronounced for the set
proposed by Alekhin.

Comparing the results computed using
Alekhin and MRST's inputs, for the LHC case ($\sqrt{S}=14$ TeV) we 
observe that at LO (see Fig.~(\ref{S1P_lhc})-b) 
the two sets give results which are
compatible within the error bands for $m_H<150$ GeV, while
for larger values of $m_H$ we observe only small differences between the two. 
At NLO, Fig.~(\ref{S1P_lhc})-d,
where the error analysis is available for both sets,
the results are compatible within the error bands for $m_H<190$ GeV,
and we have small differences for larger values of the Higgs mass.
For the NNLO case, Fig.~(\ref{S1P_lhc})-f, we notice that there
are sensible differences among the two sets.

By a similar inspection of Figs.~(\ref{S1P_tev})-b, (\ref{S1P_tev})-d and~(\ref{S1P_tev})-f,
we notice that at the Tevatron energy of $\sqrt{S}=2$ TeV,
the two PDFs sets give quite different predictions at all the three orders.

The numerical values of the total cross sections and the K-factors
as a function of the center of mass energy with the respective errors have also
been reported in Table~ \ref{table1a}, in the case of Alekhin's inputs.

\subsection{K-factors}

\begin{figure}
\subfigure[LHC]{\includegraphics[%
  width=6cm,
  angle=-90]{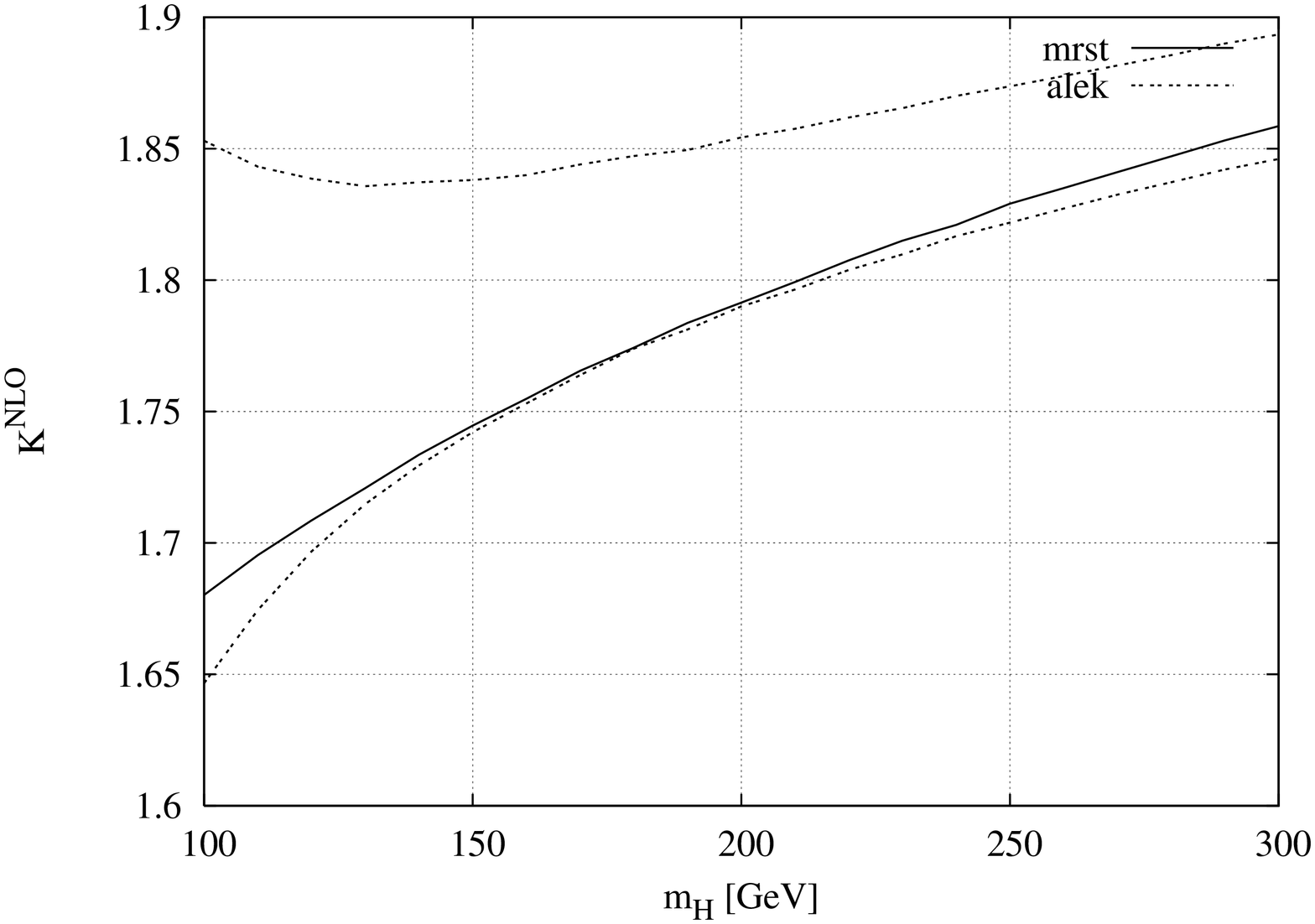}}
\subfigure[LHC]{\includegraphics[%
  width=6cm,
  angle=-90]{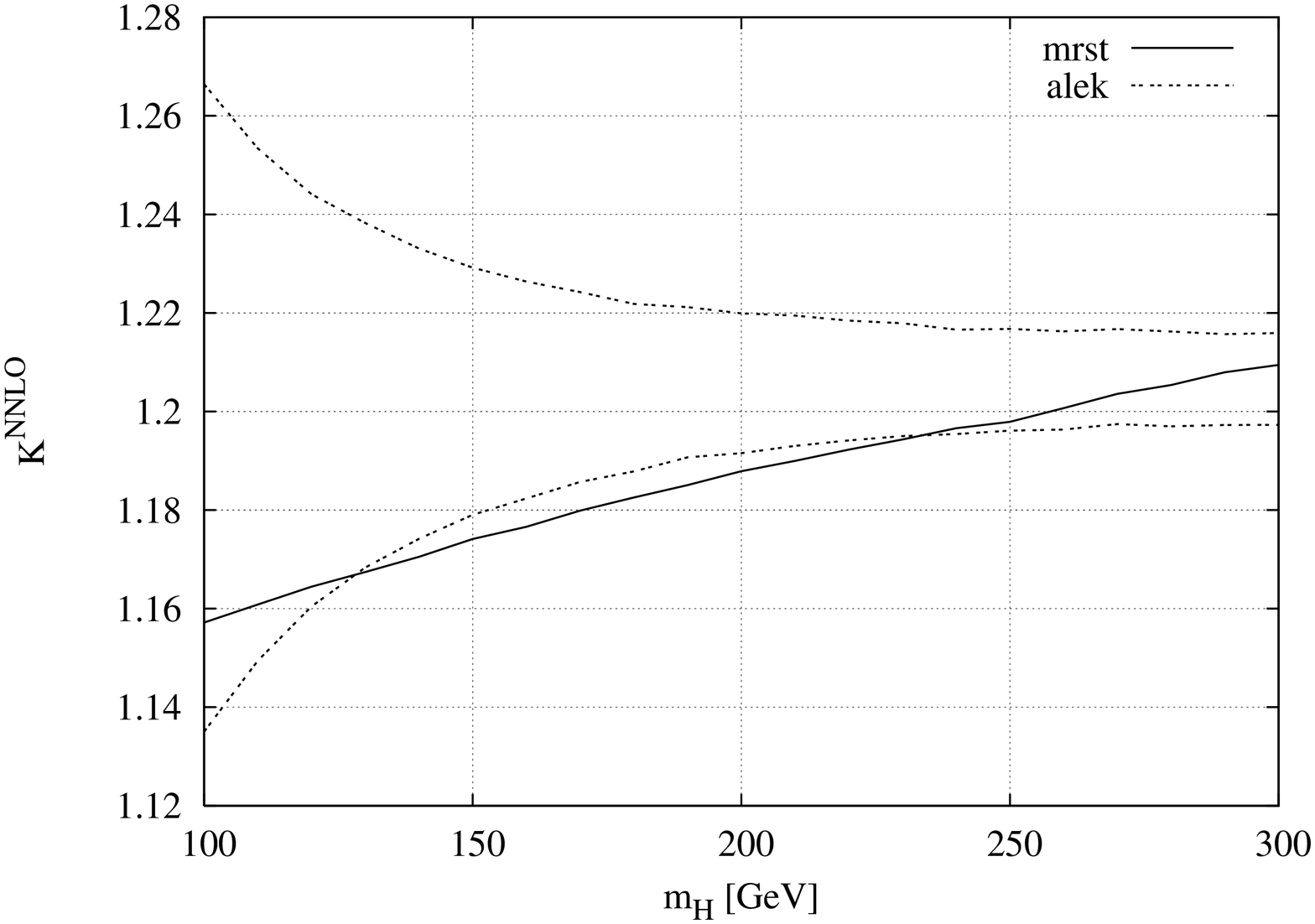}}
\subfigure[Tevatron]{\includegraphics[%
  width=6cm,
  angle=-90]{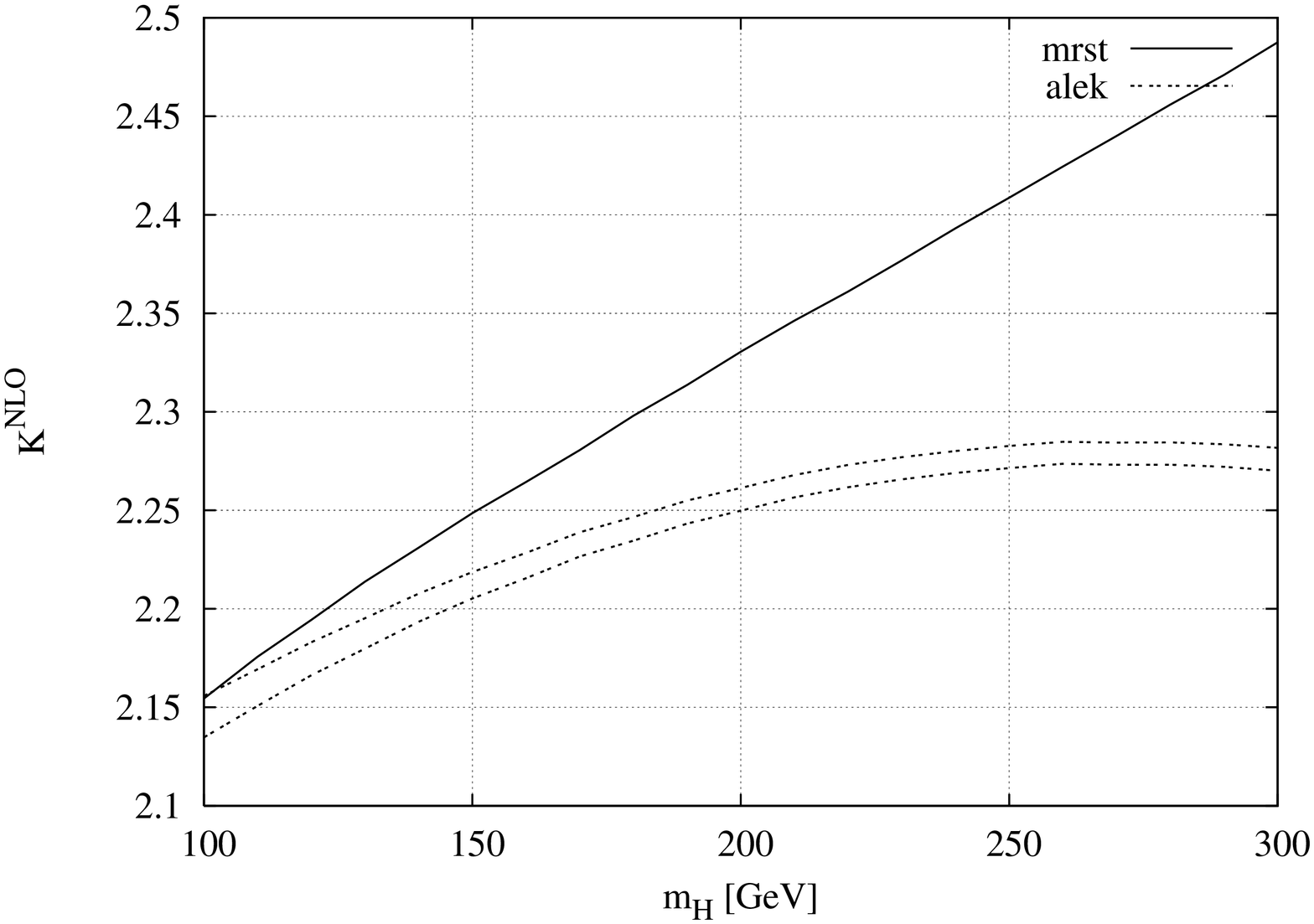}}
\subfigure[Tevatron]{\includegraphics[%
  width=6cm,
  angle=-90]{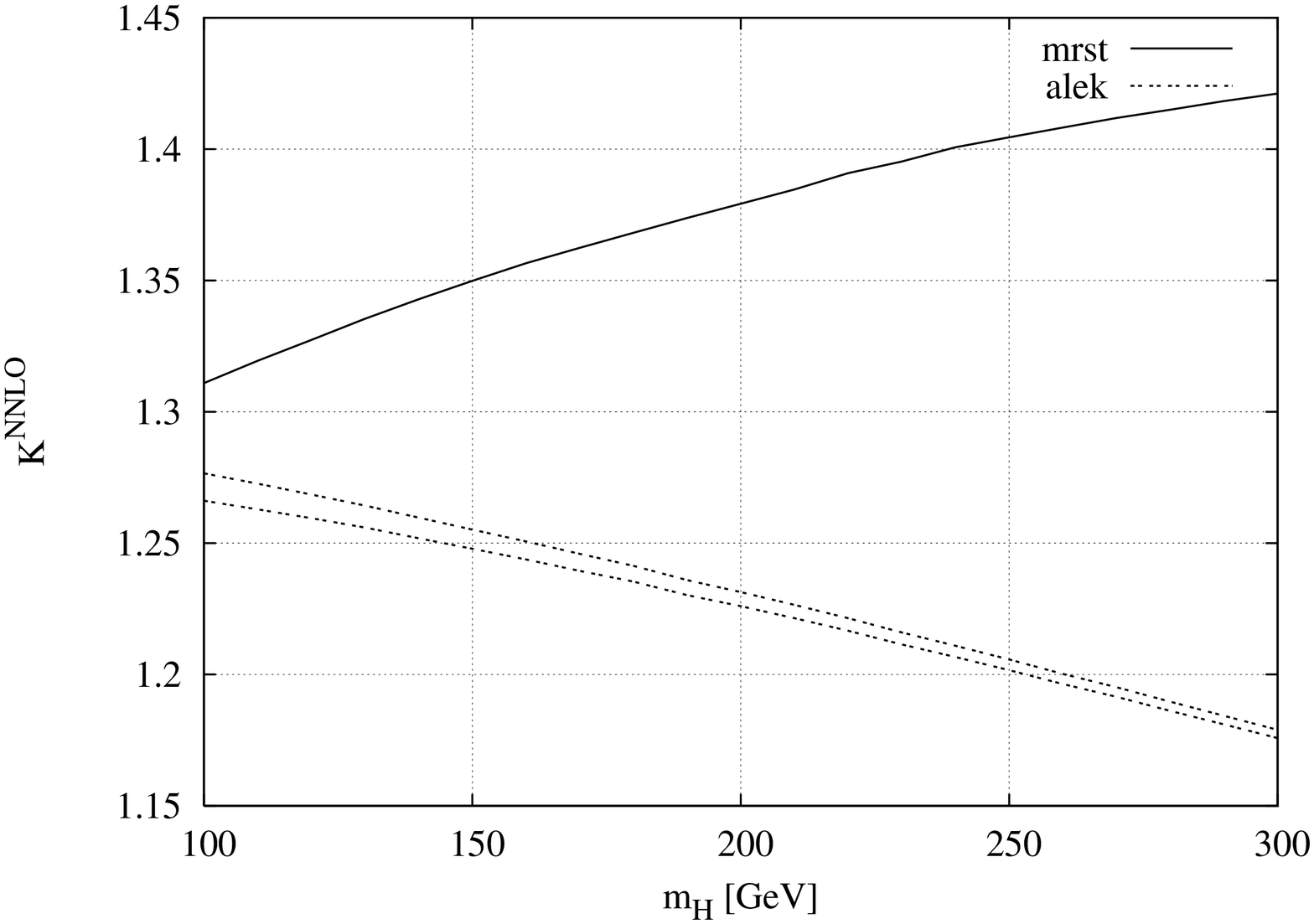}}
\caption{K-factors for the scalar Higgs at NNLO/NLO
and NLO/LO with $\mu_R^2=\mu_F^2=m_H^2$.
When available (Alekhin at all orders and MRST at NLO) the error bands are shown.}
\label{K2P}
\end{figure}

A precise indication on the impact of the NLO/NNLO corrections and the
stability of the perturbative expansion comes from a study of the
K-factors $K^{NLO}$ and $K^{NNLO}$, defined above. From the plots in
Figure \ref{K2P} the different behaviour of the predictions derived from
the two models for the parton distributions is quite evident. At 14 TeV the NLO
K-factors from both models are large, as expected, since the LO prediction 
are strongly scale dependent. The increase of $\sigma_{NLO}$
compared to the LO predictions is between 65 and 90 per cent.

In Fig.~(\ref{K2P})-a one can observe that the impact of the
NLO corrections to the LO result predicted by both sets increases
for an increasing $m_H$, with the corrections predicted by Alekhin
being the largest ones.
The trend of the MRST model in the NNLO vs NLO case Fig.~(\ref{K2P})-b
is similar, for an increasing Higgs mass, ranging from 1.16 to 1.22,
while Alekhin $K^{NNLO}$-factor is approximately constant around a
value of 1.21.

The evaluation of the
overall impact of this growth on the size of these corrections should,
however, also keep into consideration the fact that these corrections
are enhanced in a region where the cross section is sharply
decreasing (Figs.~(\ref{S1P_lhc}) and (\ref{S1P_tev})).

Moving to Tevatron energy we notice that all the K-factors are
larger than in the LHC case. For a MRST input we continue to observe
a growth of $K^{NLO}$ and $K^{NNLO}$ for an increasing $m_H$,
while for the Alekhin case this trend is slower for $K^{NLO}$, and
it is even reversed for $K^{NNLO}$. Unlike the LHC
situation, the Alekhin's K-factors are smaller than MRST ones
at Tevatron energies.

\subsection{Renormalization/factorization scale dependence}

\begin{figure}
\subfigure[LHC]{\includegraphics[%
  width=6cm,
  angle=-90]{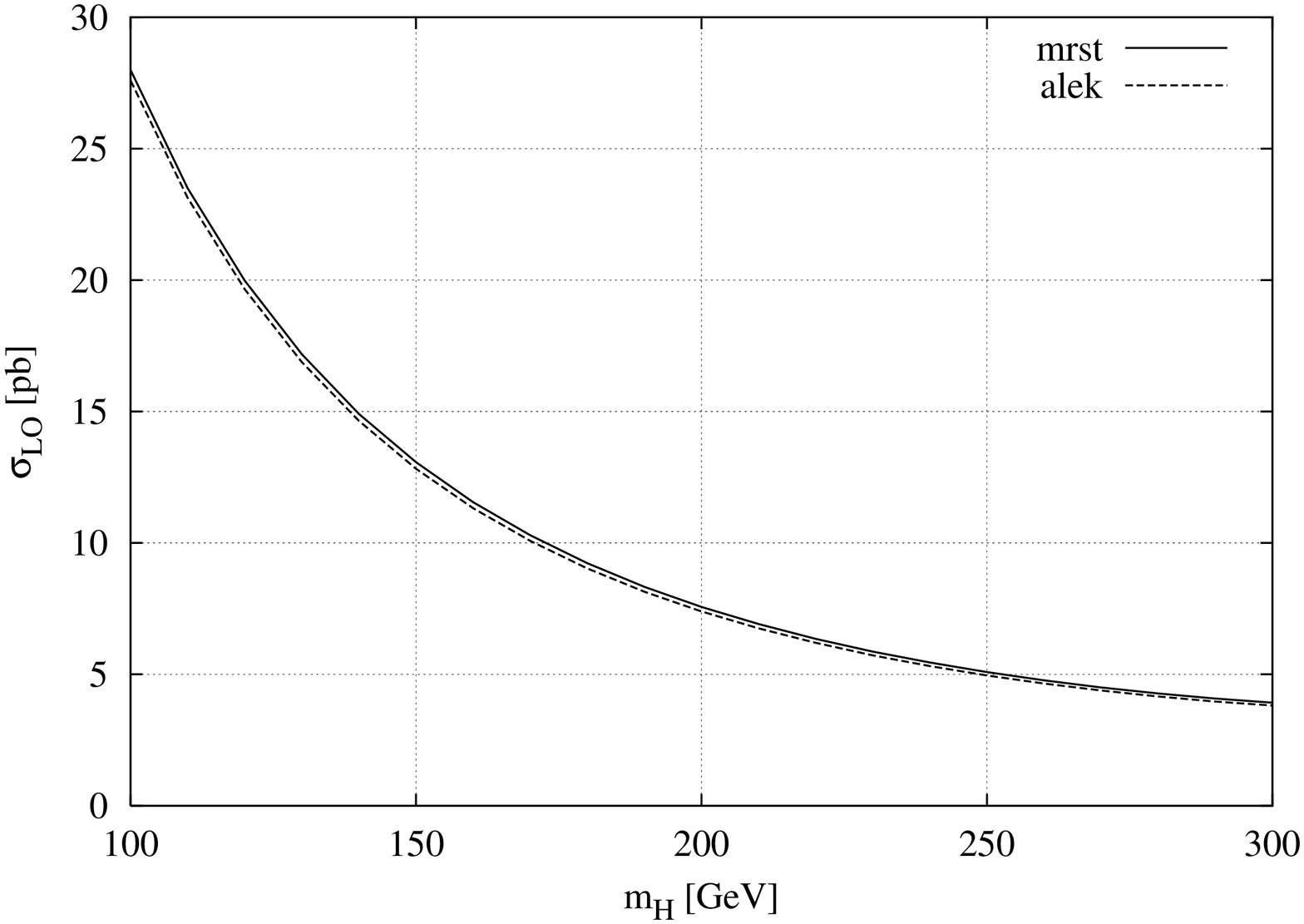}}
\subfigure[LHC]{\includegraphics[%
  width=6cm,
  angle=-90]{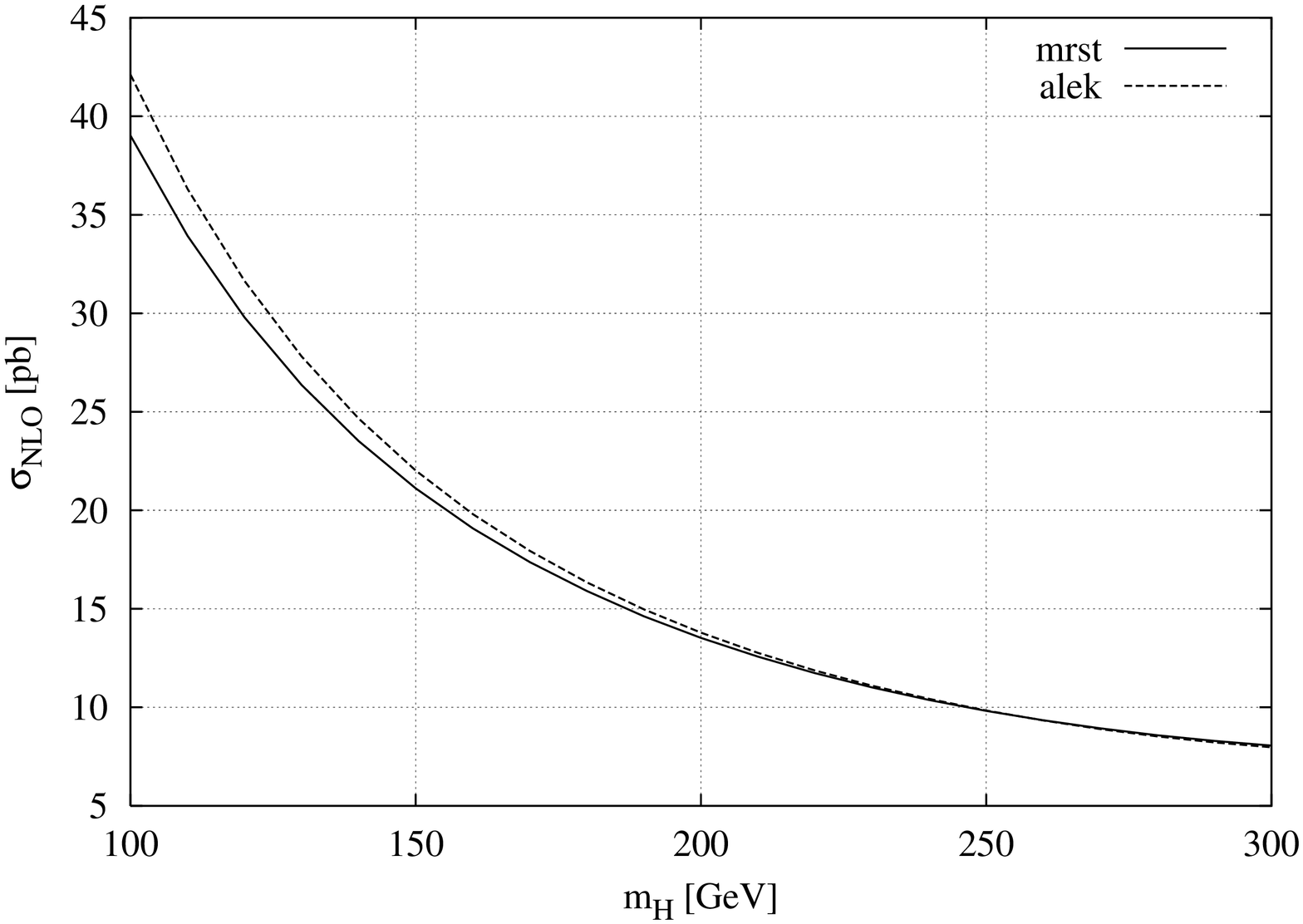}}
\subfigure[LHC]{\includegraphics[%
  width=6cm,
  angle=-90]{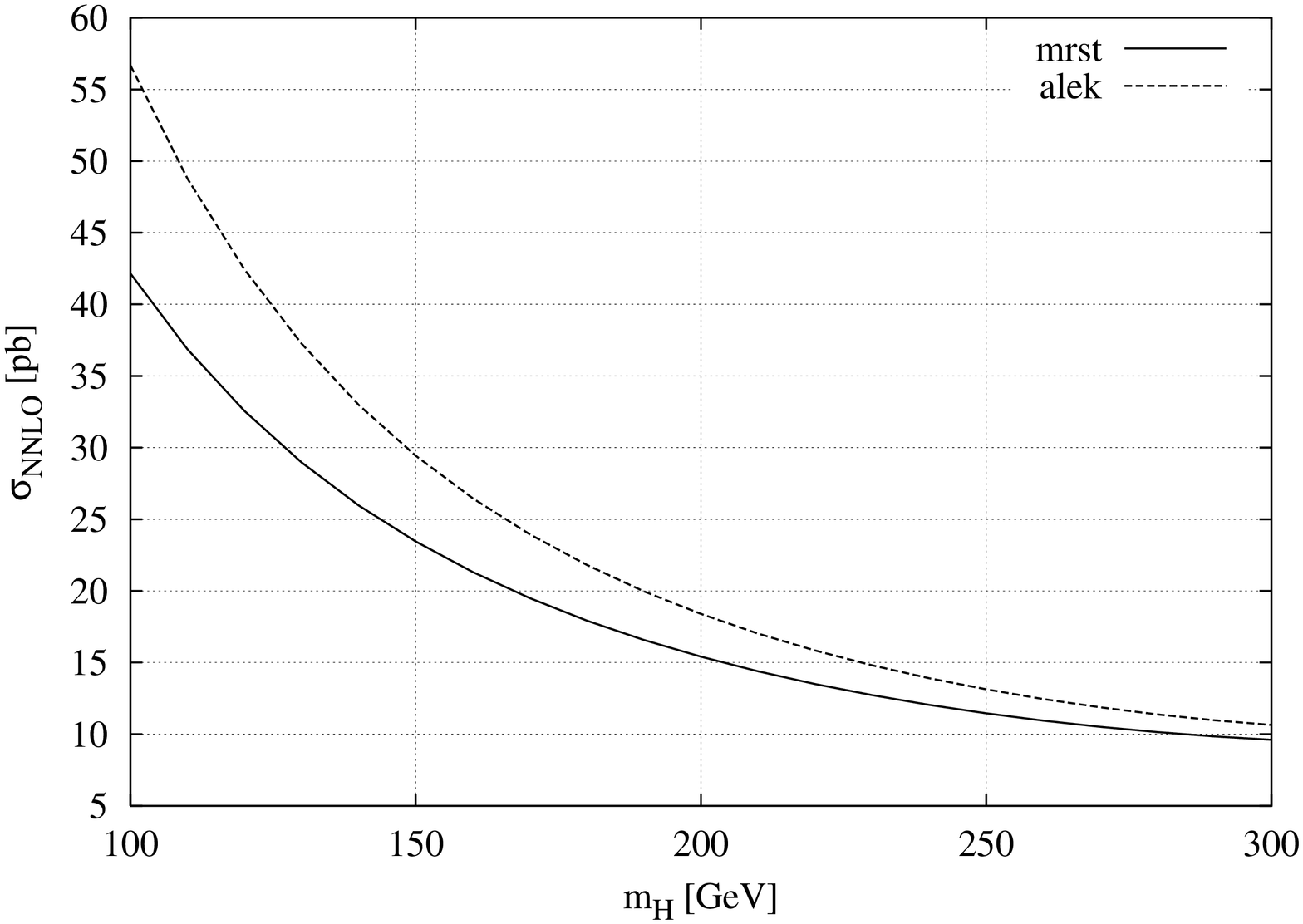}}
\subfigure[Tevatron]{\includegraphics[%
  width=6cm,
  angle=-90]{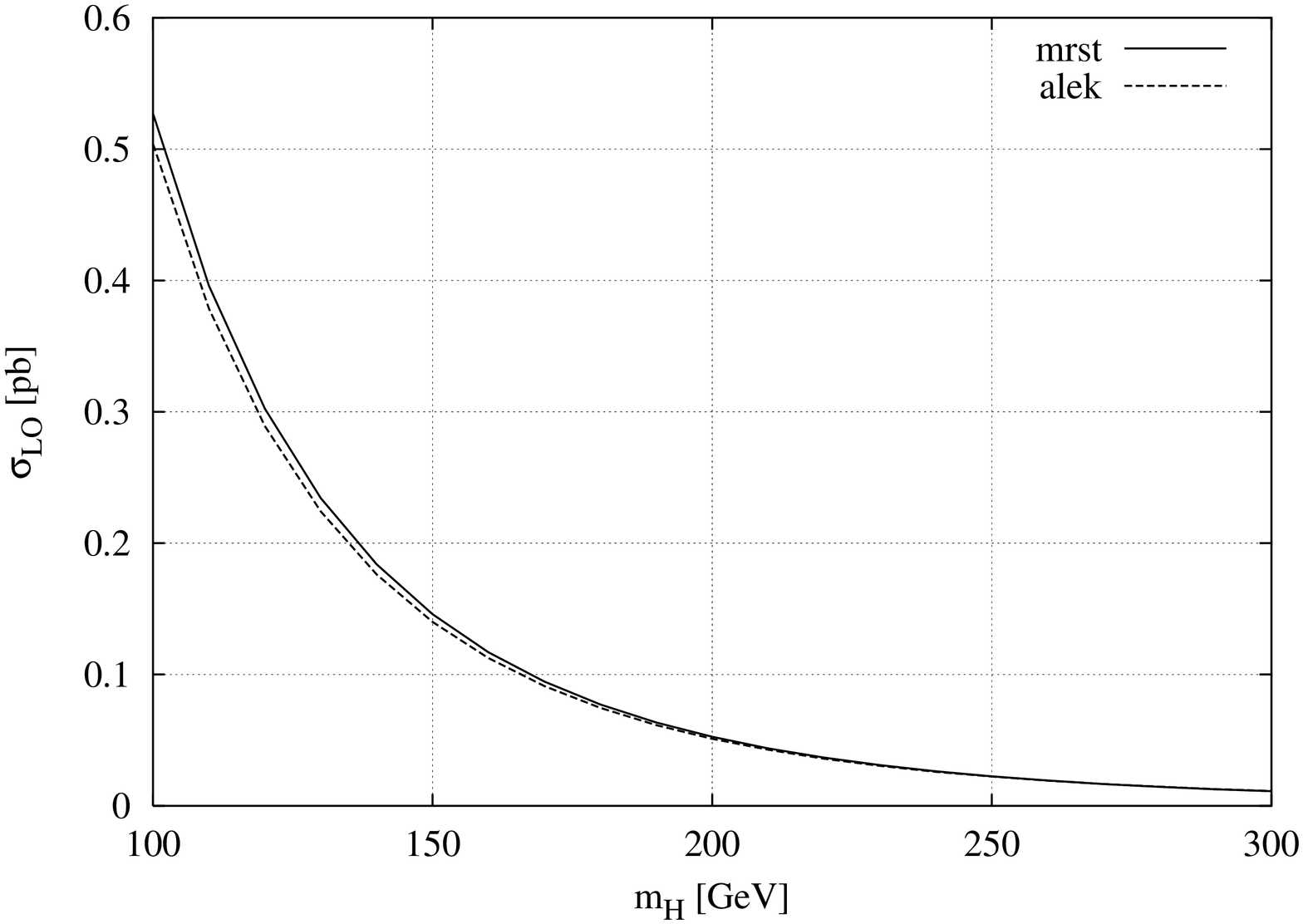}}
\subfigure[Tevatron]{\includegraphics[%
  width=6cm,
  angle=-90]{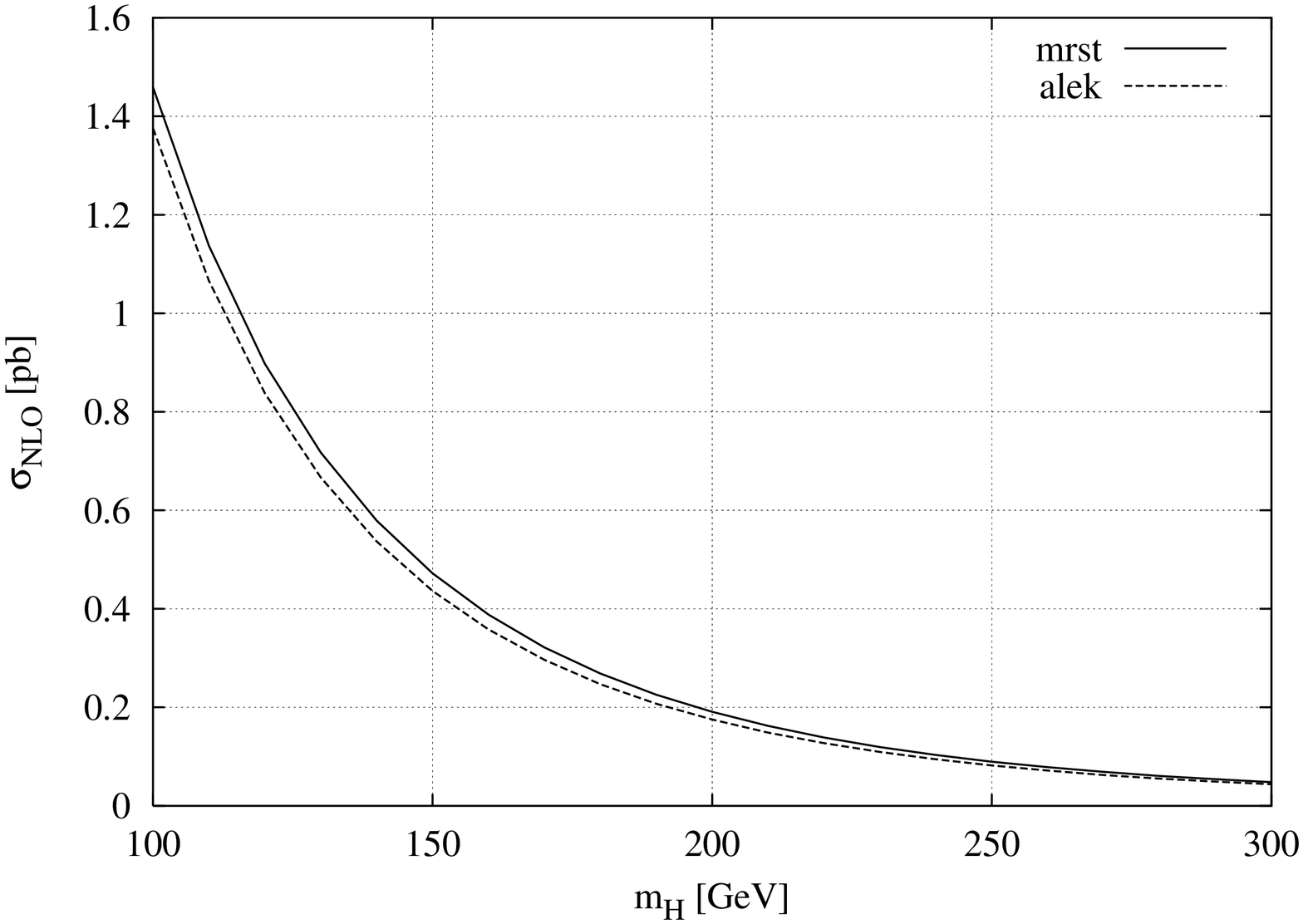}}
\subfigure[Tevatron]{\includegraphics[%
  width=6cm,
  angle=-90]{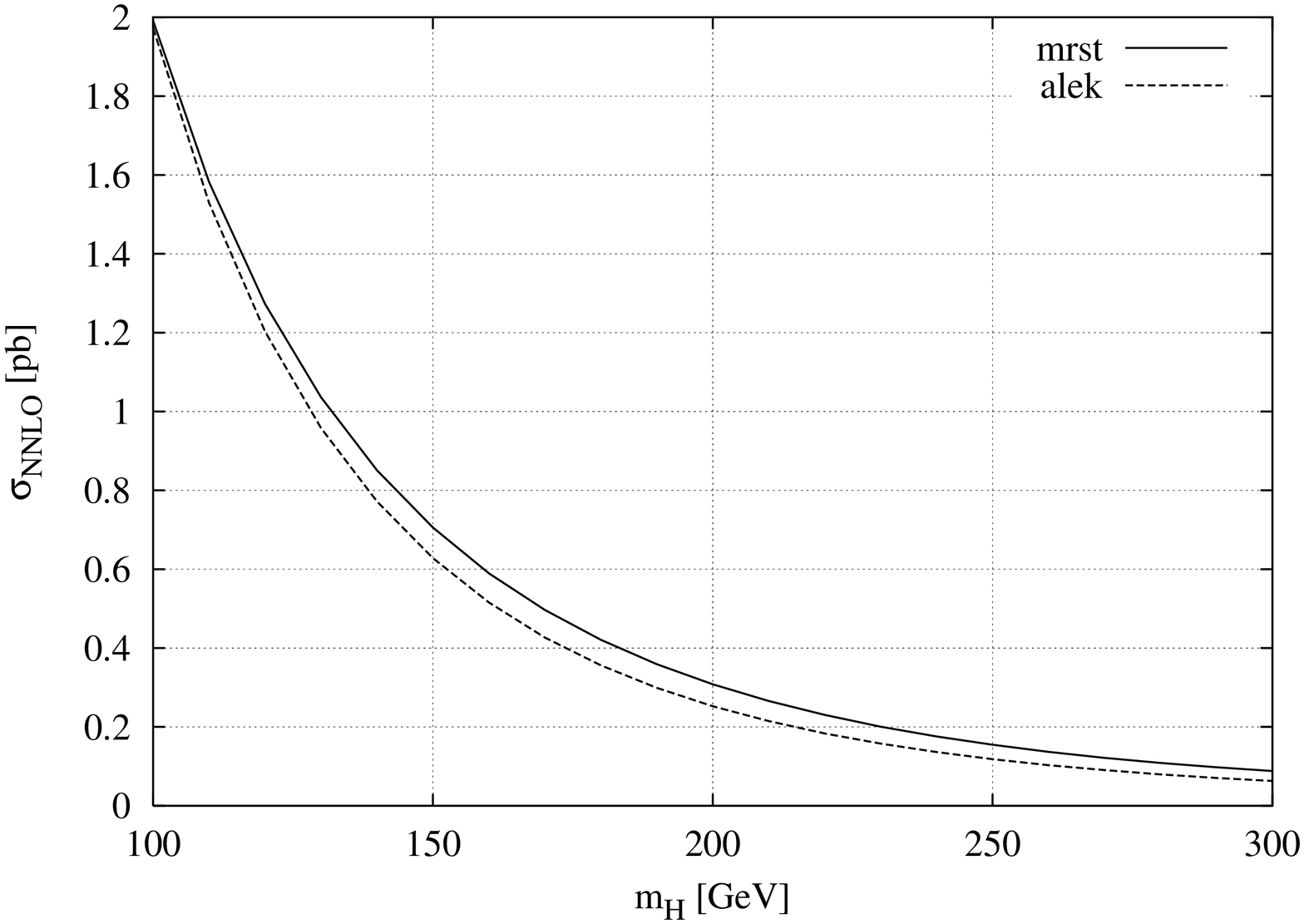}}
\caption{Cross sections for the scalar Higgs production at the LHC
and Tevatron with $\mu_R^2=(1/2)\mu_F^2$ and $\mu_F=2 m_H$}
\label{S3P}
\end{figure}

\begin{figure}
\subfigure[LHC]{\includegraphics[%
  width=6cm,
  angle=-90]{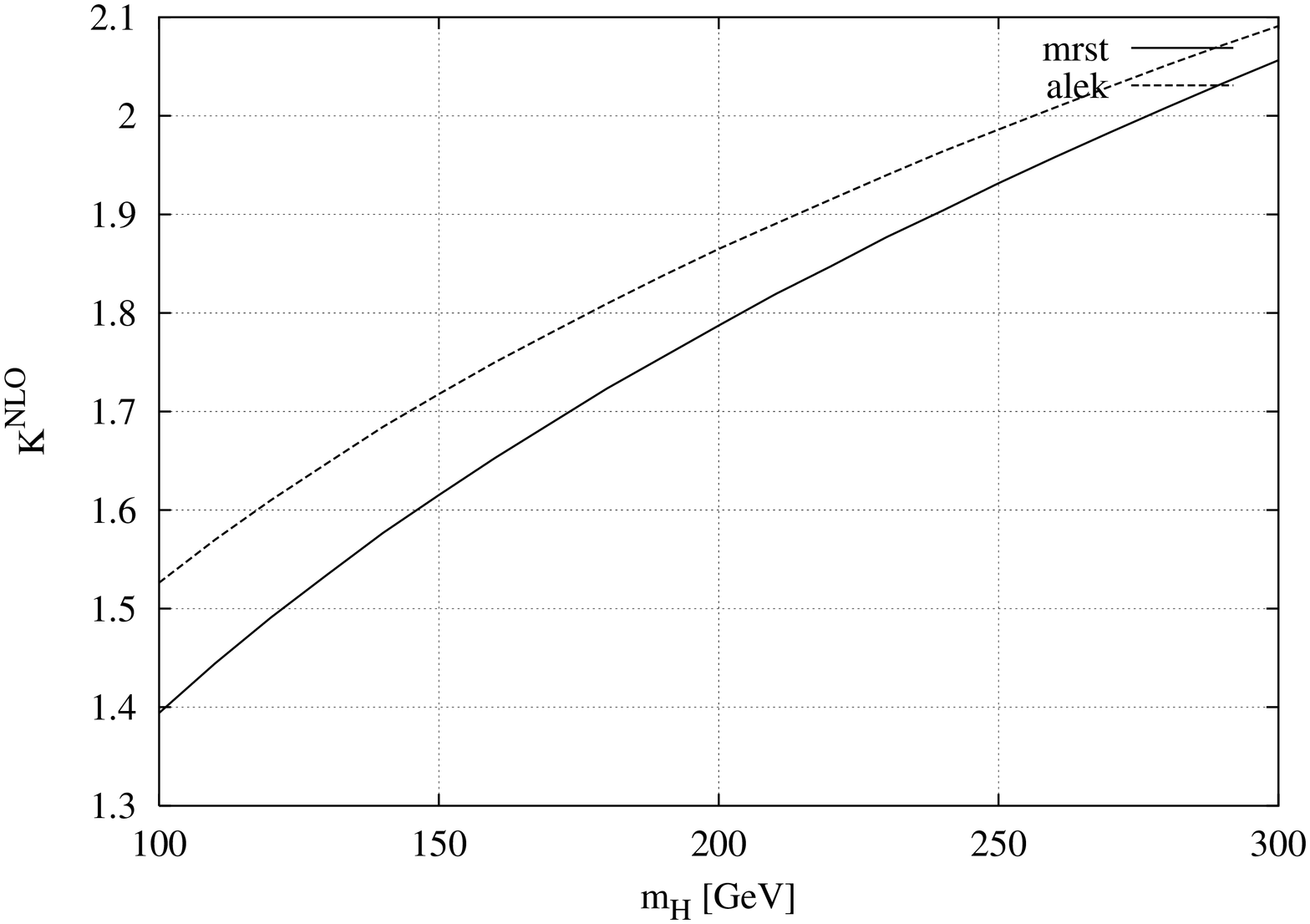}}
\subfigure[LHC]{\includegraphics[%
  width=6cm,
  angle=-90]{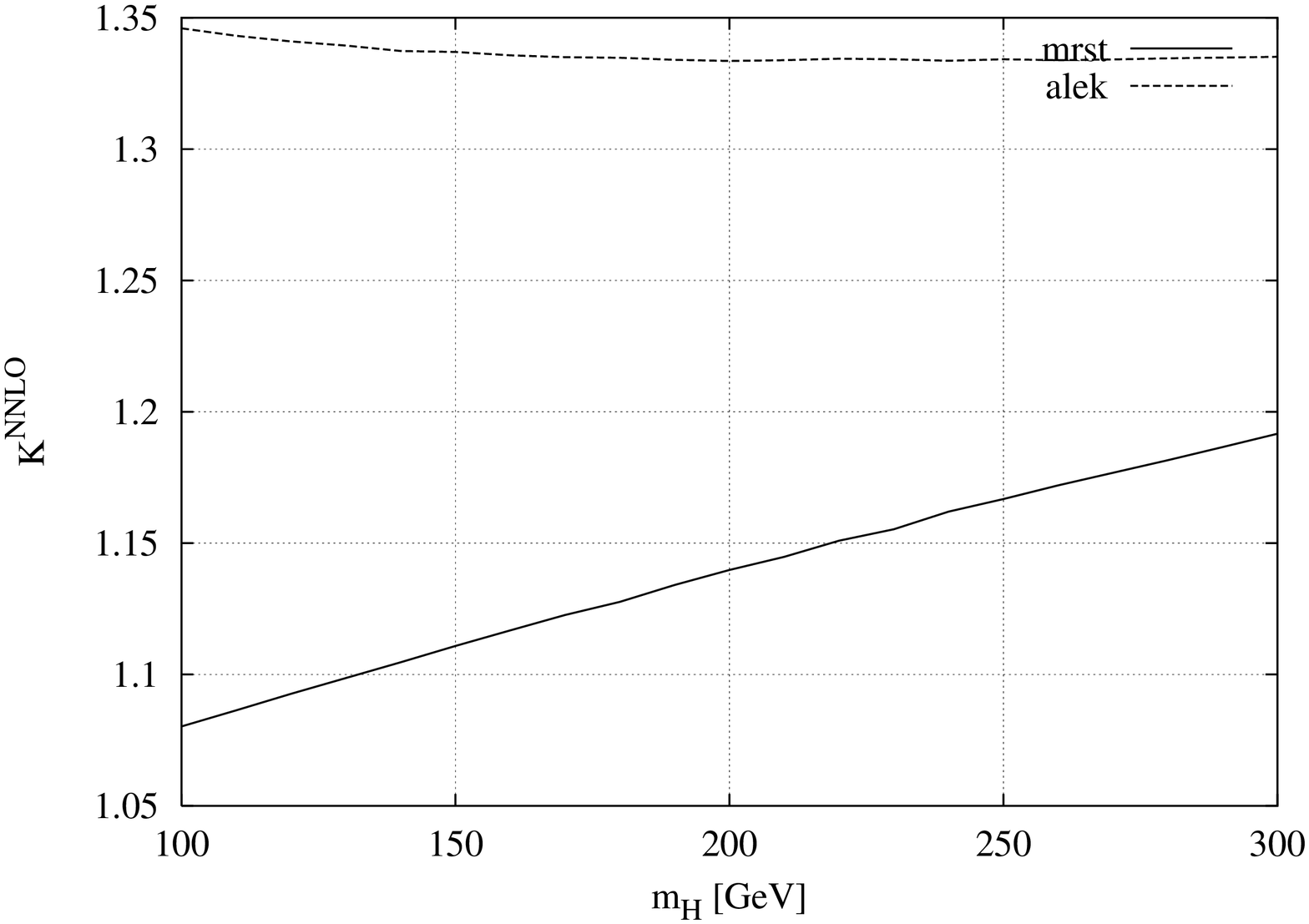}}
\subfigure[Tevatron]{\includegraphics[%
  width=6cm,
  angle=-90]{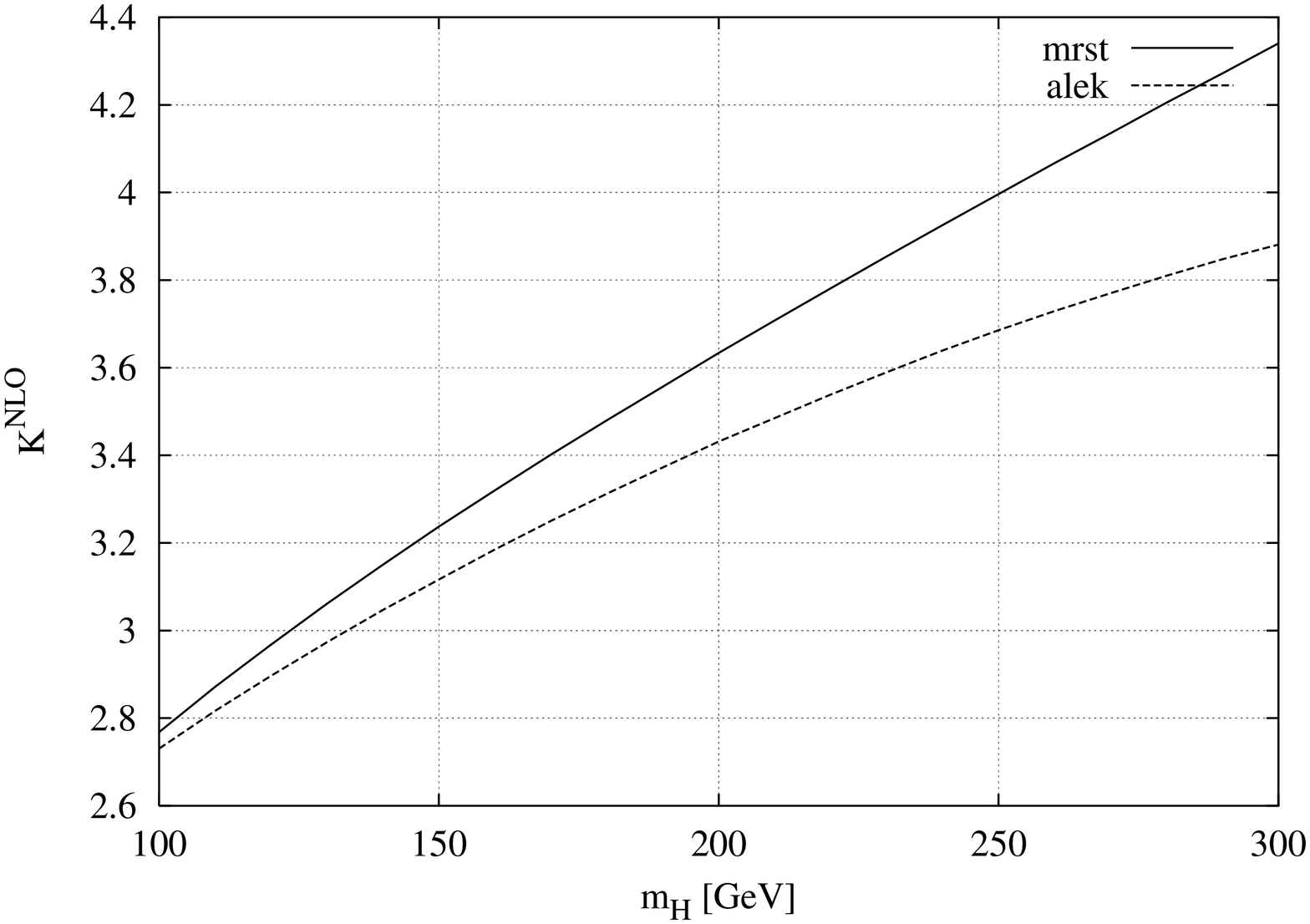}}
\subfigure[Tevatron]{\includegraphics[%
  width=6cm,
  angle=-90]{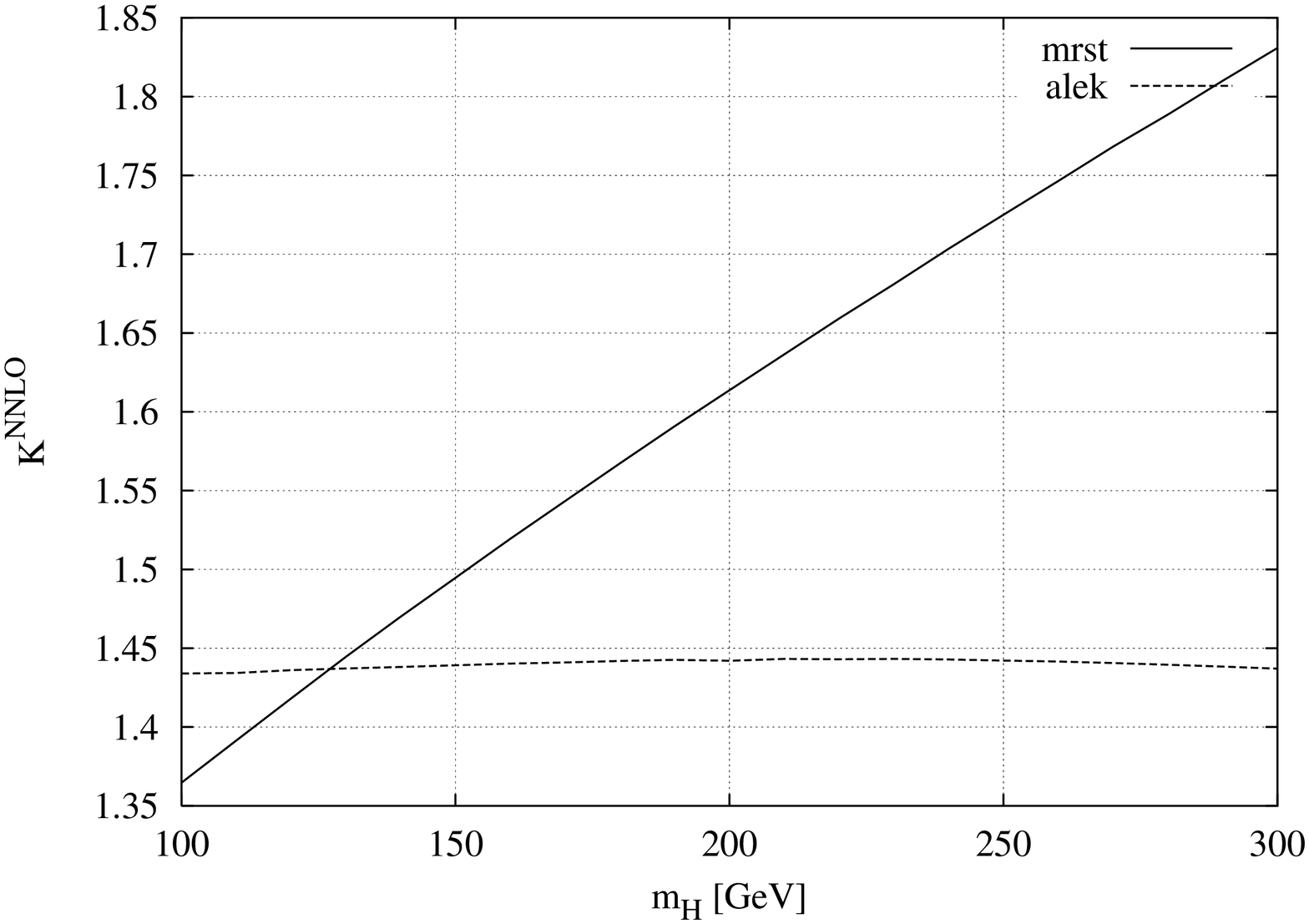}}
\caption{K-factors for the scalar Higgs at NNLO/NLO and
NLO/LO with $\mu_R^2=(1/2)\mu_F^2$ and $\mu_F=2 m_H$}
\label{K4P}
\end{figure}

Now we turn to an analysis of the dependence of our results on
$\mu_F$ and $\mu_R$.  In Figs.~\ref{S3P} we perform this study by
computing $\sigma$ as a function of the
Higgs mass for an incoming energy of 2 and 14 TeV and choose
\beq
\mu_R^2=\frac{1}{2} \mu_F^2 \qquad \mu_F=2 m_H.
\label{sel1}
\eeq
We have seen that for a typical Higgs mass around 100 GeV and 14 TeV of energy
(for  $m_H=\mu_F=\mu_R$) the cross section doubles when
we move from LO to NNLO, and a similar trend is also apparent if we fix
the relations among the scales as in (\ref{sel1}). In this case, however,
the impact of the NLO and NNLO corrections is smaller, a trend which is
apparently uniform over the whole range of the Higgs
mass explored. For $m_H=100$ GeV the scalar cross section $\sigma_{NNLO}$
is around 58 pb for coincident scales, while a different choice,
such as (\ref{sel1}) lowers it to approximately 45 pb. At Tevatron energies
the variations of the cross section with the changes of the various scales
are also sizeable. In this case for $m_H=100$ GeV the LO, NLO and NNLO
predictions ($0.6,1.2$ and 1.6 pb respectively) change approximately by
10-20 per cent if we include variations of the other scales as well.
A parallel view of this trend comes from the study of the
dependence of the K-factors. This study is presented
in Figs.~\ref{K4P}. The interval of variation of the K-factors is
substantially the same as for coincident scales, though the trends of the
two models \cite{MRST} and \cite{Alekhin} is structurally quite different
at NLO and at NNLO, with several cross-overs
among the corresponding curves taking place for $m_H$ around 200 GeV.
Another important point is that the values of $K^{NNLO}$ are, of course smaller
than $K^{NLO}$ over all the regions explored, signaling an overall stability of
the perturbative expansion. We show in Figs.~11-14 3-dimensional plots of the cross section and the 
corresponding K-factors as functions of the factorization and renormalization scales. Notice that as we move from LO to higher orders the curvature of the corresponding surfaces for the cross sections change 
from negative to positive, showing the presence of a plateau when the scales are approximately equal.

\subsection{Stability and the Choice of the Scales}
\begin{figure}
\subfigure[$C=1$]{\includegraphics[%
  width=6cm,
  angle=-90]{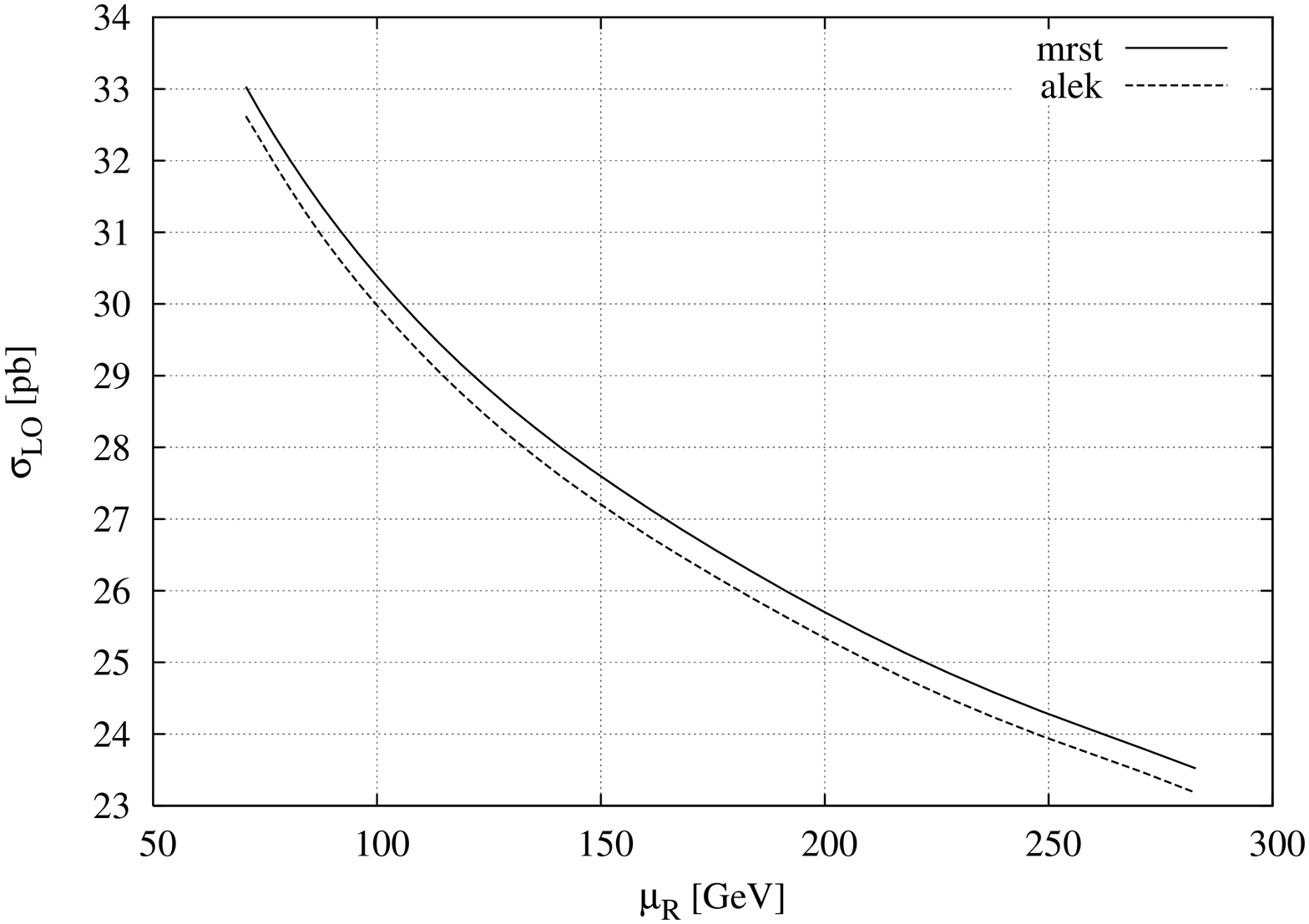}}
\subfigure[$C=1$]{\includegraphics[%
  width=6cm,
  angle=-90]{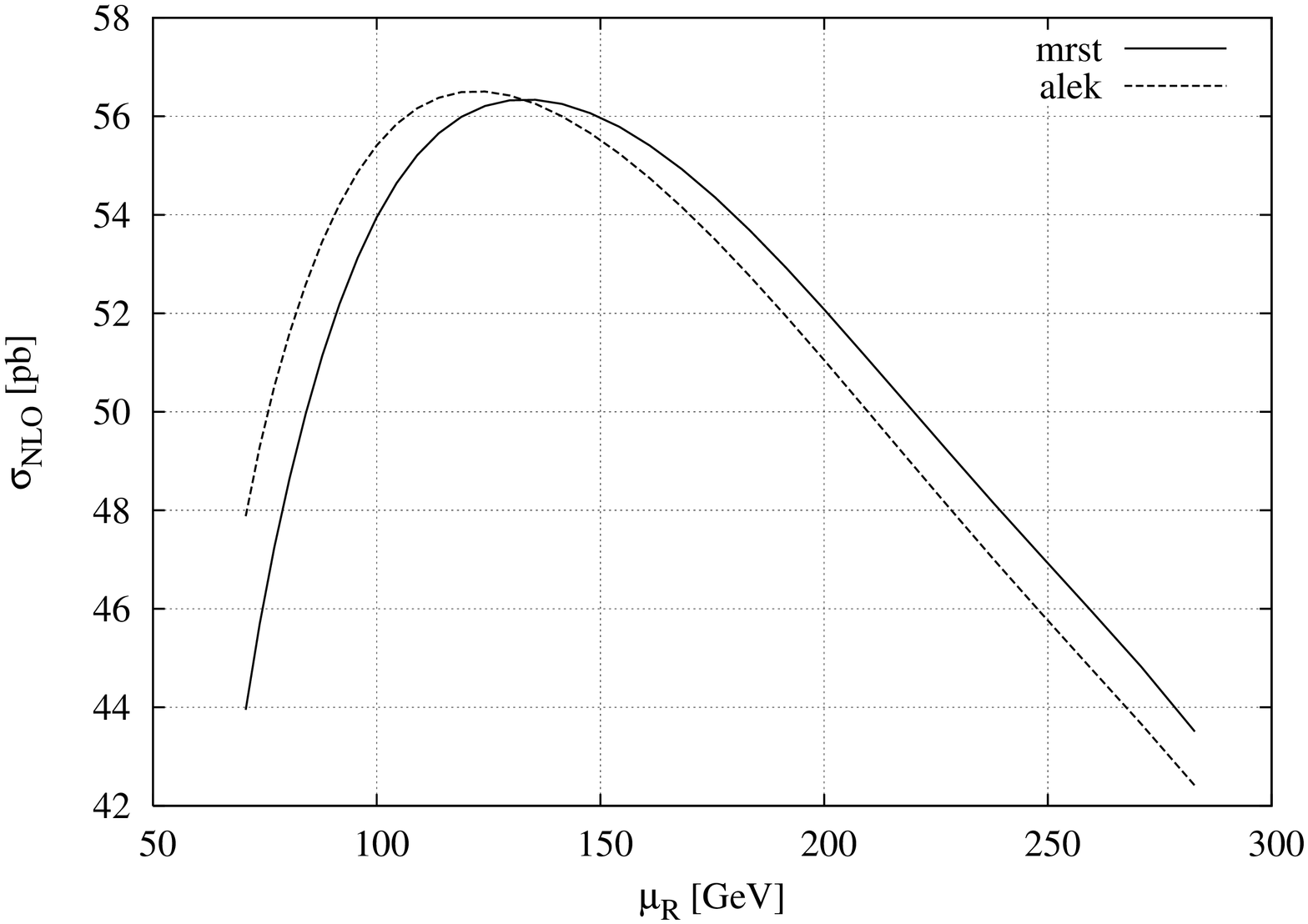}}
\subfigure[$C=1$]{\includegraphics[%
  width=6cm,
  angle=-90]{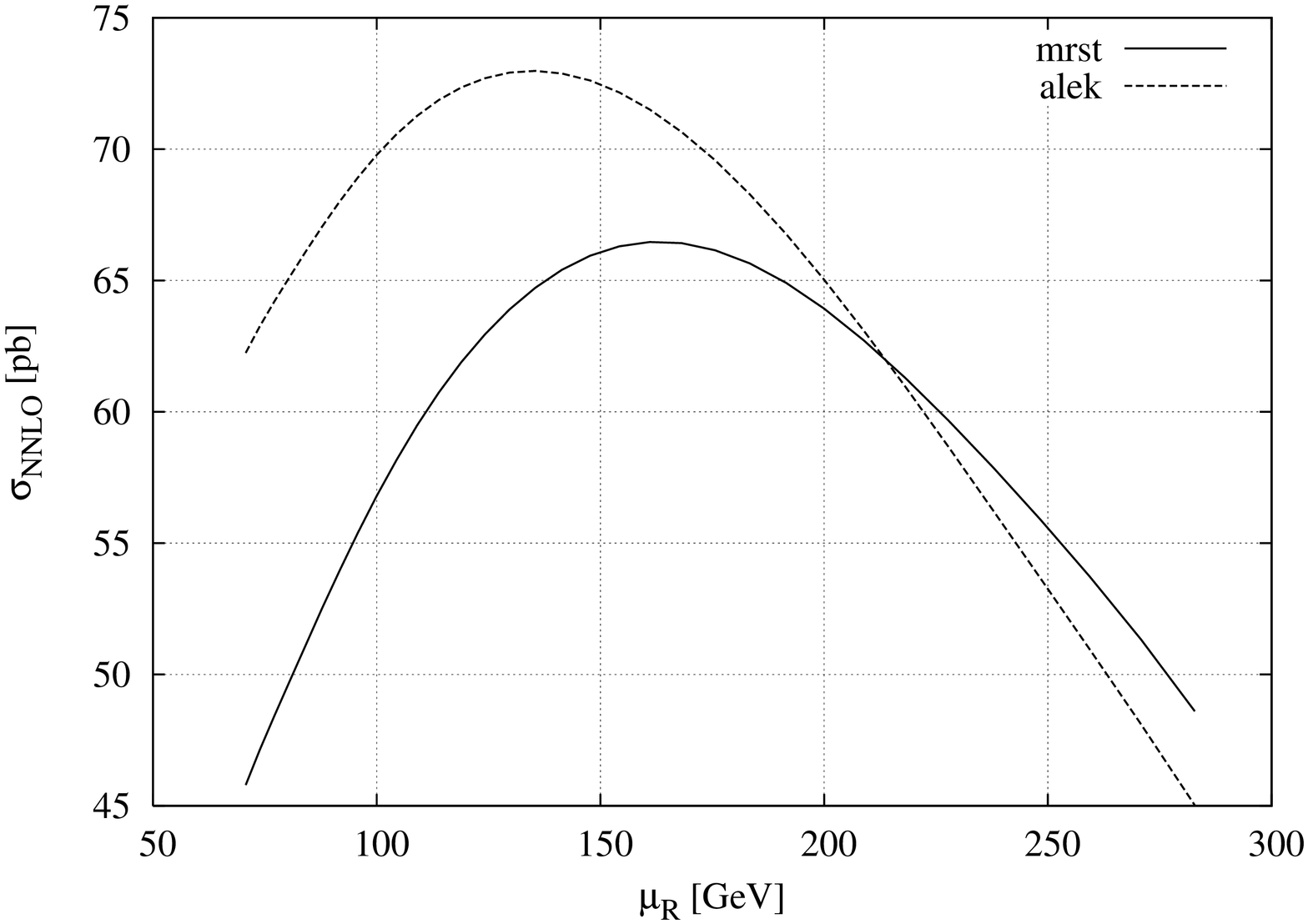}}
\subfigure[$C=1/2$]{\includegraphics[%
  width=6cm,
  angle=-90]{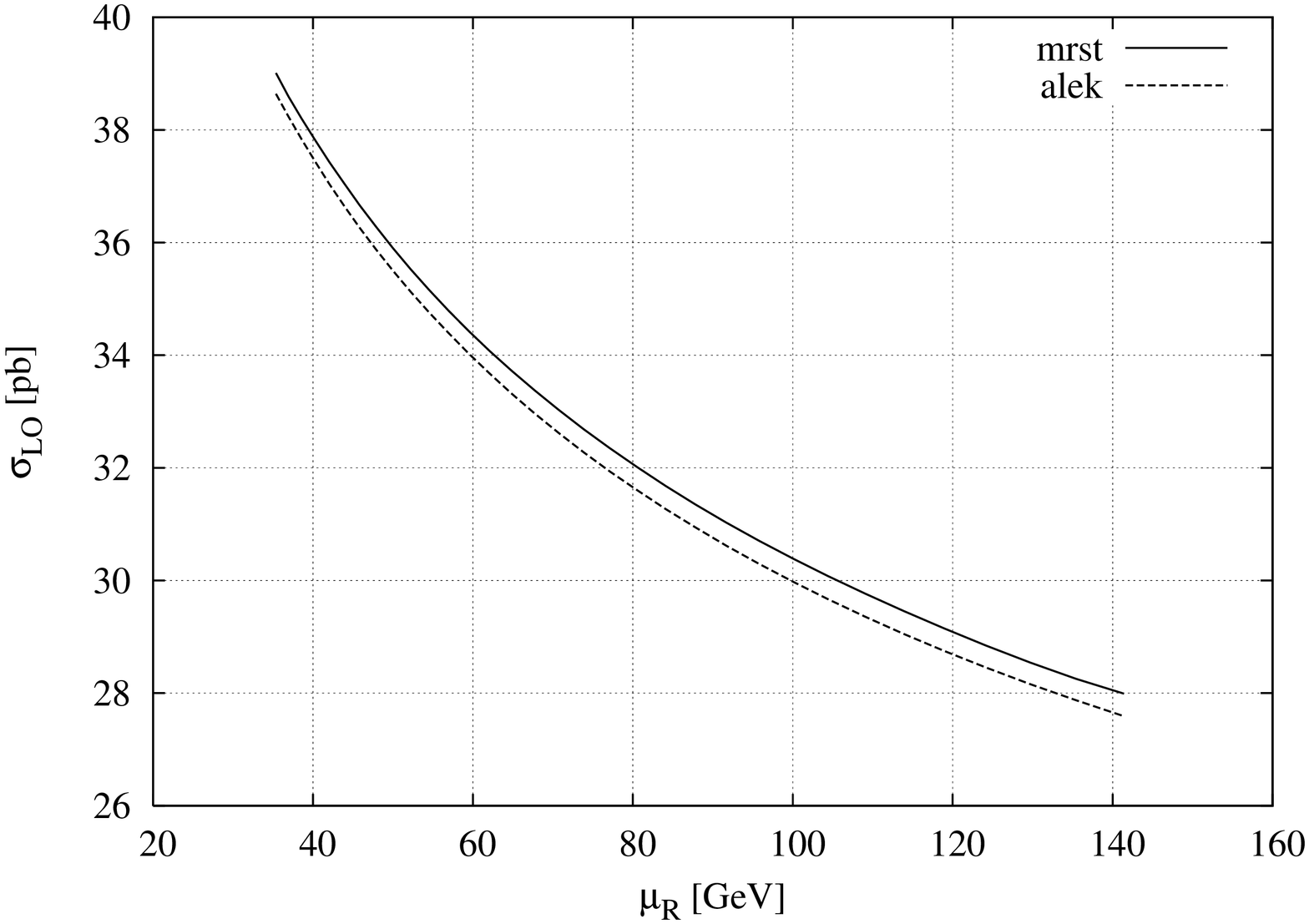}}
\subfigure[$C=1/2$]{\includegraphics[%
  width=6cm,
  angle=-90]{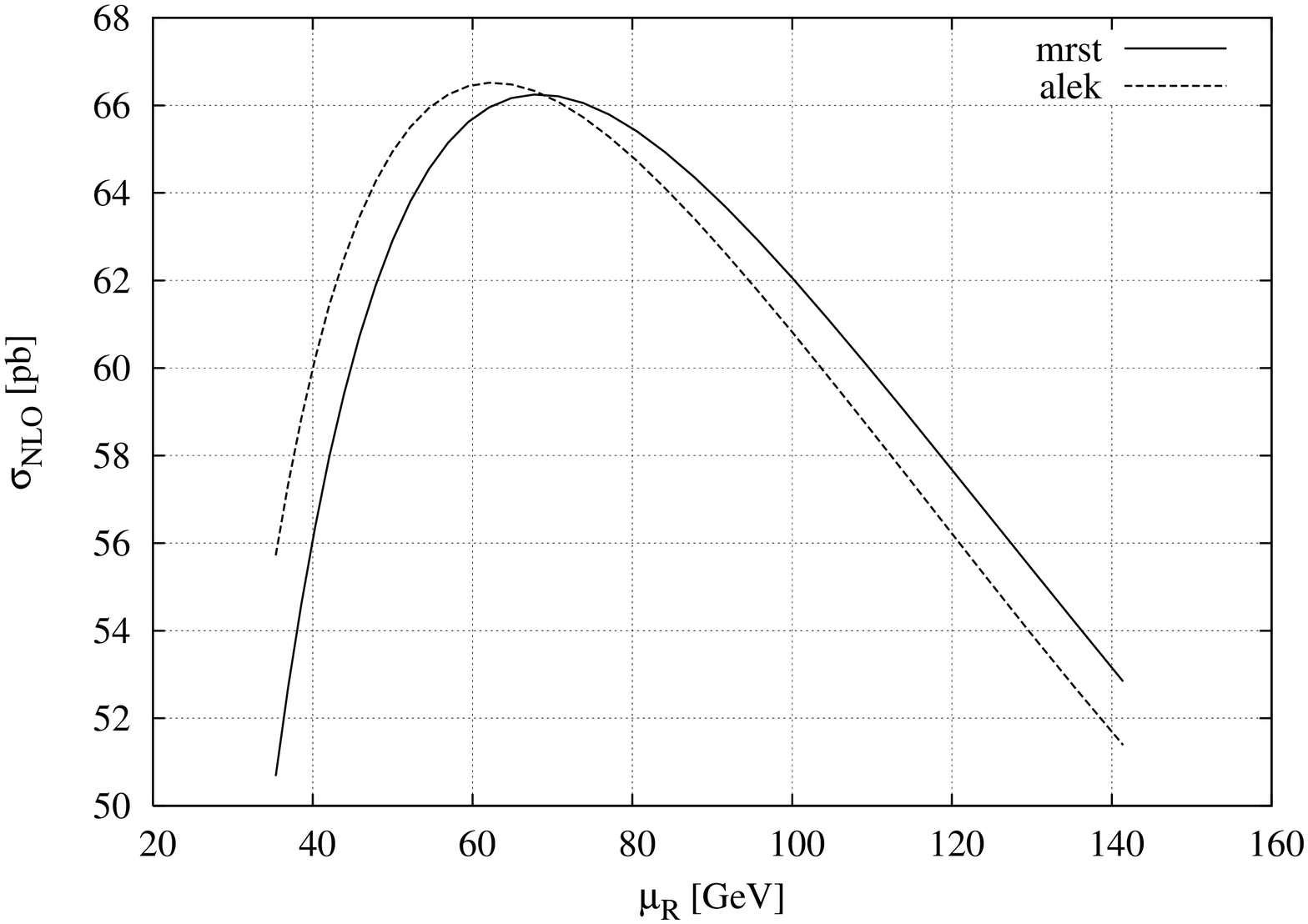}}
\subfigure[$C=1/2$]{\includegraphics[%
  width=6cm,
  angle=-90]{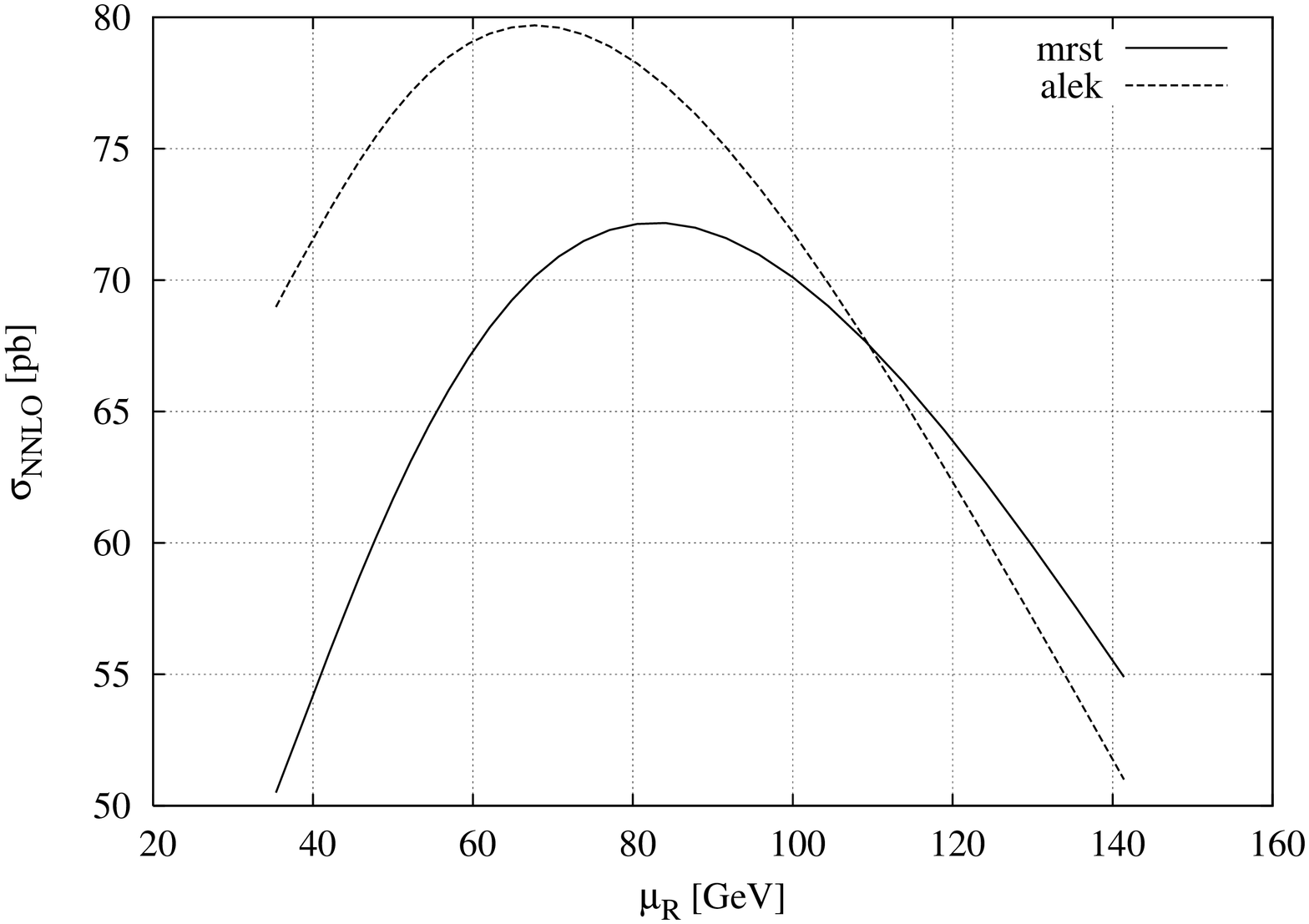}}
\caption{Cross sections for the scalar Higgs production at the LHC
as a function of $\mu_R$, with $\mu_F=C m_H$
and $m_H=100$ GeV}
\label{S5P}
\end{figure}

\begin{figure}
\subfigure[$C=1$]{\includegraphics[%
  width=6cm,
  angle=-90]{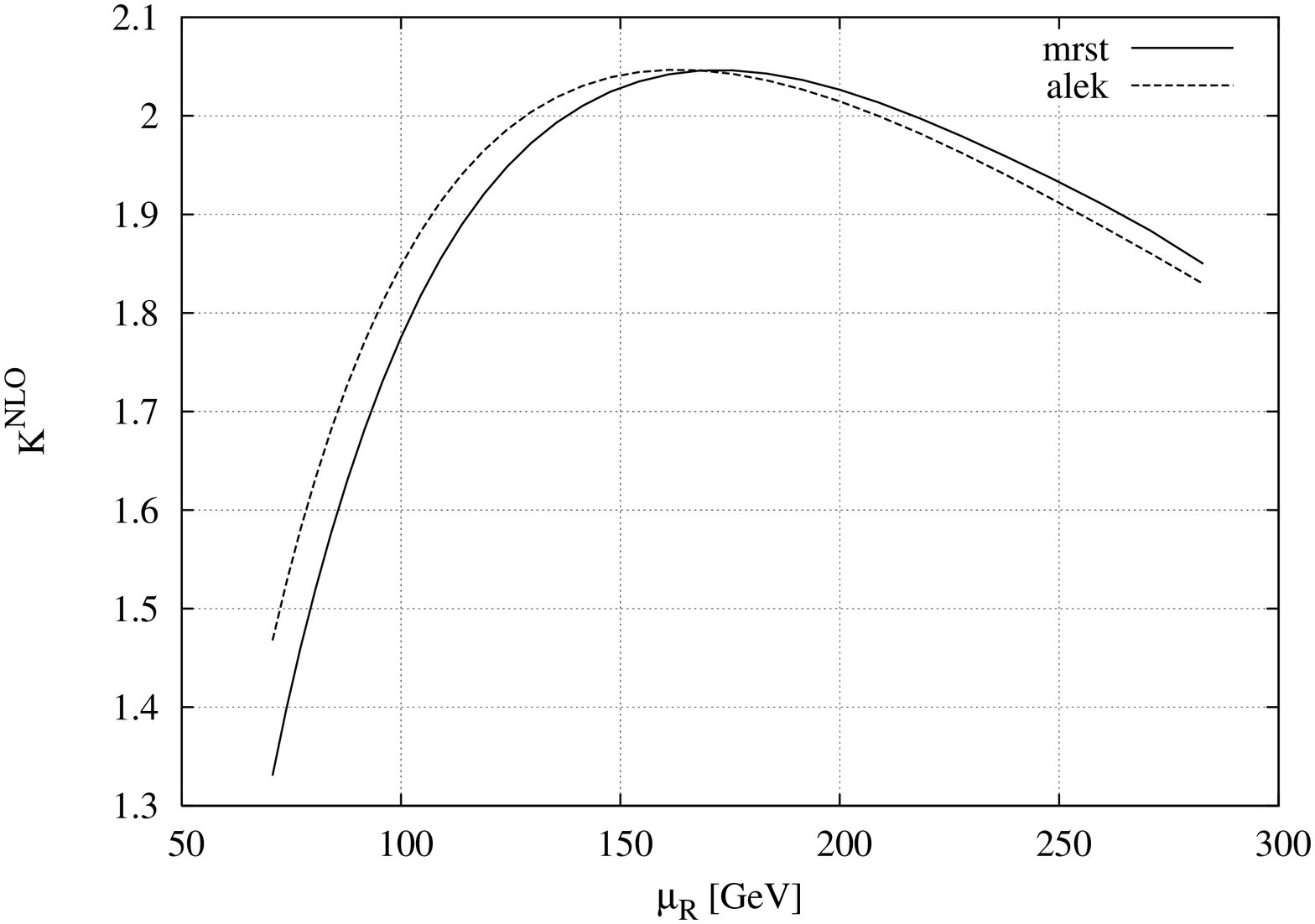}}
\subfigure[$C=1$]{\includegraphics[%
  width=6cm,
  angle=-90]{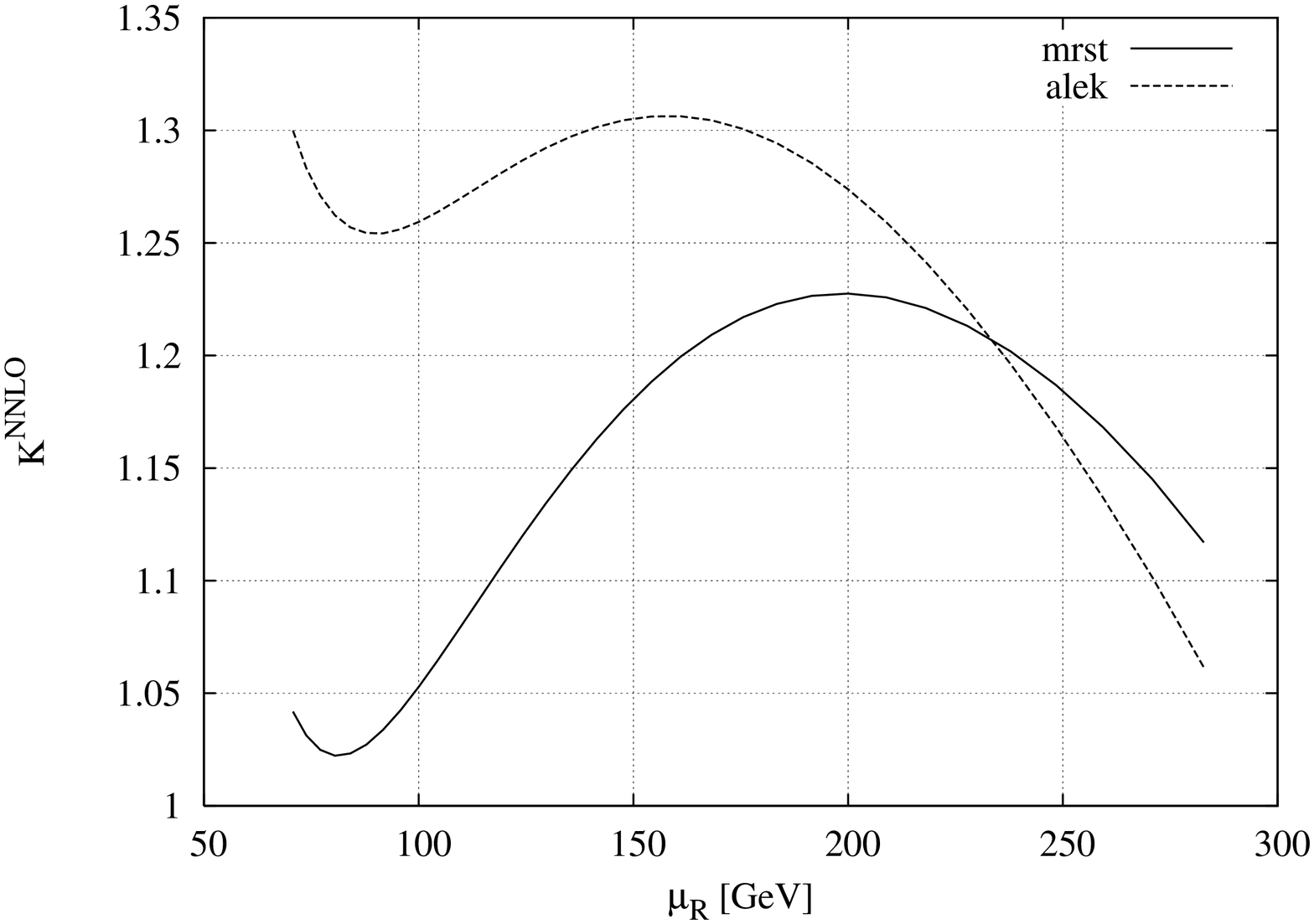}}
\subfigure[$C=1/2$]{\includegraphics[%
  width=6cm,
  angle=-90]{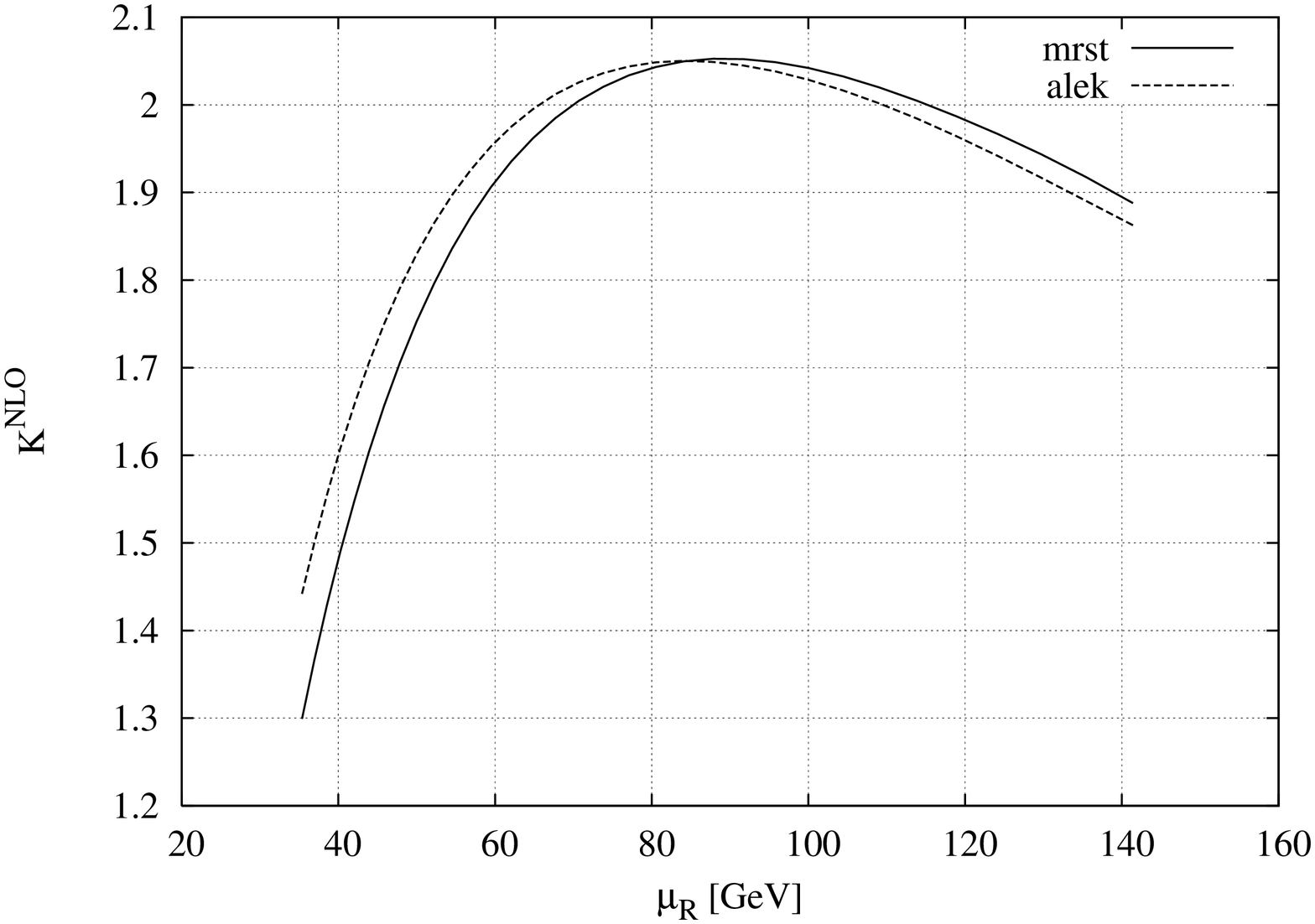}}
\subfigure[$C=1/2$]{\includegraphics[%
  width=6cm,
  angle=-90]{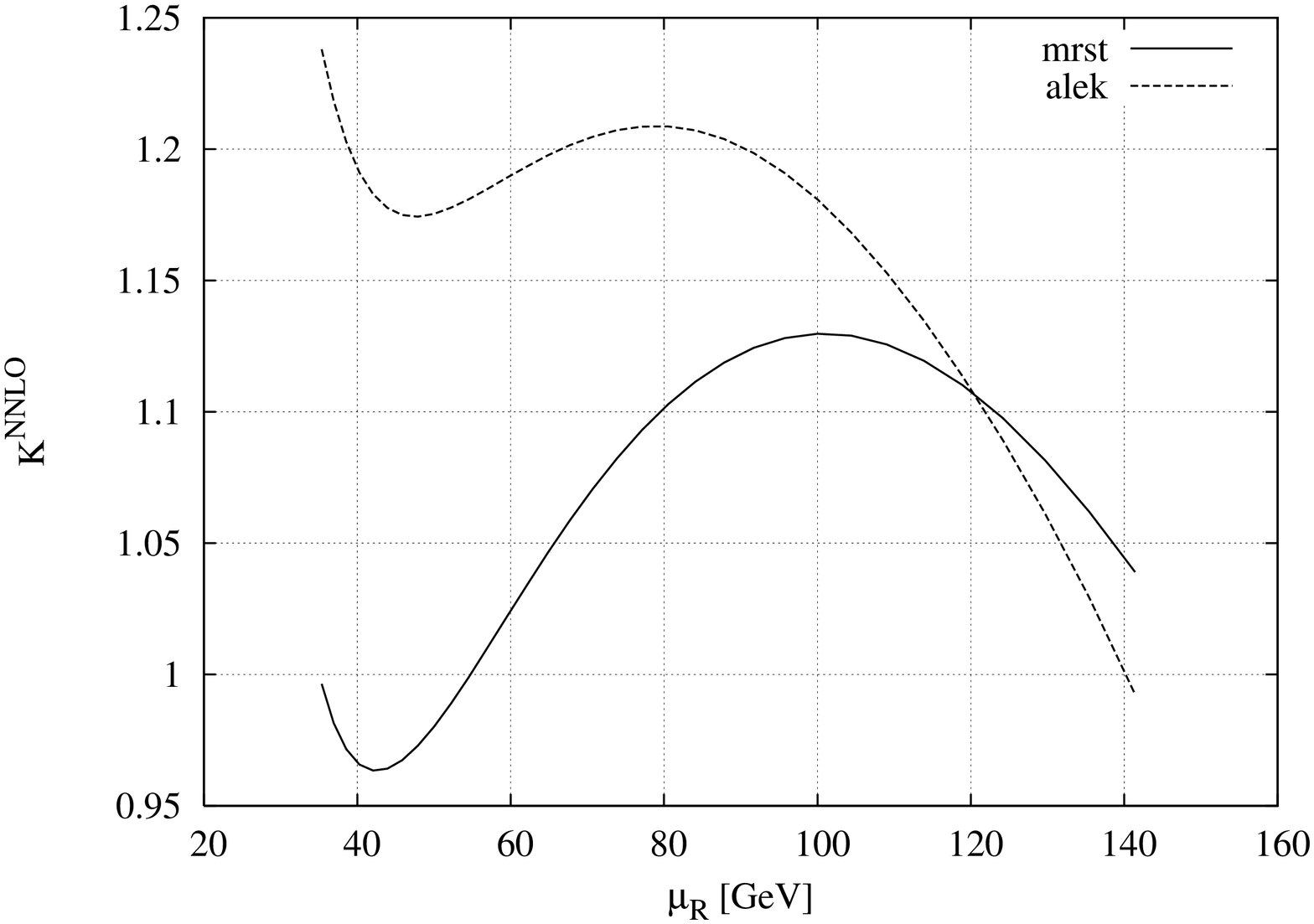}}
\caption{K-factors for the scalar Higgs production at the LHC, NNLO/NLO and
NLO/LO as a function of $\mu_R$, with $\mu_F=C m_H$
and $m_H=100$ GeV}
\label{K6P}
\end{figure}

The issue of determining the best of values of $m_H$, $\mu_F$ and $\mu_R$
in the prediction of the total cross section is a rather important one for
Higgs searches at LHC. We have therefore detailed in Figs.~\ref{S5P}
and \ref{K6P} a study of the behaviour of our results varying the
renormalization scale $\mu_R$ at a fixed value of the ratio between $\mu_F$
and $m_H$. In these figures we have chosen two values for the ratio between
these two scales. Apart from the LO behaviour of the scalar cross section,
which is clearly strongly dependent on the variation of both scales
(see Figs. \ref{S5P} (a) and (d)), and does not show any sign of stability since the
cross section can be drastically lowered by a different choice of $\mu_F$,
both the NLO and the NNLO predictions show instead a clear region of
local stability for $\mu_R > \mu_F$ but not too far away from
the ``coincidence region'' $\mu_F=\mu_R=m_H$. This can be illustrated
more simply using Fig. \ref{S5P}(b) as an example, where we have
set the incoming energy of the p-p collision at 14 TeV.
In this case, for instance, we have chosen $m_H=\mu_F=100$ GeV $(C=1)$,
and it is clear from the plots that a plateau is present in the region
of $\mu_R\sim 130$ GeV. Similar
trends are also clearly visible at NNLO, though the region of the plateau
for the scalar cross section is slightly wider. Also in this case it is
found that the condition $\mu_R > \mu_F$ generates a reduced scale
dependence. Away from this region the predictions
show a systematic scale dependence, as shown also for the choice
of $C=1/2$ in the remaining figures. In Figs.~\ref{K6P} we repeat the
same study for the K-factors, relaxing the condition on the coincidence
of all the scales and plotting the variations
of $K^{NLO}$ and $K^{NNLO}$ in terms of $\mu_R$. In the case $m_H=\mu_F=100$ GeV
the plateau is reached for $\mu_R\sim 150$ GeV for $K^{NLO}$
and $\mu_R\sim 200$ GeV for $K^{NNLO}$.  In the first case the NLO corrections
amount to an increase by 100 per cent compared to the
LO result, while the NNLO corrections modify the NLO estimates by about
20 per cent (MRST).  Similar results are obtained also for $\mu_F=50$ GeV.
In this case, at the plateau, the NLO corrections are still approximately
100 per cent compared to the LO result and the NNLO corrections increase
this value by around 15 per cent (MRST).

\subsection{Energy Dependence}
The energy dependence of the NNLO predictions for the total cross sections
and the corresponding K-factors at the LHC are shown in
Figs.~\ref{ener1}-\ref{ener3}, where we have varied the ratio
$C=\mu_F/m_H$ and $k=\mu_R^2/\mu_F^2$ in order to illustrate the variation
of the results. The cross sections increase sharply with energy and the
impact of the NNLO corrections is significant. The K-factors, in most of
the configurations chosen, vary between 1 and 2.2. We have chosen the
MRST input. The behaviour of the K-factors is influenced significantly
by the choice of the ratio $(k)$ between $\mu_R$ and $\mu_F$.
In particular, in Figs.~\ref{ener2} the NNLO K-factors increase
with $\sqrt{S}$ for $k=2$ , the center of mass energy, which is not
found for other choices of scales. The case $k=1/2$ is close in behaviour
to the coincident case $\mu_R^2=\mu_F^2$. The overall stability of the
K-factors is clearly obtained with the choice $k=1$. We have finally
included in Tables~2-10 our numerical
predictions in order to make them available to the experimental collaborations.

\section{Conclusions}

A study of the NNLO corrections to the cross section for Higgs production
has been presented.  We have implemented the exact three-loop splitting
functions in our own parton evolution code. We used as
initial conditions (at small scales) the boundary values of Martin,
Roberts, Thorne and Stirling and of Alekhin.  This study shows that the
impact of these corrections are important for the discovery of the
Higgs and for a reconstruction of its mass. The condition of stability
of the perturbative expansion is also quite evident from these studies
and suggests that the optimal choice to fix the arbitrary scales of the theory
are near the coincidence point, with $\mu_R$ in the region of a plateau.
The determination of the plateau has been performed by introducing in
the perturbative expansion and in the evolution a new independent
scale $(\mu_R)$, whose variation allows to accurately characterize
the properties of the expansion in a direct way.

While this thesis was being completed several authors have presented
studies of the total Higgs cross section based on threshold
resummation of soft or soft-plus-virtual logarithms, see \cite{mv},\cite{ravi},\cite{lm},
\cite{ij}. Our work is based on exact NNLO partonic cross sections.
Another relevant paper which recently appeared is \cite{amp}.

%%%%%%%%%%%%%%%%%%%%
%%%%%%%%%%%%%%%%%%%%%

\begin{figure}
\subfigure[$m_H=110$GeV]{\includegraphics[%
  width=12cm,
  angle=-90]{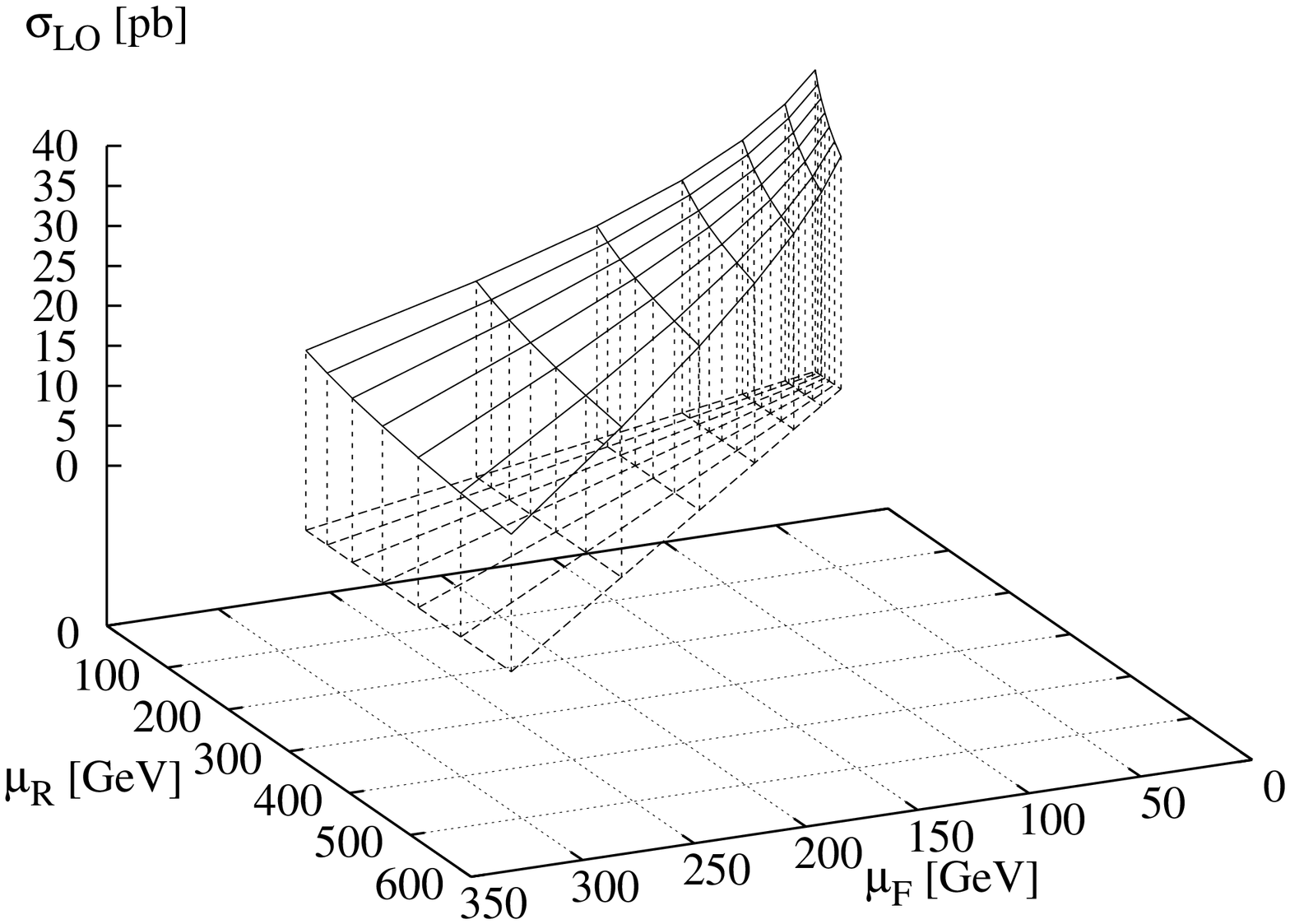}}
\caption{Three-dimensional graphs for the LO cross sections for the scalar Higgs
production at the LHC, as a function of $\mu_R$ and with $\mu_F$ with
a fixed value of $m_H$}
\label{3D1}
\end{figure}
\begin{figure}
\subfigure[$m_H=110$GeV]{\includegraphics[%
  width=12cm,
  angle=-90]{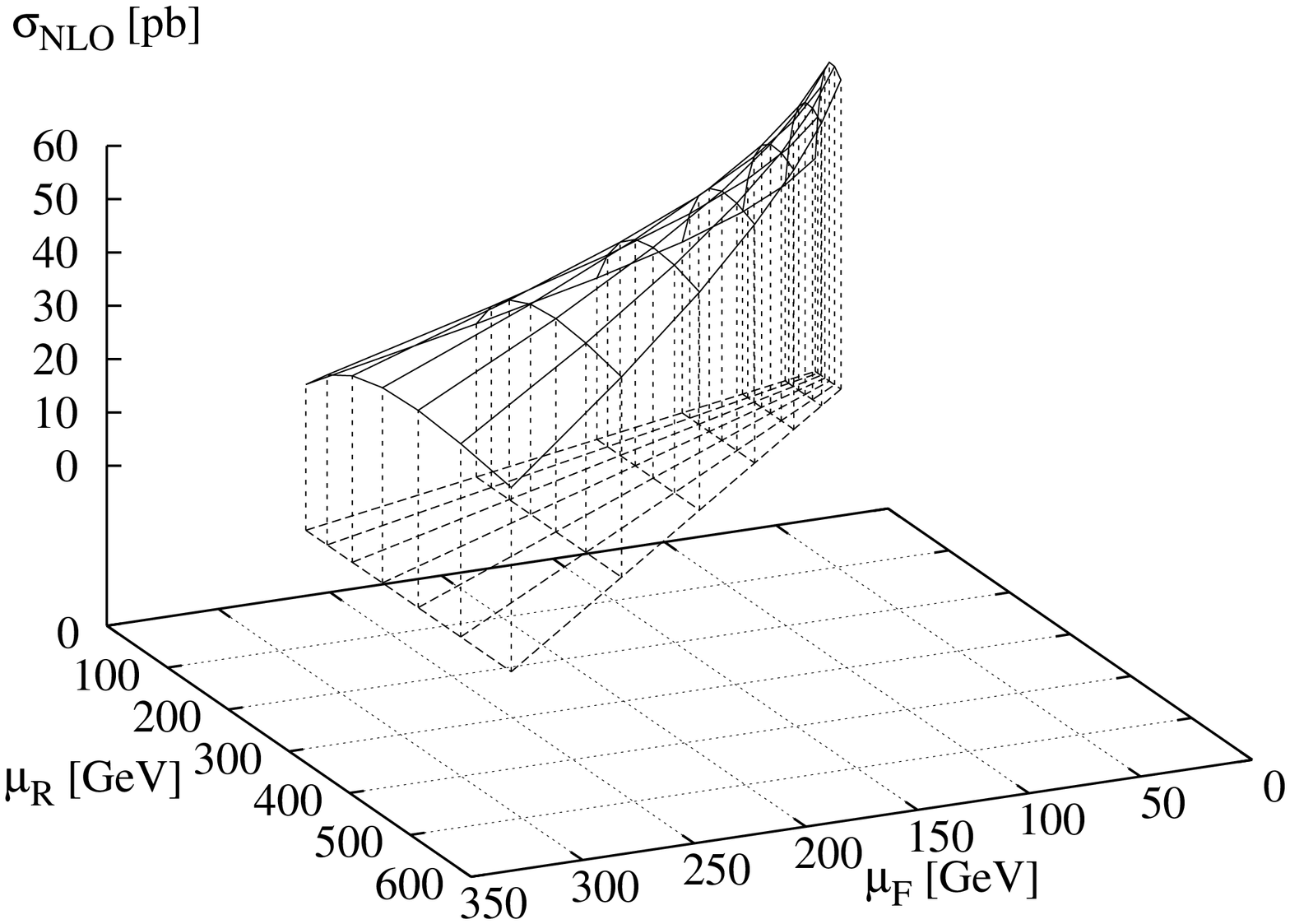}}
\caption{Three-dimensional graphs for the NLO cross sections for the scalar Higgs
production at the LHC, as a function of $\mu_R$ and with $\mu_F$ with
a fixed value of $m_H$}
\label{3D2}
\end{figure}
\begin{figure}
\subfigure[$m_H=110$GeV]{\includegraphics[%
  width=12cm,
  angle=-90]{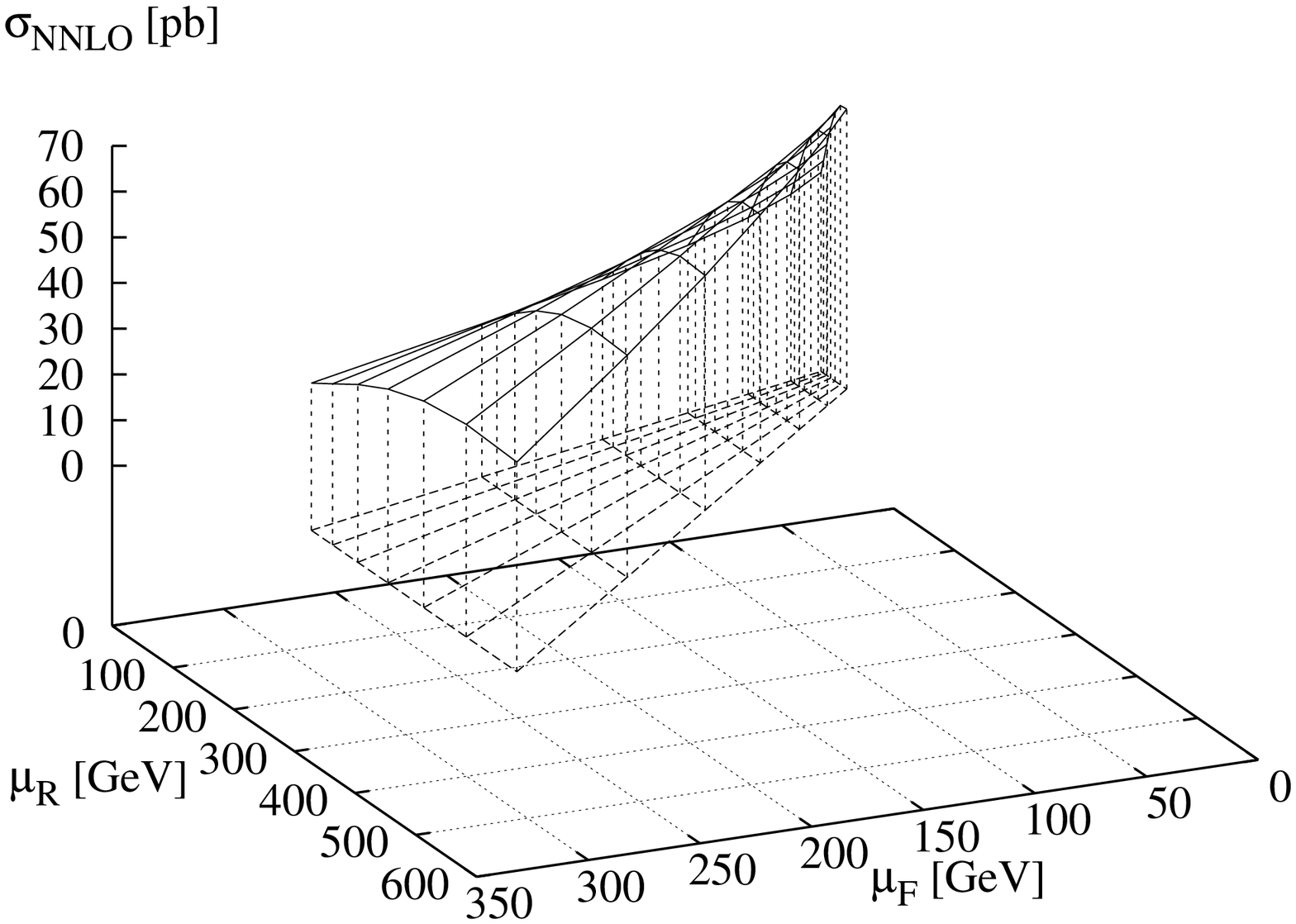}}
\caption{Three-dimensional graphs for the NNLO cross sections for the scalar Higgs
production at the LHC, as a function of $\mu_R$ and with $\mu_F$ with
a fixed value of $m_H$}
\label{3D3}
\end{figure}
\begin{figure}
\subfigure[$m_H=110$GeV]{\includegraphics[%
  width=12cm,
  angle=-90]{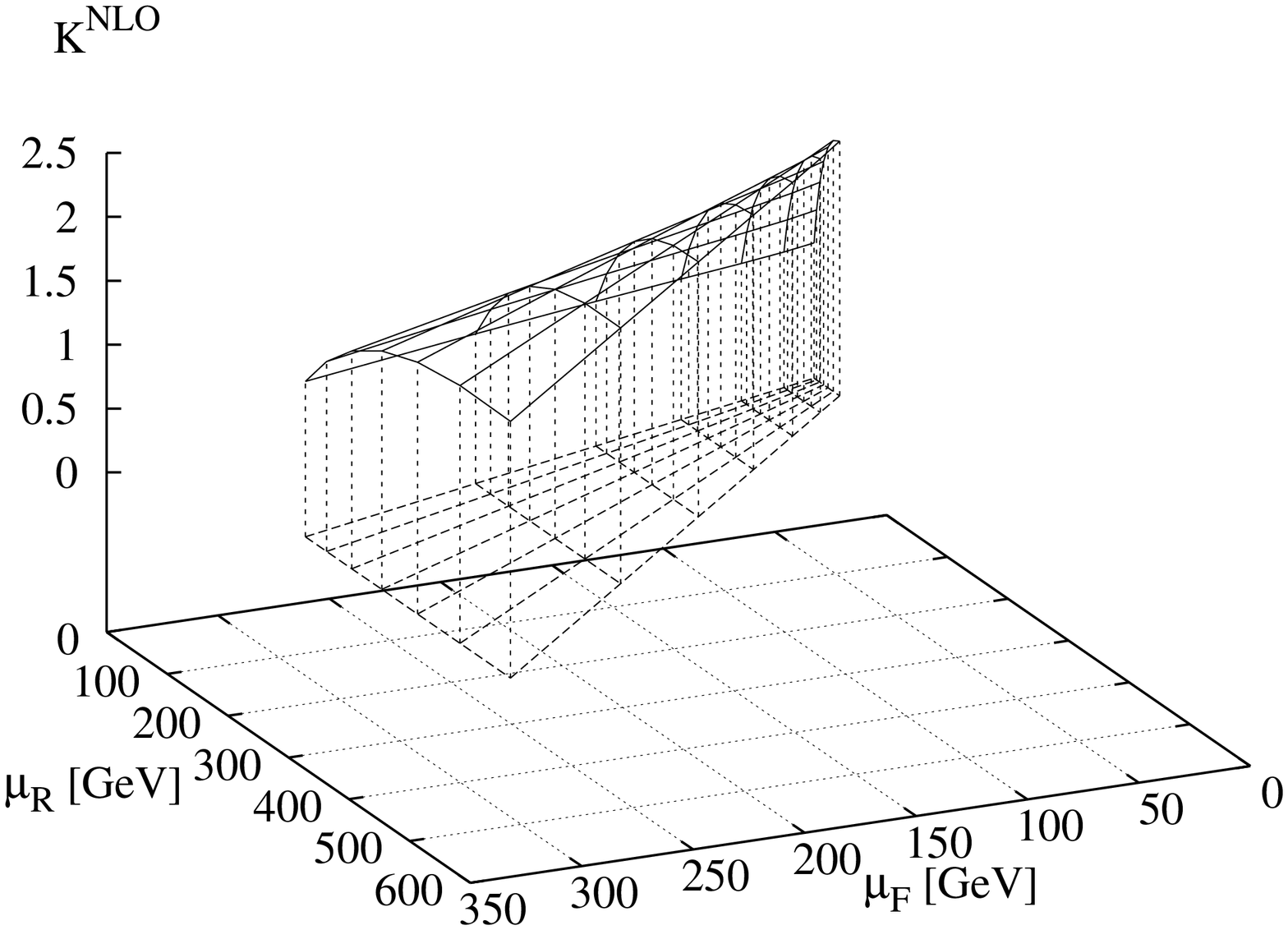}}
\caption{Three-dimensional graphs for the $K$-factor
$\sigma_{NLO}/\sigma_{LO}$ for the scalar Higgs production at the LHC,
as a function of $\mu_R$ and $\mu_F$ and at a
fixed value of $m_H$.}
\label{3D4}
\end{figure}
\begin{figure}
\subfigure[$m_H=110$GeV]{\includegraphics[%
  width=12cm,
  angle=-90]{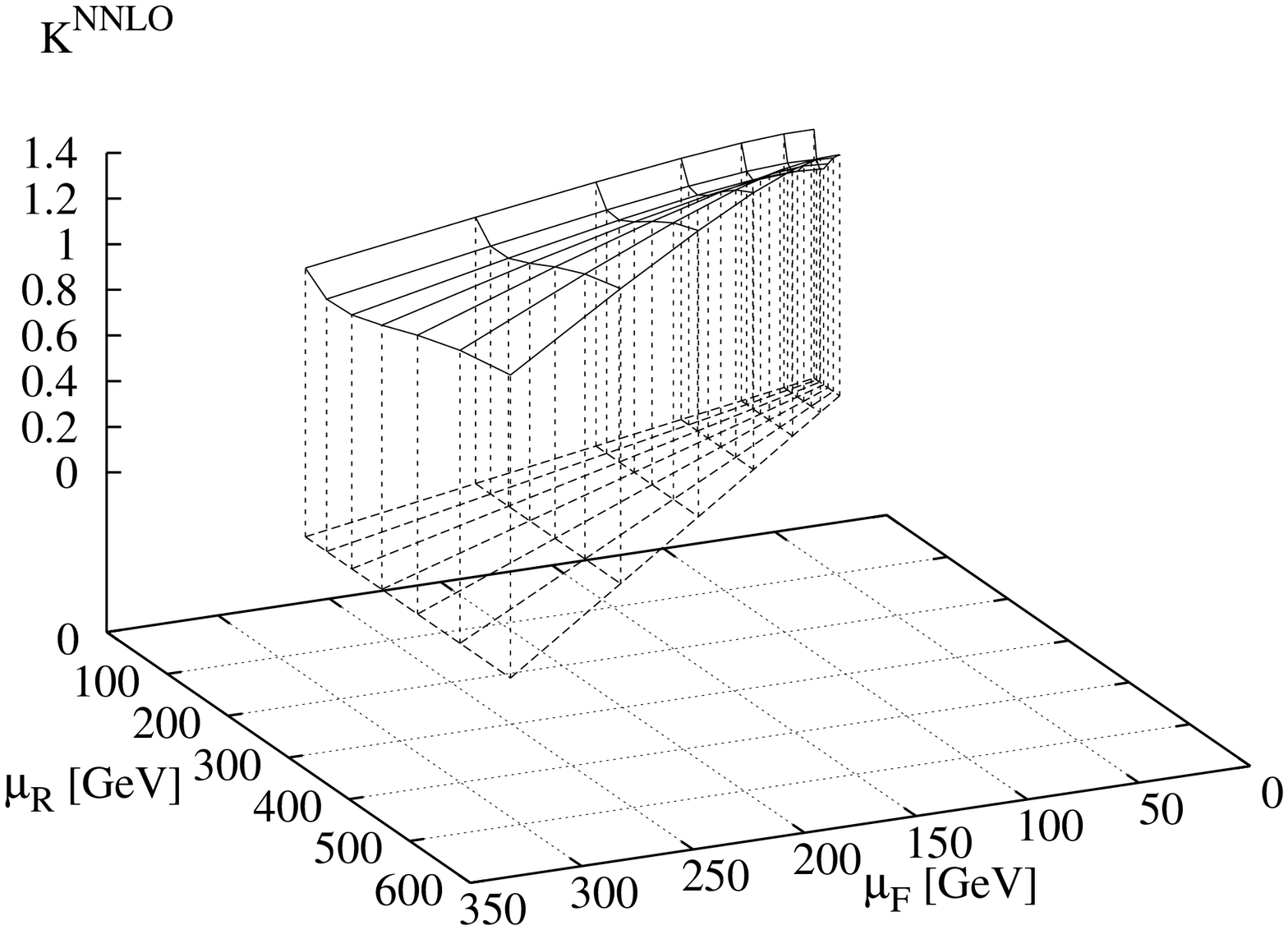}}
\caption{Three-dimensional graphs for the $K$-factor
$\sigma_{NNLO}/\sigma_{NLO}$ for the scalar Higgs production at the LHC,
as a function of $\mu_R$ and $\mu_F$ and at a
fixed value of $m_H$.}
\label{3D5}
\end{figure}
\begin{figure}
\subfigure[$C=1,\,\,k=1$]{\includegraphics[%
  width=6cm,
  angle=-90]{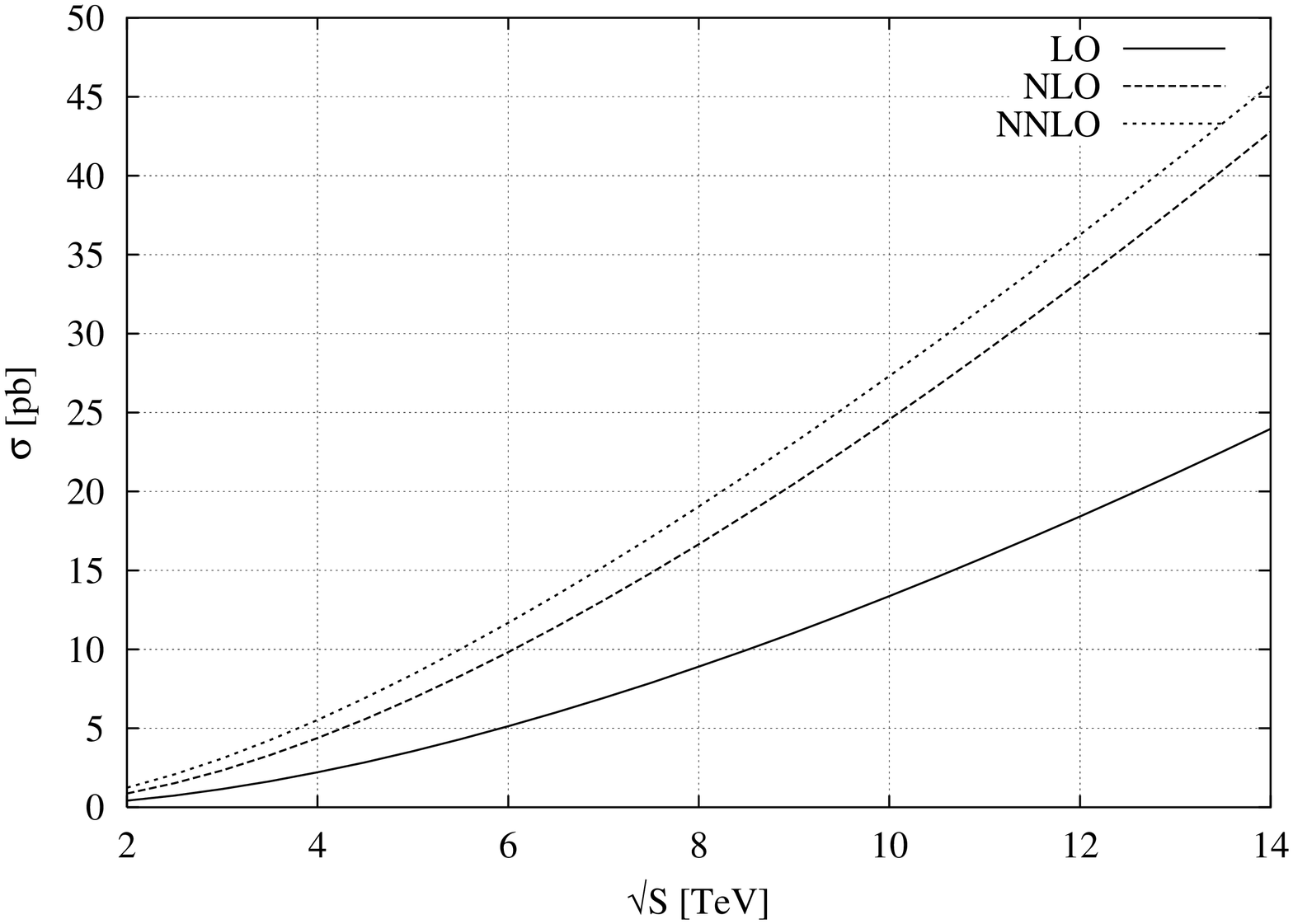}}
\subfigure[$C=1,\,\,k=1$]{\includegraphics[%
  width=6cm,
  angle=-90]{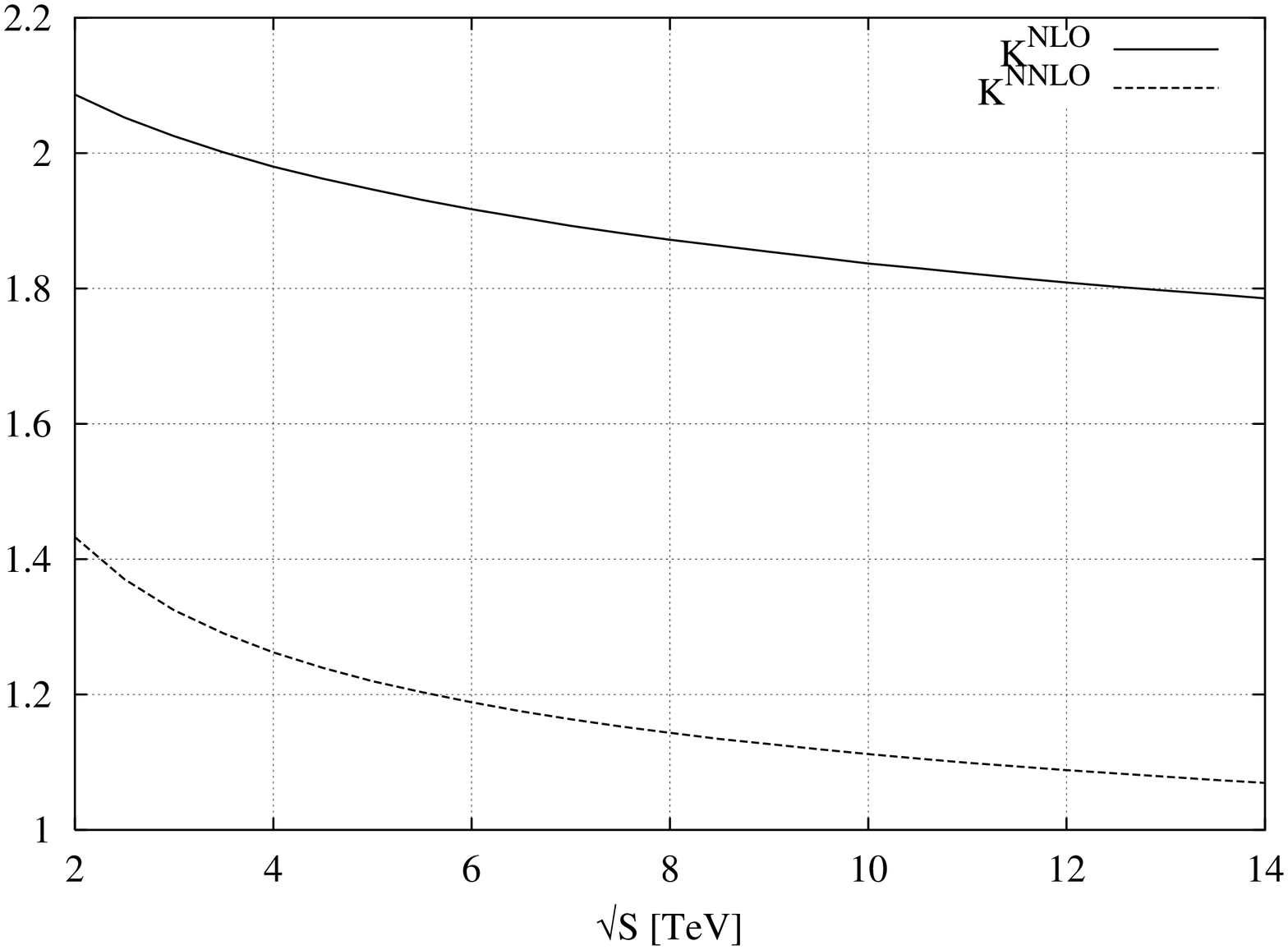}}
\subfigure[$C=2,\,\,k=1$]{\includegraphics[%
  width=6cm,
  angle=-90]{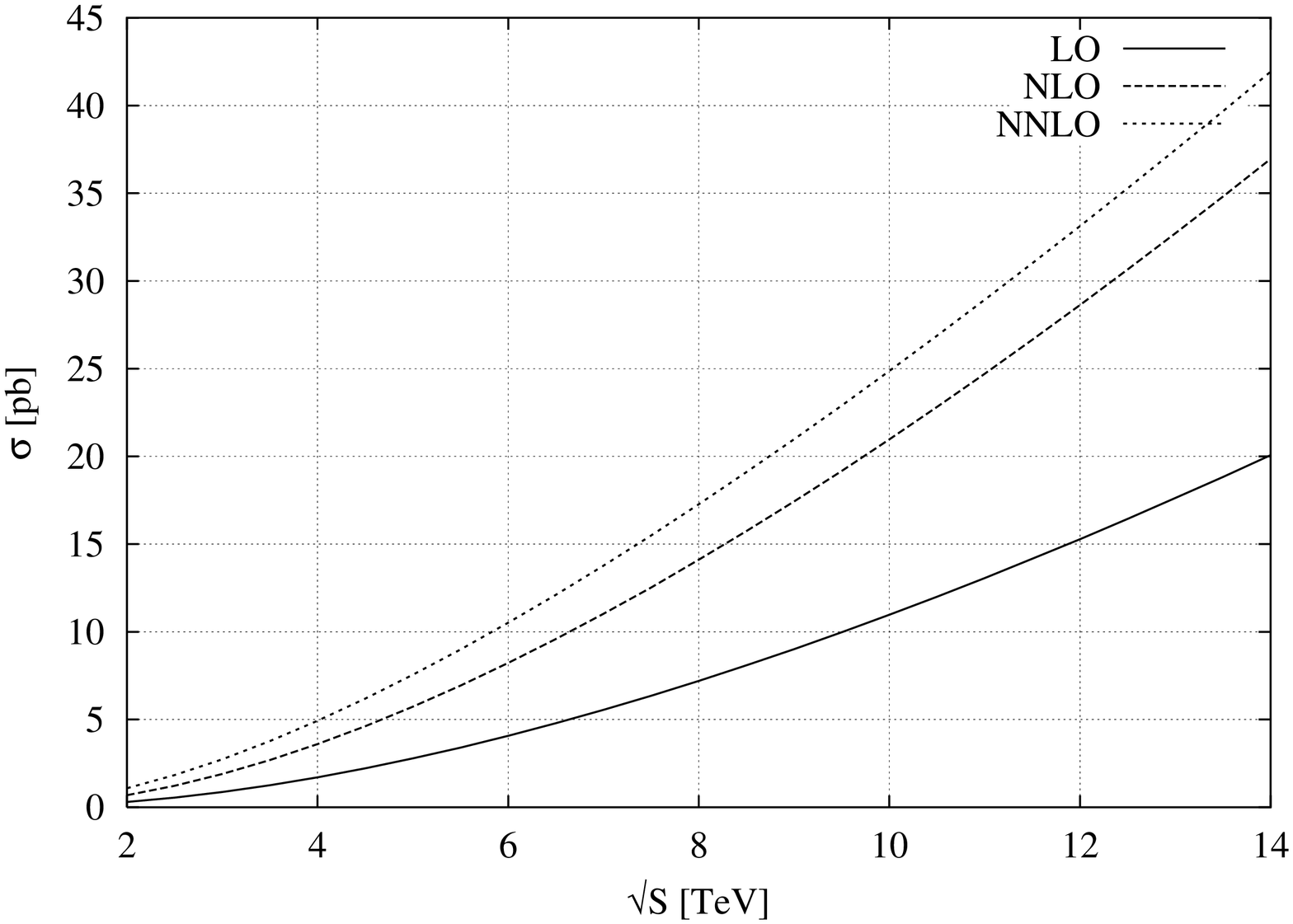}}
\subfigure[$C=2,\,\,k=1$]{\includegraphics[%
  width=6cm,
  angle=-90]{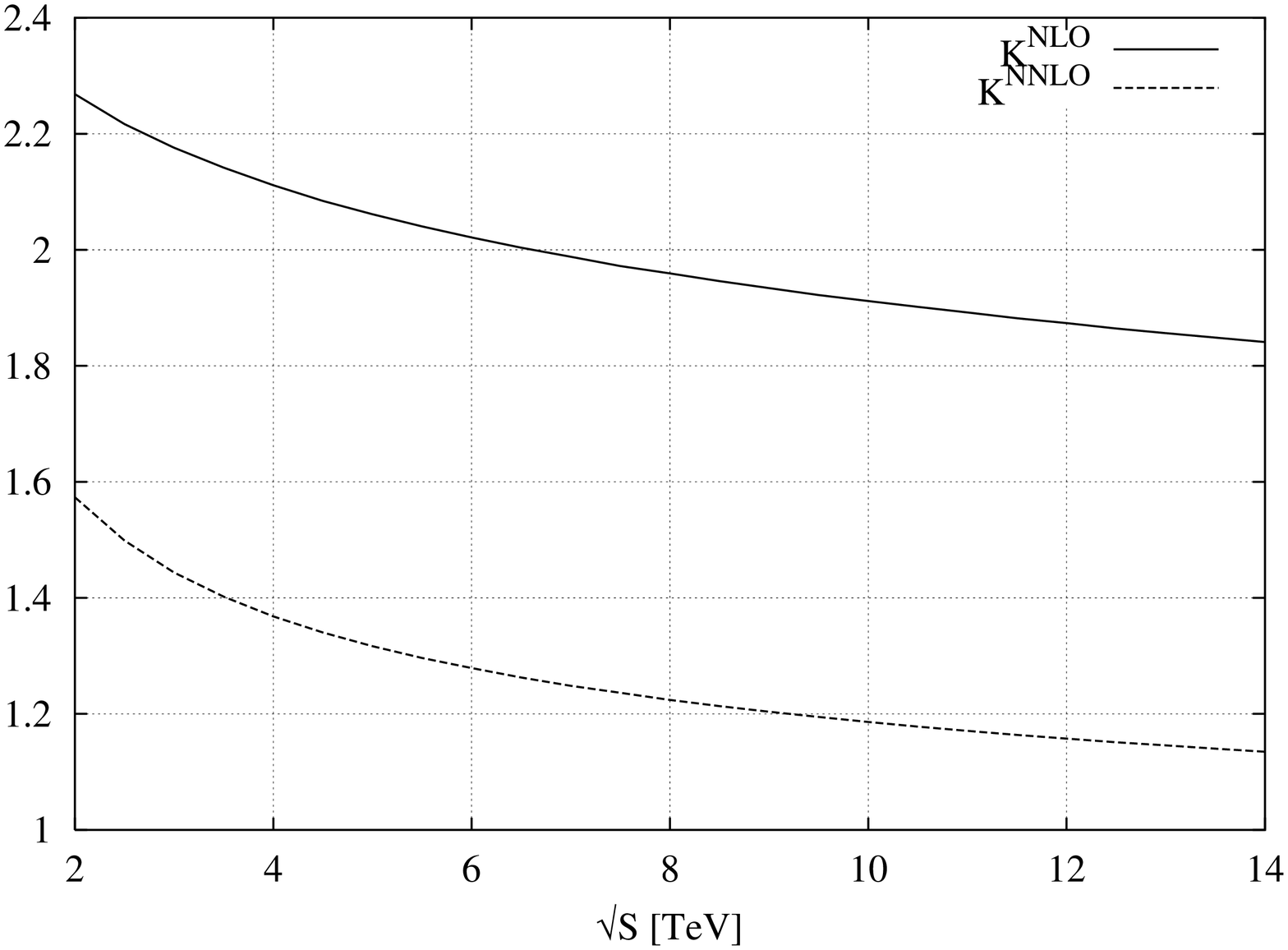}}
\subfigure[$C=1/2,\,\,k=1$]{\includegraphics[%
  width=6cm,
  angle=-90]{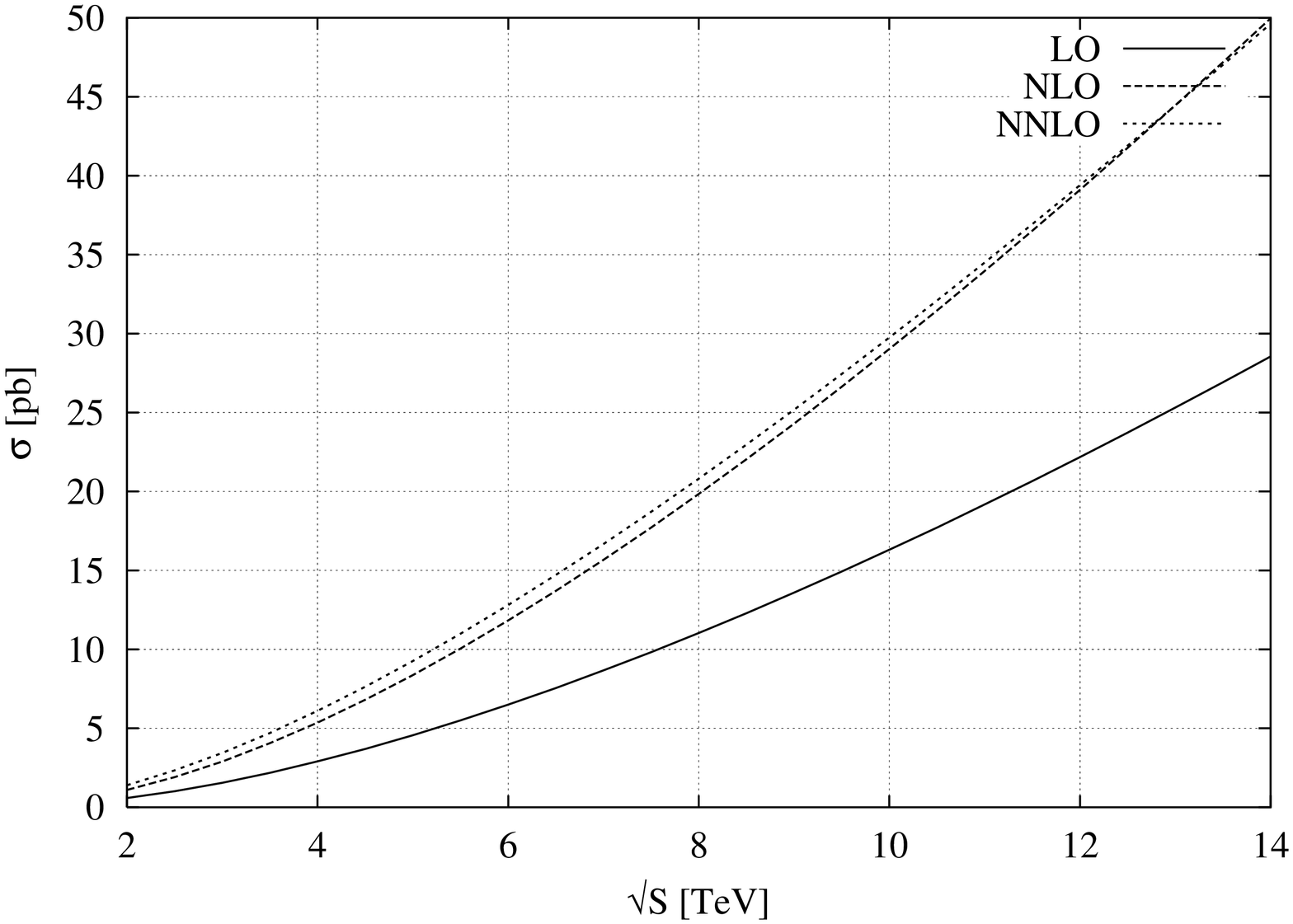}}
\subfigure[$C=1/2,\,\,k=1$]{\includegraphics[%
  width=6cm,
  angle=-90]{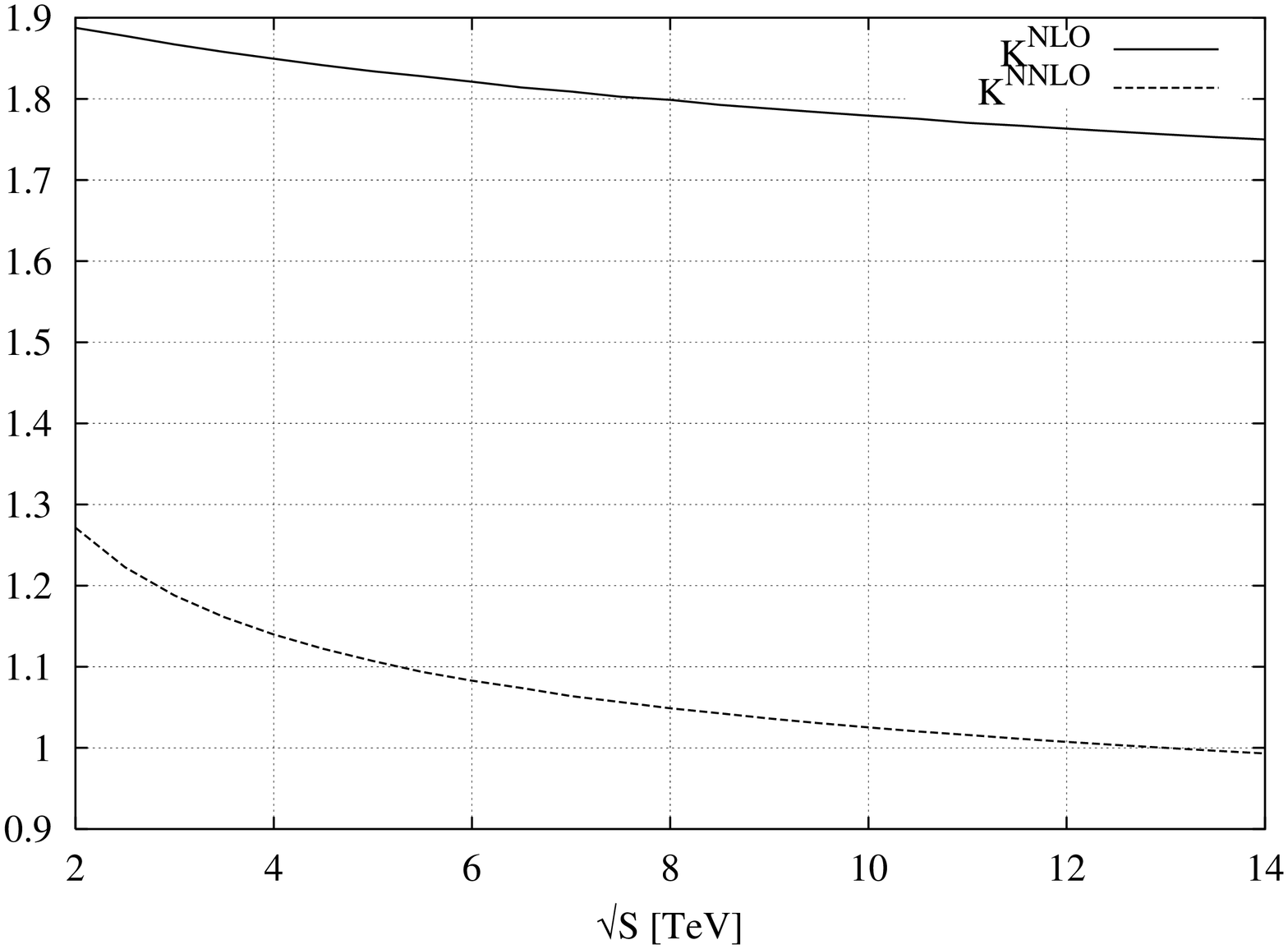}}
\caption{Cross sections and $K$-factors for the scalar Higgs
production at the LHC as a function of $\sqrt{S}$ with $\mu_F=C m_H$,
with $\mu_F^2=k\mu_R^2$ and $m_H=114$ GeV. MRST inputs have been used.}
\label{ener1}
\end{figure}
%%%%%%%%%%%%
\begin{figure}
\subfigure[$C=1,\,\,k=2$]{\includegraphics[%
  width=6cm,
  angle=-90]{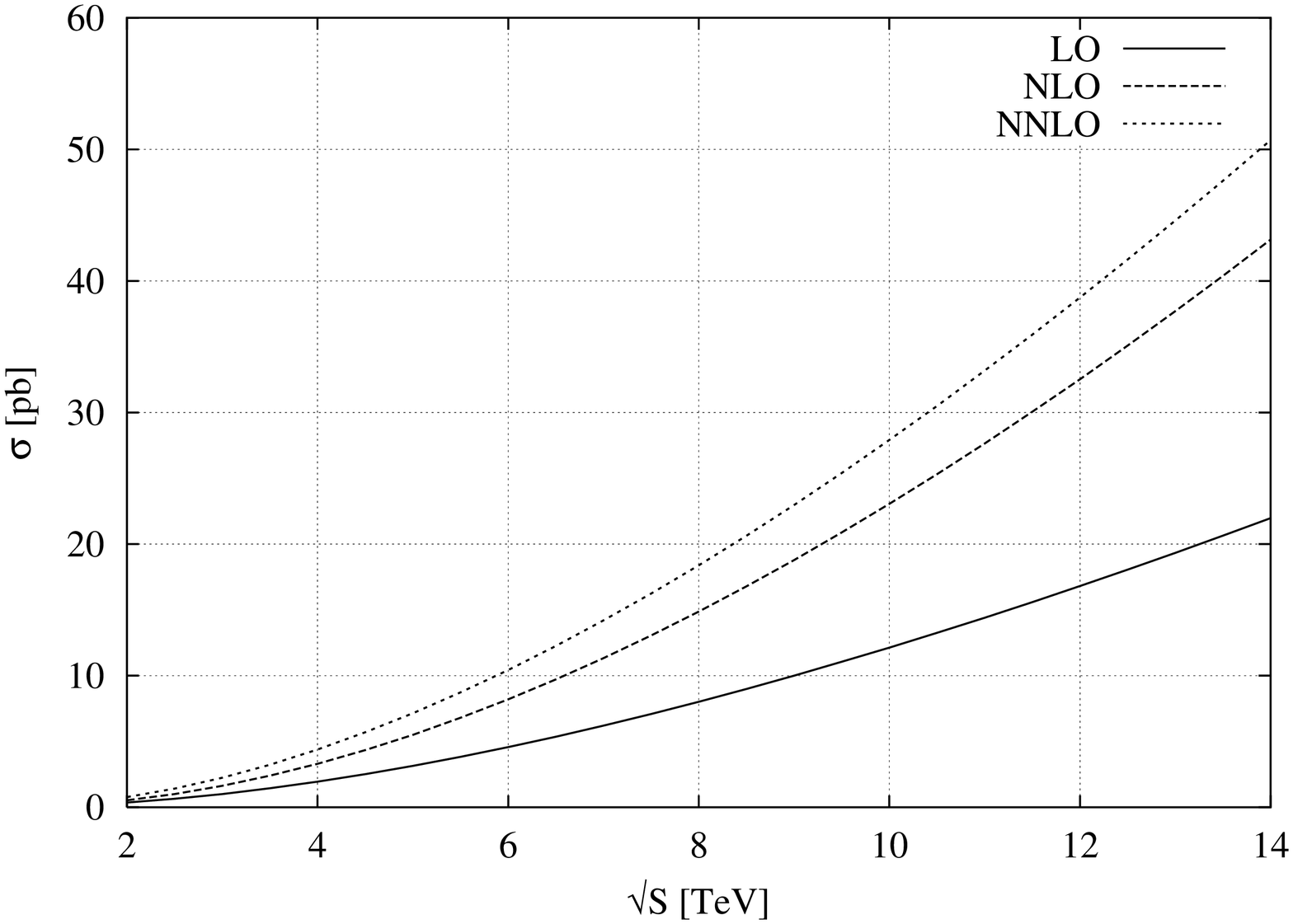}}
\subfigure[$C=1,\,\,k=2$]{\includegraphics[%
  width=6cm,
  angle=-90]{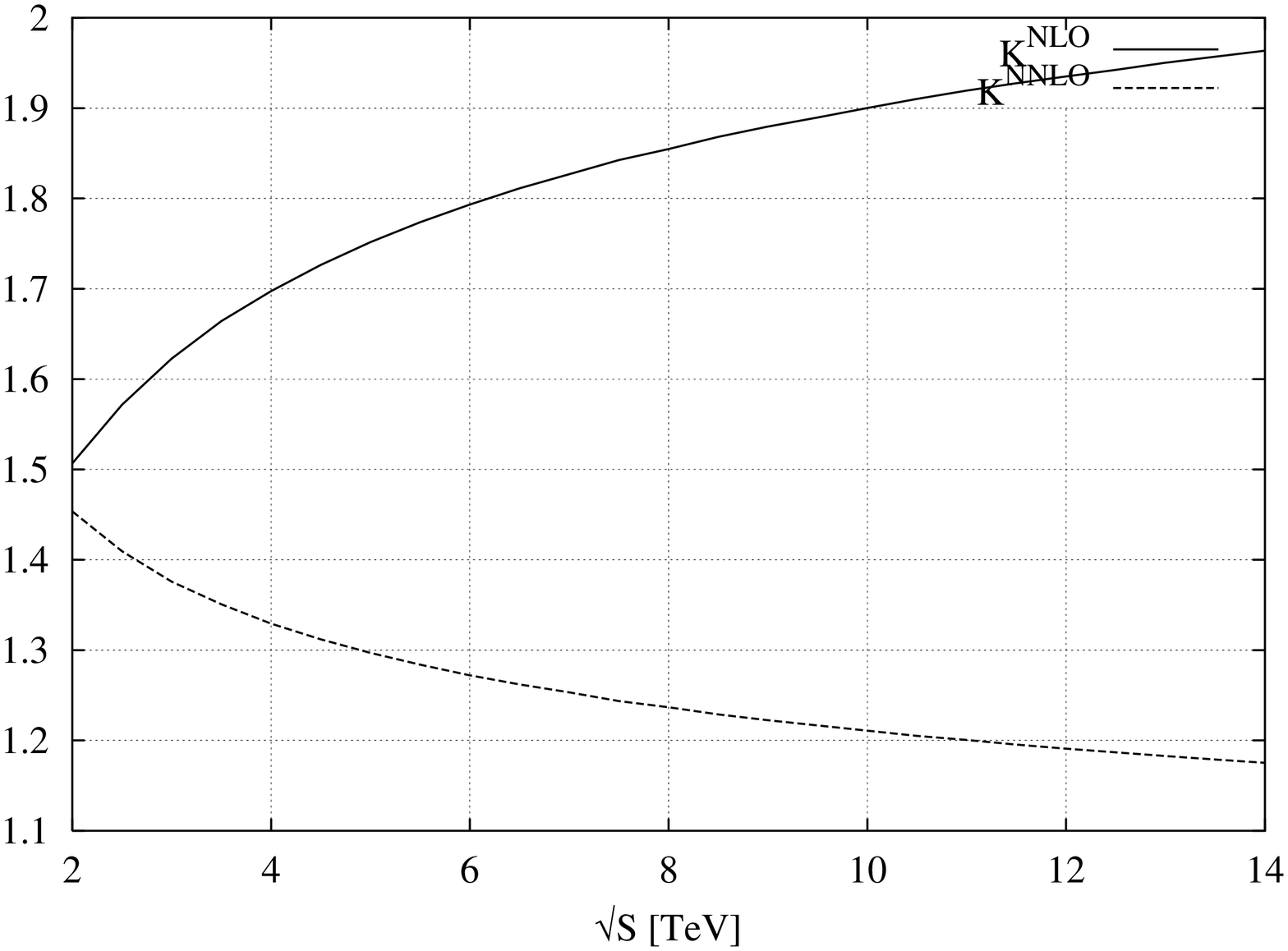}}
\subfigure[$C=2,\,\,k=2$]{\includegraphics[%
  width=6cm,
  angle=-90]{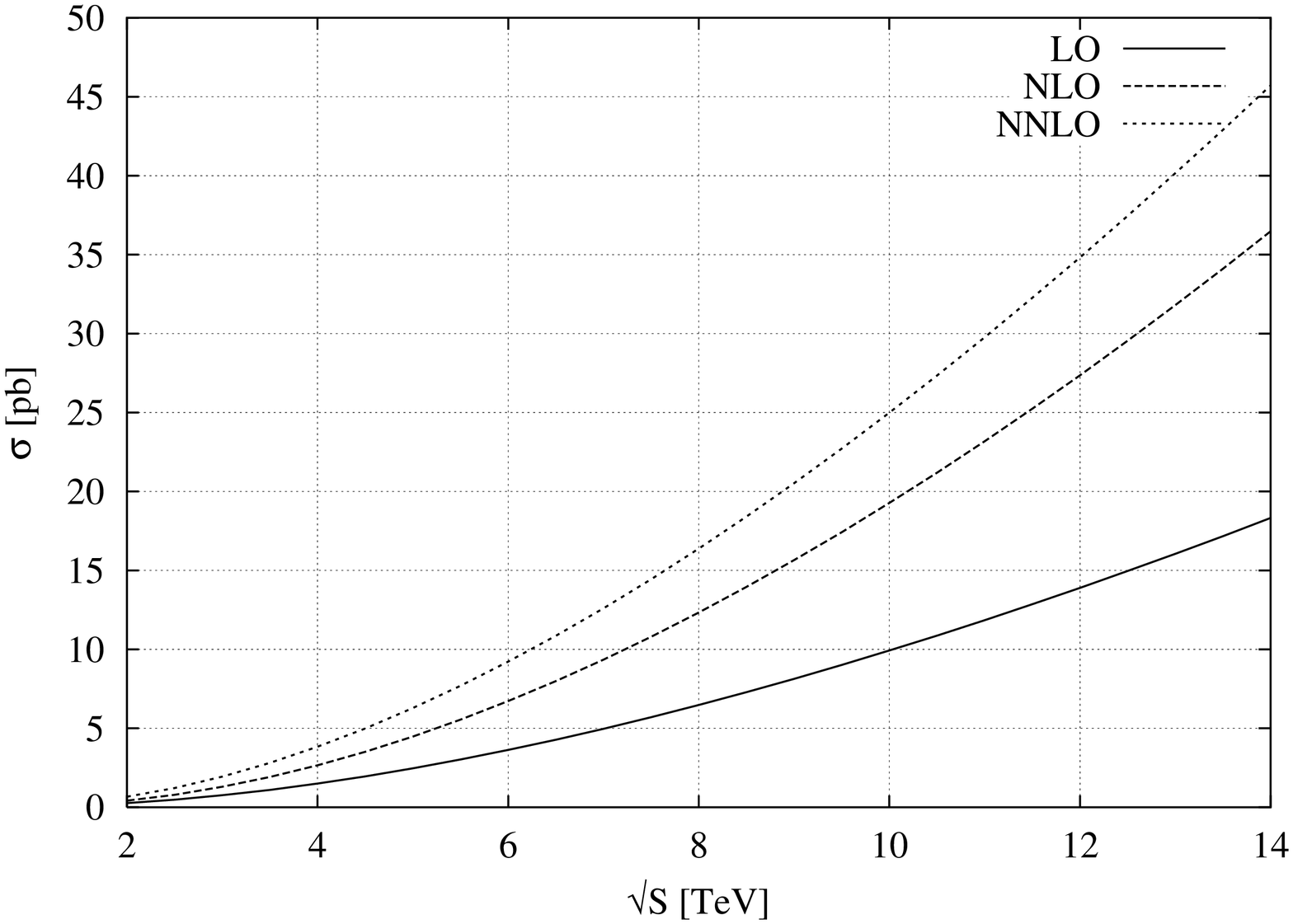}}
\subfigure[$C=2,\,\,k=2$]{\includegraphics[%
  width=6cm,
  angle=-90]{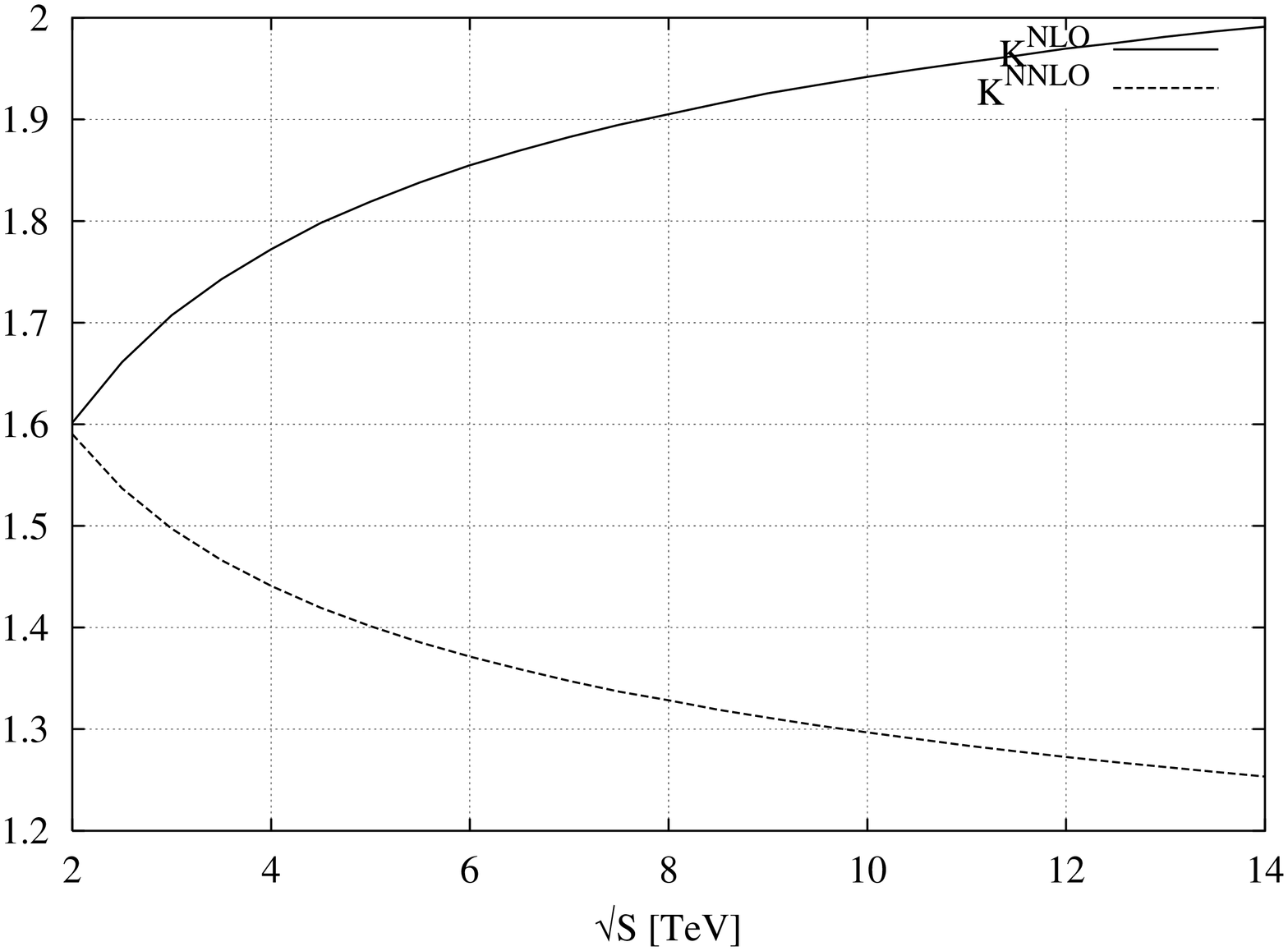}}
\subfigure[$C=1/2,\,\,k=2$]{\includegraphics[%
  width=6cm,
  angle=-90]{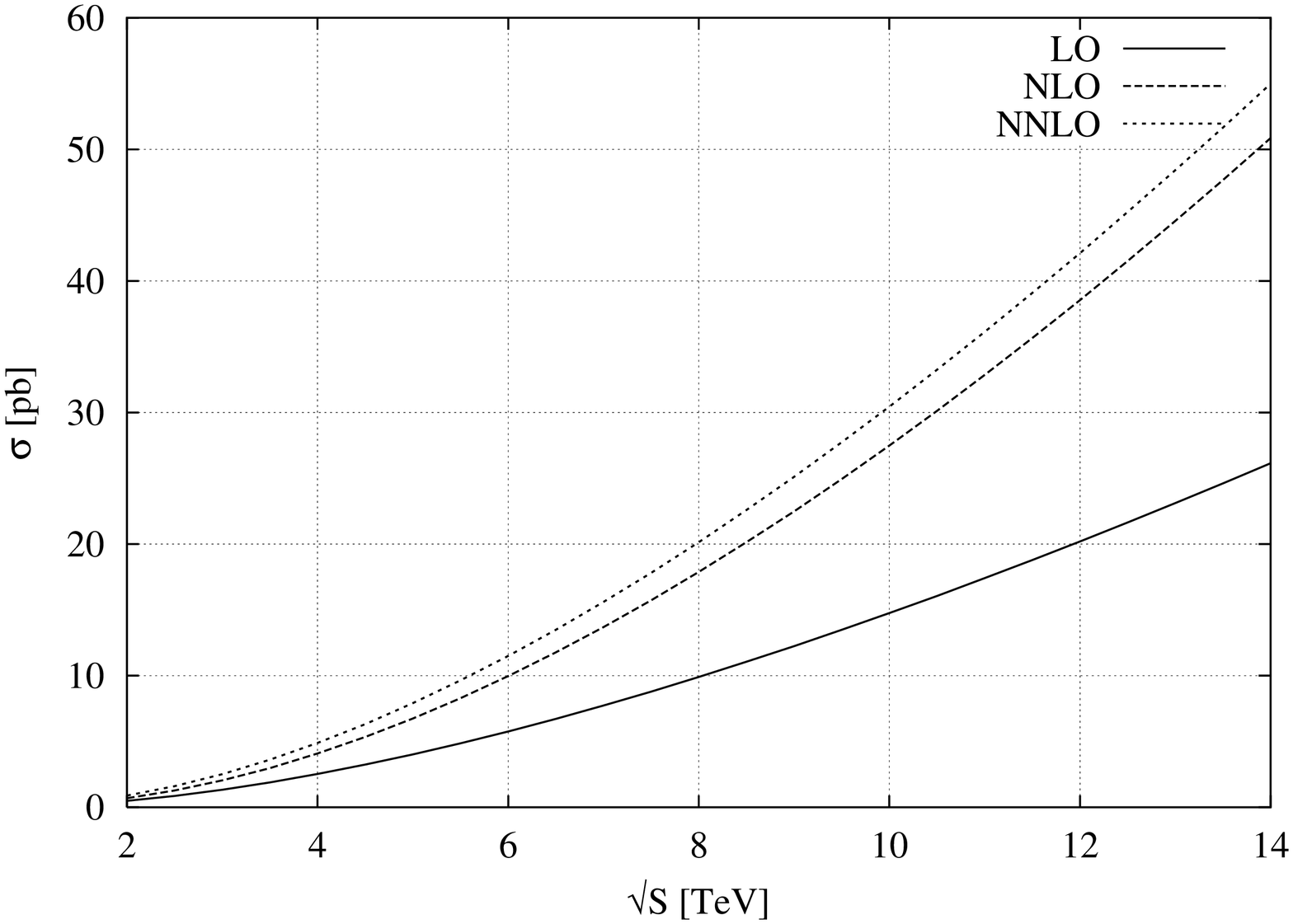}}
\subfigure[$C=1/2,\,\,k=2$]{\includegraphics[%
  width=6cm,
  angle=-90]{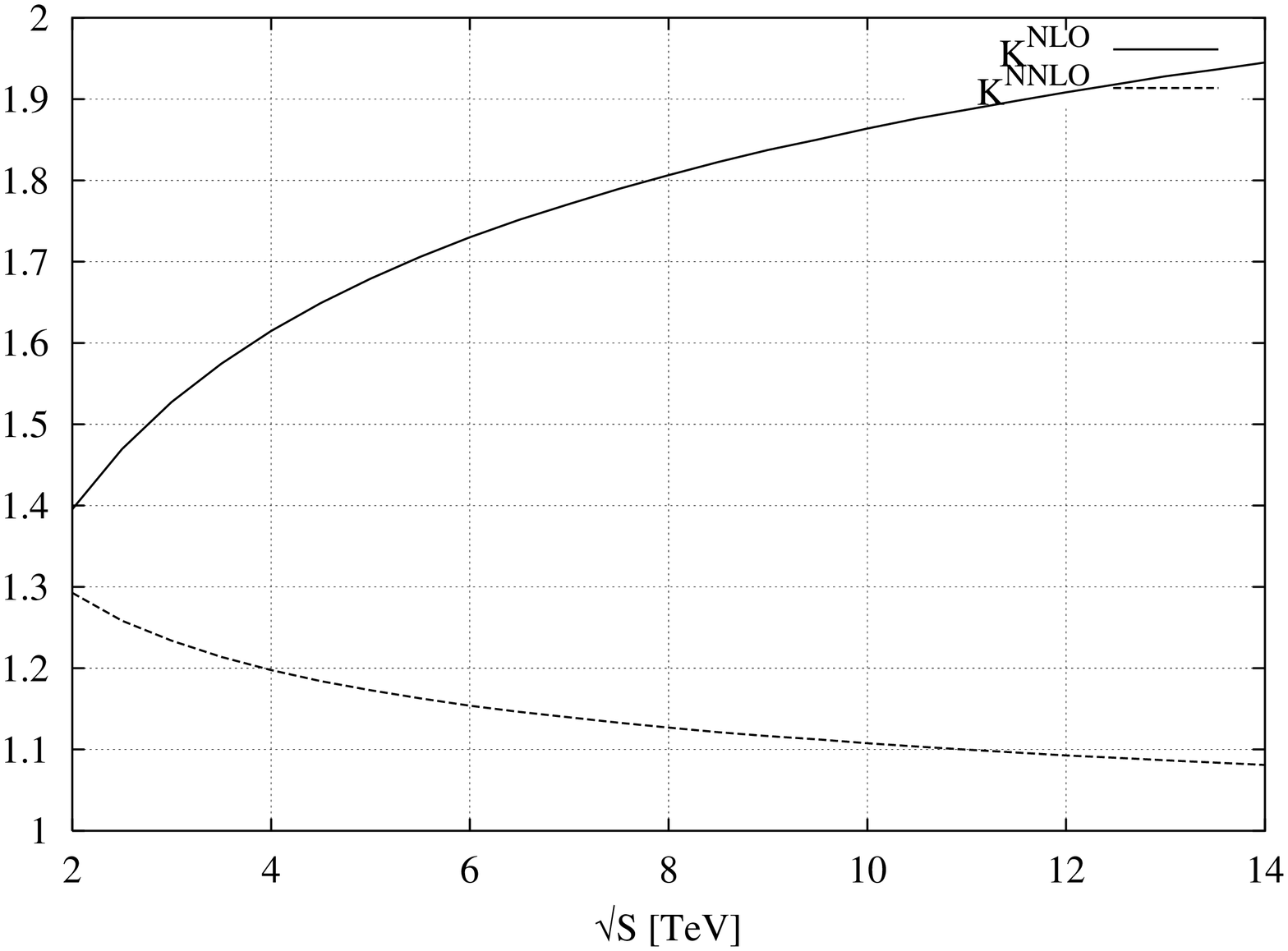}}
\caption{Cross sections and $K$-factors for the scalar Higgs
production at the LHC as a function of $\sqrt{S}$ with $\mu_F=C m_H$,
with $\mu_F^2=k\mu_R^2$ and $m_H=114$ GeV. MRST inputs have been used.}
\label{ener2}
\end{figure}

%%%%%%%%%%%%
\begin{figure}
\subfigure[$C=1,\,\,k=1/2$]{\includegraphics[%
  width=6cm,
  angle=-90]{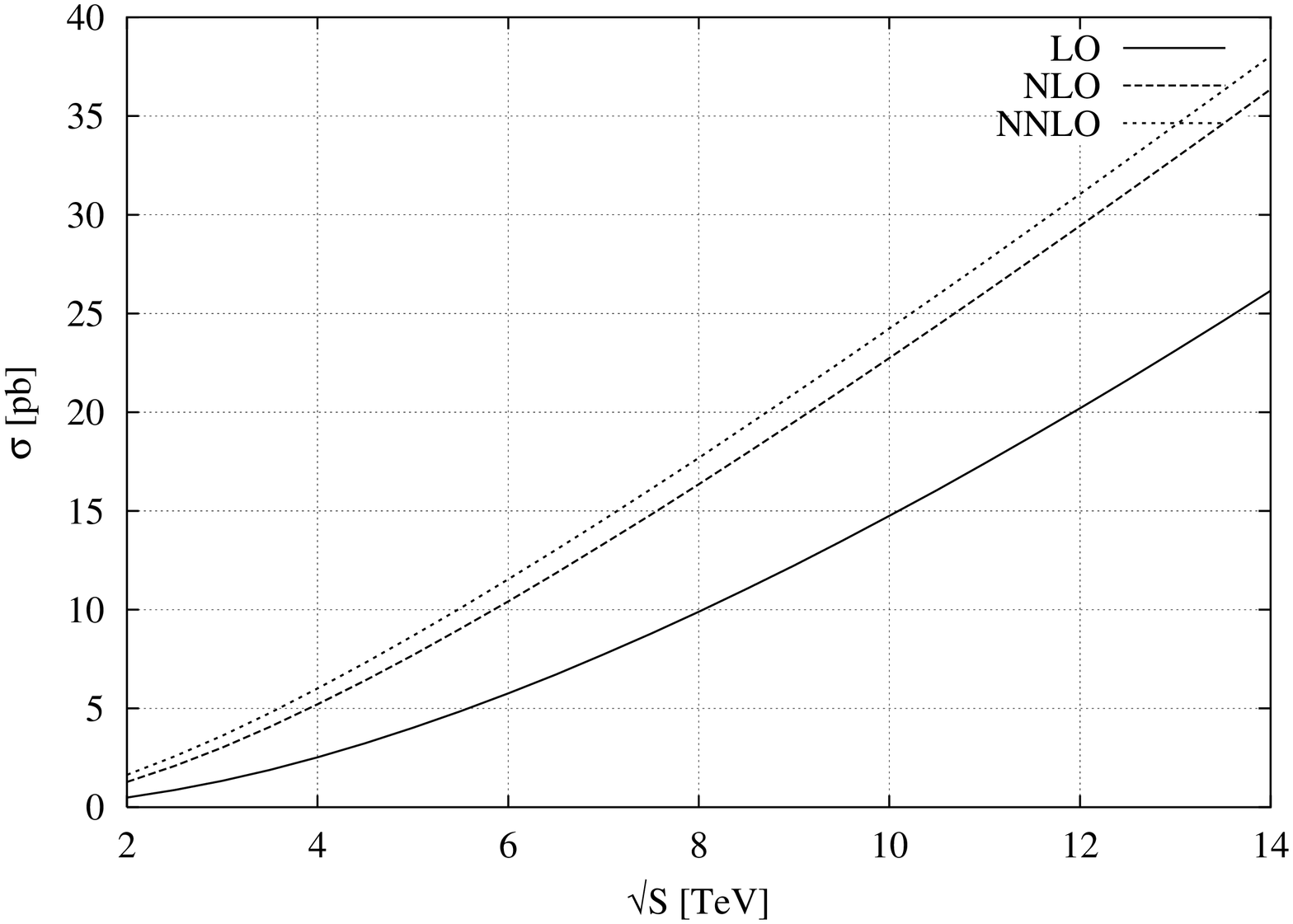}}
\subfigure[$C=1,\,\,k=1/2$]{\includegraphics[%
  width=6cm,
  angle=-90]{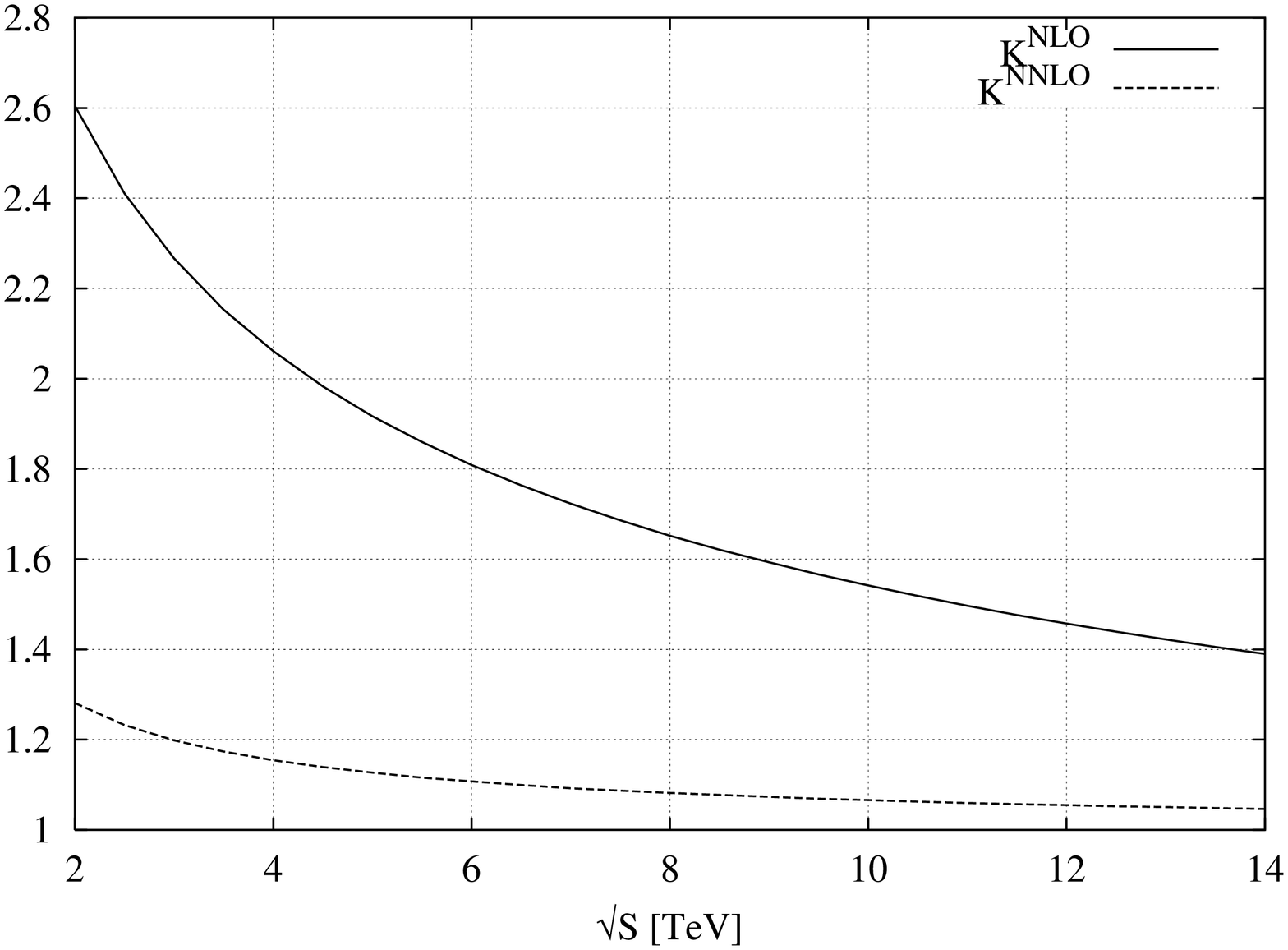}}
\subfigure[$C=2,\,\,k=1/2$]{\includegraphics[%
  width=6cm,
  angle=-90]{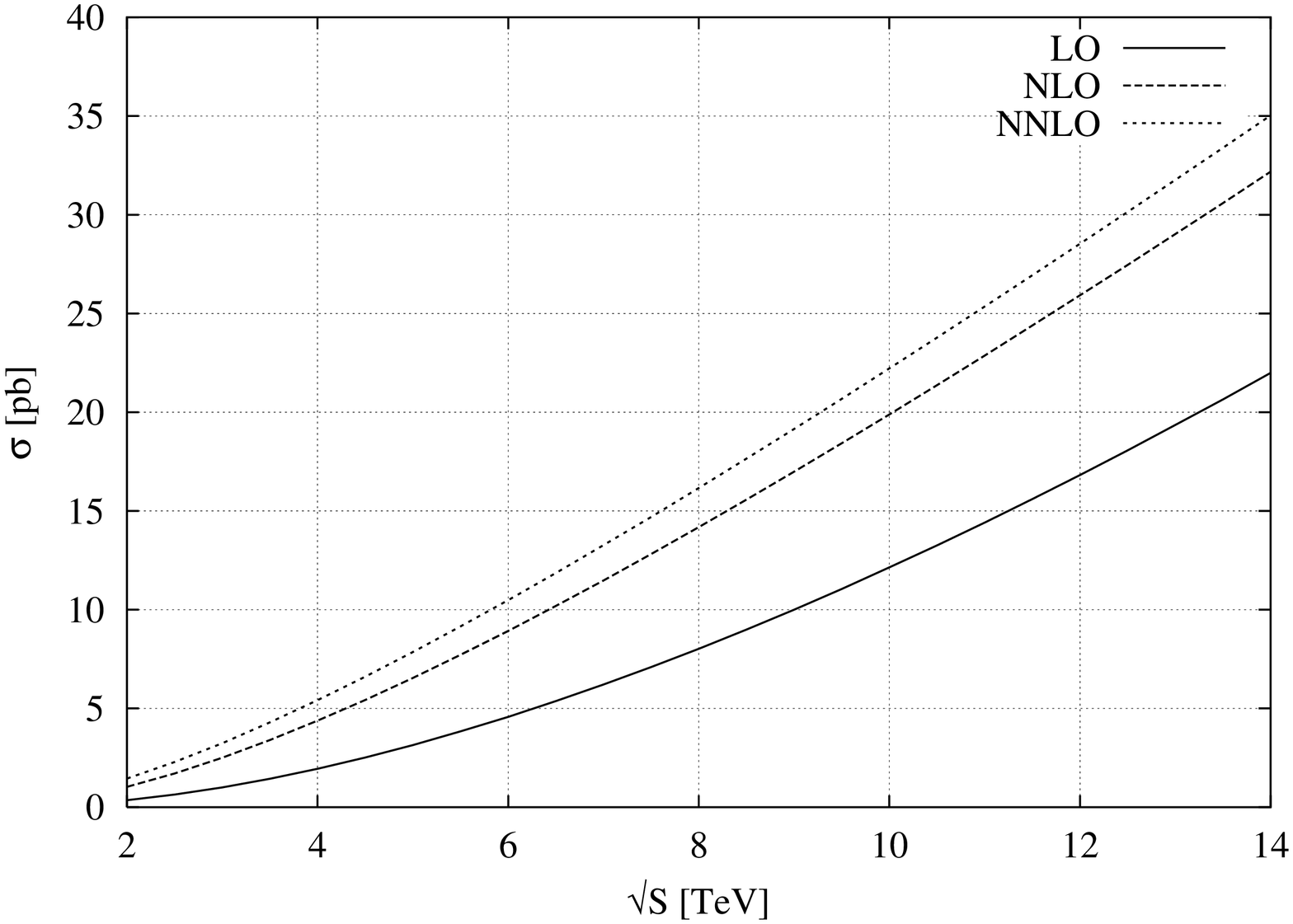}}
\subfigure[$C=2,\,\,k=1/2$]{\includegraphics[%
  width=6cm,
  angle=-90]{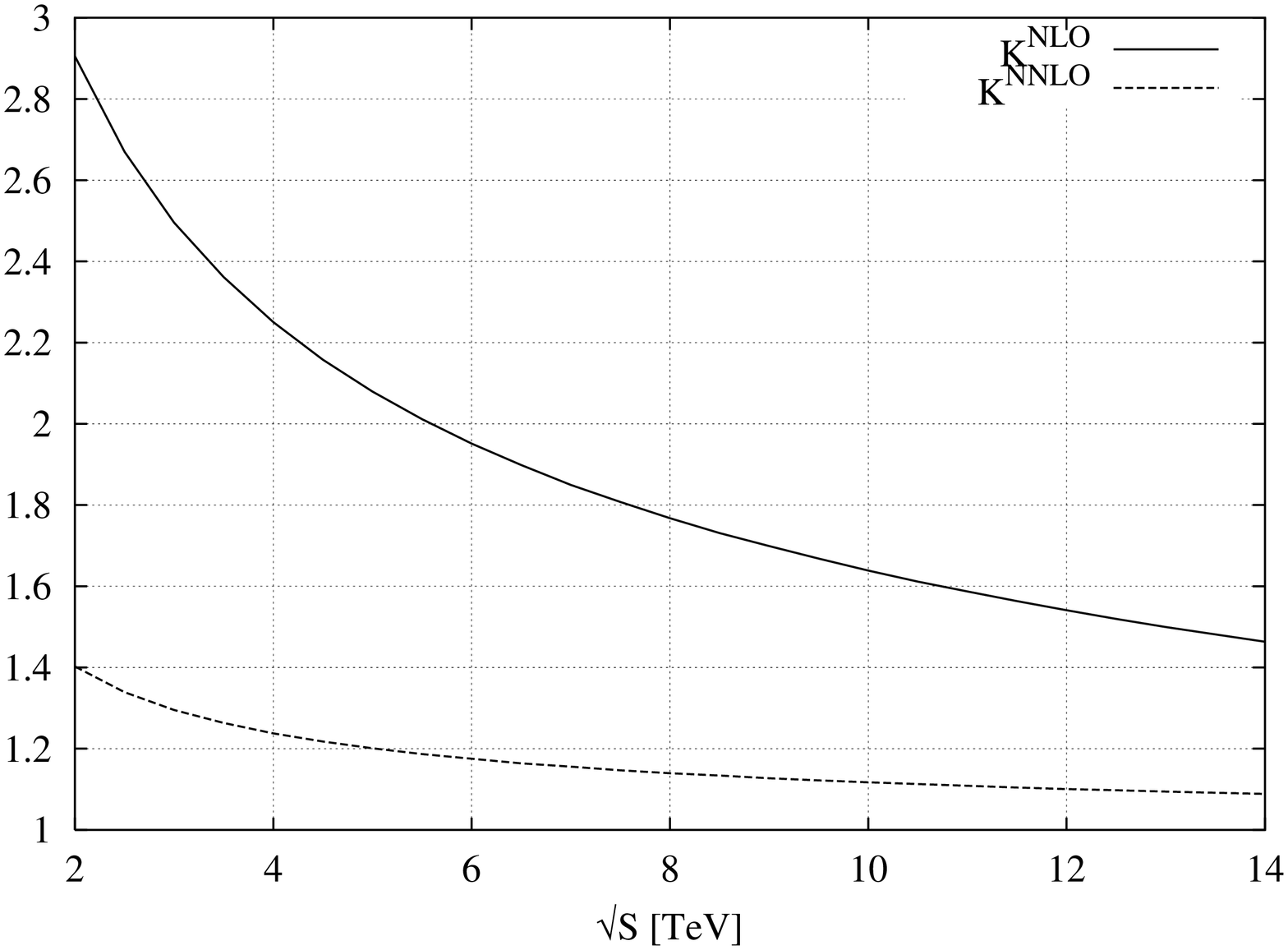}}
\subfigure[$C=1/2,\,\,k=1/2$]{\includegraphics[%
  width=6cm,
  angle=-90]{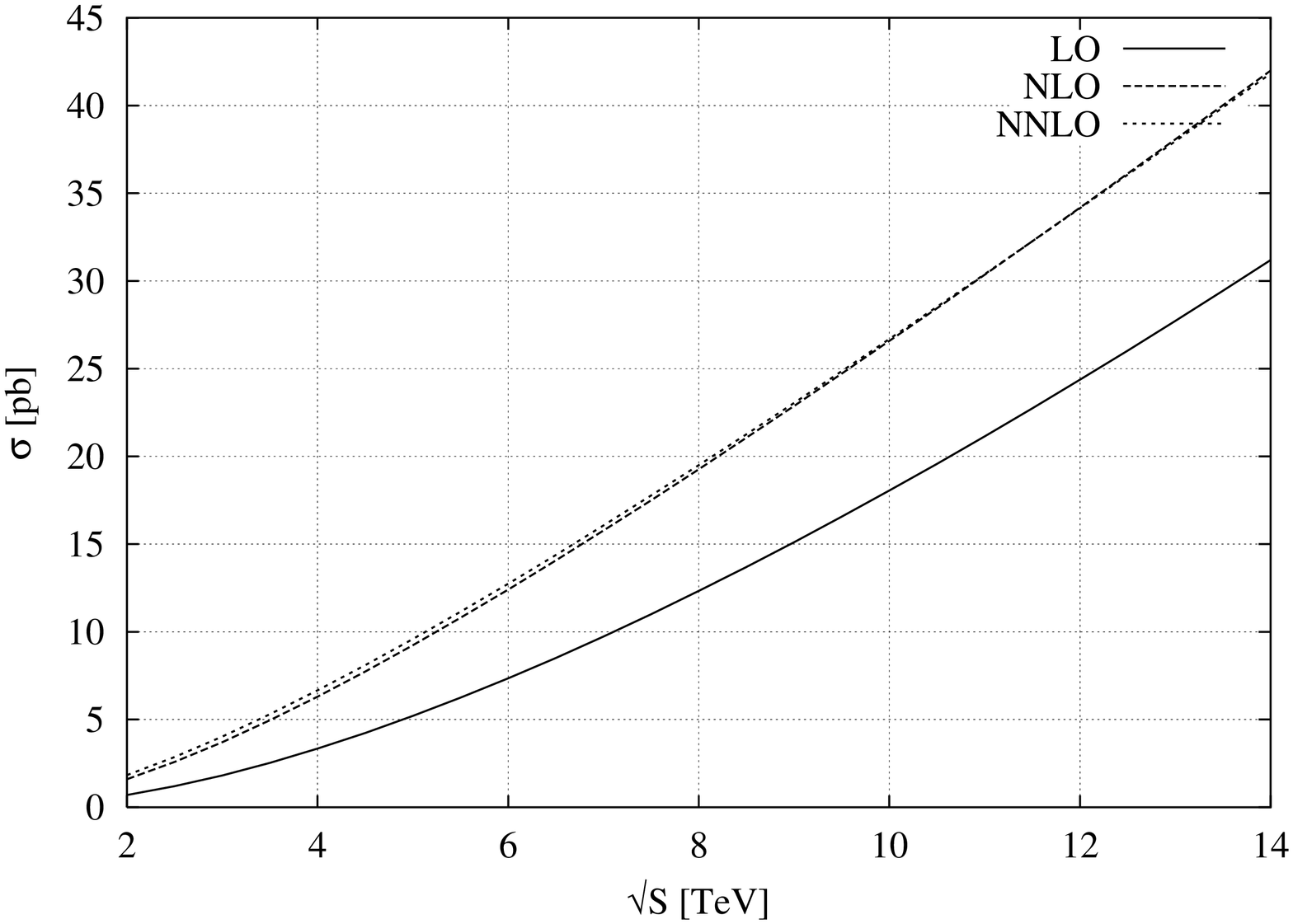}}
\subfigure[$C=1/2,\,\,k=1/2$]{\includegraphics[%
  width=6cm,
  angle=-90]{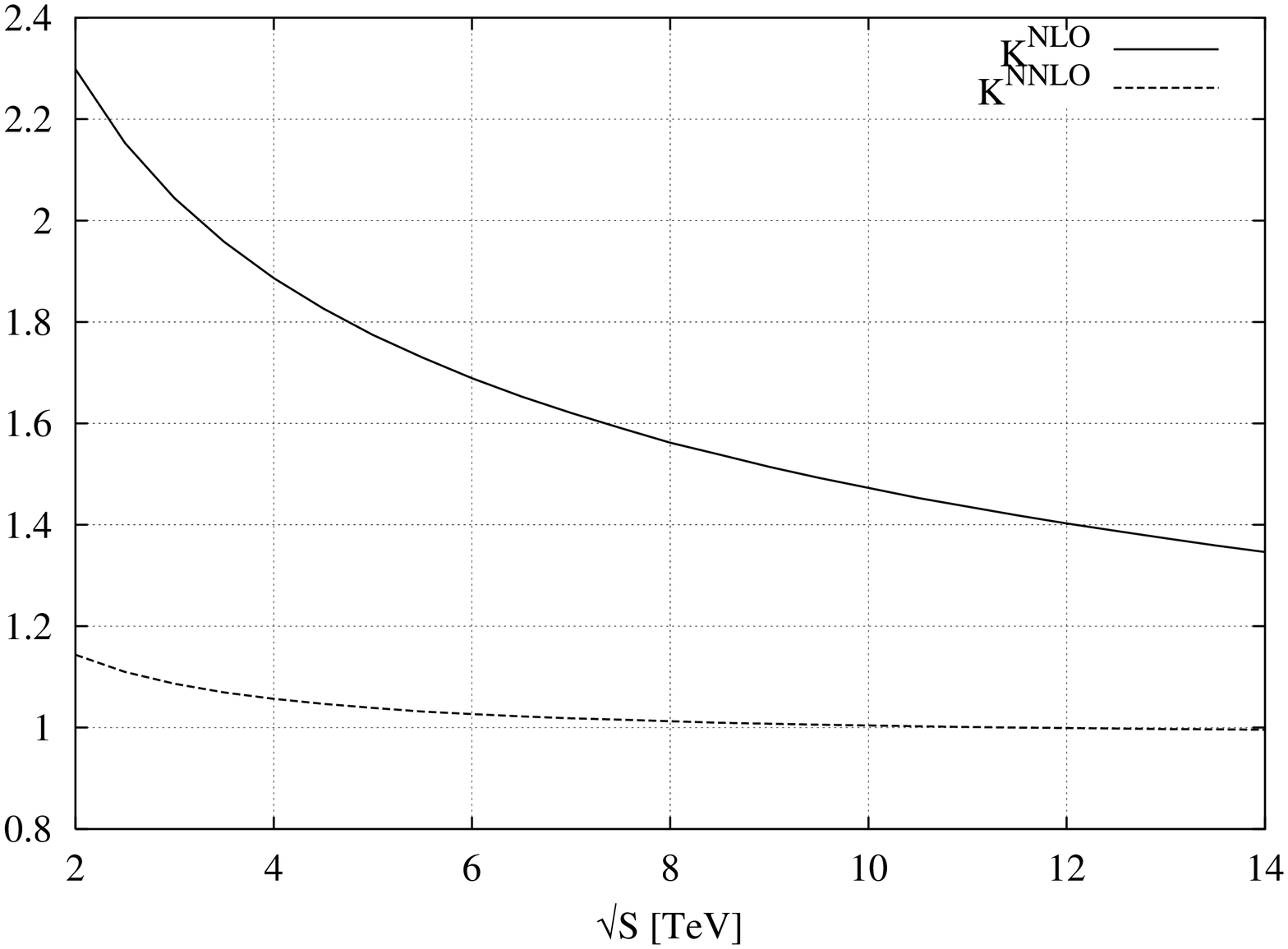}}
\caption{Cross sections and $K$-factors for the scalar Higgs
production at the LHC as a function of $\sqrt{S}$ with $\mu_F=C m_H$,
with $\mu_F^2=k\mu_R^2$ and $m_H=114$ GeV. MRST inputs have been used.}
\label{ener3}
\end{figure}

\newpage

%%%%%%%%%%%%%%%%%%%%%%%%%%%%%%%%%%%% Table 1_a  Alekhin %%%%%%%%%%%%%%%%%%%%%%%

\begin{table}
\begin{tabular}{|c|ccccc|}
\hline
$\sqrt{S}$&
$\sigma_{LO}$&
$\sigma_{NLO}$&
$\sigma_{NNLO}$&
$K^{NLO}$&
$K^{NNLO}$\tabularnewline
\hline
$2.0$&
$0.3981\pm0.0013$&
$0.861\pm0.002$&
$1.090\pm0.003$&
$2.162\pm0.009$&
$1.266\pm0.005$\tabularnewline
$2.5$&
$0.712\pm0.003$&
$1.499\pm0.004$&
$1.895\pm0.006$&
$2.106\pm0.011$&
$1.264\pm0.005$\tabularnewline
$3.0$&
$1.111\pm0.006$&
$2.291\pm0.007$&
$2.888\pm0.011$&
$2.062\pm0.013$&
$1.261\pm0.006$\tabularnewline
$3.5$&
$1.589\pm0.011$&
$3.222\pm0.010$&
$4.049\pm0.017$&
$2.028\pm0.015$&
$1.257\pm0.006$\tabularnewline
$4.0$&
$2.141\pm0.017$&
$4.279\pm0.015$&
$5.36\pm0.02$&
$1.999\pm0.018$&
$1.253\pm0.007$\tabularnewline
$4.5$&
$2.76\pm0.03$&
$5.45\pm0.02$&
$6.81\pm0.04$&
$1.974\pm0.020$&
$1.249\pm0.008$\tabularnewline
$5.0$&
$3.45\pm0.04$&
$6.73\pm0.03$&
$8.38\pm0.05$&
$1.95\pm0.02$&
$1.245\pm0.009$\tabularnewline
$5.5$&
$4.19\pm0.05$&
$8.10\pm0.04$&
$10.06\pm0.07$&
$1.93\pm0.02$&
$1.241\pm0.010$\tabularnewline
$6.0$&
$5.00\pm0.06$&
$9.57\pm0.05$&
$11.85\pm0.09$&
$1.92\pm0.03$&
$1.238\pm0.012$\tabularnewline
$6.5$&
$5.85\pm0.08$&
$11.12\pm0.07$&
$13.73\pm0.12$&
$1.90\pm0.03$&
$1.235\pm0.013$\tabularnewline
$7.0$&
$6.76\pm0.10$&
$12.74\pm0.09$&
$15.70\pm0.15$&
$1.89\pm0.03$&
$1.232\pm0.015$\tabularnewline
$7.5$&
$7.71\pm0.13$&
$14.44\pm0.11$&
$17.75\pm0.19$&
$1.87\pm0.03$&
$1.229\pm0.016$\tabularnewline
$8.0$&
$8.71\pm0.16$&
$16.21\pm0.14$&
$19.9\pm0.2$&
$1.86\pm0.04$&
$1.226\pm0.018$\tabularnewline
$8.5$&
$9.75\pm0.19$&
$18.04\pm0.18$&
$22.1\pm0.3$&
$1.85\pm0.04$&
$1.224\pm0.020$\tabularnewline
$9.0$&
$10.8\pm0.2$&
$19.9\pm0.2$&
$24.4\pm0.3$&
$1.84\pm0.04$&
$1.22\pm0.02$\tabularnewline
$9.5$&
$12.0\pm0.3$&
$21.9\pm0.3$&
$26.7\pm0.4$&
$1.83\pm0.04$&
$1.22\pm0.02$\tabularnewline
$10.0$&
$13.1\pm0.3$&
$23.9\pm0.3$&
$29.1\pm0.5$&
$1.82\pm0.05$&
$1.22\pm0.03$\tabularnewline
$10.5$&
$14.3\pm0.4$&
$25.9\pm0.4$&
$31.5\pm0.6$&
$1.81\pm0.05$&
$1.21\pm0.03$\tabularnewline
$11.0$&
$15.6\pm0.4$&
$28.1\pm0.4$&
$34.0\pm0.7$&
$1.80\pm0.06$&
$1.21\pm0.03$\tabularnewline
$11.5$&
$16.8\pm0.5$&
$30.2\pm0.5$&
$36.6\pm0.8$&
$1.80\pm0.06$&
$1.21\pm0.03$\tabularnewline
$12.0$&
$18.1\pm0.5$&
$32.4\pm0.6$&
$39.2\pm0.9$&
$1.79\pm0.06$&
$1.21\pm0.04$\tabularnewline
$12.5$&
$19.4\pm0.6$&
$34.6\pm0.7$&
$41.8\pm1.1$&
$1.78\pm0.07$&
$1.21\pm0.04$\tabularnewline
$13.0$&
$20.8\pm0.7$&
$36.9\pm0.8$&
$44.5\pm1.2$&
$1.77\pm0.07$&
$1.21\pm0.04$\tabularnewline
$13.5$&
$22.2\pm0.8$&
$39.2\pm0.9$&
$47.2\pm1.4$&
$1.77\pm0.07$&
$1.20\pm0.04$\tabularnewline
$14.0$&
$23.6\pm0.9$&
$41.6\pm1.0$&
$50.0\pm1.5$&
$1.76\pm0.08$&
$1.20\pm0.05$\tabularnewline
\hline
\end{tabular}
\caption{Values of the cross sections and $K$-factors for the scalar Higgs
production at the LHC as a function of $\sqrt{S}$ with $\mu_F=m_H$,
with $\mu_F^2=\mu_R^2$ and $m_H=114$ GeV for Alekhin, with errors.}
\label{table1a}
\end{table}

%%%%%%%%%%%%%%%%%%%%%% Table 2_a %%%%%%%%%%%%%%%%%%%%%%%%%%

\begin{table}
\begin{center}
\begin{tabular}{|c|ccccc|}
\hline
$\sqrt{S}$&
$\sigma_{LO}$&
$\sigma_{NLO}$&
$\sigma_{NNLO}$&
$K^{NLO}$&
$K^{NNLO}$\tabularnewline
\hline
$2.0$&
$0.4155$&
$0.8670$&
$1.242$&
$2.087$&
$1.433$\tabularnewline
$2.5$&
$0.7410$&
$1.521$&
$2.084$&
$2.053$&
$1.370$\tabularnewline
$3.0$&
$1.153$&
$2.335$&
$3.093$&
$2.025$&
$1.325$\tabularnewline
$3.5$&
$1.645$&
$3.292$&
$4.248$&
$2.001$&
$1.290$\tabularnewline
$4.0$&
$2.212$&
$4.380$&
$5.529$&
$1.980$&
$1.262$\tabularnewline
$4.5$&
$2.847$&
$5.587$&
$6.924$&
$1.962$&
$1.239$\tabularnewline
$5.0$&
$3.547$&
$6.903$&
$8.419$&
$1.946$&
$1.220$\tabularnewline
$5.5$&
$4.308$&
$8.318$&
$10.01$&
$1.931$&
$1.203$\tabularnewline
$6.0$&
$5.125$&
$9.826$&
$11.68$&
$1.917$&
$1.189$\tabularnewline
$6.5$&
$5.995$&
$11.42$&
$13.42$&
$1.905$&
$1.175$\tabularnewline
$7.0$&
$6.916$&
$13.09$&
$15.23$&
$1.893$&
$1.163$\tabularnewline
$7.5$&
$7.885$&
$14.84$&
$17.11$&
$1.882$&
$1.153$\tabularnewline
$8.0$&
$8.899$&
$16.66$&
$19.05$&
$1.872$&
$1.143$\tabularnewline
$8.5$&
$9.956$&
$18.55$&
$21.04$&
$1.863$&
$1.134$\tabularnewline
$9.0$&
$11.05$&
$20.49$&
$23.09$&
$1.854$&
$1.127$\tabularnewline
$9.5$&
$12.19$&
$22.50$&
$25.18$&
$1.846$&
$1.119$\tabularnewline
$10.0$&
$13.37$&
$24.56$&
$27.31$&
$1.837$&
$1.112$\tabularnewline
$10.5$&
$14.58$&
$26.68$&
$29.49$&
$1.830$&
$1.105$\tabularnewline
$11.0$&
$15.83$&
$28.85$&
$31.71$&
$1.822$&
$1.099$\tabularnewline
$11.5$&
$17.11$&
$31.06$&
$33.97$&
$1.815$&
$1.094$\tabularnewline
$12.0$&
$18.42$&
$33.32$&
$36.26$&
$1.809$&
$1.088$\tabularnewline
$12.5$&
$19.76$&
$35.62$&
$38.59$&
$1.803$&
$1.083$\tabularnewline
$13.0$&
$21.13$&
$37.97$&
$40.95$&
$1.797$&
$1.078$\tabularnewline
$13.5$&
$22.53$&
$40.36$&
$43.33$&
$1.791$&
$1.074$\tabularnewline
$14.0$&
$23.96$&
$42.78$&
$45.75$&
$1.785$&
$1.069$\tabularnewline
\hline
\end{tabular}
\end{center}
\caption{Values of the cross sections and $K$-factors for the scalar Higgs
production at the LHC as a function of $\sqrt{S}$ with $\mu_F=m_H$,
with $\mu_F^2=\mu_R^2$ and $m_H=114$ GeV. MRST inputs have been used.}
\end{table}

%%%%%%%%%%%%%%%%%%%%%%%%%%%%%%%%%%  Table 1_b    %%%%%%%%%%%%%%%%%%%%%%%

\begin{table}
\begin{center}
\begin{tabular}{|c|ccccc|}
\hline
$\sqrt{S}$&
$\sigma_{LO}$&
$\sigma_{NLO}$&
$\sigma_{NNLO}$&
$K^{NLO}$&
$K^{NNLO}$\tabularnewline
\hline
$2.0$&
$0.3029$&
$0.6871$&
$1.081$&
$2.268$&
$1.573$\tabularnewline
$2.5$&
$0.5508$&
$1.221$&
$1.830$&
$2.217$&
$1.499$\tabularnewline
$3.0$&
$0.8700$&
$1.893$&
$2.733$&
$2.176$&
$1.444$\tabularnewline
$3.5$&
$1.256$&
$2.690$&
$3.771$&
$2.142$&
$1.402$\tabularnewline
$4.0$&
$1.706$&
$3.602$&
$4.928$&
$2.111$&
$1.368$\tabularnewline
$4.5$&
$2.216$&
$4.619$&
$6.191$&
$2.084$&
$1.340$\tabularnewline
$5.0$&
$2.781$&
$5.733$&
$7.549$&
$2.061$&
$1.317$\tabularnewline
$5.5$&
$3.400$&
$6.937$&
$8.993$&
$2.040$&
$1.296$\tabularnewline
$6.0$&
$4.069$&
$8.225$&
$10.52$&
$2.021$&
$1.279$\tabularnewline
$6.5$&
$4.786$&
$9.590$&
$12.11$&
$2.004$&
$1.263$\tabularnewline
$7.0$&
$5.548$&
$11.03$&
$13.77$&
$1.988$&
$1.248$\tabularnewline
$7.5$&
$6.354$&
$12.53$&
$15.49$&
$1.972$&
$1.236$\tabularnewline
$8.0$&
$7.201$&
$14.11$&
$17.27$&
$1.959$&
$1.224$\tabularnewline
$8.5$&
$8.088$&
$15.74$&
$19.10$&
$1.946$&
$1.213$\tabularnewline
$9.0$&
$9.012$&
$17.43$&
$20.98$&
$1.934$&
$1.204$\tabularnewline
$9.5$&
$9.974$&
$19.17$&
$22.90$&
$1.922$&
$1.195$\tabularnewline
$10.0$&
$10.97$&
$20.97$&
$24.87$&
$1.912$&
$1.186$\tabularnewline
$10.5$&
$12.00$&
$22.82$&
$26.88$&
$1.902$&
$1.178$\tabularnewline
$11.0$&
$13.06$&
$24.71$&
$28.93$&
$1.892$&
$1.171$\tabularnewline
$11.5$&
$14.16$&
$26.65$&
$31.01$&
$1.882$&
$1.164$\tabularnewline
$12.0$&
$15.28$&
$28.63$&
$33.13$&
$1.874$&
$1.157$\tabularnewline
$12.5$&
$16.44$&
$30.65$&
$35.28$&
$1.864$&
$1.151$\tabularnewline
$13.0$&
$17.62$&
$32.71$&
$37.47$&
$1.856$&
$1.146$\tabularnewline
$13.5$&
$18.83$&
$34.81$&
$39.68$&
$1.849$&
$1.140$\tabularnewline
$14.0$&
$20.07$&
$36.95$&
$41.92$&
$1.841$&
$1.135$\tabularnewline
\hline
\end{tabular}
\end{center}
\caption{Values of the cross sections and $K$-factors for the scalar Higgs
production at the LHC as a function of $\sqrt{S}$ with $\mu_F=2 m_H$,
with $\mu_F^2=\mu_R^2$ and $m_H=114$ GeV. MRST inputs have been used.}
\end{table}

%%%%%%%%%%%%%%%%%%%%%%%%%%%%%%%%%%  Table 1_c    %%%%%%%%%%%%%%%%%%%%%%%
\begin{table}
\begin{center}
\begin{tabular}{|c|ccccc|}
\hline
$\sqrt{S}$&
$\sigma_{LO}$&
$\sigma_{NLO}$&
$\sigma_{NNLO}$&
$K^{NLO}$&
$K^{NNLO}$\tabularnewline
\hline
$2.0$&
$0.5817$&
$1.098$&
$1.396$&
$1.888$&
$1.271$\tabularnewline
$2.5$&
$1.014$&
$1.904$&
$2.328$&
$1.878$&
$1.223$\tabularnewline
$3.0$&
$1.551$&
$2.896$&
$3.440$&
$1.867$&
$1.188$\tabularnewline
$3.5$&
$2.182$&
$4.054$&
$4.707$&
$1.858$&
$1.161$\tabularnewline
$4.0$&
$2.899$&
$5.362$&
$6.111$&
$1.850$&
$1.140$\tabularnewline
$4.5$&
$3.695$&
$6.804$&
$7.635$&
$1.841$&
$1.122$\tabularnewline
$5.0$&
$4.563$&
$8.369$&
$9.266$&
$1.834$&
$1.107$\tabularnewline
$5.5$&
$5.498$&
$10.05$&
$10.99$&
$1.828$&
$1.094$\tabularnewline
$6.0$&
$6.496$&
$11.83$&
$12.81$&
$1.821$&
$1.083$\tabularnewline
$6.5$&
$7.552$&
$13.70$&
$14.71$&
$1.814$&
$1.074$\tabularnewline
$7.0$&
$8.662$&
$15.67$&
$16.67$&
$1.809$&
$1.064$\tabularnewline
$7.5$&
$9.824$&
$17.71$&
$18.71$&
$1.803$&
$1.056$\tabularnewline
$8.0$&
$11.03$&
$19.84$&
$20.81$&
$1.799$&
$1.049$\tabularnewline
$8.5$&
$12.29$&
$22.03$&
$22.97$&
$1.793$&
$1.043$\tabularnewline
$9.0$&
$13.59$&
$24.30$&
$25.18$&
$1.788$&
$1.036$\tabularnewline
$9.5$&
$14.93$&
$26.63$&
$27.44$&
$1.784$&
$1.030$\tabularnewline
$10.0$&
$16.31$&
$29.02$&
$29.75$&
$1.779$&
$1.025$\tabularnewline
$10.5$&
$17.72$&
$31.46$&
$32.10$&
$1.775$&
$1.020$\tabularnewline
$11.0$&
$19.18$&
$33.96$&
$34.50$&
$1.771$&
$1.016$\tabularnewline
$11.5$&
$20.66$&
$36.51$&
$36.93$&
$1.767$&
$1.012$\tabularnewline
$12.0$&
$22.18$&
$39.11$&
$39.40$&
$1.763$&
$1.007$\tabularnewline
$12.5$&
$23.73$&
$41.76$&
$41.91$&
$1.760$&
$1.004$\tabularnewline
$13.0$&
$25.31$&
$44.45$&
$44.45$&
$1.756$&
$1.000$\tabularnewline
$13.5$&
$26.92$&
$47.19$&
$47.02$&
$1.753$&
$0.9964$\tabularnewline
$14.0$&
$28.55$&
$49.96$&
$49.62$&
$1.750$&
$0.9932$\tabularnewline
\hline
\end{tabular}
\end{center}
\caption{Values of the cross sections and $K$-factors for the scalar Higgs
production at the LHC as a function of $\sqrt{S}$ with $\mu_F=(1/2) m_H$,
with $\mu_F^2=\mu_R^2$ and $m_H=114$ GeV. MRST inputs have been used.}
\end{table}

%%%%%%%%%%%%%%%%%%%%%%%%%%%%%%%%%%  Table 2_a    %%%%%%%%%%%%%%%%%%%%%%%
\begin{table}
\begin{center}
\begin{tabular}{|c|ccccc|}
\hline
$\sqrt{S}$&
$\sigma_{LO}$&
$\sigma_{NLO}$&
$\sigma_{NNLO}$&
$K^{NLO}$&
$K^{NNLO}$\tabularnewline
\hline
$2.0$&
$0.3546$&
$0.5343$&
$0.7766$&
$1.507$&
$1.453$\tabularnewline
$2.5$&
$0.6388$&
$1.004$&
$1.415$&
$1.572$&
$1.409$\tabularnewline
$3.0$&
$1.002$&
$1.626$&
$2.237$&
$1.623$&
$1.376$\tabularnewline
$3.5$&
$1.438$&
$2.393$&
$3.232$&
$1.664$&
$1.351$\tabularnewline
$4.0$&
$1.944$&
$3.300$&
$4.387$&
$1.698$&
$1.329$\tabularnewline
$4.5$&
$2.514$&
$4.340$&
$5.694$&
$1.726$&
$1.312$\tabularnewline
$5.0$&
$3.144$&
$5.507$&
$7.142$&
$1.752$&
$1.297$\tabularnewline
$5.5$&
$3.831$&
$6.795$&
$8.724$&
$1.774$&
$1.284$\tabularnewline
$6.0$&
$4.572$&
$8.199$&
$10.43$&
$1.793$&
$1.272$\tabularnewline
$6.5$&
$5.363$&
$9.714$&
$12.26$&
$1.811$&
$1.262$\tabularnewline
$7.0$&
$6.202$&
$11.33$&
$14.20$&
$1.827$&
$1.253$\tabularnewline
$7.5$&
$7.088$&
$13.06$&
$16.24$&
$1.843$&
$1.243$\tabularnewline
$8.0$&
$8.017$&
$14.87$&
$18.39$&
$1.855$&
$1.237$\tabularnewline
$8.5$&
$8.987$&
$16.79$&
$20.63$&
$1.868$&
$1.229$\tabularnewline
$9.0$&
$9.997$&
$18.79$&
$22.97$&
$1.880$&
$1.222$\tabularnewline
$9.5$&
$11.05$&
$20.88$&
$25.40$&
$1.890$&
$1.216$\tabularnewline
$10.0$&
$12.13$&
$23.05$&
$27.91$&
$1.900$&
$1.211$\tabularnewline
$10.5$&
$13.25$&
$25.31$&
$30.50$&
$1.910$&
$1.205$\tabularnewline
$11.0$&
$14.40$&
$27.64$&
$33.18$&
$1.919$&
$1.200$\tabularnewline
$11.5$&
$15.59$&
$30.05$&
$35.92$&
$1.928$&
$1.195$\tabularnewline
$12.0$&
$16.81$&
$32.53$&
$38.74$&
$1.935$&
$1.191$\tabularnewline
$12.5$&
$18.06$&
$35.08$&
$41.63$&
$1.942$&
$1.187$\tabularnewline
$13.0$&
$19.33$&
$37.70$&
$44.59$&
$1.950$&
$1.183$\tabularnewline
$13.5$&
$20.64$&
$40.39$&
$47.61$&
$1.957$&
$1.179$\tabularnewline
$14.0$&
$21.97$&
$43.14$&
$50.70$&
$1.964$&
$1.175$\tabularnewline
\hline
\end{tabular}
\end{center}
\caption{Values of the cross sections and $K$-factors for the scalar Higgs
production at the LHC as a function of $\sqrt{S}$ with $\mu_F=m_H$,
with $\mu_F^2=2\mu_R^2$ and $m_H=114$ GeV. MRST inputs have been used.}
\end{table}

\newpage
%%%%%%%%%%%%%%%%%%%%%%%%%%%%%%%%%%  Table 2_b    %%%%%%%%%%%%%%%%%%%%%%%

\begin{table}
\begin{center}
\begin{tabular}{|c|ccccc|}
\hline
$\sqrt{S}$&
$\sigma_{LO}$&
$\sigma_{NLO}$&
$\sigma_{NNLO}$&
$K^{NLO}$&
$K^{NNLO}$\tabularnewline
\hline
$2.0$&
$0.2597$&
$0.4159$&
$0.6614$&
$1.601$&
$1.590$\tabularnewline
$2.5$&
$0.4763$&
$0.7912$&
$1.216$&
$1.661$&
$1.537$\tabularnewline
$3.0$&
$0.7574$&
$1.293$&
$1.936$&
$1.707$&
$1.497$\tabularnewline
$3.5$&
$1.100$&
$1.917$&
$2.811$&
$1.743$&
$1.466$\tabularnewline
$4.0$&
$1.501$&
$2.660$&
$3.833$&
$1.772$&
$1.441$\tabularnewline
$4.5$&
$1.956$&
$3.517$&
$4.992$&
$1.798$&
$1.419$\tabularnewline
$5.0$&
$2.464$&
$4.482$&
$6.281$&
$1.819$&
$1.401$\tabularnewline
$5.5$&
$3.021$&
$5.553$&
$7.693$&
$1.838$&
$1.385$\tabularnewline
$6.0$&
$3.625$&
$6.724$&
$9.221$&
$1.855$&
$1.371$\tabularnewline
$6.5$&
$4.275$&
$7.991$&
$10.86$&
$1.869$&
$1.359$\tabularnewline
$7.0$&
$4.967$&
$9.351$&
$12.60$&
$1.883$&
$1.347$\tabularnewline
$7.5$&
$5.700$&
$10.80$&
$14.44$&
$1.895$&
$1.337$\tabularnewline
$8.0$&
$6.472$&
$12.33$&
$16.38$&
$1.905$&
$1.328$\tabularnewline
$8.5$&
$7.282$&
$13.95$&
$18.40$&
$1.916$&
$1.319$\tabularnewline
$9.0$&
$8.127$&
$15.65$&
$20.52$&
$1.926$&
$1.311$\tabularnewline
$9.5$&
$9.008$&
$17.42$&
$22.71$&
$1.934$&
$1.304$\tabularnewline
$10.0$&
$9.923$&
$19.27$&
$24.99$&
$1.942$&
$1.297$\tabularnewline
$10.5$&
$10.87$&
$21.19$&
$27.34$&
$1.949$&
$1.290$\tabularnewline
$11.0$&
$11.85$&
$23.18$&
$29.76$&
$1.956$&
$1.284$\tabularnewline
$11.5$&
$12.86$&
$25.24$&
$32.26$&
$1.963$&
$1.278$\tabularnewline
$12.0$&
$13.89$&
$27.36$&
$34.82$&
$1.970$&
$1.273$\tabularnewline
$12.5$&
$14.96$&
$29.55$&
$37.45$&
$1.975$&
$1.267$\tabularnewline
$13.0$&
$16.05$&
$31.80$&
$40.15$&
$1.981$&
$1.263$\tabularnewline
$13.5$&
$17.17$&
$34.11$&
$42.91$&
$1.987$&
$1.258$\tabularnewline
$14.0$&
$18.32$&
$36.48$&
$45.72$&
$1.991$&
$1.253$\tabularnewline
\hline
\end{tabular}
\end{center}
\caption{Values of the cross sections and $K$-factors for the scalar Higgs
production at the LHC as a function of $\sqrt{S}$ with $\mu_F=2 m_H$,
with $\mu_F^2=2\mu_R^2$ and $m_H=114$ GeV. MRST inputs have been used.}
\end{table}

%%%%%%%%%%%%%%%%%%%%%%%%%%%%%%%%%%  Table 2_c    %%%%%%%%%%%%%%%%%%%%%%%
\begin{table}
\begin{center}
\begin{tabular}{|c|ccccc|}
\hline
$\sqrt{S}$&
$\sigma_{LO}$&
$\sigma_{NLO}$&
$\sigma_{NNLO}$&
$K^{NLO}$&
$K^{NNLO}$\tabularnewline
\hline
$2.0$&
$0.4899$&
$0.6836$&
$0.8836$&
$1.395$&
$1.293$\tabularnewline
$2.5$&
$0.8642$&
$1.270$&
$1.598$&
$1.470$&
$1.258$\tabularnewline
$3.0$&
$1.333$&
$2.036$&
$2.512$&
$1.527$&
$1.234$\tabularnewline
$3.5$&
$1.890$&
$2.976$&
$3.612$&
$1.575$&
$1.214$\tabularnewline
$4.0$&
$2.526$&
$4.079$&
$4.885$&
$1.615$&
$1.198$\tabularnewline
$4.5$&
$3.237$&
$5.338$&
$6.321$&
$1.649$&
$1.184$\tabularnewline
$5.0$&
$4.016$&
$6.743$&
$7.908$&
$1.679$&
$1.173$\tabularnewline
$5.5$&
$4.858$&
$8.288$&
$9.638$&
$1.706$&
$1.163$\tabularnewline
$6.0$&
$5.761$&
$9.966$&
$11.50$&
$1.730$&
$1.154$\tabularnewline
$6.5$&
$6.719$&
$11.77$&
$13.49$&
$1.752$&
$1.146$\tabularnewline
$7.0$&
$7.730$&
$13.69$&
$15.60$&
$1.771$&
$1.140$\tabularnewline
$7.5$&
$8.790$&
$15.73$&
$17.82$&
$1.790$&
$1.133$\tabularnewline
$8.0$&
$9.898$&
$17.88$&
$20.15$&
$1.806$&
$1.127$\tabularnewline
$8.5$&
$11.05$&
$20.14$&
$22.58$&
$1.823$&
$1.121$\tabularnewline
$9.0$&
$12.24$&
$22.49$&
$25.11$&
$1.837$&
$1.116$\tabularnewline
$9.5$&
$13.48$&
$24.94$&
$27.74$&
$1.850$&
$1.112$\tabularnewline
$10.0$&
$14.75$&
$27.49$&
$30.45$&
$1.864$&
$1.108$\tabularnewline
$10.5$&
$16.06$&
$30.13$&
$33.25$&
$1.876$&
$1.104$\tabularnewline
$11.0$&
$17.41$&
$32.85$&
$36.13$&
$1.887$&
$1.100$\tabularnewline
$11.5$&
$18.79$&
$35.66$&
$39.09$&
$1.898$&
$1.096$\tabularnewline
$12.0$&
$20.20$&
$38.55$&
$42.12$&
$1.908$&
$1.093$\tabularnewline
$12.5$&
$21.64$&
$41.51$&
$45.24$&
$1.918$&
$1.090$\tabularnewline
$13.0$&
$23.11$&
$44.56$&
$48.42$&
$1.928$&
$1.087$\tabularnewline
$13.5$&
$24.62$&
$47.67$&
$51.66$&
$1.936$&
$1.084$\tabularnewline
$14.0$&
$26.15$&
$50.86$&
$54.98$&
$1.945$&
$1.081$\tabularnewline
\hline
\end{tabular}
\end{center}
\caption{Values of the cross sections and $K$-factors for the scalar Higgs
production at the LHC as a function of $\sqrt{S}$ with $\mu_F=(1/2) m_H$,
with $\mu_F^2=2\mu_R^2$ and $m_H=114$ GeV. MRST inputs have been used.}
\end{table}

%%%%%%%%%%%%%%%%%%%%%%%%%%%%%%%%%%  Table 3_a    %%%%%%%%%%%%%%%%%%%%%%%
\begin{table}
\begin{center}
\begin{tabular}{|c|ccccc|}
\hline
$\sqrt{S}$&
$\sigma_{LO}$&
$\sigma_{NLO}$&
$\sigma_{NNLO}$&
$K^{NLO}$&
$K^{NNLO}$\tabularnewline
\hline
$2.0$&
$0.4899$&
$1.276$&
$1.635$&
$2.605$&
$1.281$\tabularnewline
$2.5$&
$0.8641$&
$2.083$&
$2.567$&
$2.411$&
$1.232$\tabularnewline
$3.0$&
$1.333$&
$3.022$&
$3.622$&
$2.267$&
$1.199$\tabularnewline
$3.5$&
$1.890$&
$4.069$&
$4.775$&
$2.153$&
$1.174$\tabularnewline
$4.0$&
$2.526$&
$5.206$&
$6.009$&
$2.061$&
$1.154$\tabularnewline
$4.5$&
$3.237$&
$6.419$&
$7.312$&
$1.983$&
$1.139$\tabularnewline
$5.0$&
$4.016$&
$7.698$&
$8.673$&
$1.917$&
$1.127$\tabularnewline
$5.5$&
$4.858$&
$9.034$&
$10.08$&
$1.860$&
$1.116$\tabularnewline
$6.0$&
$5.761$&
$10.42$&
$11.54$&
$1.809$&
$1.107$\tabularnewline
$6.5$&
$6.719$&
$11.85$&
$13.03$&
$1.764$&
$1.100$\tabularnewline
$7.0$&
$7.730$&
$13.32$&
$14.55$&
$1.723$&
$1.092$\tabularnewline
$7.5$&
$8.790$&
$14.82$&
$16.11$&
$1.686$&
$1.087$\tabularnewline
$8.0$&
$9.898$&
$16.35$&
$17.69$&
$1.652$&
$1.082$\tabularnewline
$8.5$&
$11.05$&
$17.91$&
$19.30$&
$1.621$&
$1.078$\tabularnewline
$9.0$&
$12.24$&
$19.50$&
$20.93$&
$1.593$&
$1.073$\tabularnewline
$9.5$&
$13.48$&
$21.11$&
$22.57$&
$1.566$&
$1.069$\tabularnewline
$10.0$&
$14.75$&
$22.74$&
$24.24$&
$1.542$&
$1.066$\tabularnewline
$10.5$&
$16.06$&
$24.39$&
$25.92$&
$1.519$&
$1.063$\tabularnewline
$11.0$&
$17.41$&
$26.06$&
$27.61$&
$1.497$&
$1.059$\tabularnewline
$11.5$&
$18.79$&
$27.74$&
$29.32$&
$1.476$&
$1.057$\tabularnewline
$12.0$&
$20.20$&
$29.44$&
$31.05$&
$1.457$&
$1.055$\tabularnewline
$12.5$&
$21.64$&
$31.15$&
$32.78$&
$1.439$&
$1.052$\tabularnewline
$13.0$&
$23.11$&
$32.87$&
$34.53$&
$1.422$&
$1.051$\tabularnewline
$13.5$&
$24.62$&
$34.60$&
$36.28$&
$1.405$&
$1.049$\tabularnewline
$14.0$&
$26.15$&
$36.35$&
$38.04$&
$1.390$&
$1.046$\tabularnewline
\hline
\end{tabular}
\end{center}
\caption{Values of the cross sections and $K$-factors for the scalar Higgs
production at the LHC as a function of $\sqrt{S}$ with $\mu_F=m_H$,
with $\mu_F^2=(1/2)\mu_R^2$ and $m_H=114$ GeV. MRST inputs have been used.}
\end{table}

%%%%%%%%%%%%%%%%%%%%%%%%%%%%%%%%%%  Table 2_c    %%%%%%%%%%%%%%%%%%%%%%%
\begin{table}
\begin{center}
\begin{tabular}{|c|ccccc|}
\hline
$\sqrt{S}$&
$\sigma_{LO}$&
$\sigma_{NLO}$&
$\sigma_{NNLO}$&
$K^{NLO}$&
$K^{NNLO}$\tabularnewline
\hline
$2.0$&
$0.3549$&
$1.031$&
$1.446$&
$2.905$&
$1.403$\tabularnewline
$2.5$&
$0.6393$&
$1.707$&
$2.286$&
$2.670$&
$1.339$\tabularnewline
$3.0$&
$1.003$&
$2.503$&
$3.242$&
$2.496$&
$1.295$\tabularnewline
$3.5$&
$1.439$&
$3.398$&
$4.292$&
$2.361$&
$1.263$\tabularnewline
$4.0$&
$1.945$&
$4.377$&
$5.418$&
$2.250$&
$1.238$\tabularnewline
$4.5$&
$2.515$&
$5.428$&
$6.610$&
$2.158$&
$1.218$\tabularnewline
$5.0$&
$3.146$&
$6.542$&
$7.857$&
$2.079$&
$1.201$\tabularnewline
$5.5$&
$3.834$&
$7.711$&
$9.151$&
$2.011$&
$1.187$\tabularnewline
$6.0$&
$4.575$&
$8.927$&
$10.49$&
$1.951$&
$1.175$\tabularnewline
$6.5$&
$5.367$&
$10.19$&
$11.86$&
$1.899$&
$1.164$\tabularnewline
$7.0$&
$6.207$&
$11.48$&
$13.27$&
$1.850$&
$1.156$\tabularnewline
$7.5$&
$7.093$&
$12.82$&
$14.70$&
$1.807$&
$1.147$\tabularnewline
$8.0$&
$8.023$&
$14.18$&
$16.16$&
$1.767$&
$1.140$\tabularnewline
$8.5$&
$8.994$&
$15.57$&
$17.65$&
$1.731$&
$1.134$\tabularnewline
$9.0$&
$10.00$&
$16.99$&
$19.15$&
$1.699$&
$1.127$\tabularnewline
$9.5$&
$11.05$&
$18.43$&
$20.68$&
$1.668$&
$1.122$\tabularnewline
$10.0$&
$12.14$&
$19.89$&
$22.22$&
$1.638$&
$1.117$\tabularnewline
$10.5$&
$13.26$&
$21.37$&
$23.78$&
$1.612$&
$1.113$\tabularnewline
$11.0$&
$14.41$&
$22.87$&
$25.35$&
$1.587$&
$1.108$\tabularnewline
$11.5$&
$15.60$&
$24.39$&
$26.93$&
$1.563$&
$1.104$\tabularnewline
$12.0$&
$16.82$&
$25.92$&
$28.53$&
$1.541$&
$1.101$\tabularnewline
$12.5$&
$18.07$&
$27.46$&
$30.14$&
$1.520$&
$1.098$\tabularnewline
$13.0$&
$19.35$&
$29.02$&
$31.76$&
$1.500$&
$1.094$\tabularnewline
$13.5$&
$20.65$&
$30.59$&
$33.39$&
$1.481$&
$1.092$\tabularnewline
$14.0$&
$21.99$&
$32.18$&
$35.03$&
$1.463$&
$1.089$\tabularnewline
\hline
\end{tabular}
\end{center}
\caption{Values of the cross sections and $K$-factors for the scalar Higgs
production at the LHC as a function of $\sqrt{S}$ with $\mu_F=2 m_H$,
with $\mu_F^2=(1/2)\mu_R^2$ and $m_H=114$ GeV. MRST inputs have been used.}
\end{table}

%%%%%%%%%%%%%%%%%%%%%%%%%%%%%%%%%%  Table 3_c    %%%%%%%%%%%%%%%%%%%%%%%
\begin{table}
\begin{center}
\begin{tabular}{|c|ccccc|}
\hline
$\sqrt{S}$&
$\sigma_{LO}$&
$\sigma_{NLO}$&
$\sigma_{NNLO}$&
$K^{NLO}$&
$K^{NNLO}$\tabularnewline
\hline
$2.0$&
$0.6960$&
$1.600$&
$1.830$&
$2.299$&
$1.144$\tabularnewline
$2.5$&
$1.198$&
$2.579$&
$2.862$&
$2.153$&
$1.110$\tabularnewline
$3.0$&
$1.814$&
$3.708$&
$4.028$&
$2.044$&
$1.086$\tabularnewline
$3.5$&
$2.531$&
$4.956$&
$5.300$&
$1.958$&
$1.069$\tabularnewline
$4.0$&
$3.341$&
$6.303$&
$6.661$&
$1.887$&
$1.057$\tabularnewline
$4.5$&
$4.234$&
$7.734$&
$8.096$&
$1.827$&
$1.047$\tabularnewline
$5.0$&
$5.204$&
$9.234$&
$9.593$&
$1.774$&
$1.039$\tabularnewline
$5.5$&
$6.243$&
$10.80$&
$11.14$&
$1.730$&
$1.031$\tabularnewline
$6.0$&
$7.347$&
$12.41$&
$12.74$&
$1.689$&
$1.027$\tabularnewline
$6.5$&
$8.512$&
$14.07$&
$14.38$&
$1.653$&
$1.022$\tabularnewline
$7.0$&
$9.732$&
$15.77$&
$16.06$&
$1.620$&
$1.018$\tabularnewline
$7.5$&
$11.00$&
$17.50$&
$17.77$&
$1.591$&
$1.015$\tabularnewline
$8.0$&
$12.33$&
$19.26$&
$19.50$&
$1.562$&
$1.012$\tabularnewline
$8.5$&
$13.69$&
$21.06$&
$21.26$&
$1.538$&
$1.009$\tabularnewline
$9.0$&
$15.11$&
$22.88$&
$23.05$&
$1.514$&
$1.007$\tabularnewline
$9.5$&
$16.56$&
$24.72$&
$24.86$&
$1.493$&
$1.006$\tabularnewline
$10.0$&
$18.05$&
$26.58$&
$26.69$&
$1.473$&
$1.004$\tabularnewline
$10.5$&
$19.58$&
$28.45$&
$28.53$&
$1.453$&
$1.003$\tabularnewline
$11.0$&
$21.14$&
$30.35$&
$30.39$&
$1.436$&
$1.001$\tabularnewline
$11.5$&
$22.74$&
$32.26$&
$32.26$&
$1.419$&
$1.000$\tabularnewline
$12.0$&
$24.37$&
$34.18$&
$34.15$&
$1.403$&
$0.9991$\tabularnewline
$12.5$&
$26.03$&
$36.12$&
$36.05$&
$1.388$&
$0.9981$\tabularnewline
$13.0$&
$27.72$&
$38.07$&
$37.96$&
$1.373$&
$0.9971$\tabularnewline
$13.5$&
$29.44$&
$40.02$&
$39.89$&
$1.359$&
$0.9968$\tabularnewline
$14.0$&
$31.19$&
$41.99$&
$41.82$&
$1.346$&
$0.9960$\tabularnewline
\hline
\end{tabular}
\end{center}
\caption{Values of the cross sections and $K$-factors for the scalar Higgs
production at the LHC as a function of $\sqrt{S}$ with $\mu_F=(1/2) m_H$,
with $\mu_F^2=(1/2)\mu_R^2$ and $m_H=114$ GeV. MRST inputs have been used.}
\end{table}

\chapter{Deeply Virtual Neutrino Scattering (DVNS) \label{chap5}}
\fancyhead[LO]{\nouppercase{Chapter 5. Deeply Virtual Neutrino Scattering (DVNS)}}

\section{Introduction}

In this second part of the thesis, after a brief summary of the generalities of
the Deeply Virtual Compton Scattering (DVCS) process,
we discuss its generalization to the case of neutral currents and
the phenomenological implications concerning the Physics of Neutrino-Nucleon
interactions in the few GeV's range of energies.

Exclusive processes mediated by the weak force are
an area of investigation which may gather a wide interest
in the forthcoming years due to the various experimental proposals to detect
neutrino oscillations at intermediate energy using neutrino factories and superbeams
\cite{Marciano1}. These proposals require a study of the neutrino-nucleon interaction
over a wide range of energy starting from the elastic/quasi-elastic domain
up to the deep inelastic scattering (DIS) region
(see \cite{Winter},\cite{Mangano},\cite{Morfin},\cite{nuint},\cite{Quigg} for an overall overview).
However, the discussion of the neutrino nucleon interaction has, so far,
been confined either to the DIS region or to the form factor/nucleon
resonance region, while the intermediate energy region, at this time,
remains unexplored also theoretically. Clearly, to achieve a ``continuos''
description of the underlying strong interaction
dynamics, from the resonant to the perturbative regime, will require considerable effort,
since it is experimentally and
theoretically difficult to disentangle a perturbative from a non-perturbative dynamics at
intermediate energy, which
appear to be superimposed. This is best exemplified - at least in the case of
electromagnetic processes, such as Compton scattering - in the dependence of the
intermediate energy description on the momentum transfer \cite{CL}.
In this respect, the interaction of neutrinos with the constituents of the nucleon
is no different, once the partonic structure of the target is resolved. From our viewpoint,
the presence of such a gap in our knowledge well justifies any
attempt to improve the current situation.

We recall that DVCS has been extensively studied in the last few years for electromagnetic interactions.
The extension of DVCS to the case of neutral currents is presented here,
while the charged current version will be presented in the next chapter.
Also, in this chapter we will just focus on the lowest order contributions to the process,
named by us Deeply Virtual Neutrino Scattering (DVNS) in order to distinguish
it from standard DVCS, while a renormalization group analysis of the factorized amplitude,
which requires an inclusion of the modifications induced by the evolution will also be presented elsewhere.
The application of the formalism that we develop here also needs
a separate study of the isoscalar cross sections together
with a detailed analysis of the various experimental constraints at neutrino factories
in order to be applicable at forthcoming experiments.

\section{The Generalized Bjorken Region and DVCS}

Compton scattering has been investigated in the near past by several groups, since the original work by Ji and Radyushkin \cite {Ji1, Rad3, Rad1}.
Previous work on the generalized Bjorken region, which includes DVCS and predates the
``DVCS period'' can be found in \cite{Geyer}.

\begin{figure}[t]
{\par\centering \resizebox*{12cm}{!}{\includegraphics{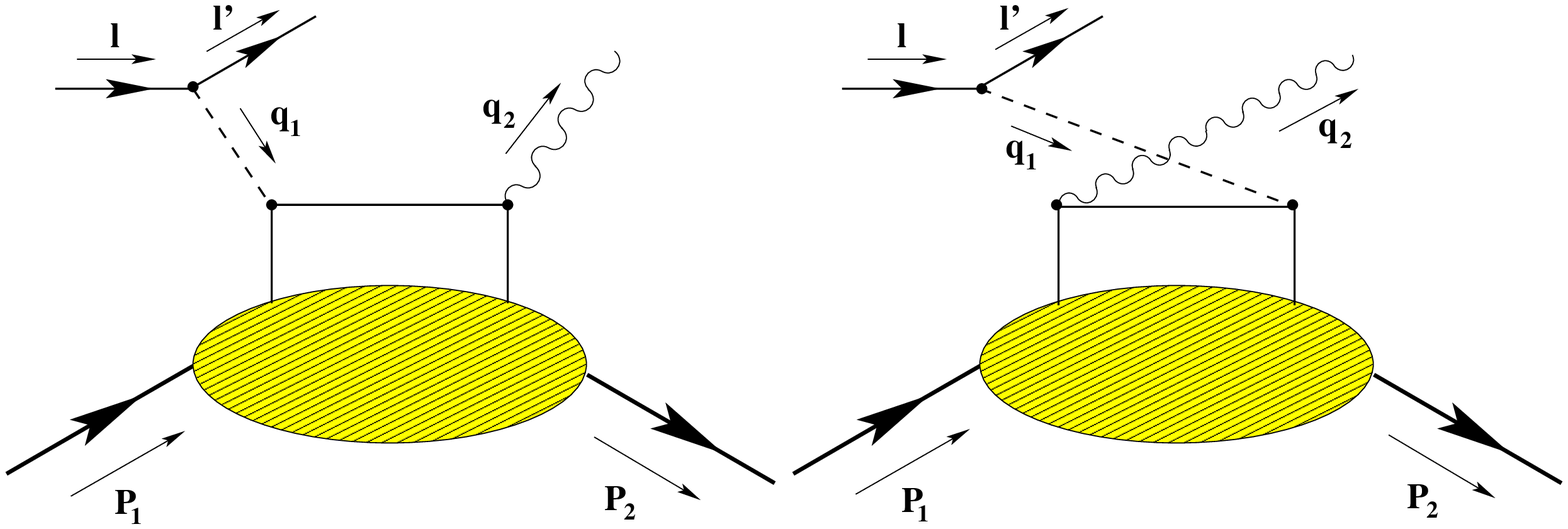}} \par}
\caption{Leading hand-bag diagrams for the process}
\label{DVCS_1.eps}
\end{figure}

A pictorial description of the process we are going to illustrate is given in Fig~\ref{DVCS_1.eps} where a neutrino of momentum $l$ scatters off a nucleon of momentum $P_1$ by an interaction with a neutral current; from the final state a photon and a nucleon emerge, of momenta $q_2$ and $P_2$ respectively, while the momenta of the final lepton is $l'$.
The process is described in terms of new constructs of the parton model termed
generalized parton distributions (GPD) or also non-forward (off-forward) parton distributions.
We recall that the regime for the study of GPD's is characterized by a deep virtuality of the exchanged photon in the initial interaction ($e +p\to e +p +\gamma$) ( $ Q^2 \approx $  2 GeV$^2$), with the final state photon kept on-shell; large energy of the hadronic system ($W^2 > 6$ GeV$^2$) above the resonance domain and small momentum transfers $|t| < 1$ GeV$^2$.
The process suffers of a severe Bethe-Heitler (BH) background, with photon emission taking place from the lepton. Therefore, in the relevant region, characterized by large $Q^2$ and small $t$, the dominant Bethe-Heitler background $(\sim 1/t)$ and the $1/Q$ behaviour of the DVCS scattering amplitude render the analysis quite complex.
From the experimental viewpoint a dedicated study of the interference BH-VCS is required in order to explore the generalized Bjorken region, and this is done by measuring asymmetries.
Opting for a symmetric choice for the defining momenta, we use as independent variables the average of the hadron and gauge bosons momenta

\beq
P_{1,2}= \bar{P} \pm \frac{\Delta}{2}\,\,\,\,\,\,\,\, q_{1,2}= \bar{q} \mp \frac{\Delta}{2}
\eeq
with $-\Delta=P_2-P_1$ being the momentum transfer. Clearly
\beq
\bar{P}\cdot \Delta=0,\,\,\,\,\, t=\Delta^2 \,\,\,\,\,\,\, \bar{P}^2=M^2 - \frac{t}{4}
\eeq
and $M$ is the nucleon mass. There are two scaling variables which are identified in the process, since 3 scalar products  can grow large in the generalized Bjorken limit: $\bar{q}^2$, $\Delta\cdot q$, $\bar{P}\cdot \bar{q}$.

The momentum transfer $t=\Delta^2$ is a small parameter in the process. Momentum asymmetries between the initial and the final state nucleon are measured by two scaling parameters, $\xi$ and $\eta$, related to ratios of the former invariants
\beq
\xi=-{\bar{q}^2\over 2 \bar{P}\cdot {\bar q}} \,\,\,\,\,\,\,\,\,\,\,\, \eta={\Delta\cdot \bar{q}\over 2 \bar{P}\cdot \bar{q}}
\eeq
where $\xi$ is a variable of Bjorken type, expressed in terms of average momenta rather than nucleon and Z-boson momenta.
The standard Bjorken variable $x= - q_1^2/( 2 P_1\cdot q_1)$ is trivially related to $\xi$ in the $t=0$ limit.
In the DIS limit $(P_1=P_2)$ $\eta=0$ and $x=\xi$, while in the DVCS limit $\eta=\xi$ and $x=2\xi/(1 +\xi)$, as one can easily deduce from the relations
\beq
q_1^2=\left(1 +\frac{\eta}{\xi}\right) \bar{q}^2 +\frac{t}{4},\,\,\,\,\,\,\,\,\,\,
q_2^2=\left(1 -\frac{\eta}{\xi}\right) \bar{q}^2 +\frac{t}{4}.
\eeq
We introduce also the inelasticity parameter $y=P_1\cdot l/(P_1\cdot q_1)$ which measures the fraction of the total energy that is transferred to the final state photon. Notice also that $\xi= \frac{\Delta^+}{2 \bar{P}^+}$ measures the ratio between the plus component of the momentum transfer and the average momentum.
A second scaling variable, related to $\xi$ is $\zeta={\Delta^+}/{{P_1}^+}$, which coincides with Bjorken $x$ $(x=\zeta)$ when $t=0$.
$\xi$, therefore, parametrizes the large component of the momentum transfer $\Delta$, which can be generically described as
\beq
\Delta= 2 \xi \bar{P} + \hat{\Delta}
\eeq
where all the components of $\hat{\Delta}$ are $O(\sqrt{t})$ \cite{Rad&Wei}.

\section{DIS versus DVNS}

\begin{figure}[t]
{\par\centering \resizebox*{6cm}{!}{\includegraphics{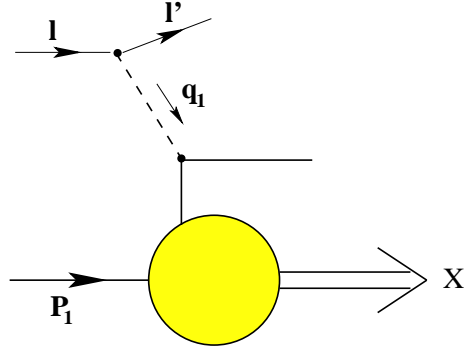}} \par}
\caption{Leading diagram for a generic DIS process}
\label{DIS}
\end{figure}

In the study of ordinary DIS scattering of neutrinos on nucleons (see Fig.~\ref{DIS}), the relevant current correlator is obtained from the T-product of two neutral currents acting on a forward nucleon state of momentum $P_1$
\ba
j_Z^\mu &\equiv& \overline{u}(l') \gamma^\mu \left( -1 + 4 \sin^{2}\theta_W + \gamma_5 \right) u(l)
\ea
where $\theta_W$ is the Weinberg angle, $l$ and $l'$ are the initial and final-state lepton. The relevant correlator is given by
\beq
T_{\mu\nu}(q_1^2,\nu)=i \int d^4 z e^{i q\cdot z}\langle P_1|T(J_Z^\mu(\xi)J_Z^{\nu}(0)
|P_1\rangle
\eeq
with $\nu=E- E'$ being the energy transfered to the nucleon, $P_1$ is the initial-state nucleon 4-momenta and $q_1 = l - l'$ is the momentum transfered.
The hadronic tensor $W_{\mu\nu}$ is related to the imaginary part of this correlator by the optical theorem. We recall that for an inclusive electroweak process mediated by neutral currents the hadronic tensor (for unpolarized scatterings) is identified in terms of 3 independent structure functions at leading twist
\beq
W_{\mu\nu}=\left(- g_{\mu\nu}+ {q_{1\mu}q_{1\nu}\over q_1^2}\right) W_1(\nu,Q^2)+
{\hat{P_1}^\mu\hat{P_1}^\nu\over P_{1}^2}{W_2(\nu,Q^2)\over M^2}
-i\epsilon_{\mu\nu\lambda\sigma}{q_1^\lambda P_1^\sigma}{W_3(\nu,Q^2)\over 2 M^2}
\eeq
where transversality of the current is obvious since $\hat{P_1}^\mu= P_1^\mu - q_1^\mu{P_1\cdot q_1}/{q_1^2}$.

The analysis at higher twists is far more involved and the total number of structure functions appearing is 14 if we include polarization effects. These are fixed by the requirements of Lorenz covariance and time reversal invariance, neglecting small CP-violating effects from the CKM matrix.
Their number can be reduced to 8 if current conservation is imposed, which is equivalent to requiring that contributions proportional to non-vanishing current quark masses can be dropped (see also \cite{Blum}).
We recall that the DIS limit is performed by the identifications
\ba
M W_1(Q^2,\nu)&=&F_1(x,Q^2) \nonumber \\
\nu W_2(Q^2,\nu)&=&F_2(x,Q^2)\nonumber \\
\nu W_3(Q^2,\nu)&=&F_3(x,Q^2),
\ea
in terms of the standard structure functions $F_1$, $F_2$ and $F_3$.
There are various ways to express the neutrino-nucleon DIS cross section, either in terms of $Q^2$ and the energy transfer, in which the scattering angle $\theta$ is integrated over, or as a triple cross section in $(Q^2,\nu,\theta)$, or yet in terms of the Bjorken variable $x$, inelasticity $y$ and the scattering angle $(x,y,\theta)$. This last case is close to the kinematical setup of our study. In this case the differential Born cross section in DIS is given by
\beq
\label{e1}
\frac{d^3 \sigma}{dx dy d\theta} = \frac{y \alpha^2}
{Q^4}\sum_{i}\eta_{i}(Q^2) L_{i}^{\mu\nu} W_{i}^{\mu\nu},
\eeq

the index $i$ denotes the different current contributions, ( $i = |\gamma|^2, |\gamma Z|, |Z|^2$ for the neutral current) and $\alpha$ denotes the fine structure constant. By $\theta$ we indicate the azimuthal angle of the final-state lepton, while $y =  (P_1\cdot q_1)/(l\cdot P_1)$ is the inelasticity parameter, and $Q^2 = -q_1^2$.
The factors $\eta_i(Q^2)$ denote the ratios of the corresponding propagator terms to the photon propagator squared,
\ba
\label{etas}
\eta^{|\gamma|^2}(Q^2)  &=& 1 ,\nonumber\\
\eta^{|\gamma Z|}(Q^2)  &=& \frac{G_F M_Z^2}{2\sqrt{2}\pi\alpha}\frac{Q^2}{Q^2+M_Z^2},\nonumber\\
\eta^{|Z|^2}(Q^2)       &=& (\eta^{|\gamma Z|})^2(Q^2).\nonumber\\
\ea
where $G_F$ is the Fermi constant and $M_Z$ is the mass of the $Z$ boson while the leptonic tensor has the form
\beq
\label{e2}
L_{\mu\nu}^i=\sum_{\lambda^\prime}\left[\bar
u(k^\prime,\lambda^\prime)\gamma_\mu(g_V^{i_1}+g_A^{i_1}
\gamma_5)u(k,\lambda)\right]^\ast\bar
u(k^\prime,\lambda^\prime)\gamma_\nu(g_V^{i_2}
+g_A^{i_2}\gamma_5)u(k,\lambda).
\eeq
In the expression above $\lambda$ and $\lambda^{\prime}$, denote the initial and final-state helicity
of the leptons. The indices $i_1$ and $i_2$ refer to the sum appearing in eq. (\ref{e1})
\begin{equation}
\label{ac}
\begin{array}{lcllcl}
g_V^{\gamma} &=& 1, & g_A^{\gamma} &=& 0, \\
g_V^Z &=& -\frac{1}{2}+2 \sin^2\theta_W,& g_A^Z &=& \frac{1}{2}, \\
\end{array}
\end{equation}
In the case of neutrino/nucleon interaction mediated by the neutral current the dominant diagram for this process appears in Fig.~\ref{DIS}. A similar diagram, with the obvious modifications, describes also charged current exchanges.
We recall that the unpolarized cross section is expressed in terms of $F_1$ and $F_2$, since $F_3$ disappears in this special case,
and in particular, after an integration over the scattering angle of the final state neutrino one obtains
\beq
\frac{d^2\sigma}{dx dy}  = 2 \pi S {\alpha^2\over Q^4} (g_V^Z)^2 \eta^{|\gamma Z|}(Q^2)
\left( 2 x y^2 F_1 + 2(1-x- {xy M^2\over S})F_2\right).
\eeq
where $S=2 M E_\nu$ is the nucleon-neutrino center of mass energy.

We also recall that in this case the cross section in the parton model is given by
\ba
\frac{d^2\sigma}{dx dy}=\frac{G_{F}^2 M E_{\nu}}{2\pi}\left( \frac{M_{Z}^2}{Q^2+M_Z^2}\right)^2\left[x q^{0}(x,Q^2)+x\bar{q}^{0}(x,Q^2)(1-y)^2\right]
\ea
where $q^{0}(x,Q^2)$ and $\bar{q}^{0}(x,Q^2)$ are linear combinations of parton distributions
\ba
q^{0}(x,Q^2)&=&\left[\frac{u_{v}(x,Q^2)+d_{v}(x,Q^2)}{2}+\frac{\bar{u}(x,Q^2)+\bar{d}(x,Q^2)}{2}\right] \left(L_u^2 + L_d^2\right)\nonumber\\
&+&\left[\frac{\bar{u}(x,Q^2)+\bar{d}(x,Q^2)}{2}\right]\left(R_u^2 + R_d^2\right)\nonumber\\
\bar{q}^{0}(x,Q^2)&=&\left[\frac{u_v(x,Q^2)+d_v(x,Q^2)}{2}+\frac{\bar{u}(x,Q^2)+\bar{d}(x,Q^2)}{2}\right]\left(R_u^2+R_d^2\right)\nonumber\\
&+&\left[\frac{\bar{u}(x,Q^2)+\bar{d}(x,Q^2)}{2}\right]\left(L_u^2+L_d^2\right)\nonumber\\
\ea
with
\ba
&&L_u =1-\frac{4}{3}\sin^2\theta_{W}\,,\hspace{1.5 cm}L_d =-1+\frac{2}{3}\sin^2\theta_{W}\nonumber\\
&&R_u =-\frac{4}{3}\sin^2\theta_{W}\,,\hspace{1.8 cm}R_d =\frac{2}{3}\sin^2\theta_{W}\nonumber\\
\ea
and we have identified the sea contributions $u_{s}$ and $d_{s}$ with $\bar{u}$ and $\bar{d}$ rispectively.

\begin{figure}[t]
{\par\centering \resizebox*{10cm}{!}{\rotatebox{-90}{\includegraphics{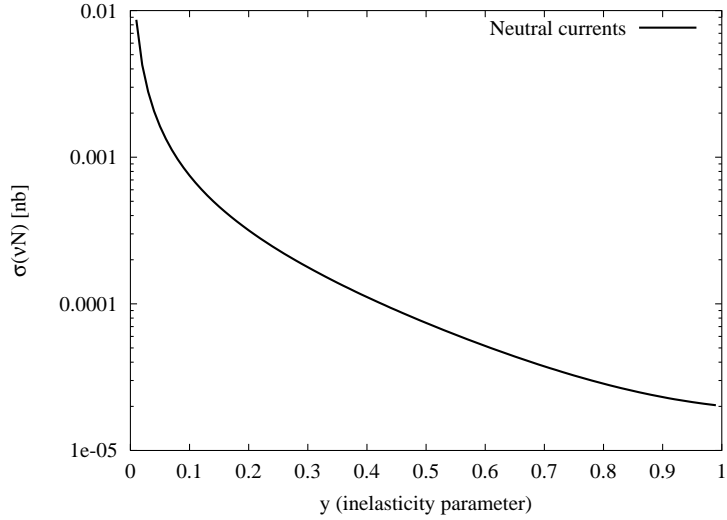}}}\par}
\caption{The cross section of a neutrino process of DIS-type
at $x\approx 0.1$ with neutral current at ultrahigh energy}
\label{sigma_Dis.ps}
\end{figure}

Let's now move to the nonforward case. Here, when a real photon is present in the final state, the relevant correlator is given by
\beq
T_{\mu\nu}(q_1^2,\nu)=i \int d^4 z e^{i q\cdot z}\langle\bar{ P} - {\Delta\over 2}|T(J_Z^\mu(-z/2)J_{\gamma}^{\nu}(z/2)|\bar{P} +{\Delta\over 2}\rangle.
\eeq
The dominant diagrams for this process appears in Fig~\ref{DVCS_1.eps}.
We impose Ward identities on both indices, which is equivalent to requiring that terms proportional to the quark masses in $\partial\cdot J_Z$ are neglected. This approximation is analogous
to the one performed in the forward case in order to reduce the structure functions from 8 to 3
(in the absence of any polarization), imposing symmetric trasversality conditions on the weak currents
\beq
\left(\bar{P}^\mu -{\Delta^\mu\over 2}\right) T_{\mu\nu}=0\,\,\,\,\,\,\,\,\,
\left(\bar{P}^\mu +{\Delta^\mu\over 2}\right) T_{\mu\nu}=0.
\eeq
The leading twist contribution to DVNS is obtained by performing a collinear expansion of the loop momentum
of the hand-bag diagram and neglecting terms of order $O(\Delta_\perp^2/Q^2)$ and $M^2/Q^2$. Transversality is satisfied at this order.
 Violation of transversality condition in the hand-bag approximation is analogous to the DVCS case, 
where it has been pointed out that one has to include systematically ``kinematical'' twist-3 operators,  which 
appear as total derivatives of twist-2 operators \cite{Rad&Wei} \cite{Penttinen} in order to restore it.  

For the parametrization of the hand-bag diagram (Fig.~\ref{DVCS_1.eps}) 
we use the light-cone decomposition in terms of 2 four-vectors $(n,\tilde{n})$, where 

\begin{eqnarray}
\tilde{n}^\mu &=& \Lambda (1,0,0,1) \nonumber \\
n^\mu &=& \frac{1}{2 \Lambda} (1,0,0,-1) \nonumber \\
\tilde{n}^2 &=& n^2 = 0 \ \ \ , \ \ \ \tilde{n}\cdot n = 1. \nonumber 
\end{eqnarray}

At the same time we set

\begin{eqnarray}
P_1^\mu&=& (1 +\xi)\tilde{n}^{\mu} +(1-\xi){\overline{M}\over 2}n^\mu - {\Delta_\perp^\mu\over 2} \nonumber \\
P_2^\mu&=& (1 +\xi)\tilde{n}^{\mu} +(1 + \xi){\overline{M}\over 2}n^\mu + {\Delta_\perp^\mu \over 2} \nonumber \\
q_1^{\mu}&=& -2 \xi\tilde{n} ^\mu + {Q^2\over 4 \xi}n^\mu \nonumber \\
k^\mu &=& \left(k\cdot n\,\right)\, \tilde{n}^\mu + \left(k\cdot \tilde{n}\right)\,\,  n^\mu + k_\perp^\mu \nonumber \\
\overline{M}^2 &=&M^2 -\frac{\Delta^2}{4}
\end{eqnarray}

with $\bar{P}^2=\overline{M}^2$. We will also use the notation $-q_1^2=Q^2$ for the invariant mass
of the virtual $Z$ boson and we will denote by $\bar{q}$ the average gauge bosons momenta respectively.

After a collinear expansion of the loop momentum we obtain

\begin{eqnarray}
&&T_{A}^{\mu \nu} = i \int \frac{d^4 k}{(2 \pi)^4} Tr\left\{ g_{u}\gamma^{\nu} \rlap/{P_D}\gamma^{\mu}
\tilde{g}\left(U_v-\gamma^{5} \right)\underline{M}^{u}(k)
+\right.\nonumber\\
&&\hspace{3.5cm}\left. g_{d}\gamma^{\nu} \rlap/{P_D}\gamma^{\mu}\tilde{g}
\left(D_v-\gamma^5\right)\underline{M}^{d}(k)\right\} \nonumber\\
&&T_{B}^{\mu \nu} = i \int \frac{d^4 k}{(2 \pi)^4} Tr\left\{ \tilde{g}\gamma^{\mu}
\left(U_v -\gamma^5\right)\rlap/{P_E}\gamma^{\nu} g_u \underline{M}^{u}(k)
+\right.\nonumber\\
&&\hspace{3.5cm}\left.\tilde{g}\gamma^{\mu} \left(D_v -\gamma^5\right)\rlap/{P_E}\gamma^{\nu}
g_d \underline{M}^{d}(k)\right\}
\label{coll}
\end{eqnarray}

where we have used the following notations

\ba
&&g_u = \frac{2}{3}e,\hspace{1cm}g_d=\frac{1}{3}e,\hspace{1cm}\tilde{g}=
\frac{g}{4 \cos{(\theta_{W})}},\nonumber\\
&&U_v=1-\frac{8}{3} \sin^2\theta_{W},\hspace{1cm}D_v=1-\frac{4}{3}\sin^2 \theta_{W},\nonumber\\
\\
&&\rlap/{P_D}=\frac{\slash{k}-\alpha\slash{\Delta}+\slash{q_1}}{\left(k - \alpha\Delta +q_1\right)^2 + i \epsilon}\,\,,\nonumber\\
&&\rlap/{P_E}=\frac{\slash{k}-\slash{q_1}+\slash{\Delta}(1-\alpha)}
{\left(k - q_1 +\Delta(1-\alpha)\right)^2 +i\epsilon} \,\,
\ea

where the constant $\alpha$ ($\alpha$ is a free parameter) ranges between $0$ and $1$.
The $\underline{M}$ matrix is the quark density matrix and is given by

\begin{eqnarray}
\underline{M}_{ab}^{(i)}(k) &=& \int d^4y e^{i k\cdot y} \langle P'|
\overline{\psi}_{a}^{(i)}(-\alpha y) \psi_{b}^{(i)}((1-\alpha)y) |P \rangle.
\label{matrix}
\end{eqnarray}

The index $i=u,d$ runs on flavours.
Using a Sudakov decomposition of the internal loop we can rewrite $T_{A}^{\mu\nu}$ and $T_{B}^{\mu\nu}$  as

\ba
&&T_{A}^{\mu\nu}=i\,\int \frac{d( k\cdot n)}{(2 \pi)^4}d(k\cdot \tilde{n}) d^2 k_{\perp}\int\frac{d\lambda}{(2\pi)}dz e^{i\lambda (z-k\cdot n)}\nonumber\\
&&\;\;\;\;\;\;\;\;\;\;\;\;Tr\left\{ g_{u}\gamma^\nu \rlap/{P_D}\gamma^{\mu}\tilde{g}\left(U_v-\gamma^5\right)\underline{M}^{u}(k)+\right.\nonumber\\
&&\hspace{2cm}\left.g_{d}\gamma^\nu \rlap/{P_D}\gamma^{\mu}\tilde{g}\left(D_v-\gamma^5\right)\underline{M}^{d}(k)\right\} \nonumber\\
\\
&&T_{B}^{\mu\nu}=i\,\int \frac{d( k\cdot n)}{(2 \pi)^4}d(k\cdot \tilde{n}) d^2 k_{\perp}\int\frac{d\lambda}{(2\pi)}dz e^{i\lambda (z-k\cdot n)}\nonumber\\
&&\;\;\;\;\;\;\;\;\;\;\;\;Tr\left\{ \tilde{g}\gamma^\mu \left(U_v -\gamma^5\right)\rlap/{P_E}\gamma^{\nu} g_u \underline{M}^{u}(k)+\right.\nonumber\\
&&\hspace{2cm}\left.\tilde{g}\gamma^\mu \left(D_v -\gamma^5\right)\rlap/{P_E}\gamma^{\nu} g_d \underline{M}^{d}(k)\right\}
\ea

to which we will refer as the direct and the exchange diagram respectively.
It is also convenient to introduce two new linear combinations $T^{\mu\nu}=T_{A}^{\mu\nu}+T_{B}^{\mu\nu}=\tilde{T}_{A}^{\mu\nu}+\tilde{T}_{B}^{\mu\nu}$ which will turn useful in order to separate Vector (V) and axial vector parts (A) of the expansion

\ba
&&\tilde{T}_{A}^{\mu\nu}=i\int \frac{d( k\cdot n)}{(2 \pi)^4}d(k\cdot\tilde{n}) d^2 k_{\perp}\int\frac{d\lambda}{(2\pi)}dz e^{i\lambda (z-k\cdot n)}\nonumber\\
&&\;\;\;\;\;\;\;\;\;\;\;\;\tilde{g} g_{u}U_v Tr\left\{\left[\gamma^\nu \rlap/{P_D}\gamma^{\mu}+\gamma^\mu \rlap/{P_E}\gamma^{\nu}\right]\underline{M}^{u}(k)\right\}+\nonumber\\
&&\hspace{1.4cm}\tilde{g} g_{d}D_vTr\left\{\left[\gamma^\nu \rlap/{P_D}\gamma^{\mu}+\gamma^\mu \rlap/{P_E}\gamma^{\nu}\right]\underline{M}^{d}(k)\right\} \nonumber\\\\
&&\tilde{T}_{B}^{\mu\nu}=-i\int \frac{d( k\cdot n)}{(2 \pi)^4}d(k\cdot \tilde{n}) d^2 k_{\perp}\int\frac{d\lambda}{(2\pi)}dz e^{i\lambda (z-k\cdot n)}\nonumber\\
&&\;\;\;\;\;\;\;\;\;\;\;\;\tilde{g} g_{u}Tr\left\{\left[\gamma^\nu \rlap/{P_D}\gamma^{\mu}+\gamma^\mu \rlap/{P_E}\gamma^{\nu}\right]\gamma^5\underline{M}^{u}(k)\right\}+\nonumber\\
&&\hspace{1.4cm}\tilde{g} g_{d}Tr\left\{\left[\gamma^\nu \rlap/{P_D}\gamma^{\mu}+\gamma^\mu \rlap/{P_E}\gamma^{\nu}\right]\gamma^5\underline{M}^{d}(k)\right\}.
\ea

with $\tilde{T}_A$ including the vector parts $(V\times V  \,\,+\,\, A\times A)$ and $\tilde{T}_B$ the axial-vector parts 

$(V\times A \,\,+ \,\,A\times V)$. After some algebraic manipulations we finally obtain

\ba
&&\tilde{T}_{A}^{\mu\nu}=\frac{i}{2}\sum_{i=u,d}\tilde{g}g_i C_i\int\frac{d\lambda dz}{(2\pi)}e^{i\lambda z} \left\{\left(\tilde{n}^{\mu} n^{\nu}+\tilde{n}^{\nu} n^{\mu} -g^{\mu\nu} \right) \alpha(z) \langle P'|\overline {\psi}^{(i)}\left(-\frac{\lambda n}{2}\right)\slash{n}\psi^{(i)}\left(\frac{\lambda n}{2}\right)|P\rangle \right.\nonumber\\
&&\hspace{5 cm}\left.+i\epsilon^{\mu\nu\alpha\beta}\tilde{n}_{\alpha} n_{\beta}\beta(z) \langle P'|\overline{\psi}^{(i)}\left(-\frac{\lambda n}{2}\right)\gamma^5 \slash{n}\psi^{(i)}\left(\frac{\lambda n}{2}\right)|P\rangle \right\}\nonumber\\
&&\tilde{T}_{B}^{\mu\nu}=-\frac{i}{2}\sum_{i=u,d}\tilde{g} g_i\int\frac{d\lambda dz}{(2\pi)}e^{i\lambda z}\left\{\left(\tilde{n}^{\mu} n^{\nu}+\tilde{n}^{\nu} n^{\mu} -g^{\mu\nu} \right) \alpha(z)\langle P'|\overline{\psi}^{(i)}\left(-\frac{\lambda n}{2}\right)\gamma^5 \slash{n}\psi^{(i)}\left(\frac{\lambda n}{2}\right)|P\rangle \right.\nonumber\\
&&\hspace{5cm}\left.+i\epsilon^{\mu\nu\alpha\beta}\tilde{n}_{\alpha} n_{\beta} \beta(z)\langle P'|\overline{\psi}^{(i)}\left(-\frac{\lambda n}{2}\right)\slash{n}\psi^{(i)}\left(\frac{\lambda n}{2}\right)|P\rangle \right\} \nonumber\\
\ea

where $C_i=U_v,\,D_v$ and 

\ba
\alpha(z)=\left(\frac{1}{z-{\xi}+i\epsilon} + \frac{1}{z+{\xi}-i\epsilon}\right),\hspace{1.5cm}
\beta(z)=\left(\frac{1}{z-{\xi} +i\epsilon} - \frac{1}{z+{\xi}-i\epsilon}\right)
\ea
are the ``first order'' propagators appearing in the factorization of the amplitude. We recall, if not obvious, that differently from 
DIS, DVCS undergoes factorization directly at amplitude level \cite{CollinsFreund}. 
 
The parameterizations of the non-forward light cone correlators in terms of GPD's is of the form given by Ji at leading twist \cite{Ji1}
\ba
&&\int\frac{d\lambda }{(2\pi)}e^{i\lambda z}\langle P'|\overline {\psi}\left(-\frac{\lambda n}{2}\right)\gamma^\mu \psi\left(\frac{\lambda n}{2}\right)|P\rangle=\nonumber\\\nonumber\\
&&H(z,\xi,\Delta^2)\overline{U}(P')\gamma^\mu U(P) + E(z,\xi,\Delta^2)\overline{U}(P')\frac{i\sigma^{\mu\nu} \Delta_{\nu}}{2M} U(P) + .....\nonumber\\
&&\int\frac{d\lambda }{(2\pi)}e^{i\lambda z}\langle P'|\overline {\psi}\left(-\frac{\lambda n}{2}\right)\gamma^\mu \gamma^5 \psi\left(\frac{\lambda n}{2}\right)|P\rangle=\nonumber\\\nonumber\\
&&\tilde{H}(z,\xi,\Delta^2)\overline{U}(P')\gamma^\mu\gamma^5 U(P) + \tilde{E}(z,\xi,\Delta^2)\overline{U}(P')\frac{\gamma^5 \Delta^{\mu}}{2M}U(P) + .....
\ea
which have been expanded in terms of functions $H,E, \tilde{H}, \tilde{E}$ \cite{Ji2} and the ellipses are meant to denote the higher-twist contributions.
It is interesting to observe that the amplitude is still described by the same light-cone correlators as in the electromagnetic case (vector, axial vector) but now parity is not conserved.
\footnote{Based on the article published in JHEP 0502, 038 (2005)}
\section{Operatorial analysis}

The operatorial structure of the T-order product of one electroweak current and one electromagnetic 
current is relevant in order 
to identify the independent amplitudes appearing in the correlator at leading twist and the study is presented here. 
We will identify four operatorial structures. For this purpose  let's start from the Fourier transform of the correlator of the two currents
\ba
T_{\mu \nu}=i\int d^{4}x e^{i qx} \langle P_2|T\left(J_{\nu}^{\gamma}(x/2) J_{\mu}^{Z_{0}}(-x/2)\right)|P_1\rangle\,, 
\ea
where for the neutral and electromagnetic currents we have the following expressions
\ba
\label{currents}
&&J^{\mu Z_{0}}(-x/2)=\frac{g}{2 \cos{\theta_W}}\overline{\psi}_{u}(-x/2)\gamma^{\mu}(g^{Z}_{u V}+g^{Z}_{u A}\gamma^5)\psi_{u}(-x/2)+\overline{\psi}_{d}(-x/2)\gamma^{\mu}(g^{Z}_{d V}+g^{Z}_{d A}\gamma^5)\psi_{d}(-x/2),\,\nonumber\\
&&J^{\nu, \gamma}(x/2)=\overline{\psi}_{d}(x/2)\gamma^{\nu}\left(-\frac{1}{3}e\right)\psi_{d}(x/2) +\overline{\psi}_{u}(x/2)\gamma^{\nu}\left(\frac{2}{3}e\right)\psi_{u}(x/2)\,.
\ea
By simple calculations one obtains
\ba
\label{timeorder}
&&\langle P_2|T\left(J_{\nu}^{\gamma}(x/2) J_{\mu}^{Z_{0}}(-x/2)\right)|P_1\rangle=\nonumber\\
&&\langle P_2|\overline{\psi}_{u}(x/2)g_u \gamma_{\nu}S(x)\gamma_{\mu}(g^{Z}_{u V}+g^{Z}_{u A}\gamma^5)\psi_{u}(-x/2)-\nonumber\\
&&\hspace{0.8 cm}\overline{\psi}_{d}(x/2)g_d \gamma_{\nu}S(x)\gamma_{\mu}(g^{Z}_{d V}+g^{Z}_{d A}\gamma^5)\psi_{d}(-x/2)+\nonumber\\
&&\hspace{0.8 cm}\overline{\psi}_{u}(x/2)\gamma_{\mu}(g^{Z}_{u V}+g^{Z}_{u A}\gamma^5)S(-x) g_u \gamma_{\nu}\psi_{u}(x/2)-\nonumber\\
&&\hspace{0.8 cm}\overline{\psi}_{d}(x/2)\gamma_{\mu}(g^{Z}_{d V}+g^{Z}_{d A}\gamma^5)S(-x) g_d \gamma_{\nu}\psi_{d}(x/2)|P_1 \rangle\,.\nonumber\\
\ea
The coefficients used in eqs. (\ref{currents}, \ref{timeorder}) $g_V^{Z}$ and $g_A^{Z}$, are
\ba
&&g^{Z}_{u V}=\frac{1}{2} + \frac{4}{3} \sin^{2}\theta_{W}\hspace{1.5 cm}g^{Z}_{u A}=-\frac{1}{2}\nonumber\\
&&g^{Z}_{d V}=-\frac{1}{2} + \frac{2}{3} \sin^{2}\theta_{W}\hspace{1.5 cm}g^{Z}_{d A}=\frac{1}{2}\,,
\ea
and
\ba
g_{u}=\frac{2}{3},&&g_{d}=\frac{1}{3}
\ea
are the absolute values of the charges of the up and down quarks in units of the electron charge.

The function $S(x)$ denotes the free quark propagator
\ba
S(x)\approx \frac{i \rlap/{x}}{2\pi^2(x^2-i\eps)^2}.
\ea

After some standard identities for the $\gamma$'s products
\ba
&&\gamma_{\mu}\gamma_{\alpha}\gamma_{\nu}=S_{\mu\alpha\nu\beta}\gamma^{\beta}+i\eps_{\mu\alpha\nu\beta}\gamma^{5}\gamma^{\beta},\nonumber\\
&&\gamma_{\mu}\gamma_{\alpha}\gamma_{\nu}\gamma^{5}=S_{\mu\alpha\nu\beta}\gamma^{\beta}\gamma^{5}-i\eps_{\mu\alpha\nu\beta}\gamma^{\beta},\nonumber\\
&&S_{\mu\alpha\nu\beta}=\left(g_{\mu\alpha}g_{\nu\beta}+g_{\nu\alpha}g_{\mu\beta}-g_{\mu\nu}g_{\alpha\beta} \right),
\ea
we rewrite the correlators as
\ba
&&T_{\mu\nu}=i\int d^{4}x\frac{e^{i q x} x^{\alpha}}{2\pi^2(x^2 -i\eps)^2}
\langle P_2|\left[g_u g_{u V}\left(S_{\mu\alpha\nu\beta}O^{\beta}_{u} -i\eps_{\mu\alpha\nu\beta}O^{5 \beta}_{u}\right)-g_u g_{u A}\left(S_{\mu\alpha\nu\beta}\tilde{O}^{5 \beta}_{u}-i\eps_{\mu\alpha\nu\beta}\tilde{O}^{\beta}_{u}\right)\right.\nonumber\\
&&\hspace{5.5 cm}\left.-g_d g_{d V}\left(S_{\mu\alpha\nu\beta}O^{\beta}_{d}-i\eps_{\mu\alpha\nu\beta}O^{5 \beta}_{d}\right)+g_d g_{d A}\left(S_{\mu\alpha\nu\beta}\tilde{O}^{5 \beta}_{d} -i\eps_{\mu\alpha\nu\beta}\tilde{O}^{\beta}_{d}\right)\right]|P_1\rangle.\,\nonumber\\
\ea
The $x$-dependence of the operators in the former equations was suppressed.

Whence the relevant operators are denoted by
\ba
&&\tilde{O}_{a}^{\beta}(x/2,-x/2)=\overline{\psi}_{a}(x/2)\gamma^{\beta}\psi_{a}(-x/2)+\overline{\psi}_{a}(-x/2)\gamma^{\beta}\psi_{a}(x/2),\nonumber\\
&&\tilde{O}_{a}^{5 \beta}(x/2,-x/2)=\overline{\psi}_{a}(x/2)\gamma^{5}\gamma^{\beta}\psi_{a}(-x/2)-\overline{\psi}_{a}(-x/2)\gamma^{5}\gamma^{\beta}\psi_{a}(x/2),\nonumber\\
&&O_{a}^{\beta}(x/2,-x/2)=\overline{\psi}_{a}(x/2)\gamma^{\beta}\psi_{a}(-x/2)-\overline{\psi}_{a}(-x/2)\gamma^{\beta}\psi_{a}(x/2),\nonumber\\
&&O_{a}^{5 \beta}(x/2,-x/2)=\overline{\psi}_{a}(x/2)\gamma^{5}\gamma^{\beta}\psi_{a}(-x/2)+\overline{\psi}_{a}(-x/2)\gamma^{5}\gamma^{\beta}\psi_{a}(x/2)\,,\nonumber\\
\ea
where $a$ is a flavour index.

\section{Phases of the Amplitude}

The numerical computation of the cross section requires a prescription for a correct handling of the singularities in the integration region at $z=\pm\xi$.
The best way to proceed is to work out explicitly the structure of the factorization formula of the amplitude using the Feynman prescription for going around the singularities, thereby isolating a principal value integral (P.V., which is real) and an imaginary contribution coming from the $\delta$ function term. A  P.V. integral is expressed in terms of ``plus'' distributions and of logarithmic terms, as illustrated below.
The expression of the factorization formula of the process in the parton model, in which $\alpha(z)$ and $\beta(z)$ appear as factors in the coefficient functions, is then given by
\ba
&&\hspace{4cm}{\cal{M}}_{f i}=J^{\mu}_{Z}(q_1)D(q_1)\eps^{\nu *}(q_1 -\Delta)\nonumber\\\nonumber\\
&&\times\left\{\frac{i}{2}\tilde{g}g_u U_v \int_{-1}^{1}dz\left(\tilde{n}^{\mu}n^{\nu}+\tilde{n}^{\nu}n^{\mu}-g^{\mu\nu} \right)\right.\nonumber\\
&&\left.\alpha(z)\left[H^u (z,\xi,\Delta^{2})\overline{U}(P_2)\slash{n}U(P_1) + E^u(z,\xi,\Delta^2)\overline{U}(P_2)\frac{i\sigma^{\mu\nu}n_{\mu}\Delta_{\nu}}{2 M}U(P_1)\right]+\right.\nonumber\\
&&\left.\beta(z)i\eps^{\mu\nu\alpha\beta}\tilde{n}_{\alpha}n_{\beta}\left[\tilde{H}^u (z,\xi,\Delta^{2})\overline{U}(P_2)\slash{n}\gamma^{5}U(P_1)+\tilde{E}^{u}(z,\xi,\Delta^{2})\overline{U}(P_2)\gamma^{5}\left(\Delta\cdot n\right)U(P_1)\right]+\right.\nonumber\\
&&\left.\frac{i}{2}\tilde{g}g_d D_v\int_{-1}^{1}dz\left\{u\rightarrow d\right\}-\right.\nonumber\\\nonumber\\
&&\left.\frac{i}{2}\tilde{g}g_u\int_{-1}^{1}dz\left(-\tilde{n}^{\mu}n^{\nu}-\tilde{n}^{\nu}n^{\mu}+ g^{\mu\nu}\right)\right.\nonumber\\
&&\left.\alpha(z)\left[\tilde{H}^u (z,\xi,\Delta^{2})\overline{U}(P_2)\slash{n}\gamma^{5}U(P_1) + \tilde{E}^u(z,\xi,\Delta^2)\overline{U}(P_2)\frac{i \gamma^{5}\Delta\cdot n}{2 M}U(P_1)\right]+\right.\nonumber\\
&&\left.\beta(z)i\eps^{\mu\nu\alpha\beta}\tilde{n}_{\alpha}n_{\beta}\left[H^u (z,\xi,\Delta^{2})\overline{U}(P_2)\slash{n}U(P_1)+E^{u}(z,\xi,\Delta^{2})\overline{U}(P_2)\frac{i\sigma^{\mu\nu}n_{\mu}\Delta_{\nu}}{2 M}(P_1)\right]-\right.\nonumber\\
&&\left.\frac{i}{2}\tilde{g}g_d\int_{-1}^{1}dz\left\{u\rightarrow d\right\}\right\}.\nonumber\\
\ea
To handle the singularity on the path of integration in the factorization formula, 
as we have already mentioned,
we use the Feynman ($i\epsilon$) prescription, thereby  generating imaginary parts.
In particular, any standard integral containing imaginary parts is then separated into real 
and imaginary contributions as
\beq
\int dz {T(z)\over z \mp \xi \pm i \epsilon}= PV \int_{-1}^1 dz {T(z)\over z \mp\xi} \,\mp \,i \pi T(\pm \xi)
\eeq
for a real coefficient $T(z)$.
We then rewrite the P.V. integral in terms of ``plus'' distributions
\ba
P.V.\int_{-1}^{1} dz {H(z)\over z - \xi}&=&
\int_{-1}^{1}dz \frac{H(z) - H(\xi)}{z - \xi}+ H(\xi)
\log\left({1-\xi\over 1 +\xi}\right) \nonumber \\
&=& \int_{-1}^1 dz {\cal Q}(z)H(z) +
\int_{-1}^1 dz \bar{{\cal Q}}(z)H(z) + H(\xi)\log\left({1-\xi\over 1 +\xi}\right) \nonumber \\
\label{rex1}
\ea
where
\ba
{\cal Q}(z)&=&\theta(-1\leq z\leq \xi) {1\over (z - \xi)_+}\nonumber \\
&=&\theta(-1\leq z\leq \xi)\left( {\theta(z < \xi)\over (z - \xi)} -
\delta(z- \xi)\int_{-1}^\xi{dz\over (z - \xi)}\right)\nonumber \\
\bar{\cal Q}(z)&=&\theta(\xi\leq z\leq 1) {1\over (z - \xi)_+}\nonumber \\
&=&\theta(-1\leq z\leq \xi)\left( {\theta(z> \xi)\over (z - \xi)} -
\delta(z- \xi)\int_{\xi}^1{dz\over (z - \xi)}\right)\nonumber \\
\label{rex2}
\ea
and the integrals are discretized using finite elements methods, in order to have high numerical accuracy.
This last point is illustrated in Appendix C, where the computations are done analytically on
a grid and then the grid spacing is sent to zero.

We can now proceed and compute the cross section. We define the scalar amplitude
\ba
{\cal M}_{f i}=J_{Lep}^{\mu}(q_1)D(q_1)T^{\mu\nu}\epsilon^{*\nu}(q_1 -\Delta)
\ea
where $D(q_1)$ is the $Z_{0}$ propagator in the Feynman gauge and $J_{Lep}^{\mu}(q_1)$ is the leptonic current
and we have introduced the polarization vector for the final state photon $\epsilon^\nu$.

In particular, for the squared amplitude we have
\ba
|{\cal M}_{f i}|^2 = -L^{\mu\lambda}D(q_1)^2 T_{\mu\nu}T^{*\nu}_{\lambda}\,
\ea
which is given, more specifically, by
\beq
{\cal |M|}^2 =\int_{-1}^1 dz \int_{-1}^1 dz' \left( K_1(z,z') \alpha(z) \alpha^*(z') + K_2(z,z') \beta(z) \beta^*(z')  \right)
\label{Fact1}
\eeq
with $K_1$ and $K_2$ real functions, combinations of the generalized distributions $(H,\tilde{H}, E,\tilde{E})$ with appropriate kinematical factors. Mixed contributions proportional to $\alpha(z)\beta^*(z')$ and $\beta(z)\alpha^*(z')$ cancel both in their real and imaginary parts and as such do not contribute to the phases. A similar result holds also for the pure electromagnetic case.

After some further manipulations, we finally rewrite the squared amplitude in terms of a P.V. contribution plus some additional
terms coming from the imaginary parts
\ba
{\cal |M|}^2 &=& P.V.\int_{-1}^1 dz \int_{-1}^1 dz' \left( K_1(z,z') \alpha(z) \alpha^*(z') + K_2(z,z') \beta(z) \beta^*(z')  \right)\nonumber \\
&&+ \pi^2 \left( K_1(\xi,\xi) - K_1(\xi,-\xi) - K_1((-\xi,\xi) + K_1(-\xi,-\xi)\right)\nonumber \\
&&+ \pi^2 \left( K_2(\xi,\xi) + K_2(\xi,-\xi) + K_2((-\xi,\xi) + K_2(-\xi,-\xi)\right)
\ea
which will be analized numerically in the sections below. In order to proceed with the 
numerical result, it is necessary to review the standard construction of the nonforward 
parton distribution functions in terms of the forward distributions, 
which is the topic of the next section.

\section{Construction of the Input Distributions}

The computation of the cross section proceeds rather straightforwardly,
though the construction of the initial conditions is more involved compared to the forward (DIS) case.
This construction has been worked out in several papers
\cite{bel1,radmus,Biernat,Freu&McD2,Initialcond,Rad1,radmus}
in the case of standard DVCS, using a diagonal input appropriately extended
to the non-diagonal kinematics. Different types of nonforward parton distribution,
all widely used in the numerical implementations have been put forward, beside Ji's original
distributions, which we will be using in order to construct the initial conditions for our process.

For our purposes it will be useful to introduce Golec-Biernat and Martin's (GBM) distributions \cite{Biernat}
at an intermediate step, which are linearly related to Ji's distributions.

\begin{figure}[t]
{\par\centering \resizebox*{10cm}{!}{\includegraphics{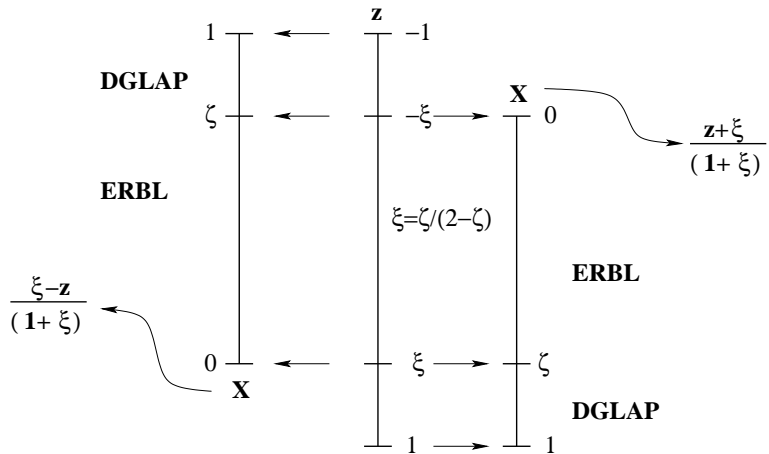}} \par}
\caption{The relationship between ${\cal F}^q (X,\zeta)$, ${\cal F}^{\bar q} (X,\zeta) $ and Ji's function $H^q (z,\xi)$.}
\label{relatio}
\end{figure}

We recall, at this point, that the quark distributions $H_{q}(z,\xi)$ have support in $z\in [-1,1]$, describing
both quark and antiquark distributions for $z>0$ and $z<0$ respectively. In terms of GBM distributions,
two distinct distributions ${\hat{\cal{F}}}^{\bar{q}}(X,\zeta)$ and ${\hat{\cal{F}}}^{q}(X,\zeta)$
with $0\leq X\leq 1$ are needed in order to cover the same information contained in Ji's distributions using only a positive
scaling variable ($X$).
In the region $X\in (\zeta,1]$ the functions ${\hat{\cal{F}}}^{q}$ and ${\hat{\cal{F}}}^{\bar{q}}$ are independent, but if $X\le\zeta$ they are related to each other, as shown in the (by now standard) plot in Fig.~\ref{relatio}.

In this new variable ($X$) the DGLAP region is described by $X > \zeta$ ($|z| > \xi$), and the ERBL region by $X<\zeta$ ($|z| < \xi$).
In the ERBL region, $\hat{{\cal F}}^q$ and $\hat{{\cal F}}^{\bar q}$ are not independent.

The relation between $H(z,\xi)$ and $\hat{{\cal{F}}}^{q}(X,\zeta)$ can be obtained explicitly \cite{Freund1} as follows:
for $z \in [-\xi,1]$ we have
\begin{equation}
\hat{{\cal F}}^{q,i} \left(X = \frac{z+\xi}{1 + \xi},\zeta\right) = \frac{H^{q,i} (z,\xi)}{1-\zeta/2} \, ,
\label{curlyq}
\end{equation}
and for $z \in [-1,\xi]$
\begin{equation}
\hat{{\cal F}}^{{\bar q},i} \left(X = \frac{\xi -z}{1 + \xi},\zeta\right) = -\frac{H^{q,i} (z,\xi)}{1-\zeta/2} \, .
\label{curlyqbar}
\end{equation}
where $i$ is a flavour index.
In our calculations we use a simplified model for the GPD's where the $\Delta^2$ dependence can be factorized as follows
\cite{Vander&guich,bel1}
\ba
H^{i}(z,\xi,\Delta^2,Q^2)&=&F^{i}_{1}\,(\Delta^2)q^{i}(z,\xi,Q^2)\nonumber\\
\tilde{H}^{i}(z,\xi,\Delta^2,Q^2)&=&G^{i}_{1}(\Delta^2)\,\Delta{q}^{i}(z,\xi,Q^2)\nonumber\\
E^{i}(z,\xi,\Delta^2,Q^2)&=&F^{i}_{2}(\Delta^2)\,r^{i}(z,\xi,Q^2)
\label{deltadropping}
\ea
where $q^{i}(z)$ and $\Delta q^{i}(z)$ are obtained from the standard non-polarized and longitudinally polarized (forward) quark distributions using a specific diagonal ansatz \cite{gluck&reya&vogt98}.
The ansatz $r^{i}(z,\xi)=q^{i}(z,\xi)$ is also necessary in order for the quark sum rule to hold \cite{Ji3}.
Analogously, in the case of the $\tilde{E}^{i}$ distributions \cite{Rad1,Man1,Fran1} one can use the
special model
\ba
\tilde{E}^{u}=\tilde{E}^{d}=\frac{1}{2\xi} \theta(\xi - |z|)\phi_{\pi}(z/\xi)g_{\pi}(\Delta^2), && g_{\pi}(\Delta^2)=\frac{4 g_{A}^{(3)}M^2}{m_{\pi}^2 -\Delta^2},\;\;\;\;\;\;\;\phi_{\pi}(x)=\frac{4}{3}(1-x^2)\nonumber\\
\ea
valid at small $\Delta^2$,
where $g_{A}^{(3)}=1.267$, $M$ is the nucleon mass and $m_{\pi}$ is the pion mass, with the normalization
\ba
F^{i}_{1}(0)=G^{i}_{1}(0)=1.
\ea
Notice that, analogously to the $H$ distributions, $q^{i}(z,\xi,Q^2)$ and $\Delta q^{i}(z,\xi,Q^2)$, which describe the
$\Delta^2=0$ limit of the H-distributions, have support in $[-1,1]$ and, again,
they describe quark distributions (for $z>0$) and antiquark distributions (for $z<0$)
\ba
&&\bar{q}^{i}(z,\xi,Q^2)=- q^{i}(-z,\xi,Q^2)\nonumber\\
&&\Delta \bar{q}^{i}(z,\xi,Q^2)=\Delta q^{i}(-z,\xi,Q^2).
\ea
Now we're going to estabilish a connection between the $q(z,\xi,\bar{Q}^2)$ and the $\hat{{\cal{F}}}^q(X,\zeta)$ functions,
which is done using Radyushkin's nonforward  ``double distributions'' \cite{Rad1}.
The construction of the input distributions,
in correspondence of an input scale $Q_0$, is performed following a standard strategy. This consists in generating nonforward double distributions $f(x,y)$ from the forward ones ($f(x)$) using a ``profile function'' $\pi(x,y)$ \cite{Rad3}
\ba
f(y,x)=\pi(y,x)f(x),
\label{doub}
\ea
where we just recall that the $\pi(y,x)$ function can be represented by
\ba
\pi(y,x)=\frac{3}{4}\frac{[1-|x|]^2 -y^2}{[1-|x|]^3}\,,
\ea
taken to be of an asymptotic shape (see ref.\cite{Rad3,radmus}) for quarks and gluons.
A more general profile is given by
\begin{equation}
\pi(x,y) = \frac{\Gamma(2b + 2)}{2^{2b+1} \Gamma^2 (b+1)} \frac{[(1 -|x|)^2 - y^2]^b}{(1 -|x|)^{2b+1}}\,\,
\label{profile}
\end{equation}
and normalized so that
\ba
&&\int^{1-|x|}_{-1+|x|} dy \, \pi(x,y)= 1 \, .
\label{profile2}
\ea
$b$ parameterizes the size of the skewing effects starting from the diagonal input. Other choices of
the profile function are also possible.
For instance, the double distributions (DD) defined above have to satisfy a symmetry constraint
based on hermiticity. This demands that these must be symmetric with respect to the exchange $y\longleftrightarrow1-x-y$, and a profile function which respects this symmetry constraint is given by \cite{Piller}
\ba
\label{constraint}
\pi(x,y)=\frac{6y(1-x-y)}{(1-x)^3}\,.
\ea
This symmetry is crucial for establishing proper analytical properties of meson production amplitudes. 
We will be using below this profile and
compare the cross section obtained with it against the one obtained with (\ref{profile}).

Now we are able to generate distributions $q(z,\xi,Q^2)$ in the $z$ variable at $\Delta^2=0$,
$q(z,\xi,Q^2)$, by integrating over the
longitudinal fraction of momentum exchange $y$ characteristic of the double distributions
\ba
q(z,\xi,Q^2)=\int_{-1}^{1}dx'\int_{-1+|x'|}^{1-|x'|}dy'\delta(x'+\xi y'-z)f(y',x',Q^2).
\label{reduction}
\ea
Using (\ref{reduction}) and the expression of the profile functions introduced above, 
the GBM distributions are generated by the relation
\ba
&&\hat{{\cal F}}^{q,a} (X,\zeta) = \frac{2}{\zeta} \int^{X}_{\frac{X-\zeta}{1-\zeta}} dx'
\pi^q \left (x', \frac{2}{\zeta} (X - x') + x' -1 \right) q^a (x') \, .
\ea
with a similar expression for the anti-quark distributions in the DGLAP region $X > \zeta ~(z < - \xi)$
\ba
&&\hat{{\cal F}}^{\bar q,a} (X,\zeta) = \frac{2}{\zeta} \int^{\frac{-X+\zeta}{1-\zeta}}_{-X} dx' \pi^q \left(x', -\frac{2}{\zeta} (X + x') + x' +1 \right) {\bar q}^a(|x'|).
\label{DGLAPqbar}
\ea
\begin{figure}[t]
{\par\centering \resizebox*{10cm}{!}{\rotatebox{-90}{\includegraphics{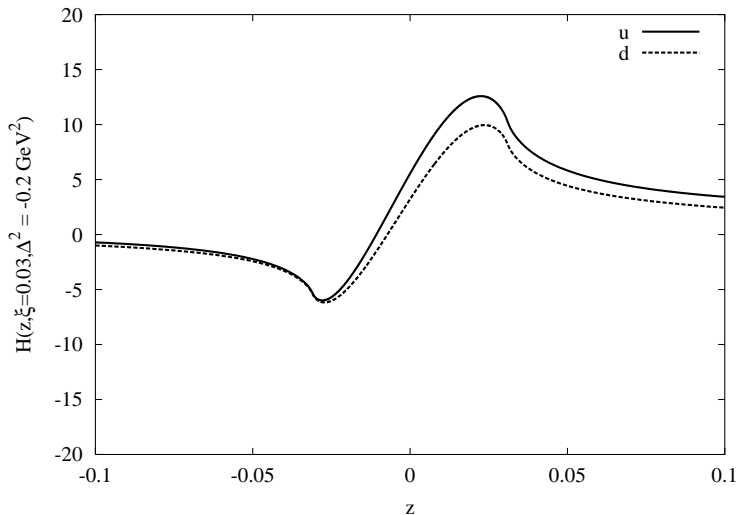}}} \par}
\caption{GPD's $H_u$ and $H_d$ generated by the diagonal parton distribution with a profile function (\ref{profile}) at an initial $0.26$ GeV$^2$}
\label{Hud}
\end{figure}
In the ERBL region, $ X < \zeta ~(|z| < \xi$), after the integration over $y$, 
we are left with the sum of two integrals
\ba
\hspace{0cm}\hat{{\cal F}}^{q,a} (X,\zeta) = \frac{2}{\zeta} \left[\int^{X}_{0} dx' \pi^q \left(x', \frac{2}{\zeta} (X - x') + x' -1 \right) q^a (x') \, - \right. \nonumber \\
\hspace{0cm}\left. \int^{0}_{X-\zeta} dx' \pi^q \left(x', \frac{2}{\zeta} (X - x') + x' -1 \right) {\bar q}^a (|x'|) \right] \, , \nonumber \\ \\
\hspace{0cm}\hat{{\cal F}}^{{\bar q},a} (X,\zeta) = - \frac{2}{\zeta} \left[ \int^{\zeta - X}_{0} dx' \pi^q \left(x', -\frac{2}{\zeta} (X + x') + x' + 1 \right) q^a (x') \, - \right. \nonumber \\
\hspace{0cm}\left. \int^{0}_{-X} dx' \pi^q \left(x', -\frac{2}{\zeta} (X + x') + x' + 1  \right) {\bar q}^a (|x'|) \right] \, .
\label{erblqqbar1}
\ea
\begin{figure}[t]
{\par\centering \resizebox*{10cm}{!}{\rotatebox{-90}{\includegraphics{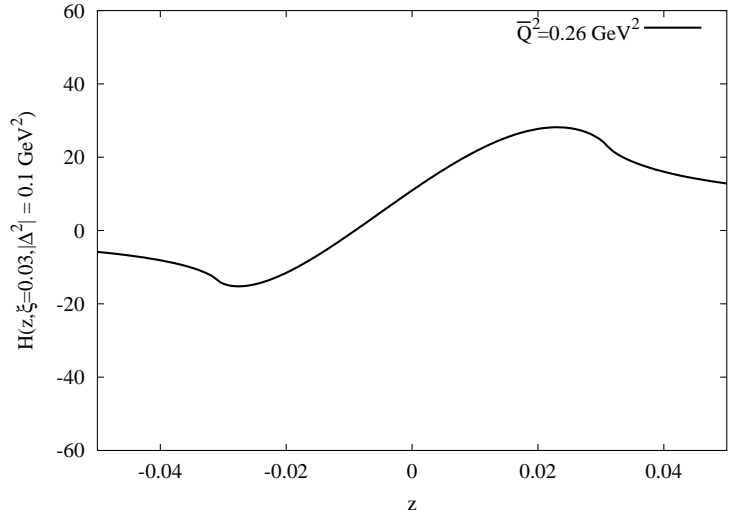}}}\par}
\caption{GPD's flavour singlet combination at $0.26$ GeV$^2$ generated with a profile (\ref{profile})}
\label{HS}
\end{figure}

Solving numerically the integrals we obtain the value of the function ${\cal{F}}^{q}$s on a grid, and using eqs.~(\ref{curlyq}) and \ref{curlyqbar} we end up with the numerical form of the H-distributions.
We have used diagonal parton distribution functions at $0.26$ GeV$^2$ \cite{gluck&reya&vogt98}
and the results of our numerical implementation can be visualized in Figs.~\ref{Hud} and \ref{HS}.

\section{The Differential Cross Section}
Our kinematical setup is illustrated in Fig.~\ref{kinematic.eps}, and we choose momenta in the target frame with the
following parameterizations
\ba
l=\left(E,0,0,E\right)\,,&& l'=\left(E',E'\cos{\phi_{\nu}}\sin{\theta_{\nu}},E'\sin{\phi_{\nu}}\sin{\theta_{\nu}},E'\cos{\theta_{\nu}}\right)\,,\nonumber\\
P_1 =\left(M,0,0,0\right)\,,&& P_2 =\left(E_2,|P_2|\cos{\phi_{N}}\sin{\theta_{N}},|P_2|\sin{\phi_{N}}\sin{\theta_{N}},|P_2|\cos{\theta_{N}}\right)\,
\ea
where the incoming neutrino is taken in the positive $\hat{z}$-direction and the nucleon is originally at rest. The first plane
is identified by the momenta of the final state nucleon and of the incoming neutrino, while the second plane is spanned by the
final state neutrino and the same $\hat{z}$-axis. $\phi_\nu$ is the angle between the $\hat{x}$ direction and the second plane, while
$\phi_{N}$ is taken between the plane of the scattered nucleon and the same $\hat{x}$ axis.
We recall that the general form of a differential cross section is given by
\ba
d\sigma =\frac{1}{4(l\cdot P_1)}|{\cal M}_{fi}|^2 (2\pi)^4 \delta^{(4)} (l+ P_1 -P_2 - l'-q_2 )\frac{d^3 \vec{l'}}{2l^{'0}(2\pi)^3}\frac{d^3 \vec{P_2}}{2P_{2}^{0}(2\pi)^3}\frac{d^3 \vec{q_2}}{2q_{2}^{0}(2\pi)^3}\,\nonumber\\
\ea
and it will be useful to express it in terms of standard quantities appearing in a standard
DIS process such as Bjorken variable $x$,
inelasticity parameter $y$, the momentum transfer plus some additional kinematical variables
typical of DVCS such as the asymmetry parameter $\xi$ and $\Delta^2$.

\begin{figure}[t]
{\par\centering \resizebox*{10cm}{!}{\rotatebox{0}{\includegraphics{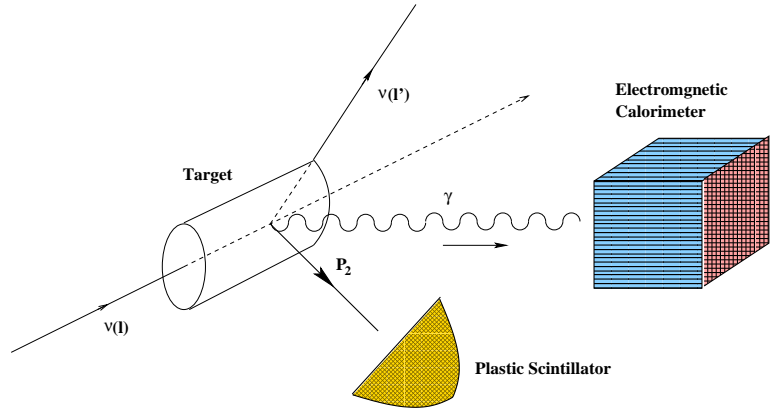}}}\par}
\caption{A pictorial description of the DVNS experimental setup, where the recoiled nucleon is
detected in coincidence with a final state photon.}
\label{Target.eps}
\end{figure}

\begin{figure}
{\centering \resizebox*{9cm}{!}{\rotatebox{0}{\includegraphics{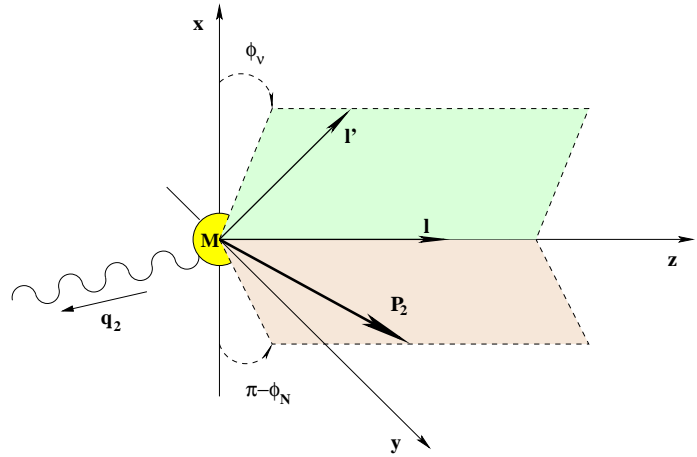}}}\par}
\caption{Kinematics of the process $\nu(l)N(P_1)\rightarrow\nu(l')N(P_2)\gamma(q_2)$}
\label{kinematic.eps}
\end{figure}
We will be using the relations
\ba
&&\bar{q}^2 = -\frac{1}{2} q_1^2 \left(1-\frac{\Delta^2}{2q_1^2} \right) \approx \frac{1}{2} Q^2\nonumber\\
&&\xi=\frac{x\left(1 -\frac{\Delta^2}{2q_1^2}\right)}{2-x\left(1 -\frac{\Delta^2}{2q_1^2}\right)}\approx \frac{2 x}{2-x}\nonumber\\
\ea
in the final computation of the cross section.
It is also important to note that $\Delta^2$ has to satisfy a kinematical constraint
\ba
\Delta^{2}_{min}=-\frac{M^2 x^2}{1-x +\frac{x M^2}{Q^2}}\left(1+ {\cal O}\left(\frac{M^2}{Q^2}\right)\right)\,.
\ea
One possible cross section to study, in analogy to the DVCS case \cite{Freu&McD}, is the following
\ba
\frac{d\sigma}{dx d Q^2 d|\Delta^2| d\phi_{r}}=\frac{y}{Q^2}\frac{d\sigma}{dx dy d|\Delta^2| d\phi_{r}}=\frac{x y^2}{8\pi Q^4}\left(1+ \frac{4M^2 x^2}{Q^2}\right)^{-\frac{1}{2}}|{\cal M}_{fi}|^2.
\ea
where $\phi_{r}$ is the angle between the lepton and the hadron scattering planes.
To proceed we also need the relations
\ba
&&l\cdot n=\left[\frac{Q^2}{2 x y} -\frac{(1+\xi)}{2\xi}\frac{Q^2}{2}\right]\chi,\nonumber\\
&&l\cdot \tilde{n}=\frac{Q^2}{2\xi}+\frac{Q^2}{4\xi^2}\left[\frac{Q^2}{2xy}-\frac{(1+\xi)}{2\xi}\frac{Q^2}{2}\right]\chi,\nonumber\\
&&n\cdot q_1 = \frac{2x}{x-2},\hspace{0.5cm}\tilde{n}\cdot q_1= Q^2 \frac{(2-x)}{8x},\nonumber\\
\ea
where $\chi$ is given by
\ba
\chi=\frac{\xi}{\frac{1+\xi}{2} \frac{Q^2}{4\xi} +\frac{\xi(1-\xi)}{2} \frac{\overline{M}^2}{2}}\,\,.
\ea
After some manipulations we obtain a simplified expression for $|{\cal M}_{fi}|^2$, similarly to eq.~(\ref{Fact1})
\ba
|{\cal M}_{fi}|^2 &=& \int_{-1}^1 dz \int_{-1}^1 dz' \left[A_1(z,z',x,t,Q^2)\alpha(z)\alpha^*(z') + A_2(z,z',x,t,Q^2)\beta(z)\beta^*(z')\right]\nonumber\\
& \times &\left[-2 Q^2 \,(4 M^2 - t)\,(x-2)^2 \,(x-1) \,x^2 y + (t-4 M^2)^2 \,(x-1)^2 \,x^4 y^2\right.\nonumber\\
&+& \left.Q^4 (x-2)^4 \,(2 - 2 y + y^2)\right]\nonumber\\
&\times&\left[2 M^2 ({M_Z}^2 + Q^2)^2 \, (x-2)^2\, [Q^2 (x-2)^2 - (4 M^2 -t) (-1 + x) x^2]^2 y^2\right]^{-1}
\label{amplicompact}
\ea
where $A_1(z,z',x,t,Q^2)$ and $A_2(z,z',x,t,Q^2)$ are functions of the invariants of the
process and of the entire set of GPD's. Their explicit form is given in Appendix D.
As we have already mentioned, the $(z,z')$ integration is be done by using the Feynman prescription to extract the phases and then
using the distributional identities (\ref{rex1})and (\ref{rex2}). For numerical accuracy we have discretized the final integrals by finite element
methods, as shown in Appendix C.

\begin{figure}
{\centering \resizebox*{9cm}{!}{\rotatebox{-90}{\includegraphics{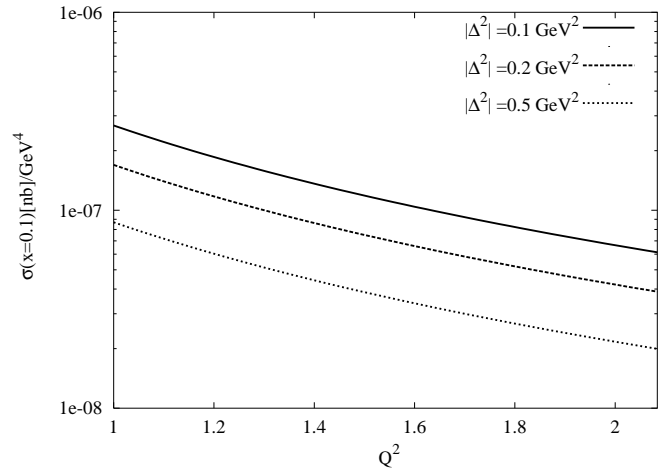}}} \par}
\caption{ DVCS cross section at $x=0.1$ and center of mass energy $M E=10$ GeV$^2$ using the profile (\ref{profile}).}
\label{set2.ps}
\end{figure}

\begin{figure}
{\centering \resizebox*{9cm}{!}{\rotatebox{-90}{\includegraphics{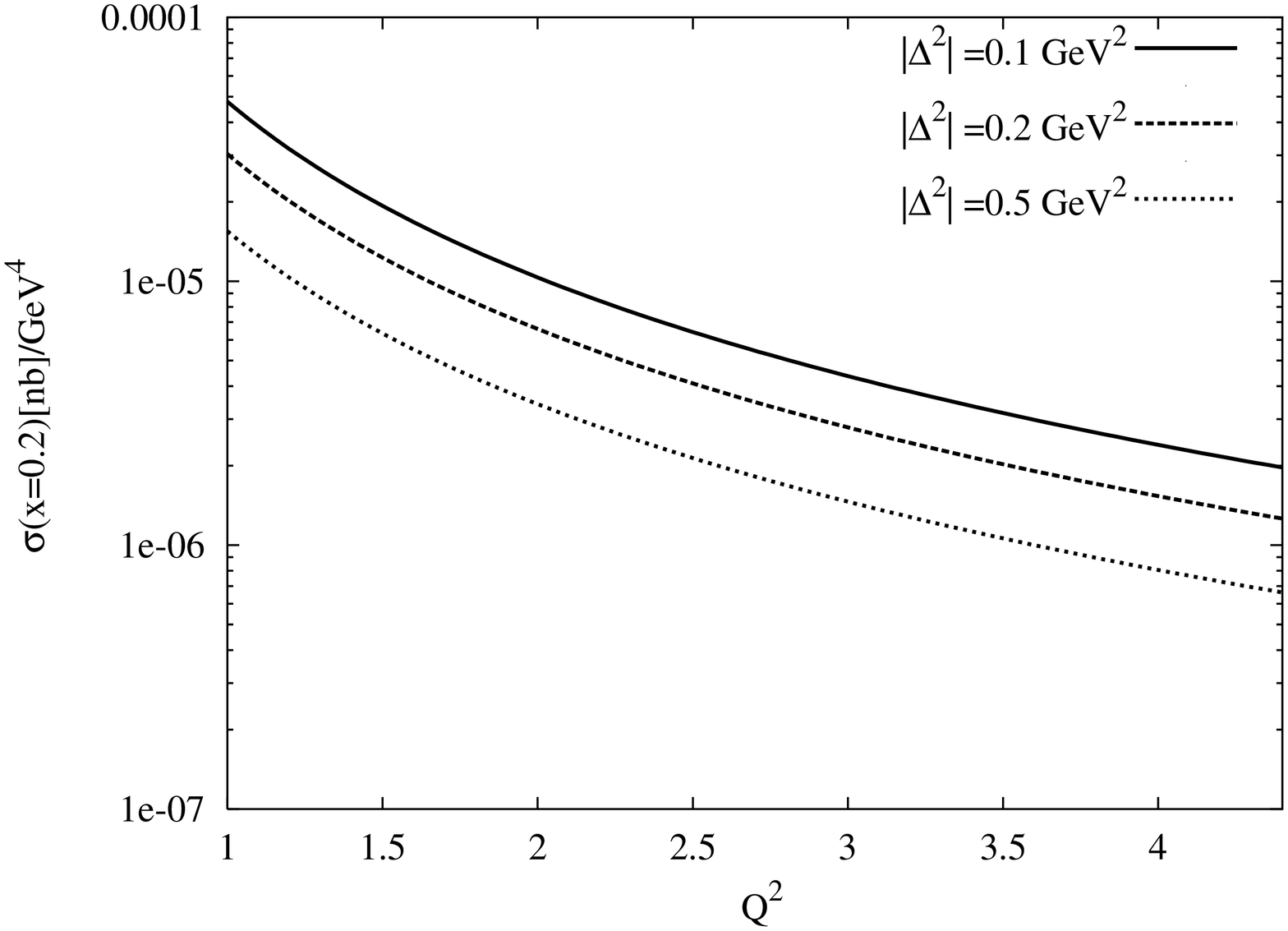}}} \par}
\caption{ DVCS cross section at $x=0.2 $ and center of mass energy $M E=10$ GeV$^2$ using the profile (\ref{profile}).}
\label{set3.ps}
\end{figure}

\begin{figure}
{\centering \resizebox*{9cm}{!}{\rotatebox{-90}{\includegraphics{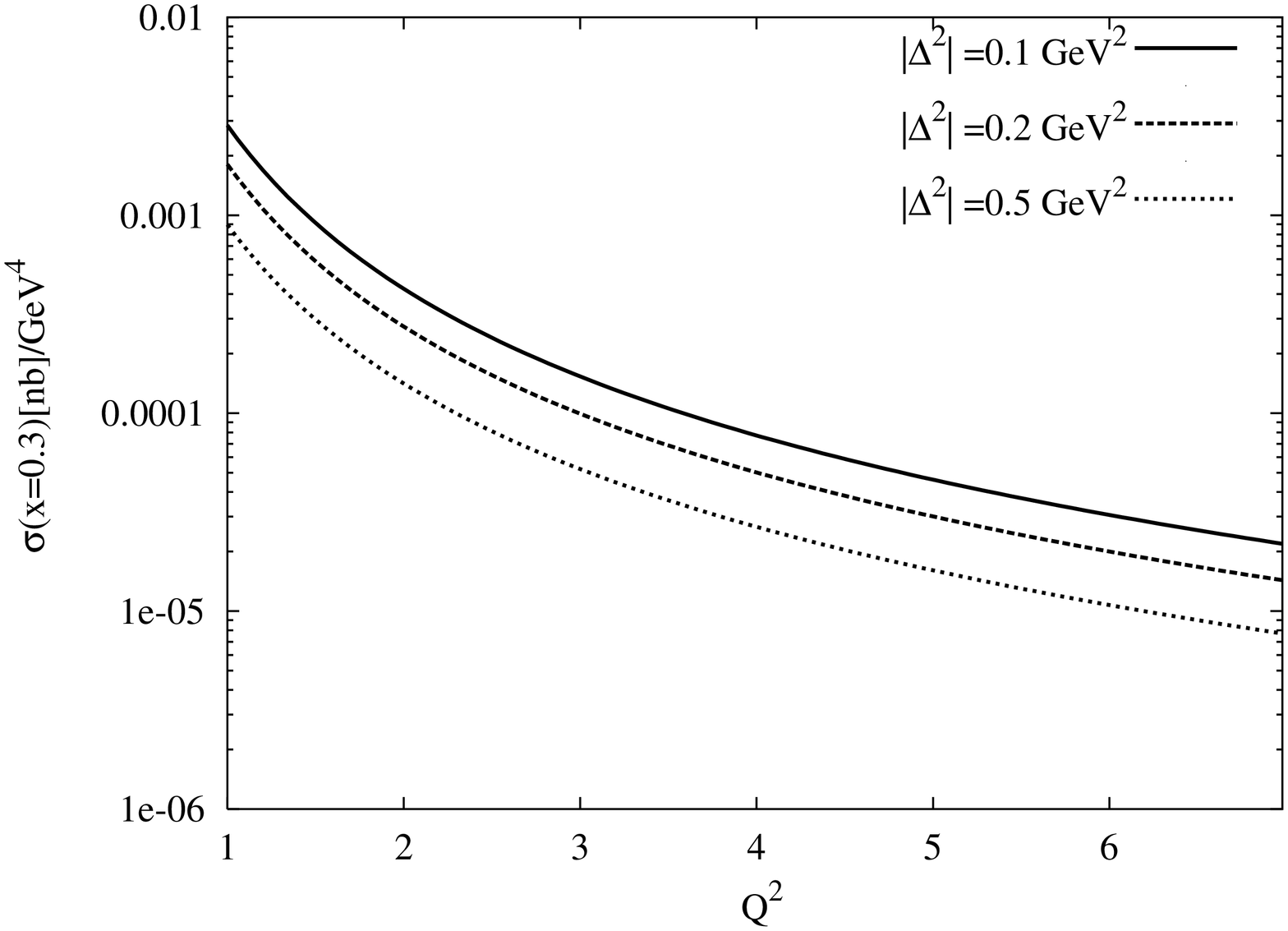}}} \par}
\caption{ DVCS cross section at $x=0.3$ and center of mass energy $M E=10$ GeV$^2$ using the profile (\ref{profile}).}
\label{set4.ps}
\end{figure}

\begin{figure}
{\centering \resizebox*{9cm}{!}{\rotatebox{-90}{\includegraphics{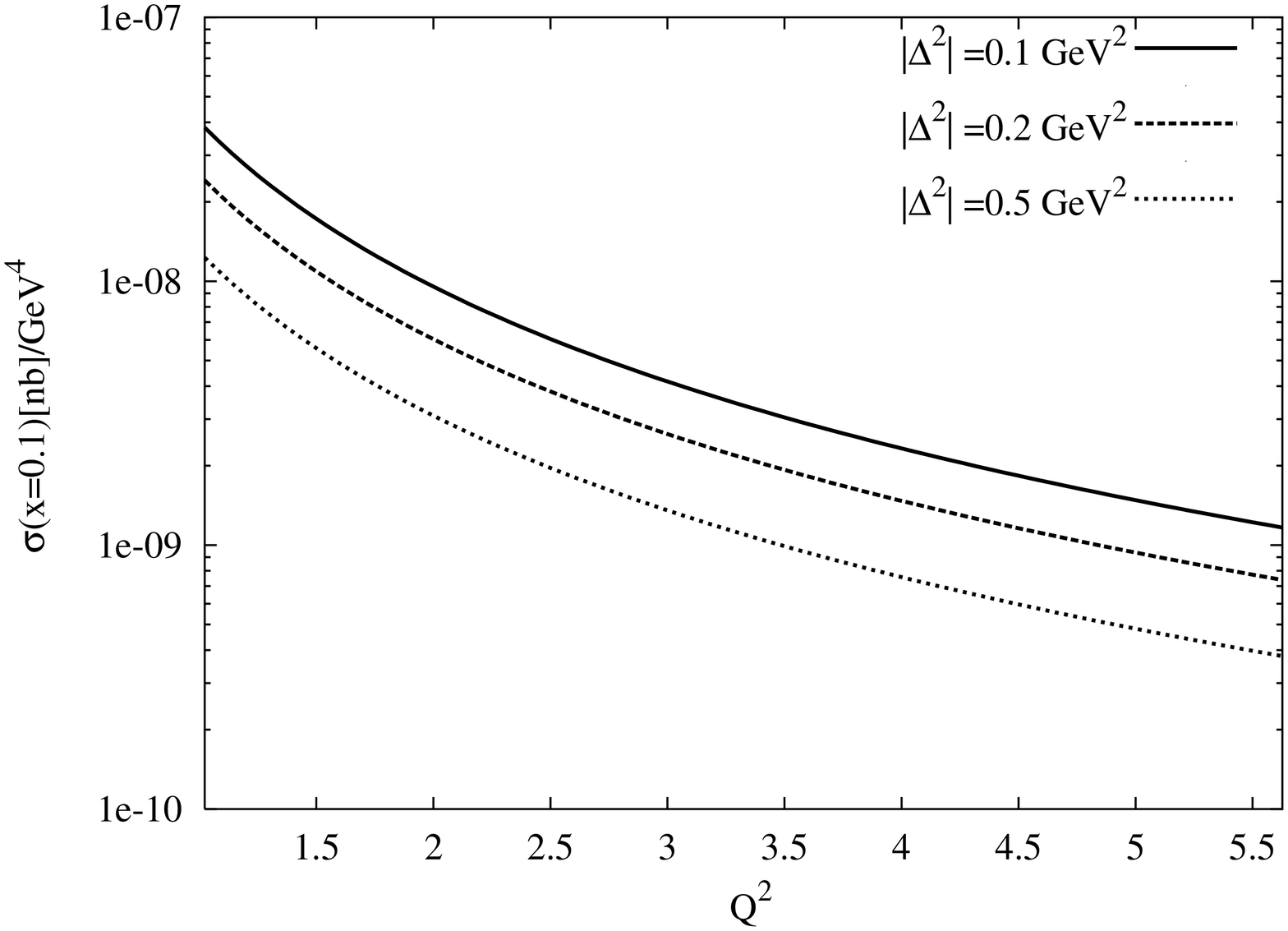}}} \par}
\caption{ DVCS cross section at $x=0.1$ and center of mass energy $M E=27$ GeV$^2$ using the profile (\ref{profile}).}
\label{set5.ps}
\end{figure}

\begin{figure}
{\centering \resizebox*{9cm}{!}{\rotatebox{-90}{\includegraphics{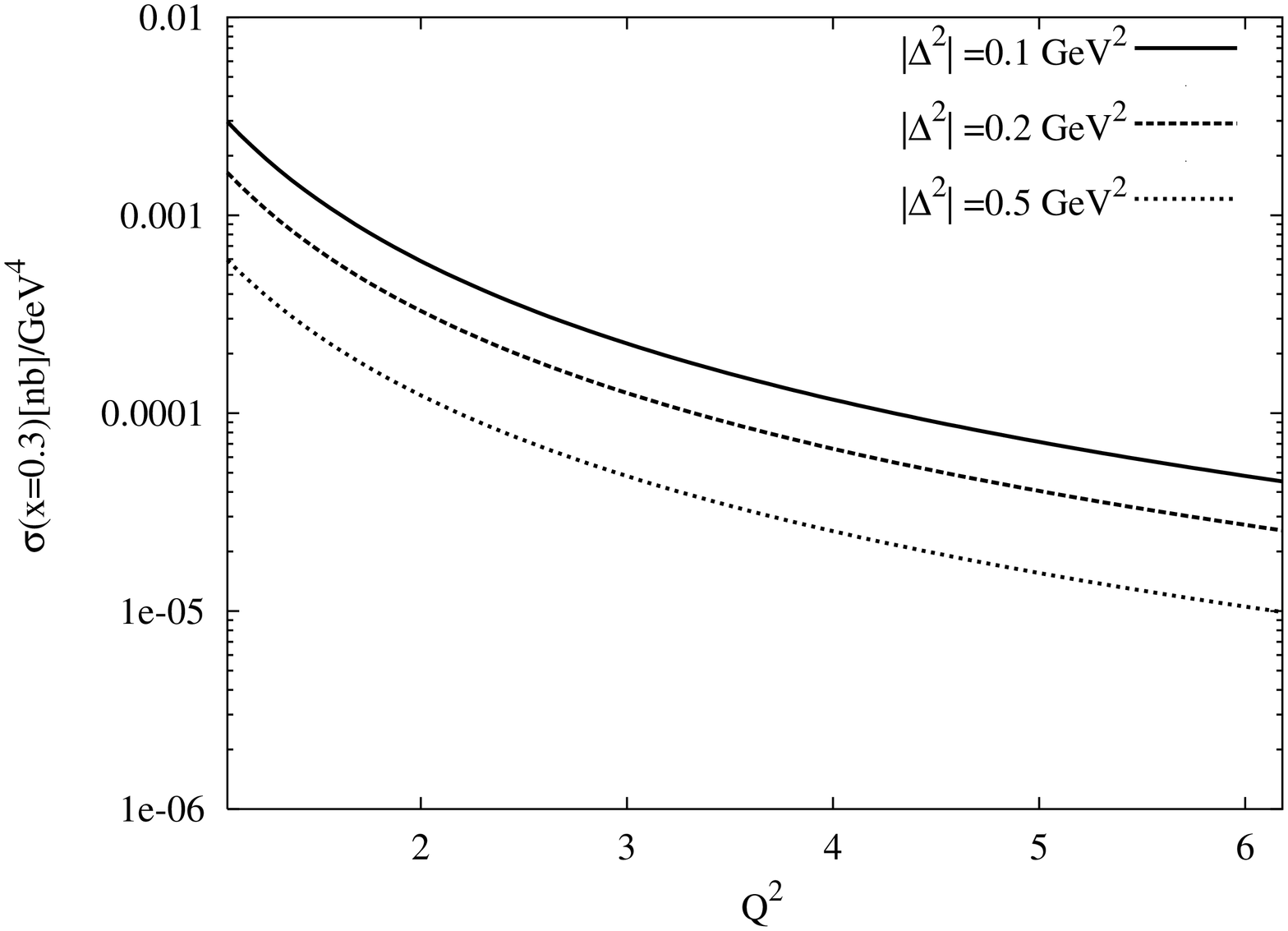}}} \par}
\caption{ DVCS cross section at $x=0.3$ and center of mass energy $M E=10$ GeV$^2$ with NPD functions generated by the profile function
(\ref{constraint}).}
\label{new_set10.ps}
\end{figure}

\begin{figure}
{\centering \resizebox*{9cm}{!}{\rotatebox{-90}{\includegraphics{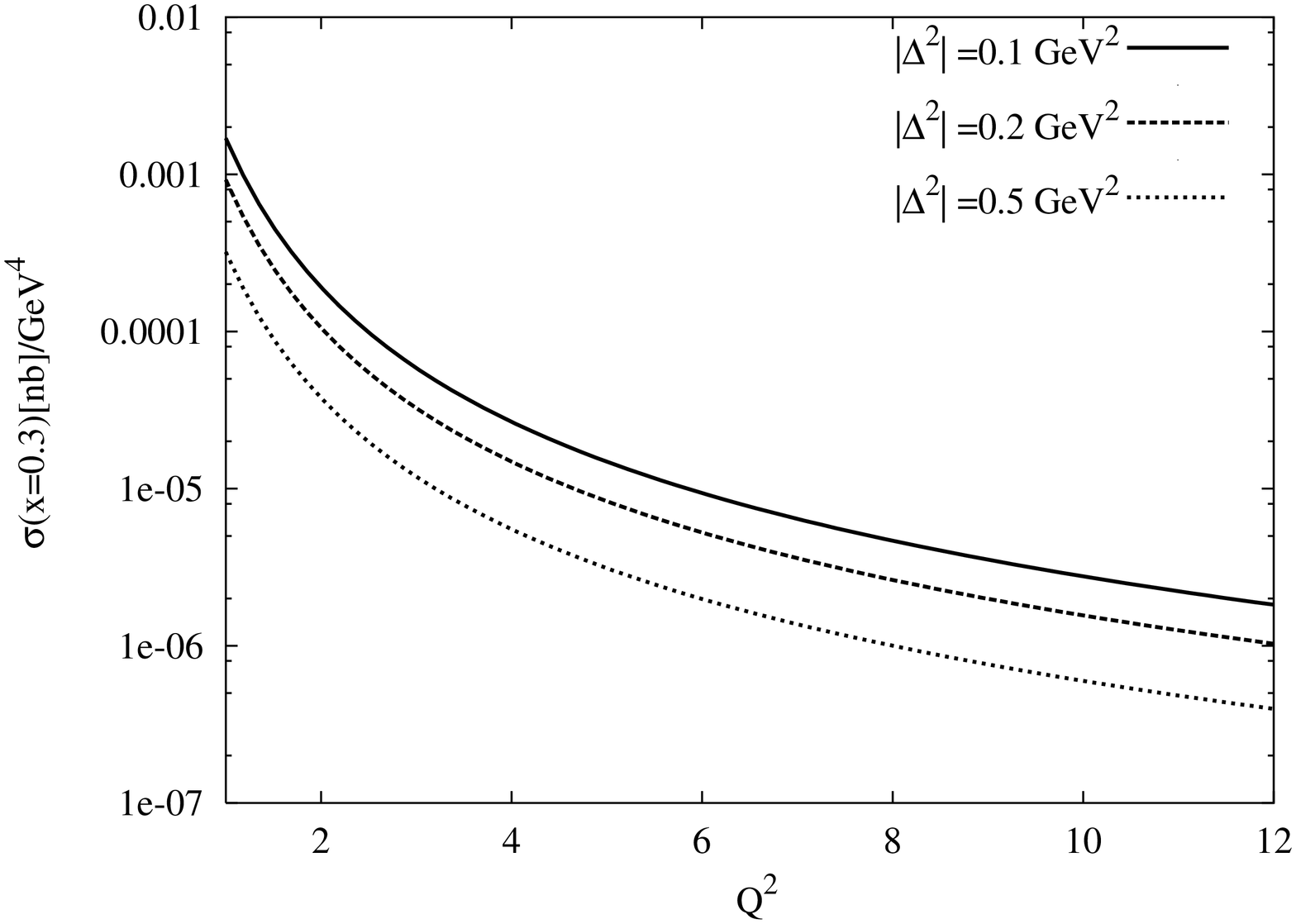}}} \par}
\caption{ DVCS cross section at $x=0.3$ and center of mass energy $M E=27$ GeV$^2$ with NPD functions generated by the profile function
(\ref{constraint}).}
\label{new_set27.ps}
\end{figure}

\begin{figure}
{\centering \resizebox*{9cm}{!}{\rotatebox{-90}{\includegraphics{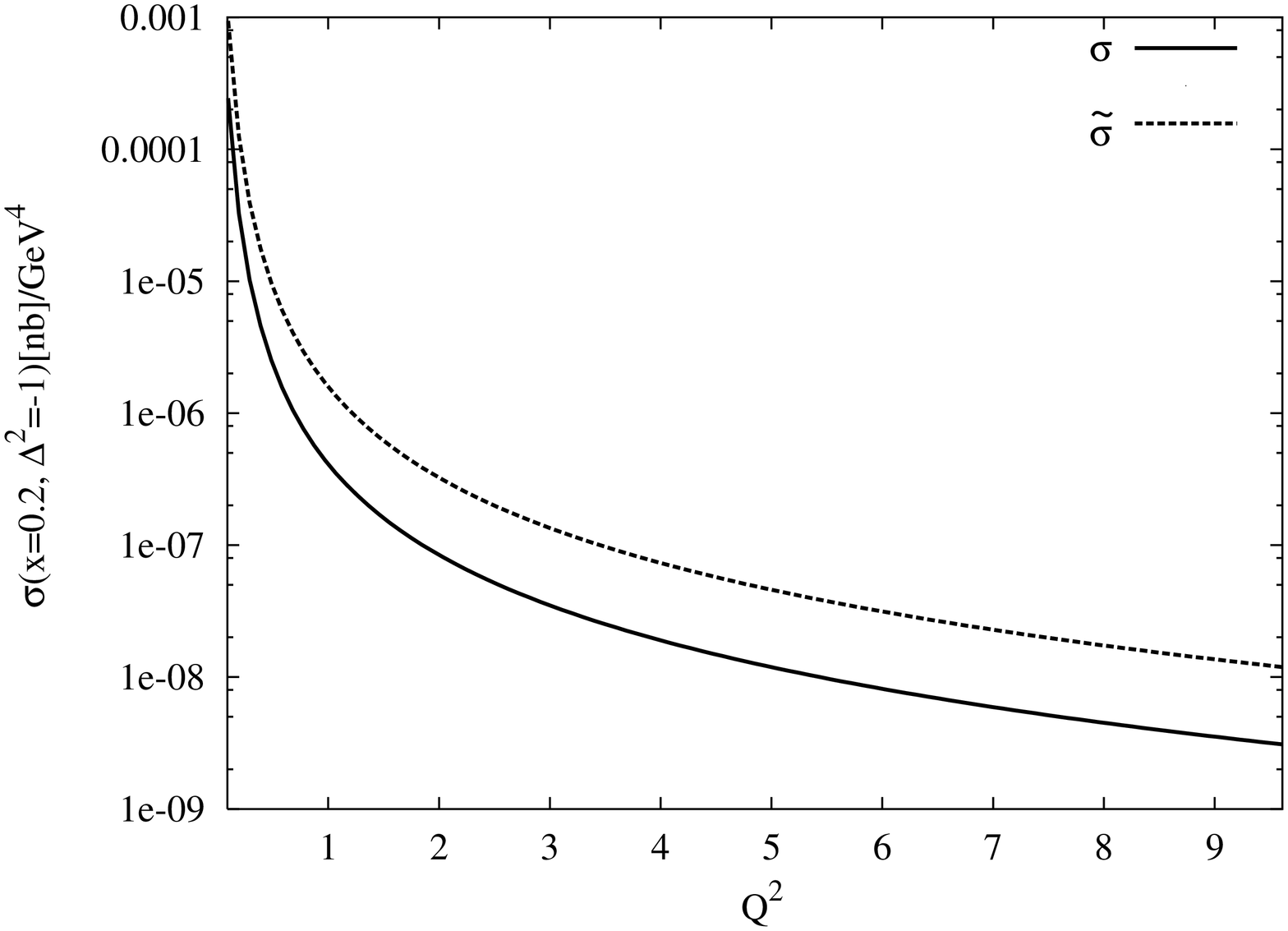}}} \par}
\caption{DVCS cross sections at $x=0.2$ and center of mass energy $M E=27$ GeV$^2$.
$\tilde{\sigma}$ using profile (\ref{constraint}).}
\label{new_sez.ps}
\end{figure}

We have plotted the differential cross section as a function of $Q^2$, for various values of
$\Delta^2$ and at fixed $x$ values. We have used both the profile given by (\ref{profile}) (Figs.~\ref{set2.ps}-\ref{set5.ps}) and the
one given in eq.~(\ref{constraint}) (Figs.~\ref{new_set10.ps} and \ref{new_set27.ps}) and their direct comparison  in a specific kinematical
region (Fig.~\ref{new_sez.ps}). The different profiles generate differences in the cross sections especially for larger
$\Delta^2$ values. Notice also that the DVNS cross section decreases rather sharply with $\Delta^2$, at the same time it increases
appreciably with $x$. The results shown are comparable with other cross sections evaluated in the quasi-elastic region
($\approx 10^{-5}$ nb) for charged and neutral current interactions, and appear to be sizeable.
Coherence effects due to neutral current interactions with heavy nuclei, in particular with the neutron component
may substantially increase
the size of the cross sections, with an enhancement proportional to $N^2$, where $N$ is the number of neutrons
\cite{PK}, though an accurate quantification of these effects requires a special study \cite{CCV} which is underway.
It is worth to emphasize that in the past this contribution
had never been included in the study of neutrino-nucleon interactions since very little was  known about the intermediate
energy kinematics in QCD from the point of view of factorization. It seems obvious to us that with the new developments now taking place in the study of QCD
at intermediate energy, especially in the case of the generalized Bjorken region, of which the deeply virtual scattering limit
is just a special case, it will be of wide interest to quantify with accuracy the role of these new contributions for neutrino factories.
In general, one expects that electromagnetic effects are suppressed compared to the standard (hadronic) deeply inelastic cross section,
and this has led in the past to a parameterization of the intermediate energy cross section as either dominated by the quasi elastic region and/or by the DIS region at higher energies, as we have mentioned in our introduction.
However, the exclusive cross section has some special positive features, one of them being to provide
a clean signal  for the detection of weakly interacting particles,
and we expect that this aspect is going to be of relevance at experimental level.

\section{Conclusions}

We have presented an extension of the standard DVCS process to the case of one neutral current exchange,
describing the scattering of a neutrino off a proton in the parton model.
We have described the leading twist behaviour of the cross section; we have found that this is
comparable to other typical
neutrino cross sections and discussed its forward or DIS limit. We have presented a complete
formalism for the study
of these processes in the parton model. The process is the natural generalization of DIS with
neutral currents and relies on
the notion of Generalized Parton Distributions, new constructs in the parton model which
have received considerable attention in recent years.
The possible applications of these new processes are manifold and we hope to return in the
near future with a discussion of some of the issues not addressed in this thesis.

\section{Appendix A}

The collinear expansion of the internal loop momentum $k$ allows to identify the light
cone operators appearing in the process at leading twist.
We recall at this point that the analysis of the hand-bag contribution is carried
out exactly as in the electromagnetic case.

To perform the collinear expansion and isolate the light-cone correlators of
DVNS from the hand-bag contribution
we use the relation
\begin{eqnarray}
\int \frac{d\lambda \ dx}{2 \pi} e^{i \lambda (x - k\cdot n)} = 1
\end{eqnarray}
inside the expression of $T^{\mu \nu}$ in order to obtain
\begin{eqnarray}
T^{\mu \nu} &=& - \int \frac{d^4k}{(2 \pi)^4} \int \frac{d\lambda \ dx}{2 \pi}
e^{i \lambda (x - k\cdot n)}  \nonumber \\
&& Tr\left\{ \left[ \gamma^\nu
\frac{i}{\rlap/k - \alpha \rlap/\Delta + \rlap/q_1 + i \epsilon} \gamma^\mu +
\gamma^\mu \frac{i}{\rlap/k + (1-\alpha) \rlap/\Delta - \rlap/q_1 + i \epsilon} \gamma^\nu\right] \underline{M}(k) \right\} \nonumber
\end{eqnarray}
and therefore
\begin{eqnarray}
T^{\mu \nu} &=& - \int \frac{d k\cdot n \ d k\cdot \tilde{n} \ d k_\perp^2}{(2 \pi)^4} \int \frac{d\lambda \ dx}{2 \pi}e^{i \lambda (x - k\cdot n)}  \int d^4z \ e^{i k\cdot z} \nonumber \\
&& Tr\left\{ \left[ \gamma^\nu
\frac{i}{\rlap/k - \alpha \rlap/\Delta + \rlap/q_1 + i \epsilon} \gamma^\mu +\gamma^\mu \frac{i}{\rlap/k + (1-\alpha) \rlap/\Delta - \rlap/q_1 + i \epsilon} \gamma^\nu\right] \right. \nonumber \\
&\cdot& \left. \langle P'|
\overline{\psi} (-\alpha z) \psi ((1-\alpha)z) |P \rangle \right\}. \nonumber
\end{eqnarray}
Keeping the leading terms for the loop momenta
\begin{eqnarray}
k^\mu - \alpha \Delta^\mu + q_1^\mu &=& \tilde{n}^\mu \left(k\cdot n + 2\alpha \xi - 2\xi\right) + n^\mu \left(k\cdot \tilde{n} - {\alpha}\xi \overline{M}^2 + \frac{Q^2}{4 \xi} \right) \nonumber  \\
k^\mu +(1-\alpha) \Delta^\mu - q_1^\mu &=& \tilde{n}^\mu \left( k\cdot n -2(1-\alpha) \xi + 2\xi \right) +
n^\mu \left( k\cdot \tilde{n} +2(1-\alpha) \xi \overline{M}^2 - \frac{Q^2}{4 \xi} \right) \nonumber
\end{eqnarray}
we thus obtain
\begin{eqnarray}
T^{\mu \nu} &=&  - \int \frac{d (k\cdot n) \ d (k\cdot\tilde{n}) \ d k_\perp^2}{(2 \pi)^4} \int \frac{d\lambda \ dx}{2 \pi}e^{i \lambda (x - k\cdot n)}  \int d^4z \ e^{i k\cdot z} \nonumber \\
&& Tr\left\{ \left[ \gamma^\nu \frac{\rlap/\tilde{n}}{2 \left(k\cdot\tilde{n}  - {\alpha} \xi \overline{M}^2 +
\frac{Q^2}{4 \xi} \right)} \gamma^\mu +\gamma^\mu \frac{\rlap/\tilde{n}}{2 \left( k\cdot\tilde{n}  +(1-\alpha){\xi}
\overline{M}^2 - \frac{Q^2}{4 \xi} \right)} \gamma^\nu \right. \right.\nonumber \\
&+& \left. \left.\gamma^\nu \frac{\rlap/n}{2 \left( k\cdot n + 2\alpha \xi - 2\xi
\right)} \gamma^\mu +
\gamma^\mu \frac{\rlap/n}{2 \left( k\cdot n -2(1-\alpha) \xi + 2\xi\right)} \gamma^\nu \right] \right. \nonumber \\
&\cdot& \left. \langle P'|
\overline{\psi} (-\alpha z) \psi ((1-\alpha)z) |P \rangle \right\}. \nonumber
\end{eqnarray}
We expand
\begin{eqnarray}
k\cdot z &=& (k\cdot n) \ (\tilde{n}\cdot z) + (k\cdot\tilde{n}) \ (n\cdot z) - \vec{k}_\perp\cdot \vec{z}_\perp \nonumber
\end{eqnarray}
and choose $\alpha = 1/2$. These expansions are introduced in eq.~\ref{coll} and after some manipulations the tensor $T$ now becomes
\begin{eqnarray}
T^{\mu \nu} |_{\alpha=1/2} &=&  -
\int \frac{d (k\cdot n) \ d (k\cdot\tilde{n} ) \ d k_\perp^2}{(2 \pi)^4} \int \frac{d\lambda \ dx}{2 \pi}e^{i \lambda (x - k\cdot n)}  \int d^4z \ e^{i k\cdot z} \nonumber \\
&& Tr\left\{ \left[ \gamma^\nu\frac{\rlap/\tilde{n}}{2 \left(k\cdot\tilde{n} - \frac{\alpha}{2} \xi \overline{M}^2 +
\frac{Q^2}{2 \xi} \right)} \gamma^\mu +\gamma^\mu \frac{\rlap/\tilde{n}}{2 \left( k\cdot \tilde{n} +(1-\alpha){\xi}
\overline{M}^2 - \frac{Q^2}{4 \xi} \right)} \gamma^\nu \right. \right.\nonumber \\
&+& \left. \left.\gamma^\nu \frac{\rlap/n}{2 \left( k\cdot n + 2\alpha \xi - 2\xi
\right)} \gamma^\mu +\gamma^\mu \frac{\rlap/n}{2 \left(
k\cdot n -2(1-\alpha) \xi + 2\xi\right)} \gamma^\nu
\right] \right. \nonumber \\
&\cdot& \left. \langle P'|
\overline{\psi} (-\alpha z) \psi ((1-\alpha)z) |P \rangle \right\}. \nonumber
\end{eqnarray}
We also recall that the expansion of the matrix element $\underline{M}(k)$ proceeds also
in this case as in the electromagnetic case
\begin{eqnarray}
\underline{M}_{ab}^{(i)}(k) &=& \int d^4y e^{i k\cdot y} \langle P'|
\overline{\psi}_{a}^{(i)}(-\alpha y) \psi_{b}^{(i)}((1-\alpha)y) |P \rangle =
A_1 \slash{\tilde{n}} + A_2\gamma_5 \slash{\tilde{n}} +...
\label{Mmatrix}
\end{eqnarray}
where the ellipses refer to terms which are of higher twist or disappear in the trace of the diagram.

\section{Appendix B}

The last ingredients needed in the construction of the input distribution
functions are the form factors $F^i_1$ and $F^i_2$. From experimental measurements we know,
by a dipole parametrization in the small $\Delta^2$ region, that
\begin{equation}
G_E^p (\Delta^2)= (1 + \kappa_p)^{-1} G_M^p (\Delta^2)= \kappa_n^{-1} G_M^n (\Delta^2)
= \left( 1 - \frac{\Delta^2}{m_V^2} \right)^{- 2} ,\qquad
G_E^n (\Delta^2) = 0,
\end{equation}
where the electric, $G_E^i (\Delta^2) = F_1^i (\Delta^2) + \frac{\Delta^2}{4 M^2} F_2^i (\Delta^2)$, and magnetic
form factors $G_M^i (\Delta^2) = F_1^i (\Delta^2) + F_2^i (\Delta^2)$ are usually parametrized in terms
of a cutoff mass $m_V = 0.84\, {\rm GeV}$. \newline
For non-polarized GPD's the valence $u$ and $d$ quark form factors in the proton can be easily extracted from
$F_I^{({p \atop n})} = 2 \left( {Q_u\atop Q_d} \right) F_I^u + \left( {Q_d \atop Q_u} \right) F_I^d$  and
given exclicitely by
\begin{equation}
2 F^u_I (\Delta^2) = 2 F_I^p (\Delta^2) + F_I^n (\Delta^2),
\qquad
F^d_I (\Delta^2) = F_I^p (\Delta^2) + 2 F_I^n (\Delta^2) ,
\quad\mbox{for}\quad I = 1,2.
\end{equation}
This exploits the fact that proton and neutron form an iso-spin doublet.

At the scale $m_{A}=0.9$ GeV one can get
\ba
G^{i}_{1}(\Delta^2)=\left(1-\frac{\Delta^2}{m_{A}^{2}}\right)^{-2}
\ea
for the valence quarks.
For the form factors $F$ we obtain
\ba
&&F_1^u=-\frac{A \Delta^2}{M^2 (1 - B \Delta^2)^2 \left(-1 + \frac{\Delta^2}{4 M^2}\right)} + \frac{1 - \frac{C \Delta^2}{M^2\left(1 - \frac{\Delta^2}{4 M^2}\right)}}{(1 - B \Delta^2)^2},\nonumber\\
&&F_2^u=\frac{D}{(1 - B \Delta^2)^2\left(1 - \frac{\Delta^2}{4 M^2}\right)},\nonumber\\
&&F_1^d=-\frac{E \Delta^2}{M^2 (1 - B \Delta^2)^2\left(-1 + \frac{\Delta^2}{4 M^2}\right)}+\frac{1 - \frac{C \Delta^2}{M^2\left(1 - \frac{\Delta^2}{4 M^2}\right)}}{(1 - B \Delta^2)^2},\nonumber\\
&&F_2^d=\frac{F}{(1 - B \Delta^2)^2\left(1 - \frac{\Delta^2}{4 M^2}\right)},\nonumber\\
\ea
where
\ba
A=0.238\hspace{1cm} B=1.417\hspace{1cm} C=0.447,\nonumber\\
D=0.835\hspace{1cm} E=0.477\hspace{1cm} F=0.120.
\ea
\section{Appendix C}

In this section we illustrate an analytical computation of the integrals by discretization,
using finite elements method.
We want to approximate with high numerical accuracy integrals of the form
\ba
P.V.\int_{-1}^{1}\frac{H(z)dz}{z-\xi}=\int_{-1}^{\xi}dz\frac{H(z)-H(\xi)}{z-\xi}
+\int_{\xi}^{1}dz\frac{H(z)-H(\xi)}{z-\xi} + H(\xi)\ln\left|\frac{\xi-1}{\xi+1}\right|.
\ea
For this purpose we start by choosing a grid on the interval  $(-1=x_0,.....,x_{n+1}=\xi)$ and define
\ba
J_1=\int_{-1}^{\xi}dz\frac{H(z)-H(\xi)}{z-\xi}=\sum_{j=0}^{n}\int_{x_j}^{x_{j+1}}dx\frac{H(x)-H(\xi)}{x-\xi}\,.
\ea
Performing a simple linear interpolation we get
\ba
J_1&=&\sum_{j=0}^{n-1}\int_{x_j}^{x_{j+1}} \left\{H(x_j)\left[\frac{x_{j+1}-x}{x_{j+1}-x_j}\right]+ H(x_{j+1})\left[\frac{x-x_{j}}{x_{j+1}-x_j}\right]\right\} \frac{dx}{x-\xi}\,\,\nonumber\\
&+&\int_{x_n}^{\xi}\left\{H(x_n)\left[\frac{\xi-x}{\xi-x_n}\right]+ H(\xi)\left[\frac{x-x_{n}}{\xi-x_n}\right]\right\} \frac{dx}{x-\xi} -\int_{-1}^{\xi}dx\frac{H(\xi)}{x-\xi}\,\,.
\ea
After the integration we are left with
\ba
J_1 &=& \sum_{j=0}^{n-1} H(x_j)\left[-1 +\left(\frac{x_{j+1}-\xi}{x_{j+1}-x_j}\right)\ln\left|\frac{x_{j+1}-\xi}{x_j-\xi}\right|\right]\nonumber\\
&+&\sum_{j=0}^{n-1}H(x_{j+1})\left[1 +\left(\frac{\xi-x_{j}}{x_{j+1}-x_j}\right)\ln\left|\frac{x_{j+1}-\xi}{x_j-\xi}\right|\right]\nonumber\\
&-& \sum_{j=0}^{n-1}H(\xi)\ln\left|\frac{x_{j+1}-\xi}{x_j-\xi}\right|-H(x_n)+H(\xi)\,\,.
\ea
Now, moving to the integral in the interval $(\xi,1)$, we introduce a similar grid of equally spaced points
$(\xi=y_0,......,y_{n+1}=1)$ and define the integral
\ba
J_2=\int_{\xi}^{1}dz\frac{H(z)-H(\xi)}{z-\xi}=\sum_{j=0}^{n}\int_{y_j}^{y_{j+1}}dy\frac{H(y)-H(\xi)}{y-\xi}.
\ea
As above, after isolating the singularity we obtain
\ba
J_2&=&\sum_{j=1}^{n}\int_{y_j}^{y_{j+1}} \left\{H(y_j)\left[1-\frac{y-y_{j}}{y_{j+1}-y_j}\right]+ H(y_{j+1})\left[\frac{y-y_{j}}{y_{j+1}-y_j}\right]\right\} \frac{dy}{y-\xi}\,\,\nonumber\\
&+&H(y_1) + H(\xi)\int_{\xi}^{y_1}\left[\frac{y_1-y}{y_1-\xi}\right]dy\,\,-\,\,\int_{\xi}^{1}dy\frac{H(\xi)}{y-\xi}\,\,.
\ea
Again, performing the integrations we obtain
\ba
J_2 &=& \sum_{j=1}^{n} H(y_j)\left[-1 +\left(\frac{y_{j+1}-\xi}{y_{j+1}-y_j}\right)\ln\left|\frac{y_{j+1}-\xi}{y_j-\xi}\right|\right]\nonumber\\
&+&\sum_{j=1}^{n}H(y_{j+1})\left[1 +\left(\frac{\xi-y_{j}}{y_{j+1}-y_j}\right)\ln\left|\frac{y_{j+1}-\xi}{y_j-\xi}\right|\right]\nonumber\\
&-&\sum_{j=1}^{n}H(\xi)\ln\left|\frac{y_{j+1}-\xi}{y_j-\xi}\right|+H(y_1)-H(\xi)\,\,.
\ea
Collecting our results, at the end we obtain
\ba
P.V.\int_{-1}^{1}\frac{H(z)dz}{z-\xi}= J_1 + J_2 +H(\xi)\ln\left|\frac{\xi-1}{\xi+1}\right|\,.
\ea
We can use the same strategy for the integrals of ``$+$'' type defined as follows
\ba
P.V.\int_{-1}^{1}\frac{H(z)dz}{z+\xi}=\int_{-1}^{-\xi}dz\frac{H(z)-H(-\xi)}{z+\xi}
+\int_{-\xi}^{1}dz\frac{H(z)-H(-\xi)}{z+\xi} + H(-\xi)\ln\left|\frac{\xi+1}{\xi-1}\right|.
\ea
This time we call our final integrals  $X_1$ and $X_2$. They are given by the expressions
\ba
X_1&=&\sum_{j=0}^{n-1} H(x_j)\left[-1 +\left(\frac{x_{j+1}+\xi}{x_{j+1}-x_j}\right)\ln\left|\frac{x_{j+1}+\xi}{x_j+\xi}\right|\right]\nonumber\\
&+&\sum_{j=0}^{n-1}H(x_{j+1})\left[1 +\left(\frac{-\xi-x_{j}}{x_{j+1}-x_j}\right)\ln\left|\frac{x_{j+1}+\xi}{x_j+\xi}\right|\right]\nonumber\\
&-& \sum_{j=0}^{n-1}H(-\xi)\ln\left|\frac{x_{j+1}+\xi}{x_j+\xi}\right|-H(x_n)+H(-\xi)\,\,,
\ea
with a discretization supported in the $(-1=x_0,.....,x_{n+1}=\xi)$ grid, and
\ba
X_2&=&\sum_{j=1}^{n} H(y_j)\left[-1 +\left(\frac{y_{j+1}+\xi}{y_{j+1}-y_j}\right)\ln\left|\frac{y_{j+1}+\xi}{y_j+\xi}\right|\right]\nonumber\\
&+&\sum_{j=1}^{n}H(y_{j+1})\left[1 +\left(\frac{-\xi-y_{j}}{y_{j+1}-y_j}\right)\ln\left|\frac{y_{j+1}+\xi}{y_j+\xi}\right|\right]\nonumber\\
&-&\sum_{j=1}^{n}H(-\xi)\ln\left|\frac{y_{j+1}+\xi}{y_j+\xi}\right|+H(y_1)-H(-\xi)\,\,.
\ea
on the $(-\xi=y_0,.....,y_{n+1}=1)$ grid. As a final result for the ``$+$'' integral we get
\ba
P.V.\int_{-1}^{1}\frac{H(z)dz}{z+\xi}= X_1 + X_2 +H(-\xi)\ln\left|\frac{\xi+1}{\xi-1}\right|\,.
\ea

\section{Appendix D}

In this section we will present the full expression of the functions $A_1$ and $A_2$ which appear in the squared amplitude
\ba
A_1(z,z'x,t,Q^2)&=&\tilde{g}^{4} Q^{2}\left.\left[4 g_{d}^{2}[\tilde{E}_{d}' (4 \tilde{H}_{d}M^{2} + \tilde{E}_{d}t)x^{2} \right.\right.\nonumber\\
&&+\left.\left.4 \tilde{H}_{d}'M^{2} (4\tilde{H}_{d}(x - 1)+ \tilde{E}_{d}x^{2})] \right.\right.\nonumber \\
&&+\left.\left.4g_{d} g_{u}[(4 \tilde{E}_{u}' \tilde{H}_{d} M^{2} + 4 \tilde{E}_{d}' \tilde{H}_{u}M^{2} +  \tilde{E}_{u}' \tilde{E}_{d} t + \tilde{E}_{d}' \tilde{E}_{u}t)x^{2} \right.\right.\nonumber \\
&&+\left.\left.4 \tilde{H}_{u}'M^{2}(4 \tilde{H}_{d}(x - 1) +  \tilde{E}_{d}x^{2}) + 4\tilde{H}_{d}'M^{2}(4 \tilde{H}_{u} (x- 1) + \tilde{E}_{u}x^{2})]\right.\right.\nonumber \\
&&+ \left.\left.D_{v} U_{v}g_{d}g_{u}[4 E_{u}' E_{d} t +4 E_{d}' E_{u} t - 4  E_{u}' E_{d}tx - 4  E_{d}' E_{u}tx + 4  E_{u}' E_{d}M^{2}x^{2}\right.\right.\nonumber \\
&&+ \left.\left.4  E_{d}' E_{u} M^{2}x^{2}  + 4 E_{u}' H_{d}M^{2}x^{2} + 4 E_{d}' H_{u}M^{2}x^{2} + E_{u}' E_{d}tx^{2} +E_{d}'E_u t x^2 \right.\right.\nonumber \\
&&+ \left.\left.4 H_{u}'M^{2}(4 H_{d}(x - 1) + E_{d}x^{2}) + 4 H_{d}'M^{2}(4 H_{u} (x - 1)+E_{u}x^{2})] \right.\right.\nonumber \\
&&+\left.\left.g_{u}^{2}[4E_{u}'E_{u}t U_{v}^{2} - 4 E_{u}'E_{u} t U_{v}^{2}x +16 \tilde{E}_{u}'\tilde{H}_{u}M^{2}x^{2} + 4\tilde{E}_{u}'\tilde{E}_{u} t x^{2} \right.\right.\nonumber \\
&&+\left.\left.4  E_{u}' E_{u}M^{2} U_{v}^{2} x^{2} + 4 E_{u}' H_{u}M^{2} U_{v}^{2} x^{2} +E_{u}'E_{u} t U_{v}^{2} x^{2} + 16\tilde{H}_{u}'M^{2}(4\tilde{H}_{u}(x - 1) + \tilde{E}_{u}x^{2})\right.\right.\nonumber \\
&&+ \left.\left.4H_{u}'M^{2} U_{v}^{2}(4H_{u}(x - 1) + E_{u}x^{2})] \right.\right.\nonumber\\
&&+\left.\left.D_{v}^{2} g_{d}^{2} [4H_{d}' M^{2}(4H_{d}(x - 1) + E_{d}x^{2}) + E_{d}'(4H_{d}M^{2}x^{2}+ E_{d}(t \,(x -2)^{2} + 4M^{2} x^{2}))]\right]\right.\nonumber\\
\ea
and for $A_2(z,z',x,t)$ we get a similar result
\ba
A_2(z,z',x,t,Q^2) &=&\left.4\tilde{g}^4 Q^2\left[g_d  g_u [4 E'_{u} E_d  t + 4 E'_{d} E_u  t - 16 D_v \tilde{H}'_u \tilde{H}_d  M^2  U_v - 16  D_v \tilde{H}'_d  \tilde{H}_u  M^2 U_v - 4 E'_u E_d t x \right.\right.\nonumber\\
&-&\left.\left.4 E'_d E_u t x + 16 D_v \tilde{H}_u \tilde{H}_d M^2 U_v x + 16 D_v \tilde{H}'_d \tilde{H}_u M^2 U_v x + 4 E'_u E_d M^2 x^2 + 4 E'_d E_u M^2 x^2 \right. \right.\nonumber\\
&+&\left.\left.4 E'_u H_d M^2 x^2 + 4 E'_d H_u M^2 x^2 + E'_u E_d t x^2 + E'_d E_u t x^2 + 4 D_v \tilde{E}_u \tilde{H}'_d M^2 U_v x^2 \right.\right.\nonumber\\
&+&\left.\left.4 D_v \tilde{E}_d \tilde{H}'_u M^2 U_v x^2 + 4 D_v \tilde{E}'_u \tilde{H}_d M^2 U_v x^2 +
4 D_v \tilde{E}'_d \tilde{H}_u M^2 U_v x^2 + D_v \tilde{E}'_u \tilde{E}_d t U_v x^2 \right.\right.\nonumber\\
&+&\left.\left.D_v \tilde{E}'_d \tilde{E}_u t U_v x^2 + 4 H'_u M^2 (4 H_d ( x-1) + E_d x^2) + 4 H'_d M^2 (4 H_u (x-1)  + E_u x^2)]\right.\right.\nonumber\\
&+&\left.\left.g_d^2 [4H'_d M^2 (4 H_d(x-1) + E_d x^2) + D_{v}^{2}(\tilde{E}'_d (4\tilde{H}_d M^2 + \tilde{E}_d t) x^2 \right.\right.\nonumber\\
&+&\left.\left.4\tilde{H}'_d M^2 (4 \tilde{H}_d (x-1) + \tilde{E}_d x^2)) + E'_d (4 H_d M^2 x^2 + E_d(t (x-2)^2 + 4 M^2 x^2))]\right.\right.\nonumber\\
&+&\left.\left.g_{u}^{2}[4 H'_u M^2 (4 H_u (x-1) + E_u x^2) + U_{v}^{2} (\tilde{E}'_u (4 \tilde{H}_u M^2 + \tilde{E}_u t) x^2 \right.\right.\nonumber\\
&+&\left.\left.4 \tilde{H}'_{u} M^2 (4 \tilde{H}_u(x-1) + \tilde{E}_u x^2)) + E'_{u} (4 H_u M^2 x^2 + E_u (t(x-2)^2 + 4 M^2 x^2))] \right] \right..\nonumber\\
\ea

\chapter{Leading Twist Amplitudes for Exclusive Neutrino Interactions
in the Deeply Virtual Limit\label{chap6}}
\fancyhead[LO]{\nouppercase{Chapter 6. Leading Twist Amplitudes for Exclusive Neutrino Interactions}}

\section{Introduction and Motivations}

In the prevoius chapter we have pointed out \cite{ACG}
that exclusive processes of DVCS-type (Deeply Virtual Compton Scattering)
could be relevant also in the theoretical study of the exclusive neutrino/nucleon interaction.
Thanks to the presence
of an on-shell photon emitted in the final state, this particle
could be tagged together with the recoiling nucleon
in a large underground detector in order to trigger on the process and
exclude contamination from other backgrounds.
With these motivations,
a study of the $\nu N\to \nu N \gamma$ process has been performed
in \cite{ACG}. The process is mediated by a neutral current and
is particularly clean since there is no
Bethe-Heitler contribution. It has been termed
{\em Deeply Virtual Neutrino Scattering} or DVNS and requires in its partonic
description the electroweak
analogue of the ``non-forward parton distributions'', previously
introduced in the study of DVCS.

In this section we extend that analysis and provide, in part,
a generalization of those results to the charged current case.
Our treatment, here, is purposely short.
The method that we use for the study of the charged processes is based on the
formalism of the non-local operator product expansion and the technique of the
harmonic polynomials, which allows to classify the various contributions
to the interaction in terms of operators of a definite geometrical twist
\cite{Lazar}.
We present here a classification of the leading twist amplitudes
of the charged process while a detailed phenomenological analysis useful
for future experimental searches will be given elsewhere.

\section{The Generalized Bjorken Region and DVCS}

Fig.~\ref{DVCS_1.eps} illustrates the process that we are going to study,
where a neutrino of momentum $l$ scatters off a nucleon of momentum $P_1$
via a neutral or a charged current interaction;
from the final state a photon and a nucleon emerge, of momenta $q_2$ and $P_2$ respectively, while the momenta of the final lepton is $l'$.
We recall that Compton scattering has been investigated in the near past by several groups, since the original works \cite {Ji1, Rad3, Rad1}.
A previous study of the Virtual Compton process in the generalized Bjorken region, of which DVCS
is just a particular case, can be found in \cite{Geyer}.
From the hadronic side, the description of the interaction proceeds
via new constructs of the parton model termed
{ \em generalized parton distributions} (GPD) or also
{\em non-forward parton distributions}. The kinematics for the study of GPD's is characterized by a deep virtuality of the exchanged photon in the initial interaction
($\nu +p\to \nu +p +\gamma$) ( $ Q^2 \approx $  2 GeV$^2$), with the final state photon kept on-shell; large energy of the hadronic system ($W^2 > 6$ GeV$^2$) above the resonance domain and small momentum transfers $|t| < 1$ GeV$^2$.
In the electroweak case, photon emission can occur from the final state
electron (in the case of charged current interactions) and provides an additional contribution to the virtual Compton
amplitude. We choose symmetric defining momenta and use as independent variables the average of the hadron and gauge bosons momenta

% \begin{figure}[t]
% {\par\centering \resizebox*{12cm}{!}{\includegraphics{DVCS_1.eps}} \par}
% \caption{Leading hand-bag diagrams for the process}
% \label{DVCS_1.eps}
% \end{figure}

\beq
P_{1,2}= \bar{P} \mp \frac{\Delta}{2}\,\,\,\,\,\,\,\, q_{1,2}= q \pm \frac{\Delta}{2},
\eeq
with $\Delta=P_2-P_1$ being the momentum transfer. Clearly
\beq
\bar{P}\cdot \Delta=0,\,\,\,\,\, \bar{P}^2=M^2 - \frac{\Delta^2}{4}
\eeq
and $M$ is the nucleon mass. There are two scaling variables which are identified in the process, 
since 3 scalar products  can grow large in the generalized Bjorken limit: $q^2$, 
$\Delta\cdot q$, $\bar{P}\cdot q$.

The momentum transfer $\Delta^2$ is a small parameter in the process. 
Momentum asymmetries between the initial and the final state nucleon are measured by 
two scaling parameters, $\xi$ and $\eta$, related to ratios of the former invariants
\ba
\xi=-\frac{q^2}{2 \bar{P}\cdot q}&& \eta=\frac{\Delta\cdot q}{2 \bar{P}\cdot q}
\ea
where $\xi$ is a variable of Bjorken type, expressed in terms of average momenta rather 
than nucleon and  gauge bosons momenta.
The standard Bjorken variable $x= - q_1^2/( 2 P_1\cdot q_1)$ is trivially related 
to $\xi$ in the $t=0$ limit and in the DVCS case $\eta=-\xi$.

Notice also that the parameter $\xi$ measures the ratio between the plus component 
of the momentum transfer and the average momentum.

$\xi$, therefore, parametrizes the large component of the momentum transfer $\Delta$, 
which can be generically described as
\beq
\label{delta}
\Delta= -2 \xi \bar{P} -{\Delta}_{\perp}
\eeq
where all the components of ${\Delta}_{\perp}$ are $O\left(\sqrt{|\Delta^2|}\right)$.

\section{Bethe-Heitler Contributions}
Prior to embark on the discussion of the virtual Compton contribution,
we quote the result for the Bethe-Heitler (BH) subprocess, which makes its first
appearance in the charged
current case, since a real photon can be radiated off the
leg of the final state lepton.
The amplitude of the BH contribution for a $W^+$ exchange is as follows
\ba
&&T_{BH}^{W^+}=-|e| \frac{g}{2\sqrt{2}}\frac{g}{\sqrt{2}}\overline{u}(l')\left[\gamma^{\mu}\frac{(\slash{l}-\slash{\Delta})}{(l-\Delta)^2 +i\eps}\gamma^{\nu}(1-\gamma^5)\right]u(l)\frac{D^{\nu\delta}(q_1)}{\Delta^2 -M_{W}^{2}+i\eps}\eps^{*}_{\mu}(q_2)\times\nonumber\\
&&\hspace{2 cm}\overline{U}(P_2)\left[\left(F_{1}^{u}(\Delta^2)-F_{1}^{d}(\Delta^2)\right)\gamma^{\delta}+\left(F_{2}^{u}(\Delta^2)-F_{2}^{d}(\Delta^2)\right)i\frac{\sigma^{\delta\alpha}\Delta_{\alpha}}{2 M}\right]U(P_1),\nonumber\\
\ea
where $\eps$ is the polarization vector of the photon
and
\ba
&&T_{BH}^{W^-}=|e| \frac{g}{2\sqrt{2}}\frac{g}{\sqrt{2}}\overline{v}(l)\left[\gamma^{\mu}\frac{(\slash{l}-\slash{\Delta})}{(l-\Delta)^2 +i\eps}\gamma^{\nu}(1-\gamma^5)\right]v(l')
\frac{D^{\nu\delta}(q_1)}{\Delta^2 -M_{W}^{2}+i\eps}\eps^{*}_{\mu}(q_2)\times\nonumber\\
&&\hspace{2 cm}\overline{U}(P_2)\left[\left(F_{1}^{u}(\Delta^2)-F_{1}^{d}(\Delta^2)\right)\gamma^{\delta}+\left(F_{2}^{u}(\Delta^2)-F_{2}^{d}(\Delta^2)\right)i\frac{\sigma^{\delta\alpha}\Delta_{\alpha}}{2 M}\right]U(P_1)\,\nonumber\\
\ea
for the $W^-$ case, with
$D^{\nu\delta}(q_1)/{(\Delta^2 -M_{W}^{2}+i\eps)}$ being the propagator of the W's and
$F_{1,2}$ the usual nucleon form factors (see also
\cite{ACG}).
\section{Structure of the Compton amplitude for charged and neutral currents}
Moving to the Compton amplitude for charged and neutral currents,
this can be expressed in terms of the correlator of currents
\ba
T_{\mu \nu}=i\int d^{4}x e^{i qx} \langle P_2|T\left(J_{\nu}^{\gamma}(x/2) J_{\mu}^{W^{\pm},Z_{0}}(-x/2)\right)|P_1\rangle\,,
\ea
where for the charged and neutral currents we have the following expressions
\ba
&&J^{\mu Z_{0}}(-x/2)=\frac{g}{2 \cos{\theta_W}}\overline{\psi}_{u}(-x/2)\gamma^{\mu}(g^{Z}_{u V}+g^{Z}_{u A}\gamma^5)\psi_{u}(-x/2)+\overline{\psi}_{d}(-x/2)\gamma^{\mu}(g^{Z}_{d V}+g^{Z}_{d A}\gamma^5)\psi_{d}(-x/2),\,\nonumber\\
&&J^{\mu W^{+}}(-x/2)=\frac{g}{2 \sqrt{2}}\overline{\psi}_{u}(-x/2)\gamma^{\mu}(1-\gamma^5)U^{*}_{u d}\psi_{d}(-x/2),\,\nonumber\\
&&J^{\mu W^{-}}(-x/2)=\frac{g}{2 \sqrt{2}}\overline{\psi}_{d}(-x/2)\gamma^{\mu}(1-\gamma^5)U_{d u}\psi_{u}(-x/2),\,\nonumber\\
&&J^{\nu, \gamma}(x/2)=\overline{\psi}_{d}(x/2)\gamma^{\nu}\left(-\frac{1}{3}e\right)\psi_{d}(x/2) +\overline{\psi}_{u}(x/2)\gamma^{\nu}\left(\frac{2}{3}e\right)\psi_{u}(x/2)\,.
\ea
Here we have chosen a simple representation of
the flavour mixing matrix $U^{*}_{u d}=U_{u d}=U_{d u}=\cos{\theta_{C}}$, where $\theta_{C}$ is the Cabibbo angle.

The coefficients $g_V^{Z}$ and $g_A^{Z}$ are
\ba
&&g^{Z}_{u V}=\frac{1}{2} + \frac{4}{3} \sin^{2}\theta_{W}\hspace{1.5 cm}g^{Z}_{u A}=-\frac{1}{2}\nonumber\\
&&g^{Z}_{d V}=-\frac{1}{2} + \frac{2}{3} \sin^{2}\theta_{W}\hspace{1.5 cm}g^{Z}_{d A}=\frac{1}{2}\,,
\ea
and 
\ba
g_{u}=\frac{2}{3},&&g_{d}=\frac{1}{3}
\ea
are the absolute 
values of the charges of the up and down quarks in units 
of the electron charge. A short computation gives
\ba
&&\langle P_2|T\left(J_{\nu}^{\gamma}(x/2) J_{\mu}^{Z_{0}}(-x/2)\right)|P_1\rangle=\nonumber\\
&&\langle P_2|\overline{\psi}_{u}(x/2)g_u \gamma_{\nu}S(x)\gamma_{\mu}(g^{Z}_{u V}+g^{Z}_{u A}\gamma^5)\psi_{u}(-x/2)-\nonumber\\
&&\hspace{0.8 cm}\overline{\psi}_{d}(x/2)g_d \gamma_{\nu}S(x)\gamma_{\mu}(g^{Z}_{d V}+g^{Z}_{d A}\gamma^5)\psi_{d}(-x/2)+\nonumber\\
&&\hspace{0.8 cm}\overline{\psi}_{u}(x/2)\gamma_{\mu}(g^{Z}_{u V}+g^{Z}_{u A}\gamma^5)S(-x) g_u \gamma_{\nu}\psi_{u}(x/2)-\nonumber\\
&&\hspace{0.8 cm}\overline{\psi}_{d}(x/2)\gamma_{\mu}(g^{Z}_{d V}+g^{Z}_{d A}\gamma^5)S(-x) g_d \gamma_{\nu}\psi_{d}(x/2)|P_1 \rangle,\, \nonumber\\
\\
&&\langle P_2|T\left(J_{\nu}^{\gamma}(x/2) J_{\mu}^{W^{+}}(-x/2)\right)|P_1\rangle=\nonumber\\
&&\langle P_2|\overline{\psi}_{u}(-x/2)\gamma_{\mu}(1-\gamma^5)U_{u d}S(-x)\gamma_{\nu}\left(-g_{d}\right)\psi_{d}(x/2)+\nonumber\\
&&\hspace{0.8 cm}\overline{\psi}_{u}(x/2)\gamma_{\nu}\left(g_{u}\right)S(x)\gamma_{\mu}(1-\gamma^5)U_{u d}\psi_{d}(-x/2) |P_1\rangle,\,\nonumber\\
\\
&&\langle P_2|T\left(J_{\nu}^{\gamma}(x/2) J_{\mu}^{W^{-}}(-x/2)\right)|P_1\rangle=\nonumber\\
&&\langle P_2|-\overline{\psi}_{d}(x/2)g_d\gamma_{\nu}S(x)\gamma_{\mu}(1-\gamma^5)U_{d u}\psi_{u}(-x/2)+\nonumber\\
&&\hspace{0.8 cm}\overline{\psi}_{d}(-x/2)\gamma_{\mu}(1-\gamma^5)S(-x)U_{d u}\psi_{u}(x/2) |P_1\rangle\,,
\ea
where all the factors $g/{2\sqrt{2}}$ and $g/2\cos{\theta_{W}}$, 
for semplicity, have been suppressed  and we have defined
\ba
S^{u}(x)=S^{d}(x)\approx \frac{i \rlap/{x}}{2\pi^2(x^2-i\eps)^2}.
\ea
Using the following identities
\ba
&&\gamma_{\mu}\gamma_{\alpha}\gamma_{\nu}=S_{\mu\alpha\nu\beta}\gamma^{\beta}+i\eps_{\mu\alpha\nu\beta}\gamma^{5}\gamma^{\beta},\nonumber\\
&&\gamma_{\mu}\gamma_{\alpha}\gamma_{\nu}\gamma^{5}=S_{\mu\alpha\nu\beta}\gamma^{\beta}\gamma^{5}-i\eps_{\mu\alpha\nu\beta}\gamma^{\beta},\nonumber\\
&&S_{\mu\alpha\nu\beta}=\left(g_{\mu\alpha}g_{\nu\beta}+g_{\nu\alpha}g_{\mu\beta}-g_{\mu\nu}g_{\alpha\beta} \right),
\ea
we rewrite the correlators as
\ba
&&T_{\mu\nu}^{Z_{0}}=i\int d^{4}x\frac{e^{i q x} x^{\alpha}}{2\pi^2(x^2 -i\eps)^2}
\langle P_2|\left[g_u g_{u V}\left(S_{\mu\alpha\nu\beta}O^{\beta}_{u} -i\eps_{\mu\alpha\nu\beta}O^{5 \beta}_{u}\right)-g_u g_{u A}\left(S_{\mu\alpha\nu\beta}\tilde{O}^{5 \beta}_{u}-i\eps_{\mu\alpha\nu\beta}\tilde{O}^{\beta}_{u}\right)\right.\nonumber\\
&&\hspace{5.5 cm}\left.-g_d g_{d V}\left(S_{\mu\alpha\nu\beta}O^{\beta}_{d}-i\eps_{\mu\alpha\nu\beta}O^{5 \beta}_{d}\right)+g_d g_{d A}\left(S_{\mu\alpha\nu\beta}\tilde{O}^{5 \beta}_{d} -i\eps_{\mu\alpha\nu\beta}\tilde{O}^{\beta}_{d}\right)\right]|P_1\rangle,\,\nonumber\\
\\
&&T^{W^{+}}_{\mu \nu}=i\int d^4 x \frac{e^{iqx}x^{\alpha}U_{u d}}{2\pi^2(x^2-i\eps)^2}\langle P_2|\left[i S_{\mu\alpha\nu\beta}\left(\tilde{O}_{u d}^{\beta}+O_{u d}^{5 \beta}\right)+\eps_{\mu\alpha\nu\beta}\left(O_{u d}^{\beta}+\tilde{O}_{u d}^{5 \beta}\right)\right] |P_1\rangle, \nonumber\\
\\
&&T^{W^{-}}_{\mu \nu}=i\int d^4 x \frac{e^{iqx}x^{\alpha}U_{d u}}{2\pi^2(x^2-i\eps)^2}\langle P_2|\left[-i S_{\mu\alpha\nu\beta}\left(\tilde{O}_{d u}^{\beta}+O_{d u}^{5 \beta}\right)-\eps_{\mu\alpha\nu\beta}\left(O_{d u}^{\beta}+\tilde{O}_{d u}^{5 \beta}\right)\right] |P_1\rangle\,. \nonumber\\
\ea
We have suppressed the $x$-dependence of the operators in the former equations. The relevant operators are denoted by
\ba
&&\tilde{O}_{a}^{\beta}(x/2,-x/2)=\overline{\psi}_{a}(x/2)\gamma^{\beta}\psi_{a}(-x/2)+\overline{\psi}_{a}(-x/2)\gamma^{\beta}\psi_{a}(x/2),\nonumber\\
&&\tilde{O}_{a}^{5 \beta}(x/2,-x/2)=\overline{\psi}_{a}(x/2)\gamma^{5}\gamma^{\beta}\psi_{a}(-x/2)-\overline{\psi}_{a}(-x/2)\gamma^{5}\gamma^{\beta}\psi_{a}(x/2),\nonumber\\
&&O_{a}^{\beta}(x/2,-x/2)=\overline{\psi}_{a}(x/2)\gamma^{\beta}\psi_{a}(-x/2)-\overline{\psi}_{a}(-x/2)\gamma^{\beta}\psi_{a}(x/2),\nonumber\\
&&O_{a}^{5 \beta}(x/2,-x/2)=\overline{\psi}_{a}(x/2)\gamma^{5}\gamma^{\beta}\psi_{a}(-x/2)+\overline{\psi}_{a}(-x/2)\gamma^{5}\gamma^{\beta}\psi_{a}(x/2),\,\nonumber\\
\\
&&\tilde{O}_{u d}^{\beta}(x/2,-x/2)=g_u\overline{\psi}_{u}(x/2)\gamma^{\beta}\psi_{d}(-x/2)+g_d\overline{\psi}_{u}(-x/2)\gamma^{\beta}\psi_{d}(x/2),\nonumber\\
&&\tilde{O}_{u d}^{5 \beta}(x/2,-x/2)=g_u\overline{\psi}_{u}(x/2)\gamma^{5}\gamma^{\beta}\psi_{d}(-x/2)-g_d\overline{\psi}_{u}(-x/2)\gamma^{5}\gamma^{\beta}\psi_{d}(x/2),\nonumber\\
&&O_{u d}^{\beta}(x/2,-x/2)=g_u\overline{\psi}_{u}(x/2)\gamma^{\beta}\psi_{d}(-x/2)-g_d\overline{\psi}_{u}(-x/2)\gamma^{\beta}\psi_{d}(x/2),\nonumber\\
&&O_{u d}^{5 \beta}(x/2,-x/2)=g_u\overline{\psi}_{u}(x/2)\gamma^{5}\gamma^{\beta}\psi_{d}(-x/2)+g_d\overline{\psi}_{u}(-x/2)\gamma^{5}\gamma^{\beta}\psi_{d}(x/2),\,\nonumber\\
\ea
and similar ones with $u \leftrightarrow d$ interchanged.

We use isospin symmetry to relate flavour nondiagonal operators
$(\hat{O}_{f f'})$ to flavour diagonal ones $(\hat{O}_{f f})$
\ba
\langle p|\hat{O}^{ud}(x)|n\rangle&=&\langle p|\hat{O}^{ud}(x)\tau^{-}|n\rangle=\langle p|\left[\hat{O}^{ud}(x),\tau^{-}\right]|n\rangle=
\langle p|\hat{O}^{uu}(x)|p\rangle-\langle p|\hat{O}^{dd}(x)|p\rangle\,,\nonumber\\
&&\langle p|\hat{O}^{ud}(x)|n\rangle=\langle n|\hat{O}^{dd}(x)|n\rangle-\langle n|\hat{O}^{uu}(x)|n\rangle\,,\nonumber\\
&&\langle n|\hat{O}^{du}(x)|p\rangle=\langle p|\hat{O}^{uu}(x)|p\rangle-\langle p|\hat{O}^{dd}(x)|p\rangle\,,\nonumber\\
&&\langle n|\hat{O}^{du}(x)|p\rangle=\langle n|\hat{O}^{dd}(x)|n\rangle-\langle n|\hat{O}^{uu}(x)|n\rangle,\,
\ea
where \ba
\tau^{\pm}=\tau^{x}\pm\tau^{y}
\ea
are isospin raising/lowering operators expressed in terms of Pauli matrices.

\section{Parameterization of nonforward matrix elements}
The extraction of the leading twist contribution to the handbag diagram
is performed using the geometrical twist expansion, as developed in
\cite{blum&rob,lazar1,lazar2,Ji2,Blum,bemuevolve,belis&muller,Beli&geyer,Pol_Weiss}, 
adapted to our case. We set the twist-2 expansions on the light cone (with $x^2=0$)
and we choose the light-cone gauge to remove the gauge link
\ba
&&\langle P_2|\overline{\psi}_{a}(-k x)\gamma^{\mu}\psi_{a}(k x)|P_1\rangle^{tw.2}=\nonumber\\
&&\int Dz e^{-ik(x\cdot P_z)}F^{a (\nu)}(z_{1},z_{2},P_{i}\cdot P_{j}x^2,P_{i}\cdot P_{j})\overline{U}(P_2)\left[\gamma^{\mu}-i k {P_{z}}^{\mu}\slash{x}\right]U(P_1)+\nonumber\\
&&\int Dz e^{-ik(x\cdot P_z)}G^{a (\nu)}(z_{1},z_{2},P_{i}\cdot P_{j}x^2,P_{i}\cdot P_{j})\overline{U}(P_2)\left[\frac{\left(i\sigma^{\mu \alpha}\Delta_{\alpha}\right)}{M} -ik{P_{z}}^{\mu}\frac{\left(i\sigma^{\alpha \beta}x_{\alpha}\Delta_{\beta}\right)}{M}\right]U(P_1)\,,\nonumber\\
\ea
with $0< k <1$ a scalar parameter, with
\ba
P_z=P_1 z_1 + P_2 z_2,\,
\ea
and
\ba
&&\langle P_2|\overline{\psi}_{a}(-k x)\gamma^{5}\gamma^{\mu}\psi_{a}(k x)|P_1\rangle^{tw.2}=\nonumber\\
&&\int Dz e^{-ik(x\cdot P_z)}F^{5 a (\nu)}(z_{1},z_{2},P_{i}\cdot P_{j}x^2,P_{i}\cdot P_{j})\overline{U}(P_2)\left[\gamma^{5}\gamma^{\mu}-i k {P_{z}}^{\mu}\gamma^{5}\slash{x}\right]U(P_1)+\nonumber\\
&&\int Dz e^{-ik(x\cdot P_z)}G^{5 a(\nu)}(z_{1},z_{2},P_{i}\cdot P_{j}x^2,P_{i}\cdot P_{j})\overline{U}(P_2)\gamma^{5}\left[\frac{\left(i\sigma^{\mu \alpha}\Delta_{\alpha}\right)}{M} -ik{P_{z}}^{\mu}\frac{\left(i\sigma^{\alpha \beta}x_{\alpha}\Delta_{\beta}\right)}{M}\right]U(P_1).\,\nonumber\\
\ea
The index $(\nu)$ in the expressions of the distribution functions $F, G$ has been introduced in order to distinguish
them from the parameterization given in \cite{blum&rob,Geyer}, which are related to linear combinations of electromagnetic
correlators. In the expressions above $a$ is a flavour index and we have introduced both a vector (Dirac) and a Pauli-type
form factor contribution with nucleon wave functions (U(P)).
The product $P_i\cdot P_j$ denotes all the possible products of the two momenta $P_1$ and $P_2$,
and the measure of integration is defined by \cite{blum&rob}
\ba
Dz=\frac{1}{2} d z_1 d z_2\,\theta(1-z_1)\,\theta(1+z_1)\,\theta(1-z_2)\,
\theta(1+z_2).\,
\ea
In our parameterization of the correlators we are omitting the so called ``trace-terms'' (see ref.~\cite{Geyer}), since these terms vanish on shell.
In order to arrive at a partonic interpretation  one introduces
variables $z_{+}$ and $z_{-}$ conjugated to $2 \bar{P}$ and $\Delta$ and defined as
\ba
&&z_{+}=1/2(z_1 +z_2),\nonumber\\
&&z_{-}=1/2(z_2 -z_1),\nonumber\\
&&Dz=dz_{+} dz_{-}\theta(1+z_+ + z_-)\theta(1+z_+ - z_-)\theta(1-z_+ + z_-)\theta(1-z_+ - z_-).\,
\ea
In terms of these new variables
$P_z=2\bar{P}z_{+}+\Delta z_{-}$, which will be used below.
At this stage, we can proceed to calculate the hadronic tensor
by performing the $x$-space integrations. This will be illustrated in the
case of the $W^+$ current, the others being similar. We define
\ba
&&\int dx^4 \frac{e^{iq x}x^{\alpha}}{2\pi^2\left(x^2-i\eps\right)^2}\langle P_2|S_{\mu\alpha\nu\beta}\tilde{O}^{a\beta}|P_1\rangle=\tilde{S}^{a}_{\mu\nu},\nonumber\\
&&\int dx^4 \frac{e^{iq x}x^{\alpha}}{2\pi^2\left(x^2-i\eps\right)^2}\langle P_2|S_{\mu\alpha\nu\beta} O^{5 a\beta}|P_1\rangle=S^{5 a}_{\mu\nu},\nonumber\\
&&\int dx^4 \frac{e^{iq x}x^{\alpha}}{2\pi^2\left(x^2-i\eps\right)^2}\langle P_2|\eps_{\mu\alpha\nu\beta} O^{a\beta}|P_1\rangle={\bf \varepsilon}^{a}_{\mu\nu},\nonumber\\
&&\int dx^4 \frac{e^{iq x}x^{\alpha}}{2\pi^2\left(x^2-i\eps\right)^2}\langle P_2|\eps_{\mu\alpha\nu\beta} \tilde{O}^{5 a\beta}|P_1\rangle=\tilde{\bf\varepsilon}^{5 a}_{\mu\nu},
\ea
and introduce the variables
\ba
&&Q_1^{\alpha}(z)=q^{\alpha}+\frac{1}{2}P^{\alpha}_z,\nonumber\\
&&Q_2^{\alpha}(z)=q^{\alpha}-\frac{1}{2}P^{\alpha}_z,
\ea
where $(z)$ is now meant to denote both variables $(z_+,z_-)$. The presence of a new variable $Q_2$, compared
to \cite{blum&rob}, is related to the fact that we are parameterizing each single bilinear
covariant rather then linear combinations of them, as in the electromagnetic case.

After some re-arrangements we get
\ba
\label{equa1}
&&\tilde{S}^{a}_{\mu\nu}=g_{u}\int Dz\frac{F^{a(\nu)}(z)}{(Q_1^2+i\eps)}\left\{\left[-g_{\mu\nu}\slash{q}+q_{\nu}\gamma_{\mu}+q_{\mu}\gamma_{\nu}\right]+\left[{P_z}_{\mu}\gamma_{\nu}+{P_z}_{\nu}\gamma_{\mu}\right]\right.\nonumber\\
&&\;\;\;\;\;\;\;\;\;\;\;\;\;\;\;\;\;\;\;\;\;\;\;\;\;\;\;\;\;\;\;\;\;\;\;\left.-\frac{\slash{q}}{(Q_1^2+i\eps)}\left[{P_z}_{\mu}{P_z}_{\nu}+{P_z}_{\mu}q_{\nu}+{P_z}_{\nu}q_{\mu}- g_{\mu\nu}(P_z\cdot q)\right]\right\}+\nonumber\\
&&\;\;\;\;\;\;\;\;g_{d}\int Dz\frac{F^{a(\nu)}(z)}{(Q_2^2+i\eps)}\left\{\left[-g_{\mu\nu}\slash{q}+q_{\nu}\gamma_{\mu}+q_{\mu}\gamma_{\nu}\right]-\left[{P_z}_{\mu}\gamma_{\nu}+{P_z}_{\nu}\gamma_{\mu}\right]\right.\nonumber\\
&&\;\;\;\;\;\;\;\;\;\;\;\;\;\;\;\;\;\;\;\;\;\;\;\;\;\;\;\;\;\;\;\;\;\;\;\left.+\frac{\slash{q}}{(Q_2^2+i\eps)}\left[-{P_z}_{\mu}{P_z}_{\nu}+{P_z}_{\mu}q_{\nu}+{P_z}_{\nu}q_{\mu}- g_{\mu\nu}(P_z\cdot q)\right]\right\}+\nonumber\\
&&\;\;\;\;\;g_{u}\int Dz\frac{G^{a(\nu)}(z)}{(Q_1^2+i\eps)}\left\{\left[-g_{\mu\nu}\frac{i\sigma^{\alpha\beta}q_{\alpha}\Delta_{\beta}}{M}+q_{\nu}\frac{i\sigma^{\mu\beta}\Delta_{\beta}}{M}+q_{\mu}\frac{i\sigma^{\nu\beta}\Delta_{\beta}}{M}\right]+\left[{P_z}_{\mu}\frac{i\sigma^{\nu\beta}\Delta_{\beta}}{M}+{P_z}_{\nu}\frac{i\sigma^{\mu\beta}\Delta_{\beta}}{M}\right]\right.\nonumber\\
&&\;\;\;\;\;\;\;\;\;\;\;\;\;\;\;\;\;\;\;\;\;\;\;\;\;\;\;\;\;\;\;\;\;\;\;\left.-\frac{i\sigma^{\alpha\beta}q_{\alpha}\Delta_{\beta}}{M(Q_1^2+i\eps)}\left[{P_z}_{\mu}{P_z}_{\nu}+{P_z}_{\mu}q_{\nu}+{P_z}_{\nu}q_{\mu}- g_{\mu\nu}(P_z\cdot q)\right]\right\}+\nonumber\\
&&\;\;\;\;\;g_{d}\int Dz\frac{G^{a(\nu)}(z)}{(Q_2^2+i\eps)}\left\{\left[-g_{\mu\nu}\frac{i\sigma^{\alpha\beta}q_{\alpha}\Delta_{\beta}}{M}+q_{\nu}\frac{i\sigma^{\mu\beta}\Delta_{\beta}}{M}+q_{\mu}\frac{i\sigma^{\nu\beta}\Delta_{\beta}}{M}\right]-\left[{P_z}_{\mu}\frac{i\sigma^{\nu\beta}\Delta_{\beta}}{M}+{P_z}_{\nu}\frac{i\sigma^{\mu\beta}\Delta_{\beta}}{M}\right]\right.\nonumber\\
&&\;\;\;\;\;\;\;\;\;\;\;\;\;\;\;\;\;\;\;\;\;\;\;\;\;\;\;\;\;\;\;\;\;\;\;\left.+\frac{i\sigma^{\alpha\beta}q_{\alpha}\Delta_{\beta}}{M(Q_2^2+i\eps)}\left[-{P_z}_{\mu}{P_z}_{\nu}+{P_z}_{\mu}q_{\nu}+{P_z}_{\nu}q_{\mu}- g_{\mu\nu}(P_z\cdot q)\right]\right\},\,\nonumber\\
\ea
with an analogous expressions for $S^{5 a}_{\mu\nu}$, that we omit,
since it can be recovered by performing the substitutions
\ba
\label{repla}
\gamma_{\mu}\rightarrow \gamma^{5}\gamma_{\mu} \hspace{.4cm}\sigma^{\mu\nu}\rightarrow \gamma^5 \sigma^{\mu\nu}, \nonumber \\
F^{a(\nu)}, G^{a(\nu)}\rightarrow F^{5 a(\nu) }, G^{5 a(\nu)}
\ea
in (\ref{equa1}).

Similarly, for ${\bf \varepsilon}^{a}_{\mu\nu}$ we get
\ba
\label{equa2}
&&{\bf \varepsilon}^{a}_{\mu\nu}=g_{u}\int Dz F^{a(\nu)}(z)\left\{ \frac{1}{(Q_1^2+i\eps)}\eps_{\mu\alpha\nu\beta}\left[q^{\alpha}\gamma^{\beta} -\frac{P_z^{\beta}q^{\alpha}\slash{q}}{(Q_1^2+i\eps)}\right]\right\}-\nonumber\\
&&\;\;\;\;\;\;\;\;\;\;g_{d}\int Dz F^{a(\nu)}(z)\left\{ \frac{1}{(Q_2^2+i\eps)}\eps_{\mu\alpha\nu\beta}\left[q^{\alpha}\gamma^{\beta} +\frac{P_z^{\beta}q^{\alpha}\slash{q}}{(Q_2^2+i\eps)}\right]\right\}+\nonumber\\
&&\;\;\;\;\;\;\;\;\;\;g_{u}\int Dz G^{a(\nu)}(z)\left\{ \frac{1}{(Q_1^2+i\eps)}\eps_{\mu\alpha\nu\beta}\left[q^{\alpha}\frac{i\sigma^{\beta\delta}\Delta_{\delta}}{M} -\frac{P_z^{\beta}q^{\alpha}(i\sigma^{\lambda\delta}q_{\lambda}\Delta_{\delta})}{M\,(Q_1^2+i\eps)}\right]\right\}-\nonumber\\
&&\;\;\;\;\;\;\;\;\;\;g_{d}\int Dz G^{a(\nu)}(z)\left\{ \frac{1}{(Q_2^2+i\eps)}\eps_{\mu\alpha\nu\beta}\left[q^{\alpha}\frac{i\sigma^{\beta\delta}\Delta_{\delta}}{M} +\frac{P_z^{\beta}q^{\alpha}(i\sigma^{\lambda\delta}q_{\lambda}\Delta_{\delta})}{M\,(Q_2^2+i\eps)}\right]\right\}\,.
\ea
The expression of $\tilde{\bf \varepsilon}^{5 a}_{\mu\nu}$
can be obtained in a similar way.
To compute the $T^{Z_{0}}_{\mu\nu}$ tensor
we need to include the following operators, which are related to the previous ones by $g_u\,, g_d\rightarrow 1$
\ba
&&\int dx^4 \frac{e^{iq x}x^{\alpha}}{2\pi^2\left(x^2-i\eps\right)^2}\langle P_2|S_{\mu\alpha\nu\beta}O^{a\beta}|P_1\rangle=S^{a}_{\mu\nu},\nonumber\\
&&\int dx^4 \frac{e^{iq x}x^{\alpha}}{2\pi^2\left(x^2-i\eps\right)^2}\langle P_2|S_{\mu\alpha\nu\beta} \tilde{O}^{5 a\beta}|P_1\rangle=\tilde{S}^{5 a}_{\mu\nu},\nonumber\\
&&\int dx^4 \frac{e^{iq x}x^{\alpha}}{2\pi^2\left(x^2-i\eps\right)^2}\langle P_2|\eps_{\mu\alpha\nu\beta} \tilde{O}^{a\beta}|P_1\rangle=\tilde{{\bf \varepsilon}}^{a}_{\mu\nu},\nonumber\\
&&\int dx^4 \frac{e^{iq x}x^{\alpha}}{2\pi^2\left(x^2-i\eps\right)^2}\langle P_2|\eps_{\mu\alpha\nu\beta} O^{5 a\beta}|P_1\rangle={\bf\varepsilon}^{5 a}_{\mu\nu}.
\ea
In this case a simple manipulation of (\ref{equa1}) gives
\ba
\label{equaZ0}
&&S^{a}_{\mu\nu}=\int Dz\frac{F^{a(\nu)}(z)}{(Q_1^2+i\eps)}\left\{\left[-g_{\mu\nu}\slash{q}+q_{\nu}\gamma_{\mu}+q_{\mu}\gamma_{\nu}\right]+\left[{P_z}_{\mu}\gamma_{\nu}+{P_z}_{\nu}\gamma_{\mu}\right]\right.\nonumber\\
&&\;\;\;\;\;\;\;\;\;\;\;\;\;\;\;\;\;\;\;\;\;\;\;\;\;\;\;\;\;\;\;\;\;\;\;\left.-\frac{\slash{q}}{(Q_1^2+i\eps)}\left[{P_z}_{\mu}{P_z}_{\nu}+{P_z}_{\mu}q_{\nu}+{P_z}_{\nu}q_{\mu}- g_{\mu\nu}(P_z\cdot q)\right]\right\}-\nonumber\\
&&\;\;\;\;\;\;\;\;\int Dz\frac{F^{a(\nu)}(z)}{(Q_2^2+i\eps)}\left\{\left[-g_{\mu\nu}\slash{q}+q_{\nu}\gamma_{\mu}+q_{\mu}\gamma_{\nu}\right]-\left[{P_z}_{\mu}\gamma_{\nu}+{P_z}_{\nu}\gamma_{\mu}\right]\right.\nonumber\\
&&\;\;\;\;\;\;\;\;\;\;\;\;\;\;\;\;\;\;\;\;\;\;\;\;\;\;\;\;\;\;\;\;\;\;\;\left.+\frac{\slash{q}}{(Q_2^2+i\eps)}\left[-{P_z}_{\mu}{P_z}_{\nu}+{P_z}_{\mu}q_{\nu}+{P_z}_{\nu}q_{\mu}- g_{\mu\nu}(P_z\cdot q)\right]\right\}+\nonumber\\
&&\;\;\;\;\;\int Dz\frac{G^{a(\nu)}(z)}{(Q_1^2+i\eps)}\left\{\left[-g_{\mu\nu}\frac{i\sigma^{\alpha\beta}q_{\alpha}\Delta_{\beta}}{M}+q_{\nu}\frac{i\sigma^{\mu\beta}\Delta_{\beta}}{M}+q_{\mu}\frac{i\sigma^{\nu\beta}\Delta_{\beta}}{M}\right]+\left[{P_z}_{\mu}\frac{i\sigma^{\nu\beta}\Delta_{\beta}}{M}+{P_z}_{\nu}\frac{i\sigma^{\mu\beta}\Delta_{\beta}}{M}\right]\right.\nonumber\\
&&\;\;\;\;\;\;\;\;\;\;\;\;\;\;\;\;\;\;\;\;\;\;\;\;\;\;\;\;\;\;\;\;\;\;\;\left.-\frac{i\sigma^{\alpha\beta}q_{\alpha}\Delta_{\beta}}{M(Q_1^2+i\eps)}\left[{P_z}_{\mu}{P_z}_{\nu}+{P_z}_{\mu}q_{\nu}+{P_z}_{\nu}q_{\mu}- g_{\mu\nu}(P_z\cdot q)\right]\right\}-\nonumber\\
&&\;\;\;\;\;\int Dz\frac{G^{a(\nu)}(z)}{(Q_2^2+i\eps)}\left\{\left[-g_{\mu\nu}\frac{i\sigma^{\alpha\beta}q_{\alpha}\Delta_{\beta}}{M}+q_{\nu}\frac{i\sigma^{\mu\beta}\Delta_{\beta}}{M}+q_{\mu}\frac{i\sigma^{\nu\beta}\Delta_{\beta}}{M}\right]-\left[{P_z}_{\mu}\frac{i\sigma^{\nu\beta}\Delta_{\beta}}{M}+{P_z}_{\nu}\frac{i\sigma^{\mu\beta}\Delta_{\beta}}{M}\right]\right.\nonumber\\
&&\;\;\;\;\;\;\;\;\;\;\;\;\;\;\;\;\;\;\;\;\;\;\;\;\;\;\;\;\;\;\;\;\;\;\;\left.+\frac{i\sigma^{\alpha\beta}q_{\alpha}\Delta_{\beta}}{M(Q_2^2+i\eps)}\left[-{P_z}_{\mu}{P_z}_{\nu}+{P_z}_{\mu}q_{\nu}+{P_z}_{\nu}q_{\mu}- g_{\mu\nu}(P_z\cdot q)\right]\right\}\,.\nonumber\\
\ea
The expression of $\tilde{S}^{5 a}_{\mu\nu}$ is obtained from
(\ref{equaZ0}) by the replacements (\ref{repla}).

For the $\tilde{\bf \varepsilon}^{a}_{\mu\nu}$ case, a re-arrangement of
(\ref{equa2}) gives
\ba
\label{equaZ02}
&&\tilde{\bf \varepsilon}^{a}_{\mu\nu}=\int Dz F^{a(\nu)}(z)\left\{ \frac{1}{(Q_1^2+i\eps)}\eps_{\mu\alpha\nu\beta}\left[q^{\alpha}\gamma^{\beta} -\frac{P_z^{\beta}q^{\alpha}\slash{q}}{(Q_1^2+i\eps)}\right]\right\}+\nonumber\\
&&\;\;\;\;\;\;\;\;\;\;\int Dz F^{a(\nu)}(z)\left\{ \frac{1}{(Q_2^2+i\eps)}\eps_{\mu\alpha\nu\beta}\left[q^{\alpha}\gamma^{\beta} +\frac{P_z^{\beta}q^{\alpha}\slash{q}}{(Q_2^2+i\eps)}\right]\right\}+\nonumber\\
&&\;\;\;\;\;\;\;\;\;\;\int Dz G^{a(\nu)}(z)\left\{ \frac{1}{(Q_1^2+i\eps)}\eps_{\mu\alpha\nu\beta}\left[q^{\alpha}\frac{i\sigma^{\beta\delta}\Delta_{\delta}}{M} -\frac{P_z^{\beta}q^{\alpha}(i\sigma^{\lambda\delta}q_{\lambda}\Delta_{\delta})}{M\,(Q_1^2+i\eps)}\right]\right\}+\nonumber\\
&&\;\;\;\;\;\;\;\;\;\;\int Dz G^{a(\nu)}(z)\left\{ \frac{1}{(Q_2^2+i\eps)}\eps_{\mu\alpha\nu\beta}\left[q^{\alpha}\frac{i\sigma^{\beta\delta}\Delta_{\delta}}{M} +\frac{P_z^{\beta}q^{\alpha}(i\sigma^{\lambda\delta}q_{\lambda}\Delta_{\delta})}{M\,(Q_2^2+i\eps)}\right]\right\}\,.
\ea
Also in this case, the expression of the ${\bf \varepsilon}^{5 a}_{\mu\nu}$ tensor is obtained  by the replacements (\ref{repla}).

\section{The partonic interpretation}
At a first sight,
the functions $F^{(\nu)}, G^{(\nu)}, F^{5(\nu)}, G^{5(\nu)}$ do not have a simple partonic interpretation.
To progress in this direction it is useful to perform the expansions
of the propagators
\ba
&&\frac{1}{Q_1^2+i \eps}\approx \frac{1}{2(\bar{P}\cdot q)}\frac{1}{\left[z_+ -\xi +\eta z_- +i\eps\right]},\nonumber\\
&&\frac{1}{Q_2^2+i \eps}\approx -\frac{1}{2(\bar{P}\cdot q)}\frac{1}{\left[z_+ +\xi +\eta z_- -i\eps\right]}
\ea
which are valid only in the asymptotic limit. In this limit only
the large kinematical invariants and their (finite) ratios are kept.
In this expansion the physical scaling variable $\xi$ appears quite naturally and one is led to
introduce a new linear combination
\ba
t=z_+ +\eta z_-\,,
\ea
to obtain
\ba
&&\frac{1}{Q_1^2+i \eps}\approx \frac{1}{2(\bar{P}\cdot q)}\frac{1}{\left[t-\xi +i\eps\right]},\nonumber\\
&&\frac{1}{Q_2^2+i \eps}\approx -\frac{1}{2(\bar{P}\cdot q)}\frac{1}{\left[t+\xi -i\eps\right]}.
\ea
Analogously, we rewrite $P_z$ using the variables $\left\{t,z_-\right\}$
\ba
P_z=2\bar{P}t +\pi z_-\,,
\ea
in terms of a new 4-vector, denoted by $\pi$,
which is a direct measure of the exchange of transverse momentum with respect
to $\bar{P}$
\ba
\pi=\Delta+ 2 \xi \bar{P}\,.
\ea
This quantity is strictly related to $\Delta_\perp$, as given in (\ref{delta}).
The dominant (large) components of the process
are related to the collinear contributions, and in our calculation the contributions proportional to the vector $\pi$ will be neglected. This, of course, introduces
a violation of the transversality of the process of $O(\Delta_\perp/2\bar{P}\cdot q)$.

Adopting the new variables $\left\{t,z_-\right\}$ and the conjugate ones $\left\{\bar{P},\pi\right\}$, the relevant integrals that we need to ``reduce'' to a single (partonic) variable
are contained in the expressions
\ba
\label{integrals}
&&H_{Q_1}(\xi)=\int dz_+ dz_- \frac{H(z_+,z_-)}{(Q_1^2+i\eps)}=\frac{1}{2\bar{P}\cdot q}\int Dz \frac{H(t+\xi z_-,z_-)}{(t-\xi+i\eps)}\nonumber\\
&&H_{Q_1}^{\mu}(\xi)=\int dz_+ dz_- \frac{H(z_+,z_-)}{(Q_1^2+i\eps)}\left[2\bar{P}^{\mu}z_+ +\Delta^{\mu}z_-\right]=\frac{1}{2\bar{P}\cdot q}\int Dz \frac{H(t+\xi z_-,z_-)}{(t-\xi+i\eps)}\left[ 2\bar{P}^{\mu}t+\pi^{\mu}z_-\right]\nonumber\\
&&H_{Q_1}^{\mu\nu}(\xi)=\int dz_+ dz_-\frac{H(z_+,z_-)}{(Q_1^2 +i\eps)^2}\left[P_z^{\mu}P_z^{\nu}+q^{\mu}P_z^{\nu}+q^{\nu}P_z^{\mu}-g^{\mu\nu}q\cdot P_z\right]\nonumber\\
&&\hspace{1.3 cm}=\frac{1}{(2\bar{P}\cdot q)^2}\int Dz\frac{H(t+\xi z_-,z_-)}{(t-\xi+i\eps)^2}\left[4\bar{P}^{\mu}\bar{P}^{\nu}t^2 + \left(2q^{\mu}\bar{P}^{\nu}+2q^{\nu}\bar{P}^{\mu}\right)t-g^{\mu\nu}(q\cdot P_z)\right.\nonumber\\
&&\hspace{6 cm}\left.+\pi^{\mu}\pi^{\nu}z_-^{2}+\left(q^{\mu}\pi^{\nu}+q^{\nu}\pi^{\mu}\right)z_- + \left(2\bar{P}^{\mu}\pi^{\nu}+2\bar{P}^{\nu}\pi^{\mu}\right)t z_-\right]\,.\nonumber\\
\ea
Here $H(z_+,z_-)$ is a generic symbol for any of the functions.
We have similar expressions for the integrals depending on the momenta $Q_2$.

The integration over the $z_-$ variable in the integrals shown above
is performed by introducing a suitable spectral representation of the function $H(t,+\xi z_-,z_-)$.
As shown in \cite{blum&rob},
we can classify these representations by the $n=0,1,...,$ powers of the variable $z_-$ ,
\ba
\hat{h}_{n}(t/\tau,\xi)=\int dz_- {z_-}^{n}\hat{h}(\frac{t}{\tau}+\xi z_-,z_-).\,
\ea
With the help of this relation one obtains
\ba
\label{repres}
&&\hat{H}_{n}(t,\xi)=\frac{1}{t^n}\int dz_- {z_-}^n H(t+\xi z_-,z_-)=\frac{1}{t^n}\int^{1}_{0}\frac{d\tau}{\tau} \tau^n \,\hat{h}_{n}(t/\tau,\xi)\nonumber\\
&&\hspace{1.5 cm}=\int_{t}^{sign(t)}\frac{d\lambda}{\lambda}\lambda^{-n}\hat{h}_{n}(\lambda,\xi)\,.
\ea
The result of this manipulation is to generate single-valued distribution amplitudes from double-valued ones.
In the single-valued distributions $\hat{h}_{n}(t,\xi)$ the new scaling variables $t$ and $\xi$
have a partonic interpretation. $\xi$ measures the asymmetry between the momenta of the initial and final states, while
it can be checked that the support of the variable $t$ is the interval $[-1,1]$. The twist-2 part of the Compton amplitude is related only to the $n=0$ moment of $z_-$.
Before performing the $z_-$ integration in each integral of Eq.~(\ref{integrals}) using Eq.~(\ref{repres})
 - a typical example is $H_{Q_1}^{\mu\nu}(\xi)$ -
we reduce such integrals to the sum of two terms using the identity
\ba
\label{purpose}
\int_{-1}^{1}dt\frac{t^m}{(t\pm\xi\mp i\eps)^2}\hat{H}_{n}(t,\xi)=
\int_{-1}^{1}dt\frac{t^{m-1}}{(t\pm\xi\mp i\eps)}\left[\hat{H}_{n}(t,\xi)
-\frac{1}{t^n}\hat{h}_{n}(t,\xi)\right]\,.
\ea
As shown in \cite{Geyer}, after the $z_-$ integration, the integrals in (\ref{integrals}) can be re-written in the form
\ba
&&H_{Q_1}(\xi)=\frac{1}{2\bar{P}\cdot q}\int_{-1}^{1}dt\frac{\hat{H}_{0}(t,\xi)}{(t-\xi+i\eps)},\nonumber\\
&&H_{Q_1}^{\mu}(\xi)=\frac{2\bar{P}^{\mu}}{2\bar{P}\cdot q}\int_{-1}^{1}dt\frac{t\hat{H}_{0}(t,\xi)}{(t-\xi+i\eps)}+ O(\pi^{\mu}),\nonumber\\
&&H_{Q_1}^{\mu\nu}(\xi)=\frac{1}{(2\bar{P}\cdot q)^2}\int_{-1}^{1}dt\frac{\left[2\hat{H}_{0}(t,\xi)-\hat{h}_{0}(t,\xi)\right]}{(t-\xi+i\eps)}4\bar{P}^{\mu}\bar{P}^{\nu}t\nonumber\\
&&\hspace{1.5 cm}+\frac{1}{(2\bar{P}\cdot q)^2}\int_{-1}^{1}dt\frac{\left[\hat{H}_{0}(t,\xi)-\hat{h}_{0}(t,\xi)\right]}{(t-\xi+i\eps)}\left\{(2q^{\mu}\bar{P}^{\nu}+2q^{\nu}\bar{P}^{\mu}-g^{\mu\nu}2q\cdot \bar{P})\right\}+O(\pi^{\mu}\pi^{\nu}),\nonumber\\
\ea
where, again, we are neglecting contributions from the terms proportional
to $\pi^\mu$, subleading in the deeply virtual limit.
The quantities that actually have a strict partonic interpretation are the $\hat{h}^{a}_{0}(t,\xi)$ functions, as
argued in ref.~\cite{Roba&Horej}.
The identification of the leading twist contributions is
performed exactly as in \cite{blum&rob}. We use a suitable form of the
polarization vectors (for the gauge bosons)
to generate the helicity components of the amplitudes
and perform the asymptotic (DVCS) limit in order to identify the
leading terms. Terms of $O(1/\sqrt{2\bar{P}\cdot q})$ are suppressed
and are not kept into account. Below we will show only
the tensor structures which survive after this limit.

\section{Organizing the Compton amplitudes}

In order to give a more compact expression for the amplitudes of our processes we define
\ba
&&g^{T \mu\nu}=-g^{\mu\nu}+\frac{q^{\mu}\bar{P}^{\nu}}{(q\cdot \bar{P})}+\frac{q^{\nu}\bar{P}^{\mu}}{(q\cdot \bar{P})},\nonumber\\
&&\alpha(t)=\frac{g_{u}}{(t-\xi+i\eps)}-\frac{g_{d}}{(t+\xi-i\eps)},\nonumber\\
&&\beta(t)=\frac{g_{u}}{(t-\xi+i\eps)}+\frac{g_{d}}{(t+\xi-i\eps)}\,.\nonumber\\
\ea
Calculating all the integrals in the Eqs. (\ref{equa1}), (\ref{equaZ0}) and
(\ref{equa2}), (\ref{equaZ02}), we rewrite the expressions of the amplitudes as follows
\ba
\label{T}
&&T^{W^{+}}_{\mu\nu}=i U_{u d}\overline{U}(P_2)\left[i\left(\tilde{S}^{u}_{\mu\nu}+ S^{5 u}_{\mu\nu}\right) + {\bf \varepsilon}^{u}_{\mu\nu}+\tilde{{\bf \varepsilon}}^{5 u}_{\mu\nu}-i\left(\tilde{S}^{d}_{\mu\nu}+ S^{5 d}_{\mu\nu}\right) -{\bf \varepsilon}^{d}_{\mu\nu}-\tilde{{\bf \varepsilon}}^{5 d}_{\mu\nu}\right]U(P_1)\,,\nonumber\\
&&T^{W^{-}}_{\mu\nu}=-i U_{d u}\overline{U}(P_2)\left[i\left(\tilde{S}^{u}_{\mu\nu}+ S^{5 u}_{\mu\nu}\right) + {\bf \varepsilon}^{u}_{\mu\nu}+\tilde{{\bf \varepsilon}}^{5 u}_{\mu\nu}-i\left(\tilde{S}^{d}_{\mu\nu}+ S^{5 d}_{\mu\nu}\right) -{\bf \varepsilon}^{d}_{\mu\nu}-\tilde{{\bf \varepsilon}}^{5 d}_{\mu\nu}\right]_{g_u \rightarrow g_d}U(P_1)\,,\nonumber\\
&&T^{Z_{0}}_{\mu\nu}= i\,\overline{U}(P_2)\left[g_u g_{u V}\left(S_{\mu\nu}^{u}-i{\bf \varepsilon}_{\mu\nu}^{5 u}\right) -g_u g_{u A}\left(\tilde{S}_{\mu\nu}^{5 u}-i\tilde{\bf \varepsilon}_{\mu\nu}^{u}\right)\right.\nonumber\\
&&\hspace{3.8 cm}\left.-g_d g_{d V}\left(S_{\mu\nu}^{d}-i{\bf \varepsilon}_{\mu\nu}^{5 d}\right)+g_d g_{d A}\left(\tilde{S}_{\mu\nu}^{5 d}-i\tilde{\bf \varepsilon}_{\mu\nu}^{d}\right)\right]U(P_1)\,,
\ea
where, suppressing all the subleading terms in the tensor structures, we get
\ba
\label{S}
&&\overline{U}(P_2)\tilde{S}^{a \mu\nu}U(P_1)=\int_{-1}^{1}dt\,\alpha(t)\frac{g^{T\mu\nu}}{2\bar{P}\cdot q}\left[\overline{U}(P_2)\slash{q}U(P_1) \hat{f}_{0}^{a}(t,\xi) +\overline{U}(P_2)(i\frac{\sigma^{\alpha\beta}q_{\alpha}\Delta_{\beta}}{M})U(P_1)\hat{g}_{0}^{a}(t,\xi)\right]+\nonumber\\
&&\hspace{3.4 cm}\int_{-1}^{1}dt\,\beta(t) \frac{\bar{P}^{\mu}\bar{P}^{\nu}}{(\bar{P}\cdot q)^2}\left[\overline{U}(P_2)\slash{q}U(P_1) t \hat{f}_{0}^{a}(t,\xi) +\overline{U}(P_2)(i\frac{\sigma^{\alpha\beta}q_{\alpha}\Delta_{\beta}}{M})U(P_1) t \hat{g}_{0}^{a}(t,\xi)\right]\,\nonumber\\
\ea
while for the ${\bf \varepsilon}^{a \mu\nu}$ expression we obtain
\ba
\label{E}
&&\overline{U}(P_2){\bf \varepsilon}^{a \mu\nu}U(P_1)=\eps^{\mu\alpha\nu\beta}\frac{2q_{\alpha}\bar{P}_{\beta}}{(2\bar{P}\cdot q)^2}\int^{1}_{-1}dt\,\beta(t)\left[\overline{U}(P_2)\slash{q}U(P_1)\hat{f}_{0}^{a}(t,\xi)\right.+\nonumber\\
&&\left.\hspace{8.1 cm}\overline{U}(P_2)(i\frac{\sigma^{\alpha\beta}q_{\alpha}\Delta_{\beta}}{M})U(P_1)\hat{g}_{0}^{a}(t,\xi)\right]\,.
\ea
Passing to the $S^{a \mu\nu}$ and $\tilde{\bf \varepsilon}^{a \mu\nu}$ tensors, which appear in the $Z_0$ neutral current exchange, we get the following formulas
\ba
\label{SP}
&&\overline{U}(P_2)S^{a \mu\nu}U(P_1)=\int_{-1}^{1}dt\left(\frac{1}{t-\xi+i\eps}+\frac{1}{t+\xi-i\eps}\right)\times\nonumber\\
&&\left\{\frac{g^{T\mu\nu}}{2\bar{P}\cdot q}\left[\overline{U}(P_2)\slash{q}U(P_1) \hat{f}_{0}^{a}(t,\xi)+\overline{U}(P_2)(i\frac{\sigma^{\alpha\beta}q_{\alpha}\Delta_{\beta}}{M})U(P_1)\hat{g}_{0}^{a}(t,\xi)\right]\right\}+\nonumber\\
&&\int_{-1}^{1}dt\left(\frac{1}{t-\xi+i\eps}-\frac{1}{t+\xi-i\eps}\right)\cdot
\frac{\bar{P}^{\mu}\bar{P}^{\nu}}{(\bar{P}\cdot q)^2}\left[\overline{U}(P_2)\slash{q}U(P_1) t \hat{f}_{0}^{a}(t,\xi) +\overline{U}(P_2)(i\frac{\sigma^{\alpha\beta}q_{\alpha}\Delta_{\beta}}{M})U(P_1) t \hat{g}_{0}^{a}(t,\xi)\right]\nonumber\\
\ea
\ba
\label{EP}
&&\overline{U}(P_2)\tilde{\bf \varepsilon}^{a \mu\nu}U(P_1)=\int_{-1}^{1}dt\left(\frac{1}{t-\xi+i\eps}-\frac{1}{t+\xi-i\eps}\right)\eps^{\mu\alpha\nu\beta}\frac{2q_{\alpha}\bar{P}_{\beta}}{(2\bar{P}\cdot q)^2}\times\nonumber\\
&&\hspace{4 cm}\left\{\left[\overline{U}(P_2)\slash{q}U(P_1)\hat{f}_{0}^{a}(t,\xi)+\overline{U}(P_2)(i\frac{\sigma^{\alpha\beta}q_{\alpha}\Delta_{\beta}}{M})U(P_1)\hat{g}_{0}^{a}(t,\xi)\right]\right\}\,.\nonumber\\
\ea
Obviously the $\tilde{S}^{a\,5 \mu\nu}$, $S^{a\,5 \mu\nu}$, $\tilde{{\bf \varepsilon}}^{a\, 5\, \mu\nu}$ and ${\bf \varepsilon}^{a\, 5\, \mu\nu}$ expressions are obtained by the substitution (\ref{repla}).

At this stage, to square the amplitude, we need to calculate the following quantity, separately for the two charged processes
\ba
T^2= |T_{DVNS}|^2 + T_{DVNS}T_{BH}^* +T_{BH}T_{DVNS}^{*}+|T_{BH}|^2\,,
\ea
which simplifies in the neutral case, since it reduces
$|T_{DVNS}|^2$ \cite{ACG}.
Eqs.~(\ref{T})-(\ref{EP}) and their axial counterparts
are our final result and provide
a description of the deeply virtual amplitude in the electroweak
sector for charged and neutral currents. The result can be expressed
in terms of a small set of parton distribution functions
which can be easily related to generalized parton distributions,
as in standard DVCS.
\footnote{Based on the article~\cite{CG_PRD}}

\section{Conclusions}
We have presented an application/extension of a method, which has been formulated
in the past for the identification of the leading twist contributions to the parton amplitude in the generalized Bjorken region,
to the electroweak case. We have considered the special case of a deeply virtual kinematics.
We have focused our attention on processes initiated by neutrinos.
From the theoretical and experimental viewpoints
the study of these processes is of interest,
since very little is known of the neutrino interaction at intermediate energy
in these more complex kinematical domains.


\begin{thebibliography}{100}
\bibitem{Sterman1} G. Sterman, Nucl.Phys.Proc.Suppl. 108, 49 (2002), G. Sterman and W. Vogelsang
 proceedings of Physics at Run II: QCD and Weak Boson Physics, hep-ph/0002132.
\bibitem{Vogt3} A. Vogt Comput. Phys. Commun. {\bf 170} 65, (2005)
\bibitem{cafacor} A. Cafarella and C. Corian\`{o}, Comput. Phys. Commun.{\bf 160}: 213, (2004).
\bibitem{nostri} A. Cafarella, C. Corian\`o and M. Guzzi, JHEP {\bf 0311}:059, (2003);
\bibitem{gordon} L.E. Gordon and G.P. Ramsey Phys.Rev.{\bf D59}:074018, (1999).
\bibitem{Rossi} G.Rossi, Phys.Rev. {\bf D 29} 852, (1984).
\bibitem{Storrow} J.H. Da Luz Vieira and J. K. Storrow, Z.Phys. {\bf C 51} 241, (1991).
\bibitem{Valle} J.W.F. Valle, hep-ph/0310125; hep-ph/0301061, Proc.Indian Natl.Sci.Acad.70A:189.
\bibitem{minos} J. Hurheim for the MINOS Coll., Amsterdam 2002,(31st ICHEP 2002), Amsterdam, Jul 2002.
\bibitem{Marciano1} M.V. Diwan et al., Phys. Rev. D {\bf 68} 012002,2003;
W.T. Weng et al, J. Phys. G {\bf 29} 1735, (2003);
W. Marciano, talk at the KITP Conference:{\em  Neutrinos: Data, Cosmos, and Planck Scale} (Mar 3-7, 2003)
\bibitem{Mangano} M. Mangano et al., Report of the nuDIS Working Group for
the ECFA-CERN Neutrino-Factory study,  hep-ph/0105155.
\bibitem{Sterman} G. Sterman and P. Stoler, Ann. Rev. Nucl. Part. Sci.
{\bf 47} 193, (1997).
\bibitem{CorianoSavkli} C. Corian\`o and C. Savkli, JHEP 9807:008,1998.
\bibitem{betafunction1} O.V. Tarasov, A.A. Vladimirov and A.Y. Zharkov, Phys. Lett. {\bf 93B} 429, (1980).
\bibitem{betafunction2} S.A. Larin and J.A.M. Vermaseren, Phys. Lett. {\bf B303} 334, (1993).
\bibitem{betafunction3} T. van Rtbergen, J. Vermaseren and S. Larin, Phys. Lett, {\bf B400} 379, (1997).
\bibitem{vogt1} S. Moch, J. Vermaseren and A. Vogt, Nucl. Phys.\, {\bf B 688}, 101, (2004);
Nucl. Phys.\, {\bf B 691}, 129, (2004).
\bibitem{Coriano} C. Corian\`o, Nucl.Phys.B627:66-94, (2002),
C. Corian\`o and A. Faraggi, Phys. Rev. {\bf D 65}, 075001, (2002).
\bibitem{ellis} R.K. Ellis, Z. Kunszt, E.M. Levin, Nucl.Phys. {\bf B 420} 517, (1994),
Erratum-ibid. {\bf B 433} 498, (1995)
\bibitem{Buras} A. Buras, Rev. Mod. Phys. {\bf 52}, 199 (1980).
\bibitem{Petronzio1} W. Furmanski and R. Petronzio, Nucl. Phys. {\bf B 195} 237, (1982).
\bibitem{Petronzio2} W. Furmanski, R. Petronzio Z. Phys. {\bf C 11} 293, (1982).
\bibitem{kosower} D. Kosower, Nucl.Phys.  {\bf B 506}, 439 (1997).
\bibitem{leshouches} W. Giele $et$ $al$. hep-ph/0204316
\bibitem{parton2005}M.~Dittmar et al., hep-ph/0511119
\bibitem{Catani} S. Catani, ``Aspects of QCD, from the Tevatron to the LHC'',
in Proceedings of Workshop on Physics at TeV Colliders, Les Houches, France, 7-18 Jun 1999, hep-ph/0005233.
%%%%%%%%%%%%%%%%%%%%%%%%%%%%%%%%%%%%%%%%%%%%%%%%%%%%%%%%%%%%%%%%%%%%%%%%%%%%%%%%%%%%%
%%%%%%%%%%%%%%%%%%%%%%%%%%%%%%%%%%%%%%%%%%%%%%%%%%%%%%%%%%%%%%%%%%%%%%%%%%%%%%%%%%%%%


% \bibitem{moch1} S. Moch, J. Vermaseren Nucl. Phys. {\bf B 573} 853, (2000)
% \bibitem{MRST} A.D. Martin, R.G. Roberts, W.J. Stirling and R.S. Thorne, Eur.Phys.J. {\bf C 23}, 73 (2002);
% Phys.Lett.{\bf B 531}, 216 (2002).
% \bibitem{Alekhin} S.I.~Alekhin, Phys.Rev.{\bf D 68}, 014002 (2003).
%%%%%%%%%%%%%%%%%%%%%%%%%%%%%%%%%%%%%%%%%%%%%%%%%%%%%%%%%%%%%%%%%%%%%%%%%%%%%%%%%
\bibitem{Ji} X. Ji, Phys.Rev.D55 (1997) 7114.
\bibitem{Radyushkin}A.V. Radyushkin, Phys.Rev.D56 (1997) 5524.
\bibitem{CafaCor} A. Cafarella and C. Corian\`{o}, Int. J. Mod. Phys. {\bf A 20} 4863, (2005)
\bibitem{Teryaev1} C. Bourrely, J. Soffer, and O.V. Teryaev Phys.Lett.B420 (1998) 375.
\bibitem{Teryaev} C. Bourrely, E. Leader, O.V. Teryaev,  hep-ph/9803238.
\bibitem{JJ} R.L. Jaffe and X. Ji, Phys. Rev. Lett. 67 (1991) 552.
\bibitem{Barone} E. Barone, Phys.Lett.B409 (1997) 499.
\bibitem{Gordon} L.E. Gordon and G. P. Ramsey, Phys.Rev.D59 (1999) 074018.
\bibitem{ERBL} A.V. Efremov and A.V. Radyushkin, Phys. Lett. B94 (1980) 245;
G. P. Lepage and S.J. Brodsky, Phys. Rev. D22 (1980) 2157.
\bibitem{CL} C. Corian\`{o} and H. N. Li, JHEP 9807 (1998) 008; Nucl. Phys. {\bf B 434}, 535 (1995).
\bibitem{Golec} K.J. Golec-Biernat and A. D. Martin, Phys.Rev.D59 (1999) 014029.
\bibitem{GRSV}M.Glck, E.Reya, M.Stratmann and W.Vogelsang, Phys.Rev.D 63 (2001) 094005
\bibitem{GRV}M.Glck, E.Reya and A.Vogt, Eur.Phys.J.C 5 (1998) 461
\bibitem{MSSV} O. Martin, A. Schafer , M. Stratmann, W. Vogelsang, Phys.Rev. {\bf D 57}: 117502, (1998)
\bibitem{CTEQ}H.L.Lai \emph{et al.}, Phys.Rev.D 55 (1997) 1280.
\bibitem{GGR}L.E.Gordon, M.Goshtasbpour and G.P.Ramsey, Phys.Rev.D 58 (1998) 094017.
%%%%%%%%%%%%%%%%%%%%%%%%%%%%%%%%%%%%%%%%%%%%%%%%%%%%%%%%%%%%%%%%%%%%%%%%%%%%%%%%%%%%
%%%%%%%%%%%%%%%%%%%%%%%%%%%%%%%%%%%%%%%%%%%%%%%%%%%%%%%%%%%%%%%%%%%%%%%%%%%%%%%%%%%%


%\bibitem{CollinsQiu} J.C. Collins and J. Qiu, Phys.Rev.D39 (1989) 1398.
%\bibitem{teryaevsoffer} C. Bourrely, J. Soffer, O.V. Teryaev, Phys.Lett.B420 (1998) 375.
%%%%%%%%%%%%%%%%%%%%%%%%%%%%%%%%%%%%%%%%%%%%%%%%%%%%%%%%%%%%%%%%%%%%%%%%%%%%%%%%%%%%
\bibitem{BCCGR} V. Barone, A. Cafarella, C. Corian\`o, M. Guzzi, P.G. Ratcliffe, Phys. Lett. {\bf B 639}:483, (2006).
\bibitem{brodsky} S.J.~Brodsky, ``Testing Quantum Chromodynamics with Antiprotons'', hep-ph/0411046.
\bibitem{bdr} For a review on the transverse polarisation of quarks in hadrons, see
V.~Barone, A.~Drago and P.G.~Ratcliffe,\emph{Phys. Rep.} \textbf{359} (2002)~1.
\bibitem{rs} J.~Ralston and D.E.~Soper, \emph{Nucl. Phys.} \textbf{B152} (1979)~109;
J.L.~Cortes, B.~Pire and J.P.~Ralston, \emph{Z. Phys.} \textbf{C55} (1992)~409;
R.L.~Jaffe and X.~Ji, \emph{Nucl. Phys.} \textbf{B375} (1992)~527.
\bibitem{collins} J.C.~Collins, \emph{Nucl. Phys.} \textbf{B396} (1993)~161.
\bibitem{rhic} G.~Bunce, N.~Saito, J.~Soffer and W.~Vogelsang, \emph{Ann. Rev. Nucl. Part. Phys.} \textbf{50} (2000)~525.
\bibitem{bcd} V.~Barone, T.~Calarco and A.~Drago, \emph{Phys. Rev.} \textbf{D56} (1997)~527.
\bibitem{mssv} O.~Martin, A.~Sch\"{a}fer, M.~Stratmann and W.~Vogelsang,
\emph{Phys. Rev.} \textbf{D57} (1998) 3084; \emph{Phys. Rev.} \textbf{D60} (1999) 117502.
\bibitem{ratcliffe} P.G.~Ratcliffe, \emph{Eur. Phys. J.} \textbf{C41} (2005)~319.
\bibitem{Mukherjee:2003pf} A.~Mukherjee, M.~Stratmann and W.~Vogelsang, \emph{Phys. Rev.} \textbf{D67}
  (2003) 114006.
\bibitem{Mukherjee:2005rw} A.~Mukherjee, M.~Stratmann and W.~Vogelsang,
\emph{Phys. Rev.} \textbf{D72} (2005) 034011.
\bibitem{bcd2} V.~Barone, T.~Calarco and A.~Drago, \emph{Phys. Lett.} \textbf{B390} (1997)~287.
\bibitem{Barone:1997fh} V.~Barone, \emph{Phys. Lett.} \textbf{B409} (1997)~499.
\bibitem{Anselmino:2004ki}
  M.~Anselmino, V.~Barone, A.~Drago and N.N.~Nikolaev,
  \emph{Phys. Lett.} \textbf{B594} (2004)~97.
\bibitem{Efremov:2004qs}
  A.V.~Efremov, K.~Goeke and P.~Schweitzer,
  \emph{Eur. Phys. J.} \textbf{C35} (2004)~207.
\bibitem{pax}
  $\mathcal{P\!AX}$ Collab.,
  V.~Barone \emph{et al.},
  ``Antiproton--Proton Scattering Experiments with Polarization'',
  Technical Proposal, hep-ex/0505054.
\bibitem{pax_prl}
  F.~Rathmann \emph{et al}., \emph{Phys. Rev. Lett.} \textbf{94} (2005) 014801.
\bibitem{guzzi}
  M.~Guzzi, in proc. of the Int. Workshop ``Transversity 2005'' (Como, 2005),
  V.~Barone and P.G.~Ratcliffe eds., World Scientific, in press.
\bibitem{sutton}
  P.J.~Sutton, A.D.~Martin, R.G.~Roberts and W.J.~Stirling,
  \emph{Phys. Rev.} \textbf{D45} (1992) 2349.
\bibitem{soffer}
  J.~Soffer, \emph{Phys. Rev. Lett.} \textbf{74} (1995) 1292.
\bibitem{grv}
  M.~Gl\"{u}ck, E.~Reya and A.~Vogt, \emph{Eur. Phys. J.} \textbf{C5} (1998)~461;
  M.~Gl\"{u}ck, E.~Reya, M.~Stratmann and W.~Vogelsang, \emph{Phys. Rev.}
  \textbf{D63} (2001) 094005.
\bibitem{h1evol}
  F.~Baldracchini, N.S.~Craigie, V.~Roberto and M.~Socolovsky, \emph{Fortschr.
  Phys.} \textbf{30} (1981) 505;
  X.~Artru and M.~Mekhfi, \emph{Z. Phys.} \textbf{C45} (1990) 669;
  A.~Hayashigaki, Y.~Kanazawa and Y.~Koike, \emph{Phys. Rev.} \textbf{D56}
  (1997) 7350;
  S.~Kumano and M.~Miyama, \emph{Phys. Rev.} \textbf{D56} (1997) R2504;
  W.~Vogelsang, \emph{Phys. Rev.} \textbf{D57} (1998) 1886;
  J.~Bl\"{u}mlein, \emph{Eur. Phys. J.} \textbf{C20} (2001) 683.
\bibitem{shimizu}
  H.~Shimizu, G.~Sterman, W.~Vogelsang and H.~Yokoya,
  \emph{Phys. Rev.} \textbf{D71} (2005) 114007.
\bibitem{sps}
  M.J.~Corden \emph{et al.}, \emph{Phys. Lett.} \textbf{B68} (1977)~96;
  \emph{Phys. Lett.} \textbf{B96} (1980)~411;
  \emph{Phys. Lett.} \textbf{B98} (1981)~220.
\bibitem{carlson}
  C.E.~Carlson and R.~Suaya, \emph{Phys. Rev.} \textbf{D18} (1978)~760.
%%%%%%%%%%%%%%%%%%%%%%%%%%%%%%%%%%%%%%%%%%%%%%%%%%%%%%%%%%%%%%%%%%%%%%%%%%%%%%%%%%%
%%%%%%%%%%%%%%%%%%%%%%%%%%%%%%%%%%%%%%%%%%%%%%%%%%%%%%%%%%%%%%%%%%%%%%%%%%%%%%%%%%%
\bibitem{RSVN} V. Ravindran, J. Smith and W.L. van Neerven,
Nucl. Phys. {\bf B665}, 325, (2003).
\bibitem{VN1} W. Van Neerven and A. Vogt,
Nucl.Phys.\, {\bf B 603}, 42, (2001); Nucl.Phys. {\bf B 588}, 345, (2000).
\bibitem{Nason} S. Catani, D. de Florian, M. Grazzini and P. Nason, JHEP {\bf 07}, 028 (2003).
\bibitem{Harl&Kil} R.V. Harlander and W.B. Kilgore, Phys. Rev. Lett. {\bf 88}, 201801 (2002).
\bibitem{Anas&Melni} C. Anastasiou and K. Melnikov, Nucl. Phys. {\bf B 646}, 220 (2002).
\bibitem{dawson} S. Dawson, hep-ph/9411325. Published in Boulder TASI 1994:0445-506 (QCD161:T45:1994). 
\bibitem{Stein} K. G. Chetyrkin, Bernd A. Kniehl and M. Steinhauser, Nucl. Phys. {\bf B 510}, 61 (1998).
\bibitem{RSVN1} V. Ravindran , J. Smith, W.L. Van Neerven, Nucl.Phys.\,{\bf B 704}, 332, (2005)
\bibitem{Laenen&Spira} M. Kr\"amer, E. Laenen and M. Spira, Nucl. Phys. {\bf B 511}, 523 (1998).
\bibitem{Chet&Bard} K.G. Chetyrkin, B.A. Kniehl, M. Steinhauser and W.A. Bardeen,
Nucl. Phys. {\bf B 535},3 (1998).
\bibitem{Giele} W. Giele $et~al$ hep-ph/0204316
\bibitem{CCG} A. Cafarella, C Corian\`{o} and M. Guzzi, Nucl. Phys. {\bf B 748}:253, (2006).
\bibitem{remiddi1} T. Gehrmann, E. Remiddi, Comput.
Phys. Commun. {\bf 141}, 296, (2001)
\bibitem{remiddi2} E. Remiddi, J.A.M. Vermaseren,
Int. J. Mod. Phys. {\bf A 15} 725, (2000)
\bibitem{Tarasov} O. V. Tarasov, A.A. Vladimirov and A. Yu. Zharkov, Phys. Lett. {\bf B 93}, 429 (1980).
\bibitem{Larin} S.A. Larin and J. Vermaseren, Phys. Lett. {\bf B 303}, 334 (1993) .
\bibitem{Ritbergen} T. van Ritbergen, J.A.M. Vermaseren and S.A. Larin, Phys. Lett. {\bf B 400}, 153 (1997).
\bibitem{bmsn}
M. Buza, Y. Matiounine, J. Smith and W.L. van Neerven,
Eur. Phys. J. {\bf C1}, 301 (1998).
\bibitem{cs}A. Chuvakin and J. Smith, Comput. Phys. Commun. {\bf 143}, 257-286 (2002).
\bibitem{CCG_Eur} A. Cafarella, C. Corian\`o, M. Guzzi, J. Smith, Eur. Phys. J. {\bf C 47}:703, (2006).
\bibitem{MRST}A.D.~Martin, R.G.~Roberts, W.J.~Stirling and R.S.~Thorne,
Eur.Phys.J.C \textbf{23} (2002) 73; Phys.Lett.B\textbf{531} (2002) 216.
\bibitem{Alekhin} S.I.~Alekhin, Phys.Rev.D \textbf{68} (2003) 014002.
\bibitem{MRST1}A.D.~Martin, R.G.~Roberts, W.J.~Stirling and R.S.~Thorne,
Eur.Phys.J.C \textbf{28} 455, (2003).
\bibitem{mv}
S. Moch and A. Vogt, Higher order soft corrections to lepton pair production
and Higgs boson production, published in Phys. Lett. {\bf B 631}:48, (2005)
\bibitem{ravi} V. Ravindran, J. Smith, W.L. van Neerven, hep-ph/0608308;
V. Ravindran, Nucl. Phys. {\bf B 746}:58, (2006);
V. Ravindran, Nucl. Phys. {\bf B 752}:173,(2006);
A. Idilbi, X.Ji and F. Yuan, hep-ph/0605068;
V. Ravindran, J. Smith, W.L. van Neerven hep-ph/0608308, to appear in Nuclear Physics {\bf B}.

\bibitem{lm}E. Laenen and L. Magnea, Phys. Lett. {\bf B 632}:270, (2006)
\bibitem{ij} A. Idilbi, X. Ji, J-P. Ma and F. Yuan, Phys. Rev. {\bf D 73}:077501, (2006)
\bibitem{amp} C. Anastasiou, K. Melnikov and F. Petriello, Phys.Rev. {\bf D 72}:097302, (2005).
%%%%%%%%%%%%%%%%%%%%%%%%%%%%%%%%%%%%%%%%%%%%%%%%%%%%%%%%%%%%%%%%%%%%%%%%%%%%%%%%%%%%%%%%%%%%%%%%%
\bibitem{Winter} K. Winter, {\em Neutrino Physics}, 2nd ed. Cambridge Univ. Press, Cambridge 2000.
\bibitem{Morfin} J.G. Morfin, M. Sakuda, Y, Suzuki, (ed.) Proc. of the
1st Workshop on Neutrino-Nucleus interactions in the Few GeV Region
(Nuint01), Nucl. Phys. B, Proc. Suppl. {\bf 112}, (2002).
\bibitem{nuint} F. Cavanna, C. Keppel, P. Lipari, M. Sakuda (ed.), Proceedings of the Intl. Workshop on Neutrino
Nucleus interactions in the Few GeV Region (NuInt04), Nucl. Phys. B, Proc. Suppl. {\bf 139}, (2005).
\bibitem{Quigg} R. Gandhi, C. Quigg, M.H. Reno ad I. Sarcevic, Phys. Rev. {\bf D58}:093009, (1998).
\bibitem{Ji1} X.~Ji, Phys. Rev. D {\bf 55} 7114 (1997)
\bibitem{Rad3} A.~V.~Radyushkin, Phys.\ Rev.\ D {\bf 56}, 5524 (1997)
\bibitem{Rad1} A.~V.~Radyushkin, Phys.\ Rev.\ D {\bf 59} 014030 (1999)
\bibitem{Geyer} J. Bl{\"u}mlein, B. Geyer, D. Robaschik, Phys. Rev. D {\bf 65} 054029 (2002)
\bibitem{Rad&Wei} A.V. Radyushkin and C. Weiss, Phys. Rev. D {\bf 63}, 114012, (2001)
\bibitem{Penttinen} M. Penttinen, M.V. Polyakov, A.G. Shuvaev, M. Strikman, Phys.Lett. B{\bf 491} 96 (2000)
\bibitem{CollinsFreund} J. C. Collins and A. Freund, Phys.Rev.D {\bf 59} 074009, (1999)
\bibitem{bel1} A. V. Belitsky, D. Muller, L. Niedermeier, A. Sch\"afer Nucl. Phys. B {\bf 593}: 289-310 (2001)
\bibitem{radmus} I. V. Musatov and A. V. Radyushkin, Phys. Rev. D {\bf 61} 074027 (2000)
\bibitem{Biernat} K.J. Golec-Biernat and A.D. Martin Phys. Rev. D {\bf 59} 014029, (1999)
\bibitem{Freu&McD2} A.~Freund and M.~McDermott, Eur. Phys. J.~C {\bf 23} 651-674 (2002)
\bibitem{Initialcond} K.J. Golec-Biernat, A.D. Martin and M.G. Ryskin, Nucl. Phys. Proc. Suppl. {\bf 79}: 365-367 (1999)
\bibitem{Freund1} A.~Freund, M.~McDermott and M. Strikman Phys.~Rev.~D {\bf 67}, 036001 (2003).
\bibitem{Vander&guich} M. Vanderhaeghen and P.A.M. Guichon, Phys. Rev. D {\bf 60} 094017 (1999)
\bibitem{gluck&reya&vogt98} M. Gluck, E. Reya, A. Vogt, Eur. Phys. J. C {\bf 5} 461-470 (1998)
\bibitem{Ji3} X.~Ji, Nucl. Phys. Proc. Suppl. {\bf 119}: 41-49 (2003)
\bibitem{Man1} L. Mankiewicz, G. Piller, A.V. Radyushkin, Eur. Phys. J. C {\bf 10} 307 (1999)
\bibitem{Fran1} L. L. Frankfurt, P. V. Pobylitsa, M. V. Polyakov, M. Strikman, Phys.Rev. D {\bf 60} 0140010 (1999)
\bibitem{Piller} L. Mankiewicz, G. Piller, T. Weigl Eur. Phys. J. C {\bf 5} (1998)
\bibitem{Freu&McD} A.~Freund and M.~McDermott, Phys. Rev. D {\bf 65} 074008 (2002)
\bibitem{PK} E. A. Paschos and A. Kartavtsev, hep-ph/0309148.
\bibitem{CCV} C. Corian\`{o}, M. Guzzi and J.D. Vergados, in preparation.
\bibitem{ACG} P. Amore, C. Corian\`{o} and M. Guzzi, JHEP {\bf 0502}: 038, (2005)
\bibitem{Lazar} M. Lazar, hep-ph/0308049 Ph.D. Thesis.
\bibitem{blum&rob} J.Bl\"umlein and D.Robaschik Nuc. Phys B {\bf 581} 449 473 (2000)
\bibitem{lazar1} J.Bl\"umlein, B. Geyer, M. Lazar and D.Robaschik, Nucl. Phys. Proc.Suppl. {\bf 89} 155-161 (2000)
\bibitem{lazar2} J.Eilers, B.Geyer and M.Lazar, Phys.Rev.D {\bf 69} 034015 (2004)
\bibitem{Ji2} X.~Ji, J. Phys. G {\bf 24}: 1181-1205 (1998)
\bibitem{Blum} J. Bl\"umlein and N. Kochelev, Nucl. Phys. B {\bf 498}: 285, (1997)
\bibitem{bemuevolve} A.~V.~ Belitsky, D.~ M{\"u}ller, L.~ Niedermeier, A.~ Sch{\"a}fer, Nucl. Phys. B {\bf 546} 279 (1999)
\bibitem{belis&muller} A. V. Belitsky, D. M{\"u}ller, L. Niedermeier, A. Sch{\"a}fer, Phys. Lett. B {\bf 437} 160 (1998)
\bibitem{Beli&geyer} A. V. Belitsky, B. Geyer, D. M{\"u}ller, A. Sch{\"a}fer, Phys. Lett. B {\bf 421} 312 (1998)
\bibitem{Pol_Weiss} M.V. Polyakov and C.Weiss Phys. Rev D {\bf 60} 114017 (1999)
\bibitem{Roba&Horej} D. Robaschik and J.Horejsi, Fortsch. Phys. {\bf 42}: 101 (1994)
\bibitem{CG_PRD} C. Corian\`o and M. Guzzi Phys. Rev. {\bf D 71}: 053002, (2005)
%%%%%%%%%%%%%%%%%%%%%%%%%%%%%%%%%%%%%%%%%%%%%%%%%%%%%%%%%%%%%%%%%%%%%%%%%%%%%%%%%%%%%%%%%%%%%%%%%%%%%%%%%%%%%%%
%%%%%%%%%%%%%%%%%%%%%%%%%%%%%%%%%%%%%%%%%%%%%%%%%%%%%%%%%%%%%%%%%%%%%%%%%%%%%%%%%%%%%%%%%%%%%%%%%%%%%%%%%%%%%%%%%%%%
\end{thebibliography}
\end{document}